\documentclass[a4paper,11pt,times,numbered,print,index]{Classes/PhDThesisPSnPDF}
\usepackage{quotchap}
\usepackage[most]{tcolorbox}
\newtcolorbox{examplebox}[1][]{%
  enhanced,
  breakable,
  colback=white,
  colframe=cyan,
  boxrule=1.5pt,
  left=6pt,right=6pt,top=6pt,bottom=6pt,
  title=#1
}
\usepackage[shortlabels]{enumitem}
\DeclareMathOperator{\sgn}{sgn}
\usepackage{amsmath}
\usepackage{amssymb}
\usepackage{mathtools}
\input{Preamble/preamble}

\title{The alternative scalar field dark sector of the Universe}

\subtitle{Cosmological inflation, quintessence and ekpyrotic model in the framework of scalar-tensor cosmology}

\author{Marcin Postolak}

\dept{Faculty of Physics and Astronomy}

\university{University of Wrocław}
\crest{\includegraphics[width=0.8\textwidth]{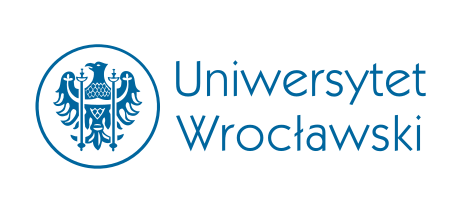}}

\supervisor{prof. dr hab. Andrzej Borowiec} 

\supervisorlinewidth{0.45\textwidth}

\degreetitle{Doctor of Philosophy \\
in the field of physical sciences}

\degreedate{June 2026}

\ifdefineAbstract
\fi

\ifdefineChapter
\fi

\begin{document}

\frontmatter

\maketitle


\begin{dedication} 

I would like to dedicate this dissertation to my loved ones, without whom it certainly would not have been possible to complete: my girlfriend, \textbf{\textit{Roksana}}; my parents, \textit{\textbf{Bożena}} and \textit{\textbf{Krzysztof}}; my sister, \textit{\textbf{Dominika}}; Antonio, Bob I, Bob II, Groovy, Milena, Stuart; and my late grandmother, \textit{\textbf{Stefania}}.\\

I would also like to dedicate this dissertation to all those struggling with mental health issues, so that they never give up in their own personal battles they fight every day.\\

Chciałbym zadedykować tę rozprawę moim bliskim, bez których z pewnością nie udałoby mi się jej ukończyć: mojej dziewczynie, \textbf{\textit{Roksanie}}; moim rodzicom, \textit{\textbf{Bożenie}} i \textit{\textbf{Krzysztofowi}}; mojej siostrze, \textit{\textbf{Dominice}}; Antonio, Bobowi I, Bobowi II, Groovy'emu, Milenie, Stuartowi; oraz świętej pamięci babci, \textit{\textbf{Stefanii}}.\\

Chciałbym również zadedykować tę rozprawę wszystkim osobom zmagającym się z problemami zdrowia psychicznego, aby nigdy nie poddawały się w swoich codziennych osobistych zmaganiach.\\

\chapter*{Preprints, publications, presentations, and participation in conferences and popular science events}
\section*{Preprints and publications}
\begin{itemize}
    \item \textbf{M. Postolak}, \href{https://cosmoversetensions.eu/learn-cosmology/did-the-big-bang-and-cosmic-inflation-really-happen/}{“\textit{Did the Big Bang and cosmic inflation really happen? (A tale of alternative cosmological models)}”}, \href{https://arxiv.org/abs/2404.18503}{arXiv:2404.18503 [physics.pop-ph], 4 2024.}
    \item A. Borowiec and \textbf{M. Postolak}, “\textit{Is it possible to separate baryonic from dark matter within the $\Lambda$-CDM formalism?}”, \href{https://doi.org/10.1016/j.physletb.2024.139176}{Phys. Lett. B \textbf{860}, 139176 (2025)}, \href{https://arxiv.org/abs/2309.10364}{arXiv:2309.10364 [gr-qc]}.
    \item \textbf{M. Postolak}, “\textit{Non-minimally coupled scalar field dark sector of the universe: in-depth (Einstein frame) case study}”, \href{https://doi.org/10.1088/1361-6382/ae7c4f}{Class. Quant. Grav. \textbf{43} (12) (2026) 125018}, \href{https://arxiv.org/abs/2505.07456}{arXiv:2505.07456 [gr-qc]}.
    \item \textbf{M. Postolak}, “\textit{Phase-resolved field-space distance criteria in ekpyrotic, bouncing, and cyclic cosmologies}”, \href{https://doi.org/10.1103/qjfz-drq6}{Phys. Rev. D \textbf{114} (4) (2026) 044016}, \href{https://arxiv.org/abs/2605.12579}{arXiv:2605.12579 [gr-qc]}.
    \item E. Di Valentino et al. (\href{https://cosmoversetensions.eu/}{\textbf{CosmoVerse}}), \textit{"The CosmoVerse White Paper: Addressing observational tensions in cosmology with systematics and fundamental physics"}, \href{https://doi.org/10.1016/j.dark.2025.101965}{Phys. Dark Univ. \textbf{49} (2025) 101965}, \href{https://arxiv.org/abs/2504.01669}{arXiv:2504.01669 [astro-ph.CO]}.\\
    Contribution:
    \begin{itemize}
        \item Revision of the paper;
        \item 4.1.3. Extra relativistic degrees of freedom;
        \item 4.3.1. Modified gravity in light of cosmic tensions;
        \item 4.4.1. Cold dark matter;
        \item 4.4.3. Interacting and decaying dark matter.
    \end{itemize}
\end{itemize}
\section*{Presentations}
\begin{itemize}
    \item \textbf{M. Postolak}, "\textit{Witamy po ciemnej stronie (Wszechświata) - o tym jak kosmologia próbuje zrozumieć nieznane}" - \href{https://wfa.uwr.edu.pl/wydarzenia/kosmos-zaczyna-sie-na-ziemi/}{Drzwi otwarte "Kosmos zaczyna się na Ziemi!"},\\
    \href{https://drive.google.com/file/d/1FRM-XB5iumB2pq9K7AJuWDrt0xIn5wHK/view?usp=sharing}{(prezentacja)}.
    \item \textbf{M. Postolak}, "\textit{Is it possible to separate baryonic from dark matter within the $\Lambda$-CDM formalism?}" - SKCM2 Symposium for Young Scientists in Europe (Wrocław),\\
    \href{https://doi.org/10.13140/RG.2.2.26884.91521}{DOI: 10.13140/RG.2.2.26884.91521}.
    \item \textbf{M. Postolak}, "\textit{Dark Matter in scalar-tensor cosmology: Is it possible to separate baryonic from dark matter within the $\Lambda$-CDM formalism?}" - PhD seminar,\\
    \href{https://doi.org/10.13140/RG.2.2.16818.58568}{DOI: 10.13140/RG.2.2.16818.58568}.
    \item \textbf{M. Postolak}, "\textit{Dark Matter in scalar-tensor cosmology. Is it possible to separate baryonic from dark matter within the $\Lambda$-CDM formalism?}" - CosmoVerse Journal Club, \href{https://doi.org/10.13140/RG.2.2.29428.16004}{DOI: 10.13140/RG.2.2.29428.16004}.
    \item \textbf{M. Postolak}, "\textit{Scalar Field Dark Matter Models. Ultralight Bosonic Dark Matter}" - PhD seminar, \href{https://doi.org/10.13140/RG.2.2.22573.09440}{DOI: 10.13140/RG.2.2.22573.09440}.
    \item \textbf{M. Postolak}, "\textit{Energy conditions and the separation between baryonic and dark matter in scalar-tensor cosmology}" - CosmoVerseSchool@Corfu “From Fundamental Physics to Data Analysis in Cosmology”, \href{https://doi.org/10.13140/RG.2.2.23676.55688}{DOI: 10.13140/RG.2.2.23676.55688}.
    \item \textbf{M. Postolak}, "\textit{Physics of the cosmological inflation - Facts and doubts}" - PhD seminar, \href{https://doi.org/10.13140/RG.2.2.34141.63202}{DOI: 10.13140/RG.2.2.34141.63202}.
    \item \textbf{M. Postolak}, "\textit{Dynamical systems applied to cosmology - How can we "look inside" the cosmological models?}" - PhD seminar, \href{https://doi.org/10.13140/RG.2.2.33224.64002}{DOI: 10.13140/RG.2.2.33224.64002}.
    \item \textbf{M. Postolak}, "\textit{Dynamical systems applied to non-minimally coupled scalar field dark sector}" - The 11th Conference of the Polish Society on Relativity, \href{https://doi.org/10.13140/RG.2.2.16430.63049}{DOI: 10.13140/RG.2.2.16430.63049}.
\end{itemize}

\end{dedication}


\begin{declaration}

I hereby declare that except where specific reference is made to the work of 
others, the contents of this dissertation are original and have not been 
submitted in whole or in part for consideration for any other degree or 
qualification in this, or any other university. This dissertation is my own 
work and contains nothing which is the outcome of work done in collaboration 
with others, except as specified in the text and Acknowledgements.


\end{declaration}

\begin{declaration}

Niniejszym oświadczam, że z wyjątkiem przypadków, w których wyraźnie powołuję się na prace innych autorów, treść niniejszej rozprawy doktorskiej jest oryginalna i nie została przedłożona, w całości ani w części, do rozpatrzenia w celu uzyskania jakiegokolwiek innego stopnia naukowego lub kwalifikacji na tej ani żadnej innej uczelni. Niniejsza praca doktorska jest moim własnym dziełem i nie zawiera żadnych elementów będących wynikiem współpracy z innymi osobami, z wyjątkiem przypadków wskazanych w tekście oraz w podziękowaniach.

\end{declaration}

\begin{acknowledgements}      

I would like to express my sincere gratitude to my loved ones: my girlfriend, \textbf{\textit{Roksana}}; my parents, \textit{\textbf{Bożena}} and \textit{\textbf{Krzysztof}}; my sister, \textit{\textbf{Dominika}}; and my late grandmother, \textit{\textbf{Stefania}}.\\
I would also like to express my gratitude to Professor \textit{\textbf{Arkadiusz Błaut}} for our conversations and his support during the difficult period between my bachelor's and master's studies, and to Professor \textit{\textbf{Maciej Matyka}} for showing me support when no one else had the courage to do so.\\
I would like to thank my supervisor, Professor \textit{\textbf{Andrzej Borowiec}}, for his collaboration on our joint article.\\
I would also like to express my gratitude to Professor \textbf{Carsten van de Bruck}, Dr. \textbf{Eleonora di Valentino}, Dr. \textbf{William Giarè}, and \textbf{Dong Ha Lee} for their time and support during my week-long stay at the University of Sheffield as part of the STSM (\href{https://cosmoversetensions.eu/}{COST Action CA21136 “CosmoVerse”}).\\
I would also like to thank \textbf{Mahdi Najafi} for every word of support he has given me.\\

Chciałbym wyrazić swoją szczerą wdzięczność moim bliskim: mojej dziewczynie, \textbf{\textit{Roksanie}}; moim rodzicom, \textit{\textbf{Bożenie}} i \textit{\textbf{Krzysztofowi}}; mojej siostrze, \textit{\textbf{Dominice}}; oraz mojej zmarłej babci, \textit{\textbf{Stefanii}}.\\
Chciałbym również podziękować profesorowi \textit{\textbf{Arkadiuszowi Błautowi}} za nasze rozmowy i wsparcie w trudnym okresie między studiami licencjackimi a magisterskimi, a także profesorowi \textit{\textbf{Maciejowi Matyce}} za okazanie mi wsparcia, gdy nikt inny nie miał na to odwagi.\\
Chciałbym podziękować mojemu promotorowi, profesorowi \textit{\textbf{Andrzejowi Borowcowi}}, za współpracę przy napisaniu naszego wspólnego artykułu.\\
Chciałbym również wyrazić swoją wdzięczność dla profesora \textbf{Carstena van de Brucka}, dr \textbf{Eleonory di Valentino}, dr. \textbf{Williama Giarè'a} oraz \textbf{Dong Ha Lee} za poświęcony czas i wsparcie podczas mojego tygodniowego pobytu na Uniwersytecie w Sheffield w ramach STSM (\href{https://cosmoversetensions.eu/}{COST Action CA21136 „CosmoVerse”}).\\
Chciałbym również podziękować \textbf{Mahdiemu Najafiemu} za każde słowo wsparcia, które mi przekazał.\\

\end{acknowledgements}

\begin{abstract}
Although the 'standard' cosmological model (LCDM/$\Lambda$CDM) provides a successful description of the large-scale evolution of the Universe, its two dominant components, dark matter and dark energy, remain fundamentally unknown. This dissertation investigates scalar field descriptions of the dark sector (of the Universe), with emphasis on scalar-tensor cosmology, inflationary effective field theory, quintessence, and ekpyrotic/cyclic alternatives to inflation.

The thesis first reviews scalar-tensor theories of gravity, starting from Brans-Dicke theory and extending to general non-minimally coupled scalar fields. Particular attention is paid to the interpretation of the scalar field, the relation between the Jordan and Einstein conformal frames, and the role of conformal transformations in cosmology. This provides the theoretical basis for treating scalar fields either as modified matter or as additional gravitational degrees of freedom.

The second part discusses cosmological inflation from the viewpoint of effective field theory. The spontaneous breaking of time translations, the Goldstone description of adiabatic perturbations, higher-dimensional operators, radiative corrections and Planck-suppressed terms are analyzed in the context of ultraviolet sensitivity and the eta problem.

The third part is devoted to quintessence as the simplest dynamical dark energy model. 'Tracking', 'scaling', 'thawing' and 'freezing' behaviors are discussed, together with observational diagnostics such as the statefinder parameters, the $(\omega_{\phi},\omega'_{\phi})$ plane and the $Om(z)$ diagnostic. The DESI DR2 results are used as motivation for considering time-dependent effective dark energy beyond a strict cosmological constant.

The final part studies ekpyrotic and cyclic cosmologies. The ekpyrotic phase is analyzed as a smoothing mechanism based on an ultra-stiff scalar field equation of state. Perturbations, non-Gaussianity, tensor modes, bounce conditions, entropy production, geodesic incompleteness and phase-resolved field-space distance bounds are discussed.

\end{abstract}
\begin{abstract}
Pomimo tego iż „standardowy” model kosmologiczny (LCDM/$\Lambda$CDM) z powodzeniem opisuje ewolucję Wszechświata w dużej skali, jego dwa dominujące składniki, ciemna materia i ciemna energia, pozostają zasadniczo nieznane. Niniejsza rozprawa doktorska poświęcona jest badaniu opisów ciemnego sektora (Wszechświata) za pomocą pól skalarnych, ze szczególnym uwzględnieniem kosmologii skalarno-tensorowej, inflacyjnej efektywnej teorii pola, kwintesencji oraz ekpyrotycznych i cyklicznych alternatyw dla inflacji.

W pracy dokonano przeglądu teorii grawitacji skalarno-tensorowych. Szczególną uwagę poświęcono interpretacji pola skalarnego, relacji między układami konforemnymi Jordana i Einsteina oraz roli transformacji konforemnych w kosmologii. Stanowi to podstawę teoretyczną do traktowania pól skalarnych jako zmodyfikowanej materii albo jako dodatkowych grawitacyjnych stopni swobody.

W drugiej części omówiono inflację kosmologiczną z punktu widzenia efektywnej teorii pola. Spontaniczne łamanie translacji czasowych, opis Goldstone’a zaburzeń adiabatycznych, operatory wyższych wymiarów, poprawki radiacyjne oraz wyrazy stłumione przez stałą Plancka zostały przeanalizowane w kontekście wrażliwości UV oraz problemu eta.

Trzecia część poświęcona jest kwintesencji jako najprostszemu dynamicznemu modelowi ciemnej energii. Omówiono zachowania typu 'tracking', 'scaling', 'thawing' oraz 'freezing', wraz z diagnostyką obserwacyjną, taką jak parametry statefinder, przestrzeń fazowa $(\omega_{\phi},\omega'_{\phi})$ oraz diagnostyka $Om(z)$. Wyniki DESI DR2 posłużą jako motywacja do rozważenia zależnej od czasu efektywnej ciemnej energii wykraczającej poza ścisłą stałą kosmologiczną.

W ostatniej części badane są kosmologie ekpyrotyczne i cykliczne. Faza ekpyrotyczna jest analizowana jako mechanizm wygładzający oparty na ultra-sztywnym równaniu stanu pola skalarnego. Omówiono zaburzenia, niegaussowość, mody tensorowe, warunki odbicia, produkcję entropii, niekompletność geodezyjną oraz granice odległości w przestrzeni pól rozdzielonych fazowo.
\end{abstract}


\tableofcontents

\listoffigures

\listoftables


\printnomenclature

\mainmatter

\begin{savequote}
"We call dark matter dark matter because we don't know what it is (laughter). And we call dark energy dark energy because we don't know what it is, either. So other than the fact that we don't quite understand 95 percent of the universe, we're doing really well."
\qauthor{\textbf{Charles L. Bennett}, \textit{\href{https://www.npr.org/2016/12/28/507208573/what-is-the-universe-made-of-scientists-respond}{What Is The Universe Made Of? Scientists Respond}}}
"The problem with general relativity is that the principles are pretty simple and the computations are always ugly."
\qauthor{\textbf{Leonard Susskind}}
\end{savequote}

\chapter{Introduction}  

\ifpdf
    \graphicspath{{Chapter1/Figs/Raster/}{Chapter1/Figs/PDF/}{Chapter1/Figs/}}
\else
    \graphicspath{{Chapter1/Figs/Vector/}{Chapter1/Figs/}}
\fi

\nomenclature[g-pi]{$\pi$}{$\simeq 3.14159\ldots$}
\nomenclature[x-hbarc]{$\hslash=c$}{$=1$}
\nomenclature[x-signature]{$\mathrm{signature}$}{$=\left(-,+,+,+\right)$}
\nomenclature[g-kappa]{$\kappa^{2}$}{$\equiv 8\pi\,G$}
\nomenclature[x-MPl]{$M_{\mathrm{Pl}}^{2}$}{$\equiv\kappa^{-2}$\quad\text{(reduced Planck mass)}}
\nomenclature[x-GF]{$G_{\mathrm{F}}$}{$\equiv \frac{\sqrt{2}}{8} \frac{g^2}{M_{\mathrm{W}}^2 c^4}=1.1663787(6) \times 10^{-5} \mathrm{GeV}^{-2}$\quad\text{(Fermi coupling constant)}}
\nomenclature[x-mPl]{$m_{\mathrm{Pl}}^{2}$}{$\equiv G^{-1}$\quad\text{(Planck mass)}}
\nomenclature[a-G]{$G$}{\text{(Newtonian gravitational constant)}}
\nomenclature[x-sym]{$B_{(ab)}$}{$\equiv\frac{1}{2}\left(B_{ab}+B_{ba}\right)$\quad\text{(symmetrization)}}
\nomenclature[x-syma]{$B_{[ab]}$}{$\equiv\frac{1}{2}\left(B_{ab}-B_{ba}\right)$\quad\text{(antisymmetrization)}}
\nomenclature[x-dAlembert]{$\Box$}{$\equiv g^{\mu\nu}\nabla_{\mu}\nabla_{\nu}$\quad\text{(d'Alembert operator)}}
\nomenclature[x-Riemann]{$R_{a b c}{ }^d$}{$=\Gamma_{a c, b}^d-\Gamma_{b c, a}^d+\Gamma_{a c}^e \Gamma_{e b}^d-\Gamma_{b c}^e \Gamma_{e a}^d$\quad\text{(Riemann tensor)}}
\nomenclature[x-Ricci]{$R_{a c}$}{$\equiv R_{a b c}{ }^b=\Gamma_{a c, b}^b-\Gamma_{b c, a}^b+\Gamma_{a c}^d \Gamma_{d e}^e-\Gamma_{b c}^d \Gamma_{d a}^b$\quad\text{(Ricci tensor)}}
\nomenclature[x-Christoffel]{$\Gamma_{i j}^m$}{$=\frac{1}{2} g^{m k}\left(\frac{\partial}{\partial x^j} g_{k i}+\frac{\partial}{\partial x^i} g_{k j}-\frac{\partial}{\partial x^k} g_{i j}\right)=\frac{1}{2} g^{m k}\left(g_{k i, j}+g_{k j, i}-g_{i j, k}\right)$\quad\text{(Christoffel symbol)}}
\nomenclature[z-MC]{MC}{Minimal coupling}
\nomenclature[z-MC]{NMC}{Non-minimal coupling}
\nomenclature[z-EFT]{EFT}{Effective field theory}
\nomenclature[z-QG]{QG}{Quantum gravity}
\nomenclature[z-ESB]{ESB}{Explicit symmetry breaking}
\nomenclature[z-SSB]{SSB}{Spontaneous symmetry breaking}
\nomenclature[z-QCD]{QCD}{Quantum chromodynamics}
\nomenclature[z-SF]{SF}{Scalar field}
\nomenclature[z-BD]{BD theory}{Brans-Dicke theory (Jordan-Fierz-Brans-Dicke theory)}
\nomenclature[z-EoM]{EoM}{Equations of motion}
\nomenclature[z-dof]{dof}{Degrees of freedom}
\nomenclature[z-GR]{GR}{General relativity (General theory of relativity)}
\nomenclature[z-STT]{STT}{Scalar–tensor theory}
\nomenclature[z-SFDM]{SFDM}{Scalar field dark matter}
\nomenclature[z-DM]{DM}{Dark matter}
\nomenclature[z-DE]{DE}{Dark energy}
\nomenclature[z-CC]{CC}{Cosmological constant}
\nomenclature[z-CCC]{CCC}{Conformal cyclic cosmology}
\nomenclature[z-LQG]{LQG}{Loop quantum gravity}
\nomenclature[z-LQC]{LQC}{Loop quantum cosmology}
\nomenclature[z-QC]{QC}{Quantum cosmology}
\nomenclature[z-BH]{BH}{Black hole}
\nomenclature[z-EC(s)]{EC(s)}{Energy condition(s)}
\nomenclature[z-NEC]{NEC}{Null energy condition}
\nomenclature[z-WEC]{WEC}{Weak energy condition}
\nomenclature[z-DEC]{DEC}{Dominant energy condition}
\nomenclature[z-SEC]{SEC}{Strong energy condition}
\nomenclature[z-QEI]{QEI}{Quantum energy inequality}
\nomenclature[z-EoS]{EoS}{Equation of state}
\nomenclature[z-MG]{MG}{Modified gravity}
\nomenclature[z-UBDM]{UBDM}{Ultralight bosonic dark matter}
\nomenclature[z-BM]{BM}{Baryonic matter}
\nomenclature[z-BBN]{BBN}{Big Bang nucleosynthesis}
\nomenclature[z-CMB]{CMB}{Cosmic microwave background}
\nomenclature[z-CMB]{BAO}{Baryon acoustic oscillations}
\nomenclature[z-PPN]{PPN}{Parameterized post-Newtonian}
\nomenclature[z-FLRW]{FLRW}{Friedmann-Lemaître-Robertson-Walker}
\nomenclature[z-FE(s)]{FE(s)}{Field equation(s)}
\nomenclature[z-KK]{KK theory}{Kaluza-Klein theory}
\nomenclature[z-MST]{MST theory}{Multi-scalar-tensor theory}
\nomenclature[z-SUGRA]{SUGRA}{Supergravity}
\nomenclature[z-LNH]{LNH}{(Dirac) large numbers hypothesis}
\nomenclature[z-WEP]{WEP}{Weak equivalence principle}
\nomenclature[z-RG]{RG}{Renormalization group}
\nomenclature[z-1PI]{1PI}{One-particle irreducible}
\nomenclature[z-UV]{UV}{Ultraviolet}
\nomenclature[z-IR]{IR}{Infrared}
\nomenclature[z-VEV]{VEV}{Vacuum expectation value}
\nomenclature[z-PGWs]{PGWs}{Primordial gravitational waves}
\nomenclature[z-LHC]{LHC}{Large Hadron Collider}
\nomenclature[x-$\pi\left(t,\vec{x}\right)$]{$\pi\left(t,\vec{x}\right)$}{Goldstone boson}
\nomenclature[z-KE]{KE}{Kinetic energy}
\nomenclature[z-PE]{PE}{Potential energy}
\nomenclature[z-PDL]{PDL}{Phantom divide line}
\nomenclature[z-LSS]{LSS}{Large scale structure}
\nomenclature[z-BKL]{BKL}{Belinsky-Khalatnikov-Lifshitz}

\label{ch:introduction}

\section{Scientific motivation}

Modern cosmology is based on an extremely successful but conceptually incomplete description of the Universe. In the standard $\Lambda$CDM model, the homogeneous and isotropic background evolution is described by the Friedmann equation:
\begin{equation}
    E^{2}(a)\equiv\frac{H^{2}(a)}{H_{0}^{2}}=\Omega_{r}^{(0)}a^{-4}+\Omega_{m}^{(0)}a^{-3}+\Omega_{k}^{(0)}a^{-2}+\Omega_{\Lambda}^{(0)}\,,
\end{equation}
where the present Universe is dominated by two components whose fundamental nature remains unknown: dark matter (DM) and dark energy (DE). Baryonic matter (BM), radiation and neutrinos are embedded in a well-tested particle physics framework, whereas DM and DE are introduced phenomenologically through their gravitational effects. This is not a weakness of cosmology, but rather a precise statement of the current frontier: the \textit{\textbf{standard model of cosmology fits an enormous amount of data, yet it leaves open the microscopic origin of most of the cosmic energy budget}} \cite{Planck:2018vyg,Planck:2018jri,CosmoVerseNetwork:2025alb}.

The dark sector is therefore not a single problem. It contains several interconnected questions. Is DM a new particle, a scalar field, a modification of gravity, or an effective description of a more complex sector? Is DE a true cosmological constant, a slowly evolving field, an interaction within the dark sector, or a sign that GR must be modified on cosmological scales? Are the observed cosmological tensions caused by unknown systematic errors, by an incomplete modeling of astrophysical data, or by new fundamental physics? These questions motivate the theoretical and phenomenological framework developed in this dissertation.

The first important lesson is that the split between different dark components is not always as straightforward as may appear from the background Friedmann equation. At the homogeneous level, baryonic and cold dark matter both contribute as pressureless components:
\begin{equation}
    \rho_{b}\propto a^{-3}\quad\land\quad\rho_{c}\propto a^{-3}\,.
\end{equation}
Their distinction is therefore not encoded in the background expansion alone, but in their microphysical properties, interactions, perturbations and observational signatures. These issues were discussed in the article \cite{Borowiec:2023kmq}, where the possibility of separating baryonic matter from dark matter within the LCDM formalism was analyzed. The conclusion is not that the LCDM description is useless, but rather, its decomposition into physical sectors contains assumptions that must be tested and generalized where necessary.

A second lesson comes from the status of inflation. Inflation is a highly successful framework for explaining the approximate spatial flatness, homogeneity and isotropy of the observable Universe, as well as the origin of nearly scale-invariant primordial perturbations \cite{Starobinsky:1980te,Guth:1980zm,Linde1982,Baumann:2009ds,Baumann_2022inflation}. However, inflation is not a single theory. It is a class of models whose ultraviolet completion, initial conditions, embedding in quantum gravity and robustness against Planck-suppressed operators remain open problems. The preprint \cite{Postolak:2024xtm} presented this broader conceptual landscape and emphasized that the Big Bang and inflationary paradigms should be understood not as final answers, but as parts of a wider family of possible cosmological scenarios.

The common theoretical language connecting these problems is the language of scalar fields (SFs). Scalar degrees of freedom appear in models of inflation, quintessence, scalar field dark matter, scalar-tensor theories of gravity, extra-dimensional theories, string-inspired effective actions and ekpyrotic/cyclic cosmologies. They are sufficiently simple to allow detailed mathematical analysis, but sufficiently general to capture a wide range of possible departures from the minimal $\Lambda$CDM picture. For this reason, scalar fields provide a natural organizing principle for the present dissertation.

\section{Scalar fields and scalar-tensor cosmology}

This thesis explores the possibility that part of the dark sector of the universe can be described by scalar degrees of freedom. Depending on their coupling structure, scalar fields can be interpreted as either new matter components or additional gravitational degrees of freedom. The distinction between the two is especially important in scalar-tensor theories (STTs), where gravitational interaction is mediated not only by the metric tensor, $g_{\mu\nu}$, but also by one or more scalar fields \cite{Brans:1961sx,Dicke:1961gz,Faraoni2004chapter,Fujii_Maeda_2003,Clifton:2011jh}.

The distinction between modified matter and modified gravity can be expressed in the most basic form as follows:
\begin{equation}
    G_{\mu\nu}=\kappa^{2}\left(T_{\mu\nu}^{(m)}+T_{\mu\nu}^{(\mathrm{new})}\right)\,,
\end{equation}
or:
\begin{equation}
    G_{\mu\nu}+\Delta G_{\mu\nu}=\kappa^{2}T_{\mu\nu}^{(m)}\,.
\end{equation}
These two forms can sometimes be rearranged in relation to each other at the background level. However, at the level of perturbations, local tests of gravity, fifth forces, gravitational waves, and matter couplings, the distinction becomes physically meaningful. A minimally coupled scalar field is usually interpreted as a matter component, while a non-minimally coupled scalar field is naturally interpreted as part of the gravitational sector.

This can be written as a general class of scalar-tensor theories in the Jordan frame \cite{Faraoni2004chapter,Fujii_Maeda_2003,Clifton:2011jh,Nojiri:2017ncd}:
\begin{equation}
    S_{J}=\int d^{4}x\sqrt{-g}\left[\frac{F(\phi)}{2\kappa^{2}}R-\frac{1}{2}K(\phi)g^{\mu\nu}\nabla_{\mu}\phi\nabla_{\nu}\phi-U(\phi)\right]+S_{m}\left[g_{\mu\nu},\psi_{m}\right]\,.
\end{equation}
The function $F(\phi)$ determines the effective gravitational coupling, $K(\phi)$ controls the normalization of the scalar kinetic term, and $U(\phi)$ plays the role of the scalar potential. Such theories naturally arise in attempts to generalize GR, in dimensional reduction, in string-inspired scenarios, and in effective descriptions of modified gravity \cite{Faraoni2004chapter,Fujii_Maeda_2003,Clifton:2011jh,Nojiri:2017ncd}.

The question of whether a scalar field belongs to the matter sector or the gravitational sector is also related to the issue of conformal frames. A Weyl transformation can map a Jordan-frame theory with a non-minimal coupling into an Einstein-frame theory with a canonical Einstein-Hilbert term, at the price of introducing scalar-dependent matter couplings. The two descriptions may be mathematically equivalent when the transformation is regular, but the physical interpretation of matter couplings, geodesics, energy conditions and quantum corrections requires care \cite{Faraoni:1998qx,Faraoni:2006fx,Catena:2006bd,Chiba:2013mha,Domenech:2016yxd}.

This issue is not merely formal. It is directly connected with the author’s article \cite{Postolak:2025qmv}, where a non-minimally coupled scalar field dark sector was studied in the Einstein frame. In such a formulation, the scalar field can interact with matter through an effective conformal coupling. The resulting dynamics can be analyzed using autonomous systems, critical points and phase-space methods. This provides a concrete example of how scalar-tensor cosmology can interpolate between modified gravity, dark sector interactions and dynamical dark energy.

\section{Inflation, quintessence and ekpyrosis as SF cosmologies}

The dissertation is organized around three major SF applications in cosmology: inflation, quintessence and ekpyrotic/cyclic cosmology.

The first application is inflation. In the simplest models, inflation is driven by a scalar field whose potential energy dominates over its kinetic energy \cite{Starobinsky:1980te,Guth:1980zm,Linde1982,Liddle2000,Baumann:2009ds,Baumann_2022inflation,Mukhanov_2005a}. The slow-roll regime requires \cite{Liddle:1994dx}:
\begin{equation}
    \epsilon_{V}\equiv\frac{M_{\mathrm{Pl}}^{2}}{2}\left(\frac{V'(\phi)}{V(\phi)}\right)^{2}\ll 1 \quad\land\quad\left|\eta_{V}\right|\equiv\left|M_{\mathrm{Pl}}^{2}\frac{V''(\phi)}{V(\phi)}\right|\ll 1\,.
\end{equation}
However, from the effective field theory (EFT) point of view, these conditions are sensitive to radiative corrections and Planck-suppressed operators \cite{Donoghue:1994dn,Burgess_2020,Cheung:2007st,Weinberg:2008hq,Baumann_McAllister_2015EFT-inflation}. A generic correction of the form:
\begin{equation}
    \Delta V\sim c\,V\left(\frac{\phi}{\Lambda}\right)^{\delta-4}
\end{equation}
may induce corrections to $\eta_{V}$ that spoil slow roll unless protected by a symmetry or controlled by a UV completion \cite{Baumann_McAllister_2015EFT-inflation,Kallosh:1995hi,Harlow:2018tng}. Thus, inflation is not only a successful phenomenological mechanism, but also a probe of high-energy physics and quantum gravity \cite{Donoghue:1994dn,Burgess_2020,Cheung:2007st,Weinberg:2008hq,Kallosh:1995hi,Harlow:2018tng}.

The second application is quintessence. In this case, the scalar field is responsible for late-time accelerated expansion rather than primordial inflation \cite{Ratra:1987rm,Wetterich:1987fm,Caldwell:1997ii,Peebles:2002gy,Copeland:2006wr,Frieman:2008sn,Tsujikawa:2013fta}. A canonical quintessence field is described by:
\begin{equation}
    \rho_{\phi}=\frac{1}{2}\dot{\phi}^{2}+V(\phi) \quad\land\quad p_{\phi}=\frac{1}{2}\dot{\phi}^{2}-V(\phi)\,,
\end{equation}
and therefore:
\begin{equation}
    \rho_{\phi}+p_{\phi}=\dot{\phi}^{2}\geq 0\,.
\end{equation}
Consequently, a minimally coupled canonical quintessence field with positive energy density cannot cross the phantom divide line (PDL) $\omega=-1$ \cite{Caldwell:1999ew,Cline:2003gs,Vikman:2004dc,Copeland:2006wr,Tsujikawa:2013fta}. This simple observation becomes important in light of DESI DR2, whose combined analyses suggest that the late-time expansion history may be more flexibly described by a time-dependent effective dark energy equation of state than by a strict cosmological constant \cite{DESI:2025zgx,DESI:2025fii}. Such results do not prove quintessence or scalar-tensor gravity, but they motivate a systematic analysis of dynamical dark energy models \cite{Caldwell:2005tm,Linder:2007wa,Copeland:2006wr,Tsujikawa:2013fta,Amendola_Tsujikawa_2010q}.

The third application is ekpyrotic and cyclic cosmology. In ekpyrotic models, a scalar field rolling down a steep negative potential produces an ultra-stiff equation of state:
\begin{equation}
    \omega_{\mathrm{ekp}}\gg 1\,.
\end{equation}
During contraction, the ekpyrotic energy density grows faster than anisotropy, spatial curvature and ordinary matter. This makes ekpyrosis a powerful smoothing mechanism and an alternative to inflationary smoothing \cite{Khoury:2001wf,Lehners:2008vx,Battefeld:2014uga}. However, ekpyrotic models face their own difficulties: the bounce, perturbation matching, non-Gaussianity, entropy, geodesic completeness and the construction of a controlled EFT through the non-singular transition.

These issues are directly related to the preprint \cite{Postolak:2026okk}, where phase-resolved field-space distance bounds were analyzed in ekpyrotic, bouncing and cyclic cosmologies. The key idea is to separate the total SF distance into physically distinct phases and to ask whether each phase remains under theoretical control. In ekpyrotic contraction with approximately constant $\epsilon_{\mathrm{ek}}$, one obtains a characteristic relation of the form:
\begin{equation}
    d_{\mathrm{ek}}\simeq\frac{\sqrt{2\epsilon_{\mathrm{ek}}}}{\epsilon_{\mathrm{ek}}-1}N_{\mathrm{sm}}\,,
\end{equation}
where $N_{\mathrm{sm}}$ measures the smoothing in $|H|$. In the ultra-stiff limit, this gives:
\begin{equation}
    d_{\mathrm{ek}}\simeq\sqrt{\frac{2}{\epsilon_{\mathrm{ek}}}}
    N_{\mathrm{sm}}\,,
\end{equation}
so that a sub-Planckian distance budget requires sufficiently large $\epsilon_{\mathrm{ek}}$. This illustrates how scalar field cosmology can be constrained not only by observations, but also by internal EFT and field-space consistency conditions.

\section{Original contributions and scope of the thesis}

The original work connected with this dissertation can be summarized as follows:
\begin{examplebox}[Research contributions connected with the dissertation]
    \begin{enumerate}[label=(\alph*)]
        \item In \cite{Borowiec:2023kmq}, the separation between baryonic and dark matter within the $\Lambda$CDM formalism was analyzed. This work motivates a careful treatment of the dark sector and emphasizes that the physical interpretation of cosmological components depends on more than the background expansion history;
        \item In \cite{Postolak:2024xtm}, alternative cosmological models were presented in a broader conceptual context. This work motivates the comparative approach adopted in the dissertation, where inflation is analyzed together with non-inflationary early-Universe scenarios;
        \item In \cite{Postolak:2025qmv}, an Einstein frame non-minimally coupled scalar field dark sector model was studied in detail. This work provides the direct link between scalar-tensor formalism, dark sector interactions and dynamical systems methods;
        \item In \cite{CosmoVerseNetwork:2025alb}, the author contributed to the discussion of cosmological tensions, including extra relativistic degrees of freedom, modified gravity, cold dark matter, and interacting or decaying dark matter. This broader context motivates the thesis perspective that observational tensions should be addressed simultaneously through systematics and fundamental physics;
        \item In \cite{Postolak:2026okk}, phase-resolved field-space distance bounds were developed for ekpyrotic, bouncing and cyclic cosmologies. This work connects early-Universe alternatives with EFT control and field-space consistency.
    \end{enumerate}
\end{examplebox}
The dissertation \textit{\textbf{does not aim to prove that one specific scalar field model is the final description of the dark sector}}. Rather, its purpose is to develop a coherent framework in which scalar fields can be used to organize, compare and constrain possible deviations from the minimal $\Lambda$CDM picture.

\section{Structure of the dissertation}

Chapter \ref{Sec:STTs} introduces the formalism of scalar-tensor theories. It begins with Brans-Dicke theory as the prototype of scalar-tensor gravity, derives the field equations, discusses the GR limit, and explains the connection with Kaluza-Klein theory. It then presents the general scalar-tensor framework, motivations from dark energy and high-energy physics, conformal transformations, and the question of mathematical and physical equivalence between Jordan and Einstein frames.

Chapter \ref{Sec:inflation} discusses inflation from the viewpoint of EFT and UV sensitivity. It explains how the EFT approach parameterizes unknown high-energy physics, how time translations are spontaneously broken in an inflationary background, how the Goldstone mode describes adiabatic perturbations, and why scalar field masses are sensitive to radiative and Planck-suppressed corrections.

Chapter \ref{Sec:quintessence} is dedicated to quintessence. It presents the canonical scalar field formalism, tracking and scaling solutions, dynamical systems methods, and observational diagnostics such as the statefinder parameters, the $(\omega_{\phi},\omega'_{\phi})$ plane and the $Om(z)$ diagnostic. Special attention is given to DESI DR2 and to the interpretation of time-dependent dark energy.

Chapter \ref{Sec:ekpyrotic} discusses ekpyrotic and cyclic cosmology. It analyzes ekpyrotic smoothing, negative scalar field potentials, multi-field perturbations, non-Gaussianity, tensor perturbations, cyclic evolution, bounce conditions, entropy and geodesic incompleteness. This chapter discusses the strengths and open problems of cyclic ekpyrotic scenarios.

Chapter \ref{Sec:conclusions} summarizes the chapter's conclusions and presents perspectives for future work. The chapter identifies common lessons from different scalar field scenarios and outlines possible directions for scalar-tensor dark sector cosmology, early-Universe models controlled by EFT, and observational tests beyond the LCDM model.

\begin{savequote}
"Could it be, nevertheless, that Einstein's theory is wrong? Might it be necessary to modify it—to find a new theory of gravity that can explain both the stronger gravity and the apparent antigravity being observed today—rather than simply throwing in invisible things to make the standard model work?"
\qauthor{\textbf{John Moffat}, \textit{Reinventing Gravity: A Physicist Goes Beyond Einstein}}
\end{savequote}

\chapter{The formalism of scalar-tensor theories}  
\label{Sec:STTs}
\ifpdf
    \graphicspath{{Chapter2/Figs/Raster/}{Chapter2/Figs/PDF/}{Chapter2/Figs/}}
\else
    \graphicspath{{Chapter2/Figs/Vector/}{Chapter2/Figs/}}
\fi


\section{Brans-Dicke theory}
\label{BDtheory-section}
The \textit{\textbf{Brans-Dicke theory}}\footnote{Also known as Jordan-Fierz-Brans-Dicke theory.} (BD theory) \cite{Brans:1961sx,Fierz:1956zz,Faraoni2004chapter,Fujii_Maeda_2003} provides a modern prototype for modified theories of gravity based on the scalar-tensor formalism. In the so-called \textit{Jordan conformal frame}, where the scalar field is non-minimally coupled to gravity (for more details, see Sec.~\ref{EJframes-section}) the action takes the following form\footnote{From a theoretical perspective (\textit{\textbf{ghost-free condition}}), the minimum requirement for the free parameter is $\omega_{\mathrm{BD}}>-\frac{3}{2}$.} \cite{Faraoni2004chapter}:
\begin{equation}\label{BDaction}
    \begin{aligned}
    S_{\mathrm{BD}} & =\frac{1}{16 \pi} \int d^4 x \sqrt{-g}\left[\phi R-\frac{\omega_{\mathrm{BD}}}{\phi} g^{\mu\nu} \nabla_\mu \phi \nabla_\nu \phi-V(\phi)\right]+S_{\mathrm{m}} \\
    & \equiv S_{\phi}+S_{\mathrm{m}}\,.
    \end{aligned}
\end{equation}
The first segment, $S_{\phi}$, describes the gravitational part of the action, and the second one, $S_{\mathrm{m}}$, describes the matter sector (i.e., all forms of matter other than the scalar field):
\begin{equation}
    S_{\mathrm{m}}=\int d^4 x \sqrt{-g}\,\mathcal{L}_{\mathrm{m}}\,.
\end{equation}
In such a case, the matter content is minimally coupled (MC) to the SF $\phi$ - there is no direct coupling between these two constituents.

Furthermore, from the above information concerning the physical formalism under consideration, it follows that in this particular theory, the gravitational field is described not only by the metric tensor $g_{\mu\nu}$, but also by the BD SF $\phi$. Concerning the physical interpretation of the potential of this scalar field, it could be recognized as a sort of generalization of the cosmological constant (dark energy). It is also worth pointing out that the original motivation for BD theory was to find a formalism that satisfied \textit{\textbf{Mach's principle}}\footnote{This principle can be summarized by saying that \textit{inertia and local inertial frames are determined by the global distribution and motion of matter in the universe}.} \cite{Mach_2013}.

\subsection{General equations of motion}
\subsubsection{Metric sector}
We begin the procedure of deriving the general form of the equations of motion for BD theory with a variation of the Ricci scalar using the Palatini identity:
\begin{equation}\label{generalRvariation}
    \delta R=R_{ab}\,\delta g^{ab}+\underbrace{\nabla_{a}\left(\nabla_{b}\,\delta g^{ab}-g^{cd}\nabla^{a}\,\delta g_{cd}\right)}_{\text{Total divergence}}\,,
\end{equation}
where the 2nd term is the total divergence and could be dropped by assuming appropriate boundary conditions. The next step is the variation of the segment describing the non-minimal coupling (NMC) in the action \eqref{BDaction}, namely:
\begin{equation}\label{NMC-variation}
    \delta\left(\sqrt{-g}\,\phi R\right)=\phi\,\delta\left(\sqrt{-g}R\right)+R\,\delta\left(\sqrt{-g}\,\phi\right)\,.
\end{equation}
Expanding the relation:
\begin{equation}
    \delta\left(\sqrt{-g}R\right)=\sqrt{-g}\,\underbrace{\left(R_{ab}-\frac{1}{2}g_{ab}R\right)}_{G_{ab}}\,\delta g^{ab}=\sqrt{-g}\,G_{ab}\,\delta g^{ab}\,,
\end{equation}
where:
\begin{equation}\label{Einstein-tensor}
    G_{ab}\equiv R_{ab}-\frac{1}{2}g_{ab}R
\end{equation}
is the \textit{\textbf{Einstein tensor}}, and:
\begin{equation}
    \delta\left(\sqrt{-g}\,\phi\right)=\delta\phi\sqrt{-g}+\phi\,\delta\left(\sqrt{-g}\right)=\sqrt{-g}\left(\delta\phi-\frac{1}{2}\phi\,g_{ab}\,\delta g^{ab}\right)
\end{equation}
gives the final form of \eqref{NMC-variation}:
\begin{equation}
    \delta\left(\sqrt{-g}\,\phi R\right)=\sqrt{-g}\left(\phi\,G_{ab}+g_{ab}\,\Box\phi-\nabla_{a}\nabla_{b}\,\phi\right)\,\delta g^{ab}\,.
\end{equation}
Consequently, one could write the following:
\begin{equation}
    \int{d^{4}x\,\sqrt{-g}\left(\delta\phi\,R\right)}=\int{d^{4}x\,\sqrt{-g}\left(g_{ab}\,\Box\phi-\nabla_{a}\nabla_{b}\,\phi\right)\,\delta g^{ab}}\,.
\end{equation}
The next step is to address the variation of the kinetic term in \eqref{BDaction}:
\begin{equation}\label{kinetic-variation}
    \delta\left(\sqrt{-g}\,\frac{\omega_{\mathrm{BD}}}{\phi}\nabla^{c}\phi\,\nabla_{c}\phi\right)\,.
\end{equation}
First of all, let us note that, in fact, we perform a variation with respect to the metric tensor, so:
\begin{equation}
    \delta\left(\nabla^{c}\phi\,\nabla_{c}\phi\right)=-\delta g^{ab}\,\nabla_{a}\phi\,\nabla_{b}\phi
\end{equation}
which yields:
\begin{equation}
    \delta\left(\frac{\omega_{\mathrm{BD}}}{\phi}\,g^{cd}\,\nabla_{c}\phi\,\nabla_{d}\phi\right)=-\frac{\omega_{\mathrm{BD}}}{\phi}\,\nabla_{a}\phi\,\nabla_{b}\phi\,\delta g^{ab}\,.
\end{equation}
Then, using the identity for the variation of the square root of the metric tensor determinant:
\begin{equation}
    \delta\left(\sqrt{-g}\right)=-\frac{1}{2}\sqrt{-g}\,g_{ab}\,\delta g^{ab}
\end{equation}
we can write down the final form of the expression \eqref{kinetic-variation}:
\begin{equation}\label{kinetic-variation-final}
    \delta\left(\sqrt{-g}\,\frac{\omega_{\mathrm{BD}}}{\phi}\nabla^{c}\phi\,\nabla_{c}\phi\right)=-\sqrt{-g}\,\frac{\omega_{\mathrm{BD}}}{\phi}\left(\frac{1}{2}g_{ab}\,\nabla^{c}\phi\,\nabla_{c}\phi+\nabla_{a}\phi\,\nabla_{b}\phi\right)\,\delta g^{ab}\,.
\end{equation}
Variation of the SF potential term produces:
\begin{equation}
    \delta\left[\sqrt{-g}\,V(\phi)\right]=\frac{1}{2}\sqrt{-g}\,V(\phi)\,g_{ab}\,\delta g^{ab}\,.
\end{equation}
and of the matter part:
\begin{equation}
    \delta S_{\mathrm{m}}=-\frac{1}{2}\int{d^{4}x\,\sqrt{-g}\,T_{ab}^{(m)}\,\delta g^{ab}}\,,
\end{equation}
where $T_{ab}^{(m)}$ denotes the \textit{\textbf{energy-momentum tensor}} for the matter sector:
\begin{equation}
    T_{ab}^{(m)}\equiv -\frac{2}{\sqrt{-g}}\,\frac{\delta}{\delta g^{ab}}\left(\sqrt{-g}\,\mathcal{L}_{\mathrm{m}}\right)\,.
\end{equation}
By considering all the performed calculations, the total variation of the Brans-Dicke action \eqref{BDaction} with respect to the metric is expressed as follows:
\begin{equation}\label{total-BD-variation}
    \begin{aligned}
    \delta S_{\mathrm{BD}} = & \frac{1}{16\pi}\int d^{4}x\,\sqrt{-g}\Biggl[\phi\,G_{ab}+g_{ab}\,\Box\phi-\nabla_{a}\nabla_{b}\phi-\frac{\omega_{BD}}{\phi}\Bigl(\nabla_{a}\phi\,\nabla_{b}\phi-\frac{1}{2}g_{ab}\,\nabla^{c}\phi\,\nabla_{c}\phi\Bigr) \\
    & -\frac{1}{2}V(\phi)\,g_{ab}\Biggr]\,\delta g^{ab}-\frac{1}{2}\int d^{4}x\,\sqrt{-g}\,T_{ab}^{(m)}\,\delta g^{ab}=0\,,
    \end{aligned}
\end{equation}
and produces the following explicit form of the \textit{\textbf{general field equations of the BD theory for the metric sector}}:
\begin{equation}\label{BD-general-EoM}
        \boxed{G_{ab}=\frac{8\pi}{\phi}\,T_{ab}^{(m)}+\frac{\omega_{\mathrm{BD}}}{\phi^{2}}\left(\nabla_{a}\phi\,\nabla_{b}\phi-\frac{1}{2}g_{ab}\,\nabla^{c}\phi\,\nabla_{c}\phi\right)+\frac{1}{\phi}\left(\nabla_{a}\nabla_{b}\phi-g_{ab}\,\Box\phi\right)-\frac{V(\phi)}{2\phi}\,g_{ab}}\,.
\end{equation}
\subsubsection{Scalar field sector}
Variation of the gravitational part of \eqref{BDaction} gives:
\begin{equation}\label{SF-gravitational-variation}
    \delta S_{\phi}=\frac{1}{16\pi}\int d^{4}x\,\sqrt{-g}\left[\delta\phi\, R+\omega_{\mathrm{BD}}\frac{\nabla^{c}\phi\,\nabla_{c}\phi}{\phi^{2}}\,\delta\phi-2\frac{\omega_{\mathrm{BD}}}{\phi}\nabla^{c}\phi\,\nabla_{c}\delta\phi-V'(\phi)\delta\phi\right]\,.
\end{equation}
The calculations for most of the terms in \eqref{SF-gravitational-variation} are straightforward, but it is worth taking a closer look at the 3rd term. Let us introduce a vector field defined as follows:
\begin{equation}
    A^{c}\equiv 2\frac{\omega_{\mathrm{BD}}}{\phi}\nabla^{c}\phi\,
\end{equation}
so that we must calculate the following integral:
\begin{equation}
    \mathcal{I}_{1}\equiv -\int d^{4}x\,\sqrt{-g}\,A^{c}\,\nabla_{c}\delta\phi\,.
\end{equation}
Using the covariant integration by parts one obtains:
\begin{equation}
    \int d^{4}x\,\sqrt{-g} \,\nabla_c\left(A^c\,\delta \phi\right)=\int d^{4}x\,\sqrt{-g}\left[\nabla_{c} A^c\,\delta \phi+A^{c}\,\nabla_{c}\delta \phi\right]\,.
\end{equation}
Moreover, neglecting the boundary term (assuming $\delta\phi=0$ on $\partial\mathcal{M}$):
\begin{equation}
    \int d^{4}x\,\sqrt{-g}\,A^{c}\, \nabla_{c}\delta\phi=-\int d^{4}x\,\sqrt{-g}\,\delta\phi\, \nabla_{c}A^c\,,
\end{equation}
yields:
\begin{equation}\label{I1-simplified}
    \mathcal{I}_{1}=\int d^{4}x\,\sqrt{-g}\,\delta\phi\,\nabla_{c}A^{c}\,.
\end{equation}
The formula in the integral \eqref{I1-simplified} could be written as:
\begin{equation}
    \nabla_{c}A^{c}=\nabla_{c}\left(2\frac{\omega_{\mathrm{BD}}}{\phi}\nabla^{c}\phi\right)=2\omega_{\mathrm{BD}}\,\nabla_{c}\left(\frac{1}{\phi}\nabla^{c}\phi\right)\,.
\end{equation}
The next step is to use the Leibniz rule:
\begin{equation}
    \nabla_{c}\left(\frac{1}{\phi}\nabla^{c}\phi\right)=\left(\nabla_{c}\frac{1}{\phi}\right)\nabla^{c}\phi+\frac{1}{\phi}\nabla_{c}\nabla^{c}\phi=-\frac{1}{\phi^{2}}\nabla_{c}\phi\nabla^{c}\phi+\frac{1}{\phi}\Box\phi\,,
\end{equation}
therefore:
\begin{equation}
    \nabla_{c}A^{c}=2\omega_{\mathrm{BD}}\left(-\frac{1}{\phi^{2}}\nabla_{c}\phi\nabla^{c}\phi+\frac{1}{\phi}\Box\phi\right)\,.
\end{equation}
Finally, the integral takes the form of:
\begin{equation}
    \mathcal{I}_{1}=\int d^{4}x\,\sqrt{-g}\,\delta\phi\left(-2\omega_{\mathrm{BD}}\frac{\nabla_{c}\phi\nabla^{c}\phi}{\phi^{2}}+2\frac{\omega_{\mathrm{BD}}}{\phi}\Box\phi\right)\,.
\end{equation}
Taking everything into account, the variation principle leads to:
\begin{equation}
    \delta S_{\phi}=\frac{1}{16\pi}\int d^{4}x\,\sqrt{-g}\,\delta\phi\left[R-\frac{\omega_{\mathrm{BD}}}{\phi^{2}}\nabla^{c}\phi\nabla_{c}\phi+2\frac{\omega_{\mathrm{BD}}}{\phi}\Box\phi-V'(\phi)\right]=0\,,
\end{equation}
which consequently provides:
\begin{equation}
    R-\frac{\omega_{\mathrm{BD}}}{\phi^{2}}\nabla^{c}\phi\nabla_{c}\phi+2\frac{\omega_{\mathrm{BD}}}{\phi}\Box\phi-V'(\phi)=0\,.
\end{equation}
It is possible to obtain a simpler form of the equations of motion by eliminating the Ricci scalar using the field equations for the metric sector \eqref{BD-general-EoM}:
\begin{equation}\label{Ricci-scalar-BD}
    R=\frac{1}{\phi}\left[8\pi\,T_{(\mathrm{m})}+3\Box\phi+\frac{\omega_{\mathrm{BD}}}{\phi}\left(\nabla\phi\right)^{2}+2V(\phi)\right]\,.
\end{equation}
Finally, the \textbf{equation of motion for the scalar field in BD theory} takes the form:
\begin{equation}\label{BD-equation-SF}
    \boxed{\Box\phi=\frac{1}{2\omega_{\mathrm{BD}}+3}\left[8\pi\,T_{\mathrm{m}}+\phi\,V'(\phi)-2V(\phi)\right]}\,.
\end{equation}
Several important conclusions could be drawn from the above equation.

First of all, the BD SF $\phi$ has \textbf{non-conformal matter} (matter with non-vanishing trace):
\begin{equation}\label{non-conformal-matter-BD}
    T_{\mathrm{m}}\equiv T^{(\mathrm{m})\mu}_{\quad\;\;\;\;\;\mu}\neq 0
\end{equation}
\textbf{as the source}. However, the \textbf{SF is not directly coupled} to either the \textbf{matter energy-momentum tensor} or the \textbf{matter Lagrangian}. In fact, \textbf{$\phi$ acts back on matter only via the metric tensor} in the manner prescribed by the field equations \eqref{BD-general-EoM}.

Secondly, the term proportional to the derivative of the scalar field potential and the potential itself vanishes in the case of a standard massive field, namely for:
\begin{equation}
    V(\phi)=\frac{1}{2}m^{2}\phi^{2}\,.
\end{equation}

The most important physical implication resulting from the form of action \eqref{BDaction} and equations of motion \eqref{BD-general-EoM} is the observation that in BD theory, the \textit{\textbf{scalar field plays the role of the inverse of the effective gravitational coupling}}:
\begin{equation}
    G_{\mathrm{eff}}(\phi)\equiv\frac{1}{\phi}\,,
\end{equation}
which is essentially a function dependent on a point in spacetime. It is usually assumed that the scalar field only takes positive values:
\begin{equation}
    \phi>0\,.
\end{equation}
This is because of the attractive nature of gravity.
\subsubsection{Brans-Dicke parameter}
The BD parameter $\omega_{\mathrm{BD}}$ is a \textbf{free parameter} of the theory. From a theoretical/phenomenological point of view, some scientists believe that the natural range of values for free parameters is values of the order of unity\footnote{The issue of \textit{\textbf{naturalness}} and \textit{\textbf{fine-tuning}} is quite controversial in the scientific community. More information on this topic can be found, for example, in \cite{tHooft:1979rat,Hossenfelder:2018ikr,Grinbaum:2009sk,Adams:2019kby,Burgess_2020,Sloan:2020zer}. Moreover, the author of this dissertation would like to point out that he categorically disagrees with the arguments behind the so-called \textit{anthropic principle} \cite{Carter1974}, as he considers it (as does a significant fraction of the scientific community) to be outside the scope of science.}. In our case, this would mean that:
\begin{equation}
    \omega_{\mathrm{BD}}\sim\mathcal{O}(1)\,.
\end{equation}
In this specific case, it can be justified on physical grounds by a value close to the strength of gravitational coupling (e.g., low-energy limit of string theories \cite{Polchinski_1998a,Polchinski_1998b,Green_Schwarz_Witten_2012a,Green_Schwarz_Witten_2012b}).

The constraints on the acceptable values of the parameter $\omega_{\mathrm{BD}}$ come mainly from observational tests of modified theories of gravity \cite{Will_2018,Will:2014kxa,Ishak:2018his}, which seem to rule out values of the order of unity and favor values of \cite{Bertotti:2003rm,Williams:2004qba}:
\begin{equation}
    \omega_{\mathrm{BD}}\gtrsim 4\times 10^{4}\,.
\end{equation}
in the case of massless SF. Another possible bound comes from cosmological (CMB) constraints (also for the massless field) \cite{Avilez:2013dxa}:
\begin{equation}
    \omega_{\mathrm{BD}}>890\,.
\end{equation}
The above implies that the value of the free parameter in BD theory should be fine-tuned in order to be consistent with experimental results. Furthermore, a significant observation is the conclusion that \textit{\textbf{the higher the values of $\boldsymbol{\omega_{\mathrm{BD}}}$, the closer Brans-Dicke theory is to the standard description known from GR}}. One can notice this in the first term on the right-hand side of the field equation \eqref{BD-equation-SF} or through analysis within the framework of PPN formalism\footnote{In fact, $\gamma_{\mathrm{PPN}}$ parameter indicates how much space curvature is produced by unit rest mass. For more details, see \cite{Misner:1973prbch39,Will_2018,Will:2014kxa}.} \cite{Perivolaropoulos:2009ak}:
\begin{equation}
    \gamma_{\mathrm{PPN}}=\frac{1+\omega_{\mathrm{BD}}}{2+\omega_{\mathrm{BD}}}\quad\implies\quad \left|\gamma_{\mathrm{PPN}}-1\right|=\frac{1}{2+\omega_{\mathrm{BD}}}\propto\mathcal{O}\left(\frac{1}{\omega_{\mathrm{BD}}}\right)\,.
\end{equation}
The renewed interest in BD theory (and its generalizations) originates from modified theories of gravity belonging to the widely-understood class of \textbf{scalar-tensor theories} (STTs). In addition, BD theory can be closely linked to the compactification (of potential extra dimensions) within \textbf{Kaluza-Klein theory} \cite{Cho1992,Overduin:1997sri} (more details in Sec.~\ref{BDandKK-section}) and the \textbf{low-energy limit} of the gravitational sector of \textbf{bosonic (super)string theory} \cite{Maeda1988,Blaschke:2004wa,Gasperini_2007ch2}, in which:
\begin{equation}
    \omega_{\mathrm{BD}}=-1\,.
\end{equation}
\subsection{Equations of motion in the FLRW Universe}
A homogeneous and isotropic model of the universe is described using the concept of \textbf{FLRW metric} which, in comoving polar coordinates $\left(t,r,\theta,\varphi\right)$, takes the form:
\begin{equation}\label{FLRW-metric-general}
    ds^{2}=-dt^{2}+a^{2}(t)\left[\frac{dr^{2}}{1-k\,r^{2}}+r^{2}\left(d\theta^{2}+\sin^{2}{\theta}\,d\varphi^{2}\right)\right]\,,
\end{equation}
where:
\begin{equation}
    k=0,\pm 1
\end{equation}
denotes the spatial curvature and:
\begin{equation}
    d\Omega^{2}\equiv d\theta^{2}+\sin^{2}{\theta}\,d\varphi^{2}
\end{equation}
is a metric on the unit 2-sphere (angular part of the spatial line element).

The above metric allows one to derive Friedmann equations, which provide a description of the evolution of the universe. The 1st \textbf{Friedmann equation} is essentially a formula describing \textbf{constraints}:
\begin{equation}\label{IFriedmann-general}
    \left(\frac{\dot{a}}{a}\right)^{2}\equiv H^{2}=\frac{\kappa^{2}}{3}\rho_{\mathrm{tot}}-\frac{k}{a^{2}}
\end{equation}
and the 2nd Friedmann equation shows how the scale factor evolves over time:
\begin{equation}\label{IIFriedmann-general}
    \frac{\ddot{a}}{a}\equiv \dot{H}+H^{2}=-\frac{\kappa^{2}}{6}\left(\rho_{\mathrm{tot}}+3p_{\mathrm{tot}}\right)\,,
\end{equation}
where:
\begin{equation}
    \rho_{\mathrm{tot}}\equiv\sum_{i}\rho_{i}\quad\land\quad p_{\mathrm{tot}}\equiv\sum_{i}p_{i}
\end{equation}
denotes \textbf{total energy density} and \textbf{total pressure}, respectively.
Moreover, one can easily insert the explicit dependence on the Hubble parameter $H$ from \eqref{IFriedmann-general} into the equation for $\dot{H}$ \eqref{IIFriedmann-general} and obtain a compact form of the EoM:
\begin{equation}
    \frac{a\,\ddot{a}-\dot{a}^{2}}{a^{2}}\equiv\dot{H}=-\frac{\kappa^{2}}{2}\left(\rho+p\right)+\frac{k}{a^{2}}\,.
\end{equation}
\subsubsection{Scalar field sector}
The Brans-Dicke scalar field depends only on the cosmic time $t$, so we can write:
\begin{equation}
    \nabla^{c}\phi\nabla_{c}\phi=-\dot{\phi}^{2}
\end{equation}
and:
\begin{equation}\label{Box-phi-general}
    \Box\phi=\nabla^{\mu}\nabla_{\mu}\phi=\frac{1}{\sqrt{-g}}\partial_{\mu}\left(\sqrt{-g}\,g^{\mu\nu}\partial_{\nu}\phi\right)\,.
\end{equation}
Furthermore, the square root of the metric determinant takes the form:
\begin{equation}
    \sqrt{-g}=a^{3}(t)\,r^{2}\sin{\theta}\,\frac{1}{\sqrt{1-k\,r^{2}}}\,.
\end{equation}
In the case of a homogeneous SF:
\begin{equation}
    \phi=\phi(t)\,.
\end{equation}
Consequently, differentiation will only be performed with respect to cosmic time. This means that equation \eqref{Box-phi-general} will contain only a $0-0$ component and can therefore be simplified to the following expression:
\begin{equation}\label{Box-homogeneous-SF}
    \Box\phi=\frac{1}{\sqrt{-g}}\partial_{0}\left(\sqrt{-g}\,g^{00}\partial_{0}\phi\right)\quad\land\quad\partial_{0}\equiv\partial_{t}\,.
\end{equation}
The above also implies that only the cubic term of the scale factor contributes to the $\sqrt{-g}$. Taking all the relevant aspects into account enables the d'Alembert operator \eqref{Box-homogeneous-SF} to be calculated directly in the case of the FLRW metric:
\begin{equation}
    \Box\phi=a^{-3}\partial_{t}\left(-a^{3}\,\dot{\phi}\right)=-a^{-3}\left(3a^{2}\,\dot{a}\,\dot{\phi}+a^{3}\,\ddot{\phi}\right)=-\left(\ddot{\phi}+3H\,\dot{\phi}\right)\,.
\end{equation}
Under the assumption that the matter content of the universe is described by the \textbf{perfect fluid}\footnote{No heat flow in the fluid rest frame (no energy flux orthogonal to $u^{\mu}$) and no anisotropic stresses (pressure is the same in all spatial directions in the rest frame).} stress-energy tensor:
\begin{equation}\label{perfect-fluid}
    T_{\mu\nu}^{(\mathrm{tot})}=\left(\rho_{\mathrm{tot}}+p_{\mathrm{tot}}\right)\,u_{\mu}u_{\nu}+p_{\mathrm{tot}}\,g_{\mu\nu}\quad\land\quad u^{\mu}u_{\mu}=-1
\end{equation}
with the trace:
\begin{equation}
    T_{(\mathrm{tot})}\equiv g^{\mu\nu}\,T_{\mu\nu}=-\rho_{\mathrm{tot}}+3p_{\mathrm{tot}}\,,
\end{equation}
allows one to obtain the \textbf{final form of the scalar sector BD theory EoM} (\textit{Klein-Gordon equation in Brans-Dicke theory}) \eqref{BD-equation-SF}:
\begin{equation}\label{BD-final-EoM-SF}
    \boxed{\ddot{\phi}+3H\dot{\phi}=\frac{1}{2\omega_{\mathrm{BD}}+3}\Bigl[8\pi\left(\rho_{\mathrm{tot}}-3p_{\mathrm{tot}}\right)+\phi\,V'(\phi)-2V(\phi)\Bigr]}\,.
\end{equation}
that can be further simplified by assuming a \textbf{barotropic\footnote{Barotropic fluids are fluids whose (energy) density depends only on pressure.} equation of state}\footnote{For radiation (relativistic matter) the EoS parameter takes the value of $\omega_{r}=\frac{1}{3}$, for non-relativistic matter (dark and baryonic) $\omega_{m}=0$, and for the cosmological constant $\omega_{\Lambda}=-1$.} (EoS) for the matter content of the Universe:
\begin{equation}\label{barotropic-EoS}
    p_{i}\equiv\omega_{i}\rho_{i}\quad\land\quad \omega_{i}=\mathrm{const}
\end{equation}
into the following:
\begin{equation}
    \ddot{\phi}+3H\dot{\phi}=\frac{1}{2\omega_{\mathrm{BD}}+3}\left[8\pi\left(\sum_{i}\rho_{i}(1-3\omega_{i})\right)+\phi\,V'(\phi)-2V(\phi)\right]\,.
\end{equation}
Using the barotropic EoS \eqref{barotropic-EoS} in the principle of covariant conservation of the energy-momentum tensor enables the explicit calculation of the dependence of energy density on the scale factor for the individual components of the Universe's energy balance:
\begin{equation}
    \nabla^{\nu}T_{\mu\nu}^{(i)}=0\quad\implies\quad \dot{\rho}_{i}+3H\rho_{i}\left(1+\omega_{i}\right)=0\,,
\end{equation}
which yields the following relation for the energy densities:
\begin{equation}
    \rho_{i}(a)=\rho_{i,0}\,a^{-3\left(1+\omega_{i}\right)}\quad\land\quad \rho_{i}(a=1)=\rho_{i,0}=\mathrm{const}\,.
\end{equation}
\subsubsection{Metric sector}
In order to obtain the 1st Friedmann equation we need to take into account the following $0-0$ and $0$ components, respectively:
\begin{equation}
    G_{00}=3\left(H^{2}+\frac{k}{a^{2}}\right)\quad\land\quad T_{00}=\rho_{\mathrm{tot}}\quad\land\quad \nabla_{0}\phi=\dot{\phi}\,.
\end{equation}
Using \eqref{BD-general-EoM} and performing some simple calculations, we obtain the following formula:
\begin{equation}
    \boxed{H^{2}=\frac{8\pi}{3\phi}\rho_{\mathrm{tot}}+\frac{\omega_{\mathrm{BD}}}{6}\left(\frac{\dot{\phi}}{\phi}\right)^{2}-H\left(\frac{\dot{\phi}}{\phi}\right)+\frac{V(\phi)}{6\phi}-\frac{k}{a^{2}}}\,.
\end{equation}
In the case of the $i-i$ and $i$ components, one obtains:
\begin{equation}
    G_{ii}=-\left(2\dot{H}+3H^{2}+\frac{k}{a^{2}}\right)\,a^{2}\quad\land\quad T_{ii}^{(\mathrm{tot})}=a^{2}\,p_{(\mathrm{tot})}\quad\land\quad \nabla_{i}\phi=0
\end{equation}
and therefore:
\begin{equation}
    2\dot{H}+3H^{2}=-\left[\frac{8\pi}{\phi}p_{(\mathrm{tot})}+\frac{\omega_{\mathrm{BD}}}{2}\left(\frac{\dot{\phi}}{\phi}\right)^{2}+\frac{1}{\phi}\left(\ddot{\phi}+2H\dot{\phi}\right)\right]+\frac{V(\phi)}{2\phi}-\frac{k}{a^{2}}\,.
\end{equation}
Furthermore, using the equation of motion \eqref{BD-final-EoM-SF} for the BD SF and the fact that:
\begin{equation}
    \ddot{\phi}+2H\dot{\phi}=\ddot{\phi}+3H\dot{\phi}-H\dot{\phi}
\end{equation}
yields the \textbf{final form of the EoM for the metric sector in the BD theory}:
\begin{equation}\label{BD-final-EoM-metric}
    \boxed{\begin{aligned}
    \dot{H}= & -\frac{8\pi}{\left(2\omega_{\mathrm{BD}}+3\right)\phi}\Bigl[\left(\omega_{\mathrm{BD}}+2\right)\rho_{\mathrm{tot}}+\omega_{\mathrm{BD}}\,p_{\mathrm{tot}}\Bigr]-\frac{\omega_{\mathrm{BD}}}{2}\left(\frac{\dot{\phi}}{\phi}\right)^{2}+2H\left(\frac{\dot{\phi}}{\phi}\right) \\
    & +\frac{1}{2\left(2\omega_{\mathrm{BD}}+3\right)\phi}\Bigl[\phi\,V'(\phi)-2V(\phi)\Bigr]+\frac{k}{a^{2}}
    \end{aligned}}
\end{equation}
In a manner analogous to the Klein-Gordon equation \eqref{BD-final-EoM-SF}, the field equation \eqref{BD-final-EoM-metric} can be simplified by means of the EoS \eqref{barotropic-EoS}, which consequently leads to:
\begin{equation}
    \begin{aligned}
    \dot{H}= & -\frac{8\pi}{\left(2\omega_{\mathrm{BD}}+3\right)\phi}\Biggl\{\sum_{i}\Bigl[\omega_{\mathrm{BD}}\left(1+\omega_{i}\right)+2\Bigr]\rho_{i}\Biggr\}-\frac{\omega_{\mathrm{BD}}}{2}\left(\frac{\dot{\phi}}{\phi}\right)^{2}+2H\left(\frac{\dot{\phi}}{\phi}\right) \\
    & +\frac{1}{2\left(2\omega_{\mathrm{BD}}+3\right)\phi}\Bigl[\phi\,V'(\phi)-2V(\phi)\Bigr]+\frac{k}{a^{2}}
    \end{aligned}
\end{equation}
By introducing the so-called \textbf{total effective EoS parameter}:
\begin{equation}\label{effective-EoS}
    \omega_{\mathrm{tot}}\equiv\frac{p_{\mathrm{tot}}}{\rho_{\mathrm{tot}}}=\frac{\sum_{i}\omega_{i}\rho_{i}}{\sum_{i}\rho_{i}}\quad\implies\quad p_{\mathrm{tot}}\equiv\omega_{\mathrm{tot}}\rho_{\mathrm{tot}}
\end{equation}
the equations of motion \eqref{BD-final-EoM-SF} and \eqref{BD-final-EoM-metric} can be expressed as:
\begin{equation}
    \boxed{\ddot{\phi}+3H\dot{\phi}=\frac{1}{2\omega_{\mathrm{BD}}+3}\Bigl[8\pi\left(1-3\omega_{\mathrm{tot}}\right)\rho_{\mathrm{tot}}+\phi\,V'(\phi)-2V(\phi)\Bigr]}
\end{equation}
and:
\begin{equation}
    \boxed{\begin{aligned}
    \dot{H}= & -\frac{8\pi}{\left(2\omega_{\mathrm{BD}}+3\right)\phi}\Bigl[\omega_{\mathrm{BD}}\left(1+\omega_{\mathrm{tot}}\right)+2\Bigr]\rho_{\mathrm{tot}}-\frac{\omega_{\mathrm{BD}}}{2}\left(\frac{\dot{\phi}}{\phi}\right)^{2}+2H\left(\frac{\dot{\phi}}{\phi}\right) \\
    & +\frac{1}{2\left(2\omega_{\mathrm{BD}}+3\right)\phi}\Bigl[\phi\,V'(\phi)-2V(\phi)\Bigr]+\frac{k}{a^{2}}
    \end{aligned}}\,,
\end{equation}
respectively.
\begin{examplebox}[Massive or free Brans-Dicke SF without ordinary matter (with or without radiation) in the spatially flat Universe]
\begin{itemize}
    \item SF potential:
    \begin{equation}
        V(\phi)=\frac{1}{2}m^{2}\phi^{2}\quad\lor\quad V(\phi)=0
    \end{equation}
    \item EoM for the scalar sector:
    \begin{equation}\label{BD-KG-vanishing}
        \ddot{\phi}+3H\dot{\phi}=0
    \end{equation}
    $\implies$ Exactly the same equation as in the \textbf{vacuum case}.
    \item In fact, the \textit{\textbf{BD scalar only couples to the trace of the matter energy–momentum tensor}}:
    \begin{equation}
        T_{(\mathrm{i})}=-\rho_{i}+3p_{i}=\rho_{i}\left(3\omega_{i}-1\right)
    \end{equation}
    that vanishes in the case of radiation ($\omega_{r}=\frac{1}{3}$):
    \begin{equation}
        T_{(\mathrm{r})}=0\,.
    \end{equation}
    \item Therefore, it can be concluded that the BD SF:
    \begin{enumerate}
        \item \textbf{Is not sourced by radiation (relativistic matter)};
        \item Evolves only through the \textbf{Hubble friction} and its own \textbf{initial conditions}.
    \end{enumerate}
    \item Important subclass of solutions:
    \begin{equation}
        \dot{\phi}=0\quad\implies\quad \phi(t)=\phi_{0}=\mathrm{const}\implies\quad \dot{H}=-\frac{16\pi}{3\phi_{0}}\rho_{\mathrm{r}}
    \end{equation}
    $\implies$ \textit{\textbf{BD theory EoM reduces to the usual GR field equations (FEs) with constant gravitational coupling}}:
    \begin{equation}
        G_{\mathrm{eff}}\equiv\frac{1}{\phi_{0}}\,.
    \end{equation}
    \item In such a case, one obtains the standard \textbf{scaling solutions} for the \textbf{radiation-domination (RD) era}:
    \begin{equation}\label{scaling-RD}
        a(t)\propto\sqrt{t}\quad\land\quad H(t)\propto\frac{1}{2t}
    \end{equation}
    $\implies$ \textit{\textbf{The radiation-dominated FLRW background is indistinguishable from GR at the homogeneous background level, with the BD scalar remaining at a constant value.}}\footnote{This was first demonstrated in the work of \textit{\textbf{Hidekazu Nariai}} \cite{Nariai:1968ncy}.}
    \item Multiplying \eqref{BD-KG-vanishing} by the factor $a^{3}$ and noting that the resulting formula is equivalent to the following derivative:
    \begin{equation}
        \frac{d}{dt}\left(a^3\,\dot{\phi}\right)=3a^{2}\,\dot{a}\,\dot{\phi}+a^3\,\ddot{\phi}=3a^3 H\dot{\phi}+a^3\,\ddot{\phi}
    \end{equation}
    one could obtain \textbf{another class of solutions} characterized by the following condition:
    \begin{equation}
        \frac{d}{dt}\left(a^3\,\dot{\phi}\right)=0\quad\implies\quad \dot{\phi}=C_{1}a^{-3}\,.
    \end{equation}
    \item Using the scaling solution \eqref{scaling-RD} for RD epoch one obtains:
    \begin{equation}
        \dot{\phi}(t)=C_{1}\,t^{-3/2}\quad\implies\quad \phi(t)=C_{1}\int{t^{-3/2}dt}=C_{2}-\frac{2C_{1}}{\sqrt{t}}\,,
    \end{equation}
    where the \textbf{BD SF asymptotically approaches a constant value}:
    \begin{equation}
        \lim_{t->{}^{+}\infty}{\phi(t)}=C_{2}\equiv\phi_{\infty}=\mathrm{const}
    \end{equation}
    and therefore the final relationship is:
    \begin{equation}
        \phi(t)=\phi_{\infty}-\frac{2C_{1}}{\sqrt{t}}
    \end{equation}
    $\equiv$ \textbf{'decaying BD mode' on top of the GR-like radiation background}.
\end{itemize}
\textit{\textbf{\underline{Conclusion:}}}\\
Even when the \textbf{BD field is dynamical}, \textbf{radiation does not act as a source}, it only \textbf{damps the time derivative of the BD field via expansion}.
\end{examplebox}
\subsubsection{Interpretation of the Brans-Dicke scalar field}
The interpretation of the scalar field in BD theory is a matter of some controversy. This is attributed to the fact that this field can be considered from two perspectives:
\begin{enumerate}[(a)]
    \item As an another \textit{\textbf{cosmological fluid}} (part of the \textbf{matter sector}) \cite{Clifton:2011jh,Nesseris:2022hhc};
    \item As a part of the \textit{\textbf{gravitational field}} (constituent of the \textbf{gravitational sector} and \textbf{not a form of matter}) \cite{Faraoni:2004pi,Fujii:2003pa}.
\end{enumerate}
It is important to acknowledge that this issue is intimately connected to the notion of \textbf{conformal frames} and their dual mathematical and physical interpretations (see Sec.~\ref{EJframes-section} for further details). In his book \cite{Faraoni:2004pi}, Faraoni highlights that this phenomenology (interpretation) of the scalar field, i.e. the connection (or lack thereof) between BD SF and the matter sector, distinguishes the interpretation of FEs in BD theory as \textbf{effective Einstein field equations (GR)} from the recognition in the FEs of scalar-tensor a theory that is \textbf{different from Einstein's theory}.

To a certain extent, this is a matter of phenomenology (or perhaps even the philosophy of physics and cosmology \cite{Ellis:2006fy,Rovelli:2007uwt,Batterman2013,Routledge2021,Baerdemaeker_2025}), whereby one must select between modifications in the geometry of spacetime (the gravitational sector) or modifications in the matter content of the Universe (the right-hand side of the field equations). \textit{\textbf{The crucial question is whether these approaches are equivalent}} $\ldots$

\subsubsection{Spatially flat vacuum model of Brans-Dicke cosmology}
\begin{examplebox}[Duality symmetry]
    \begin{itemize}
        \item Vacuum stress-energy tensor:
        \begin{equation}
            T_{\mu\nu}^{(\mathrm{m})}=0\,.
        \end{equation}
        \item For vanishing or linear SF potential (CC term in the dilaton action):
        \begin{equation}
            V(\phi)=0\quad\lor\quad V(\phi)=\Lambda\phi\quad\land\quad \Lambda=\mathrm{const}\,,
        \end{equation}
        the vacuum EoM for BD theory become:
        \begin{subequations}\label{Vacuum-BD-EoM}
            \begin{align}
                \dot{H} &= -\frac{\omega_{\mathrm{BD}}}{2}\left(\frac{\dot{\phi}}{\phi}\right)^{2}+2H\left(\frac{\dot{\phi}}{\phi}\right)\,, \\
                \ddot{\phi} & =-3H\dot{\phi}
            \end{align}
        \end{subequations}
        and:
        \begin{subequations}\label{Linear-BD-EoM}
            \begin{align}
                \dot{H} &= -\left[\frac{\Lambda}{2\left(3+2\omega_{\mathrm{BD}}\right)}+\frac{\omega_{\mathrm{BD}}}{2}\left(\frac{\dot{\phi}}{\phi}\right)^{2}\right]+2H\left(\frac{\dot{\phi}}{\phi}\right)\,, \\
                \ddot{\phi} & =-\left(\frac{\Lambda\phi}{3+2\omega_{\mathrm{BD}}}+3H\dot{\phi}\right)\,,
            \end{align}
        \end{subequations}
        respectively.
        \item Taking into account the following redefinitions \cite{Lidsey:1995ft,Lidsey:1996yf}:
        \begin{equation}
            \alpha\equiv\ln{a}\quad\land\quad \Phi\equiv -\ln{\left(G\phi\right)}
        \end{equation}
        in \eqref{Vacuum-BD-EoM} and \eqref{Linear-BD-EoM} leads to the \textbf{duality transformation} (\textbf{\textit{duality symmetry}}) of the form \cite{Lidsey:1995ft,Lidsey:1996yf}:
        \begin{subequations}\label{Duality-transformation}
            \begin{align}
                \alpha & \mapsto \left(1-\frac{2}{3\omega_{\mathrm{BD}}+4}\right)\,\alpha-2\left(\frac{\omega_{\mathrm{BD}}+1}{3\omega_{\mathrm{BD}}+4}\right)\,\Phi \,, \\
                \Phi & \mapsto -\left(\frac{6}{3\omega_{\mathrm{BD}}+4}\right)\,\alpha+\left(\frac{2}{3\omega_{\mathrm{BD}}+4}-1\right)\,\Phi \,.
            \end{align}
        \end{subequations}
        \item Furthermore, the transformation \eqref{Duality-transformation} can be interpreted as a \textbf{generalization of the scale factor duality} that is characteristic of the \textbf{effective action of string theory} \cite{Gasperini:1991ak,Giveon:1994fu,Tseytlin:1991xk,Giveon:1991jj}:
        \begin{equation}\label{string-duality}
            \alpha\mapsto -\alpha \quad\land\quad \Phi\mapsto \Phi-6\alpha\,.
        \end{equation}
        \item Essentially, taking into consideration:
        \begin{equation}
            \omega_{\mathrm{BD}}=-1
        \end{equation}
        in duality transformation \eqref{Duality-transformation} yields the string duality \eqref{string-duality}.
    \end{itemize}
\end{examplebox}
\subsection{GR limit}
As mentioned previously, it is widely accepted that the \textbf{Brans-Dicke theory is indistinguishable from Albert Einstein's general theory of relativity} in the limit of \cite{Weinberg:1972kfs,Will_2018}:
\begin{equation}\label{omegaBD-limit}
    \omega_{\mathrm{BD}}\to\infty\,.
\end{equation}
However, it should be noted that this is \textit{\textbf{only justifiable in the majority of cases, and not, as a general rule}}. It has been demonstrated that there exist exact solutions for BD theory \cite{Matsuda:1972zp,Romero_Barros_1992,Romero:1992ci,Romero:1992bu,Paiva:1993qa,Scheel:1994yn,Anchordoqui:1997du} that \textit{\textbf{do not reduce to Einstein field equations solutions}} in the limit specified in \eqref{omegaBD-limit}. This phenomenon, named the \textit{\textbf{"anomalous" asymptotic behavior}}, has been observed in the context of solutions that incorporate the \textit{\textbf{traceless matter energy-momentum tensor}} \cite{Banerjee:1996iy} for which:
\begin{subequations}\label{anomalous-BD}
            \begin{align}
                G_{\mu\nu} & =\frac{\omega_{\mathrm{BD}}}{\phi^{2}}\left(\nabla_{\mu}\phi\,\nabla_{\nu}\phi-\frac{1}{2}g_{\mu\nu}\nabla_{\rho}\phi\,\nabla^{\rho}\phi\right)+\frac{1}{\phi}\Bigl(\nabla_{\mu}\nabla_{\nu}\phi-g_{\mu\nu}\,\Box\phi\Bigr) \,, \\
                \Box\phi & =0 \,.
            \end{align}
\end{subequations}
In point of fact, the relations \eqref{anomalous-BD} have been shown to possess \textbf{exact static spherical solutions} - \textit{\textbf{class I (exterior vacuum)}} \cite{Brans:1962zz,Bhadra:2005mc}.
\begin{examplebox}[Class I (exterior vacuum) solutions]
\begin{itemize}
    \item The class I solutions:
\begin{subequations}\label{classI-exterior-vacuum}
            \begin{align}
                ds^{2} & = -\left(\frac{r-k}{r+k}\right)^{2/\lambda}\,dt^{2}+\left(1+\frac{k}{r}\right)^{4}\,\left(\frac{r-k}{r+k}\right)^{-2\left(1+C\right)/\lambda} \,, \\
                \phi(r) & = \phi_{0}\,\left(\frac{r-k}{r+k}\right)^{C/\lambda} \,,
            \end{align}
\end{subequations}
where:
\begin{subequations}\label{constants-classI-vacuum}
            \begin{align}
                k & = \mathrm{const}>0\quad\text{(length scale)} \,, \\
                C & = \mathrm{const}\quad\text{(integration constant - scalar charge)} \,,
            \end{align}
\end{subequations}
and\footnote{For all $\omega_{\mathrm{BD}}>-\frac{3}{2}$.}:
\begin{equation}\label{lambda-squared-classI}
    \lambda^{2}=\left(C+1\right)^{2}-C\left(1-\frac{1}{2}\,\omega_{\mathrm{BD}}\,C\right)=\frac{1}{2}C^{2}\left(\omega_{\mathrm{BD}}+2\right)+C+1\,.
\end{equation}
\item Within the astrophysics context \cite{Horbatsch:2010hj}, the value of the scalar charge $C$ is determined through the process of \textbf{matching to the regular interior solution} (no a priori $\omega_{\mathrm{BD}}$-dependence) \cite{Bhadra:2005mc,Bhadra:2001bda}.
\item Using \eqref{constants-classI-vacuum}, one can conclude that the dominant term in the relation \eqref{lambda-squared-classI} for $\lambda$ behaves as follows:
\begin{equation}\label{lambda-classI-approximation}
    \lambda=\left|C\right|\sqrt{\frac{\omega_{\mathrm{BD}}}{2}}\left[1+\mathcal{O}\left(\frac{1}{\omega_{\mathrm{BD}}}\right)\right]\,,
\end{equation}
and therefore, the ratio of parameters become:
\begin{equation}\label{ratio-classI-approximate}
    \frac{C}{\lambda}=\sgn{\left(C\right)}\sqrt{\frac{2}{\omega_{\mathrm{BD}}}}\left[1+\mathcal{O}\left(\frac{1}{\omega_{\mathrm{BD}}}\right)\right]\,.
\end{equation}
\item Inserting \eqref{lambda-classI-approximation} and the ratio \eqref{ratio-classI-approximate} into the explicit exterior vacuum solution \eqref{classI-exterior-vacuum}, and Taylor expanding the exponential gives:
\begin{equation}
    \begin{aligned}
        \phi(r) & =\phi_{0}\,\exp{\left[\frac{C}{\lambda}\ln{\left(\frac{r-k}{r+k}\right)}\right]} \\
        & = \phi_{0}\left[1+\frac{C}{\lambda}\ln{\left(\frac{r-k}{r+k}\right)}+\mathcal{O}\left(\frac{C^{2}}{\lambda^{2}}\right)\right]\,.
    \end{aligned}
\end{equation}
\item Due to the fact that the ratio behaves as:
\begin{equation}
    \frac{C}{\lambda}\sim\mathcal{O}\left(\frac{1}{\sqrt{\omega_{\mathrm{BD}}}}\right)
\end{equation}
while the logarithm is $\mathcal{O}(1)$ for any finite $r>k$, one concludes that the BD SF behaves asymptotically as follows:
\begin{equation}
    \lim_{\omega_{\mathrm{BD}}\to{}^{+}\infty}{\phi(r)}=\phi_{0}+\mathcal{O}\left(\frac{1}{\sqrt{\omega_{\mathrm{BD}}}}\right)
\end{equation}
$\implies$ \textit{\textbf{The deviation from a constant BD scalar dies off as}} $\boldsymbol{\frac{1}{\sqrt{\omega_{\mathrm{BD}}}}}$ \textit{\textbf{rather than}} $\boldsymbol{\frac{1}{\omega_{\mathrm{BD}}}}$.
\end{itemize}
\textbf{\textit{\underline{Conclusion:}}}\\
In the context of this particular class of models, the so-called "\textit{\textbf{large Brans-Dicke parameter anomaly}}" \cite{Faraoni:1998yq} is of relevance. This anomaly is characterized by the fact that, while the \textbf{kinetic energy of the SF is finite}, the \textbf{metric does not smoothly approach the Schwarzschild solution}.
\end{examplebox}

\subsection{Connection to the Kaluza-Klein theory}\label{BDandKK-section}
\begin{figure}[h]
    \centering
    \includegraphics[width=1\linewidth]{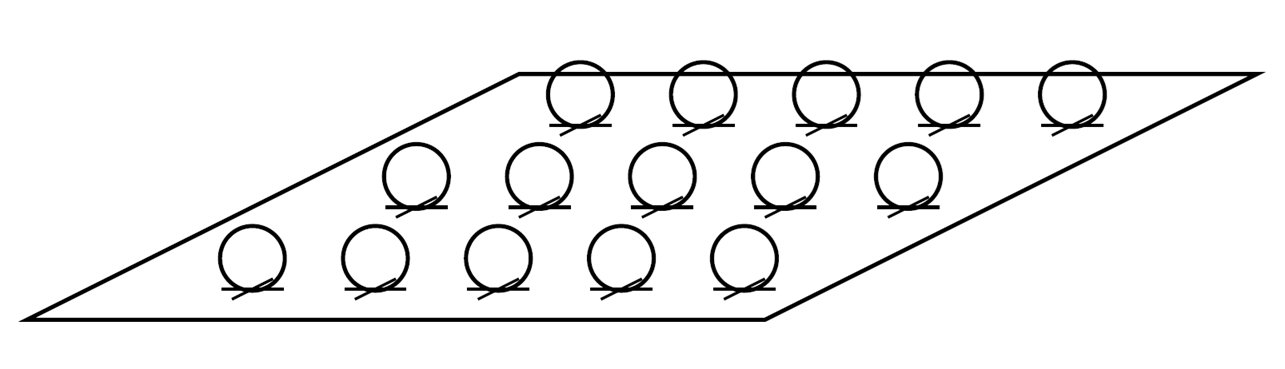}
    \caption{In KK theory, every point in spacetime is assumed to have a small extra dimension \cite{Gron2007}.}
    \label{fig:KK-figure}
\end{figure}

The \textbf{Kaluza}\footnote{Interesting fact: Theodor Kaluza's family originates from the same town as the author of this dissertation -\textbf{ Racibórz}. At the time of Kaluza's birth, the city (then called \textit{Ratibor}) belonged to the \textit{German Empire's Prussian Province of Silesia}, while today it is a city in the \textit{Silesian Voivodeship in Poland}.}\textbf{-Klein theory}\footnote{From a historical point of view, it is worth mentioning that in \textbf{1914} (before the publication of Einstein's general relativity), \textit{\textbf{Gunnar Nordström}} proposed in his article \cite{Nordstrom:1914ejq} a 5D scalar-tensor theory attempting to unify gravity with electromagnetism. Then, in \textbf{1919}, \textit{\textbf{Theodor Kaluza}} developed a theory that unified electromagnetism and gravity by applying linearized GR field equations \cite{Kaluza1921}. It was finally completed in \textbf{1926} by \textit{\textbf{Oskar Klein}} \cite{Klein1926}.} (KK theory) \cite{Kaluza1921,Klein1926} is an attempt to develop a unified (classical) field theory of gravitation and electromagnetism, based on the concept of an extra spatial dimension beyond the four dimensions of GR. The proposal states that there are three spatial dimensions and one dimension of time, as well as an additional spatial dimension shaped in the form of a tiny circle (see, Fig.~\ref{fig:KK-figure}).

It is interesting to note that BD theory can be derived from the field equations of Kaluza-Klein theory. This has been proven in many modern studies that discuss KK theory in detail \cite{Gron2007,Overduin:1997sri,Bailin:1987jd,Duff:1994tn,Wesson1999,Fujii_Maeda_2003KK,Plebanski_Krasinski_2024}. In fact, BD SF is mathematically (geometrically) related to the \textbf{determinant of the metric} defined on the manifold that describes \textbf{additional spatial dimensions}.

In the most elementary version of classical KK theory with a single \textbf{dilaton} field (SF), one has the following spacetime:
\begin{equation}\label{KK-spacetime}
    \Bigl(\mathcal{M}\otimes\mathcal{K},\,\hat{g}_{AB}\Bigr)\,,
\end{equation}
where:
\begin{itemize}
    \item $\mathcal{M}$ is the $4D$ manifold with one time-like dimension;
    \item $\mathcal{K}$ is the submanifold with $d\geq 1$ spatial dimensions;
\end{itemize}
and $(4+d)$-dimensional quantities are denoted by a hat (e.g., $\hat{g}_{AB}$).\\
In the case under consideration, the $(4+d)$-dimensional metric can be expressed as follows \cite{Faraoni2004chapter}:
\begin{equation}
    \Bigl(\hat{g}_{AB}\Bigr)\equiv \left(\begin{array}{cc}
\hat{g}_{ab} & 0 \\
0 & \hat{\phi}_{\alpha\beta}
\end{array}\right)\,,
\end{equation}
where:
\begin{subequations}\label{KK-metric-indices}
  \begin{alignat}{2}
    A,B         & = 0,1,\ldots,3+d   &\quad& \text{(full metric indices)} \,, \\
    a,b         & = 0,1,2,3          &\quad& \text{(indices associated with $\mathcal{M}$)} \,, \\
    \alpha,\beta& = 4,5,\ldots,3+d   &\quad& \text{(indices associated with $\mathcal{K}$)} \,.
  \end{alignat}
\end{subequations}
The original proposal by Kaluza and Klein can be seen to differ significantly at first glance. In a purely gravitational approach (\textit{\textbf{Kaluza-Klein gravity/cosmology}}) that takes into account additional dimensions \textbf{without unifying with Maxwell's equations}, there are \textbf{no off-diagonal terms} in the metric. which were present in the \textbf{original proposal} (for $d=1$) \cite{Kaluza1921,Klein1926,Plebanski_Krasinski_2024,Overduin:1997sri}:
\begin{equation}\label{KK-original-metric}
  \Bigl(\hat{g}_{AB}^{\mathrm{(KK)}}\Bigr)
  =
  \begin{pmatrix}
    \hat{g}_{ab}+\Phi\,A_{a}A_{b} & \Phi\,A_{a} \\
    \Phi\,A_{b} & \Phi
  \end{pmatrix}\,,
\end{equation}
where:
\begin{equation}
    \Phi=\hat{\phi}_{44}
\end{equation}
parametrizes the \textbf{size} of the compact extra dimension:
\begin{equation}
    \mathcal{K}\simeq S^{1}\,,
\end{equation}
and $A_{a}$ is responsible for the \textbf{'tilt'} of the extra dimension:
\begin{equation}
    A_{a}\equiv\frac{\hat{g}_{a4}}{\hat{g}_{44}}=\frac{\hat{g}_{a4}}{\Phi}\,,
\end{equation}
which in classical electrodynamics reduces to the well-known form for the \textbf{electromagnetic 4-potential}:
\begin{equation}
    A_{a}=\left(-\Phi,\vec{A}\right)
\end{equation}
with the electric (scalar) potential $\Phi$ and the magnetic (vector) potential $\vec{A}$.

\subsubsection{Kaluza-Klein cosmology}
From the above formalism, it follows that $\hat{g}_{ab}$ is the FLRW metric defined on the manifold $\mathcal{M}$ and $\hat{\phi}_{ab}$ is the Riemannian metric defined on the submanifold $\mathcal{K}$.\\
Firstly, let us consider the vacuum FEs of GR in the $(4+d)$-dimensional case:
\begin{equation}\label{KK-action-cosmology}
    S_{\mathrm{KK}}=\int d^{(4+d)}x\,\sqrt{-\hat{g}}\,\mathcal{L}^{(4+d)}=\frac{1}{16\pi\hat{G}}\int d^{(4+d)}x\,\sqrt{-\hat{g}}\,\left(\hat{R}+\hat{\Lambda}\right)\,,
\end{equation}
where:
\begin{equation}
    \hat{g}\equiv\det{\left(\hat{g}_{AB}\right)}\,,
\end{equation}
and:
\begin{itemize}
    \item $\hat{R}$ is the Ricci curvature associated with the metric $\hat{g}_{AB}$,
    \item $\hat{\Lambda}$ is the $\left(4+d\right)$-dimensional CC,
    \item $\hat{G}$ is the gravitational coupling.
\end{itemize}
In order to consider a regime (e.g. inflation) in which the dynamics of the universe is dominated by a single SF (Kaluza-Klein scalar obtained from pure geometry), the higher-dimensional energy-momentum tensor describing higher-dimensional matter (different from $\hat{\Lambda}$) is assumed to vanish.\\
We can also introduce the extra dimensions metric determinant:
\begin{equation}
    \varphi\equiv\left|\det{\left(\hat{\phi}_{\alpha\beta}\right)}\right|
\end{equation}
together with the following symmetric tensor of the form:
\begin{equation}
    \rho_{\alpha\beta}\equiv \varphi^{-1/d}\,\phi_{\alpha\beta}\,,
\end{equation}
for which by the definition:
\begin{equation}
    \Bigl|\det{\left(\rho_{\alpha\beta}\right)}\Bigr|=1
\end{equation}
condition is satisfied.

The hypothesis is postulated that \textit{\textbf{extra spatial dimensions are curled up on a microscopic scale}} of size $\ell$; therefore, the integral over the $(4+d)$ dimensions in equation \eqref{KK-action-cosmology} is split into the product of an integral over the 4 spacetime dimensions and of an integral over the remaining $d$ dimensions. The final integration yields the following result \cite{Faraoni2004chapter}:
\begin{equation}\label{KK-action-cosmology-reduced}
    S_{\mathrm{KK}}^{(\mathrm{reduced})}=\frac{V^{(\ell)}}{16\pi \hat{G}}\int d^{4}x\,\sqrt{-g}\,\sqrt{\varphi}\left[\left(R+R_{\mathcal{K}}+\hat{\Lambda}\right)+\frac{(d-1)}{4d}g^{ab}\frac{\nabla_{a}\varphi\nabla_{b}\varphi}{\varphi^{2}}\right]\,,
\end{equation}
where:
\begin{itemize}
    \item $V^{(\ell)}$ is the volume of the compact manifold $\mathcal{K}$,
    \item $R_{\mathcal{K}}$ is the Ricci curvature of the manifold $\mathcal{K}$.
\end{itemize}
Then, by defining a new scalar field $\phi$ of the form:
\begin{equation}
    \phi\equiv\sqrt{\varphi}
\end{equation}
and rescaling the coupling constant by the compact manifold volume:
\begin{equation}
    G\equiv\frac{\hat{G}}{V^{(\ell)}}
\end{equation}
yields the final form of the action for the \textit{\textbf{vacuum Kaluza-Klein cosmology}}\footnote{In such a case the BD-like SF $\phi$ is a dimensionless quantity.} (with CC) \cite{Faraoni2004chapter}:
\begin{equation}\label{KK-action-cosmology-final}
    \boxed{S_{\mathrm{KK}}^{(\mathrm{BD})}=\frac{1}{16\pi G}\int d^{4}x\,\sqrt{-g}\left[\phi\left(R+R_{\mathcal{K}}+\hat{\Lambda}\right)+\frac{(d-1)}{d}g^{ab}\frac{\nabla_{a}\phi\nabla_{b}\phi}{\phi}\right]}\,.
\end{equation}
The above action describes the \textit{\textbf{BD theory}}, in which the BD parameter takes the following explicit form:
    \begin{equation}\label{KK-omega-BD-general}
        \omega_{\mathrm{BD}}=-\frac{(d-1)}{d}\,.
    \end{equation}
\begin{examplebox}[The possible implications of a 5th extra dimension from Kaluza-Klein theory]
    \begin{itemize}
        \item The \textbf{Kaluza–Klein theory} may provide a \textbf{geometric justification} for the existence of a \textbf{cosmological SF}. In this case, the field would originate from a \textbf{\textit{higher-dimensional (vacuum) GR (cosmology)}};
        \item \textbf{Gravitation} and \textbf{electromagnetism} are reduced to \textit{\textbf{geometric properties of a 5D spacetime}};
        \item The 5D world is \textit{\textbf{neutral}} and \textit{\textbf{free from electromagnetic fields}} \cite{Gron2007};
        \item The \textbf{nature of the 5D SF} can be identified by observing that what is perceived as \textbf{charge} is the \textbf{\textit{motion of a neutral particle around a closed fifth dimension}} which generates an \textbf{\textit{inertial dragging field}} \cite{Gron2007};
        \item The \textbf{Coulomb field} could be the \textbf{projection of the inertial dragging field} into 4D spacetime \cite{Gron:1986yd};
        \item There would be \textbf{no electromagnetic fields} if \textbf{gravity} is correctly described by a \textbf{Newtonian theory}, which involves \textbf{no inertial dragging field} \cite{Gron2007};
        \item From a \textit{5D perspective}, \textbf{electromagnetism} could be a \textbf{gravitational effect that disappears in the Newtonian limit} \cite{Gron2007}.
    \end{itemize}
\end{examplebox}

\section{Scalar-tensor theories}
Scalar-tensor theories of gravity \cite{Bergmann:1968ve,Nordtvedt:1970uv,Wagoner:1970vr,Damour:1992kf,Esposito-Farese:2000pbo,Capozziello:2011et,Shankaranarayanan:2022wbx,Goenner:2012cq,Quiros:2019ktw,Fujii:2003pa,Faraoni:2004pi,Clifton:2011jh} represent a highly extensive class of modified gravity models. The relatively simple structure of these theories, which is usually the result of the presence of an additional scalar field (or fields), allows for an effective description of many modifications of general relativity resulting from much more complex theories/hypotheses (e.g., LQG, string theory, models beyond the Standard Model etc.). Thanks to the 'useful' structure of mathematical formalism, this is one of the most widely studied and fastest-growing areas of theoretical physics.
\subsection{History of scalar-tensor theories}
In its simplest form, modern scalar-tensor gravity can be characterized by a metric $g_{\mu\nu}$ supplemented by a scalar field $\phi$, which changes the effective strength and/or composition of gravitational interactions. Goenner’s historical analysis\footnote{More details on the chronology of the development of scalar-tensor theories of gravity can also be found in \cite{Brans:2005ra}.} \cite{Goenner:2012cq} emphasizes that the \textbf{emergence of STTs} was not a single linear progression, but rather a \textbf{convergence of several research frameworks} shaped by conceptual motivations and scientific sociology, including communication barriers, publication venues and post-war scientific networks. The summary can be found in Table~\ref{tab:STTs_timeline}.

\begin{table}[!htbp]
\centering
\caption{A selection of milestones in the development of ST gravity (cosmology), along with an explanation of their relevance.}
\label{tab:STTs_timeline}
\small
\setlength{\tabcolsep}{6.2pt}
\renewcommand{\arraystretch}{1.28}

\rowcolors{2}{cyan!6}{white}
\begin{tabularx}{\linewidth}{p{2.05cm}p{3.2cm}XX}
\hline
\rowcolor{cyan!15}
\textbf{Year(s)} & \textbf{Author(s)} & \textbf{Conceptual contribution} & \textbf{Relevance for later STTs} \\
\hline
1921--1926 &
Kaluza \cite{Kaluza1921}; Klein \cite{Klein1926} &
$5D$ unification: $g_{\mu\nu}$ and a vector field are present in the higher-dimensional metric, along with a scalar sector (after compactification). &
Provides a geometric origin of scalar companions to gravity and is later a key structural ingredient in Jordan-Thiry-type approaches \cite{Goenner:2012cq}. \\

1937--1938 &
Dirac \cite{Dirac:1937ti,Dirac:1938mt} &
The variability of 'constants' on cosmological timescales, particularly $G$, is motivated by large-number reasoning. &
It gives cosmological validity to a dynamically effective gravitational coupling, which is naturally represented by the SF \cite{Goenner:2012cq}. \\

1941 &
Scherrer \cite{Scherrer1941theorie} &
An action principle in which $R$ is multiplied by a scalar prefactor (a proto-scalar–tensor structure). &
This historically underappreciated precursor demonstrates that ST Lagrangian structures emerged independently of the Jordan framework \cite{Goenner:2012cq}. \\

1946--1947 &
Jordan (et al.) \cite{Jordan1946,JordanMuller1947} &
The variable $G$ approach is tied to the $5D$/projective unification traditions and is explicitly related to the arguments about cosmological variability. &
Introduces a coherent research programme in which the scalar encodes an effective gravitational coupling and has a strong influence on subsequent ST formulations \cite{Goenner:2012cq}. \\

1951 &
Thiry \cite{Thiry1951} &
The Jordan-Thiry approach has been mathematically systematized in a unitary-field setting. &
It clarifies consistency, solution structure and the formal foundations, and helps stabilize the framework at the level of mathematical physics \cite{Goenner:2012cq}. \\

1956 &
Fierz \cite{Fierz:1956zz} &
A critique of Jordan's framework focusing on the physical interpretation, particularly with regard to matter coupling. &
It draws attention to the differences between formulations and highlights the problems that can arise from interpreting them, which in turn makes for clearer $4D$ representations \cite{Goenner:2012cq}. \\

1961--1962 &
Brans-Dicke \cite{Brans:1961sx,Brans:1962zz}; Dicke \cite{Dicke:1961gz} &
A simple $4D$ modification of GR with a Machian motivation that can be interpreted as a local unit transformation of conformal rescalings. &
Due to its clarity and orientation towards testing the theory, it becomes the prototype ST model. It also initiates subsequent discussions on the (in)equivalence of conformal representations. \cite{Goenner:2012cq}. \\

1968--1970 &
Bergmann \cite{Bergmann:1968ve}; Nordtvedt \cite{Nordtvedt:1970uv}; Wagoner \cite{Wagoner:1970vr} &
General STT space, as well as systematic weak-field (PPN) phenomenology and GWs aspects. &
It transforms ST gravity from a single model into a class of parameters and functions tied to observational constraints, establishing the modern experimental framework \cite{Faraoni:2004pi,Fujii:2003pa,Goenner:2012cq}. \\
\hline
\end{tabularx}
\end{table}

\subsubsection{Scalar sectors from unification and cosmological variability - (Pre)history}
From a mathematical perspective, an important development in scalar-tensor theories of gravity is the emergence of the \textit{\textbf{Kaluza–Klein theory}} \cite{Kaluza1921,Klein1926}. As a result of dimensional compactification, this theory generates the following degrees of freedom:
\begin{itemize}
    \item $4D$ metric tensor,
    \item vector field,
    \item \textit{additional \textbf{scalar field}}.
\end{itemize}
Although this scalar did not constitute a well-established alternative to GR on its own, it naturally provided a \textbf{geometric origin} for a companion scalar to gravity. This would later become central to the \textbf{Jordan–Thiry} ideas \cite{Thiry1951}.

Meanwhile, Dirac's thoughts on large numbers - \textit{\textbf{Dirac large numbers hypothesis}} (LNH)\footnote{The original version of the hypothesis assumes that Newton's gravitational constant is inversely proportional to the age of the universe ($G\propto T_{U}^{-1}$), while the mass of the universe is directly proportional to the square of its age ($M_{U}\propto T_{U}^{2}$).} \cite{Dirac:1937ti,Dirac:1938mt} implied that \textit{physical 'constants'} (especially $G$) could change over \textit{cosmological timescales}, which makes an effective scalar gravitational coupling particularly plausible in cosmology.

\subsubsection{Several closely related proposals - Genesis}
Between 1941 and the early 1960s, ST formalism was proposed independently by multiple scientists \cite{Goenner:2012cq}. \textbf{Scherrer} \cite{Scherrer1941theorie} introduced an early action principle involving a scalar prefactor multiplying the Ricci scalar, which is conceptually similar to later (Jordan frame) \textit{scalar-tensor Lagrangians} but has been cited less frequently in the literature. \textbf{Jordan} developed a \textbf{variable} $\boldsymbol{G}$ \textbf{formalism} \cite{Jordan1946,JordanMuller1947} closely associated with \textbf{5D unification} and explicitly connected to the \textbf{cosmological motivation for variability}, while \textbf{Thiry} \cite{Thiry1951} provided a \textbf{mathematically systematic development} in this field.

Conceptual and interpretational issues, especially regarding the \textbf{physically meaningful coupling of the scalar sector to the matter sector}, were highlighted by critiques such as those of \textbf{Fierz} \cite{Fierz:1956zz}.

\subsubsection{Brans-Dicke theory as a prototype}
The (Jordan-Fierz-)\textbf{Brans–Dicke} formulation of the STTs \cite{Brans:1961sx,Fierz:1956zz} has become the \textit{\textbf{dominant paradigm}} in its field of gravity (cosmology). This follows from a 'relatively minor'\footnote{The most general formulation of the gravity theory in $4D$ spacetime using a metric tensor and a scalar field leading to second-order EoM is the \textit{\textbf{Horndeski's theory}} \cite{Horndeski:1974wa,Charmousis:2011bf,Deffayet:2011gz,Kobayashi:2019hrl,Horndeski:2024sjk}. In the case of arbitrary-dimensional spacetime, the generalization is the \textbf{Lovelock gravity} \cite{Lovelock:1971yv,Padmanabhan:2013xyr}.} modification of Einstein's theory of gravity in the $4D$ formalism closely related to Mach's principle, focusing on the possibility of testing theories experimentally to determine their accuracy and possible validity.

Furthermore, \textbf{Dicke's} analysis of the \textbf{local transformation of units} \cite{Dicke:1961gz} aims to clarify how different conformal representations can represent the same physics in different representations. This can be considered the original seed of the \textbf{debate on conformal frames} (see, Sec.~\ref{EJframes-section}).

Thereafter, \textbf{Bergmann's} generalization \cite{Bergmann:1968ve} and \textbf{Nordtvedt's} systematics concerning phenomenology in the weak field limit (PPN) \cite{Nordtvedt:1970uv}, and \textbf{Wagoner's} mechanism for generating GWs \cite{Wagoner:1970vr} established STT as a \textbf{broad testable class of gravitational theories} rather than a single simple model. This evolution of ST models resulted in a formalism that formed the basis for attempts to describe the \textbf{early} (e.g. \textit{cosmological inflation} \cite{Martin:2013tda}) and \textbf{late} (e.g. \textit{quintessence} \cite{Tsujikawa:2013fta,Amendola_Tsujikawa_2010q}) stages of the Universe's evolution.

\subsection{Mathematical framework}
In the aforementioned \textit{\textbf{Jordan conformal frame}}, STTs can be described using the following action \cite{Faraoni2004chapter}:
\begin{equation}\label{STT-action-general}
    S_{(\mathrm{STT})}=\int d^{4}x\,\sqrt{-g}\left[\frac{f(\phi)}{2}R-\frac{\omega_{\mathrm{BD}}(\phi)}{2}\nabla^{c}\phi\,\nabla_{c}\phi-V(\phi)\right]+S_{\mathrm{m}}\,.
\end{equation}
In this case (contrary to the \textbf{\textit{Einstein frame}}), the component of the action that is responsible for matter \textit{\textbf{does not depend explicitly on the scalar field}}:
\begin{equation}
    S_{\mathrm{m}}=\int d^{4}x\,\sqrt{-g}\,\mathcal{L}_{\mathrm{m}}\,.
\end{equation}
It is evident from the form of the action \eqref{STT-action-general} that the free parameter $\omega_{\mathrm{BD}}$ known from the Brans-Dicke theory becomes a function explicitly dependent on the scalar field $\phi$ and varies with the spacetime point, i.e.:
\begin{equation}\label{STT-omega}
    \omega_{\mathrm{BD}}\equiv\omega_{\mathrm{BD}}(\phi)\,.
\end{equation}
Furthermore, it is straightforward to demonstrate that the action \eqref{STT-action-general} contains a special case that describes the \textbf{BD theory}. This occurs for the following functional relationships of the theory parameters:
\begin{equation}
    f(\phi)=\frac{\phi}{8\pi}\quad\land\quad \omega_{\mathrm{BD}}(\phi)=\frac{\omega_{0}}{8\pi}\frac{1}{\phi}\quad\land\quad V(\phi)\mapsto\frac{V(\phi)}{16\pi}\,,
\end{equation}
where:
\begin{equation}
    \omega_{0}\equiv\mathrm{const}\,.
\end{equation}
A natural extension of the standard class of STTs are the so-called \textit{\textbf{multi-scalar-tensor theories}} (MST theories) \cite{Damour:1992we,Berkin:1993bt,Rainer:1996gw,Kuusk:2014sna,Kuusk:2015dda}, which take into account the existence of \textit{\textbf{multiple cosmological scalar fields}} (of different origins, depending on the type of theory) in their structure.

Turning back to the standard version of STTs, the variation of \eqref{STT-action-general} with respect to the metric tensor and the scalar field enables us to derive the direct forms of the equations of motion \cite{Faraoni2004chapter}:
\begin{subequations}\label{STT-EoM-general}
            \begin{align}
                f(\phi)\,H^{2} & = \frac{1}{3}\left[\rho_{\mathrm{tot}}+\frac{\omega_{\mathrm{BD}}(\phi)}{2}\dot{\phi}^{2}+V(\phi)\right]-\dot{f}(\phi)H \,, \\
                f(\phi)\,\dot{H} & = -\frac{1}{2}\left[\rho_{\mathrm{tot}}+p_{\mathrm{tot}}+\omega_{\mathrm{BD}}(\phi)\,\dot{\phi}^{2}+\ddot{f}(\phi)-\dot{f}(\phi)H\right] \,, \\
                \ddot{\phi}+3H\dot{\phi} & = -\frac{1}{2\omega_{\mathrm{BD}}(\phi)}\left[\omega'_{\mathrm{BD}}(\phi)\,\dot{\phi}^{2}-f'(\phi)R+2V'(\phi)\right] \,,
            \end{align}
\end{subequations}
where:
\begin{equation}\label{STT-dot-phi}
    \dot{f}(\phi)=f'(\phi)\,\dot{\phi}\quad\land\quad \ddot{f}(\phi)=f''(\phi)\,\dot{\phi}^{2}+f'(\phi)\,\ddot{\phi}\,,
\end{equation}
and the \textit{\textbf{total stress-energy tensor is covariantly conserved}}, namely:
\begin{equation}
    \nabla^{b}\,T_{ab}^{(\mathrm{tot})}=0\,.
\end{equation}
Using identities from \eqref{STT-dot-phi}, the Ricci scalar formula for a spatially flat FLRW universe:
\begin{equation}
    R=6\left(2H^{2}+\dot{H}\right)
\end{equation}
and substituting the $\dot{f}(\phi)H$ dependence from the 1st equation in \eqref{STT-EoM-general} into the 2nd one yields the \textbf{final form of the STT FEs} derived from \eqref{STT-action-general}:
\begin{subequations}\label{STT-EoM-compact}
            \begin{align}
                \Aboxed{f(\phi)\,\dot{H} & = -\frac{1}{2}\left[\left(\frac{2}{3}+\omega_{\mathrm{tot}}\right)\rho_{\mathrm{tot}}+\left\{\frac{5\omega_{\mathrm{BD}}(\phi)}{6}+f''(\phi)\right\}\,\dot{\phi}^2+f'(\phi)\,\ddot{\phi}+f(\phi)\,H^2-\frac{V(\phi)}{3}\right]} \,, \\
                \Aboxed{\ddot{\phi}+3H\dot{\phi} & = \frac{f'(\phi)\Bigl[\left(1-3 \omega_{\mathrm{tot}}\right)\rho_{\mathrm{tot}}+4V(\phi)-\Bigl\{\omega_{\mathrm{BD}}(\phi)+3f''(\phi)\Bigr\}\,\dot{\phi}^2\Bigr]-f(\phi)\left[\omega'_{\mathrm{BD}}(\phi)\,\dot{\phi}^2+2V'(\phi)\right]}{2f(\phi)\,\omega_{\mathrm{BD}}(\phi)+3f'(\phi)^2}} \,,
            \end{align}
\end{subequations}
where $\omega_{\mathrm{tot}}$ was defined in \eqref{effective-EoS} and the following \textbf{(conformal frame SF redefinition) condition must be satisfied}\footnote{The precise control of the Jordan-to-Einstein frame SF redefinition is enabled by this combination.}:
\begin{equation}\label{cf-SF-redefinition-condition}
    \Delta(\phi)\equiv 2f(\phi)\,\omega_{\mathrm{BD}}(\phi)+3f'(\phi)^2\neq 0\quad\implies\quad \boxed{\omega_{\mathrm{BD}}(\phi)\neq -\frac{3}{2}\frac{f'(\phi)^{2}}{f(\phi)}} \,.
\end{equation}
\begin{examplebox}[Phenomenological (mathematical) examples]
    \begin{itemize}
        \item \textit{\textbf{Minimal coupling}}:
        \begin{equation}
            f(\phi)=f_{0}\equiv\mathrm{const}\quad\implies\quad \Delta(\phi)=2f_{0}\,\omega_{\mathrm{BD}}(\phi)\quad\iff\quad \omega_{\mathrm{BD}}(\phi)\neq 0\,;
        \end{equation}
        \item \textit{\textbf{Brans–Dicke-like}}:
        \begin{equation}
            f(\phi)=\phi \quad\implies\quad \Delta(\phi)=2\phi\,\omega_{\mathrm{BD}}(\phi)+3 \quad\iff\quad \left(\phi>0\;\land\;\omega_{\mathrm{BD}}(\phi)>0\right)\,;
        \end{equation}
        \item \textit{\textbf{Quadratic coupling}}:
        \begin{equation}
            f(\phi)=\xi\phi^{2}\;\land\;\xi\neq 0\;\implies\; \Delta(\phi)=2\xi\phi^{2}\left[\omega_{\mathrm{BD}}(\phi)+6\xi\right]\;\iff\;\left(\omega_{\mathrm{BD}}(\phi)\neq -6\xi\;\land\;\phi\neq 0\right)\,;
        \end{equation}
        \item \textbf{\textit{Exponential coupling}}:
        \begin{equation}
        \begin{aligned}
            & f(\phi)=f_{0}\,e^{\alpha\phi}\;\land\;f_{0}\neq 0\;\implies\; \Delta(\phi)=f(\phi)\left[2\omega_{\mathrm{BD}}(\phi)+3\alpha^{2}f(\phi)\right] \\
            & \iff\;\omega_{\mathrm{BD}}(\phi)\neq -\frac{3}{2}\alpha^{2}f(\phi)
        \end{aligned}
        \end{equation}
        $\implies$ \textbf{The general degeneracy curve} (safe subcases):
        \begin{equation}
            f_{0}>0\quad\land\quad\omega_{\mathrm{BD}}\geq 0 \quad\implies\quad \Delta(\phi)>0\quad\iff\quad \phi\in\mathbb{R}
        \end{equation}
        \begin{equation}
            f_{0}>0\;\land\;\omega_{\mathrm{BD}}=\omega_{0}\equiv\mathrm{const}>0 \;\implies\; \Delta(\phi)=f(\phi)\left[2\omega_{0}+3\alpha^{2}f(\phi)\right]>0\;\iff\;\phi\in\mathbb{R}
        \end{equation}
    \end{itemize}
\end{examplebox}
In fact, the action from which the equations of motion for scalar-tensor theories are derived could also take the other forms, for example:
\begin{equation}\label{STT-action-v2}
    S_{(\mathrm{STT})}=\frac{1}{16\pi}\int d^{4}x\,\sqrt{-g}\left[\phi R-\frac{\omega_{\mathrm{BD}}(\phi)}{\phi}g^{ab}\nabla_{a}\phi\,\nabla_{b}\phi-V(\phi)\right]+S_{\mathrm{m}}\,,
\end{equation}
or the most usual with a \textbf{canonical kinetic energy segment} for a \textit{\textbf{redefined scalar field}} $\varphi$ \cite{Nordtvedt:1970uv}\footnote{This is possible if the function $f(\varphi)$ has a regular inverse, $f^{-1}(\varphi)$. Examples that do not meet this condition include functions such as a series with even powers of the SF $\phi$ \cite{Liddle:1991am,Torres:1996hv}.}:
\begin{equation}\label{STT-action-canonical-KE}
    \boxed{S_{(\mathrm{STT})}=\frac{1}{16\pi}\int d^{4}x\,\sqrt{-g}\left[f(\varphi)R-\frac{1}{2}g^{ab}\nabla_{a}\varphi\,\nabla_{b}\varphi-U(\varphi)\right]+S_{\mathrm{m}}}\,,
\end{equation}
where:
\begin{equation}
    \phi=f(\varphi)\quad\land\quad \omega_{\mathrm{BD}}=\frac{f(\varphi)}{2f'(\varphi)^{2}}\quad\land\quad U(\varphi)=V\bigl[f(\varphi)\bigr]\,.
\end{equation}
In such a case, the FEs take the following general form:
\begin{subequations}
\begin{align}
    f(\varphi)\,G_{ab} & = 8\pi\,T_{ab}^{(\mathrm{tot})}+\frac{1}{2}\left[\nabla_{a}\phi\,\nabla_{b}\phi-g_{ab}\left\{\frac{1}{2}\bigl(\nabla\varphi\bigr)^{2}+U(\varphi)\right\}\right]+\Bigl(\nabla_{a}\nabla_{b}-g_{ab}\,\Box\Bigr)\,f(\varphi) \,, \\
    \Box\varphi & = -f'(\varphi)\,R+U'(\varphi) \,, 
\end{align}
\end{subequations}
and therefore, explicitly:
\begin{subequations}\label{STT-EoM}
    \begin{align}
        f(\varphi)\,H^{2} & = \frac{1}{3}\left[8\pi\rho_{\mathrm{tot}}+\frac{1}{2}\left\{\frac{1}{2}\dot{\varphi}^{2}+U(\varphi)\right\}-3H\dot{f}(\varphi)\right] \,, \\
        f(\varphi)\,\dot{H} & = -\left[4\pi\bigl(1+\omega_{\mathrm{tot}}\bigr)\rho_{\mathrm{tot}}+\frac{1}{2}\left\{\frac{1}{2}\dot{\varphi}^{2}+\ddot{f}(\varphi)-\dot{f}(\varphi)H\right\}\right] \,, \\
        \ddot{\varphi}+3H\dot{\varphi} & = U'(\varphi)-\frac{f'(\varphi)}{f(\varphi)}\left[8\pi\bigl(1-3\omega_{\mathrm{tot}}\bigr)\rho_{\mathrm{tot}}-\frac{1}{2}\dot{\varphi}^{2}+2U(\varphi)-3\Bigl\{\ddot{f}(\varphi)+3\dot{f}(\varphi)H\Big\}\right] \,,
    \end{align}
\end{subequations}
where we have used the fact that:
\begin{equation}
    R=\frac{1}{f(\varphi)}\left[8\pi\bigl(1-3\omega_{\mathrm{tot}}\bigr)\rho_{\mathrm{tot}}-\frac{1}{2}\dot{\varphi}^{2}+2U(\varphi)-3\Bigl\{\ddot{f}(\varphi)+3\dot{f}(\varphi)H\Big\}\right]\,.
\end{equation}
Eliminating the constraint equation from \eqref{STT-EoM} yields the \textit{\textbf{final form of the EoM for the scalar-tensor theory in the Jordan frame}}:
\begin{subequations}\label{STT-EoM-simplified}
            \begin{align}
                \Aboxed{\dot{H} & = \frac{f'(\varphi)\,U'(\varphi)+\frac{1}{2}U(\varphi)+H f'(\varphi)\,\dot{\varphi}-8\pi\,\omega_{\mathrm{tot}}\rho_{\mathrm{tot}}-3H^2\Bigl[f(\varphi)+4f'(\varphi)^{2}\Bigr]-\Bigl[f''(\varphi)+\frac{1}{4}\Bigr] \dot{\varphi}^2}{2\Bigl[f(\varphi)+3 f'(\varphi)^2\Bigr]}} \,, \\
                \Aboxed{\ddot{\phi} & = \frac{f'(\varphi)\Bigl[8 \pi\bigl(1-3 \omega_{\mathrm{tot}}\bigr) \rho_{\mathrm{tot}}+2 U(\varphi)-\bigl\{\frac{1}{2}+3 f''(\varphi)\bigr\} \dot{\varphi}^2\Bigr]-f(\varphi) U'(\varphi)}{f(\varphi)+3 f'(\varphi)^2}-3H\dot{\phi}} \,,
            \end{align}
\end{subequations}
under the condition:
\begin{equation}
    \boxed{f(\varphi)+3 f'(\varphi)^2\neq 0}\quad\implies\quad f(\varphi)\neq \frac{1}{12}\left(-\varphi^{2}\pm 2i\sqrt{3}\,C_{1}\varphi+3C_{1}^2\right)\quad\land\quad C_{1}\equiv\mathrm{const}\,.
\end{equation}
\begin{examplebox}[Observational constraints on the STTs]
    \begin{itemize}
        \item \textit{\textbf{BBN constraints on the variation of}} $\boldsymbol{G}$ (95.4\% CL) \cite{Alvey:2019ctk}:
        \begin{equation}
        \boxed{\begin{array}{ll}
            \dfrac{G_{\mathrm{BBN}}}{G_0} \simeq \dfrac{f\left(\varphi_0\right)}{f\left(\varphi_{\mathrm{BBN}}\right)}=0.98_{-0.06}^{+0.06} & (\mathrm{BBN})\,, \\
            \dfrac{G_{\mathrm{BBN}}}{G_0} \simeq \dfrac{f\left(\varphi_0\right)}{f\left(\varphi_{\mathrm{BBN}}\right)}=0.99_{-0.05}^{+0.06} & \left(\mathrm{BBN}+\Omega_{\mathrm{b}} h^2\right)\,.
\end{array}}
\end{equation}
        For a slow-evolving linear function of time for $G$ \cite{Uzan:2002vq,Uzan:2024ded}:
        \begin{equation}
            G(t)=G_{\mathrm{BBN}}+\dot{G} \left(t-t_{\mathrm{BBN}}\right)\quad\implies\quad G_0=G_{\mathrm{BBN}}+\dot{G}\left(t_0-t_{\mathrm{BBN}}\right)
        \end{equation}
        one obtains \cite{Alvey:2019ctk}:
        \begin{equation}
            \boxed{\begin{aligned}
            & \left|\frac{\dot{G}}{G_0}\right|\simeq\left|\frac{\dot{f}\left(\varphi_{0}\right)}{f\left(\varphi_{0}\right)}\right|=1.4_{-4.7}^{+4.4} \times 10^{-12}\, \mathrm{yr}^{-1} \quad (\mathrm{BBN})\,, \\
            & \left|\frac{\dot{G}}{G_0}\right|\simeq\left|\frac{\dot{f}\left(\varphi_{0}\right)}{f\left(\varphi_{0}\right)}\right|=0.7_{-4.3}^{+3.8} \times 10^{-12}\, \mathrm{yr}^{-1} \quad \left(\mathrm{BBN}+\Omega_{\mathrm{b}} h^2\right)\,.
            \end{aligned}}
        \end{equation}
        \item \textit{\textbf{PPN constraints}} \cite{Bertotti:2003rm,Williams:2004qba}:
        \begin{equation}
            \bigl|\gamma_{\mathrm{PPN}}-1\bigr|=\left|\frac{2f'\left(\varphi_{0}\right)^{2}}{f\left(\varphi_{0}\right)+4f'\left(\varphi_{0}\right)^{2}}\right|\simeq 2\left|\frac{f'\left(\varphi_{0}\right)^{2}}{f\left(\varphi_{0}\right)}\right|\,,
        \end{equation}
        \begin{equation}
            \bigl|\beta_{\mathrm{PPN}}-1\bigr|=\frac{1}{2}\left|\frac{f\left(\varphi_{0}\right)f'\left(\varphi_{0}\right)^{2}\left[f'\left(\varphi_{0}\right)^{2}-2f\left(\varphi_{0}\right)f''\left(\varphi_{0}\right)\right]}{\left[f\left(\varphi_{0}\right)+3f'\left(\varphi_{0}\right)^{2}\right]\left[f\left(\varphi_{0}\right)+4f'\left(\varphi_{0}\right)^{2}\right]^{2}}\right|\simeq\left|\frac{f'\left(\varphi_{0}\right)^{2}}{f\left(\varphi_{0}\right)}\frac{f''\left(\varphi_{0}\right)}{f\left(\varphi_{0}\right)}\right|
        \end{equation}
        \begin{equation}
            \bigl|\gamma_{\mathrm{PPN}}-1\bigr|=\left(2.1 \pm 2.3\right) \times 10^{-5}\quad\implies\quad \boxed{\frac{f'\left(\varphi_{0}\right)^{2}}{f\left(\varphi_{0}\right)}\lesssim \frac{1}{2}\bigl|\gamma_{\mathrm{PPN}}-1\bigr|\simeq 10^{-5}}
        \end{equation}
        \item \textit{\textbf{GWs constraints}} (\textit{GW230529}) \cite{Wang:2025ehy}:
        \begin{equation}
            \omega_{\mathrm{BD}} \gtrsim \mathcal{O}\left(10^1-10^2\right)\quad\land\quad \omega_{\mathrm{BD}}(\varphi)=\frac{1}{2}\frac{f(\varphi)}{f'(\varphi)^{2}}\,,
        \end{equation}
        \begin{equation}
            \implies\quad \boxed{\frac{f'(\varphi)^{2}}{f(\varphi)}=\frac{1}{2}\frac{1}{\omega_{\mathrm{BD}}}\lesssim 5\times 10^{-3}}\,.
        \end{equation}
    \end{itemize}
\end{examplebox}

\begin{examplebox}[Convergence of STTs to GR]
    \begin{itemize}
        \item For general STTs the PPN parameters become of the form \cite{Clifton:2011jh}:
        \begin{equation}
            \gamma_{\mathrm{PPN}}=\frac{1+\omega_{\mathrm{BD}}(\phi)}{2+\omega_{\mathrm{BD}}(\phi)}\quad\land\quad \beta_{\mathrm{PPN}}=1+\frac{\omega'_{\mathrm{BD}}(\phi)}{\bigl[4+2\omega_{\mathrm{BD}}(\phi)\bigr]\bigl[3+2\omega_{\mathrm{BD}}(\phi)\bigr]^{2}}\,.
        \end{equation}
        \item Large $\omega_{\mathrm{BD}}(\phi)$ expansion yields:
        \begin{equation}
            \gamma_{\mathrm{PPN}}-1=\frac{1+\omega_{\mathrm{BD}}(\phi)}{2+\omega_{\mathrm{BD}}(\phi)}-1=-\frac{1}{2+\omega_{\mathrm{BD}}(\phi)}\sim -\frac{1}{\omega_{\mathrm{BD}}(\phi)}\,,
        \end{equation}
        \begin{equation}
            \beta_{\mathrm{PPN}}-1\sim\frac{\omega'_{\mathrm{BD}}(\phi)}{\bigl[2\omega_{\mathrm{BD}}(\phi)\bigr]\bigl[2\omega_{\mathrm{BD}}(\phi)\bigr]^{2}}=\frac{1}{8}\frac{\omega'_{\mathrm{BD}}(\phi)}{\omega_{\mathrm{BD}}(\phi)^{3}}\sim\frac{\omega'_{\mathrm{BD}}(\phi)}{\omega_{\mathrm{BD}}(\phi)^{3}}\,,
        \end{equation}
        therefore:
        \begin{subequations}
        \begin{align}
            \gamma_{\mathrm{PPN}} &\to 1
            \quad\implies\quad
            \omega_{\mathrm{BD}}(\phi)\to\infty\,,
            \\
            \beta_{\mathrm{PPN}} &\to 1
            \quad\implies\quad
            \frac{\omega'_{\mathrm{BD}}(\phi)}{\omega_{\mathrm{BD}}(\phi)^{3}}\to 0
            \,\Bigl(\implies\,\omega_{\mathrm{BD}}\to\mathrm{const}\Bigr)\,.
        \end{align}
        \end{subequations}
    \end{itemize}
    \textbf{\textit{\underline{Conclusion:}}}\\
    \textit{\textbf{General STT converges to GR under the condition}}\footnote{This was originally demonstrated by \textit{\textbf{Kenneth Nordtvedt}} in \cite{Nordtvedt:1970uv}.} \cite{Nordtvedt:1970uv,Damour:1992kf}:
        \begin{equation}
            \boxed{\omega_{\mathrm{BD}}(\phi)\to\infty\quad\land\quad \frac{\omega'_{\mathrm{BD}}(\phi)}{\omega_{\mathrm{BD}}(\phi)^{3}}\to 0}\,.
        \end{equation}
\end{examplebox}

\subsection{Motivations}
\subsubsection{Dynamical dark energy from DESI DR2(?)}
\label{sec:DESI-DR2-STT-motivation}
The detailed observational discussion of the DESI DR2 constraints on dynamical dark energy, including the CPL parameter space, the $Om(z)$ and statefinder diagnostics, and the reconstruction of quintessence-like scenarios, is presented in Sec.~\ref{sec:DESI-DR2-ch4}. In this section, the same observational results will be used only in a more limited sense: as a \textit{\textbf{phenomenological motivation for considering STTs as a natural extension of GR}}.

The main point is not that DESI DR2 proves a STT. Rather, the DESI DR2 analyses indicate that the \textbf{late-time expansion history may be more flexibly described by an effective DE sector with a time-dependent EoS}\footnote{Another way of interpreting the data obtained within DESI DR2 in relation to \textit{\textbf{dynamical (evolving) dark energy}} is to consider the concept of \textit{\textbf{interacting dark energy}} \cite{Caldera-Cabral:2008yyo,vanderWesthuizen:2025vcb,vanderWesthuizen:2025mnw,vanderWesthuizen:2025rip,Pan:2025qwy,Guedezounme:2025wav,Li:2026xaz}.} than by a cosmological constant \cite{DESI:2025zgx,DESI:2025fii}.

In the commonly used CPL parameter space \cite{Chevallier:2000qy,Linder:2002et}:
\begin{equation}
    \omega_{\mathrm{DE}}(a)=\omega_{0}+\omega_{a}\left(1-a\right) \quad\iff\quad \omega_{\mathrm{DE}}(z)=\omega_{0}+\omega_{a}\frac{z}{1+z}\,,
\end{equation}
the region favored by the combined DESI DR2 analyses, is characterized by:
\begin{equation}
    \omega_{0}>-1 \quad\land\quad \omega_{a}<0\,.
\end{equation}
However, the condition $\omega_{a}<0$ alone is not sufficient to guarantee a phantom-like regime in the past. Since:
\begin{equation}
    \lim_{z\to\infty}\omega_{\mathrm{DE}}(z)=\omega_{0}+\omega_{a}\,,
\end{equation}
the effective EoS evolves from a phantom-like value in the past to a non-phantom value today only if:
\begin{equation}
    \omega_{0}>-1 \quad\land\quad \omega_{0}+\omega_{a}<-1\,.
\end{equation}
Equivalently, the crossing of the phantom divide line (PDL) occurs at:
\begin{equation}
    z_{\mathrm{PDL}}=-\frac{1+\omega_{0}}{1+\omega_{0}+\omega_{a}}\,,
\end{equation}
provided that this redshift is positive. This behavior should be interpreted as an \textbf{effective reconstruction of the background expansion history, not as a direct measurement of a fundamental SF EoS}.

This distinction is important for the present section. A minimally coupled canonical scalar field in GR can be described by:
\begin{equation}
    \rho_{\phi}=\frac{1}{2}\dot{\phi}^{2}+V(\phi) \quad\land\quad
    p_{\phi}=\frac{1}{2}\dot{\phi}^{2}-V(\phi)\,,
\end{equation}
and therefore:
\begin{equation}
    \rho_{\phi}+p_{\phi}=\dot{\phi}^{2}\geq 0\,.
\end{equation}
For a positive SF energy density, this implies:
\begin{equation}
    \omega_{\phi}\equiv\frac{p_{\phi}}{\rho_{\phi}}\geq -1\,.
\end{equation}
Consequently, a single canonical quintessence field cannot cross the phantom divide line (PDL) of:
\begin{equation}
    \omega=-1
\end{equation}
without either modifying the matter sector, introducing additional degrees of freedom, allowing non-canonical kinetic terms, or changing the gravitational sector.

The STTs provide one of the most economical gravitational ways of realizing such an effective behavior. In the Jordan frame, a broad class of STTs may be written as \cite{Faraoni2004chapter,Fujii_Maeda_2003,Clifton:2011jh,Tsujikawa:2013fta,Nojiri:2017ncd}:
\begin{equation}
    \boxed{S_{J}=\int d^{4}x\sqrt{-g}\left[\frac{F(\phi)}{2\kappa^{2}}R-\frac{1}{2}K(\phi)\,g^{\mu\nu}\nabla_{\mu}\phi\nabla_{\nu}\phi-U(\phi)\right]+S_{m}\left[g_{\mu\nu},\psi_{m}\right]}\,,
    \label{Jordan-STT-action-DESI-motivation}
\end{equation}
where $F(\phi)$ determines the effective gravitational coupling, $K(\phi)$ characterizes the normalization of SF kinetic energy and $U(\phi)$ is the scalar self-interaction potential. The corresponding field equations can be written as:
\begin{equation}
    F(\phi)G_{\mu\nu}=\kappa^{2}\left(T_{\mu\nu}^{(m)}+T_{\mu\nu}^{(\phi)}\right)+\nabla_{\mu}\nabla_{\nu}F-g_{\mu\nu}\Box F\,.
\end{equation}
For a spatially flat FLRW background, this gives:
\begin{align}
    3F H^{2}&=\kappa^{2}\left(\rho_{m}+\rho_{r}+\frac{1}{2}K\dot{\phi}^{2}+U\right)-3H\dot{F}\,,
    \label{STT-Friedmann-1-DESI-motivation}
    \\
    -2F\dot{H}&=\kappa^{2}\left(\rho_{m}+\frac{4}{3}\rho_{r}+K\dot{\phi}^{2}
    \right)+\ddot{F}-H\dot{F}\,.
    \label{STT-Friedmann-2-DESI-motivation}
\end{align}
Thus, even if the SF has a canonical kinetic term, the non-minimal coupling generates additional contributions proportional to $\dot{F}$ and $\ddot{F}$. These terms are absent in minimally coupled quintessence.

To see why this is relevant for DESI-like reconstructions, let us rewrite \eqref{STT-Friedmann-1-DESI-motivation} and \eqref{STT-Friedmann-2-DESI-motivation} in a GR-like form with a fixed reference value $F_{0}\equiv F(\phi_{0})$:
\begin{align}
    3H^{2}&=\frac{\kappa^{2}}{F_{0}}\left(\rho_{m}+\rho_{r}+\rho_{\mathrm{DE}}^{(\mathrm{eff})}\right)\,,
    \\
    -2\dot{H}&=\frac{\kappa^{2}}{F_{0}}\left(\rho_{m}+\frac{4}{3}\rho_{r}+\rho_{\mathrm{DE}}^{(\mathrm{eff})}+p_{\mathrm{DE}}^{(\mathrm{eff})}\right)\,.
\end{align}
Then, the effective DE density and pressure are given by:
\begin{align}
    \kappa^{2}\rho_{\mathrm{DE}}^{(\mathrm{eff})}&=\kappa^{2}\left(\frac{1}{2}K\dot{\phi}^{2}+U\right)-3H\dot{F}+3H^{2}\left(F_{0}-F\right)\,,
    \\
    \kappa^{2}p_{\mathrm{DE}}^{(\mathrm{eff})}&=\kappa^{2}\left(\frac{1}{2}K\dot{\phi}^{2}-U\right)+\ddot{F}+2H\dot{F}+\left(F_{0}-F\right)\left(-2\dot{H}-3H^{2}\right)\,.
\end{align}
In a consequence, one may write:
\begin{equation}
    \boxed{\kappa^{2}\left[\rho_{\mathrm{DE}}^{(\mathrm{eff})}+p_{\mathrm{DE}}^{(\mathrm{eff})}\right]=\kappa^{2}K\dot{\phi}^{2}+\ddot{F}-H\dot{F}-2\dot{H}\left(F_{0}-F\right)}\,.
    \label{rho-plus-p-effective-STT}
\end{equation}
In the case of minimally coupled limit:
\begin{equation}
    F(\phi)=F_{0}=\mathrm{const}\,,
\end{equation}
one recovers the relation:
\begin{equation}
    \rho_{\mathrm{DE}}^{(\mathrm{eff})}+p_{\mathrm{DE}}^{(\mathrm{eff})}=K\dot{\phi}^{2}\geq 0\,,
\end{equation}
so that the effective EoS cannot cross $\omega=-1$ for $K>0$. In a genuine STT, however, the derivative terms in \eqref{rho-plus-p-effective-STT} can change the sign of the effective combination $\rho_{\mathrm{DE}}^{(\mathrm{eff})}+p_{\mathrm{DE}}^{(\mathrm{eff})}$. Therefore:
\begin{equation}
    \boxed{\omega_{\mathrm{DE}}^{(\mathrm{eff})}=\frac{p_{\mathrm{DE}}^{(\mathrm{eff})}}{\rho_{\mathrm{DE}}^{(\mathrm{eff})}}}
\end{equation}
\textit{\textbf{may cross the PDL as an effective GR description}}, even when the underlying theory contains a healthy scalar degree of freedom.

\begin{examplebox}[DESI DR2 as a motivation for STTs]
    The DESI DR2 results do not uniquely select a STT. However, they motivate STTs for three reasons:
    \begin{enumerate}[label=(\alph*)]
        \item They suggest that the late-time cosmic expansion may prefer an effective, time-dependent DE sector rather than a strict cosmological constant;
        \item A minimally coupled canonical quintessence field cannot realize a genuine phantom crossing;
        \item In STTs, the non-minimal coupling $F(\phi)R$ makes the effective EoS reconstructed under GR assumptions differ from the fundamental SF EoS.
    \end{enumerate}
\end{examplebox}

\subsubsection{String Theory (Superstring theory, M-theory, Supergravity (SUGRA))}
\textit{\textbf{String theory}}\footnote{It is worth mentioning that, to this day, there is an ongoing debate within the scientific community as to \textit{whether string theory fits within the scientific method} of physics. The main concerns are its possible \textit{lack of verifiability} and the associated philosophical issues \cite{Dawid:2013maa,Ellis:2014sjr,Castelvecchi2015}.} and its various extensions (such as \textbf{Superstring theory} \cite{Polchinski_1998b,Green_Schwarz_Witten_2012a,Green_Schwarz_Witten_2012b}, \textbf{M-theory} \cite{Becker:2006dvp}, \textbf{Supergravity} \cite{VanNieuwenhuizen:1981ae,Freedman:2012zz,Nath:2016qzm,DallAgata:2021uvl,Nastase:2024itd}) are characterized by an important aspect that is inseparably connected with the motivation for STTs. Namely, an \textit{inherent feature} of these theories is the presence of a \textit{SF coupled with gravity} - the \textit{\textbf{dilaton}}\footnote{It is worth noting that, apart from dilaton motivation, SFs also appear in many other areas of physics, especially in \textit{particle physics}. Examples include the \textbf{Higgs boson} and non-fundamental fields such as \textbf{composite bosons} and \textbf{fermion condensates}. However, in cosmology, \textbf{inflaton} and \textbf{quintessence} can serve as examples.} \cite{Wetterich:1987fm,Taylor:1988nw,Brandenberger:1998zs,Gasperini2008}.

Another important aspect is the fact that the gravitational sector of BD theory is characterized by \textbf{conformal invariance} due to the following transformations \cite{Faraoni:2004pi}:
\begin{subequations}\label{BD-invariance}
            \begin{align}
                g_{\mu\nu}\mapsto\tilde{g}_{\mu\nu} & = \left(G\phi\right)^{2\alpha}\,g_{\mu\nu}\quad\land\quad \alpha\neq \frac{1}{2} \,, \\
                G\phi\mapsto G\tilde{\phi} & = \left(G\phi\right)^{1-2\alpha} \,,
            \end{align}
\end{subequations}
with:
\begin{equation}\label{BD-invariance1}
    \tilde{\omega}_{\mathrm{BD}}=\frac{\omega_{\mathrm{BD}}-6\alpha\,(\alpha-1)}{\left(1-2\alpha\right)^{2}}\,.
\end{equation}
This is reminiscent of the \textbf{conformal invariance of string theories at high energies} \cite{Cho1992}.

Notably, the \textit{most compelling motivation for STTs in the context of string theory phenomenology} emerges from the aforementioned \textbf{low-energy limit of bosonic string theory}. This corresponds to the BD theory, characterized by a constant free parameter value \cite{Callan:1985ia,Fradkin:1985ys}:
\begin{equation}\label{omega-BD-1}
    \omega_{\mathrm{BD}}=-1\,.
\end{equation}

The low-energy action of the bosonic string theory in the \textit{string frame} at the tree level takes the following form \cite{Polchinski_1998a}:
\begin{equation}\label{low-energy-string-frame}
S=\frac{1}{2 \kappa_{D}^2} \int d^{D}x\,\sqrt{-g}\,e^{-2\Phi}\left[R[g]-4g^{\mu\nu}\partial_{\mu}\Phi\,\partial_{\nu}\Phi-\frac{1}{12}H_{\mu\nu\sigma}H^{\mu\nu\sigma}-\frac{D-26}{l_{s}^2}\right]\,,
\end{equation}
where:
\begin{equation}
    \kappa_{D}\equiv 8\pi G_{(D)}
\end{equation}
is the gravitational coupling constant in $D$ target space dimensions, and:
\begin{equation}
    H_{\mu\nu\sigma}=\partial_{[\mu} b_{\nu\sigma]}
\end{equation}
denotes the strength of the \textbf{Kalb-Ramond field}, $b_{\nu\sigma}$. Moreover, $\Phi$ represents the \textit{dimensionless \textbf{string dilaton}} and $l_{s}$ is the \textbf{string scale}.

It is usually assumed that:
\begin{equation}
    H_{\mu\nu\sigma}H^{\mu\nu\sigma}=0\,,
\end{equation}
and therefore, only the string dilaton is considered. This is related to the fact that in a spatially homogeneous and isotropic Universe, the field $H_{\mu\nu\sigma}$ reduces to the form \cite{Faraoni:2004pi}:
\begin{equation}
    H_{0\nu\sigma}=0\quad\land\quad H_{123}=h(t)\,.
\end{equation}
and can also be modeled using a \textbf{perfect fluid} approximation \cite{Kolitch:1994qa}. Therefore, the \textbf{inclusion of matter} models the term associated with the \textbf{Kalb-Ramond 3-form} in the FLRW models.

In the case of $4D$ spacetime, a simple redefinition of the dilaton field:
\begin{equation}
    \varphi=e^{-2\Phi}
\end{equation}
yields the BD theory described by the free parameter of the form \eqref{omega-BD-1}, namely:
\begin{equation}
    S=\frac{1}{2\kappa^{2}}\int d^{4}x\,\sqrt{-g}\,\left[\varphi R+\frac{1}{\varphi}\,g^{\mu\nu}\partial_{\mu}\varphi\,\partial_{\nu}\varphi\right]\,.
\end{equation}
In the general case of the $\boldsymbol{p}$\textbf{-brane in the} $\boldsymbol{d}$\textbf{-dimensional spacetime} (after \textit{compactification}), the BD parameter becomes \cite{Duff:1994an}:
\begin{equation}
    \omega_{\mathrm{BD}}=-\frac{(d-1)\,(p-1)-(p+1)^{2}}{(d-2)\,(p-1)-(p+1)^{2}}=\frac{p-1}{d\,(1-p)+p\,(4+p)-1}-1\,.
\end{equation}
Other motivations related to string theory include:
\begin{itemize}
    \item \textbf{Extended} \cite{La:1989pn} and \textbf{hyperextended} \cite{Steinhardt1990} \textbf{inflation};
    \item \textbf{Randall-Sundrum brane-world} models \cite{Randall:1999ee,Randall:1999vf};
    \item \textbf{Domain wall} ($4D$ brane) in a \textit{higher-dimensional space} \cite{Garriga:1999yh,Barcelo:2000ta}.
\end{itemize}

\subsection{Conformal transformations}
\label{Conf-transf-subsection}
\textit{\textbf{Conformal transformations}}\footnote{From a mathematical perspective, a conformal transformation preserves angles locally, but not necessarily lengths.} \cite{Wald:1984rg_conformal,Fujii_Maeda_2003_conformal,Kroon:2016ink,Carroll_2019_conformal,Hawking_Ellis_2023_conformal} are a widely used mathematical technique related to STTs. They allow for the transformation of the metric tensor and the redefinition of the SF so that new dynamical variables can be obtained:
\begin{equation}\label{EF-variables}
    \left(\tilde{g}_{\mu\nu},\tilde{\phi}\right)\,.
\end{equation}
This implies the existence of a gravitational sector in the considered theory that is similar to that known from GR, as well as a new SF, denoted by ${\tilde{\phi}}$, which is characterized by a canonical kinetic energy density. This new set of dynamical variables \eqref{EF-variables} is called the \textbf{\textit{Einstein (conformal) frame}}, as opposed to the \textit{\textbf{Jordan (conformal) frame}}:
\begin{equation}\label{JF-variables}
    \left(g_{\mu\nu},\phi\right)\,.
\end{equation}
As was stated previously, interpretations regarding mathematical and physical (non)equivalence are the subject of much controversy among physicists and cosmologists. The following subsection is entirely dedicated to this issue (see Sec.~\ref{EJframes-section}).
\subsubsection{Mathematical details}
To conduct a comprehensive analysis of the mathematical aspects of conformal transformations, let us consider the following \textit{\textbf{spacetime}}:
\begin{equation}
    \left(\mathcal{M},g_{\mu\nu}\right)
\end{equation}
which consists of:
\begin{itemize}
    \item Smooth $n\geq2$-dimensional manifold $\mathcal{M}$;
    \item Lorentzian or Riemannian metric $g_{\mu\nu}$ on manifold $\mathcal{M}$.
\end{itemize}
The following (spacetime) \textit{point-dependent metric rescaling} is referred to as a \textit{\textbf{conformal transformation}}\footnote{We use the notation $\Omega^{2}(x)$ because we are limiting ourselves to transformations that \textit{leave the sign of the line interval unchanged}.}:
\begin{equation}\label{conf-transf}
    g_{\mu\nu}\mapsto \tilde{g}_{\mu\nu}=\Omega^{2}(x)\,g_{\mu\nu}\quad\iff\quad ds^{2}\mapsto d\tilde{s}^{2}=\Omega^{2}(x)\,ds^{2} \,,
\end{equation}
and the term '\textit{\textbf{conformal factor}}'\footnote{Regular, nowhere vanishing function.} is used to refer to the function $\Omega(x)$.
\begin{examplebox}[Properties of conformal transformations]
    \begin{itemize}
        \item The \textbf{lengths} of \textit{spacelike} and \textit{timelike} intervals \textbf{change};
        \item The \textbf{norms} of \textit{spacelike} and \textit{timelike} vectors are \textbf{changed};
        \item The \textit{rescaled metric} $\tilde{g}_{\mu\nu}$ leaves \textit{null intervals} and \textit{null vectors} \textbf{unchanged};
        \item The \textit{light cones} remain \textbf{unchanged};
        \item The \textit{causal structure} of spacetimes $\left(\mathcal{M},g_{\mu\nu}\right)$ and $\left(\mathcal{M},\tilde{g}_{\mu\nu}\right)$ is \textbf{identical}.
    \end{itemize}
\end{examplebox}
\subsubsection{Conformal transformations of geometric quantities}
Using the definition of the inverse metric tensor:
\begin{equation}
    \tilde{g}_{\mu\sigma}\tilde{g}^{\sigma\nu}=\delta_{\mu}^{\nu}
\end{equation}
and the definition of the conformal transformation \eqref{conf-transf} one obtains:
\begin{equation}
    \Omega^{2}(x)\tilde{g}_{\mu\sigma}\tilde{g}^{\sigma\nu}=\delta_{\mu}^{\nu}\,,
\end{equation}
so the \textbf{inverse transformed metric tensor} takes the form:
\begin{equation}\label{inv-metr-transf}
    \tilde{g}^{\mu\nu}=\Omega^{-2}(x)\,g^{\mu\nu}\,.
\end{equation}
Using the algebraic identity for a scalar $c$ and $n\times n$ matrix $A$:
\begin{equation}
    \det{\left(cA\right)}=c^{n}\det{\left(A\right)}
\end{equation}
yields:
\begin{equation}
    \tilde{g}\equiv\det{\left(\tilde{g}_{\mu\nu}\right)}=\det{\left[\Omega^{2}(x)\,g_{\mu\nu}\right]}=\Omega^{2n}(x)\det{\left(g_{\mu\nu}\right)}\,,
\end{equation}
and therefore, the \textbf{determinant of the transformed metric tensor} is given by the formula:
\begin{equation}
    \tilde{g}=\Omega^{2n}(x)\,g\,.
\end{equation}
The standard form of the Levi-Civita connection is:
\begin{equation}
    \Gamma_{\nu\sigma}^{\mu}=\frac{1}{2}g^{\mu\rho}\Bigl(\partial_{\nu}g_{\rho\sigma}+\partial_{\sigma}g_{\rho\nu}-\partial_{\rho}g_{\nu\sigma}\Bigr)=\frac{1}{2}g^{\mu\rho}\bigl(g_{\rho\sigma,\nu}+g_{\rho\nu,\sigma}-g_{\nu\sigma,\rho}\bigr)\,.
\end{equation}
Using \eqref{conf-transf} and the fact that $\Omega(x)$ is a scalar, we can show that:
\begin{equation}
    \partial_{\mu}\Omega(x)=\nabla_{\mu}\Omega(x)\,,
\end{equation}
and therefore, it implies the following form of the \textbf{transformed Levi-Civita connection}:
\begin{equation}
\begin{aligned}
    \tilde{\Gamma}_{\nu\sigma}^{\mu} & = \Gamma_{\nu\sigma}^{\mu}+g^{\mu\rho}\,\Omega^{-1}\Bigl[\bigl(\nabla_{\nu}\Omega\bigr)\,g_{\rho\sigma}+\bigl(\nabla_{\sigma}\Omega\bigr)\,g_{\rho\nu}-\bigl(\nabla_{\rho}\Omega\bigr)\,g_{\nu\sigma}\Bigr] \\
    & = \Gamma_{\nu\sigma}^{\mu}+\underbrace{\Omega^{-1}\Bigl[\delta_{\nu}^{\mu}\,\nabla_{\sigma}\Omega+\delta_{\sigma}^{\mu}\,\nabla_{\nu}\Omega-g_{\nu\sigma}\nabla^{\mu}\Omega\Bigr]}_{\kappa^{\mu}{}_{\nu\sigma}}\,.
\end{aligned}
\end{equation}
Then, by defining the \textit{difference tensor} between the original and transformed connection as:
\begin{equation}\label{diff-tensor}
    \kappa^{\mu}{}_{\nu\sigma}\equiv\tilde{\Gamma}_{\nu\sigma}^{\mu}-\Gamma_{\nu\sigma}^{\mu}=\Omega^{-1}\Bigl[\delta_{\nu}^{\mu}\,\nabla_{\sigma}\Omega+\delta_{\sigma}^{\mu}\,\nabla_{\nu}\Omega-g_{\nu\sigma}\nabla^{\mu}\Omega\Bigr]\,,
\end{equation}
one could show that for two \textit{torsionless connections} the \textbf{transformed Riemann curvature tensor} satisfies the relation\footnote{From this point onwards, we will assume that we are operating in $4D$ spacetime.}:
\begin{equation}
    \tilde{R}^{\mu}{}_{\nu\sigma\rho}=R^{\mu}{}_{\nu\sigma\rho}+2\Bigl(\nabla_{[\sigma}\,\kappa^{\mu}{}_{\rho]\nu}+\kappa^{\mu}{}_{[\sigma|\eta|}\,\kappa^{\eta}{}_{\rho]\nu}\Bigr)\,.
\end{equation}
Moreover, inserting the explicit form of the difference tensor \eqref{diff-tensor} and using the identity:
\begin{equation}
   \nabla_{\mu}\bigl(\ln{\Omega}\bigr)=\Omega^{-1}\nabla_{\mu}\Omega \,,
\end{equation}
gives the \textit{other form} of the \textbf{transformed Riemann curvature tensor}:
\begin{equation}\label{Riem-transf}
\begin{aligned}
    \tilde{R}_{\mu\nu\sigma}{}^{\rho} = &\,R_{\mu\nu\sigma}{}^{\rho}+2\Bigl[\delta^{\rho}_{[\mu}\nabla_{\nu]}\nabla_{\sigma}\bigl(\ln{\Omega}\bigr)-g^{\rho\eta}g_{\sigma[\mu}\nabla_{\nu]}\nabla_{\eta}\bigl(\ln{\Omega}\bigr)+\nabla_{[\mu}\bigl(\ln{\Omega}\bigr)\delta^{\rho}_{\nu]}\nabla_{\sigma}\bigl(\ln{\Omega}\bigr) \\
    &\,-\nabla_{[\mu}\bigl(\ln{\Omega}\bigr)g_{\nu]\sigma}g^{\rho\eta}\nabla_{\eta}\bigl(\ln{\Omega}\bigr)-g_{\sigma[\mu}\delta^{\rho}_{\nu]}g^{\eta\zeta}\nabla_{\eta}\bigl(\ln{\Omega}\bigr)\nabla_{\zeta}\bigl(\ln{\Omega}\bigr)\Bigr]\,.
\end{aligned}
\end{equation}
Contracting \eqref{Riem-transf} on $\rho$ with $\sigma$ yields the \textbf{transformed Ricci tensor}:
\begin{equation}\label{Ric-transf}
    \tilde{R}_{\mu\nu}=R_{\mu\nu}-2\Bigl[\nabla_{\mu}\nabla_{\nu}\bigl(\ln{\Omega}\bigr)-\nabla_{\mu}\bigl(\ln{\Omega}\bigr)\nabla_{\nu}\bigl(\ln{\Omega}\bigr)+g_{\mu\nu}\bigl(\nabla\ln{\Omega}\bigr)^{2}\Bigr]-g_{\mu\nu}\nabla^{2}\bigl(\ln{\Omega}\bigr)\,.
\end{equation}
According to the definition of Ricci scalar:
\begin{equation}
    \tilde{R}=\tilde{g}^{\mu\nu}\tilde{R}_{\mu\nu}
\end{equation}
together with \eqref{inv-metr-transf} and \eqref{Ric-transf} one obtains the \textbf{transformed Ricci scalar}:
\begin{equation}
    \tilde{R}=\Omega^{-2}\Bigl[R-6\Bigl\{\Box\ln{\Omega}+\bigl(\nabla\ln{\Omega}\bigr)^{2}\Bigr\}\Bigr]\,.
\end{equation}
Furthermore, using the following identity:
\begin{equation}
    \bigl(\nabla\ln{\Omega}\bigr)^{2}=g^{\mu\nu}\bigl(\nabla_{\mu}\ln{\Omega}\bigr)\,\bigl(\nabla_{\nu}\ln{\Omega}\bigr)
\end{equation}
this gives \textit{another form} of the \textbf{transformed Ricci scalar}:
\begin{equation}
    \tilde{R}=\Omega^{-2}\left(R-6\frac{\Box\Omega}{\Omega}\right)\,.
\end{equation}
The \textit{\textbf{Weyl (conformal) tensor}}\footnote{In GR, the Weyl tensor is the only component of curvature that exists in free space (i.e. solution of the vacuum FEs). It controls the way that GWs propagate through regions of space that are free of matter \cite{Hawking_Ellis_2023_conformal,Penrose_2005_Reality,Danehkar:2007mj}. Moreover, the Weyl tensor illustrates how tidal forces distort the shape of a body. For $D=2,3$, the Weyl curvature tensor vanishes identically, and for $D\geq 4$, it is generally non-zero.} \cite{Weyl1918} is the traceless part of the Riemann tensor:
\begin{equation}
    C^{\mu}{}_{\nu\sigma\rho}\equiv R^{\mu}{}_{\nu\sigma\rho}-4 S_{[\sigma}{}^{[\mu}\delta^{\nu]}_{\rho]}\,,
\end{equation}
where:
\begin{equation}
    S_{\mu\nu}\equiv\frac{1}{2}\left(R_{\mu\nu}-\frac{1}{6}R g_{\mu\nu}\right)
\end{equation}
is the \textit{\textbf{Schouten tensor}}\footnote{In the case of $D=4$, the trace of the Schouten tensor is: $S^{\mu}{}_{\mu}=\frac{1}{6}R$.} \cite{Besse1987} that transforms as follows:
\begin{equation}\label{Schouten-transf}
    S_{\mu\nu}\mapsto S_{\mu\nu}-\Bigl[\nabla_{\mu}\nabla_{\nu}\ln{\Omega}-\bigl(\nabla_{\mu}\ln{\Omega}\bigr)\bigl(\nabla_{\nu}\ln{\Omega}\bigr)\Bigr]\,.
\end{equation}
Using \eqref{Schouten-transf} with \eqref{Riem-transf} yields that all extra terms cancel, therefore the \textbf{transformed (1,3) Weyl tensor} becomes:
\begin{equation}\label{Weyl-inv}
    \tilde{C}^{\mu}{}_{\nu\sigma\rho}=C^{\mu}{}_{\nu\sigma\rho}\,.
\end{equation}
\begin{examplebox}[Conformal invariance of the Weyl tensor]
    \begin{itemize}
        \item The version of the \textit{\textbf{Weyl tensor with mixed indices}}, namely \textit{(1,3)} \eqref{Weyl-inv}, is the \textit{\textbf{only one that is conformally invariant}};
        \item The \textit{\textbf{other versions of the Weyl tensor}} (with different index positions) \textbf{are not conformally invariant}\footnote{$\Omega\neq\mathrm{const}$.}:
        \begin{subequations}
        \begin{equation}
            \tilde{C}_{\mu\nu\sigma\rho}=\Omega^{2}\,C_{\mu\nu\sigma\rho}\neq C_{\mu\nu\sigma\rho}\,,
        \end{equation}
        \begin{equation}
            \tilde{C}_{\mu\nu}{}^{\sigma\rho}=\Omega^{-2}\,C_{\mu\nu}{}^{\sigma\rho}\neq C_{\mu\nu}{}^{\sigma\rho}\,,
        \end{equation}
        \begin{equation}
            \tilde{C}^{\mu\nu\sigma\rho}=\Omega^{-6}\,C^{\mu\nu\sigma\rho}\neq C^{\mu\nu\sigma\rho}\,.
        \end{equation}
        \end{subequations}
    \end{itemize}
    \textbf{\textit{\underline{Conclusion:}}}\\
    Each raised index contributes a factor of $\Omega^{-2}$.
\end{examplebox}
The covariant conservation law for a (symmetric) energy-momentum tensor:
\begin{equation}
    \nabla^{\nu}T_{\mu\nu}^{(\mathrm{m})}=0
\end{equation}
\textit{\textbf{is not conformally invariant in general}}, and transforms as follows \cite{Borowiec:2023kmq}:
\begin{equation}\label{cons-transf}
    \tilde{\nabla}^{\nu}\tilde{T}_{\mu\nu}^{(\mathrm{m})}=-\frac{\Omega'(\phi)}{\Omega(\phi)}\tilde{T}^{(\mathrm{m})}\tilde{\nabla}_{\mu}\phi\equiv Q_{\mu}\,.
\end{equation}
For a spatially homogeneous SF:
\begin{equation}
    \phi=\phi(t)
\end{equation}
\eqref{cons-transf} becomes \cite{Borowiec:2023kmq}:
\begin{equation}\label{cons-transf-hom}
    \tilde{\nabla}^{\nu}\tilde{T}_{\mu\nu}^{(\mathrm{m})}=-\frac{\Omega'(\phi)}{\Omega(\phi)}\tilde{T}^{(\mathrm{m})}\dot{\phi}\,.
\end{equation}
\begin{examplebox}[Conformal invariance of the covariant conservation law for the energy-momentum tensor]
    \begin{itemize}
        \item In the case of matter characterized by a \textit{\textbf{non-vanishing trace of the energy-momentum tensor}}:
        \begin{equation}\label{nonzero-trace}
            T^{(\mathrm{m})}\neq 0\,,
        \end{equation}
        the conservation law \textit{\textbf{is not conformally invariant}} and obeys the relation given in equation \eqref{cons-transf} or \eqref{cons-transf-hom};
        \item For a matter with vanishing trace (e.g., \textit{radiation/relativistic matter} or \textit{Maxwell field}):
        \begin{equation}
            T^{(\mathrm{m})}=0
        \end{equation}
        the covariant conservation law is \textit{\textbf{conformally invariant}}.
    \end{itemize}
    \textbf{\textit{\underline{Conclusion:}}}\\
    In the general case, as described by equation \eqref{nonzero-trace}, a conformal transformation implies the \textit{\textbf{transfer of energy and momentum between the matter sector and the SF}}.
\end{examplebox}
\begin{examplebox}[5th force]
    The transformation law described by formula \eqref{cons-transf} has the following important implications:
    \begin{itemize}
        \item \textit{Timelike geodesics} of the metric $g_{\mu\nu}$ \textbf{are not geodesics} of the transformed metric $\tilde{g}_{\mu\nu}$;
        \item Considering a \textit{\textbf{non-relativistic particle in the Newtonian limit}}, geodesic equation becomes:
        \begin{equation}
            \ddot{x}^{i}+\Gamma^{i}_{00}=-\frac{\Omega'(\phi)}{\Omega(\phi)}\tilde{\nabla}^{i}\phi\,;
        \end{equation}
        \item The Christoffel symbol $\Gamma^{i}_{00}$ contains the \textit{\textbf{Newtonian force}}:
        \begin{equation}
            \Gamma_{00}^{i}=\partial^{i}\Phi_{\mathrm{N}}
        \end{equation}
        and therefore, one could interpret the following as a \textit{\textbf{5th force}}\footnote{Assuming that we are dealing with a \textit{standard kinetic term} of the SF in the action.}:
        \begin{equation}
            F_{5}\equiv -\frac{\Omega'(\phi)}{\Omega(\phi)}\nabla\phi\,.
        \end{equation}
    \end{itemize}
    \textbf{\textit{\underline{Conclusion:}}}\\
    Particles in \textit{free fall} in the Universe $(\mathcal{M},g_{\mu\nu})$ are subject to a \textit{\textbf{5th force}} proportional to $\tilde{\nabla}^{\mu}\phi$ in the conformally rescaled Universe $(\mathcal{M},\tilde{g}_{\mu\nu})$.
\end{examplebox}
Furthermore, the \textbf{transformed} relationships for the \textbf{energy density} and \textbf{pressure} of material components take the following explicit forms \cite{Borowiec:2023kmq}:
\begin{subequations}
    \begin{equation}\label{en-den-transf}
        \tilde{\rho}^{(\mathrm{m})}(a,\phi)=\Omega^{-4}\rho^{(\mathrm{m})}\left(a\, \Omega^{-1}\right)\,,
    \end{equation}
    \begin{equation}\label{p-transf}
        \tilde{p}^{(\mathrm{m})}(a,\phi)=\Omega^{-4}p^{(\mathrm{m})}\left(a\,\Omega^{-1}\right)\,.
    \end{equation}
\end{subequations}
and satisfy the following \textbf{transformed (non)continuity equation}:
\begin{equation}
    \dot{\tilde{\rho}}^{(\mathrm{m})}+3H\left(\tilde{\rho}^{(\mathrm{m})}+\tilde{p}^{(\mathrm{m})}\right)=\frac{\Omega'(\phi)}{\Omega(\phi)}\dot{\phi}\left(3\tilde{p}^{(\mathrm{m})}-\tilde{\rho}^{(\mathrm{m})}\right)
\end{equation}
which, in the case of a \textit{barotropic fluid}:
\begin{equation}
    p^{(\mathrm{m})}=\omega^{(\mathrm{m})}\rho^{(\mathrm{m})}\quad\land\quad \omega^{(\mathrm{m})}=\mathrm{const} \,,
\end{equation}
takes the form:
\begin{equation}
    \dot{\tilde{\rho}}^{(\mathrm{m})}+3H\tilde{\rho}^{(\mathrm{m})}\left(1+\omega^{(\mathrm{m})}\right)=\frac{\Omega'(\phi)}{\Omega(\phi)}\dot{\phi}\,\tilde{\rho}^{(\mathrm{m})}\left(3\omega^{(\mathrm{m})}-1\right)\,.
\end{equation}
\section{Conformal frames: Mathematical and Physical (Non)equivalence}
\label{EJframes-section}
Different conformal frames, most notably the Jordan and Einstein frames, are admitted by STTs and higher-derivative gravity models. These frames are connected by a Weyl (rescaling) transformation of the metric. Dicke previously stressed that only dimensionless quantities are physical observables and that such transformations can be understood as local changes of units \cite{Dicke:1961gz}. As long as units and matter couplings are handled consistently, this observation serves as the foundation for the contemporary belief that various conformal frames can be thought of as distinct parametrizations of the same physics \cite{Magnano:1993bd,Faraoni:2006fx,Catena:2006bd,Postma:2014vaa}.

Nevertheless, there is a \textit{\textbf{noticeable difference in viewpoint}} in the literature. Some authors argue that the \textit{\textbf{frames are physically and mathematically equivalent}} \cite{Magnano:1993bd,Faraoni:2006fx,Catena:2006bd,Postma:2014vaa,Flanagan:2004bz,Chiba:2013mha,Domenech:2016yxd}, whereas others claim that there is \textit{\textbf{physical non-equivalence at the classical or quantum level}} \cite{Faraoni:1999hp,Capozziello:2010sc,Quiros:2012rnn,Kamenshchik:2014waa}. Some even characterize the entire controversy as a '\textit{\textbf{pseudo-issue}}' resulting from the \textit{improper application of conformal transformations} \cite{Faraoni:2006fx,Quiros:2012rnn,Faraoni:1998qx}.

In this subsection, we examine the issue of (non-)equivalence of conformal frames based on the existing literature \cite{Dicke:1961gz,Magnano:1993bd,Faraoni:2006fx,Catena:2006bd,Postma:2014vaa,Flanagan:2004bz,Chiba:2013mha,Domenech:2016yxd,Faraoni:1999hp,Capozziello:2010sc,Quiros:2012rnn,Kamenshchik:2014waa,Faraoni:1998qx}. We structure our discussion by considering three main aspects:
\begin{enumerate}[label=(\alph*)]
  \item \textit{\textbf{Mathematical (non)equivalence}} of the formulations;
  \item \textit{\textbf{Physical (non)equivalence}} at the \textit{\textbf{classical}} and \textit{\textbf{quantum}} levels;
  \item Divergence of opinions regarding \textit{\textbf{which frame, if any, should be regarded as 'physical' one}}.
\end{enumerate}
\subsection{STTs in the Jordan frame}
\subsubsection{Action and FEs}
The general action for the STT in the \textit{\textbf{Jordan (conformal) frame}} can be written as follows \cite{Faraoni:2006fx,Catena:2006bd,Flanagan:2004bz,Faraoni:1998qx}:
\begin{equation}
  S_{J}\bigl[g_{\mu\nu},\phi,\psi_{\mathrm{m}}\bigl]
  = \int d^4x\,\sqrt{-g}\left[\frac{1}{2\kappa^2}\mathcal{A}(\phi) R-\frac{1}{2}\mathcal{B}(\phi) g^{\mu\nu}\partial_{\mu}\phi\,\partial_{\nu}\phi-\mathcal{V}(\phi)\right]+S_{\mathrm{m}}\bigl[g_{\mu\nu},\psi_{\mathrm{m}}\bigr],
  \label{eq:SJ-def}
\end{equation}
and yields the following field equations for the metric and SF sectors:
\begin{equation}
  \mA(\phi) G_{\mu\nu}= \kappa^2\,T^{(\mm)}_{\mu\nu}+ \mB(\phi)\left(\partial_{\mu}\phi\,\partial_{\nu}\phi-\frac{1}{2} g_{\mu\nu}\,\partial_{\alpha}\phi\,\partial^{\alpha}\phi\right)-g_{\mu\nu} \mV(\phi)+\nabla_{\mu}\nabla_{\nu}\mA(\phi)-g_{\mu\nu}\Box \mA(\phi)\,,
  \label{eq:Jordan-Einstein-eq}
\end{equation}
\begin{equation}
    \mB(\phi)\,\Box\phi+\frac{1}{2}\mB'(\phi)\,\partial_{\alpha}\phi\,\partial^{\alpha}\phi-\frac{1}{2\kappa^2}\mA'(\phi)\,R+\mV'(\phi)= 0\,.
\label{eq:Jordan-scalar-eq}
\end{equation}
Diffeomorphism invariance implies:
\begin{equation}
  \nabla_{\mu}T^{(\mm)\mu}{}_{\nu}=0\,,
\end{equation}
This means that the \textit{\textbf{matter is covariantly conserved in the Jordan frame}}.
\subsubsection{Geodesics and WEP}
When non-gravitational forces are absent, the \textit{test particles follow the geodesics} of the Jordan frame, i.e. matter is MC to $g_{\mu\nu}$:
\begin{equation}
  S_{\mathrm{tp}}^{(\mathrm{J})}=-m\int ds_{J}\quad\land\quad ds_{J}^{2}=g_{\mu\nu}dx^{\mu}dx^{\nu}\,,
\end{equation}
and leads to the geodesic equation:
\begin{equation}
    \frac{d^{2}x^{\mu}}{d\tau_{J}^2}+\Gamma^{\mu}_{\nu\rho}\frac{dx^{\nu}}{d\tau_{J}}\frac{dx^{\rho}}{d\tau_{J}}=0\,.
\end{equation}
Consequently, the Jordan frame exhibits the \textit{\textbf{weak equivalence principle}} (WEP) by construction, which is often used to support the idea that it is the '\textit{physical}' frame. However, Dicke noted that a local change of units can be understood as a conformal rescaling of the metric \cite{Dicke:1961gz}. According to this view, the definition of the unit system determines whether matter is MC, and the concept of a physical frame must be expressed in terms of dimensionless observables rather than metric couplings alone \cite{Magnano:1993bd,Faraoni:2006fx,Catena:2006bd,Postma:2014vaa}.
\subsection{Conformal transformation to the Einstein frame}
\subsubsection{Weyl rescaling and curvature transformation}
Let us consider a general Weyl (conformal) rescaling:
\begin{equation}
    \tilde{g}_{\mu\nu}= \Omega^{2}(x)\,g_{\mu\nu},\quad\land\quad \Omega(x)>0\,.
\label{eq:weyl-rescaling}
\end{equation}
We know that the inverse metric, its determinant, and Ricci scalar transform in the following manner (see, Sec.~\ref{Conf-transf-subsection}):
\begin{subequations}
    \begin{equation}
        \tilde{g}^{\mu\nu}=\Omega^{-2}\,g^{\mu\nu}\,,
    \end{equation}
    \begin{equation}
        \sqrt{-\tilde{g}}=\Omega^{4}\,\sqrt{-g}\,,
    \end{equation}
    \begin{equation}\label{eq:R-transform}
        \tilde{R}=\Omega^{-2}\Bigl[R-6\Bigl\{\Box\ln\Omega+g^{\mu\nu}\bigl(\partial_{\mu}\ln{\Omega}\bigr)\,\bigl(\partial_{\nu}\ln{\Omega}\bigr)\Bigr\}\Bigr]\,.
    \end{equation}
\end{subequations}
By substituting \eqref{eq:R-transform} into \eqref{eq:SJ-def} and choosing the conformal factor $\Omega(x)$ appropriately, the non-minimal coupling described by the $\mA(\phi)R$ term can be mapped into a \textit{\textbf{canonical Einstein-Hilbert term}}.
\subsubsection{Choice of the conformal factor}
In the new (Einstein) frame, we require the gravitational sector to have the canonical form, namely:
\begin{equation}
    S_{E}\supset \frac{\MPl^{2}}{2}\int d^{4}x\,\sqrt{-\tilde{g}}\,\tilde{R}=\frac{1}{2\kappa^2}\int d^{4}x\,\sqrt{-\tilde{g}}\,\tilde{R}\,.
\end{equation}
This can be accomplished by choosing the following form of $\Omega(\phi)$:
\begin{equation}\label{eq:Omega-choice}
    \Omega^{2}(\phi)=\mA(\phi)\,,
\end{equation}
and therefore:
\begin{equation}
    \sqrt{-g}\,\frac{\mA(\phi)}{2\kappa^2}R=\frac{\sqrt{-\tilde g}}{\kappa^{2}}\,\left[\frac{1}{2}\tilde{R}-\frac{3}{2}\tilde{g}^{\mu\nu}\Bigl(\partial_{\mu}\ln{\mA(\phi)}\Bigr)\,\Bigl(\partial_{\nu}\ln{\mA(\phi)}\Bigr)-3\,\tilde\Box \ln{\mA(\phi)}\right]\,,
\end{equation}
The last term can be removed from the action since it is a \textit{total derivative} \cite{Faraoni:1998qx,Flanagan:2004bz,Faraoni:2006fx}.
\subsubsection{Kinetic term and canonical normalization}
The SF kinetic term transforms as follows:
\begin{equation}\label{kinetic-transf}
\begin{aligned}
    -\frac{1}{2}\sqrt{-g}\,\mB(\phi)\,g^{\mu\nu}\partial_{\mu}\phi\,\partial_{\nu}\phi
  &= -\frac{1}{2}\sqrt{-\tilde{g}}\,\Omega^{-2}(\phi)\,\mB(\phi)\,
     \tilde{g}^{\mu\nu}\,\partial_{\mu}\phi\,\partial_{\nu}\phi \\
  &= -\frac{1}{2}\sqrt{-\tilde{g}}\,\frac{\mB(\phi)}{\mA(\phi)}
     \tilde{g}^{\mu\nu}\,\partial_{\mu}\phi\,\partial_{\nu}\phi\,.
\end{aligned}
\end{equation}
Combining \eqref{kinetic-transf} with the derivative term from the curvature transformation gives the explicit form of the \textit{\textbf{Einstein frame kinetic prefactor}} \cite{Faraoni:1998qx,Catena:2006bd,Postma:2014vaa}:
\begin{equation}\label{eq:Kphi-def}
    \mathcal{K}(\phi)\equiv\frac{\mB(\phi)}{\mA(\phi)}+\frac{3}{2\kappa^2}\left(\frac{\mA'(\phi)}{\mA(\phi)}\right)^2\,.
\end{equation}
This results in the following expression for the \textit{\textbf{scalar sector}} in the Einstein frame:
\begin{equation}
    S_{\phi}^{(\mathrm{E})}=-\int d^{4}x\,\sqrt{-\tilde{g}}\,\left[\frac{1}{2}\mathcal{K}(\phi)\,\tilde{g}^{\mu\nu}\partial_{\mu}\phi\,\partial_{\nu}\phi-\tilde{\mV}(\phi)\right]
\end{equation}
with the SF potential:
\begin{equation}\label{eq:potential-transform}
    \tilde{\mV}(\phi)=\Omega^{-4}(\phi)\,\mV(\phi)=\frac{\mV(\phi)}{\mA^{2}(\phi)}\,.
\end{equation}
If $\mathcal{K}(\phi)>0$, a \textit{\textbf{canonically normalized Einstein frame SF}} $\chi$ could be defined as:
\begin{equation}\label{eq:chi-phi-redef}
  \frac{d\chi}{d\phi}=\sqrt{\mathcal{K}(\phi)}\,.
\end{equation}
and yields the \textit{\textbf{final form of the Einstein frame action for the STT}}:
\begin{equation}\label{eq:SE-grav-scalar}
    S_{\mathrm{E}}=\int d^{4}x\,\sqrt{-\tilde{g}}\left[\frac{\MPl^2}{2}\tilde{R}-\frac{1}{2}\tilde{g}^{\mu\nu}\partial_{\mu}\chi\,\partial_{\nu}\chi-\mV(\chi)\right]\,,
\end{equation}
with:
\begin{equation}
    \mV(\chi)\equiv\tilde{\mV}\bigl[\phi\left(\chi\right)\bigr]
\end{equation}
given by \eqref{eq:potential-transform}.
\subsubsection{Matter sector and non-geodesic motion}
The matter action undergoes the following transformation:
\begin{equation}
    S_{\mm}\left[g_{\mu\nu},\psi_{\mm}\right] = S_{\mm}\left[\Omega^{-2}(\phi)\,\tilde{g}_{\mu\nu},\psi_{\mm}\right]=S_{\mm}\left[\mA(\phi)^{-1}\tilde{g}_{\mu\nu},\psi_{\mm}\right]\,.
\end{equation}
For a test particle, one finds:
\begin{equation}\label{SF-dep-mass1}
    S_{\mathrm{tp}}^{(\mathrm{E})}=-m \int ds_{J}=-m \int\sqrt{-g_{\mu\nu}\,dx^{\mu} dx^{\nu}}=-m \int \Omega^{-1}(\phi)\, d\tilde{s}\,,
\end{equation}
where:
\begin{equation}
    d\tilde{s}^2=-\tilde{g}_{\mu\nu}dx^{\mu} dx^{\nu}\,.
\end{equation}
\begin{examplebox}[(Varying) SF-dependent mass and 5th force in the Einstein frame]
\begin{itemize}
    \item The relation \eqref{SF-dep-mass1} could be interpreted as the existence of a \textit{\textbf{(varying) SF-dependent mass}} of the test particle in the Einstein frame described by the explicit form:
    \begin{equation}
        \tilde{m}\left(\chi\right)=m\,\Omega^{-1}\left[\phi\left(\chi\right)\right]=\frac{m}{\sqrt{\mA\left[\phi\left(\chi\right)\right]}}\,;
    \end{equation}
    \item Consequently, \textit{\textbf{massive particles are accelerated by an effective '5th force'}} sourced by the $\chi$-dependence and \textit{\textbf{do not follow the geodesics}} of the $\tilde{g}_{\mu\nu}$ metric \cite{Faraoni:1999hp,Faraoni:2006fx}.
\end{itemize}
\end{examplebox}
As previously stated, the energy-momentum tensor for matter, $T_{\mu\nu}^{(\mm)}$, satisfies the (non)conservation equation of the form:
\begin{equation}
    \tilde{\nabla}_{\mu}\tilde{T}^{(\mm)\mu}{}_{\nu}=\alpha\left(\chi\right)\tilde{T}^{(\mm)}\tilde{\nabla}_{\nu}\chi\quad\land\quad \alpha\left(\chi\right)\equiv\frac{d\ln{\mA(\chi)}}{d\chi}\,.
\end{equation}
The \textit{\textbf{total energy-stress tensor}} (including both SF and matter) \textit{\textbf{is covariantly conserved}}, indicating the presence of underlying \textit{diffeomorphism invariance}. However, only its \textit{splitting} into 'geometry', 'scalar', and 'matter' components \textit{is frame-dependent}.
\subsubsection{Mathematical equivalence as field redefinition}
One can show that we can relate the actions \eqref{eq:SJ-def} and \eqref{eq:SE-grav-scalar} together with the transformed matter sector by an \textit{\textbf{invertible field redefinition}}:
\begin{equation}
    \Bigl(g_{\mu\nu},\phi,\psi_{\mm}\Bigr)\quad\longleftrightarrow\quad\Bigl(\tilde{g}_{\mu\nu},\chi,\psi_{\mm}\Bigr)
\end{equation}
as long as:
\begin{equation}
    \mA(\phi)>0\quad\land\quad \mathcal{K}(\phi)>0\,,
\end{equation}
in order for $\Omega(\phi)$ and $\chi(\phi)$ to be \textit{smooth} and \textit{monotonic} functions \cite{Magnano:1993bd,Faraoni:2006fx,Catena:2006bd,Postma:2014vaa,Flanagan:2004bz,Faraoni:1998qx}.\\
Solutions to the Jordan frame FEs correspond \textit{one-to-one} with solutions to the Einstein frame EoM, provided that singularities of the conformal factor are avoided. This establishes the \textit{\textbf{mathematical equivalence between the two formulations}}.

The central question is \textit{\textbf{whether physical observables}} - that is, \textit{relations between dimensionless quantities that are measured experimentally} - \textit{\textbf{are also invariant under this mapping}}. This requires careful consideration of units and what is held fixed upon comparing frames.
\subsection{Classical physical (non)equivalence}
\subsubsection{Conformal rescalings as changes of units (Dicke’s interpretation)}
Dicke's analysis demonstrates that a conformal rescaling can be regarded as a local change of units \cite{Dicke:1961gz}. Under the rescaling given by \eqref{eq:weyl-rescaling}, suppose that the \textit{unit length}, \textit{time}, and \textit{mass} are also rescaled:
\begin{equation}
    \tilde\ell_{\mathrm{u}}=\Omega\,\ell_{\mathrm{u}}\quad\land\quad\tilde{t}_{\mathrm{u}}=\Omega\,t_{\mathrm{u}}\quad\land\quad\tilde{m}_{\mathrm{u}}=\Omega^{-1}\,m_{\mathrm{u}}\,.
\end{equation}
Then the \textit{physical length} $L$ transforms as:
\begin{equation}
    \tilde{L}=\Omega\,L\,,
\end{equation}
but the \textit{dimensionless ratio}:
\begin{equation}
    \frac{\tilde{L}}{\tilde{\ell}_\mathrm{u}}=\frac{L}{\ell_{\mathrm{u}}}
\end{equation}
\textit{remains invariant}. The same applies to time and mass.

The energy densities scale as given in \eqref{en-den-transf}. The unit energy density also scales in this way \cite{Faraoni:2006fx}, namely:
\begin{equation}
    \frac{\tilde{\rho}}{\tilde{\rho}_{\mathrm{u}}}=\frac{\rho}{\rho_{\mathrm{u}}}\,.
\end{equation}
\textit{Catena}, \textit{Pietroni} and \textit{Scarabello} \cite{Catena:2006bd}, as well as \textit{Postma} and \textit{Volponi} \cite{Postma:2014vaa}, systematically developed this viewpoint. They introduced \textbf{frame-invariant combinations} of the metric and the local unit scale, showing that \textit{\textbf{all physical observables can be expressed in terms of such invariants}}, \textit{independently of the chosen conformal frame}. In this conceptual framework, the \textit{\textbf{Einstein and Jordan frames represent distinct unit conventions rather than different physics}}.
\subsubsection{Cosmological observables}
In a spatially flat FLRW spacetime, the Jordan frame line element can be written as follows:
\begin{equation}
    ds_{\mathrm{J}}^{2}=-dt^{2}+a^2(t)\,d\vec{x}^{2}\,.
\end{equation}
According to the selected form of the conformal factor \eqref{eq:Omega-choice} one obtains:
\begin{equation}
    \tilde{g}_{\mu\nu}=\mA(\phi)\,g_{\mu\nu}\,,
\end{equation}
which yields:
\begin{equation}
    ds_{\mathrm{E}}^2=-d\tilde{t}^{2}+\tilde{a}^2(\tilde{t})\,d\vec{x}^2\,,
\end{equation}
where:
\begin{equation}
    d\tilde{t}=\sqrt{\mA(\phi)}\,dt \quad\land\quad \tilde{a}(\tilde{t})=\sqrt{\mA(\phi)}\,a(t)\,.
\end{equation}
Null geodesics are \textit{invariant under conformal transformations}, meaning that the \textit{comoving radial coordinate of light rays} and the \textit{conformal time} remain unchanged. \textit{Chiba} and \textit{Yamaguchi} explicitly derive the relation between the background and perturbation variables in the Jordan and Einstein frames \cite{Chiba:2013mha}, and demonstrate that \textit{\textbf{cosmological observables}} such as redshift, luminosity, angular diameter distance, and CMB temperature anisotropy \textit{\textbf{are frame-independent}} when expressed in terms of \textit{suitably defined physical parameters and units}.

Similarly, \textit{Catena et al.} \cite{Catena:2006bd} reformulate the Boltzmann equation, particle freeze-out and CMB photon evolution in a frame-invariant way. Moreover \textit{Dom\`enech} and \textit{Sasaki} review these results and conclude that classical cosmological observables are frame-invariant for STTs with universal matter coupling once units and couplings are properly translated \cite{Domenech:2016yxd}.

\subsubsection{Einstein frame as 'physical' frame in nonlinear gravity}
The analysis by \textit{Magnano} and \textit{Sokołowski} \cite{Magnano:1993bd} shows that, in certain conditions, nonlinear $f(R)$ gravity is equivalent to GR plus a SF in the Einstein frame. They point out that in this frame, the scalar possesses a standard kinetic term and positive energy.

In contrast, in the Jordan frame, the effective energy associated with higher-derivative terms could be negative or violate the energy conditions. Therefore, they regard the \textit{\textbf{Einstein frame as the most 'physical' representation of this class of theories}}, despite the underlying mathematical equivalence.

\subsubsection{Conformal frame 'pseudoissue' and singularities}
\textit{Faraoni}, \textit{Gunzig} and \textit{Nardone} \cite{Faraoni:1998qx} provide a detailed discussion of conformal transformations and highlight that many apparent differences between frames arise from the \textit{comparison of dimensionful quantities under different unit conventions}. Moreover, \textit{Faraoni} and \textit{Nadeau} \cite{Faraoni:2006fx} further support this argument by analyzing examples in which a cosmological singularity appears to be present in one frame but absent in another. They demonstrate that, \textit{once running units are implemented in the Einstein frame, the presence or absence of singularities in dimensionless observables becomes frame-independent}. They refer to the debate concerning the conformal frames as a \textit{\textbf{'pseudoissue' at the classical level}}.

\subsubsection{Explicit non-equivalence in $f(R)$ cosmology}
\textit{Capozziello}, \textit{Martin-Moruno} and \textit{Rubano} study a specific $f(R)$ model \cite{Capozziello:2010sc}, comparing its cosmological evolution in the Jordan and Einstein frames. They found that \textit{if the same matter sector is identified and units are treated as fixed in both frames}, the \textit{background dynamics can differ} to the extent that \textbf{only the Jordan frame model fits supernova data}, while the \textbf{Einstein frame model does not}. They interpret this as \textbf{\textit{physical non-equivalence of the frames for that theory}}.

From another perspective \cite{Faraoni:2006fx,Catena:2006bd,Postma:2014vaa,Chiba:2013mha}, this conclusion arises from the \textit{inconsistent identification of physical parameters and units between the conformal frames}. However, when matter couplings and unit conventions are systematically converted, the \textit{dimensionless distance modulus becomes frame-invariant}, thereby \textbf{eliminating the apparent non-equivalence}. Therefore, many 'counterexamples' can be understood as \textit{comparisons between different theories rather than different frames of the same theory}.

\subsection{Geodesics, energy conditions and singularities}
Since conformal transformations preserve \textit{null cones}, \textit{null geodesics} and the \textit{causal structure} \textit{are also invariant}. Nevertheless, timelike geodesics are not generally mapped to timelike geodesics of the transformed metric due to the \textit{\textbf{effective scalar force in the Einstein frame}} \cite{Faraoni:1999hp,Faraoni:2006fx}. This results in a \textit{\textbf{dependence on the conformal frame of the naive formulation of the WEP}} if \textit{masses and units are not transformed simultaneously}.

The \textit{energy conditions}, which are formulated in terms of the energy-momentum and Einstein tensors, \textit{transform non-trivially under conformal rescalings} \cite{Magnano:1993bd,Faraoni:1998qx}. Depending on how the total energy-momentum is split between the scalar and matter sectors, \textbf{apparent violations of the WEC or NEC may be found in one frame but not another}. \textit{Faraoni} and \textit{Nadeau} \cite{Faraoni:2006fx} demonstrate that, when focusing on physical, dimensionless quantities, \textbf{energy conditions and singularity structure are frame-independent}. \textit{Capozziello et al.} \cite{Capozziello:2010sc} and \textit{Quiros et al.} \cite{Quiros:2012rnn} interpret some of these differences as \textbf{evidence of genuine non-equivalence}.

\subsection{Quantum aspects and frame dependence}
\subsubsection{Path integrals, measure and off-shell quantities}
At the quantum level, a conformal transformation corresponds to a nonlinear field redefinition in the path integral. According to the equivalence theorem, $S$-matrix elements should be invariant under such redefinitions if the Jacobian is incorporated correctly. However, quantum corrections are typically examined using the background-field method within an \textit{off-shell effective action} formalism. \textit{Off-shell quantities}, such as the \textit{one-particle irreducible effective action} depend on the choice of field variables, which can lead to \textbf{frame dependence non-equivalence} \cite{Kamenshchik:2014waa,Domenech:2016yxd}.

\textit{Kamenshchik} and \textit{Steinwachs} \cite{Kamenshchik:2014waa} calculate \textit{one-loop divergences} in general STTs in both the Jordan and Einstein frames using the \textit{generalized Schwinger–DeWitt technique}. They find that:
\begin{enumerate}[label=(\alph*)]
    \item The off-shell one-loop counterterms and $\beta$ functions \textbf{are different between frames};
    \item In agreement with the \textit{equivalence theorem}, the \textbf{divergent parts coincide on-shell}.
\end{enumerate}
Therefore, \textbf{quantum corrections are frame-dependent at the off-shell level but frame-independent for on-shell observables}, at least to \textit{one loop}.
\subsubsection{RG $\beta$ functions from one-loop divergences}
In the context of \textit{Kamenshchik} and \textit{Steinwachs}' one-loop analysis of STTs \cite{Kamenshchik:2014waa}, the (perturbative) renormalization group (RG) functions that control the scale dependence of renormalized couplings (or, more generally, coupling \textit{functionals}) are referred to as $\beta$ functions.

A set of renormalized couplings at the renormalization scale $\mu$ is represented by $\bigl\{g_i\bigr\}$. The \textbf{\textit{beta functions}} \textit{encode their RG running} \cite{Collins:1984xc,Peskin:1995ev,Bagnuls:2000ae}:
\begin{equation}\label{beta-RG-eq1}
    \beta_{g_{i}}\Bigl(\bigl\{g\bigr\}\Bigr)\equiv \mu\frac{d g_{i}\left(\mu\right)}{d\mu}=\frac{d g_{i}}{d\ln\mu}\,.
\end{equation}
Alternatively, if we set:
\begin{equation}
    t\equiv\ln\mu\,,
\end{equation}
then one can write the another form of \eqref{beta-RG-eq1} \cite{Collins:1984xc,Peskin:1995ev,Bagnuls:2000ae}:
\begin{equation}
    \frac{d g_{i}}{dt}=\beta_{g_i}\,.
\end{equation}

\subsubsection{Extraction from one-loop divergences}
Using dimensional regularization:
\begin{equation}
    d=4-\epsilon\,,
\end{equation}
the divergent part of the one-loop effective action takes the following generic local form:
\begin{equation}\label{eq:GammaDivGeneric}
    \Gamma^{(1)}_{\mathrm{div}}=\frac{1}{\epsilon}\int d^{4}x\,\sqrt{-g}\,
  \sum_{i}{c_{i}\Bigl(\bigl\{g\bigr\}\Bigr)\,\mathcal{O}_i}\,,
\end{equation}
where:
\begin{itemize}
    \item $\bigl\{\mathcal{O}_i\bigr\}$ is the basis of \textit{local operators} consistent with the symmetries (e.g. $R^{2}$, $R_{\mu\nu}R^{\mu\nu}$, $R\left(\nabla\phi\right)^{2}$ etc.);
    \item $c_i$ are functions of the \textit{renormalized couplings}. 
\end{itemize}
Renormalization introduces counterterms that cancel out the divergence \eqref{eq:GammaDivGeneric}. The RG running and, consequently, the $\beta$ functions in minimal subtraction schemes are determined by the residues of the simple $\frac{1}{\epsilon}$ poles \cite{Collins:1984xc,Peskin:1995ev,Bagnuls:2000ae}.

\subsubsection{ST specific form: $\beta$ functionals}
In general, the couplings in STTs are often SF-dependent functions (e.g. $\mA(\phi)$, $\mB(\phi)$, $\mV(\phi)$). For the considered Jordan frame action for STT \eqref{eq:SJ-def}, the RG running can be described by the following beta functionals:
\begin{equation}
    \beta_{\mA}(\phi)\equiv \mu\frac{d\mA(\phi)}{d\mu}\quad\land\quad\beta_{\mB}(\phi)\equiv\mu\frac{d\mB(\phi)}{d\mu}\quad\land\quad\beta_{\mV}(\phi)\equiv\mu\frac{d\mV(\phi)}{d\mu}\,.
\end{equation}
\subsubsection{Off-shell frame dependence vs. on-shell equivalence}
We extract conventional beta functions from \textit{\textbf{off-shell}}. Due to their \textit{\textbf{UV divergences}}, they can change under nonlinear field redefinitions \cite{Kamefuchi:1961sb} (e.g. \textit{Jordan to Einstein frame}), even when \textit{on-shell observables remain invariant}. This is consistent with the equivalence theorem for field redefinitions and the underlying on-shell invariance. A classic approach to eliminating off-shell parameterization dependence at the level of the effective action is the \textit{'unique' effective action} construction of \textit{Vilkovisky} \cite{Vilkovisky:1984st}.

\subsubsection{Effective actions, renormalization and cosmology}
In cosmological applications, one often works with an effective action evaluated on backgrounds that \textit{\textbf{do not exactly solve the classical equations}} (e.g. during \textit{slow-roll inflation}), and one is interested in RG improved potentials and beta functions. In such contexts, the dependence of the effective action on the off-shell frame implies that \textit{\textbf{RG flows and apparent fine-tuning issues may appear different in different frames}} \cite{Kamenshchik:2014waa,Domenech:2016yxd}.

\textit{Dom\`enech} and \textit{Sasaki} \cite{Domenech:2016yxd} emphasize that cosmological observables remain frame-equivalent at the classical level. Nevertheless, when quantum corrections are included, it is important to ensure that renormalization conditions and the identification of physical parameters are specified in a frame-independent manner. \textit{Kamenshchik} and \textit{Steinwachs} \cite{Kamenshchik:2014waa} propose the use of a geometrical, \textit{field-space covariant version of the background-field method} (\textit{\textbf{Vilkovisky–DeWitt effective action}}) in order to minimize such ambiguities.

\subsubsection{Conformal transformations vs. Weyl invariance}
\textit{Quiros et al.} \cite{Quiros:2012rnn} draw attention to the difference between the \textit{\textbf{generic conformal transformations}} that are used to relate the \textit{Jordan} and \textit{Einstein} frames, and the \textit{\textbf{genuine local Weyl invariance}} of a theory. Brans-Dicke and related ST models \textit{are not invariant under Weyl transformations}, therefore, \textit{\textbf{one cannot rely on Weyl symmetry to claim frame equivalence}}.

Quantum anomalies further complicate the situation, even in theories with \textit{classical scale invariance}. Based on this perspective, they consider the \textit{\textbf{conformal frame debate to be a consequence of employing symmetry arguments for transformations that do not represent actual symmetries of the microscopic theory}}.

\subsection{Divergent viewpoints on the 'physical' frame}
\subsubsection{Supporters of equivalence}
The first group of scientists suggests that the \textit{\textbf{Jordan and Einstein frames are physically equivalent}} and that any \textit{\textbf{apparent discrepancy arises from the inconsistent identification of physical quantities}}:
\begin{itemize}
    \item \textit{Magnano} and \textit{Soko{\l}owski} \cite{Magnano:1993bd} demonstrate the \textit{\textbf{dynamical equivalence}} between $f(R)$ gravity and Einstein gravity with a SF. Although they favour the \textit{Einstein frame as being physically more transparent}, they acknowledge the underlying \textit{equivalence with the Jordan frame};
    \item According to \textit{Faraoni} and \textit{Nadeau} \cite{Faraoni:2006fx}, when running units are adopted and the focus moves towards dimensionless observables, the \textit{\textbf{problem of conformal frames becomes a 'pseudoissue'}};
    \item \textit{Catena et al.} \cite{Catena:2006bd} and \textit{Postma et al.} \cite{Postma:2014vaa} construct explicit, \textit{\textbf{frame-invariant formulations of ST cosmology}}, including \textit{Boltzmann equations} and \textit{perturbations}. Thus, they make \textit{\textbf{frame equivalence}} manifest;
    \item \textit{Dom\`enech} and \textit{Sasaki} \cite{Domenech:2016yxd} review the use of conformal frames in cosmology, concluding that \textit{\textbf{classical cosmological observables are frame-equivalent}} under \textit{broadly defined conditions}.
\end{itemize}

\subsubsection{Supporters of non-equivalence}
The second group emphasizes the \textit{\textbf{physical non-equivalence of frames}}, either in the \textit{classical} or \textit{quantum} sense:
\begin{itemize}
    \item \textit{Faraoni} and \textit{Gunzig} \cite{Faraoni:1999hp} point out that the \textit{\textbf{WEP is not satisfied in the Einstein frame}}, and that \textit{\textbf{matter does not move along geodesics of the metric tensor}} $\tilde{g}_{\mu\nu}$. They suggest that this is \textit{\textbf{evidence against the physicality of the Einstein frame}};
    \item \textit{Capozziello}, \textit{Martin-Moruno} and \textit{Rubano} \cite{Capozziello:2010sc} present an explicit $f(R)$ cosmological model that demonstrates how the Jordan and Einstein frames \textit{produce different fits to supernova data} when interpreted using \textit{fixed units}. They interpret such differences as \textit{\textbf{indicating physical non-equivalence}};
    \item \textit{Quiros et al.} \cite{Quiros:2012rnn} highlight that the \textit{\textbf{literature on conformal transformations frequently confuses a simple change of variables with Weyl invariance}}, emphasizing that \textit{\textbf{there is no compelling reason to assume frame equivalence in STTs}};
    \item \textit{Kamenshchik} and \textit{Steinwachs} \cite{Kamenshchik:2014waa} demonstrate that \textit{\textbf{off-shell quantum corrections differ between the conformal frames}}. They interpret such differences as evidence that \textit{\textbf{quantum theories formulated in different conformal frames are strictly inequivalent}}, unless one limits oneself to \textit{on-shell observables}.
\end{itemize}

\subsubsection{Interpretation vs. observables}
In a cosmological context, \textit{Dom\`enech} and \textit{Sasaki} \cite{Domenech:2016yxd} point out that \textit{\textbf{different conformal frames correspond to different splits of the system into 'gravity' and 'matter' sectors}}. In one frame, the inflaton is \textit{minimally coupled and drives an accelerating phase}, whereas in another, the same physical evolution is described as a \textit{modified matter sector in a different geometry}. While observables such as the \textit{CMB power spectrum} remain unchanged, the interpretation of their causes differs between the frames. This interpretational flexibility contributes to the ongoing debate over which, if any, frame should be considered '\textit{\textbf{physical}}'.

\subsection{Pragmatic viewpoint for cosmology}
A \textit{\textbf{pragmatic viewpoint regarding the issue of conformal frames}} emerges for the purposes of practical cosmological model building and data analysis:
\begin{enumerate}[label=(\alph*)]
    \item At the \textit{\textbf{classical level}}, the \textit{\textbf{Jordan and Einstein frames are equivalent}} \textit{provided one works with dimensionless, frame-invariant observables and translates matter couplings and units consistently} \cite{Faraoni:2006fx,Catena:2006bd,Postma:2014vaa,Chiba:2013mha,Domenech:2016yxd};
    \item \textit{\textbf{Different interpretations}} (e.g. which field 'drives' acceleration, the localization of energy, and the matter vs. gravity sector) \textit{\textbf{reflect conventions rather than different physics}}, provided one uses \textit{universal matter couplings};
    \item \textit{\textbf{At the quantum level}}, \textit{off-shell effective actions} and \textit{RG flows} \textit{\textbf{depend on the chosen frame}}. It is important to \textit{\textbf{take care to ensure that renormalization conditions and definitions of physical parameters are formulated consistently}} \cite{Kamenshchik:2014waa,Domenech:2016yxd}.
\end{enumerate}

\subsection{Open directions}
There are several issues that remain open and require further investigation:
\begin{enumerate}[label=(\roman*)]
    \item The \textit{\textbf{role of conformal anomalies}} and \textit{\textbf{explicit scale symmetry breaking}} in establishing the \textit{\textbf{quantum status associated with frame transformations}} \cite{Quiros:2012rnn,Kamenshchik:2014waa};
    \item The extension of \textit{\textbf{frame-invariant formulation}} to \textit{\textbf{multi-field}} \cite{Damour:1992we,Berkin:1993bt,Rainer:1996gw,Kuusk:2014sna,Kuusk:2015dda} and \textit{\textbf{higher-derivative STTs}} \cite{Clifton:2011jh,Burgess:2003jk,Nojiri:2010wj,Nojiri:2017ncd,Sotiriou:2008rp,DeFelice:2010aj} and \textit{\textbf{their implementation in realistic scenarios of dark energy and the early Universe}} \cite{Catena:2006bd,Domenech:2016yxd};
    \item The behavior of \textit{\textbf{energy conditions}} (\textit{classical} \cite{Visser:1999de,Barcelo:2002bv,Cattoen:2006yh,Curiel2017,MartinMoruno2017,Kontou:2020bta,Hawking_Ellis_2023ECs}, \textit{semi-classical} \cite{Ford:1994bj,MartinMoruno2017}, and \textit{quantum} \cite{Ford:1994bj,Kontou:2020bta}), \textit{\textbf{singularity theorems}} \cite{Penrose:1964wq,Hawking:1970zqf,Hawking1996nature,Senovilla:1998oua,Hawking_Ellis_2023Sing1,Hawking_Ellis_2023Sing2,Hawking_Ellis_2023Sing3,Tipler:1978zz,Cattoen:2005dx} and \textit{\textbf{horizon thermodynamics}} \cite{Bekenstein:1972tm,Bekenstein:1973ur,Bardeen:1973gs,Bekenstein:1974ax,Hawking:1975vcx,Wald:1999vt,Padmanabhan:2002sha,Padmanabhan:2003gd,Bousso:2002ju} under \textit{conformal transformations}, particularly when \textit{\textbf{non-trivial matter couplings are present}} \cite{Magnano:1993bd,Faraoni:2006fx,Capozziello:2010sc,Faraoni:1998qx}.
\end{enumerate}
These topics connect the debate about the conformal frame to more general questions concerning the geometrical foundations of gravity, the practical definition of observables in cosmology and the structure of low-energy EFTs of gravity \cite{Burgess:2003jk,Donoghue:1994dn}.

\begin{savequote}
"Although we have remarkable observational evidence that something like inflation occurred in the early universe, inflation cannot yet be considered a part of the standard model of cosmology, with the same level of confidence as, for example, BBN is a fact about the early universe."
\qauthor{\textbf{Daniel Baumann} \cite{Baumann:2022mni}}
"You know how sometimes you meet somebody and they’re really nice, so you invite them over to your house and you keep talking with them and they keep telling you more and more cool stuff? But then at some point you’re like, maybe we should we call it a day, but they just won’t leave and they keep talking and as more stuff comes up it becomes more and more disturbing and you’re like, just stop already? That’s kind of what happened with inflation."
\qauthor{\textbf{Max Tegmark} \cite{Gefter2012}}
\end{savequote}
\chapter{Issues with cosmological inflation(?): UV sensitivity}
\label{Sec:inflation}

\ifpdf
    \graphicspath{{Chapter3/Figs/Raster/}{Chapter3/Figs/PDF/}{Chapter3/Figs/}}
\else
    \graphicspath{{Chapter3/Figs/Vector/}{Chapter3/Figs/}}
\fi

\section{Introduction}
Cosmological inflation remains one of the most influential ideas in modern theoretical cosmology. In its simplest form, it posits an initial period of accelerated expansion that can account for the approximate homogeneity, isotropy, and spatial flatness of the observable universe. It also provides a mechanism for generating primordial scalar and tensor perturbations. Starobinsky, Guth, and Linde first introduced this concept in their pioneering works \cite{Starobinsky:1980te,Guth:1980zm,Linde1982}, and it has since been developed into a broad framework. Its theoretical foundations, phenomenological implications, and observational tests have been extensively reviewed in the literature \cite{Guth1984,Liddle2000,Brandenberger2000,Mukhanov_2005a,Mukhanov_2005b,Lemoine2007,Baumann:2009ds,Baumann:2014nda,Baumann_2022inflation,Senatore:2016aui,Martin:2013tda}. However, inflation is not a single theory but rather a class of models. The questions of its ultraviolet completion, initial conditions, and embedding in a more fundamental theory remain unanswered.

This chapter discusses inflation from the perspectives of effective field theory and ultraviolet sensitivity. Adopting this viewpoint is particularly useful, given that inflation involves physics at energy scales far above those that can be tested in laboratory experiments. The philosophy of effective field theory is to separate low-energy degrees of freedom from unknown high-energy physics, encoding the latter in a tower of higher-dimensional operators suppressed by a cutoff scale, $\Lambda$ \cite{Burgess_2020,Georgi:1993mps,Kaplan:2005es,Meissner:2022cbi}. In this sense, an EFT is a systematic way of parametrizing ignorance about short-distance physics, not merely an approximation scheme. Furthermore, gravity itself can be treated as an effective quantum field theory at energies below its cutoff, which is expected to be at or below the Planck scale unless additional lighter gravitational degrees of freedom appear \cite{Donoghue:1994dn,Burgess:2003jk,Donoghue2024EFT1,Donoghue2024EFT2,Donoghue2024EFT3,Donoghue2024EFT4,Donoghue2024EFT5}. This observation is central to building inflationary models, since Planck-suppressed corrections may modify the scalar potential and disrupt slow roll.

A second organizing principle of the chapter is symmetry. In particle physics, spontaneous symmetry breaking is most familiar from gauge theories, where one must carefully distinguish physical global symmetries from gauge redundancies \cite{Weinberg:1996kr}. In cosmology, the time-dependent background selects a preferred clock and therefore spontaneously breaks time translations or, equivalently, time diffeomorphisms in the sense of their non-linear realization after the choice of a preferred slicing. The associated Goldstone mode, $\pi$, provides an efficient language for describing adiabatic scalar perturbations, since it non-linearly realizes the broken time reparametrizations and is related at leading order to the comoving curvature perturbation. This is the conceptual basis of the effective field theory of inflation, in which the most general action compatible with the unbroken spatial diffeomorphisms is constructed in unitary gauge and the Goldstone boson is restored by the St{\"u}ckelberg trick \cite{Lyth:1998xn,Cheung:2007st,Weinberg:2008hq,Ashoorioon:2018uey,Baumann_McAllister_2015appEFTinfl,Baumann_McAllister_2015EFT-inflation}.

This symmetry-based formulation is powerful because it separates universal consequences of the inflationary background from model-dependent details. The background evolution fixes the universal part of the action, while higher-order operators control the dynamics of fluctuations, the sound speed, non-Gaussianities and possible deviations from the simplest slow-roll predictions. At the same time, expectations from quantum gravity suggest that exact global symmetries should not survive in a consistent theory of gravity. This has important consequences for the naturalness of shift symmetries and for the stability of inflationary potentials against Planck-suppressed corrections \cite{Kallosh:1995hi,Baumann_McAllister_2015EFT-inflation}.

A complete inflationary scenario must also explain the transition from the accelerated phase to the hot radiation-dominated Universe. Since inflation dilutes any pre-existing matter and radiation, the energy stored in the inflationary sector must be transferred to ordinary particles through reheating or preheating \cite{Albrecht:1982mp,Allahverdi:2010xz,Amin:2014eta,Dai:2014jja}. Reheating is not a secondary detail: it determines the thermal history after inflation, affects the relation between inflationary scales and observable modes, and may constrain otherwise viable models. Therefore, EFT control of inflation must be supplemented by a consistent description of the exit from inflation and of the couplings between the inflaton and matter.

The simplest inflationary scenarios can explain the nearly scale-invariant and approximately Gaussian primordial perturbations. However, this does not establish inflation as a UV-complete theory. There are several theoretical issues, including the measure problem, the degree of fine-tuning required in the initial conditions and SF potential, trans-Planckian sensitivity, the role of quantum gravity corrections, and the question of whether inflation removes or merely postpones the initial singularity \cite{Penrose:1988mg,Earman1999,Martin:2000xs,Gibbons:2006pa,Steinhardt2011,Brandenberger:2012aj,Ijjas:2013vea,Ijjas:2014nta,Ijjas:2015hcc,Ijjas2017,Martin2020,Postolak:2024xtm}. Observational constraints on inflation include measurements of the CMB and large-scale structure, as well as bounds on primordial gravitational waves \cite{Brandenberger:2011eq,Guth:2013sya,Planck:2018jri,Planck:2019kim,Chowdhury:2019otk,Martin:2024qnn,Ferreira:2025lrd}. Moreover, according to theorems on geodesic incompleteness, inflationary spacetimes are generally not past-eternal \cite{Borde:2001nh,DiTucci:2019xcr}. Nonetheless, causality-based arguments and analyses of initial data continue to clarify what inflation does and does not explain \cite{Vachaspati:1998dy,Bunch:1978yq,Goldwirth:1991rj,Garfinkle:2023vzf}. Related singularity results further emphasize that accelerated expansion alone does not automatically provide a complete theory of the origin of the Universe \cite{Borde:1993xh,Borde:1996pt}.

For these reasons, it is useful to compare inflationary EFT with effective descriptions of non-singular cosmologies, including bouncing scenarios and other alternatives in which the initial singularity may be avoided or replaced by a transition between contraction and expansion \cite{Cai:2016thi,Cai:2017tku}. Such models face their own difficulties, including stability, gradient instabilities, anisotropy growth and the construction of a controlled EFT through the non-singular phase. Nevertheless, they provide an important contrast: they show that the central questions of early-universe cosmology are not exhausted by fitting the scalar spectral index and the tensor-to-scalar ratio, but also include UV sensitivity, symmetry protection, causality, geodesic completeness and the consistency of the effective description.

The chapter is organized as follows. First, we recall the bottom-up EFT logic and illustrate how gravity differs from renormalizable gauge theories because the gravitational coupling is dimensionful. We then discuss spontaneous breaking of time translations in cosmology and its relation to the Goldstone description of adiabatic perturbations. Next, we construct the EFT of inflation in unitary gauge, restore the Goldstone mode, and identify the universal and model-dependent parts of the action. We subsequently discuss reheating as the necessary bridge between inflation and the hot Big Bang. Finally, we analyze the ultraviolet sensitivity of scalar masses and Planck-suppressed operators, emphasizing why controlling radiative corrections is essential for the naturalness of slow-roll inflation.

\section{Bottom up: parameterizing ignorance}
The EFT formalism is based on a simple separation of scales. If the physical processes under consideration are characterized by the energy scale $E$, while the microscopic completion contains heavier degrees of freedom with masses of order $\Lambda$, then for:
\begin{equation}
    E\ll\Lambda\,,
\end{equation}
the heavy modes do not have to be included explicitly. Their effects are encoded in a tower of local operators compatible with the symmetries of the low-energy theory \cite{Burgess_2020,Georgi:1993mps,Kaplan:2005es,Meissner:2022cbi}. In this sense, the bottom-up EFT construction is a systematic way of \textit{\textbf{parameterizing ignorance}} about the UV theory. For a generic light field $\phi$, the effective Lagrangian may be organized in the following form:
\begin{equation}
    \mathcal{L}_{\mathrm{eff}}[\phi]=\mathcal{L}_{l}[\phi]+\sum_{i}{c_{i} \frac{\mathcal{O}_{i}[\phi]}{\Lambda^{\delta_{i}-4}}}\,,
\end{equation}
where $\mathcal{L}_{l}$ contains the leading relevant and marginal operators, $\mathcal{O}_{i}$ are higher-dimensional operators of mass dimension $\delta_{i}>4$, $c_i$ are dimensionless Wilson coefficients and $\Lambda$ is the cut-off scale. The contribution of an operator with dimension $\delta_i$ to an amplitude is suppressed by powers of:
\begin{equation}
    \left(\frac{E}{\Lambda}\right)^{\delta_i-4}\,.
\end{equation}
Therefore, the EFT expansion is predictive at low energies even if the UV completion is unknown.

Gravity provides one of the most important examples of this logic. The leading low-energy gravitational action is the Einstein-Hilbert action:
\begin{equation}
    S_{\mathrm{EH}}=\frac{M_{\mathrm{Pl}}^2}{2}\int d^{4}x\,\sqrt{-g}\,R\,.
\end{equation}
Expanding the metric around Minkowski spacetime:
\begin{equation}
    g_{\mu\nu}\equiv\eta_{\mu\nu}+\frac{1}{M_{\mathrm{Pl}}}\,h_{\mu\nu}
\end{equation}
one obtains:
\begin{equation}
    S_{\mathrm{EH}}=\int d^{4}x\left[\left(\partial h\right)^2+\frac{1}{M_{\mathrm{Pl}}}\,h\,\left(\partial h\right)^2+\frac{1}{M_{\mathrm{Pl}}^2}\,h^2\,\left(\partial h\right)^2+\ldots\right]\,.
\end{equation}
This expansion shows that graviton self-interactions are controlled by inverse powers of $M_{\mathrm{Pl}}$. Since the gravitational coupling has a negative mass dimension, perturbative GR is not renormalizable in the traditional sense. However, this is precisely the situation in which the EFT interpretation is useful: GR is predictive as a low-energy QFT, provided one includes all operators allowed by diffeomorphism invariance order by order in the derivative expansion \cite{Donoghue:1994dn,Burgess:2003jk,Donoghue2024EFT1,Donoghue2024EFT2,Donoghue2024EFT3,Donoghue2024EFT4,Donoghue2024EFT5}.
This should be contrasted with Yang-Mills theory \cite{Yang:1954ek,Weinberg_1996YM,Paschos_2023YM}:
\begin{equation}
    S_{\mathrm{YM}}=\int d^{4}x\left[\left(\partial A\right)^2+g\,A^2\, \partial A+g^2\,A^4\right]\,,
\end{equation}
where the gauge coupling $g$ is dimensionless in four spacetime dimensions. Thus, Yang-Mills self-interactions are not suppressed by inverse powers of a heavy scale. In GR, by contrast, every additional graviton interaction comes with further powers of $1/M_{\mathrm{Pl}}$. This is the origin of the statement that perturbative quantum gravity is naturally organized as an EFT.

At the schematic level, the gravitational cutoff is expected to be of order of the Planck scale:
\begin{equation}
    \Lambda=M_{\mathrm{Pl}}\,,
\end{equation}
although in a concrete UV completion the actual cutoff may be lower, for example because of additional heavy states, compactification scales, or strong-coupling effects. The most general purely gravitational EFT action compatible with diffeomorphism invariance can be written as:
\begin{equation}
    S_g=\int d^{4}x\,\sqrt{-g}\left[M_{\Lambda}^4+\frac{M_{\mathrm{Pl}}^2}{2}R +c_{1}R^{2}+c_{2}R_{\mu\nu}R^{\mu\nu}+\frac{1}{M^2}\left(d_{1}R^{3}+\ldots\right)+\ldots\right]
\end{equation}
The first term corresponds to the cosmological constant contribution, the second one is the Einstein-Hilbert term, while the remaining terms represent higher-curvature corrections. In four spacetime dimensions, one may choose different quadratic curvature bases because the Gauss-Bonnet combination is topological. Thus, an explicit $R_{\mu\nu\rho\sigma}R^{\mu\nu\rho\sigma}$ term can be exchanged for a combination of $R^2$ and $R_{\mu\nu}R^{\mu\nu}$ up to a topological invariant.

When matter fields are included, the complete low-energy action has the form of:
\begin{equation}
    S_{\mathrm{eff}}\left[\phi,g\right]=S_{g}+S_{\mathrm{eff}}[\phi]+S_{g,\phi}\,,
\end{equation}
where $S_{g}$ contains purely gravitational operators, $S_{\mathrm{eff}}[\phi]$ contains the SF operators in the absence of curvature, and $S_{g,\phi}$ contains mixed operators involving both $\phi$ and the metric. In the bottom-up expansion one may write that:
\begin{equation}
    S_{g,\phi}=\int d^{4}x\,\sqrt{-g}\left[\sum_{i}{c_{i}\frac{\mathcal{O}_{i}[g,\phi]}{\Lambda^{\delta_{i}-4}}}\right]\,.
\end{equation}

\section{Symmetry breaking in cosmology}
\begin{examplebox}[What does it mean to break a gauge symmetry?]
\begin{itemize}
    \item The \textit{set of gauge symmetries} $G$ can be divided into symmetries that:
            \begin{enumerate}[label=(\alph*)]
                \item \textit{\textbf{Tend towards identity}} at \textit{spatial infinity} ($G_{*}$);
                \item \textit{\textbf{Do not approach the identity}} at \textit{spatial infinity} ($G/G_{*}$) - \textit{\textbf{global part}} of the gauge transformations;
            \end{enumerate}
    \item The \textit{only part of the gauge symmetry that is broken} is the \textit{\textbf{global part}};
    \item '\textit{\textbf{SSB of a gauge symmetry}}' $\equiv$ \textbf{SSB of} $\boldsymbol{G/G_{*}}$ \textbf{symmetry} associated with the \textit{Noether current} and the \textit{conserved charge} ($G_{*}$ symmetry \textbf{is not broken}).
\end{itemize}
\end{examplebox}
\subsection{SSB of time translations}
\begin{examplebox}[SSB in cosmology]
    \begin{itemize}
        \item \textit{Metric tensor} $g_{\mu\nu}$ - \textit{\textbf{gauge field}};
        \item \textit{Invariance of GR with respect to spacetime diffeomorphisms} - \textit{\textbf{gauge symmetry}}:
        \begin{equation}
            x^{\mu}\mapsto x'^{\mu}\left(x^{\nu}\right)
        \end{equation}
        \item \textit{SSB of time translations} - global part of the time diffeomorphisms:
        \begin{equation}
            t \mapsto t'\left(x^{\nu}\right)
        \end{equation}
        \item If spacetime \textbf{does not have} at least a \textit{local timelike Killing vector} $\implies$ \textbf{Breaking of time translations};
        \item Example - \textit{\textbf{de Sitter spacetime}} \cite{Griffiths_Podolsky_2009dS}:
        \begin{equation}
            ds^2=-dt^2+e^{2Ht}\,d\vec{x}^2 \quad\implies\quad \begin{dcases}
                t \mapsto t+\lambda \\
                \vec{x} \mapsto e^{-H\lambda}\,\vec{x}
            \end{dcases}
        \end{equation}
        $\implies$ Existence of a \textit{\textbf{timelike Killing vector}}
        \item Ideal de Sitter space - no preferred time slicing (all slices are related by gauge transformations);
        \item \textit{\textbf{Cosmological inflation}} - \textbf{deviation} from the \textit{ideal de Sitter} model;
        \item Inflationary '\textit{\textbf{clock}}' (\textit{\textbf{order parameter}}) informs about \textit{how much exponential expansion remains} (e.g., Hubble parameter, $H$);
        \item Nevertheless, \textit{it is more convenient to use} the \textbf{time-dependent EVs} $\boldsymbol{\psi_{m}(t)}$ \textbf{of the matter fields} $\boldsymbol{\psi_{m}}$ (e.g., inflaton $\phi$ or inflationary vacuum energy density $\rho$)\\
        $\implies$ \textit{\textbf{Preferred time slicing}} (\textit{defined and labeled by values of} $\psi_{m}$)\\
        $\implies$ \textbf{Breaking of the time translation invariance};
        \item Despite the fact that the \textit{spacetime-independent transformation} of a time variable:
        \begin{equation}
            t \mapsto t+\xi \quad\land\quad \xi=\mathrm{const}\,,
        \end{equation}
        is a system redundancy (even for a preferred time slicing), nonetheless, the \textit{\textbf{spacetime-dependent transformation}}:
        \begin{equation}\label{time-shift-1}
            t \mapsto t+\pi\left(t,\vec{x}\right)
        \end{equation}
        \textit{\textbf{is not invariant}} unless:
        \begin{equation}
            \pi\left(t,\vec{x}\right)=\mathrm{const}\,;
        \end{equation}
        $\implies$ \textbf{SSB} of time translation $\implies$ \textit{\textbf{Existence of a Goldstone boson}}\footnote{More details concerning symmetry breaking in a gauge theory one may find in the textbook \cite{Weinberg:1996kr}.} $\boldsymbol{\pi\left(t,\vec{x}\right)}$;
        \item This corresponds to a \textit{spacetime-dependent transformation} along the \textbf{broken generator}:
        \begin{equation}
            U\left(t,\vec{x}\right)\equiv t+\pi\left(t,\vec{x}\right)\,.
        \end{equation}
    \end{itemize}
\end{examplebox}
\subsection{Adiabatic perturbations}
From a physical perspective, an important point is that Goldstone boson fluctuations are related to cosmological observables. Namely, perturbations induced by \textit{local time shifts}:
\begin{equation}
    \delta\psi_{m}\left(t,\vec{x}\right)\equiv \psi_{m}\left[t+\pi\left(t,\vec{x}\right)\right]-\psi_m(t)
\end{equation}
correspond to the \textit{\textbf{adiabatic fluctuations}} \cite{Kodama:1984ziu,Mukhanov:1990me,Malik:2008im,Liddle2000}. This means that, under a linear approximation, these fluctuations are proportional to the Goldstone mode:
\begin{equation}
    \delta\psi_{m}=\dot{\psi}_{m}(t)\,\pi\left(t,\vec{x}\right)\,.
\end{equation}
As observational data show \textit{no signs of deviation from adiabatic initial conditions} \cite{Planck:2018jri,Planck:2019kim}, the \textit{Goldstone boson} provides the \textbf{most efficient way} to describe cosmological data. Moreover, in the \textit{spatially flat gauge} for which:
\begin{equation}
    g_{ij}=a^{2}(t)\,\delta_{ij}
\end{equation}
all metric perturbations are related to the Goldstone mode via the Einstein field equations. In order to obtain purely adiabatic fluctuations, a time shift in the opposite direction with respect to \eqref{time-shift-1} can be performed:
\begin{equation}\label{time-shift-2}
    t \mapsto t-\pi\left(t,\vec{x}\right)
\end{equation}
and therefore, remove all matter fluctuations, namely:
\begin{equation}
    \delta\psi_{m} \mapsto 0\,.
\end{equation}
The result of applying the transformation described by equation \eqref{time-shift-2} is the \textit{generation of isotropic perturbations for the spatial components of the metric tensor} \cite{Baumann_McAllister_2015appEFTinfl}:
\begin{equation}\label{pi-R-relation}
    \delta g_{ij}=a^{2}(t)\,e^{2\mathcal{R}\left(t,\vec{x}\right)}\,\delta_{ij} \quad\land\quad \mathcal{R}\left(t,\vec{x}\right)=-H\pi\left(t,\vec{x}\right)\,.
\end{equation}
\begin{examplebox}[Relation between perturbations in the \textit{comoving} and \textit{spatially flat} gauge]
\begin{itemize}
    \item The equation \eqref{pi-R-relation} leads to an important conclusion that the \textit{\textbf{Goldstone boson}} $\boldsymbol{\pi\left(t,\vec{x}\right)}$ \textit{in spatially flat gauge} \textbf{is proportional} to the \textit{\textbf{curvature perturbation}} $\boldsymbol{\mathcal{R}\left(t,\vec{x}\right)}$ in the \textit{comoving (unitary) gauge};
    \item Moreover, in the case of:
    \begin{equation}
        H\simeq\mathrm{const}
    \end{equation}
    one could use the quantities \textit{\textbf{interchangeably}}.
\end{itemize}
\end{examplebox}
\subsection{Effective action in comoving (unitary) gauge}
Earlier, we demonstrated that matter perturbations can be removed by a local time shift when one uses a comoving gauge:
\begin{equation}
    \pi\left(t,\vec{x}\right)=0\,.
\end{equation}
This simply means the 'absorption' of the Goldstone boson by the metric tensor. Once we have decided on a specific gauge, the action does not necessarily have to be invariant due to full diffeomorphism, but only due to time-dependent spatial diffeomorphisms:
\begin{equation}
    x^i \mapsto x^i+\xi^i\left(t,x^j\right)\,.
\end{equation}
For this reason, our model is characterized by reduced symmetry, which in turn can generate many new segments in the action.

\begin{figure}[htbp]
    \centering
    \includegraphics[width=1\linewidth]{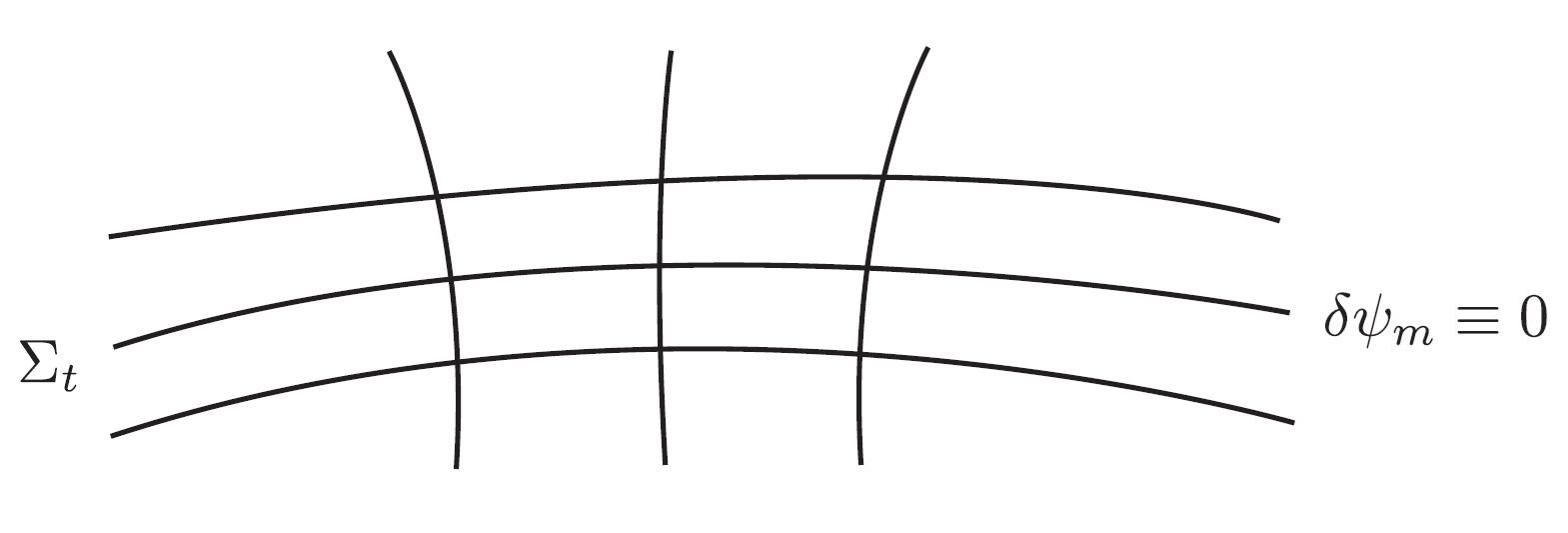}
    \caption{The foliation of spacetime into a series of spacelike hypersurfaces in the unitary gauge under consideration \cite{Baumann_McAllister_2015appEFTinfl}.}
    \label{fig:hypersurfaces-unitary-gauge}
\end{figure}

From a geometrical point of view, our gauge describes metric perturbations defined on $t=\mathrm{const}$ hypersurfaces $\Sigma_{t}$ (Fig.~\ref{fig:hypersurfaces-unitary-gauge}).
Therefore, a convenient step is to introduce an \textit{orthogonal unit four-vector} to $\Sigma_{t}$ and denoted by $n_{\mu}$:
\begin{equation}
    n_{\mu}\equiv -\frac{\delta_{\mu}{}^{0}}{\sqrt{-g^{00}}}\,.
\end{equation}
in order to construct (by contraction of the covariant tensors) mathematical objects such as $g^{00}$ and $R^{00}$, present in the effective action. It is also necessary to define the quantities characterizing the 3D geometry of the $\Sigma_{t}$ hypersurfaces.
\begin{examplebox}[Geometry of $\Sigma_{t}$ hypersurfaces]
    \begin{itemize}
        \item \textit{\textbf{Induced metric}}:
        \begin{equation}
            h_{\mu\nu}\equiv g_{\mu\nu}+n_{\mu}\,n_{\nu}\,;
        \end{equation}
        \item \textit{\textbf{Extrinsic curvature tensor}}:
        \begin{equation}
            K_{\mu\nu}\equiv h_{\mu}{}^{\rho}\,\nabla_{\rho}\,n_{\nu}\,;
        \end{equation}
        \item \textit{\textbf{Riemann curvature tensor of the induced metric}} defined by the \textit{Gauss–Codazzi equation}\footnote{It defines the relation between quantities defined in 3D and 4D space.} \cite{Poisson_2004Codazzi,Baumgarte_Shapiro_2010Codazzi}:
        \begin{equation}
            \hat{R}_{\alpha\beta\gamma\delta}\equiv h_{\alpha}{}^{\mu}\,h_{\beta}{}^{\nu}\,h_{\gamma}{}^{\rho}\,h_{\delta}{}^{\sigma}\,R_{\mu\nu\rho\sigma}-K_{\alpha\gamma}\,K_{\beta\delta}+K_{\alpha\delta}\,K_{\beta\gamma}\,.
        \end{equation}
    \end{itemize}
\end{examplebox}
According to the given analysis, the \textit{most general form of the action} is expressed as follows \cite{Cheung:2007st,Baumann_McAllister_2015appEFTinfl}:
\begin{equation}\label{action-unitary-general}
    S=\int d^{4}x\,\sqrt{-g}\,\mathcal{L}\Bigl[R_{\mu\nu\rho\sigma},g^{00},K_{\mu\nu},\hat{R}_{\mu\nu},\nabla_{\mu};t\Bigr]\,.
\end{equation}
\subsection{Universal part of the action}
Taking the \textit{\textbf{spatially flat FLRW background}} into consideration: 
\begin{equation}
    ds^{2}=-dt^{2}+a^{2}(t)\,\delta_{ij}\,dx^{i}dx^{j} \,,
\end{equation}
we find that:
\begin{equation}
    \bar{g}^{00}=-1\quad\land\quad \bar{R}\equiv\bar{R}^{\mu}{}_{\mu}=6\left(2H^{2}+\dot{H}\right)\quad\land\quad\bar{K}\equiv\bar{K}^{\mu}{}_{\mu}=3H\,,
\end{equation}
and therefore, at linear order in fluctuations, for \eqref{action-unitary-general} one obtains \cite{Baumann_McAllister_2015appEFTinfl}:
\begin{equation}\label{action-FLRW-unitary}
    S=\int d^{4}x\,\sqrt{-g}\left[\Lambda_{0}(t)+c_{1}\left(g^{00}+1\right)+c_{2}\left(K-3H\right)+c_{3}\left(R-\bar{R}(t)\right)\right]\,,
\end{equation}
where the coefficients:
\begin{equation}
    c_{1}=c_{1}(t)\quad\land\quad c_{2}=c_{2}(t)\quad\land\quad c_{3}=c_{3}(t)
\end{equation}
are \textbf{explicitly time-dependent}, due to the \textit{breaking of time diffeomorphisms}.
\begin{examplebox}[Remark]
One could:
    \begin{itemize}
        \item Absorb $\bar{R}(t)$ into $\Lambda_{0}(t)$ (0th-order segment);
        \item Remove $c_{3}(t)$ by applying a \textit{conformal transformation} of the metric $g^{\mu\nu}$;
        \item Integrate by parts the extrinsic curvature term:
        \begin{equation}
            \int d^{4}x\,\sqrt{-g}\,c_{2}(t)\,K=-\int d^{4}x\,\sqrt{-g}\,n^{\mu}\nabla_{\mu}\,c_{2}(t)=\int d^{4}x\,\sqrt{-g}\,\sqrt{-g^{00}}\,\dot{c}_{2}(t)\,.
        \end{equation}
    \end{itemize}
\end{examplebox}
Based on the points mentioned above, the action \eqref{action-FLRW-unitary} can be simplified into the following form:
\begin{equation}\label{action-FLRW-unitary-simplified}
    S=\int d^{4}x\,\sqrt{-g}\left[\frac{1}{2}M_{\mathrm{Pl}}^{2}\,R-\Lambda(t)-c(t)\,g^{00}\right]+\Delta S\,,
\end{equation}
where $\Delta S$ denotes the higher-order terms. Moreover, the functions $\Lambda(t)$ and $c(t)$ are determined by the background, namely varying the linear part of \eqref{action-FLRW-unitary-simplified} with respect to $g^{\mu\nu}$ we obtain:
\begin{equation}
\begin{aligned}
H^2 & =\frac{1}{3M_{\mathrm{Pl}}^2}\Bigl[c(t)+\Lambda(t)\Bigr]\,, \\
\dot{H} & =-\frac{1}{3 M_{\mathrm{Pl}}^2}\Bigl[2c(t)-\Lambda(t)\Bigr]-H^2\,,
\end{aligned}
\end{equation}
so that:
\begin{equation}
    \dot{H}=-\frac{1}{M_{\mathrm{Pl}}^{2}}c(t)\,,
\end{equation}
and therefore fixes the functions:
\begin{equation}
    \boxed{\Lambda(t)=M_{\mathrm{Pl}}^{2}\left(3H^{2}+\dot{H}\right) \quad\land\quad c(t)=-M_{\mathrm{Pl}}^{2}\,\dot{H}}\,.
\end{equation}
Hence, we can write down the final form of the \textit{\textbf{universal part of the action}} \eqref{action-FLRW-unitary-simplified} for the \textit{FLRW background}:
\begin{examplebox}[Universal part of the action (for the FLRW background)]
\begin{itemize}
    \item \textit{Universal part of the action}:
    \begin{equation}\label{action-FLRW-unitary-final}
        S=M_{\mathrm{Pl}}^{2}\int d^{4}x\,\sqrt{-g}\underbrace{\left[\frac{1}{2}R-\left(3H^{2}+\dot{H}\right)+\dot{H}\,g^{00}\right]}_{\text{Universal part of the action}}+\underbrace{\Delta S}_{\text{higher-order corrections}}
    \end{equation}
    is \textit{\textbf{completely determined by the FLRW background}} (Hubble parameter $H(t)$);
    \item The \textit{\textbf{differences between individual models are covered by higher-order corrections}} ($\Delta S$).
\end{itemize}
\end{examplebox}
\begin{examplebox}[Higher-order corrections]
    One may write the \textit{\textbf{higher-order terms}} in the action \eqref{action-FLRW-unitary-final} as powers of fluctuations \cite{Baumann_McAllister_2015appEFTinfl}:
    \begin{equation}\label{unitary-higher-order}
    \begin{aligned}
        \Delta S= & \int d^{4}x\,\sqrt{-g}\Biggl[\frac{1}{2}M_{2}^{4}(t)\left(\delta g^{00}\right)^{2}+\frac{1}{3!}M_{3}^{4}(t)\left(\delta g^{00}\right)^{3}+\frac{1}{4!}M_{4}^{4}(t)\left(\delta g^{00}\right)^{4}+\ldots \\
        & -\frac{1}{2}\bar{M}_{1}^{3}(t)\delta g^{00}\delta K-\frac{1}{2}\bar{M}_{2}^{2}(t)\delta K^{2}-\frac{1}{2}\bar{M}_{3}^{2}(t)\left(\delta K^{\mu}{}_{\nu}\right)^{2}+\ldots -\frac{1}{2}\hat{M}_{1}^{2}(t)\delta g^{00}\hat{R}+\ldots\Biggr]
    \end{aligned}
    \end{equation}
    where:
\begin{equation}
    \delta g^{00}\equiv g^{00}+1 \quad\land\quad \delta K_{\mu\nu}\equiv K_{\mu\nu}-a^{2}H\, h_{\mu\nu}\,.
\end{equation}
\end{examplebox}
\subsection{Goldstone boson}
In order to better understand the essence of the action \eqref{unitary-higher-order}, we can now introduce the Goldstone boson associated with SSB of the time translation. The use of the so-called ‘\textit{\textbf{St{\"u}ckelberg trick}}’ \cite{Stueckelberg1938} allows the gauge invariance of the theory to be restored for the boson. For this purpose, the following (spacetime-dependent) transformation should be performed for the time variable:
\begin{equation}\label{Stuckelberg-trick}
    t \mapsto t'=t+\pi\left(t,x^{i}\right)\,.
\end{equation}
The transformation law described by \eqref{Stuckelberg-trick} \textit{does not affect} the \textit{4D Ricci scalar} and the \textit{volume element} either. Meanwhile, all coefficients explicitly dependent on time will transform according to the following manner \cite{Baumann_McAllister_2015appEFTinfl}:
\begin{equation}
    f(t)\mapsto f\left(t+\pi\right)=f(t)+\dot{f}(t)\,\pi+\frac{1}{2}\ddot{f}(t)\,\pi^{2}+\ldots\,.
\end{equation}
Moreover, the covariant and contravariant tensors shall transform as \cite{Baumann_McAllister_2015appEFTinfl}:
\begin{subequations}
    \begin{equation}
        t_{\mu\nu}\mapsto\frac{\partial x^{\alpha}}{\partial x^{\mu\prime}}\frac{\partial x^{\beta}}{\partial x^{\nu\prime}}\,t_{\alpha\beta}=\frac{1}{\Bigl(\delta^{\alpha}_{\mu}+\delta^{\alpha}_{0}\,\partial_{\mu}\pi\Bigr)\left(\delta^{\beta}_{\nu}+\delta^{\beta}_{0}\,\partial_{\nu}\pi\right)}\,t_{\alpha\beta}\,,
    \end{equation}
    \begin{equation}
        t^{\mu\nu}\mapsto\frac{\partial x^{\mu\prime}}{\partial x^{\alpha}}\frac{\partial x^{\nu\prime}}{\partial x^{\beta}}\,t^{\alpha\beta}=\Bigl(\delta^{\mu}_{\alpha}+\delta^{\mu}_{0}\partial_{\alpha}\pi\Bigr)\Bigl(\delta^{\nu}_{\beta}+\delta^{\nu}_{0}\partial_{\beta}\pi\Bigr)\,t^{\alpha\beta}\,,
    \end{equation}
\end{subequations}
respectively. Therefore, the contravariant elements of the metric tensor are subject to the following transformation laws:
\begin{subequations}
\begin{align}
    g^{00} & \mapsto g^{00}+2\,\partial_{\mu}\pi\,g^{0\mu}+\partial_{\mu}\pi\,\partial_{\nu}\pi\,g^{\mu\nu}\,, \\
    g^{0i} & \mapsto g^{0i}+\partial_{\mu}\pi\,g^{\mu i}\,, \\
    g^{ij} & \mapsto g^{ij}\,.
\end{align}
\end{subequations}
On the other hand, the covariant components transform as follows:
\begin{subequations}
\begin{align}
    g_{00} & \mapsto \frac{g_{00}}{\left(1+\dot\pi\right)^2}\simeq g_{00}-2\dot{\pi}\,g_{00} \,, \\
    g_{0i} & \mapsto \frac{g_{0i}}{\left(1+\dot{\pi}\right)} -\frac{g_{00}\,\partial_i\pi}{\left(1+\dot{\pi}\right)^2}\simeq g_{0i}-\dot{\pi}\,g_{0i}-g_{00}\,\partial_{i}\pi \,, \\
    g_{ij} & \mapsto g_{ij} -\frac{g_{0i}\,\partial_{j}\pi+g_{0j}\,\partial_{i}\pi}{\left(1+\dot{\pi}\right)}+\frac{g_{00}\,\partial_{i}\pi\,\partial_{j}\pi}{\left(1+\dot{\pi}\right)^2}\simeq g_{ij}-g_{0i}\,\partial_{j}\pi-g_{0j}\,\partial_{i}\pi \,.
\end{align}
\end{subequations}
When it comes to quantities defined on hypersurfaces $\Sigma_{t}$, the situation is a bit more complicated because coordinate transformation actually changes the hypersurfaces themselves. For this reason, it is more useful to express quantities defined on $\Sigma_{t}$ through the 4D metric tensor \cite{Baumann_McAllister_2015appEFTinfl}:
\begin{equation}
    K_{\mu\nu}=\frac{\delta^{0}_{\nu}\,\partial_{\mu}g^{00}}{2\left(-g^{00}\right)^{3/2}}+\frac{\delta^{0}_{\mu}\delta^{0}_{\nu}\,g^{0\sigma}\,\partial_{\sigma}g^{00}}{2\left(-g^{00}\right)^{5/2}}-g^{0\sigma}\frac{\Bigl(\partial_{\mu}g_{\sigma\nu}+\partial_{\nu}g_{\sigma\mu}-\partial_{\sigma}g_{\mu\nu}\Bigr)}{2\sqrt{-g^{00}}}\,.
\end{equation}
\begin{examplebox}[Explicit transformation laws]
    \begin{itemize}
        \item The metric fluctuation:
        \begin{equation}\label{g00-relation}
            \delta g^{00}=-2\dot{\pi}-\dot{\pi}^{2}+\frac{\left(\partial_{i}\pi\right)^{2}}{a^{2}}\,;
        \end{equation}
        \item The extrinsic curvature:
        \begin{subequations}
        \begin{align}
            K_{ij} & \mapsto a^{2}H\,\delta_{ij}-\partial_{i}\partial_{j}\pi+\ldots\,, \\
            K^{i}{}_{j} & \mapsto H\,\delta^{i}{}_{j}-a^{-2}\,\partial^{i}\partial_{j}\pi+\ldots\,, \\
            K & \mapsto 3H-a^{-2}\,\partial^{2}\pi+\ldots \quad\land\quad \partial^{2}\equiv\delta^{ij}\partial_{i}\partial_{j}\,;
        \end{align}
        \end{subequations}
        \item The fluctuations of extrinsic curvature:
        \begin{subequations}\label{K-relation}
        \begin{align}
            \delta K^{i}{}_{j} & \mapsto -a^{-2}\,\partial^{i}\partial_{j}\pi+\ldots\,, \\
            \delta K & \mapsto a^{-2}\,\partial^{2}\pi+\ldots\,;
        \end{align}
        \end{subequations}
        \item The intrinsic Ricci curvature:
        \begin{subequations}
        \begin{align}
            \hat{R}_{ij} & \mapsto H\Bigl(\partial_{i}\partial_{j}\pi+\delta_{ij}\,\partial^{2}\pi\Bigr)+\ldots \,, \\
            \hat{R} & \mapsto 4\frac{H}{a^{2}}\partial^{2}\pi+\ldots \,;
        \end{align}
        \end{subequations}
        \item The fluctuations of intrinsic Ricci curvature:
        \begin{subequations}
        \begin{align}
            \delta\hat{R}_{ij} & \mapsto \hat{R}_{ij} \,, \\
            \delta\hat{R} & \mapsto \hat{R} \,.
        \end{align}
        \end{subequations}
    \end{itemize}
\end{examplebox}
\subsection{Decoupling limit}
\begin{examplebox}[Goldstone action]
    Up to terms involving powers of $\delta g^{00}$, the \textit{\textbf{Goldstone action}} becomes \cite{Baumann_McAllister_2015appEFTinfl}:
    \begin{equation}\label{Goldstone-action1}
    \begin{aligned}
    S= & \int d^{4}x\,\sqrt{-g}\Biggl[\frac{1}{2}M_{\mathrm{Pl}}^{2}\,R-M_{\mathrm{Pl}}^{2}\Bigl\{3H^{2}(t+\pi)+\dot{H}(t+\pi)\Bigr\}+M_{\mathrm{Pl}}^{2}\,\dot{H}\Bigl(g^{00}+2\,\partial_{\mu}\pi\,g^{0\mu} \\
    & +\partial_{\mu}\pi\,\partial_{\nu}\pi\,g^{\mu\nu}\Bigr)+\sum_{n=2}^{\infty}{\frac{M_{n}^{4}(t+\pi)}{n!}}\Bigl(1+g^{00}+2\,\partial_{\mu}\pi\,g^{0\mu}+\partial_{\mu}\pi\,\partial_{\nu}\pi\,g^{\mu\nu}\Bigr)^{n}\Biggr]\,.
    \end{aligned}
    \end{equation}
    $\implies$ \textit{\textbf{The Goldstone boson is mixed with metric perturbations in a rather complicated manner}}.
\end{examplebox}
By analogy with the standard formalism known from QFT, in this case too, above a certain frequency $\omega_{\mathrm{mix}}$, the \textit{\textbf{Goldstone boson decouples from}} $\boldsymbol{\delta g^{\mu\nu}}$. Such a situation corresponds to the so-called decoupling limit:
\begin{examplebox}[Decoupling limit]
    \begin{equation}
        \begin{dcases}
            M_{\mathrm{Pl}}\to\infty \\
            \dot{H}\to 0
        \end{dcases}
        \quad\land\quad M_{\mathrm{Pl}}\,\dot{H}=\mathrm{const}
    \end{equation}
    \begin{itemize}
        \item \textit{\textbf{The gravitational perturbations decouple from the Goldstone boson above the limiting frequency of}}:
        \begin{equation}
            \omega_{\mathrm{mix}}^{2}\equiv\left|\dot{H}\right|\,;
        \end{equation}
        \item In the case of:
        \begin{equation}
            \omega >\omega_{\mathrm{mix}}\,,
        \end{equation}
        the Goldstone action \eqref{Goldstone-action1} can be evaluated in the \textit{unperturbed spacetime} $\bar{g}^{\mu\nu}$ and takes a \textit{simplified form}:
        \begin{equation}\label{Goldstone-action2}
        \begin{aligned}
            S= & \int d^{4}x\,\sqrt{-g}\Biggl[\frac{1}{2}M_{\mathrm{Pl}}^{2}\,R-M_{\mathrm{Pl}}^{2}\Bigl\{3H^{2}(t+\pi)+\dot{H}(t+\pi)\Bigr\}+M_{\mathrm{Pl}}^{2}\,\dot{H}\Bigl\{-1-2\dot{\pi}+\left(\partial_{\mu}\pi\right)^{2}\Bigr\} \\
            & +\sum_{n=2}^{\infty}{\frac{M_{n}^{4}(t+\pi)}{n!}}\Bigl(-2\,\dot{\pi}+\left(\partial_{\mu}\pi\right)^{2}\Bigr)^{n}\Biggr]\,.
        \end{aligned}
        \end{equation}
    \end{itemize}
\end{examplebox}
\section{EFT of cosmological inflation}
\begin{examplebox}[Quasi-de Sitter approximation]
    \begin{itemize}
        \item In the \textit{quasi-de Sitter limit}, it is postulated that the fractional change in the Hubble parameter and its time derivative are negligible, namely:
        \begin{equation}
            \left|\frac{\dot{H}}{H^2}\right| \ll 1 \quad\land\quad \left|\frac{\ddot{H}}{H\dot{H}}\right| \ll 1\,;
        \end{equation}
        \item Similar restrictions are also usually assumed to apply to all coefficients in the effective action under consideration\footnote{In practice, this means that the action for metric fluctuations remains invariant under time translation.}:
        \begin{equation}
            \left|\frac{\dot{c}_{i}}{H\,c_{i}}\right| \ll 1\,,
        \end{equation}
        for example:
        \begin{equation}
            \left|\frac{\dot{M}_{2}}{H\,M_{2}}\right| \ll 1 \,.
        \end{equation}
    \end{itemize}
\end{examplebox}
\subsection{Universal part of the action}
The Friedmann equations may be written in the form of:
\begin{subequations}\label{Friedmann-SF}
        \begin{align}
            3M_{\mathrm{Pl}}^{2}\,H^{2} & =\rho_{\phi} \,, \\
            -2M_{\mathrm{Pl}}^{2}\,\dot{H} & =\rho_{\phi}+p_{\phi}=\dot{\bar{\phi}}^{2} \,,
        \end{align}
\end{subequations}
where:
\begin{equation}
    \rho_{\phi}\equiv\frac{1}{2}\dot{\bar{\phi}}^{2}+V(\bar{\phi})\quad\land\quad p_{\phi}\equiv\frac{1}{2}\dot{\bar{\phi}}^{2}-V(\bar{\phi})\,.
\end{equation}
Furthermore, the Lagrangian for a SF is defined as:
\begin{equation}
    \mathcal{L}_{\phi}=-\frac{1}{2}g^{\mu\nu}\partial_{\mu}\phi\,\partial_{\nu}\phi-V(\phi)\,,
\end{equation}
and in the comoving gauge:
\begin{equation}
    \phi=\bar{\phi}(t)\,,
\end{equation}
the following identities are satisfied:
\begin{equation}
    \frac{1}{2}\dot{\bar{\phi}}^{2}=-M_{\mathrm{Pl}}^{2}\,\dot{H}\quad\land\quad V(\bar{\phi})=M_{\mathrm{Pl}}^{2}\left(3H^{2}+\dot{H}\right)\,,
\end{equation}
therefore:
\begin{equation}
    \mathcal{L}_{\phi}=-\frac{1}{2}\dot{\bar{\phi}}^{2}\,g^{00}-V(\bar{\phi})=M_{\mathrm{Pl}}^{2}\left[\dot{H}\,g^{00}-\left(3H^{2}+\dot{H}\right)\right]
\end{equation}
corresponds to \eqref{action-FLRW-unitary-final} under the assumption:
\begin{equation}
    \left|\dot{H}\right|\ll H^{2}\,.
\end{equation}

\subsection{Quadratic effective action}
The unitary gauge action hides the scalar fluctuation inside the metric. To make the scalar degree of freedom explicit, one restores time diffeomorphisms by applying the St{\"u}ckelberg replacement \cite{Cheung:2007st,Baumann_McAllister_2015appEFTinfl}:
\begin{equation}
    t\mapsto t+\pi(t,\vec{x})\,.
\end{equation}
The field $\pi$ is the Goldstone boson associated with the spontaneous breaking of time translations by the inflationary background. Under this replacement, every time-dependent coefficient in the unitary gauge action becomes a function of $t+\pi$, while $g^{00}$ transforms as:
\begin{equation}\label{g00-epsilon}
    \delta g^{00}=2\varepsilon H\pi\,.
\end{equation}
In a consequence, the Goldstone part of the universal action may be written as:
\begin{equation}\label{quadratic-effective-Goldstone}
    \mathcal{L}_{\pi}=M_{\mathrm{Pl}}^{2}\Bigl[-\left(3H^{2}(t+\pi)+\dot{H}(t+\pi)\right)+\dot{H}\Bigl(g^{00}+2\partial_{\mu}\pi\,g^{0\mu}+\partial_{\mu}\pi\,\partial_{\nu}\pi\,g^{\mu\nu}\Bigr)\Bigr]\,.
\end{equation}
Now, we can write that:
\begin{equation}
    2M_{\mathrm{Pl}}^{2}\,\dot{H}\,\dot{\pi}\,\delta g^{00}=4M_{\mathrm{Pl}}^{2}\,\dot{H}\left(\varepsilon H\,\dot{\pi}\,\pi\right)\,,
\end{equation}
which implies the following relations:
\begin{equation}
    S_{\mathrm{mix}}=4M_{\mathrm{Pl}}^{2}\int dt\,d^{3}x\,a^{3}(t)\dot{H}\left(\varepsilon H\,\dot{\pi}\,\pi\right)\,,
\end{equation}
\begin{equation}
    \dot{\pi}\,\pi=\frac{1}{2}\partial_{t}\left(\pi^{2}\right)\,,
\end{equation}
\begin{equation}
    S_{\mathrm{mix}}=2M_{\mathrm{Pl}}^{2}\int dt\,d^{3}x\,a^{3}\dot{H}\,\varepsilon\,H\,\partial_{t}\left(\pi^{2}\right)\,.
\end{equation}
Integrating by parts over cosmic time:
\begin{equation}
    \int dt\,F(t)\,\partial_{t}\left(\pi^{2}\right)=\underbrace{\Bigl[F(t)\,\pi^{2}\Bigr]\Biggl|_{t_{i}}^{t_{f}}}_{\text{boundary term}}-\int dt\,\dot{F}(t)\,\pi^{2}\,,
\end{equation}
where:
\begin{equation}
    F(t)=a^{3}\dot{H}\varepsilon H\,,
\end{equation}
and neglecting the boundary term, one gets:
\begin{equation}
    S_{\mathrm{mix}}=-2M_{\mathrm{Pl}}^{2}\int dt\,d^{3}x\,\partial_{t}\left(a^{3}\dot{H}\varepsilon H\right)\pi^{2}\,.
\end{equation}
In such a case, the time derivative:
\begin{equation}
    \partial_{t}\left(a^{3}\dot{H}\varepsilon H\right)=a^{3}\left[3H^{2}\dot{H}\varepsilon+\partial_{t}\left(\dot{H}\varepsilon H\right)\right]
\end{equation}
yields:
\begin{equation}
    S_{\mathrm{mix}}=-2M_{\mathrm{Pl}}^{2}\int dt\,d^{3}x\,a^{3}\left[3H^{2}\dot{H}\varepsilon+\partial_{t}\left(\dot{H}\varepsilon H\right)\right]\pi^{2}\,,
\end{equation}
so that:
\begin{equation}
    4M_{\mathrm{Pl}}^{2}\,\dot{H}\left(\varepsilon H\,\dot{\pi}\,\pi\right)\to -6M_{\mathrm{Pl}}^{2}\,\dot{H}\varepsilon H^{2}\pi^{2}-2M_{\mathrm{Pl}}^{2}\,\partial_{t}\left(\dot{H}\varepsilon H\right)\pi^{2}\,.
\end{equation}
In similar manner one can obtain that:
\begin{equation}
    -3M_{\mathrm{Pl}}^{2}\,H^{2}(t+\pi)=3M_{\mathrm{Pl}}^{2}\,\dot{H}\left(\varepsilon H^{2}\pi^{2}\right)
\end{equation}
\begin{equation}
    \frac{\mathrm{mixing}}{\mathrm{kinetic}}=\frac{3\varepsilon H^{2}\pi^{2}}{\dot{\pi}^{2}}=\frac{3\varepsilon H^{2}}{\omega^{2}}
\end{equation}
and the mixing is negligible in the case of decoupling limit:
\begin{equation}\label{decoupling-frequency}
    \omega\gg\omega_{\mathrm{mix}}\equiv\sqrt{\varepsilon}\,H\,.
\end{equation}

\begin{examplebox}[Decoupling in quasi-de Sitter spacetime]
    \begin{itemize}
        \item The special feature of quasi-de Sitter spacetime described by:
        \begin{equation}
            \varepsilon\ll 1
        \end{equation}
        is the fact that \textit{\textbf{decoupling occurs at relatively low frequencies (energy scales)}};
        \item Moreover, the \textit{\textbf{horizon crossing}}:
        \begin{equation}
            \omega=H
        \end{equation}
        \textit{\textbf{falls in the decoupling limit}}.
    \end{itemize}
\end{examplebox}
For the unperturbed spacetime, one obtains the following form of the quadratic Goldstone action:
\begin{equation}\label{quadratic-Goldstone-action1}
    \mathcal{L}_{\pi}^{(2)}=M_{\mathrm{Pl}}^{2}\,\left|\dot{H}\right|\left(\dot{\pi}^{2}-\frac{\left(\partial_{i}\pi\right)^{2}}{a^{2}}\right)\quad\land\quad \left(\partial_{i}\pi\right)^{2}\equiv\delta^{ij}\partial_{i}\pi\,\partial_{j}\pi\,.
\end{equation}
Moreover, the symmetry breaking scale - Noether current for time translations are given by:
\begin{equation}\label{Noether-current-Goldstone}
    J^{\mu}=T^{0\mu}=-\sqrt{2M_{\mathrm{Pl}}^{2}\,\left|\dot{H}\right|}\,\partial^{\mu}\pi_{c}+\mathcal{O}\left(\pi_{c}^{2}\right)\quad\land\quad \pi_{c}\equiv\sqrt{2M_{\mathrm{Pl}}^{2}\,\left|\dot{H}\right|}\,\pi=f_{\pi}^{2}\,\pi
\end{equation}
\begin{examplebox}[Noether current and SSB scale for the quadratic Goldstone action]
    \begin{itemize}
        \item In the decoupling approximation one could rewrite the action \eqref{quadratic-Goldstone-action1} in the form known from the case of \textit{free SF in flat (Minkowski) space}:
        \begin{equation}
            \mathcal{L}_{\pi}^{(2)}=M_{\mathrm{Pl}}^{2}\,\left|\dot{H}\right|\left(\dot{\pi}^{2}-\left(\nabla\pi\right)^{2}\right)=-M_{\mathrm{Pl}}^{2}\,\left|\dot{H}\right|\,\eta^{\mu\nu}\partial_{\mu}\pi\,\partial_{\nu}\pi\,;
        \end{equation}
        \item By introducing the \textit{\textbf{canonically normalized Goldstone field}} \eqref{Noether-current-Goldstone} one can write:
        \begin{equation}
            \partial_{\mu}\pi=\frac{\partial_{\mu}\pi_{c}}{f_{\pi}^{2}}\quad\land\quad \mathcal{L}_{\pi}^{(2)}=-\frac{1}{2}\eta^{\mu\nu}\partial_{\mu}\pi_{c}\,\partial_{\nu}\pi_{c}\,;
        \end{equation}
        \item The Goldstone field $\pi$ realizes the time translation symmetry nonlinearly as a shift:
        \begin{equation}
            \delta\pi=-\xi^{0}+\mathcal{O}\left(\pi\,\xi^{0},\partial\xi^{0}\right)\,,
        \end{equation}
        up to linear order one may promote the transformation parameter into a spacetime-dependent function:
        \begin{equation}
            \delta\pi(x)=-\alpha(x)\,;
        \end{equation}
        \item Variation of the Lagrangian is:
        \begin{equation}
            \delta\mathcal{L}=\frac{\partial\mathcal{L}}{\partial\left(\partial_{\mu}\pi\right)}\partial_{\mu}\left(\delta\pi\right)=-f_{\pi}^{4}\,\partial^{\mu}\pi\,\partial_{\mu}\left(-\alpha\right)=f_{\pi}^{4}\,\partial^{\mu}\pi\,\partial_{\mu}\alpha\,;
        \end{equation}
        \item Comparing with the standard Noether relation:
        \begin{equation}
            \delta\mathcal{L}=J^{\mu}\partial_{\mu}\alpha
        \end{equation}
        yields the explicit form for the \textit{\textbf{Noether current}}:
        \begin{equation}
            J^{\mu}=f_{\pi}^{4}\,\partial^{\mu}\pi\,;
        \end{equation}
        \item Furthermore, using \eqref{Friedmann-SF} with \eqref{Noether-current-Goldstone} immediately yields the \textit{\textbf{relation between the SSB scale and the inflaton kinetic energy}} in the decoupling limit:
        \begin{equation}\label{SSB-scale1}
            f_{\pi}^{4}=2M_{\mathrm{Pl}}^{2}\left|\dot{H}\right|=\dot{\bar{\phi}}^{2}\,.
        \end{equation}
    \end{itemize}
    $\implies$ \textit{\textbf{The time variation of the inflaton field is responsible for the breaking of the (time translation) symmetry}}.
\end{examplebox}
Using \eqref{quadratic-effective-Goldstone} and \eqref{g00-epsilon} one can obtain that:
\begin{equation}\label{massive-Goldstone-action}
    \mathcal{L}_{\pi}^{(2)}=M_{\mathrm{Pl}}^{2}\left|\dot{H}\right|\left(\dot{\pi}^{2}-\frac{\left(\partial_{i}\pi\right)^{2}}{a^{2}}+3\varepsilon H^{2}\pi^{2}\right)\,,
\end{equation}
so that:
\begin{equation}\label{pi-mass-tachyonic}
    m_{\pi}^{2}\equiv -3\varepsilon H^{2}=3\dot{H}=-\frac{3}{2}\frac{f_{\pi}^{4}}{M_{\mathrm{Pl}}^{2}}=-\frac{3}{2}\frac{\dot{\bar{\phi}}^{2}}{M_{\mathrm{Pl}}^{2}}
\end{equation}
During inflationary stage, such a mass is negative and slow-roll suppressed:
\begin{equation}
    \dot{H}<0\quad\implies\quad \frac{\left|m_{\pi}^{2}\right|}{H^{2}}=3\varepsilon\ll 1
\end{equation}
$\implies$ \textit{\textbf{The Goldstone boson has a small tachyonic mass}}.
\begin{examplebox}[(Massless) Physical adiabatic mode]
    \begin{itemize}
        \item Nevertheless, \eqref{pi-mass-tachyonic} \textit{\textbf{does not mean that the physical adiabatic mode is massive}};
        \item Using the action \eqref{massive-Goldstone-action} with the definition of the comoving curvature perturbation \eqref{pi-R-relation} yields the exact result in the form of the action for the \textit{\textbf{massless and conserved outside the horizon comoving curvature perturbation}}:
        \begin{equation}
            \mathcal{L}_{\mathcal{R}}^{(2)}=\frac{1}{2}\frac{f_{\pi}^{4}}{H^{2}}\left(\dot{\mathcal{R}}^{2}-\frac{\left(\partial_{i}\mathcal{R}\right)^{2}}{a^{2}}\right)
        \end{equation}
    \end{itemize}
    $\implies$ \textit{\textbf{The small tachyonic mass term is specific to the Goldstone description}}.
\end{examplebox}

\subsection{Single-derivative corrections}
The action for single-derivative corrections could be written as:
\begin{equation}
    \Delta\mathcal{L}=\sum_{n=2}^{\infty}{\frac{M_{n}^{4}(t)}{n!}\left(\delta g^{00}\right)^{n}}\,,
\end{equation}
which yields the so called $\boldsymbol{P(X,\phi)}$ \textit{\textbf{theories}} \cite{Armendariz-Picon:1999hyi,Garriga:1999vw,Armendariz-Picon:2000ulo,Arkani-Hamed:2003pdi,Alishahiha:2004eh,Chen:2006nt,Copeland:2006wr,Babichev:2007dw,Amendola_Tsujikawa_2010kessence}:
\begin{equation}
    \mathcal{L}_{P(X)}=P(X,\phi)\quad\land\quad X\equiv -\frac{1}{2}\left(\partial\phi\right)^{2}\,.
\end{equation}
In unitary (comoving) gauge one obtains that:
\begin{equation}
    \mathcal{L}_{P(X)}\equiv P\left(-\frac{1}{2}\dot{\bar{\phi}}^{2}g^{00},\bar{\phi}\right)=M_{\mathrm{Pl}}^{2}\Bigl[\dot{H}g^{00}-\left(3H^{2}+\dot{H}\right)\Bigr]+\sum_{n=2}^{\infty}{\frac{\left(-\frac{1}{2}\dot{\bar{\phi}}^{2}\right)^{n}\dfrac{\partial^{n}P}{\partial X^{n}}}{n!}\Bigl(\delta g^{00}\Bigr)^{n}}\,,
\end{equation}
where:
\begin{equation}
    M_{n}^{4}(t)\equiv\left(-\frac{1}{2}\dot{\bar{\phi}}^{2}\right)^{n}\frac{\partial^{n}P}{\partial X^{n}}\,.
\end{equation}
\subsubsection{Quadratic action}
Taking the Goldstone boson into account up to the quadratic operator gives:
\begin{equation}\label{quadratic-derivative-term}
    \frac{1}{2}M_{2}^{4}\Bigl(\delta g^{00}\Bigr)^{2}\mapsto\frac{1}{2}M_{2}^{4}\Bigl(1+g^{00}+2\,\partial_{\mu}\pi\,g^{0\mu}+\partial_{\mu}\pi\,\partial_{\nu}\pi\,g^{\mu\nu}\Bigr)^{2}
\end{equation}
Thus, in the regime of:
\begin{equation}
    M_{2}^{4}\gg M_{\mathrm{Pl}}^{2}\left|\dot{H}\right|
\end{equation}
the dominant mixing comes from the $M_{2}^{4}\,\dot{\pi}\delta g^{00}$ interaction term.

The decoupling limit is the same as in \eqref{decoupling-frequency}, therefore, \eqref{quadratic-derivative-term} in the unperturbed spacetime becomes:
\begin{equation}
    \frac{1}{2}M_{2}^{4}\Bigl(\delta g^{00}\Bigr)^{2}\mapsto 2M_{2}^{4}\left(\dot{\pi}^{2}+\dot{\pi}^{3}-\dot{\pi}\frac{\left(\partial_{i}\pi\right)^{2}}{a^{2}}\right)+\mathcal{O}\left(\pi^{4}\right)\,.
\end{equation}
\begin{examplebox}[Speed of sound from the quadratic operator]
\begin{itemize}
    \item The quadratic operator $\left(\delta g^{00}\right)^{2}$ \textit{\textbf{modifies the Goldstone kinetic segment}} but \textit{\textbf{not the spatial gradient term}} $\left(\partial_{i}\pi\right)^{2}$ (\textit{fixed by the FLRW background symmetries});
    \item Implementation of the quadratic operator into the Goldstone action generates a \textit{\textbf{nontrivial speed of sound for the fluctuations}}:
    \begin{equation}
        \mathcal{L}_{\pi}^{(2)}=M_{\mathrm{Pl}}^{2}\frac{\left|\dot{H}\right|}{c_{s}^{2}}\left(\dot{\pi}^{2}-c_{s}^{2}\frac{\left(\partial_{i}\pi\right)^{2}}{a^{2}}\right)\,,
    \end{equation}
    where:
    \begin{equation}
        c_{s}^{2}\equiv\frac{M_{\mathrm{Pl}}^{2}\,\dot{H}}{M_{\mathrm{Pl}}^{2}\,\dot{H}-2M_{2}^{4}}=1+\frac{2M_{2}^{4}}{M_{\mathrm{Pl}}^{2}\,\dot{H}-2M_{2}^{4}}\quad\land\quad M_{2}^{4}>0\,;
    \end{equation}
    \item In the regime of:
    \begin{equation}
        M_{2}^{4}\gg M_{\mathrm{Pl}}^{2}\left|\dot{H}\right|
    \end{equation}
    one obtains the physical values for the speed of sound:
    \begin{equation}
        c_{s}^{2}\ll 1\,.
    \end{equation}
\end{itemize}
\end{examplebox}
\begin{examplebox}[Redefinition of the spatial coordinates - SSB scale]
    \begin{itemize}
        \item Absorption of the speed of sound in redefined spatial coordinates:
        \begin{equation}
            x^{i}\mapsto \tilde{x}^{i}=\frac{x^{i}}{c_{s}}
        \end{equation}
        allows us to restore the \textit{(fake) Lorentz invariance} in the action:
        \begin{equation}
            \tilde{\mathcal{L}}_{\pi}^{(2)}\equiv c_{s}^{3}\,\mathcal{L}_{\pi}^{(2)}=M_{\mathrm{Pl}}^{2}\left|\dot{H}\right|\,c_{s}\left(\dot{\pi}^{2}-\frac{\left(\tilde{\partial}_{i}\pi\right)^{2}}{a^{2}}\right)\,;
        \end{equation}
        \item Therefore, one can \textit{\textbf{read off the SSB scale directly from the normalization of the kinetic part}}, namely:
        \begin{equation}
            f_{\pi}^{4}\equiv 2M_{\mathrm{Pl}}^{2}\left|\dot{H}\right|\,c_{s}
        \end{equation}
        and reduces into \eqref{SSB-scale1} in the limiting case of:
        \begin{equation}
            c_{s}\to 1\,.
        \end{equation}
    \end{itemize}
\end{examplebox}
\subsubsection{Cubic action}
The cubic segment of the Goldstone action:
\begin{equation}\label{cubic-Goldstone-action1}
    \mathcal{L}_{\pi}^{(3)}=M_{\mathrm{Pl}}^{2}\left(c_{s}^{-2}-1\right)\left|\dot{H}\right|\left(\dot{\pi}^{3}-\dot{\pi}\frac{\left(\partial_{i}\pi\right)^{2}}{a^{2}}\right)
\end{equation}
implies:
\begin{equation}
    \tilde{\mathcal{L}}_{\pi}^{(3)}=\frac{1}{2}f_{\pi}^{4}\left(1-c_{s}^{2}\right)\left(\dot{\pi}^{3}-c_{s}^{-2}\,\dot{\pi}\frac{\left(\tilde{\partial}_{i}\pi\right)^{2}}{a^{2}}\right)\,.
\end{equation}
Moreover, the cubic operator in the decoupling limit:
\begin{equation}
    \frac{1}{3!}M_{3}^{4}\Bigl(\delta g^{00}\Bigr)^{3}\mapsto \frac{1}{3!}M_{3}^{4}\left(-2\dot{\pi}+\left(\partial_{\mu}\pi\right)^{2}\right)^{3}
\end{equation}
together with:
\begin{equation}
    A\equiv \left(\partial_{\mu}\pi\right)^{2}=g^{\mu\nu}\partial_{\mu}\pi\,\partial_{\nu}\pi=-\dot{\pi}^{2}+\frac{\left(\partial_{i}\pi\right)^{2}}{a^{2}}\,,
\end{equation}
and taking into account the fact that:
\begin{equation}
    \dot{\pi}=\mathcal{O}(\pi)\quad\land\quad A=\mathcal{O}\left(\pi^{2}\right)\,,
\end{equation}
give:
\begin{equation}
    \left(-2\dot{\pi}+A\right)^{3}=-8\underbrace{\dot{\pi}^{3}}_{\mathcal{O}\left(\pi^{3}\right)}+12\underbrace{\dot{\pi}^{2}A}_{\mathcal{O}\left(\pi^{4}\right)}-6\underbrace{\dot{\pi}A^{2}}_{\mathcal{O}\left(\pi^{5}\right)}+\underbrace{A^{3}}_{\mathcal{O}\left(\pi^{6}\right)}=-8\dot{\pi}^{3}+\ldots\,.
\end{equation}
This yields the following relation:
\begin{equation}\label{cubic-term}
    \frac{1}{3!}M_{3}^{4}\Bigl(\delta g^{00}\Bigr)^{3}\mapsto \frac{1}{3!}M_{3}^{4}\left(-2\dot{\pi}+\left(\partial_{\mu}\pi\right)^{2}\right)^{3}=-\frac{4}{3}M_{3}^{4}\,\dot{\pi}^{3}+\ldots
\end{equation}
By inserting \eqref{cubic-term} into \eqref{cubic-Goldstone-action1}, we obtain:
\begin{equation}
    \mathcal{L}_{\pi}^{(3)}=M_{\mathrm{Pl}}^{2}\left(c_{s}^{-2}-1\right)\dot{H}\left[\dot{\pi}\frac{\left(\partial_{i}\pi\right)^{2}}{a^{2}}+\frac{B}{c_{s}^{2}}\dot{\pi}^{3}\right]\,,
\end{equation}
where:
\begin{equation}
    \frac{B}{c_{s}^{2}}\equiv\frac{2}{3}\frac{M_{3}^{4}}{M_{2}^{4}}-1\,,
\end{equation}
so that, he cubic Goldstone Lagrangian has two independent operators: $\dot{\pi}\left(\partial_{i}\pi\right)^{2}$ and $\dot{\pi}^{3}$.

\subsubsection{Non-Gaussianity}
The cubic Lagrangian of the form:
\begin{equation}
    \tilde{\mathcal{L}}_{\pi}^{(3)}=-\frac{1}{2}\left(\tilde{\partial}_{\mu}\pi_{c}\right)^{2}-\frac{1}{2\Lambda^{2}}\left[\dot{\pi}_{c}\frac{\left(\tilde{\partial}_{i}\pi_{c}\right)^{2}}{a^{2}}+B\,\dot{\pi}_{c}^{3}\right]\,,
\end{equation}
where:
\begin{equation}
    \Lambda^{4}\equiv f_{\pi}^{4}\frac{c_{s}^{4}}{\bigl(1-c_{s}^{2}\bigr)^{2}}=2 M_{\mathrm{Pl}}^{2}\left|\dot{H}\right|\frac{c_{s}^{5}}{\bigl(1-c_{s}^{2}\bigr)^{2}}\,,
\end{equation}
denotes the \textit{\textbf{strong coupling scale}}. Therefore, the ratio between the energy scales may be written as:
\begin{equation}
    \left(\frac{f_{\pi}}{\Lambda}\right)^{2}=\frac{1-c_{s}^{2}}{c_{s}^{2}}=c_{s}^{-2}-1\quad\land\quad \left(\frac{f_{\pi}}{\Lambda}\right)^{4}=\frac{\left(1-c_{s}^{2}\right)^{2}}{c_{s}^{4}}
\end{equation}
where:
\begin{equation}
    0<c_{s}\leq 1\,.
\end{equation}
Thus, the strength of the cubic interaction is described by the dimensionless ratio $\left(f_{\pi}/\Lambda\right)^{2}$.

In addition, an amplitude of non-Gaussianity is given by \cite{Baumann_McAllister_2015appEFTinfl,Bartolo:2004if,Chen:2010xka,Yadav:2010fz,Komatsu:2010hc,Durrer_2020non-Gaussianities}:
\begin{equation}
    f_{\mathrm{NL}}\,\mathcal{R}\equiv\frac{\mathcal{L}_{3}}{\mathcal{L}_{2}}\Bigg|_{\omega=H}\,,
\end{equation}
where:
\begin{equation}
    \frac{\left(k_{1}k_{2}k_{3}\right)^2}{\Delta_{\mathcal{R}}^4} \mathcal{B}_{\mathcal{R}}\left(k_{1},k_{2},k_{3}\right)=\frac{18}{5} f_{\mathrm{NL}}\,S\left(k_{1},k_{2},k_{3}\right)\,.
\end{equation}
At horizon crossing (in the rescaled coordinates):
\begin{equation}
    \omega\sim H \quad\land\quad \dot{\pi}_{c}\sim H\pi_{c}\quad\land\quad \frac{\tilde{\partial}_{i}\pi_{c}}{a}\sim H\pi_{c}\,,
\end{equation}
which implies:
\begin{equation}
    \mathcal{L}_{2}\sim H^{2}\pi_{c}^{2}\quad\land\quad \mathcal{L}_{3}\sim\frac{1}{2\Lambda^{2}}H^{3}\pi_{c}^{3}\,,
\end{equation}
so that:
\begin{equation}
    \frac{\mathcal{L}_{3}}{\mathcal{L}_{2}}\Bigg|_{\omega=H}\sim\frac{1}{2\Lambda^{2}}H\pi_{c}
\end{equation}
Moreover:
\begin{equation}
    \mathcal{R}=-H\pi=-H\frac{\pi_{c}}{f_{\pi}^{2}}\quad\implies\quad H \pi_{c}\sim f_{\pi}^{2}\,\mathcal{R}
\end{equation}
yields:
\begin{equation}
    f_{\mathrm{NL}}\,\mathcal{R}\sim\left(\frac{f_{\pi}}{\Lambda}\right)^{2}\mathcal{R}\sim \left(c_{s}^{-2}-1\right)\mathcal{R}\,.
\end{equation}
Therefore, small values of the speed of sound, $c_{s}$, generate enhanced interactions (enhanced equilateral non-Gaussianity). Moreover, the detailed analysis shows that \cite{Baumann_McAllister_2015appEFTinfl}:
\begin{examplebox}[Non-Gaussianity from the cubic Goldstone interaction]
    \begin{itemize}
        \item The non-Gaussianities originating from the relevant operators in the cubic Goldstone action are as follows:
        \begin{equation}
            f_{\mathrm{NL}}^{\dot{\pi}\left(\partial_{i}\pi\right)^{2}}=-\frac{85}{324}\left(\frac{f_{\pi}}{\Lambda}\right)^{2}\quad\land\quad f_{\mathrm{NL}}^{\dot{\pi}^{3}}=\frac{5}{81}B\left(\frac{f_{\pi}}{\Lambda}\right)^{2}\,;
        \end{equation}
        \item The corresponding equilateral and orthogonal amplitudes:
        \begin{equation}\label{equil-ortho}
            f_{\mathrm{NL}}^{\mathrm{equil}}=\left(-0.27+0.08\,B\right)\left(\frac{f_{\pi}}{\Lambda}\right)^{2} \quad\land\quad f_{\mathrm{NL}}^{\mathrm{ortho}}=\left(0.02+0.02\,B\right)\left(\frac{f_{\pi}}{\Lambda}\right)^{2}
        \end{equation}
    \end{itemize}
\end{examplebox}
\begin{examplebox}[Derivation of equilateral and orthogonal amplitudes]
    \begin{itemize}
        \item Let us denote the \textit{\textbf{shape functions}} as:
        \begin{subequations}
        \begin{align}
            S_{1} & \equiv S_{\dot{\mathcal{R}}\left(\partial_{i}\mathcal{R}\right)^{2}}=\frac{\hat{k}_{1}^{2}-\hat{k}_{2}^{2}-\hat{k}_{3}^{2}}{\hat{k}_{1}\hat{k}_{2}\hat{k}_{3}}\left(-1+\sum_{i>j}{\frac{1}{9}\hat{k}_{i}\hat{k}_{j}}+\frac{1}{27}\hat{k}_{1}\hat{k}_{2}\hat{k}_{3}\right)+\text{permutations}\,, \\
            S_{2} & \equiv S_{\dot{\mathcal{R}}^{3}}=\hat{k}_{1}\hat{k}_{2}\hat{k}_{3}\,,
        \end{align}
        \end{subequations}
        where:
        \begin{equation}
            \hat{k}_{i}\equiv\frac{k_{i}}{K}\quad\land\quad K\equiv\frac{1}{3}\left(k_{1}+k_{2}+k_{3}\right)\,;
        \end{equation}
        \item The amplitudes, on the other hand, let us define as:
        \begin{subequations}\label{f1f2}
            \begin{align}
                f_{1} & \equiv f_{\mathrm{NL}}^{\dot{\pi}\left(\partial_{i}\pi\right)^{2}}=-\frac{85}{324}\left(\frac{f_{\pi}}{\Lambda}\right)^{2} \,, \\
                f_{2} & \equiv f_{\mathrm{NL}}^{\dot{\pi}^{3}}=\frac{5}{81}B\left(\frac{f_{\pi}}{\Lambda}\right)^{2} \,;
            \end{align}
        \end{subequations}
        \item Then, the total bispectrum becomes:
        \begin{equation}\label{total-bispectrum}
            \mathcal{B}_{\mathcal{R}}=\mathcal{B}_{1}+\mathcal{B}_{2} \quad\land\quad S_{\mathrm{tot}}=f_{1}S_{1}+f_{2}S_{2}\,;
        \end{equation}
        \item Projection onto the equilateral and orthogonal basis yields:
        \begin{equation}\label{projection-shape}
            S_{\mathrm{tot}}\mapsto f_{\mathrm{NL}}^{\mathrm{equil}}S_{\mathrm{equil}}+f_{\mathrm{NL}}^{\mathrm{ortho}}S_{\mathrm{ortho}}
        \end{equation}
        \item Due to the linear character of projection \eqref{projection-shape} the amplitudes must be linear combinations of the amplitudes:
        \begin{equation}
            \begin{pmatrix}
                f_{\mathrm{NL}}^{\mathrm{equil}} \\
                f_{\mathrm{NL}}^{\mathrm{ortho}}
            \end{pmatrix}
            =
            \begin{pmatrix}
                \alpha_{1} & \alpha_{2} \\
                \beta_{1} & \beta_{2}
            \end{pmatrix}
            \begin{pmatrix}
                f_{1} \\
                f_{2}
            \end{pmatrix}\,;
        \end{equation}
        \item In the case under consideration, the coefficients $\alpha_{i}$ and $\beta_{i}$ satisfy the following relations:
        \begin{subequations}\label{alpha-beta}
            \begin{align}
                -\frac{85}{324}\alpha_{1} & =-0.27 \quad\implies\quad \alpha_{1} =0.27\times\frac{324}{85}\simeq 1.03\,,\\
                \frac{5}{81}\alpha_{2} & =0.08 \quad\implies\quad \alpha_{2}=0.08\times\frac{81}{5}\simeq 1.30\,,\\
                -\frac{85}{324}\beta_{1} & =0.02 \quad\implies\quad \beta_{1}=-0.02\times\frac{324}{85}\simeq -0.076\,,\\
                \frac{5}{81}\beta_{2} & =0.02 \quad\implies\quad \beta_{2}=0.02\times\frac{81}{5}=0.324\,;
            \end{align}
        \end{subequations}
        \item Substituting \eqref{f1f2} into \eqref{total-bispectrum} with the coefficients given by \eqref{alpha-beta} produces exactly the relation \eqref{equil-ortho}.
    \end{itemize}
\end{examplebox}
Furthermore, one can also observe that there is no unnatural hierarchy between the scales related with the two operators for \cite{Senatore:2010jy}:
\begin{equation}
    B\sim\mathcal{O}(1)\,.
\end{equation}

\begin{examplebox}[Observational constraints on primordial non-Gaussianities]
    \begin{itemize}
        \item \textbf{Planck PR4} \cite{Jung:2025nss}:
        \begin{equation}
            f_{\mathrm{NL}}^{\mathrm{equil}}=6 \pm 46 \quad\land\quad f_{\mathrm{NL}}^{\mathrm{ortho}}=-8 \pm 21\,;
        \end{equation}
        \item Therefore, the following \textit{\textbf{hierarchy of energy scales}} arises:
        \begin{equation}
            \left(\frac{f_{\pi}}{\Lambda}\right)^{2}\lesssim 10^{2} \quad\implies\quad \boxed{\frac{f_{\pi}}{\Lambda}\lesssim 10}
        \end{equation}
        and limitations on the \textit{\textbf{speed of sound for the comoving curvature perturbations}}:
        \begin{equation}
            c_{s}\gtrsim 0.05\,.
        \end{equation}
    \end{itemize}
\end{examplebox}

\subsubsection{Theoretical limit}
\begin{examplebox}[Hierarchy between SSB and freeze-out scales]
    \begin{itemize}
        \item \textit{\textbf{Planck PR4}} \cite{Rosenberg:2022sdy}:
        \begin{equation}
            10^{9}A_{s}=2.081\pm 0.031 \quad\implies\quad A_{s}\equiv\Delta_{\mathcal{R}}^{2}\left(k_{*}\right)=\frac{1}{4\pi^{2}}\left(\frac{H}{f_{\pi}}\right)^{4}=2.081\times 10^{-9}
        \end{equation}
        yields:
        \begin{equation}
            \frac{f_{\pi}}{H}=\left(4\pi^{2}A_{s}\right)^{-1/4}\simeq 59.07 \quad\implies\quad \boxed{f_{\pi}\simeq 59.07\,H}\,;
        \end{equation}
        \item Moreover, \textit{\textbf{PGWs would determine the Hubble scale in terms of the Planck scale}} \cite{Krauss:2010ty,Guzzetti:2016mkm,Caprini:2018mtu,Maggiore-inflation-PGWs,Maggiore-PGWs,Tasinato2020PGWs,Caprini2020stochastic,Caldwell:2022qsj}:
        \begin{equation}
            H=\pi\,M_{\mathrm{Pl}}\sqrt{\frac{r A_{s}}{2}}\simeq 3.205\times 10^{-5}\sqrt{\frac{r}{0.1}}\,M_{\mathrm{Pl}}\,.
        \end{equation}
    \end{itemize}
\end{examplebox}

It was also shown that there is a breakdown of perturbative unitarity of Goldstone boson scattering for \cite{Baumann:2011su,Baumann:2014cja}:
\begin{equation}
    \omega^{4}>\Lambda_{u}^{4}\equiv\frac{24\pi}{5}\left(1-c_{s}^{2}\right)\Lambda^{4}\,.
\end{equation}

\begin{examplebox}[Hierarchy between the breakdown of perturbative unitarity and the SSB scale]
    \begin{itemize}
        \item For:
        \begin{equation}
            \Lambda_{u}>f_{\pi}
        \end{equation}
        $\implies$ \textit{\textbf{Weakly coupled slow-roll background with perturbative higher-derivative corrections}};
        \item For:
        \begin{equation}
            \Lambda_{u}<f_{\pi}
        \end{equation}
        $\implies$ \textit{\textbf{Completion below the SSB scale may signal strongly coupled dynamics}};
        \item For the critical value:
        \begin{equation}
            \Lambda_{u}=f_{\pi}
        \end{equation}
        $\implies$ \textit{\textbf{Threshold for the speed of sound}}\footnote{Assuming that $0<c_{s*}<1$.}:
        \begin{equation}
            \frac{24\pi}{5}\left(\frac{2M_{\mathrm{Pl}}^{2}\left|\dot{H}\right|\,c_{s*}^{5}}{1-c_{s*}^{2}}\right)=2M_{\mathrm{Pl}}^{2}\left|\dot{H}\right|\,c_{s*} \implies \boxed{c_{s*}=\frac{1}{4}\sqrt{\frac{\sqrt{5\left(5+96\pi\right)}-5}{3\pi}}\simeq 0.476}
        \end{equation}
        $\implies$ \textit{\textbf{Threshold for the amplitude of the non-Gaussianity}}\footnote{For $B\sim\mathcal{O}(1)$.}:
        \begin{subequations}
        \begin{align}
            f_{\mathrm{NL}*}^{\dot{\pi}\left(\partial_{i}\pi\right)^{2}} & =-\frac{85}{324}\left(c_{s*}^{-2}-1\right)\simeq \boldsymbol{-0.896}\,, \\
            f_{\mathrm{NL}*}^{\dot{\pi}^{3}} & \simeq\frac{5}{81}\left(c_{s*}^{-2}-1\right)\simeq \boldsymbol{0.211}\,, \\
            f_{\mathrm{NL}*}^{\mathrm{equil}} & \simeq -0.19\left(c_{s*}^{-2}-1\right)\simeq\boldsymbol{-0.649} \,, \\
            f_{\mathrm{NL}*}^{\mathrm{ortho}} & \simeq 0.04\left(c_{s*}^{-2}-1\right)\simeq\boldsymbol{0.137} \,.
        \end{align}    
        \end{subequations}
    \end{itemize}
    \textit{\textbf{\underline{Conclusion:}}}\\
    \textit{\textbf{The current observations cannot rule out non-Gaussianity above the threshold and an important target for future experiments is probing equilateral non-Gaussianity down to}} $\boldsymbol{f_{\mathrm{NL}}^{\mathrm{equil}}\sim\mathcal{O}(1)}$ \cite{Komatsu:2010hc,CMB-S4:2016ple,SimonsObservatory:2018koc,Meerburg:2019qqi,Karagiannis:2019jjx,Cabass:2022epm,CMB-S4:2022ght,Euclid:2025hlc}.
\end{examplebox}

\begin{figure}[htbp]
    \centering
    \includegraphics[width=0.6\linewidth]{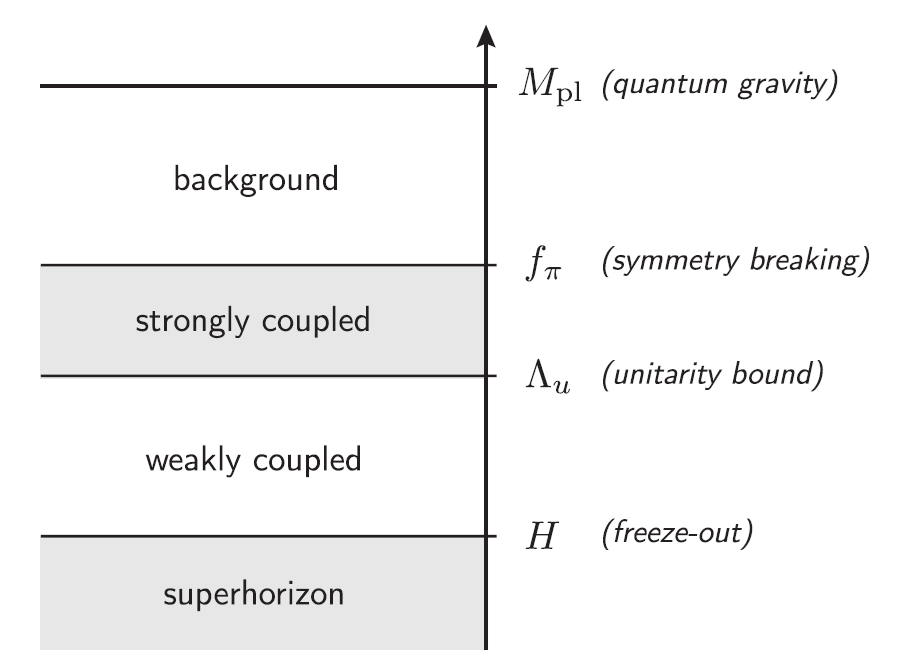}
    \caption{The relevant energy scales of the inflationary EFT. It is important to note the qualitative distinction of whether the breakdown of perturbative unitarity of Goldstone boson scattering $\Lambda_{u}$ is above or below the SSB energy scale $f_{\pi}$ \cite{Baumann_McAllister_2015appEFTinfl}.}
    \label{fig:placeholder}
\end{figure}

\subsection{Higher-derivative corrections}
Higher orders in the derivative expansion yields existence of operators
related with extrinsic and intrinsic curvature (\textit{\textbf{ghost inflation}} \cite{Arkani-Hamed:2003juy} and \textit{\textbf{Galileon inflation}} \cite{Burrage:2010cu}):
\begin{equation}\label{higher-derivative-corrections}
    \Delta\mathcal{L}=-\frac{1}{2}\left[\bar{M}_{1}^{3}(t)\,\delta g^{00}\delta K+\bar{M}_{2}^{2}(t)\,\delta K^{2}+\bar{M}_{3}^{2}(t)\bigl(\delta K^{\mu}{}_{\nu}\bigr)^{2}+\ldots\right]-\frac{1}{2}\Bigl[\hat{M}_{1}^{2}(t)\,\delta g^{00}\hat{R}+\ldots\Bigr]\,.
\end{equation}
\subsubsection{Expansion of higher-derivative operators}
Using \eqref{g00-relation} and \eqref{K-relation} the first operator in the Lagrangian \eqref{higher-derivative-corrections} becomes:
\begin{equation}
\begin{aligned}
    -\frac{1}{2}\bar{M}_{1}^{3}(t)\,\delta g^{00}\delta K & = -\frac{1}{2}\bar{M}_{1}^{3}\left(-2\dot{\pi}+\frac{\left(\partial_{i}\pi\right)^{2}}{a^{2}}+\ldots\right)\left(-\frac{\partial_{i}^{2}\pi}{a^{2}}+\ldots\right)= \\
    & = -\frac{1}{2}\bar{M}_{1}^{3}\left[2\dot{\pi}\frac{\partial_{i}^{2}\pi}{a^{2}}-\frac{\partial_{i}^{2}\pi\, \left(\partial_{j}\pi\right)^{2}}{a^{4}}+\ldots\right]\,.
\end{aligned}
\end{equation}
Integrating the 1st term by parts gives:
\begin{equation}\label{1st-high-der-corr}
    \int d^{4}x\,a^{3}\,2\dot{\pi}\frac{\partial_{i}^{2}\pi}{a^{2}}=2\int dt\,d^{3}x\,a\,\dot{\pi}\,\partial_{i}^{2}\pi=-2\int dt\,d^{3}x\,a\,\partial_{i}\dot{\pi}\,\partial_{i}\pi
\end{equation}
Moreover, one can observe that:
\begin{equation}
    \partial_{i}\dot{\pi}\,\partial_{i}\pi=\frac{1}{2}\partial_{t}\left[\left(\partial_{i}\pi\right)^{2}\right]\,,
\end{equation}
therefore, \eqref{1st-high-der-corr} yields:
\begin{equation}
    -\int dt\,d^{3}x\,a\,\partial_{t}\left(\partial_{i}\pi\right)^{2}=\int dt\,d^{3}x\,\dot{a}\left(\partial_{i}\pi\right)^{2}=\int dt\,d^{3}x\,a^{3}H\frac{\left(\partial_{i}\pi\right)^{2}}{a^{2}}
\end{equation}
so that:
\begin{equation}
    -\frac{1}{2}\bar{M}_{1}^{3}\,\delta g^{00}\delta K=-\frac{1}{2}\bar{M}_{1}^{3}\left[H\frac{\left(\partial_{i}\pi\right)^{2}}{a^{2}}-\frac{\partial_{i}^{2}\pi\,\left(\partial_{j}\pi\right)^{2}}{a^{4}}+\ldots\right]\,.
\end{equation}
Thus, the quadratic curvature operators are the following:
\begin{subequations}
    \begin{align}
        -\frac{1}{2}\bar{M}_{2}^{2}\,\delta K^{2} & = -\frac{1}{2}\bar{M}_{2}^{2}\frac{\left(\partial_{i}^{2}\pi\right)^{2}}{a^{4}}+\ldots \,,\\
        -\frac{1}{2}\bar{M}_{3}^{2}\,\left(\delta K^{\mu}{}_{\nu}\right)^{2} & = -\frac{1}{2}\bar{M}_{3}^{2}\frac{\left(\partial_{i}\partial_{j}\pi\right)^{2}}{a^{4}}+\ldots \,.
    \end{align}
\end{subequations}
Inside the action:
\begin{equation}
    \int d^{3}x\,\left(\partial_{i}\partial_{j}\pi\right)^{2}=\int d^{3}x\,\,\left(\partial_{i}^{2}\pi\right)^{2}\,,
\end{equation}
after integration by parts, we obtain:
\begin{equation}
    -\frac{1}{2}\left[\bar{M}_{2}^{2}\,\delta K^{2}+\bar{M}_{3}^{2}\,\left(\delta K^{\mu}{}_{\nu}\right)^{2}\right]=-\frac{1}{2}\left(\bar{M}_{2}^{2}+\bar{M}_{3}^{2}\right)\frac{\left(\partial_{i}^{2}\pi\right)^{2}}{a^{4}}+\ldots\,.
\end{equation}
Furthermore, the segment in the Lagrangian \eqref{higher-derivative-corrections} that contains the intrinsic Ricci curvature yields the following: 
\begin{equation}
    -\frac{1}{2}\bar{M}_{1}^{2}\,\delta g^{00}\hat{R}=-\frac{1}{2}\bar{M}_{1}^{2}\bigl(-2\dot{\pi}+\ldots\bigr)\left(4H\frac{\partial_{i}^{2}\pi}{a^{2}}+\ldots\right)=4\bar{M}_{1}^{2}H\,\dot{\pi}\frac{\partial_{i}^{2}\pi}{a^{2}}+\ldots
\end{equation}
and after integration by parts generates the term:
\begin{equation}
    -\frac{1}{2}\hat{M}_{1}^{2}\,\delta g^{00}\hat{R}=2\hat{M}_{1}^{2}\left[H^{2}\frac{\left(\partial_{i}\pi\right)^{2}}{a^{2}}+\ldots\right]\,.
\end{equation}
At the end of the day, the \textbf{higher-derivative corrections in the Goldstone Lagrangian} can be written in the following manner:
\begin{equation}\label{higher-derivative-final}
\begin{aligned}
    \Delta\mathcal{L}_{\pi} = & -\frac{1}{2}\bar{M}_{1}^{3}\left[H\frac{\left(\partial_{i}\pi\right)^{2}}{a^{2}}-\frac{\partial_{i}^{2}\pi\,\left(\partial_{j}\pi\right)^{2}}{a^{4}}+\ldots\right]-\frac{1}{2}\left(\bar{M}_{2}^{2}+\bar{M}_{3}^{2}\right)\frac{\left(\partial_{i}^{2}\pi\right)^{2}}{a^{4}}+\ldots \\
    & +2\hat{M}_{1}^{2}\left[H^{2}\frac{\left(\partial_{i}\pi\right)^{2}}{a^{2}}+\ldots\right]
\end{aligned}
\end{equation}
Phenomenology of these contributions \cite{Senatore:2009gt,Bartolo:2010bj,Creminelli:2010qf,Burrage:2010cu}

\subsubsection{Quadratic action and dispersion relation}
The higher-derivative corrections in \eqref{higher-derivative-final} yield the higher-derivative contribution:
\begin{equation}
    \propto\left(\bar{M}_{2}^{2}+\bar{M}_{3}^{2}\right)\left(\partial_{i}^{2}\pi\right)^{2}
\end{equation}
to the quadratic Goldstone boson Lagrangian:
\begin{equation}
    \mathcal{L}_{\pi}^{(2)}=\left(2M_{2}^{4}-M_{\mathrm{Pl}}^{2}\dot{H}\right)^{2}\dot{\pi}^{2}+M_{\mathrm{Pl}}^{2}\,\dot{H}\frac{\left(\partial_{i}\pi\right)^{2}}{a^{2}}-\frac{1}{2}\left(\bar{M}_{2}^{2}+\bar{M}_{3}^{2}\right)\frac{\left(\partial_{i}^{2}\pi\right)^{2}}{a^{4}}+\ldots\,.
\end{equation}
If the higher-derivative term dominates over the gradient one at horizon crossing, one may take into account the de Sitter limit:
\begin{equation}
    \dot{H}\to 0\,,
\end{equation}
which consequently yields the following form of the quadratic action:
\begin{equation}
     \mathcal{L}_{\pi}^{(2)}\simeq 2M_{2}^{4}\,\dot{\pi}^{2}-\frac{1}{2}\left(\bar{M}_{2}^{2}+\bar{M}_{3}^{2}\right)\frac{\left(\partial_{i}^{2}\pi\right)^{2}}{a^{4}}+\ldots\,.
\end{equation}
For a Fourier mode:
\begin{equation}
    \pi_{k}\propto e^{-i\omega t+i\vec{k}\cdot\vec{x}}
\end{equation}
one obtains:
\begin{equation}
    4M_{2}^{4}\,\omega^{2}=\left(\bar{M}_{2}^{2}+\bar{M}_{3}^{2}\right)\frac{k^{4}}{a^{4}}\,,
\end{equation}
so that the \textit{\textbf{dispersion relation is of the quartic (non-relativistic) character}}:
\begin{equation}\label{quartic-dispersion1}
    \omega^{2}\propto k^{4}\,.
\end{equation}
Moreover, by introducing the following energy scale (in the form of a combination of the mass scales):
\begin{equation}
    \rho^{2}\equiv\frac{4M_{2}^{4}}{\bar{M}_{2}^{2}+\bar{M}_{3}^{2}}
\end{equation}
produces the \textit{compact form of the dispersion relation}:
\begin{equation}\label{quartic-dispersion2}
    \omega=\frac{k^{2}}{a^{2}\rho}\,.
\end{equation}
\begin{examplebox}[Non-relativistic (quartic) dispersion relation - ghost inflation]
    The \textit{\textbf{quartic (non-relativistic) dispersion relation}} of the form \eqref{quartic-dispersion1} (or \eqref{quartic-dispersion2}) is a \textit{\textbf{distinctive feature of the ghost inflation}} scenario \cite{Arkani-Hamed:2003juy}.
\end{examplebox}
At the freeze-out:
\begin{equation}
    \omega\left(k,a\right)\sim H\,,
\end{equation}
therefore, we could define:
\begin{equation}\label{pf}
    p_{f}\equiv\frac{k_{f}}{a}\sim\sqrt{H\rho}\,.
\end{equation}
Moreover, using the definition of a group velocity:
\begin{equation}
    v_{g}\equiv\frac{\partial\omega}{\partial p}=\frac{\partial}{\partial p}\left(\frac{p^{2}}{\rho}\right)=2\frac{p}{\rho}
\end{equation}
with the use of \eqref{pf} yields:
\begin{equation}
    v_{g}\left(p_{f}\right)\sim 2\sqrt{\frac{H}{\rho}}\,.
\end{equation}
In order to obtain a \textit{subluminal propagation} one must take into account a hierarchy of the form:
\begin{equation}
    \rho\gtrsim 4H\,.
\end{equation}

\subsubsection{Mode equation and power spectrum}
The dependence of the scale factor on the conformal time is expressed as follows:
\begin{equation}
    a(\tau)=-\left(H \tau\right)^{-1}\,.
\end{equation}
Consequently, the mode equation takes the form:
\begin{equation}
    \pi_{k}''(\tau)-\frac{2}{\tau}\pi_{k}'(\tau)+\frac{H^{2}k^{4}\tau^{2}}{\rho^{2}}\pi_{k}(\tau)=0\,.
\end{equation}
Using the following substitution:
\begin{equation}\label{pi-substitution}
    \pi_{k}(\tau)=(-\tau)^{3/2}\,u_{k}\left(y\right)\quad\land\quad y\equiv\frac{H k^{2}\tau^{2}}{2\rho}
\end{equation}
one obtains the Bessel equation of order $\nu=\frac{3}{4}$ \cite{King_Billingham_Otto_2003Bessel,Prosperetti_2011Bessel,Whittaker_Watson_2021Bessel}:
\begin{equation}\label{Bessel1}
    y^{2}\frac{d^{2}u}{dy^{2}}+y\frac{du}{dy}+\left(y^{2}-\nu^{2}\right)\,u=0\quad\land\quad u=u(y)\,.
\end{equation}
whose solution, taking into account the \textit{\textbf{Bunch-Davies initial conditions}} \cite{Bunch:1978yq}, is:
\begin{equation}
    u_{k}(y)\propto H_{3/4}^{(1)}(y)\,,
\end{equation}
where $H_{3/4}^{(1)}(y)$ is the \textit{Hankel function of the first kind} (\textit{Bessel function of the third kind}) \cite{King_Billingham_Otto_2003Bessel,Prosperetti_2011Bessel,Whittaker_Watson_2021Bessel}. Furthermore, using an asymptotic behavior of the Hankel function at late-times (small values of the function argument $y$), one finds the \textit{\textbf{scale-invariant spectrum} of the comoving curvature perturbations} for the \textit{\textbf{ghost inflation}} scenario:
\begin{equation}
    \Delta_{\mathcal{R}}^{2}(k)=\frac{\Gamma^{2}\left(\frac{3}{4}\right)}{2\pi^{3}}\left(\frac{H}{M_{2}}\right)^{4}\left(\frac{\rho}{H}\right)^{3/2}\simeq\mathrm{const}\,.
\end{equation}
\begin{examplebox}[Derivation of the scale-invariant spectrum for ghost inflation]
Let us introduce the following quantities:
\begin{equation}
    \alpha\equiv\frac{H k^{2}}{\rho}\quad\land\quad y\equiv\frac{\alpha\,\tau^{2}}{2}=\frac{H k^{2} \tau^{2}}{2\rho}\,,
\end{equation}
so that:
\begin{equation}
    \frac{dy}{d\tau}=\alpha\,\tau=\frac{2y}{\tau}\,.
\end{equation}
Based on the substitution in \eqref{pi-substitution}, we know that during inflation:
\begin{equation}
    \pi_{k}(\tau)=(-\tau)^{3/2}\,u_{k}(y)\quad\land\quad \tau<0\,.
\end{equation}
The first derivative with respect to conformal time becomes:
\begin{equation}\label{pi-der1}
    \pi_{k}'\equiv\frac{d}{d\tau}\left[(-\tau)^{3/2}\,u_{k}(y)\right]=-\frac{3}{2}(-\tau)^{1/2}\,u_{k}+\alpha\,\tau\,(-\tau)^{3/2}\,u_{k,y}\,,
\end{equation}
where:
\begin{equation}
    u_{k,y}\equiv\frac{d u_{k}}{dy}\quad\land\quad u_{k}=u_{k}(y)\,.
\end{equation}
By using the following identity:
\begin{equation}
    \tau=-(-\tau)\quad\implies\quad \alpha\,\tau\,(-\tau)^{3/2}=-\alpha\,(-\tau)^{5/2}=-(-\tau)^{1/2}\,\alpha\,\tau^{2}
\end{equation}
we can simplify the formula \eqref{pi-der1}:
\begin{equation}
    \pi_{k}'=(-\tau)^{1/2}\left[-\frac{3}{2}u_{k}-\alpha\,\tau^{2}\,u_{k,y}\right]=(-\tau)^{1/2}\left[-\frac{3}{2}u_{k}-2y\,u_{k,y}\right]\,.
\end{equation}
In order to calculate the second derivative of the Goldstone $\pi_{k}$ mode:
\begin{equation}
    \pi_{k}''\equiv\frac{d}{d\tau}\left[(-\tau)^{1/2}\left(-\frac{3}{2}u_{k}-2y\,u_{k,y}\right)\right]
\end{equation}
let us introduce the functions:
\begin{equation}
    B(\tau)=(-\tau)^{1/2}\quad\land\quad F(y)=-\frac{3}{2}u_{k}-2y\,u_{k,y}\,
\end{equation}
that lead to:
\begin{equation}
    \pi_{k}''=B'\,F+B\,F'\,,
\end{equation}
where:
\begin{equation}
    B'=-\frac{1}{2}(-\tau)^{-1/2}
\end{equation}
and:
\begin{equation}
\begin{aligned}
    F' =-\frac{3}{2}u_{k,y}\,\frac{dy}{d\tau}-2\left(\frac{dy}{d\tau}u_{k,y}+y\,u_{k,yy}\,\frac{dy}{d\tau}\right) & =-\frac{3}{2}\frac{2y}{\tau}\,u_{k,y}-2\left(\frac{2y}{\tau}\,u_{k,y}+y\,u_{k,yy}\,\frac{2y}{\tau}\right)= \\
    & = -\frac{7y}{\tau}\,u_{k,y}-\frac{4y^{2}}{\tau}\,u_{k,yy}\,.
\end{aligned}
\end{equation}
In a consequence, the second derivative takes the form:
\begin{equation}
    \pi_{k}''=(-\tau)^{-1/2}\left[\frac{3}{4}u_{k}+8y\,u_{k,y}+4y^{2}\,u_{k,yy}\right]\,.
\end{equation}
On the other hand, the friction term can be written as:
\begin{equation}
    -\frac{2}{\tau}\pi_{k}'=-\frac{2}{\tau}(-\tau)^{1/2}\left[-\frac{3}{2}u_{k}-2y\,u_{k,y}\right]=(-\tau)^{-1/2}\Bigl[-3u_{k}-4y\,u_{k,y}\Bigr]
\end{equation}
and the frequency term as:
\begin{equation}
    \frac{H^{2}k^{4}\tau^{2}}{\rho^{2}}\pi_{k}=\alpha^{2}\tau^{2}\,(-\tau)^{3/2}\,u_{k}=4y^{2}\,(-\tau)^{-1/2}\,u_{k}\,.
\end{equation}
Using the above calculations, we thus obtain the \textit{\textbf{Bessel equation}} \eqref{Bessel1} \cite{King_Billingham_Otto_2003Bessel,Prosperetti_2011Bessel,Whittaker_Watson_2021Bessel}:
\begin{equation}\label{Bessel2}
    \boxed{y^{2}\,u_{k,yy}+y\,u_{k,y}+\left(y^{2}-\frac{9}{16}\right)\,u_{k}=0}
\end{equation}
with:
\begin{equation}
    \nu=\frac{3}{4}\,.
\end{equation}
The general solution of \eqref{Bessel2} takes the form:
\begin{equation}
    u_{k}(y)=c_{1}\,H_{3/4}^{(1)}(y)+c_{2}\,H_{3/4}^{(2)}(y)\,,
\end{equation}
so that the Goldstone mode can be written in terms of the Hankel functions of the 1st and 2nd kind:
\begin{equation}
    \pi_{k}(\tau)=(-\tau)^{3/2}\left[c_{1}\,H_{3/4}^{(1)}(y)+c_{2}\,H_{3/4}^{(2)}(y)\right]\,.
\end{equation}
At early-times:
\begin{equation}
    \tau\to{}^{-}\infty\quad\implies\quad y\to{}^{+}\infty
\end{equation}
and the large $y$ asymptotic behavior is \cite{King_Billingham_Otto_2003Bessel,Prosperetti_2011Bessel,Whittaker_Watson_2021Bessel}:
\begin{subequations}
    \begin{align}
        H_{\nu}^{(1)}(y) & \sim\sqrt{\frac{2}{\pi\,y}}\,e^{i\left(y-\frac{\pi\,\nu}{2}-\frac{\pi}{4}\right)} \,, \\
        H_{\nu}^{(2)}(y) & \sim\sqrt{\frac{2}{\pi\,y}}\,e^{-i\left(y-\frac{\pi\,\nu}{2}-\frac{\pi}{4}\right)} \,.
    \end{align}
\end{subequations}
From physical perspective, we are looking for a \textit{positive frequency solution}, therefore:
\begin{equation}
    c_{2}=0\quad\implies\quad \pi_{k}(\tau)\sim (-\tau)^{3/2}\,H_{3/4}^{(1)}(y)
\end{equation}
up to a normalization factor.

On the other hand, at late-times:
\begin{equation}
    \tau\to 0^{-}\quad\implies\quad y\to 0^{+}\,,
\end{equation}
and asymptotically the 1st Hankel function behaves in the following manner \cite{King_Billingham_Otto_2003Bessel,Prosperetti_2011Bessel,Whittaker_Watson_2021Bessel}:
\begin{equation}
    H_{\nu}^{(1)}(y)\sim -\frac{i}{\pi}\Gamma(\nu)\left(\frac{2}{y}\right)^{\nu}\quad\land\quad \nu>0\,,
\end{equation}
and in the specific case we are considering:
\begin{equation}
    H_{3/4}^{(1)}\sim -\frac{i}{\pi}\Gamma\left(\frac{3}{4}\right)\left(\frac{2}{y}\right)^{3/4}\,.
\end{equation}
Consequently, our solution takes the form:
\begin{equation}
    \pi_{k}(\tau)\sim (-\tau)^{3/2}\left[-\frac{i}{\pi}\Gamma\left(\frac{3}{4}\right)\left(\frac{2}{y}\right)^{3/4}\right]\,,
\end{equation}
where we may write explicitly the $y$-dependent segment in terms of the comoving wavenumber $k$:
\begin{equation}
    \left(\frac{2}{y}\right)^{3/4}=\left(\frac{4\rho}{H k^{2} \tau^{2}}\right)^{3/4}=2\sqrt{2}\,\left(\frac{\rho}{H}\right)^{3/4}\,k^{-3/2}\,(-\tau)^{-3/2}
\end{equation}
that yields into the \textit{\textbf{final form of the Goldstone mode function}}:
\begin{equation}
    \boxed{\pi_{k}\propto\Gamma\left(\frac{3}{4}\right)\left(\frac{\rho}{H}\right)^{3/4}\,k^{-3/2}\neq\pi_{k}(\tau)}
\end{equation}
$\implies$ The late-time mode amplitude is \textit{independent of conformal time} - the \textit{\textbf{mode freezes-out}} (approaches a \textit{constant value}).

Using the definition of curvature perturbation \eqref{pi-R-relation} one may obtain:
\begin{equation}
    \left|\mathcal{R}_{k}\right|^{2}=H^{2}\left|\pi_{k}\right|^{2}\propto H^{2}\left(\frac{\rho}{H}\right)^{3/2}\,k^{-3}\,.
\end{equation}
Furthermore, in the Fourier space we can define:
\begin{equation}
    \mathcal{R}\left(\vec{x}\right)=\int\frac{d^{3}k}{\left(2\pi\right)^{3}}\,e^{i\vec{k}\cdot\vec{x}}\,\mathcal{R}_{\vec{k}}
\end{equation}
and:
\begin{equation}
    \mathcal{R}_{\vec{k}}=\int d^{3}x\,e^{-i\vec{k}\cdot\vec{x}}\,\mathcal{R}\left(\vec{x}\right)\,,
\end{equation}
which due to the homogeneity yields the following form of the \textit{two-point correlation function}:
\begin{equation}
    \left\langle\mathcal{R}_{\vec{k}}\mathcal{R}_{\vec{k}'}\right\rangle=\left(2\pi\right)^{3}\,\delta^{(3)}(\vec{k}+\vec{k}')\,P_{\mathcal{R}}(k)\,,
\end{equation}
where:
\begin{equation}\label{dimensionful-power-spectrum-ghost}
    P_{\mathcal{R}}(k)=\left|\mathcal{R}_{k}\right|^{2}\propto k^{-3}
\end{equation}
is the (dimensionful) \textit{\textbf{power spectrum}}.

Finally, using \eqref{dimensionful-power-spectrum-ghost} one can obtain the \textit{\textbf{scale-invariant dimensionless power spectrum}} for the curvature perturbations in \textit{ghost inflation}:
\begin{equation}
    \boxed{\Delta_{\mathcal{R}}^{2}(k)\equiv\frac{k^{3}}{2\pi^{2}}P_{\mathcal{R}}(k)=\frac{\Gamma^{2}\left(\frac{3}{4}\right)}{2\pi^{3}}\left(\frac{H}{M_{2}}\right)^{4}\left(\frac{\rho}{H}\right)^{3/2}\simeq\mathrm{const}}\,.
\end{equation}
\end{examplebox}
For the purpose of estimating the significance of individual operators, one can use the so-called \textit{Lifshitz scaling}.
\begin{examplebox}[Lifshitz scaling]
\begin{itemize}
    \item The \textit{\textbf{Lifshitz scaling}} \cite{Lifshitz1941theory,Hohenberg:1977ym,Horava:2009uw} is an \textit{anisotropic scaling symmetry between time and space}:
    \begin{equation}
        t\mapsto \lambda^{z}\,t\quad\land\quad x^{i}\mapsto \lambda\,x^{i}\,,
    \end{equation}
    or equivalently, \textit{between momentum and frequency} variables:
    \begin{equation}
        k\mapsto \lambda^{-1}\,k\quad\land\quad \omega\mapsto \lambda^{-z}\,\omega\,,
    \end{equation}
    where the exponent $z$ is the \textit{dynamical critical exponent} that describes how the characteristic frequency scales with momentum:
    \begin{equation}
        \omega\sim k^{z}\,;
    \end{equation}
    \item Remark: Lifshitz scaling differs from the \textit{relativistic scaling}:
    \begin{equation}
        t\mapsto \lambda\,t\quad\land\quad x^{i}\mapsto \lambda\,x^{i}\,.
    \end{equation}
\end{itemize}
\end{examplebox}
In our case, the dynamical critical exponent takes the value of:
\begin{equation}
    z=2\,,
\end{equation}
therefore, we must apply the following explicit form of the scaling:
\begin{equation}\label{Lifhistz1}
\begin{dcases}
    t\mapsto s^{-1}\,t \\
    x^{i}\mapsto s^{-1/2}\,x^{i}
\end{dcases}
\quad\iff\quad
\begin{dcases}
    \omega\mapsto s\,\omega \\
    k\mapsto s^{1/2}\,k\,.
\end{dcases}
\end{equation}
Furthermore, one can also note that the partial derivatives scale oppositely to the left-hand side of \eqref{Lifhistz1}:
\begin{equation}
    \partial_{t}\mapsto s\,\partial_{t}\quad\land\quad \partial_{i}\mapsto s^{1/2}\,\partial_{i}\,.
\end{equation}
The next step is to determine the scaling of the Goldstone field:
\begin{equation}
    \pi\mapsto s^{\Delta}\,\pi\,,
\end{equation}
where the numerical value of the parameter $\Delta$ is determined by requiring the quadratic Goldstone action to scale homogeneously.
\begin{examplebox}[Determination of the value for the $\Delta$ parameter]
    \begin{itemize}
        \item The measure scales as:
        \begin{equation}
            dt\,d^{3}x\mapsto s^{-1}\,dt\,s^{-3/2}\,d^{3}x=s^{-5/2}\,dt\,d^{3}x\,;
        \end{equation}
        \item The kinetic term scales as:
        \begin{equation}
            \dot{\pi}^{2}\sim\left(\partial_{t}\pi\right)^{2}\mapsto\left(s\,s^{\Delta}\,\pi\right)^{2}=s^{2\left(1+\Delta\right)}\,\dot{\pi}^{2}
        \end{equation}
        and its contribution to the action becomes:
        \begin{equation}
            S_{\dot{\pi}^{2}}\mapsto s^{-5/2}\,s^{2\left(1+\Delta\right)}\,S_{\dot{\pi}^{2}}=s^{-1/2+2\Delta}\,S_{\dot{\pi}^{2}}\,;
        \end{equation}
        \item In similar manner, the higher-derivative term scales as:
        \begin{equation}
            \partial_{i}^{2}\mapsto s\,\partial_{i}^{2}\quad\implies\quad \left(\partial_{i}^{2}\pi\right)^{2}\mapsto\left(s\,s^{\Delta}\,\pi\right)^{2}=s^{2\left(1+\Delta\right)}\left(\partial_{i}^{2}\pi\right)^{2}\,,
        \end{equation}
        so that:
        \begin{equation}
            S_{\left(\partial_{i}^{2}\pi\right)^{2}}\mapsto s^{-5/2}\,s^{2\left(1+\Delta\right)}\,S_{\left(\partial_{i}^{2}\pi\right)^{2}}=s^{-1/2+2\Delta}\,S_{\left(\partial_{i}^{2}\pi\right)^{2}}\,;
        \end{equation}
        \item Finally, the condition for the quadratic action to be scale invariant takes the form:
        \begin{equation}
            -\frac{1}{2}+2\Delta=0\quad\implies\quad \Delta=\frac{1}{4}\,,
        \end{equation}
        which yields the \textit{\textbf{explicit form of the scaling for the Goldstone boson}}:
        \begin{equation}\label{Lifshitz-Goldstone}
            \boxed{\pi\mapsto s^{1/4}\,\pi}\,.
        \end{equation}
    \end{itemize}
\end{examplebox}
Using the relationship \eqref{Lifshitz-Goldstone} for the Goldstone field, we are able to determine the explicit scaling factors for all derivatives relevant to our analysis, namely:
\begin{equation}
    \dot{\pi}\sim s^{5/4}\quad\land\quad \partial_{i}\pi\sim s^{3/4}\quad\land\quad \partial_{i}^{2}\pi\sim s^{5/4}\,,
\end{equation}
so, eventually, we can also establish scaling for individual operators:
\begin{examplebox}[Scaling of the higher-derivative operators]
    The operators scale in the following manner:
    \begin{equation}
        \dot{\pi}\,\left(\partial_{i}\pi\right)^{2}\sim s^{11/4}\quad\land\quad \partial_{i}^{2}\pi\,\left(\partial_{j}\pi\right)^{2}\sim s^{11/4}\quad\land\quad \dot{\pi}^{3}\sim s^{15/4}\,.
    \end{equation}
    \textit{\textbf{\underline{Conclusion:}}}\\
    The \textbf{most relevant operators with the same scaling in the cubic Goldstone action} are $\boldsymbol{\dot{\pi}\,\left(\partial_{i}\pi\right)^{2}}$ and $\boldsymbol{\partial_{i}^{2}\pi\,\left(\partial_{j}\pi\right)^{2}}$, whereas the operator $\dot{\pi}^{3}$ is more \textit{irrelevant and therefore suppressed} in the action.
\end{examplebox}
Finally, the \textit{\textbf{final form of the cubic Goldstone Lagrangian}} is the following:
\begin{equation}\label{cubic-Goldstone-action-final}
    \boxed{\mathcal{L}_{\pi}^{(3)}=2M_{2}^{4}\frac{\dot{\pi}\,\left(\partial_{i}\pi\right)^{2}}{a^{2}}+\frac{1}{2}\bar{M}_{1}^{3}\frac{\partial_{i}^{2}\pi\,\left(\partial_{j}\pi\right)^{2}}{a^{4}}+\ldots}\,.
\end{equation}
\subsubsection{Parametric size of the bispectrum}
At the freeze-out:
\begin{equation}
    \omega\sim H \quad\land\quad p_{f}\sim\sqrt{H\rho}
\end{equation}
the quadratic Goldstone Lagrangian is of the order:
\begin{equation}\label{quadratic-estimation}
    \mathcal{L}_{2}\sim M_{2}^{4}H^{2}\pi^{2}\,.
\end{equation}
The first cubic operator in the action \eqref{cubic-Goldstone-action-final} can be approximated by the following expression:
\begin{equation}
    \mathcal{L}_{3}^{(1)}\sim 2M_{2}^{4}\,\dot{\pi}\,\left(\partial_{i}\pi\right)^{2}\sim 2M_{2}^{4}\left(H \pi\right)\left(p_{f}^{2}\,\pi^{2}\right)\sim 2M_{2}^{4} H^{2}\rho\,\pi^{3}\,.
\end{equation}
Dividing the above expression by the estimate for the quadratic action \eqref{quadratic-estimation}, we obtain the ratio:
\begin{equation}
    \frac{\mathcal{L}_{3}^{(1)}}{\mathcal{L}_{2}}\sim\rho\,\pi=\frac{\rho}{H}\mathcal{R}\,,
\end{equation}
which yields into the \textit{\textbf{amplitude of non-Gaussianity for the first cubic operator}}:
\begin{equation}
    \boxed{f_{\mathrm{NL}}^{\dot{\pi}\left(\partial_{i}\pi\right)^{2}}\sim\frac{\rho}{H}}\,.
\end{equation}
Furthermore, the second cubic operator in \eqref{cubic-Goldstone-action-final} may be estimated as:
\begin{equation}
    \mathcal{L}_{3}^{(2)}\sim\bar{M}_{1}^{3}\left(\partial_{i}^{2}\pi\right)\left(\partial_{j}\pi\right)^{2}\sim\bar{M}_{1}^{3}\,p_{f}^{4}\,\pi^{3}\sim\bar{M}_{1}^{3}H^{2}\rho^{2}\pi^{3}\,,
\end{equation}
and dividing once again by \eqref{quadratic-estimation} produces the following ratio:
\begin{equation}
    \frac{\mathcal{L}_{3}^{(2)}}{\mathcal{L}_{2}}\sim\frac{\bar{M}_{1}^{3}}{M_{2}^{4}}\rho^{2}\,\pi=\frac{\bar{M}_{1}^{3}}{\left(\bar{M}_{2}^{2}+\bar{M}_{3}^{2}\right)}\pi=\frac{\bar{M}_{1}^{3}}{\left(\bar{M}_{2}^{2}+\bar{M}_{3}^{2}\right)}\frac{\mathcal{R}}{H}\,.
\end{equation}
This implies the \textit{\textbf{amplitude of non-Gaussianity for the second cubic operator}}:
\begin{equation}
    \boxed{f_{\mathrm{NL}}^{\partial_{i}^{2}\pi\left(\partial_{j}\pi\right)^{2}}\sim\frac{\bar{M}_{1}^{3}}{\left(\bar{M}_{2}^{2}+\bar{M}_{3}^{2}\right)}\frac{1}{H}}\,.
\end{equation}
\begin{examplebox}[Exact results for the amplitudes of non-Gaussianity in the ghost inflationary scenario]
    The exact relationships take the following forms \cite{Senatore:2009gt}:
    \begin{subequations}
        \begin{align}
            f_{\mathrm{NL}}^{\dot{\pi}\left(\partial_{i}\pi\right)^{2}} & =0.25\frac{\rho}{H}\,, \\
            f_{\mathrm{NL}}^{\partial_{i}^{2}\pi\left(\partial_{j}\pi\right)^{2}} & =0.13\frac{\bar{M}_{1}^{3}}{\left(\bar{M}_{2}^{2}+\bar{M}_{3}^{2}\right)}\frac{1}{H}\,.
        \end{align}
    \end{subequations}
\end{examplebox}
\begin{examplebox}[Observational non-Gaussianity constraints on the ghost inflation]
    \begin{itemize}
        \item \textit{\textbf{Planck 2018}} ($T+E$) \cite{Planck:2019kim}:
        \begin{equation}
            f_{\mathrm{NL}}^{\mathrm{Ghost}}=-48\pm 52\,.
        \end{equation}
    \end{itemize}
\end{examplebox}
\subsection{Final form of the Goldstone Lagrangian and field equation}
\subsubsection{Goldstone Lagrangian}
Based on our considerations and derivations, we can write the \textit{\textbf{final form of the Goldstone Lagrangian at the decoupling limit}} \textit{up to a cubic order}:
\begin{examplebox}[Final form of the Goldstone Lagrangian (at the decoupling limit)]
    \begin{equation}\label{Goldstone-action-final}
    \begin{aligned}
        \mathcal{L}_{\pi} = &\,M_{\mathrm{Pl}}\left|\dot{H}\right|\left[\dot{\pi}^{2}-\frac{\left(\partial_{i}\pi\right)^{2}}{a^{2}}\right]+2M_{2}^{4}\left[\dot{\pi}^{2}-\dot{\pi}\frac{\left(\partial_{i}\pi\right)^{2}}{a^{2}}\right]+\left(2M_{2}^{4}-\frac{4}{3}M_{3}^{4}\right)\dot{\pi}^{3} \\
        & -\frac{1}{2}\left[H \bar{M}_{1}^{3}\frac{\left(\partial_{i}\pi\right)^{2}}{a^{2}}+\left(\bar{M}_{2}^{2}+\bar{M}_{3}^{2}\right)\frac{\left(\partial_{i}^{2}\pi\right)^{2}}{a^{4}}-\bar{M}_{1}^{3}\frac{\partial_{i}^{2}\pi\,\left(\partial_{j}\pi\right)^{2}}{a^{4}}\right]+\ldots\,.
    \end{aligned}
\end{equation}
\end{examplebox}
\subsubsection{Goldstone field equation}
\begin{examplebox}[Convention]
    \begin{itemize}
        \item We will use the differential operators:
        \begin{equation}
            \Delta\equiv\partial_{i}^{2}=\delta^{ij}\partial_{i}\partial_{j}\quad\land\quad \left(\partial\pi\right)^{2}\equiv\left(\partial_{i}\pi\right)^{2}\,,
        \end{equation}
        where $\Delta$ denotes the \textit{Laplace operator} (\textit{Laplacian});
        \item Furthermore, we introduce the following (time-dependent) functions, which constitute the (time-dependent) parameters in the resulting Euler–Lagrange equations:
        \begin{align}\label{Goldstone-final-parameters}
            A=A(t) & \equiv M_{\mathrm{Pl}}^{2}\left|\dot{H}\right|+2M_{2}^{4} & \land\quad & B=B(t)\equiv 2M_{2}^{4}-\frac{4}{3}M_{3}^{4} \,, \\
            C=C(t) & \equiv M_{\mathrm{Pl}}^{2}\left|\dot{H}\right|+\frac{1}{2}H \bar{M}_{1}^{3} & \land\quad & E=E(t)\equiv\bar{M}_{2}^{2}+\bar{M}_{3}^{2}\,;
        \end{align}
        \item The Goldstone action associated with \eqref{Goldstone-action-final} takes the form:
        \begin{equation}
            S_{\pi}\equiv\int d^{4}x\,\sqrt{-g}\,\mathcal{L}_{\pi}=\int dt\,d^{3}x\,a^{3}\mathcal{L}_{\pi}\,,
        \end{equation}
        where:
        \begin{equation}
            L_{\pi}\equiv a^{3}\mathcal{L}_{\pi}\,.
        \end{equation}
    \end{itemize}
\end{examplebox}
The \textbf{Euler-Lagrange equation for a cosmological Goldstone boson} takes the following form:
\begin{equation}
    \underbrace{\frac{\partial L_{\pi}}{\partial\pi}}_{=0}-\frac{d}{dt}\left(\frac{\partial L_{\pi}}{\partial\dot{\pi}}\right)-\partial_{i}\left(\frac{\partial L_{\pi}}{\partial\left(\partial_{i}\pi\right)}\right)+\partial_{i}\partial_{j}\left(\frac{\partial L_{\pi}}{\partial\left(\partial_{i}\partial_{j}\pi\right)}\right)=0\,,
\end{equation}
where the first term vanishes because $\mathcal{L}_{\pi}$ given by \eqref{Goldstone-action-final} \textit{does not depend on the Goldstone field itself}, but \textit{only on its derivatives}. Furthermore, using \eqref{Goldstone-final-parameters} allows one to rewrite the Goldstone Lagrangian as:
\begin{equation}\label{Goldstone-Lagrangian-coefficients}
    \mathcal{L}_{\pi}=A\,\dot{\pi}^{2}-C\frac{\left(\partial\pi\right)^{2}}{a^{2}}-2M_{2}^{4}\,\dot{\pi}\frac{\left(\partial\pi\right)^{2}}{a^{2}}+B\,\dot{\pi}^{3}-\frac{1}{2}E\frac{\left(\Delta\pi\right)^{2}}{a^{4}}+\frac{1}{2}\bar{M}_{1}^{3}\frac{\Delta\pi\,\left(\partial\pi\right)^{2}}{a^{4}}\,.
\end{equation}
\subsubsection{Explicit forms of the equation of motion}
\begin{equation}
    \frac{\partial L_{\pi}}{\partial\dot{\pi}}=2a^{3}A\,\dot{\pi}-2a\,M_{2}^{4}\left(\partial\pi\right)^{2}+3a^{3}B\,\dot{\pi}^{2}\,,
\end{equation}
\begin{equation}
    \frac{\partial L_{\pi}}{\partial\left(\partial_{i}\pi\right)}=-2a\,C\, \partial_{i}\pi-4a\,M_{2}^{4}\,\dot{\pi}\,\partial_{i}\pi+\frac{\bar{M}_{1}^{3}}{a}\left(\Delta\pi\right)\,\partial_{i}\pi\,,
\end{equation}
\begin{equation}
    \frac{\partial L_{\pi}}{\partial\left(\partial_{i}\partial_{j}\pi\right)}=\delta^{ij}\left[-\frac{E}{a}\Delta\pi+\frac{\bar{M}_{1}^{3}}{2a}\left(\partial\pi\right)^{2}\right]\,.
\end{equation}
\begin{equation}\label{Goldstone-EoM1}
    \boxed{\begin{aligned}
        \frac{d}{dt}\left[2a^{3}A\,\dot{\pi}-2a\,M_{2}^{4}\,\left(\partial\pi\right)^{2}+3a^{3}B\,\dot{\pi}^{2}\right]= & \,2a\,C\,\Delta\pi+4a\,M_{2}^{4}\,\partial_{i}\bigl(\dot{\pi}\,\partial_{i}\pi\bigr)-\frac{\bar{M}_{1}^{3}}{a}\partial_{i}\bigl[\left(\Delta\pi\right)\,\partial_{i}\pi\bigr] \\
        & -\frac{E}{a}\Delta^{2}\pi+\frac{\bar{M}_{1}^{3}}{2a}\Delta\Bigl[\bigl(\partial\pi\bigr)^{2}\Bigr]
    \end{aligned}}\,.
\end{equation}
Subsequently, the equation of motion \eqref{Goldstone-EoM1} can be further transformed by employing the following identities:
\begin{align}
    \Delta\Bigl[\bigl(\partial\pi\bigr)^{2}\Bigr] & = 2\bigl(\partial_{i}\partial_{j}\pi\bigr)^{2}+2\,\partial_{i}\pi\,\partial_{i}\bigl(\Delta\pi\bigr) \,, \\
    \partial_{i}\Bigl[\bigl(\Delta\pi\bigr)\,\partial_{i}\pi\Bigr] & = \partial_{i}\bigl(\Delta\pi\bigr)\,\partial_{i}\pi+\bigl(\Delta\pi\bigr)^{2}
\end{align}
which leads to the simplification of factors associated with $\bar{M}_{1}^{3}$:
\begin{equation}
    -\frac{\bar{M}_{1}^{3}}{a}\partial_{i}\bigl[\left(\Delta\pi\right)\,\partial_{i}\pi\bigr]+\frac{\bar{M}_{1}^{3}}{2a}\Delta\Bigl[\bigl(\partial\pi\bigr)^{2}\Bigr]=\frac{\bar{M}_{1}^{3}}{a}\Bigl[\bigl(\partial_{i}\partial_{j}\pi\bigr)^{2}-\bigl(\Delta\pi\bigr)^{2}\Bigr] \,.
\end{equation}
Consequently, this results in \textbf{another explicit form of the Goldstone field equation}:
\begin{equation}\label{Goldstone-EoM2}
    \boxed{\begin{aligned}
        \frac{d}{dt}\left[2a^{3}A\,\dot{\pi}-2a\,M_{2}^{4}\,\left(\partial\pi\right)^{2}+3a^{3}B\,\dot{\pi}^{2}\right]= & \,2a\,C\,\Delta\pi+4a\,M_{2}^{4}\,\partial_{i}\bigl(\dot{\pi}\,\partial_{i}\pi\bigr)-\frac{E}{a}\Delta^{2}\pi \\
        & +\frac{\bar{M}_{1}^{3}}{a}\Bigl[\bigl(\partial_{i}\partial_{j}\pi\bigr)^{2}-\bigl(\Delta\pi\bigr)^{2}\Bigr]
    \end{aligned}}\,.
\end{equation}
\subsubsection{Linearized Goldstone mode equation}
Considering only the quadratic part of the full Lagrangian \eqref{Goldstone-Lagrangian-coefficients} (or equivalently \eqref{Goldstone-action-final}):
\begin{equation}
    \mathcal{L}_{\pi}^{(2)}=A\,\dot{\pi}^{2}-C\frac{\left(\partial\pi\right)^{2}}{a^{2}}-\frac{1}{2}E\frac{\left(\Delta\pi\right)^{2}}{a^{4}}
\end{equation}
yields the \textbf{linearized equation of motion for the Goldstone boson}:
\begin{equation}
    \frac{d}{dt}\left(a^{3} A\,\dot{\pi}\right)-a\,C\,\Delta\pi+\frac{E}{2a}\Delta^{2}\pi=0\,,
\end{equation}
which, when the term involving the time derivative is expanded, takes the form similar to the \textit{Klein-Gordon equation in the FLRW Universe}:
\begin{examplebox}[Linearized Goldstone field equation]
\begin{equation}\label{linearized-Goldstone-EoM1}
    A\,\ddot{\pi}+\left(3H A+\dot{A}\right)\dot{\pi}-C\frac{\Delta\pi}{a^{2}}+\frac{1}{2}E\frac{\Delta^{2}\pi}{a^{4}}=0\,.
\end{equation}
\end{examplebox}
Furthermore, in the \textit{\textbf{quasi-de Sitter limit}}:
\begin{equation}
    A\propto\left|\dot{H}\right|\simeq\mathrm{const}\,,
\end{equation}
the equation \eqref{linearized-Goldstone-EoM1} can be expressed as:
\begin{examplebox}[Linearized Goldstone field equation in the quasi-de Sitter limit]
\begin{equation}\label{linearized-Goldstone-EoM2}
    A\left(\ddot{\pi}+3H\dot{\pi}\right)-C\frac{\Delta\pi}{a^{2}}+\frac{1}{2}E\frac{\Delta^{2}\pi}{a^{4}}=0\,.
\end{equation}
\end{examplebox}
\subsection{Conclusions on the equation of motion for the cosmological Goldstone boson}
\begin{examplebox}[Conclusions and observations regarding the EoM for the Goldstone field]
    The following conclusions and observations can be drawn based on the conducted analysis on the mechanism of SSB in the phase of cosmological inflation:
    \begin{enumerate}[label=(\Roman*)]
        \item The \textit{EFT formalism interpolates} between the standard \textit{\textbf{single-clock regime}} and the \textit{\textbf{higher-derivative ghost inflation regime}} \cite{Cheung:2007st,Arkani-Hamed:2003juy};
        \item In the \textit{\textbf{Fourier space}}:
        \begin{equation}
            \Delta\pi_{\vec{k}}=-k^{2}\,\pi_{\vec{k}}\quad\land\quad \Delta^{2}\pi_{\vec{k}}=k^{4}\,\pi_{\vec{k}}\,,
        \end{equation}
        so that the linearized EoM \eqref{linearized-Goldstone-EoM1} takes the form:
        \begin{equation}\label{Fourier-Goldstone-EoM1}
            \boxed{A\,\ddot{\pi}_{k}+\left(3H A+\dot{A}\right)\dot{\pi}_{k}+\left(C\frac{k^{2}}{a^{2}}+\frac{1}{2}E\frac{k^{4}}{a^{4}}\right)\pi_{k}=0}\,.
        \end{equation}
        In a consequence, the \textit{\textbf{Goldstone fluctuation behaves like a (time-dependent) damped oscillator}} that possesses \textbf{two distinct restoring force terms}:
        \begin{enumerate}[label=(\alph*)]
            \item An ordinary quadratic term,
            \item An additional higher-derivative quartic term;
        \end{enumerate}
        \item \textbf{\textit{Stability conditions}} \cite{Cheung:2007st}:
        \begin{enumerate}[label=(\alph*)]
            \item \textbf{No ghost}:
            \begin{equation}
                \boxed{A\equiv M_{\mathrm{Pl}}^{2}\left|\dot{H}\right|+2M_{2}^{4}>0}\,;
            \end{equation}
            \item \textbf{Gradient (for $\boldsymbol{k^{2}}$ domination)}:
            \begin{equation}
                \boxed{C\equiv M_{\mathrm{Pl}}^{2}\left|\dot{H}\right|+\frac{1}{2}H \bar{M}_{1}^{3}>0}\,,
            \end{equation}
            \item \textbf{Gradient (for $\boldsymbol{k^{4}}$ domination)}:
            \begin{equation}
                \boxed{E\equiv\bar{M}_{2}^{2}+\bar{M}_{3}^{2}>0}\,;
            \end{equation}
        \end{enumerate}
        \item Considering a \textit{\textbf{short wavelength}} (\textit{WKB}\footnote{Wentzel-Kramers-Brillouin.} \cite{Wentzel:1926aor,Kramers:1926njj,Brillouin:1926blg}) \textit{\textbf{regime}}, and \textit{neglecting the Hubble friction} together with the assumption of a \textit{slow time dependence of the coefficients}, allows one to obtain an \textbf{\textit{approximate exact solution}} of the form:
        \begin{equation}
            \pi_{k}\sim e^{-i\omega t}\,,
        \end{equation}
        which yields:
        \begin{equation}
            -A\,\omega^{2}+C\frac{k^{2}}{a^{2}}+\frac{1}{2}E\frac{k^{4}}{a^{4}}\simeq 0\,,
        \end{equation}
        so that:
        \begin{equation}\label{omega-squared}
            \boxed{\omega^{2}\simeq\frac{C}{A}\frac{k^{2}}{a^{2}}+\frac{E}{2A}\frac{k^{4}}{a^{4}}}
        \end{equation}
        $\implies$ There are \textit{\textbf{two propagation regimes}} \cite{Cheung:2007st,Arkani-Hamed:2003juy}:
        \begin{enumerate}[label=(\alph*)]
            \item \textbf{Relativistic regime} (\textit{quadratic domination}):
            \begin{equation}
                \boxed{\omega^{2}\simeq c_{s}^{2}\,p^{2}\quad\land\quad c_{s}^{2}\equiv\frac{C}{A}\quad\land\quad p\equiv\frac{k}{a}}\,,
            \end{equation}
            \item \textbf{Ghost inflation - quartic dispersion regime} (\textit{quartic domination}):
            \begin{equation}
                \boxed{\omega^{2}\simeq\frac{E}{2A}p^{4}\equiv\frac{p^{4}}{\rho_{\mathrm{eff}}^{2}}\quad\land\quad \rho_{\mathrm{eff}}^{2}\equiv\frac{2A}{E}}\,;
            \end{equation}
        \end{enumerate}
        with the transition scale coming from \eqref{omega-squared}:
        \begin{equation}
            C\,p_{*}^{2}\simeq\frac{1}{2}E\,p_{*}^{4}\quad\implies\quad\boxed{p_{*}^{2}\simeq\frac{2C}{E}}
        \end{equation}
        $\implies$ For physical momenta:
        \begin{equation}
            p \ll p_{*}\,,
        \end{equation}
        the \textbf{dynamics is effectively governed by a speed of sound} $c_{s}$ (relativistic regime), while for:
        \begin{equation}
            p \gg p_{*}\,,
        \end{equation}
        the \textbf{quartic term becomes dominant}.\\
        This is a \textit{\textbf{precise mathematical criterion that indicates when higher-derivative effects become significant}} \cite{Cheung:2007st,Arkani-Hamed:2003juy};
        \item In the \textit{\textbf{superhorizon limit}}:
        \begin{equation}
            \frac{k}{aH}\to 0 \quad\overset{\eqref{linearized-Goldstone-EoM1}}{\implies}\quad\boxed{\frac{d}{dt}\left(a^{3}A\,\dot{\pi}\right)\simeq 0}\,,
        \end{equation}
        which yields:
        \begin{equation}
            a^{3}A\,\dot{\pi}=\mathrm{const}\quad\land\quad \dot{\pi}\propto\frac{1}{a^{3}A}\,,
        \end{equation}
        so that:
        \begin{equation}\label{pi-t}
            \boxed{\pi(t)=\pi_{0}+\pi_{1}\int\frac{dt}{a^{3}(t)\,A(t)}}
        \end{equation}
        $\implies$ \textbf{Constant mode plus a decaying mode} (for a quasi-de Sitter):
        \begin{equation}
            \begin{dcases}
                a(t)\propto e^{Ht} \\
                A(t)\simeq\mathrm{const}>0 \\
                H=\mathrm{const}
            \end{dcases}
            \quad\implies\quad \int\frac{dt}{a^{3}(t)\,A(t)}\simeq\mathrm{const}-\frac{e^{-3Ht}}{3HA}=\mathrm{const}-\frac{1}{3HA}a^{-3}\,.
        \end{equation}
        If $A$ \textit{varies slowly} and the inflationary stage continues, the second term in \eqref{pi-t} \textbf{decays rapidly} (the \textbf{Goldstone mode freezes out} and the \textbf{curvature perturbation} $\boldsymbol{\mathcal{R}}$ \textbf{is conserved on superhorizon scales} \cite{Cheung:2007st,Weinberg:2003sw});
        \item The EoM \eqref{linearized-Goldstone-EoM1} identifies the following \textit{\textbf{analytical limits}}:
        \begin{enumerate}[label=(\alph*)]
            \item \textbf{Single-field} (\textit{slow-roll} \cite{Liddle:1994dx}, \textit{k-inflation} \cite{Armendariz-Picon:1999hyi}/\textit{DBI}\footnote{Dirac-Born-Infeld.} \textit{theories} \cite{Silverstein:2003hf,Peiris:2007gz,Baumann:2009ni,Baumann_McAllister_2015DBIinflation,Alishahiha:2004eh}):
            \begin{equation}
                \begin{dcases}
                    E=0 \\
                    \bar{M}_{1}=0
                \end{dcases}
                \quad\implies\quad \boxed{A\,\ddot{\pi}+\left(3HA+\dot{A}\right)\dot{\pi}-C\frac{\Delta\pi}{a^{2}}=0}\,,
            \end{equation}
            \item \textbf{Ghost inflation} \cite{Arkani-Hamed:2003juy}:
            \begin{equation}
                \begin{dcases}
                    C\simeq 0 \\
                    E>0
                \end{dcases}
            \quad\implies\quad \boxed{A\,\ddot{\pi}_{k}+\left(3HA+\dot{A}\right)\dot{\pi}_{k}+\frac{1}{2}E\frac{k^{4}}{a^{4}}\pi_{k}=0}
            \end{equation}
            \textbf{\textit{\underline{Physical (observational) implications}}}:
            \begin{itemize}
                \item A \textbf{scale-invariant spectrum with enhanced equilateral non-Gaussianity} \cite{Arkani-Hamed:2003juy}.
                \item The \textbf{bispectrum is much larger than in the conventional slow-roll scenario} \cite{Arkani-Hamed:2003juy}.
            \end{itemize}
        \end{enumerate}
        \item The \textit{\textbf{exact EoM}} \eqref{Goldstone-EoM2} \textit{\textbf{contains the cubic self-interactions}} which \textit{do not affect the free mode functions at linear order}, but \textit{\textbf{they do source mode coupling}} and therefore \textit{\textbf{generate non-Gaussianity}}. These terms are the \textit{\textbf{origin of the bispectrum and higher correlators}}.
        $\implies$ \textit{\textbf{\underline{The central message of inflationary EFT:}}}\\
        \textbf{The reduced speed of sound and the higher-derivative interactions lead to an increase in non-Gaussianity} \cite{Cheung:2007st};
        \item The \textbf{\textit{relevant freeze-out conditions}} are as follows:
        \begin{enumerate}[label=(\alph*)]
            \item \textbf{Relativistic regime}:
            \begin{equation}
                \boxed{\omega\simeq c_{s}\frac{k}{a}\sim H}\,,
            \end{equation}
            \item \textbf{Quartic regime}:
            \begin{equation}
                \boxed{\omega\simeq\frac{p^{2}}{\rho_{\mathrm{eff}}}\sim H}\quad\implies\quad p_{f}\simeq\sqrt{H\rho_{\mathrm{eff}}}\,,
            \end{equation}
        \end{enumerate}
        $\implies$ The EoM itself determines which notion of 'horizon crossing' is relevant:
        \begin{enumerate}[label=(\alph*)]
            \item \textbf{Slow-roll inflation}:
            \begin{equation}
                \boxed{\frac{k}{a}\sim H}\,,
            \end{equation}
            \item \textbf{Models with small} $\boldsymbol{c_{s}}$:
            \begin{equation}
                \boxed{\frac{k}{a}\sim\frac{H}{c_{s}}}\,,
            \end{equation}
            \item \textbf{Ghost inflation}:
            \begin{equation}
                \boxed{\frac{k}{a}\sim\sqrt{H\rho_{\mathrm{eff}}}}\,;
            \end{equation}
        \end{enumerate}
        \item In the \textit{\textbf{canonical slow-roll (single-clock) limit}}:
        \begin{equation}
            M_{2}=M_{3}=\bar{M}_{1}=\bar{M}_{2}=\bar{M}_{3}=0\,,
        \end{equation}
        one recovers the \textbf{decoupled Goldstone EoM}:
        \begin{equation}
            \boxed{\ddot{\pi}+3H\dot{\pi}-\frac{\Delta\pi}{a^{2}}=0}
        \end{equation}
        $\implies$ \textbf{A massless, free field Goldstone boson that induces nearly Gaussian curvature perturbations}.
    \end{enumerate}
\end{examplebox}

\subsection{Is this the end of the story? More degrees of freedom}
\begin{equation}
    S=S_{\pi}+S_{\psi}+S_{\mathrm{mix}}
\end{equation}
\begin{equation}
\begin{aligned}
    S= & \int d^{4}x\,\sqrt{-g}\Biggl[M_{\mathrm{Pl}}^{2}\left\{\frac{1}{2}R-\left(3H^{2}+\dot{H}\right)+\dot{H}\,g^{00}\right\}+\sum_{a}\left(-\frac{1}{2}\bigl(\partial\psi_{a}\bigr)^{2}-\frac{1}{2}m_{a}\psi_{a}^{2}\right) \\
    & +F\Bigl(\delta g^{00},\delta K_{\mu\nu},\hat{R},\psi_{a},\nabla_{\mu};t\Bigr)\Biggr]
\end{aligned}
\end{equation}
Unitary gauge interaction:
\begin{equation}
    \Delta\mathcal{L}_{\mathrm{mix}}=\beta_{a}(t)\,\delta g^{00}\,\psi_{a}
\end{equation}
induces the following term in the Goldstone action:
\begin{equation}
    \Delta\mathcal{L}_{\mathrm{mix}}=-2\beta_{a}(t)\,\dot{\pi}\,\psi_{a}+\beta_{a}(t)\left[-\dot{\pi}^{2}+\frac{\left(\partial_{i}\pi\right)^{2}}{a^{2}}\right]\psi_{a}+\ldots\,.
\end{equation}
\subsubsection{Heavy fields}
The evolution of adiabatic modes during inflation can be influenced by heavy fields \cite{Burgess:2002ub,Tolley:2009fg,Achucarro:2010da,Achucarro:2010jv,Achucarro:2012yr,Achucarro:2012sm,Cespedes:2012hu,Gong:2013sma}:
\begin{equation}
    m_{\psi}\gg H\,.
\end{equation}
The so-called \textit{\textbf{Gelaton field}} is one example of such an influence:
\begin{examplebox}[Gelaton field]
    A \textit{\textbf{gelaton field}} $\phi$ is a \textit{\textbf{heavy scalar degree of freedom}}:
    \begin{equation}
        m_{\mathrm{gel}}\gg H\,,
    \end{equation}
    that is \textbf{strongly coupled to the inflaton field} $\chi$ \cite{Tolley:2009fg}:
    \begin{equation}
        S_{\mathrm{gel}}=\int d^{4}x\,\sqrt{-g}\left[\frac{1}{2}M_{\mathrm{Pl}}^{2}\,R-\frac{1}{2}\left(\partial_{\mu}\phi\right)\left(\partial^{\mu} \phi\right)-\frac{1}{2}e^{2b(\phi)}\left(\partial_{\mu} \chi\right)\left(\partial^{\mu}\chi\right)-V\bigl(\phi,\chi\bigr)\right]\,.
    \end{equation}
    Integrating it out leaves a \textbf{significant imprint on the low-energy adiabatic dynamics}:
    \begin{equation}
        S=\int d^{4}x\,\sqrt{-g}\left[\frac{1}{2}M_{\mathrm{Pl}}^{2}\,R+p\bigl(X,\chi\bigr)+\ldots\right]\quad\land\quad X \equiv-\frac{1}{2}(\partial \chi)^2\,,
    \end{equation}
    where:
    \begin{equation}
        p\bigl(X,\chi\bigr)=e^{2b\bigl[\phi_{0}\bigl(\chi,X\bigr)\bigr]}X-V\bigl[\phi_{0}\bigl(\chi,X\bigr),\chi\bigr]
    \end{equation}
    comes from the \textit{minimum of an effective potential}:
    \begin{equation}
        \partial_{\phi}V\bigl(\phi_{0},\chi\bigr)-2b'\left(\phi_0\right) e^{2b\left(\phi_0\right)}X=0 \,;
    \end{equation}
    typically in the form of an \textbf{enhanced equilateral non-Gaussianity} and a \textbf{reduced effective speed of sound} \cite{Tolley:2009fg}:
    \begin{equation}
        c_{s}^2=\left(1+\frac{4e^{2b(\phi)}\,b'(\phi)^2\,\dot{\chi}^2}{m_{\mathrm{gel}}^2}\right)^{-1}<1\,,
    \end{equation}
    where:
    \begin{equation}
        m_{\mathrm{gel}}^{2}\equiv\partial_{\phi}^{2}\,V\bigl(\phi,\chi\bigr)-X\,\partial_{\phi}^{2}\left(e^{2b(\phi)}\right)=\partial_{\phi}^{2}\,V\bigl(\phi,\chi\bigr)-\dot{\chi}^{2}\,e^{2b(\phi)}\Bigl[2b'(\phi)^{2}+b''(\phi)\Bigr]
    \end{equation}
    is the \textit{effective gelaton mass}.
\end{examplebox}
For a quadratic toy model\footnote{Here, the symbol $\rho$ denotes the coefficient of the bilinear mixing operator $\dot{\pi}\,\psi$, which parametrizes the strength of the coupling between the Goldstone mode and the heavy field, and should not be confused with the scale of the dispersion relation, also denoted by the symbol $\rho$, which was used previously in the context of ghost inflation.} \cite{Senatore:2010wk,Achucarro:2012yr}:
\begin{equation}
    \mathcal{L}=-M_{\mathrm{Pl}}^{2}\left|\dot{H}\right|\left(\partial_{\mu}\pi\right)^{2}+\frac{1}{2}\dot{\psi}^{2}-\frac{1}{2}\frac{\left(\partial_{i}\psi\right)^{2}}{a^{2}}-\frac{1}{2}m_{\psi}^{2}\psi^{2}+2\rho\,\dot{\pi}\,\psi
\end{equation}
with:
\begin{equation}
    \omega^{2}\ll m_{\psi}^{2}\quad\land\quad p^{2}\equiv\frac{k^{2}}{a^{2}}\ll m_{\psi}^{2}\,,
\end{equation}
the heavy field $\psi$ may be integrated out and its equation of motion becomes:
\begin{equation}
    \left(m_{\psi}^{2}-\frac{\Delta}{a^{2}}+\ldots\right)\psi\simeq 2\rho\,\dot{\pi}\,,
\end{equation}
where:
\begin{equation}
    \psi\simeq\frac{2\rho}{m_{\psi}^{2}}\dot{\pi}+\ldots\,.
\end{equation}
Now, we may write that:
\begin{equation}
    \Delta\mathcal{L}_{\mathrm{eff}}\simeq\frac{2\rho^{2}}{m_{\psi}^{2}}\dot{\pi}^{2}+\ldots\,,
\end{equation}
therefore, the speed of sound takes the following explicit form \cite{Senatore:2010wk,Achucarro:2012sm}:
\begin{equation}
    c_{s}^{-2}=1+\frac{4\dot{\theta}^{2}}{M_{\mathrm{eff}}^{2}}\,,
\end{equation}
where $\dot{\theta}$ is the \textit{turn rate of the background in the field space} and $M_{\mathrm{eff}}$ is the \textit{effective mass of the entropic (isocurvature) fluctuation} that couples to the adiabatic mode \cite{Gordon:2000hv,Bassett:2005xm,Langlois:2008mn,Malik:2008im}.

\subsubsection{Light (or Hubble scale) fields}
If:
\begin{equation}
    m_{\psi}\lesssim H\,,
\end{equation}
the \textbf{extra fields cannot be integrated out}, so that the \textit{\textbf{single-clock description is insufficient}}. The simplest example of such an evolution is a quasi-single-field inflation with an additional scalar $\sigma$:
\begin{equation}
    \mathcal{L}=-M_{\mathrm{Pl}}\left|\dot{H}\right|\left(\partial_{\mu}\pi\right)^{2}+\frac{1}{2}\dot{\sigma}^{2}-\frac{1}{2}\frac{\left(\partial_{i}\sigma\right)^{2}}{a^{2}}-\frac{1}{2}m_{\sigma}^{2}\sigma^{2}+\rho\,\dot{\pi}\,\sigma+\mu\,\sigma^{3}+\ldots
\end{equation}
The \textbf{squeezed limit of the bispectrum} for:
\begin{equation}
    m_{\sigma}<\frac{3}{2}H
\end{equation}
is modified such that the \textbf{three-point correlation function} of the \textit{\textbf{primordial curvature perturbation}} \textit{\textbf{(on uniform density hypersurfaces)}}\footnote{In general, the quantities $\zeta$ and $\mathcal{R}$ \textbf{are not identical} because they are \textit{\textbf{two different gauge invariants}}. \textit{\textbf{They only coincide for adiabatic perturbations on super-Hubble scales}}, in which context they are often used interchangeably in the literature (e.g. \cite{Malik:2008im}).} $\zeta$ takes the form \cite{Chen:2009zp}:
\begin{equation}\label{squeezed-light}
    \boxed{\left\langle\zeta_{\vec{q}}\,\zeta_{\vec{k}}\,\zeta_{-\vec{k}}\right\rangle'\propto P_{\zeta}(q)\,P_{\zeta}(k) \left(\frac{q}{k}\right)^{3/2-\nu}\quad\land\quad \nu\equiv\sqrt{\frac{9}{4}-\frac{m_{\sigma}^{2}}{H^{2}}}}\,,
\end{equation}
where:
\begin{equation}
    q\ll k\,.
\end{equation}
Here, $q$ denotes the magnitude of the '\textit{soft}' (\textbf{long-wavelength}) momentum, and $k$ denotes the magnitude of the '\textit{hard}' (\textbf{short-wavelength}) momentum.

\begin{figure}[htbp]
    \centering
    \includegraphics[width=1\linewidth]{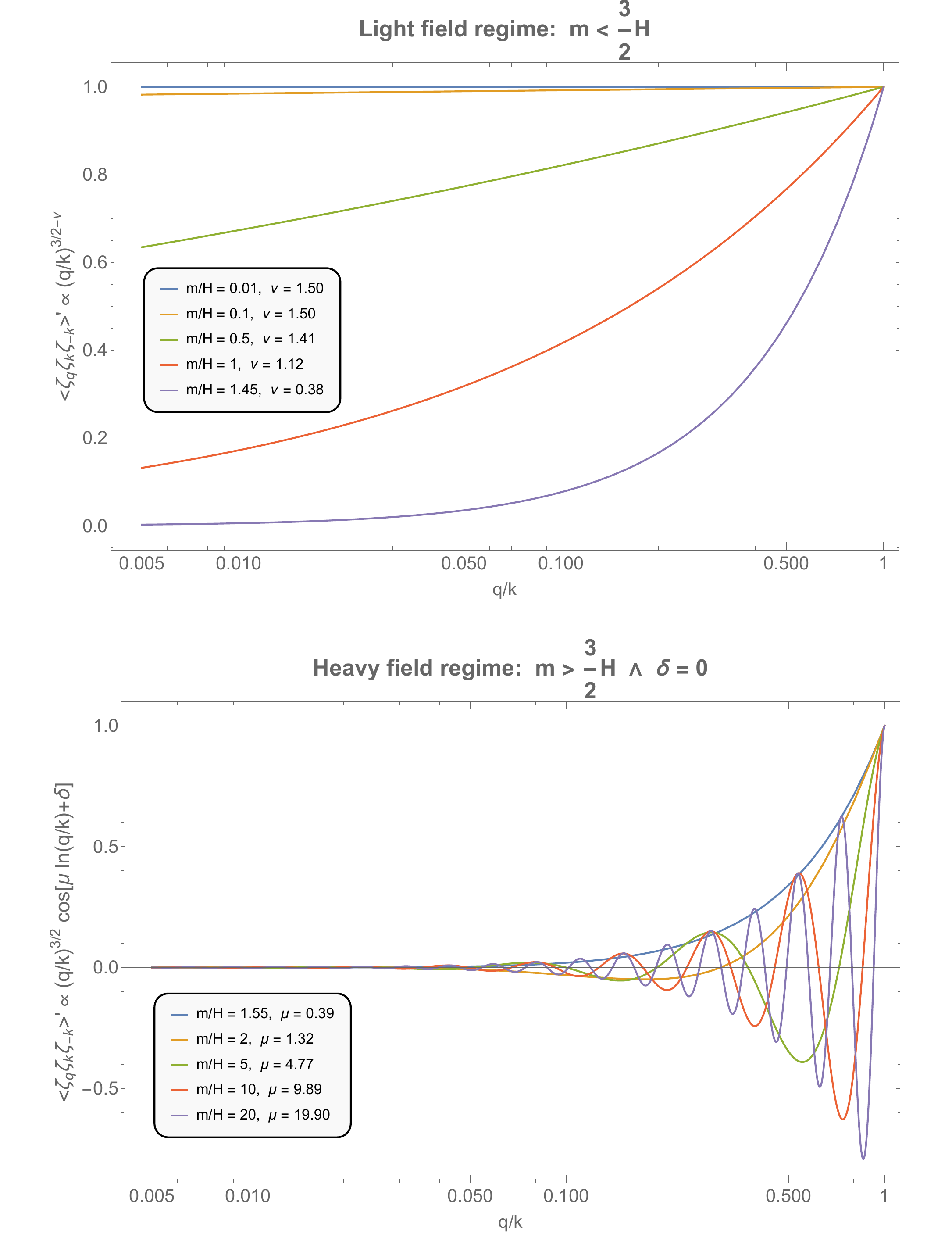}
    \caption{The reduced three-point correlation functions (light field regime \eqref{squeezed-light} at the top and heavy field regime \eqref{squeezed-heavy} at the bottom) for different mass scale hierarchies in the inflationary scenario with an additional scalar degree of freedom.}
    \label{fig:Squeezed-limit-bispectrum-inflation}
\end{figure}

On the other hand, if:
\begin{equation}
    m_{\sigma}>\frac{3}{2}H\,,
\end{equation}
the \textbf{signal becomes oscillatory} \cite{Chen:2009zp}:
\begin{equation}\label{squeezed-heavy}
    \boxed{\left\langle\zeta_{\vec{q}}\,\zeta_{\vec{k}}\,\zeta_{-\vec{k}}\right\rangle'\propto P_{\zeta}(q)\,P_{\zeta}(k) \left(\frac{q}{k}\right)^{3/2}\cos{\left[\mu\ln{\left(\frac{q}{k}\right)+\delta}\right]}\quad\land\quad \mu\equiv\sqrt{\frac{m_{\sigma}^{2}}{H^{2}}-\frac{9}{4}}}
\end{equation}
and also becomes the \textit{\textbf{potential source of the cosmological collider signal of Hubble-mass particles}} \cite{Arkani-Hamed:2015bza,Lee:2016vti}\footnote{Possible mechanisms for enhancing such signals and improving their observational prospects are explored in \cite{Wang:2019gbi}.}. Fig.~\ref{fig:Squeezed-limit-bispectrum-inflation} illustrates both cases of the squeezed bispectra.
\begin{examplebox}[Bispectrum notation]
    \begin{itemize}
        \item Instead of writing $\left\langle\zeta_{\vec{k}_{1}}\,\zeta_{\vec{k}_{2}}\,\zeta_{\vec{k}_{3}}\right\rangle$, one can define the \textit{\textbf{reduced three-point function}}:
        \begin{equation}
            \left\langle\zeta_{\vec{k}_{1}}\,\zeta_{\vec{k}_{2}}\,\zeta_{\vec{k}_{3}}\right\rangle=\left(2\pi\right)^{3}\,\delta^{(3)}\left(\vec{k}_{1}+\vec{k}_{2}+\vec{k}_{3}\right)\left\langle\zeta_{\vec{k}_{1}}\,\zeta_{\vec{k}_{2}}\,\zeta_{\vec{k}_{3}}\right\rangle'\,,
        \end{equation}
        or equivalently:
        \begin{equation}
            \left\langle\zeta_{\vec{k}_{1}}\,\zeta_{\vec{k}_{2}}\,\zeta_{\vec{k}_{3}}\right\rangle=\left(2\pi\right)^{3}\,\delta^{(3)}\left(\vec{k}_{1}+\vec{k}_{2}+\vec{k}_{3}\right)\,B_{\zeta}\bigl(k_{1},k_{2},k_{3}\bigr) \,,
        \end{equation}
        so that:
        \begin{equation}
            \boxed{\left\langle\zeta_{\vec{k}_{1}}\,\zeta_{\vec{k}_{2}}\,\zeta_{\vec{k}_{3}}\right\rangle'\equiv B_{\zeta}\bigl(k_{1},k_{2},k_{3}\bigr)}\,.
        \end{equation}
        It is the \textbf{standard bispectrum notation in the context of primordial non-Gaussianity} (e.g., \cite{Chen:2010xka});
        \item The \textit{motivation for factoring out the delta function is associated with \textbf{translational invariance}}. \textit{In position space}, statistical homogeneity leads to the conclusion that \textit{correlators depend only on separations}. However, \textit{in Fourier space}, this becomes \textit{momentum conservation}:
        \begin{equation}
            \vec{k}_{1}+\vec{k}_{2}+\vec{k}_{3}=0\,.
        \end{equation}
        The prime simply removes this universal factor, thereby allowing one to \textit{\textbf{focus on the physically relevant momentum dependence of the bispectrum itself}};
        \item Moreover, the \textbf{three external momenta used in the case of squeezed limit} are given by:
        \begin{equation}
            \vec{k}_{1}=\vec{q}\quad\land\quad \vec{k}_{2}=\vec{k}\quad\land\quad \vec{k}_{3}=-\vec{k}-\vec{q}\,,
        \end{equation}
        where:
        \begin{equation}
            q\equiv\left|\vec{q}\right|\ll k\equiv\left|\vec{k}\right|\quad\implies\quad \left|-\vec{k}-\vec{q}\right|\simeq k\,
        \end{equation}
        $\implies$ \textbf{One long mode and two short modes} $\equiv$ \textbf{\textit{The standard momentum configuration used in quasi-single-field inflation and cosmological collider physics}} \cite{Chen:2009zp,Arkani-Hamed:2015bza}.
    \end{itemize}
\end{examplebox}

\subsubsection{Symmetries and SUSY}
A \textbf{general type of multifield EFT} may include many more operators than a single-field EFT \cite{Senatore:2010wk}:
\begin{equation}\label{multifield-EFT}
    F\supset \lambda_{ab}(t)\,\psi_{a}\psi_{b}+\kappa_{abc}(t)\,\psi_{a}\psi_{b}\psi_{c}+\beta_{a}(t)\,\delta g^{00}\,\psi_{a}+\gamma_{a}(t)\,\delta g^{00}\partial^{0}\psi_{a}+\ldots\,,
\end{equation}
where $\lambda_{ab}(t)$ is a \textit{bilinear coupling matrix} (time-dependent mass/mixing matrix), $\kappa_{abc}(t)$ is a \textit{cubic cross-coupling tensor} (three-field interactions), $\beta_{a}(t)$ is a \textit{linear mixing coefficient} (strength of a derivative mixing), and $\gamma_a(t)$ is a \textit{derivative mixing coefficient} (time-dependent Wilson coefficient) controlling the coupling between the lapse fluctuation $\delta g^{00}$ and the time derivative of the extra field $\partial^0\psi_{a}$. Moreover, in \eqref{multifield-EFT} $a,b,c$ denote the field-space labels (not spacetime indices).

It is a common practice to \textit{postulate additional symmetries} to develop \textit{predictive theories} and \textit{technically natural} light fields. A concrete example would be the field, $psi_{I}$, which transforms as a vector under the \textit{\textbf{internal}} $\boldsymbol{SO(N)}$ \textit{\textbf{symmetry}}. In this case, only combinations of the field (and its derivatives) that are invariant under the given symmetry are permitted:
\begin{equation}
    \psi_{I}\psi_{I} \quad\land\quad \partial_{\mu}\psi_{I}\,\partial^{\mu}\psi_{I} \quad\land\quad \left(\psi_{I}\psi_{I}\right)^{2} \quad\land\quad \ldots\,.
\end{equation}
\begin{examplebox}[Internal $SO(N)$ symmetry]
    \begin{itemize}
        \item The \textit{\textbf{special orthogonal group}} $\boldsymbol{SO(N)}$ is the group of \textbf{orientation-preserving linear transformations} of $\mathbb{R}^{N}$ that preserve the \textit{Euclidean inner product} \cite{Pal_2019chapter}:
    \begin{equation}
        SO(N)=\bigl\{R\in GL\left(N,\mathbb{R}\right)\,\big|\,R^{T}R=\mathbb{I}_{N}\,\land\,\det R=1\bigr\}\,;
    \end{equation}
    \item If the additional fields $\psi_I$ ($I=1,\ldots,N$) transform in the \textit{\textbf{vector representation}} of this group:
    \begin{equation}
        \psi_{I}\mapsto R_{IJ}\,\psi_{J} \quad\land\quad R\in SO(N)\,,
    \end{equation}
    then the EFT is restricted only to \textbf{$SO(N)$-invariant combinations}, such as:
    \begin{equation}
        \psi_{I}\psi_{I} \quad\land\quad \partial_{\mu}\psi_{I}\,\partial^{\mu}\psi_{I} \quad\land\quad \bigl(\psi_{I}\psi_{I}\bigr)^{2} \quad\land\quad \psi_{I}\psi_{I}\,\delta g^{00} \quad\land\quad\ldots\,,
    \end{equation}
    meanwhile, all terms containing uncontracted internal indices are forbidden:
    \begin{equation}
        \psi_{I} \quad\land\quad \lambda_{I}(t)\psi_{I} \quad\land\quad \kappa_{IJK}(t)\psi_{I}\psi_{J}\psi_{K} \quad\land\quad \ldots\,.
    \end{equation}
    and this is only the case if the corresponding tensors are chosen to be $SO(N)$-invariants;
    \item The associated \textit{Lie algebra} is the space of \textit{\textbf{real antisymmetric matrices}} \cite{Pal_2019chapter}:
    \begin{equation}
        \mathfrak{so}(N)=\bigl\{X\in M_{N}\left(\mathbb{R}\right)\,\big|\,X^{T}=-X\bigr\}\,,
    \end{equation}
    so that:
    \begin{equation}
        \dim SO(N)=\frac{1}{2}N\left(N-1\right)\,;
    \end{equation}
    \item Nevertheless, in the context of SUSY, one should distinguish the \textit{internal} $SO(N)$ symmetry from the \textbf{supercharge symmetry}, since supercharges are \textit{spinorial objects} transformed under the corresponding \textit{spin group} rather than directly under $SO(N)$. Nonetheless, orthogonal groups may appear as \textbf{R-symmetry groups} in \textit{extended SUSY} \cite{Sohnius:1985qm,Martin:1997ns,Freedman:2012zz}.
    \end{itemize}
    \textit{\textbf{\underline{Conclusion:}}}\\
    Incorporating an internal $SO(N)$ symmetry \textit{\textbf{could be}} a simple and efficient method of reducing down on the number of multifield EFT operators that are allowed. This can possible make the theory more predictable and helps keep it \textit{technically natural}.
\end{examplebox}
\begin{examplebox}[R-symmetry]
    An \textit{\textbf{R-symmetry}} is an \textit{\textbf{internal automorphism symmetry of the supersymmetry algebra that acts nontrivially on the supercharges}} themselves:
    \begin{equation}
        Q^{A}\mapsto U^{A}{}_{B}\,Q^{B}\,,
    \end{equation}
    thus \textbf{constraining the allowed SUSY couplings much more strongly than an ordinary internal symmetry} \cite{Freedman:2012zz,Sohnius:1985qm,Martin:1997ns,Cecotti_2015chapter}.
\end{examplebox}
In addition, approximate shift symmetries further suppress non-derivative operators.

Theoretical (mathematical) models associated with supersymmetry and its variants \textit{may offer a possible theoretical explanation} for naturally light extra degrees of freedom. These models can also relate the mass of these SFs to the Hubble parameter in a mathematically elegant way.

The work by Baumann and Green \cite{Baumann:2011nk} is one attempt to identify \textbf{potential observational signatures of SUSY originating from the early stages of the Universe's evolution}. They highlight the possible connection between \textit{naturalness}, \textit{SUSY}, \textit{Hubble mass fields}, and \textit{squeezed-limit signatures}.
         
\begin{figure}[htbp]
    \centering
    \includegraphics[width=1\linewidth]{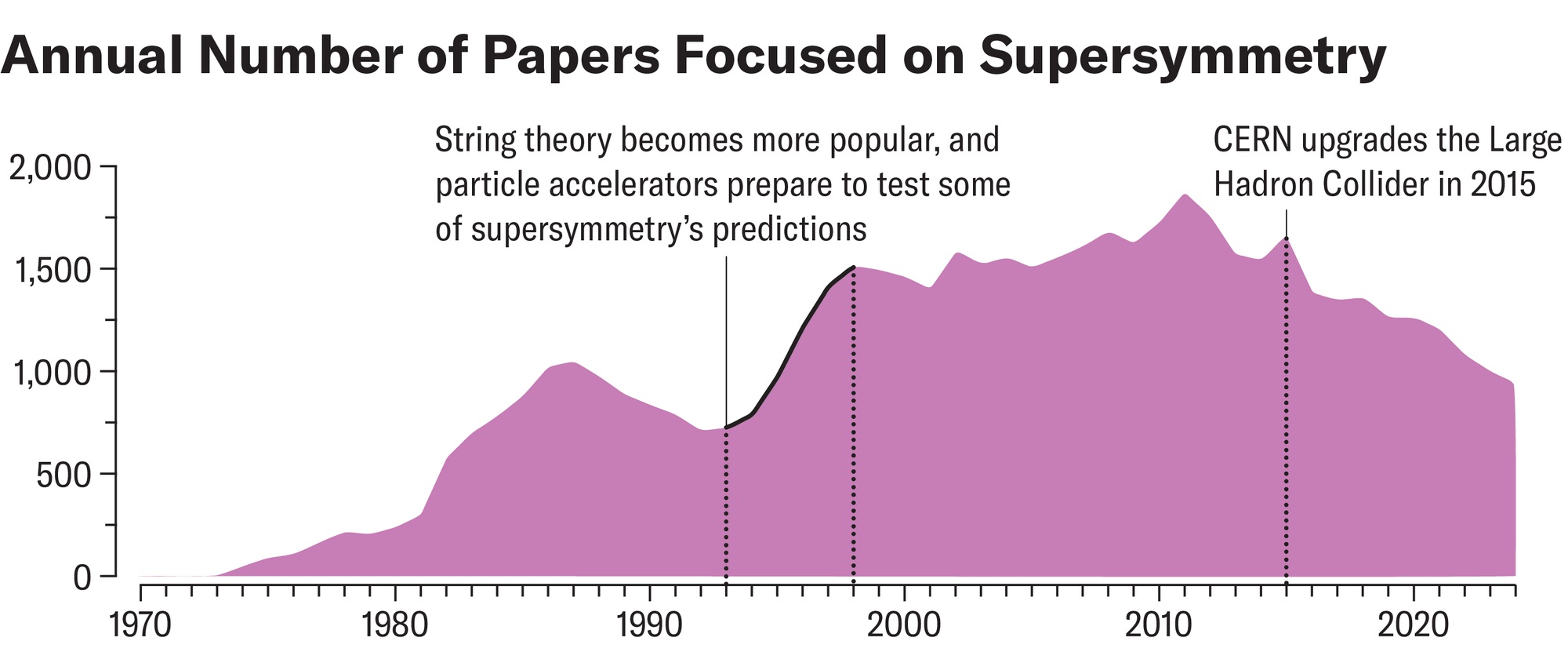}
    \caption{The annual number of research papers on supersymmetry indicates a significant trend of waning interest in this topic within the scientific community. More observational data is being collected by the LHC, but none of it seems to support the prevailing paradigm of SUSY \cite{Petrakou:2025}.}
    \label{fig:SUSY-sociology}
\end{figure}

It should be emphasized that, to this day, \textit{\textbf{no observational evidence has been found to confirm the existence of supersymmetry}} (or any of its variants) \cite{Haber:1984rc,Canepa:2019hph,DOnofrio:2025PDGSUSYExp} \textit{\textbf{in the real world}} (\textit{if such evidence even exists}). The issue of SUSY's impact on the development of modern physics reflects the broader challenges of the \textit{sociology of science} \cite{LykkenSpiropulu2014,Petrakou:2025} (see Fig.~\ref{fig:SUSY-sociology}).
The LHC was built to 'find' signs of supersymmetry. Despite providing ever-increasing amounts of data, the LHC offers no confirmation of the widely accepted SUSY hypothesis. However, the debate regarding SUSY remains still open \cite{Constantin:2025mex}$\ldots$

\subsubsection{Dissipative sectors and non-vacuum fluctuations}
The additional degrees of freedom coupled to the Goldstone mode need not behave as weakly coupled spectator fields. More generally, they may constitute an \emph{environment} whose microscopic dynamics is not resolved explicitly in the low-energy description. In the EFT language, this is encoded by coupling the Goldstone field to composite operators:
\begin{equation}
\bigl\{\mathcal O,\mathcal O_\mu,\mathcal O_{\mu\nu},\ldots\bigr\}
\label{eq:adddof_composite_ops}
\end{equation}
belonging to a \textit{dissipative sector} \cite{Cheung:2007st,LopezNacir:2011kk,LopezNacir:2012rm,Salcedo:2024smn}.

A schematic interaction consistent with the non-linearly realized time diffeomorphisms is:
\begin{equation}
    S_{\mathrm{int}}=\int d^4x\sqrt{-g}\,\Bigl[f(t+\pi)\,\mathcal{O}+f^{\mu}(t+\pi)\,\mathcal{O}_{\mu}+f^{\mu\nu}(t+\pi)\,\mathcal{O}_{\mu\nu}+\ldots\Bigr]\,.
\label{eq:adddof_dissipative_action}
\end{equation}
Expanding around the background clock:
\begin{equation}
    f(t+\pi)=f(t)+\dot{f}(t)\pi+\frac{1}{2}\ddot{f}(t)\pi^{2}+\ldots\,,
\label{eq:adddof_f_expand}
\end{equation}
one obtains, already at linear order, a coupling of the Goldstone fluctuation to the environmental operators:
\begin{equation}
    \Delta\mathcal{L}_{\mathrm{int}}\supset\dot{f}(t)\,\pi\,\mathcal{O}+\ldots\,.
\label{eq:adddof_linear_source}
\end{equation}
This is the \textbf{\textit{dissipative/open system} analogue of the usual St{\"u}ckelberg restoration of broken time diffeomorphisms}, now allowing for exchange of energy and information between the adiabatic mode and an unresolved sector \cite{Cheung:2007st,LopezNacir:2011kk,Salcedo:2024smn}.

In the full open system description, integrating out the environment generates, in general, nonlocal dissipation and noise kernels in the \textit{Schwinger-Keldysh} or \textit{closed time path} effective action \cite{Feynman:1963fq,Calzetta:2008iqa,Hu:2008rga,Kamenev_2011book}. Under suitable locality (Markovian) and near-equilibrium assumptions, however, these kernels reduce to a local \textit{\textbf{Langevin-type equation}}\footnote{See \cite{Hu:2008rga,Hu_Verdaguer_2020} for more details on \textit{semiclassical} and \textit{stochastic} gravity.} for the Fourier modes of the Goldstone boson \cite{LopezNacir:2011kk,Salcedo:2024smn}:
\begin{equation}
    \ddot{\pi}_{k}+\bigl(3H+\gamma\bigr)\dot{\pi}_{k}+\omega_{k}^{2}\left(t,k\right)\,\pi_{k}=\frac{\xi_{k}}{N_{c}}\quad\land\quad \omega_{k}^{2}\left(t,k\right)=c_{s}^{2}\frac{k^{2}}{a^{2}}+m_{\mathrm{eff}}^{2}+\ldots\,,
\label{eq:adddof_langevin}
\end{equation}
where $\gamma$ is an effective dissipation coefficient, $N_{c}$ denotes the kinetic normalization of the canonically normalized mode, and $\xi_{k}$ is a stochastic source induced by the environmental sector \cite{LopezNacir:2011kk,Salcedo:2024smn,Berera:2008ar,Kamali:2023lzq}.

\begin{examplebox}[Langevin equation]
    \begin{itemize}
        \item The \textit{\textbf{Langevin equation}} is a \textbf{stochastic differential equation} for a coarse-grained variable $q(t)$, consisting of a deterministic differential operator and a stochastic source term \cite{Langevin1908original,Langevin1908translation,Amir_2020Langevin,Livi_Politi_2025Langevin}:
        \begin{equation}\label{Langevin-equation}
            \ddot{q}(t)+\Gamma\,\dot{q}(t)+\Omega^{2}\,q(t)=\xi(t)\,.
        \end{equation}
        Here $\Gamma$ is a damping coefficient, $\Omega$ is a frequency parameter, and $\xi(t)$ is a stochastic process specified statistically rather than deterministically;
        \item More generally, one may consider a \textit{\textbf{nonlocal}} integro-differential form:
        \begin{equation}
            \ddot{q}(t)+\Omega^{2}(t)\,q(t)+\int^{t}dt'\,\Sigma_{R}\left(t,t'\right)\,q\left(t'\right)=\xi(t)\,,
        \end{equation}
        where $\Sigma_{R}\left(t,t'\right)$ is a \textbf{retarded memory kernel}. The ordinary local Langevin equation is recovered in the \textit{\textbf{Markovian limit}}, when the kernel admits a local derivative expansion:
        \begin{equation}
            \Sigma_{R}\left(t,t'\right)\mapsto 2\Gamma\,\delta'\left(t-t'\right)+\delta\Omega^{2}\,\delta\left(t-t'\right)+\ldots\,,
        \end{equation}
        where:
        \begin{equation}
            \delta'\left(t-t'\right)\equiv\frac{d\delta\left(t-t'\right)}{d\left(t-t'\right)}
        \end{equation}
        denotes the derivative of the Dirac delta distribution. The coefficients $\Gamma$ and $\delta\Omega^{2}$ arise as local approximations to a more general memory effect \cite{Amir_2020Langevin,Livi_Politi_2025Langevin};
        \item The stochastic source $\xi(t)$ is characterized by its \textit{\textbf{statistical moments}}. In the idealized \textit{\textbf{white noise}} limit, $\xi(t)$ is taken to have vanishing mean and delta-correlated covariance:
        \begin{equation}
            \left\langle\xi\left(t\right)\right\rangle=0 \quad\land\quad
            \left\langle\xi\left(t\right)\xi\left(t'\right)\right\rangle = \mathcal{N}\,\delta\left(t-t'\right)\,,
        \end{equation}
        where $\mathcal{N}$ is the noise amplitude. The adjective '\textit{white}' means that the noise has \textit{no preferred time scale} - its correlations vanish for:
        \begin{equation}
            t\neq t'\,,
        \end{equation}
        and equivalently its power spectral density is independent of frequency. If all higher connected correlators vanish, the noise is \textit{\textbf{Gaussian white noise}} \cite{Amir_2020Langevin,Livi_Politi_2025Langevin}. Strictly speaking, white noise is an idealized \textit{\textbf{generalized stochastic process}}, or distribution, rather than an ordinary function of time, and it may be regarded as the time derivative of a Wiener process \cite{Blei_2001chapter,Capinski_Kopp_Traple_2012chapter}:
        \begin{equation}
            \xi(t)=\frac{dW(t)}{dt}\,;
        \end{equation}
        \item The Langevin equation is mathematically equivalent, under the same Gaussian-Markovian assumptions, to a \textit{\textbf{Fokker-Planck equation}} for the probability density in phase space, $P\left(q,\dot{q};t\right)$. For the second order Langevin equation \eqref{Langevin-equation}, this is more precisely the \textit{\textbf{Kramers-Fokker-Planck equation}} \cite{Amir_2020Langevin,Livi_Politi_2025Langevin}. Hence, the same stochastic dynamics may be described either in terms of random trajectories or in terms of the evolution of a probability distribution;
        \item For a linear Langevin equation, the solution can be written using the \textit{\textbf{retarded Green function}} $G_{R}\left(t,t'\right)$ of the deterministic operator:
        \begin{equation}
            q(t)=q^{(\mathrm{hom})}(t)+\int^{t}dt'\,G_{R}\left(t,t'\right)\,\xi\left(t'\right)\,,
        \end{equation}
        where $q^{(\mathrm{hom})}(t)$ is the solution of the associated homogeneous equation. This decomposition separates the deterministic part of the evolution from the stochastic contribution induced by the noise term;
        \item Consequently, assuming linearity, vanishing mean noise, and no initial cross-correlation between the homogeneous solution and the stochastic source, the two-point function of $q(t)$ splits into a homogeneous part and a noise-induced part:
        \begin{equation}
            \left\langle q\left(t\right)\,q\left(t'\right)\right\rangle=
            \left\langle q^{(\mathrm{hom})}\left(t\right)\,q^{(\mathrm{hom})}\left(t'\right)\right\rangle+\int^{t}dt_{1}\int^{t'}dt_{2}\,
            G_{R}\left(t,t_{1}\right)\,G_{R}\left(t',t_{2}\right)\,
            \left\langle \xi(t_{1})\xi(t_{2})\right\rangle\,.
        \end{equation}
        Therefore, the statistical properties of the solution are determined jointly by the deterministic Green function and the covariance structure of the stochastic source.
    \end{itemize}
    \textit{\textbf{\underline{Conclusion:}}}\\
    The Langevin equation is a \textit{\textbf{mathematical effective evolution law}} in which unresolved microscopic dynamics is encoded by a deterministic differential operator together with a stochastic process specified by its correlators. Its essential mathematical content lies in the interplay between \textit{\textbf{drift}}, \textit{\textbf{damping}}, \textit{\textbf{memory kernels}}, and \textit{\textbf{noise statistics}}, together with its equivalent formulation in terms of the \textbf{Kramers-Fokker-Planck equation}.
\end{examplebox}
In this local limit the stochastic noise is often approximated as Gaussian and white, with correlator of the schematic form:
\begin{equation}
    \left\langle\xi_{\vec{k}}\left(t\right)\,\xi_{\vec{k'}}\left(t'\right)\right\rangle=\left(2\pi\right)^{3}\,\delta^{(3)}\left(\vec{k}+\vec{k'}\right)\frac{\nu_{\xi}(t)}{a^{3}(t)}\delta\left(t-t'\right)\,,
\label{eq:adddof_noise}
\end{equation}
where the amplitude $\nu_{\xi}$ is related to $\gamma$ by a fluctuation-dissipation relation in the local near-equilibrium regime by the following relation (up to convention-dependent normalization) \cite{Berera:2008ar,Kamali:2023lzq,Berera:1995wh}:
\begin{equation}
    \nu_{\xi}\propto\gamma\,T\,.
\end{equation}
The curvature perturbation then receives contributions from both the homogeneous (vacuum or quasi-vacuum) solution and the stochastic sourcing generated by the environment:
\begin{equation}
    \Delta_{\zeta}^{2}=\Delta_{\zeta,\mathrm{(hom)}}^2+\Delta_{\zeta,\mathrm{(noise)}}^2\,,
\label{eq:adddof_power_split}
\end{equation}
with the standard single-clock identification:
\begin{equation}
    \zeta\simeq -H\,\pi
\end{equation}
holding after freeze-out in the decoupling regime \cite{Cheung:2007st,LopezNacir:2011kk}. In the strongly dissipative regime:
\begin{equation}
\gamma \gg H\,,
\label{eq:adddof_gamma_large}
\end{equation}
the stochastic contribution may dominate over the intrinsic vacuum part, leading to an enhancement of the scalar power spectrum and, more generally, to distinctive non-Gaussian signatures of environmental interactions \cite{LopezNacir:2011kk,Salcedo:2024smn}.

\subsubsection{Conclusions}
The single-field (single-clock) Goldstone effective field theory (EFT) is the first step towards a more general class of descriptions of fluctuations that could arise during a potential inflationary epoch.

The description concerning additional degrees of freedom can be divided into two regimes:
\begin{enumerate}[label=(\alph*)]
    \item For a \textit{\textbf{heavy fields}} (with an adiabatic evolution):
    \begin{equation}
        m_{\psi}\gg H\,,
    \end{equation}
    the extra fields can be \textit{\textbf{integrated out}}, which simply \textbf{renormalizes the coefficients} of the Goldstone EFT;
    \item For a \textit{\textbf{light/Hubble scale fields}}:
    \begin{equation}
        m_{\psi}\lesssim H\,,
    \end{equation}
    one should \textit{\textbf{explicitly take into account in the action the additional fields}} whose presence may lead to new \textbf{squeezed-limit power spectra} dominated by \textit{non-vacuum fluctuations}. Consequently, this leads to a \textit{much more extensive phenomenology of inflation} than in the case of a single scalar field scenario.
\end{enumerate}

\section{Slow-roll inflation and the role of the inflaton mass}
The dynamics (at the background level) of the cosmological \textit{\textbf{slow-roll}}\footnote{For more details on slow-roll formalism, see \cite{Liddle:1994dx}.} inflationary scenario with a single scalar field (inflaton) is described by the following effective action for the canonical field:
\begin{equation}
    S=\int{d^{4}x\,\sqrt{-g}\left[\frac{M_{\mathrm{Pl}}^{2}}{2}R-\frac{1}{2}g^{\mu\nu}\partial_{\mu}\phi\,\partial_{\nu}\phi-V(\phi)\right]}\,.
\end{equation}
For the slow-roll regime characterized by the mathematical conditions:
\begin{equation}
    \left|\dot{\phi}^2\right| \ll V(\phi)\quad\land\quad \left|\ddot{\phi}\right| \ll 3H\dot{\phi}\sim\left|V'(\phi)\right|
\end{equation}
the key role is played by \textit{potential slow-roll parameters} describing the behavior of self-interaction potential:
\begin{equation}\label{slow-roll-parameters}
    \boxed{\epsilon_{V}\equiv\frac{M_{\mathrm{Pl}}^{2}}{2}\left(\frac{V'(\phi)}{V(\phi)}\right)^{2} \quad\land\quad \eta_{V}\equiv M_{\mathrm{Pl}}^{2}\left(\frac{V''(\phi)}{V(\phi)}\right)}\,,
\end{equation}
where:
\begin{equation}
    \boxed{\epsilon_{V}\ll 1 \quad\land\quad \left|\eta_{V}\right|\ll 1}\,.
\end{equation}
As will be shown later in this chapter, the slow roll parameter $\eta_{V}$ is especially sensitive to the inflaton mass scale (radiative corrections).

In order to perform a detailed analysis, let us consider the following general form of the potential for the inflaton:
\begin{equation}\label{inflation-potential}
    V(\phi)=V_{0}+\frac{1}{2}m^{2}\phi^{2}\,.
\end{equation}
\begin{examplebox}[Remark: Hilltop inflation]
    \begin{itemize}
        \item For the SF potential given by \eqref{inflation-potential} \textit{inflation occurs near the fixed point of a symmetry}:
        \begin{equation}\label{fixed-point-symmetry-inflation}
            V'(\phi)=0\,;
        \end{equation}
        \item In the case (taken into account in our considerations) of:
        \begin{equation}
            m^{2}>0
        \end{equation}
        symmetry is preserved and the potential \eqref{inflation-potential} may be interpreted as a \textit{\textbf{massive SF (inflaton) with a constant vacuum energy density}} ('\textit{false vacuum}' or \textit{cosmological constant}-like term);
        \item From a mathematical point of view, such a form can be obtained by expanding the unknown explicit form of $V(\phi)$ into a Taylor series around the fixed point \eqref{fixed-point-symmetry-inflation};
        \item Nevertheless, taking into account the negative values:
        \begin{equation}
            m^{2}<0
        \end{equation}
        yields the \textit{\textbf{spontaneous symmetry breaking}} (SSB) \cite{Baumann_McAllister_2015EFT-inflation}:
        \begin{equation}\label{hilltop-infl1}
            V(\phi)=V_{0}\left[1+\frac{1}{2}\eta_{0}\left(\frac{\phi}{M_{\mathrm{Pl}}}\right)^{2}\right]\,,
        \end{equation}
        where:
        \begin{equation}
            \eta_{0}\simeq\eta_{V}<0
        \end{equation}
        is the (effective) \textit{\textbf{tachyonic mass}} - value of $\eta_{V}$ at the maximum;
        \item For small values of the parameter $\eta_{0}$, the inflaton potential \eqref{hilltop-infl1} corresponds to the so-called \textit{\textbf{hilltop inflation}}\footnote{Such a form of inflaton potential appeared for the first time in the context of $\boldsymbol{\mathcal{N}=1}$ \textit{\textbf{SUGRA}} \cite{Linde:1983psb}.} scenario \cite{Boubekeur:2005zm,Kallosh:2019jnl}:
        \begin{equation}
            V(\phi)=V_{0}-\frac{1}{2}m^{2}\phi^{2}+\ldots=V_{0}\left[1-\frac{1}{2}\left|\eta_{0}\right|\left(\frac{\phi}{M_{\mathrm{Pl}}}\right)^2+\ldots\right]\quad\land\quad m^{2}>0
        \end{equation}
        that belongs to the class of \textit{\textbf{small-field models}} characterized by the following SF excursion during the inflationary phase:
        \begin{equation}
            \Delta\phi<M_{\mathrm{Pl}}\,.
        \end{equation}
    \end{itemize}
\end{examplebox}
Assuming that the constant term in \eqref{inflation-potential} dominates during inflation, we obtain:
\begin{equation}
    V''(\phi)\simeq m^{2} \quad\land\quad V(\phi)\simeq V_{0}\,,
\end{equation}
therefore:
\begin{equation}
    \eta_{V}\simeq M_{\mathrm{Pl}}^{2}\left(\frac{m^{2}}{V_{0}}\right)\,.
\end{equation}
During the inflationary epoch, a reasonable approximation of the Universe's evolution is the quasi de Sitter expansion:
\begin{equation}
    3M_{\mathrm{Pl}}^{2}\,H^{2}\simeq V_{0}\,,
\end{equation}
consequently, this leads to an explicit dependence of the slow-roll parameter on the SF mass and the Hubble parameter:
\begin{equation}\label{eta-de-Sitter}
    \eta_{V}\simeq\frac{m^{2}}{3H^{2}}\,.
\end{equation}
From \eqref{eta-de-Sitter} it follows that the following \textit{\textbf{hierarchy between}} $\boldsymbol{m}$ \textit{\textbf{and}} $\boldsymbol{H}$ \textit{\textbf{is required}} in order to satisfy the slow-roll regime:
\begin{equation}\label{inflaton-mass-hierarchy}
    m^{2}\ll H^{2}\,.
\end{equation}
Indeed, understanding the stability of this hierarchy, taking into account quantum corrections, is the essence of the \textit{\textbf{problem associated with the parameter}} $\boldsymbol{\eta_{V}}$ \cite{Lyth:1998xn,Baumann:2009ds,Baumann_McAllister_2015EFT-inflation}.

\section{Radiative corrections to a scalar mass: quadratic sensitivity}
From the perspective of \textit{\textbf{effective field theory}}\footnote{For more information on the EFT formalism in the general framework of gravity, see \cite{Donoghue:1994dn,Burgess:2003jk,Donoghue2024EFT1,Donoghue2024EFT2,Donoghue2024EFT3,Donoghue2024EFT4,Donoghue2024EFT5}.} (EFT) \cite{Burgess_2020,Georgi:1993mps,Meissner:2022cbi}, the inflaton is a \textit{light scalar particle} that is coupled to \textit{heavier degrees of freedom} \cite{Lyth:1998xn,Cheung:2007st,Weinberg:2008hq,Baumann_McAllister_2015appEFTinfl,Baumann_McAllister_2015EFT-inflation}. Therefore, integration of 'heavy' physics on the $\Lambda$ energy scale generates \textit{renormalized parameters} and \textit{higher-dimension operators}. The central aspect of SFs (e.g., inflaton) is the fact that their \textit{\textbf{masses are generically UV sensitive}} - loop corrections behave as:
\begin{equation}\label{UV-1}
    \Delta m^{2}\propto\Lambda^{2}\,.
\end{equation}
This sensitivity lies at the heart of the debate on \textit{\textbf{naturalness (hierarchy problem)}} in particle physics and cosmology \cite{Craig:2022eqo,Peskin:2025lsg,Giudice:2017pzm,Wells:2018sus,Hossenfelder:2018ikr}. Specific examples include:
\begin{enumerate}[label=(\alph*)]
    \item \textit{\textbf{Electroweak hierarchy}} (Higgs boson mass vs. Planck scale):
    \begin{equation}
        \begin{cases}
            v\equiv\left(\sqrt{2}\,G_{\mathrm{F}}\right)^{-1/2}\simeq 246\,\mathrm{GeV}\quad\text{(VEV of the Higgs field)} \\
            M_{\mathrm{Pl}}\equiv\bigl(8\pi\,G\bigr)^{-1/2}\simeq 2.4\times 10^{18}\,\mathrm{GeV}\quad\text{(reduced Planck mass)}
        \end{cases}
        \implies\boxed{\frac{v}{M_{\mathrm{Pl}}}\sim 10^{-16}}\,;
    \end{equation}
    \item \textit{\textbf{Cosmological constant problem}} (vacuum energy vs. particle physics scale):
    \begin{equation}
        \begin{cases}
            \rho_{\Lambda}\simeq 10^{-47}\,\mathrm{GeV}^{4}\quad\text{(vacuum energy density)} \\
            M_{\mathrm{Pl}}^{4}\simeq 10^{73}\,\mathrm{GeV}^{4}\quad\text{(particle physics scale)}
        \end{cases}
        \implies\boxed{\frac{\rho_{\Lambda}}{M_{\mathrm{Pl}}^{4}}\sim 10^{-120}}\,;
    \end{equation}
    \item \textit{\textbf{Strong CP problem}}:
    \begin{equation}
        \boxed{\bigl|\theta_{\mathrm{QCD}}\bigr|\lesssim 10^{-10}\ll 1}\,;
    \end{equation}
    \item \textit{\textbf{Neutrino mass hierarchy}}:
    \begin{equation}
        m_{\nu}\simeq 10^{-1}\,\mathrm{eV}\quad\text{(typical neutrino masses)}\quad\implies\quad \boxed{\frac{m_{\nu}}{v}\sim 10^{-13}}\,;
    \end{equation}
    \item \textit{\textbf{Inflaton mass hierarchy}} (inflaton mass vs. Hubble scale during inflation) given by \eqref{inflaton-mass-hierarchy}.
\end{enumerate}
We shall explicitly derive the relation \eqref{UV-1} for a simple toy model using a sharp momentum cutoff. More specifically, we will use standard methods known from QFT, including \textit{Wick rotation} and \textit{loop integrals} \cite{Peskin:1995ev,Weinberg:1995mt,Srednicki:2007qs}.

\begin{examplebox}[Convention]
Consistent with the adopted metric signature $\left(-,+,+,+\right)$, the following convention will be applied:
    \begin{itemize}
        \item Metric contraction of the gradient of the SF (inflaton):
        \begin{equation}
            \left(\nabla\phi\right)^{2}=\left(\partial\phi\right)^{2}\equiv g^{\mu\nu}\partial_{\mu}\phi\,\partial_{\nu}\phi\,;
        \end{equation}
        \item Momenta invariant:
        \begin{equation}
            k^{2}\equiv -k_{\mu}k^{\mu}=\left(k^{0}\right)^{2}-\vec{k}^{2}\,;
        \end{equation}
        \item Free scalar propagator:
        \begin{equation}
            \tilde{G}_{F}(k)=\frac{i}{(k^0)^2-\vec{k}^{\,2}-m^2+i\varepsilon}=\frac{i}{k^2-m^2+i\varepsilon}\,.
        \end{equation}
    \end{itemize}
\end{examplebox}
\subsection{Quartic interaction and one-loop 1PI self-energy}
We will examine a toy model with a real SF in flat spacetime with quartic self-interaction:
\begin{equation}\label{quartic-action}
    \mathcal{L}_{\mathrm{bare}}=-\frac{1}{2}\left(\partial\phi\right)^{2}-\frac{1}{2}m_{0}^{2}\phi^{2}-\frac{1}{4!}\lambda_{0}\phi^{4}\,,
\end{equation}
where $m_{0}^{2}$ and $\lambda_{0}$ are bare parameters.

The leading correction to the two-point function is the one-loop tadpole and the corresponding \textit{\textbf{one-particle-irreducible (1PI) self-energy}} $\boldsymbol{\Pi(p^{2})}$ is defined in such a way that the full propagator is obtained as a result of Dyson summation in terms of 1PI contributions. Therefore, at one loop we obtain the following \cite{Peskin:1995ev,Weinberg:1995mt,Srednicki:2007qs}:
\begin{equation}\label{one-loop-integral}
    i\,\Pi\left(p^{2}\right)=-\frac{i\lambda_{0}}{2}\int\frac{d^{4}k}{\left(2\pi\right)^{4}}\frac{i}{k^{2}-m_{0}^{2}+i\varepsilon}\,.
\end{equation}
In the case of action \eqref{quartic-action}, such a diagram is momentum-independent in this order, so that:
\begin{equation}
    \Pi\left(p^{2}\right)=\mathrm{const}\,.
\end{equation}
Consequently, we can define the induced one-loop contribution (shift) to the (bare) mass parameter as:
\begin{equation}
    \Delta m^{2}\equiv\Pi(0)\,.
\end{equation}

\subsection{Wick rotation: from Minkowski to Euclidean momentum}
We may calculate the loop integral \eqref{one-loop-integral} using the \textit{Wick rotation} approach:
\begin{equation}
    k^{0}=i k_{\mathrm{E}}^{0}\quad\land\quad d^{4}k=i\,d^{4}k_{\mathrm{E}}\quad\land\quad k_{\mathrm{E}}^{2}\equiv\left(k_{\mathrm{E}}^{0}\right)^{2}+\vec{k}^{2}\,,
\end{equation}
so that:
\begin{equation}
    k^{2}-m_{0}^{2}+i\varepsilon=\left(k^{0}\right)^{2}-\vec{k}^{2}-m_{0}^{2}+i\varepsilon\quad\mapsto\quad  -\left(k_{\mathrm{E}}^{2}+m_{0}^{2}\right)
\end{equation}
and the integral in \eqref{one-loop-integral} becomes:
\begin{equation}\label{Eucl-int}
    \int\frac{d^{4}k}{\left(2\pi\right)^{4}}\frac{i}{k^{2}-m_{0}^{2}+i\varepsilon}=\int\frac{d^{4}k_{\mathrm{E}}}{\left(2\pi\right)^{4}}\frac{1}{k_{\mathrm{E}}^{2}+m_{0}^{2}}\,.
\end{equation}
Substituting \eqref{Eucl-int} into \eqref{one-loop-integral} yields the following:
\begin{equation}
    \Pi(0)=\frac{\lambda_{0}}{2}\int\frac{d^{4}k_{\mathrm{E}}}{\left(2\pi\right)^{4}}\frac{1}{k_{\mathrm{E}}^{2}+m_{0}^{2}}\,.
\end{equation}
\subsection{Sharp cutoff}
Applying a \textit{sharp Euclidean cutoff} for the momenta:
\begin{equation}
    \left|k_{\mathrm{E}}\right|\leq\Lambda
\end{equation}
and using the properties of 4D spherical coordinates:
\begin{equation}
    \int_{\left|k_{\mathrm{E}}\right|\leq\Lambda}d^{4}k_{\mathrm{E}}=\Omega_{4}\int_{0}^{\Lambda}dk\,k^{3} \quad\land\quad \Omega_{4}=2\pi^{2}
\end{equation}
gives:
\begin{equation}\label{Pi-zero-cubic}
    \Pi(0)=\frac{\lambda_{0}}{2}\frac{\Omega_{4}}{\left(2\pi\right)^{4}}\int_{0}^{\Lambda}dk\,\frac{k^{3}}{k^{2}+m_{0}^{2}}=\frac{\lambda_{0}}{16\pi^{2}}\int_{0}^{\Lambda}dk\,\frac{k^{3}}{k^{2}+m_{0}^{2}}\,.
\end{equation}

\begin{examplebox}[4D spherical coordinates]
    Let:
    \begin{equation}
        k_\mathrm{E{\mu}}\in\mathbb{R}^4\quad\land\quad k\equiv\left|k_{\mathrm{E}}\right|=\sqrt{k_{\mathrm{E1}}^{2}+k_{\mathrm{E2}}^{2}+k_{\mathrm{E3}}^{2}+k_{\mathrm{E4}}^{2}}\,.
    \end{equation}
    and introduce the angles $\left(\chi,\theta,\phi\right)$ with the radius $k$:
    \begin{equation}
        \begin{aligned}
            & k_{\mathrm{E1}}=k\cos{\chi} \,, \\
            & k_{\mathrm{E2}}=k\sin{\chi}\cos{\theta} \,, \\
            & k_{\mathrm{E3}}=k\sin{\chi}\sin{\theta}\cos{\phi} \,, \\
            & k_{\mathrm{E4}}=k\sin{\chi}\sin{\theta}\sin{\phi} \,,
        \end{aligned}
    \end{equation}
    where:
    \begin{equation}
        k\in[0,\Lambda] \quad\land\quad \chi\in[0,\pi] \quad\land\quad \theta\in[0,\pi] \quad\land\quad \phi\in[0,2\pi)\,.
    \end{equation}
    \begin{itemize}
    \item The Euclidean line element takes the form:
    \begin{equation}
        ds^2=d k^2+k^2\,d\Omega_{3}^2\,,
    \end{equation}
    so that the volume element factorizes in the following manner:
    \begin{equation}
        d^{4}k_{\mathrm{E}}=k^{3}\,dk\,d\Omega_{3}\,,
    \end{equation}
    where:
    \begin{equation}
        d\Omega_{3}=\sin^{2}{\chi}\sin{\theta}\,d\chi\,d\theta\,d\phi
    \end{equation}
    is the solid angle element on the unit 3-sphere $S^{3}$;
    \item Therefore:
    \begin{equation}
        \int_{\left|k_{\mathrm{E}}\right|\leq\Lambda}d^{4}k_{\mathrm{E}}=\int_{0}^{\Lambda}dk\,k^{3}\int_{S^{3}}d\Omega_{3}\equiv\Omega_{4}\int_{0}^{\Lambda}dk\,k^{3}\,,
    \end{equation}
    where:
    \begin{equation}
        \Omega_{4}\equiv\int_{S^{3}}d\Omega_{3}
    \end{equation}
    is the 4D solid angle (surface area of the unit 3-sphere);
    \item Explicit form of $\Omega_{4}$:
    \begin{equation}
        \Omega_{4}=\int_{0}^{\pi}d\chi\,\sin^{2}{\chi}\int_{0}^{\pi}d\theta\,\sin{\theta}\int_{0}^{2\pi}d\phi=\frac{\pi}{2}\cdot 2 \cdot 2\pi=2\pi^{2}\,.
    \end{equation}
\end{itemize}
\end{examplebox}
Rewriting the integrand in \eqref{Pi-zero-cubic} as:
\begin{equation}
    \frac{k^{3}}{k^{2}+m_{0}^{2}}=k-\frac{k\,m_{0}^{2}}{k^{2}+m_{0}^{2}}
\end{equation}
yields:
\begin{equation}
    \int_{0}^{\Lambda}dk\,\frac{k^{3}}{k^{2}+m_{0}^{2}}=\int_{0}^{\Lambda}dk\,k\,- m_{0}^{2}\int_{0}^{\Lambda}dk\,\frac{k}{k^{2}+m_{0}^{2}}=\frac{\Lambda^{2}}{2}-\frac{m_{0}^{2}}{2}\,\ln{\left(\frac{\Lambda^{2}+m_{0}^{2}}{m_{0}^{2}}\right)}\,,
\end{equation}
hence:
\begin{equation}\label{Pi-zero1}
    \Pi(0)=\frac{\lambda_{0}}{32\pi^{2}}\left[\Lambda^{2}-m_{0}^{2}\ln{\left(\frac{\Lambda^{2}+m_{0}^{2}}{m_{0}^{2}}\right)}\right]\,.
\end{equation}
In the regime of:
\begin{equation}
    \Lambda\gg m_{0}
\end{equation}
\eqref{Pi-zero1} takes the form:
\begin{equation}
    \Pi(0)\simeq\frac{\lambda_{0}}{32\pi^2}\left[\Lambda^{2}-m_{0}^{2}\ln{\left(\frac{\Lambda^2}{m_{0}^2}\right)}\right]\,,
\end{equation}
leading to the conclusion that the induced correction to the (bare) mass contains an UV-sensitive square term:
\begin{equation}
    \boxed{\Delta m^2 \equiv \Pi(0)=\frac{\lambda_0}{32\pi^2}\Lambda^2+\mathcal{O}\left[\ln{\left(\frac{\Lambda^2}{m_{0}^2}\right)}\right]}
\end{equation}
consistent with the predictions of QFT and EFT \cite{Peskin:1995ev,Weinberg:1995mt,Srednicki:2007qs,Lyth:1998xn,Cheung:2007st,Weinberg:2008hq,Baumann_McAllister_2015appEFTinfl,Baumann_McAllister_2015EFT-inflation}.

\subsection{Heavy physics: EFT interpretation}
For the EFT formalism, the UV sensitivity is very often associated with integrating out heavy degrees of freedom for the energy scale:
\begin{equation}
    M\sim\Lambda\,.
\end{equation}
For a heavy scalar $\chi$ with mass $M$ described by the action \cite{Baumann_McAllister_2015EFT-inflation,Burgess_2020}:
\begin{equation}
    \mathcal{L}\supset -\frac{1}{2}\left(\partial\chi\right)^{2}-\frac{1}{2}M^{2}\chi^{2}-\frac{1}{2}g^{2}\phi^{2}\chi^{2}\,,
\end{equation}
where $g$ is the coupling constant, one finds the following relation for the quantum corrections \cite{Georgi:1993mps}:
\begin{equation}\label{delta-m-phi}
    \Delta m_{\phi}^{2}\sim\frac{g^{2}}{16\pi^{2}}M^{2}\sim\frac{g^{2}}{16\pi^{2}}\Lambda^{2}
\end{equation}
up to logarithmic terms. Therefore, the appearance of the heavy scale factor destabilizes the hierarchy of energy scales unless it is protected by \textit{symmetry} (\textit{\textbf{technical naturalness}}) \cite{Susskind:1978ms,tHooft:1979rat}.

\section{Eta problem I: radiative corrections}
\begin{equation}
    \eta_{V}\simeq\frac{m^{2}+\Delta m^{2}}{3H^{2}}=\eta_{V}^{\mathrm{(tree)}}+\Delta\eta_{V} \quad\land\quad \Delta\eta_{V}\simeq\frac{\Delta\eta_{V}}{3H^{2}}
\end{equation}
\begin{equation}
    \Delta m^{2}\sim\frac{g^{2}}{16\pi^{2}}\Lambda^{2}\quad\implies\quad \boxed{\Delta\eta_{V}\sim\frac{g^{2}}{48\pi^{2}}\frac{\Lambda^{2}}{H^{2}}\Bigg|_{\Lambda\gtrsim H}\gtrsim 1}\,.
\end{equation}
\begin{equation}
    \Delta m^{2}\sim\frac{\lambda_{0}}{32\pi^{2}}\Lambda^{2}\quad\implies\quad \boxed{\Delta\eta_{V}\sim\frac{\lambda_{0}}{96\pi^{2}}\frac{\Lambda^{2}}{H^{2}}\Bigg|_{\Lambda\gtrsim H}\gtrsim 1}\,.
\end{equation}
\begin{examplebox}[Eta problem I (Radiative corrections)]
    We know that a consistent inflationary EFT must satisfy:
    \begin{equation}
         \Lambda\gtrsim H\,,
    \end{equation}
    so that the horizon-scale modes are within the EFT domain \cite{Baumann:2014nda,Georgi:1993mps,Kaplan:2005es}, and therefore:
    \begin{equation}
        \boxed{\Delta\eta_{V}\sim\frac{\Lambda^{2}}{H^{2}}\Bigg|_{\Lambda\gtrsim H}\gtrsim 1}\,.
    \end{equation}
    \textit{\textbf{\underline{Conclusion:}}}\\
    \textbf{Long-term slow-roll inflation appears to be unnatural.}
\end{examplebox}

\subsection{SUSY: partial cancellation of radiative corrections}
One can show that \cite{Peskin:1995ev,Srednicki:2007qs}:
\begin{equation}
    \Delta m_{\mathrm{SUSY}}^{2}\propto m_{\mathrm{soft}}^{2}
\end{equation}
and therefore \cite{Baumann:2014nda}:
\begin{enumerate}[label=(\alph*)]
    \item For:
    \begin{equation}
        \omega\gg H
    \end{equation}
    flat space cancellations;
    \item For:
    \begin{equation}
        \omega\lesssim H
    \end{equation}
    sensitivity to quasi-de Sitter curvature:
    \begin{equation}
        \Delta m^{2}\sim H^{2} \quad\implies\quad \Delta\eta_{V}\sim\mathcal{O}(1) 
    \end{equation}
\end{enumerate}
If:
\begin{equation}
    \epsilon_{V}\ll \left|\eta_{V}\right|
\end{equation}
then:
\begin{equation}
    n_{s}-1\simeq 2\eta_{V}
\end{equation}
and observations imply \cite{Liddle2000,Baumann:2009ds,Baumann:2014nda}:
\begin{equation}
    n_{s}\simeq0.96 \quad\implies\quad \eta_{V}\simeq -0.02
\end{equation}

\subsection{Shift symmetry and technical naturalness}
Considering shift symmetry of the form:
\begin{equation}
    \phi\mapsto\phi+\mathrm{const}\,,
\end{equation}
we may schematically write that \cite{tHooft:1979rat,Peskin:1995ev,Weinberg:1995mt,Srednicki:2007qs}:
\begin{equation}
    m^{2}\propto g \quad\land\quad \Delta m^{2}\propto g\,\Lambda^{2}\,.
\end{equation}
The above gives rise to two regimes within the EFT:
\begin{enumerate}[label=(\alph*)]
    \item Without gravity - radiatively stable:
    \begin{equation}
        \left|\eta_{V}\right|\ll 1\,;
    \end{equation}
    \item Including gravity - Planck-suppressed operators are expected to break continuous global symmetries \cite{Kallosh:1995hi,Baumann:2014nda,Banks:2010zn,Harlow:2018jwu,Harlow:2018tng}:
    \begin{equation}
        \Delta\eta_{V}\sim\mathcal{O}(1)\,.
    \end{equation}
\end{enumerate}

\section{Eta problem II: Planck-suppressed operators}
The EFT expansion of the action takes the form of \cite{Baumann:2014nda,Georgi:1993mps,Kaplan:2005es,Burgess:2017ytm}:
\begin{equation}
    S_{\mathrm{eff}}[\phi]=\int d^{4}x\,\sqrt{-g}\,\left[\frac{1}{2}M_{\mathrm{Pl}}^{2}\,R+\mathcal{L}_{\ell}[\phi]+\sum_{i}\frac{c_{i}}{\Lambda^{\delta_{i}-4}}\hat{\mathcal{O}}_{i}[\phi]\right]\,,
\end{equation}
where $\hat{\mathcal{O}}_{i}$ have mass dimension $\delta_{i}>4$, $c_i$ are (typically) dimensionless Wilson coefficients, and $\Lambda$ is the cutoff, with $\Lambda\lesssim M_{\mathrm{Pl}}$ and $\Lambda \gtrsim H$ during inflation.
\subsection{Dimension-six correction}
Dimension-six operator:
\begin{equation}
    \hat{\mathcal{O}}_{6}=V_{\ell}(\phi)\,\phi^2
\end{equation}
that enters the potential as:
\begin{equation}
    \Delta V(\phi)=\frac{c}{\Lambda^{2}}\hat{\mathcal{O}}_{6}=c\,V_{\ell}(\phi)\frac{\phi^{2}}{\Lambda^{2}}\,,
\end{equation}
with $c$ dimensionless and expected to be \cite{Baumann:2014nda,Georgi:1993mps,Burgess:2017ytm}:
\begin{equation}
    c \sim\mathcal{O}(1)
\end{equation}
in the absence of special structure. The full potential becomes:
\begin{equation}\label{V-eta-2}
    V(\phi)=V_{\ell}(\phi)+\Delta V(\phi)=V_{\ell}(\phi)\left(1+c \frac{\phi^{2}}{\Lambda^{2}}\right)\,.
\end{equation}
Provided that:
\begin{equation}
    \frac{\phi^{2}}{\Lambda^{2}}\ll 1
\end{equation}
yields:
\begin{equation}
    \frac{\Delta V(\phi)}{V(\phi)}\sim c\,\left(\frac{\phi}{\Lambda}\right)^{2}\ll 1\,.
\end{equation}
Now, we can compute the effect on $\eta_{V}$ given by \eqref{slow-roll-parameters}. Differentiating the relation \eqref{V-eta-2} produces:
\begin{equation}
    \begin{aligned}
        V'(\phi) & =V'_{\ell}(\phi)\left(1+c\frac{\phi^{2}}{\Lambda^{2}}\right)+V_{\ell}(\phi)\left(2c\frac{\phi}{\Lambda^{2}}\right)\,, \\
        V''(\phi) & =V''_{\ell}(\phi)\left(1+c\frac{\phi^{2}}{\Lambda^{2}}\right)+4c\frac{\phi}{\Lambda^{2}}V'_{\ell}(\phi)+2c \frac{V_{\ell}(\phi)}{\Lambda^{2}}\,,
\end{aligned}
\end{equation}
so that:
\begin{equation}
    \eta_{V}(\phi)=M_{\mathrm{Pl}}^{2}\frac{V''_{\ell}(\phi)\left(1+c \frac{\phi^{2}}{\Lambda^{2}}\right)+4c\frac{\phi}{\Lambda^{2}}V'_{\ell}(\phi)+2c\frac{V_{\ell}(\phi)}{\Lambda^{2}}}{V_{\ell}(\phi)\left(1+c \frac{\phi^{2}}{\Lambda^{2}}\right)}\,.
\end{equation}
Introducing:
\begin{equation}
    \eta_{\ell}\equiv M_{\mathrm{Pl}}^{2}\frac{V''_{\ell}(\phi)}{V_{\ell}(\phi)}
\end{equation}
and the small parameter:
\begin{equation}
    \delta_{\Lambda}\equiv c\frac{\phi^{2}}{\Lambda^{2}}\ll 1
\end{equation}
together with the approximation:
\begin{equation}
    \frac{1}{1+\delta_{\Lambda}}=1-\delta_{\Lambda}+\mathcal{O}\left(\delta_{\Lambda}^{2}\right) \,,
\end{equation}
one obtains:
\begin{equation}
    \begin{aligned}
        \eta(\phi) & =\eta_{\ell}(\phi)+M_{\mathrm{Pl}}^{2}\left[4c \frac{\phi}{\Lambda^{2}}\frac{V'_{\ell}(\phi)}{V_{\ell}(\phi)}+2 \frac{c}{\Lambda^{2}}\right]\frac{1}{1+\delta_{\Lambda}} \\
        & \simeq \eta_{\ell}(\phi)+M_{\mathrm{Pl}}^2\left[4c\frac{\phi}{\Lambda^{2}}\frac{V'_{\ell}(\phi)}{V_{\ell}(\phi)}+2\frac{c}{\Lambda^{2}}\right]\left(1-\delta_{\Lambda}\right)+\mathcal{O}\left(\delta_{\Lambda}^2\right)\,.
    \end{aligned}
\end{equation}
Thus:
\begin{equation}
    \begin{aligned}
        \Delta\eta_{V}\equiv\eta_{V}-\eta_{\ell}\simeq &\,2 c\left(\frac{M_{\mathrm{Pl}}}{\Lambda}\right)^{2}+4 c\left(\frac{M_{\mathrm{Pl}}}{\Lambda}\right)^{2}\left(\frac{\phi}{\Lambda}\right)\left(M_{\mathrm{Pl}}\frac{V'_{\ell}(\phi)}{V_{\ell}(\phi)}\right) \\
        & -\delta_{\Lambda}\left[2c\left(\frac{M_{\mathrm{Pl}}}{\Lambda}\right)^{2}+4c\left(\frac{M_{\mathrm{Pl}}}{\Lambda}\right)^{2}\left(\frac{\phi}{\Lambda}\right)\left(M_{\mathrm{Pl}}\frac{V'_{\ell}(\phi)}{V_{\ell}(\phi)}\right)\right]+\mathcal{O}\left(\delta_{\Lambda}^2\right)\,.
    \end{aligned}
\end{equation}
During slow-roll:
\begin{equation}
    M_{\mathrm{Pl}}\frac{V'_{\ell}(\phi)}{V_{\ell}(\phi)}=\sqrt{2\epsilon_{V}} \quad\land\quad \left|\eta_{\ell}\right|\ll 1\,,
\end{equation}
therefore, the dominant contribution is:
\begin{examplebox}[Eta problem II (Planck-suppressed operators)]
\begin{equation}\label{delta-eta-appr}
    \Delta\eta_{V}\simeq 2c\,\left(\frac{M_{\mathrm{Pl}}}{\Lambda}\right)^{2} \quad\land\quad \frac{\phi}{\Lambda}\ll 1\,.
\end{equation}
Hence, \textit{\textbf{dimension-six corrections are generically dangerous for slow-roll}} if:
\begin{equation}
    \Lambda\lesssim M_{\mathrm{Pl}}\,.
\end{equation}
\end{examplebox}
\subsection{General operator of dimension \texorpdfstring{$\delta$}{delta}}
Consider a higher-dimension correction of the monomial form:
\begin{equation}
    \Delta V(\phi)=c\,\bar{V}\left(\frac{\phi}{\Lambda}\right)^{\delta-4} \quad\land\quad \delta>4\,,
\end{equation}
where:
\begin{equation}
    \bar{V}\sim V(\phi)\,.
\end{equation}
is approximately constant during the relevant slow-roll phase and $c$ is a dimensionless parameter. In the limit of:
\begin{equation}
    \Delta V\ll V\,,
\end{equation}
we may approximate the value of $\Delta\eta_{V}$ as:
\begin{equation}\label{V-bar}
    \Delta\eta_{V}\simeq M_{\mathrm{Pl}}^{2}\frac{\Delta V''}{\bar{V}}\,.
\end{equation}
Now, computing the derivatives:
\begin{align}
    \Delta V'(\phi) & = c\,\bar{V}(\delta-4)\,\Lambda^{-(\delta-4)}\,\phi^{\delta-5} \\
    \Delta V''(\phi) & = c\,\bar{V}(\delta-4)(\delta-5)\,\Lambda^{-(\delta-4)}\,\phi^{\delta-6}
\end{align}
and inserting them into \eqref{V-bar} produces:
\begin{equation}\label{final-corrections-eta}
    \boxed{\Delta\eta_{V}\simeq c\,(\delta-4)(\delta-5)\left(\frac{M_{\mathrm{Pl}}}{\Lambda}\right)^{2}\left(\frac{\phi}{\Lambda}\right)^{\delta-6}}\,,
\end{equation}
which is consistent with the result of \eqref{delta-eta-appr}.
\begin{examplebox}[General monomial operator]
    The monomial example demonstrates that for:
\begin{itemize}
    \item $\delta=5$ there is \textbf{no direct shift} to $\eta_{V}$:
    \begin{equation}
        \delta-5=0\quad\implies\quad \Delta V''(\phi)=0\,;
    \end{equation}
    \item $\delta=6$ there are \textbf{generically dangerous corrections} for $\Lambda\lesssim M_{\mathrm{Pl}}$:
    \begin{equation}
        \Delta\eta_{V}\sim\left(\frac{M_{\mathrm{Pl}}}{\Lambda}\right)^{2}\,;
    \end{equation}
    \item $\delta \geq 7$ and $\phi\ll\Lambda$ the \textbf{higher operators decouple more efficiently} due to the fact that the $\left(\frac{\phi}{\Lambda}\right)^{\delta-6}$ term in \eqref{final-corrections-eta} suppresses $\Delta\eta_{V}$.
\end{itemize}
\textit{\textbf{\underline{Conclusion:}}}\\
\textbf{Controlling the lowest-dimensional, Planck-suppressed operator is essential for small-field inflationary models.}
\end{examplebox}
\subsection{Interpretation and QG}
Even for \textit{radiatively stable renormalizable sector}, the Planck-suppressed operators with:
\begin{equation}
    c\sim\mathcal{O}(1) \quad\land\quad \Lambda\lesssim M_{\mathrm{Pl}}
\end{equation}
usually generates (via \eqref{delta-eta-appr} and \eqref{final-corrections-eta}):
\begin{equation}
    \Delta\eta_{V}\sim\mathcal{O}(1)\,,
\end{equation}
meaning that \textit{\textbf{slow-roll inflation is not necessarily top-down natural}}. The expectation that consistent QG theory admits no exact global symmetries explains why such operators are generic, unless prohibited by a specific UV structure \cite{Kallosh:1995hi,Banks:2010zn,Harlow:2018jwu,Harlow:2018tng}.

These considerations help to explain why \textit{\textbf{constructing fully controlled UV embeddings of slow-roll inflation is challenging}} - the \textit{eta problems} demonstrate how \textbf{radiative effects and Planck-suppressed operators can cause slow-roll to become unstable unless additional structure is present}.

\section{Summary}
The main conclusion of the chapter is twofold. First, EFT provides the most economical and model-independent language for describing inflationary perturbations and for organizing possible deviations from the simplest slow-roll scenario. Second, the same EFT logic reveals why inflation is UV sensitive: without symmetry protection, special dynamics or a controlled embedding in a more fundamental theory, radiative corrections and higher-dimensional operators tend to destabilize the slow-roll conditions. This motivates comparison with non-singular and bouncing cosmologies, where similar EFT questions arise in a different dynamical setting \cite{Cai:2016thi,Cai:2017tku}. \textit{\textbf{Thus, the issue is not merely whether inflation can fit the data, but whether the inflationary EFT can remain under theoretical control up to the scale at which new gravitational or QG physics becomes relevant.}}
\begin{savequote}
"Quintessence may take many forms. The simplest models propose a quantum field whose energy is varying so slowly that it looks, at first glance, like a constant vacuum energy. The idea is borrowed from inflationary Cosmology [\ldots]. The key difference is that quintessence is much weaker than the inflaton."
\qauthor{\textbf{Jeremiah P. Ostriker, Paul J. Steinhardt} \cite{OstrikerSciAm}}
“Quintessence theories for cosmic acceleration imbue dark energy with a non-trivial dynamics that offers hope in distinguishing the physical origin of this component.”
\qauthor{\textbf{Eric V. Linder} \cite{Linder:2007wa}}
\end{savequote}

\chapter{Quintessence}
\label{Sec:quintessence}
\ifpdf
    \graphicspath{{Chapter5/Figs/Raster/}{Chapter5/Figs/PDF/}{Chapter5/Figs/}}
\else
    \graphicspath{{Chapter5/Figs/Vector/}{Chapter5/Figs/}}
\fi

\section{Introduction}
The discovery that the late-time Universe is undergoing accelerated expansion led to the introduction of a new physical ingredient usually referred to as \textit{\textbf{dark energy}} (DE). The simplest possibility is the cosmological constant $\Lambda$, which corresponds to a vacuum component with a constant equation of state (EoS) parameter:
\begin{equation}
    \omega_{\Lambda}\equiv\frac{p_{\Lambda}}{\rho_{\Lambda}}=-1\,. \end{equation}
Although the cosmological constant is remarkably successful phenomenologically, it also raises well-known theoretical problems, especially the smallness of the observed DE density and the coincidence between the present matter and DE abundances. These difficulties motivate the study of dynamical alternatives to $\Lambda$, in which the accelerated expansion is generated by a new time-dependent degree of freedom or by a modification of the gravitational sector \cite{Peebles:2002gy,Copeland:2006wr,Frieman:2008sn,Tsujikawa:2013fta,Amendola_Tsujikawa_2010q}.

A useful \textit{\textbf{phenomenological classification separates models of DE beyond the cosmological constant into two broad classes}}:
\begin{enumerate}[label=(\alph*)]
    \item \textit{\textbf{Modified matter}} (MM) models, in which the stress-energy tensor $T_{\mu\nu}$ contains an additional exotic component with sufficiently negative pressure, $p<0$, capable of driving the accelerated expansion;
    \item \textit{\textbf{Modified gravity}} (MG) models, in which the geometrical part of the gravitational field equations is modified, for example through corrections to the Einstein tensor $G_{\mu\nu}$ or through additional gravitational degrees of freedom.
    \end{enumerate}
Schematically, one may write this distinction as (for MM):
\begin{equation}
    \boxed{G_{\mu\nu}=\kappa^{2}\left(T_{\mu\nu}^{(m)}+T_{\mu\nu}^{(\mathrm{DE})}\right)}\,,
\end{equation} or, alternatively (for MG):
\begin{equation}
    \boxed{G_{\mu\nu}+\Delta G_{\mu\nu}=\kappa^{2}T_{\mu\nu}^{(m)}}\,.
\end{equation}
At the level of the homogeneous expansion history, these two descriptions can often be rewritten into each other by moving terms between the left- and right-hand sides of the gravitational field equations. Therefore, from the perspective of the background Friedmann equations alone, the distinction between modified matter and modified gravity is not unique. However, from the point of view of field theory, the distinction is physically meaningful: MM models introduce new matter fields on a fixed gravitational framework, whereas MG models modify the propagating gravitational degrees of freedom, their couplings or the geometrical structure of the theory.

\textit{\textbf{Quintessence}}\footnote{The name comes from the Latin phrase '\textit{quinta essentia}' (fifth element).} is the canonical SF $\phi$ with a self-interaction potential $V(\phi)$ that is responsible for the late-time accelerated expansion of the Universe \cite{Ratra:1987rm,Caldwell:1997ii,Carroll:1998zi,Wetterich:1987fm}. In such a formalism, the equation of state (EoS) parameter $\omega_{\phi}$ changes dynamically with time.

In the 1980s, cosmological models were proposed that incorporated quintessence in the presence of the other components of the Universe: matter and radiation \cite{Fujii:1982ms,Ford:1987de,Wetterich:1987fm,Ratra:1987rm}. The next step was to identify the so-called “\textit{tracker}” \textit{fields} (attractor-like solutions in which the SF energy density tracks the background fluid energy density for a rather wide range of initial conditions) \cite{Steinhardt:1999nw} relevant to addressing the coincidence problem for dark energy \cite{Zlatev:1998tr}.

The main challenge for all quintessence models is the existence of a \textit{sufficiently flat scalar field potential that leads to the current slow-roll evolution}, which is characterized by an energy density \cite{Planck:2018vyg}:
\begin{equation}
    \rho_{\mathrm{DE}}\sim 10^{-123} m_{\mathrm{Pl}}^{4}
\end{equation}
and a mass:
\begin{equation}
    m_{\phi}\lesssim H_{0}\sim 10^{-33}\,\mathrm{eV}\,.
\end{equation}

\section{Quintessence formalism}
As previously mentioned, the quintessence model postulates the existence of a canonical scalar field that interacts only through gravitational interaction with other components of the Universe (i.e., a minimally coupled scalar field)\footnote{We will use the notation consistent with \cite{Amendola_Tsujikawa_2010q}. That is, the subscript 'M' denotes a general perfect fluid (including a total fluid), and the subscript 'm' denotes a non-relativistic matter.}:
\begin{equation}
    S_{q}=\int d^{4}x\,\sqrt{-g}\left[\frac{1}{2\kappa^{2}}R+\mathcal{L}_{\phi}\right]+S_{M}\,,
\end{equation}
where:
\begin{equation}
    \mathcal{L}_{\phi}=-\frac{1}{2}g^{\mu\nu}\partial_{\mu}\phi\,\partial_{\nu}\phi-V(\phi)\,.
\end{equation}
and $S_{M}$ describes the rest of the matter content of the Universe (matter and radiation).
For the matter content, let us consider a barotropic EoS for a perfect fluid:
\begin{equation}
    p_{M}=\omega_{M}\,\rho_{M}\,,
\end{equation}
that yields the standard continuity equation in cosmology produced from the covariant conservation law:
\begin{equation}
    \nabla_{\mu}T^{\mu\nu}_{M}=0 \quad\implies\quad \dot{\rho}_{M}+3H\rho_{M}\left(1+\omega_{M}\right)=0\,,
\end{equation}
with the general solution:
\begin{equation}
    \rho_{M}\propto a^{-3\left(1+\omega_{M}\right)}\,.
\end{equation}
The stress-energy tensor for the quintessence field has the form of \cite{Kolb:1990vq}:
\begin{equation}
    T_{\mu\nu}^{(\phi)}=-\frac{2}{\sqrt{-g}}\frac{\delta\left(\sqrt{-g}\,\mathcal{L}_{\phi}\right)}{\delta g^{\mu\nu}}=\partial_{\mu}\phi\,\partial_{\nu}\phi-g_{\mu\nu}\left[\frac{1}{2}g^{\alpha\beta}\partial_{\alpha}\phi\,\partial_{\beta}\phi+V(\phi)\right]\,,
\end{equation}
and in the FLRW background is described by the following quantities:
\begin{align}\label{rho-p-EoS}
    \rho_{\phi} & =-T_{0}{}^{0(\phi)}=\frac{1}{2}\dot{\phi}^{2}+V(\phi)\,, \\
    p_{\phi} & = \frac{1}{3}T_{i}{}^{i(\phi)}=\frac{1}{2}\dot{\phi}^{2}-V(\phi)\,, \\
    \omega_{\phi} & \equiv\frac{p_{\phi}}{\rho_{\phi}}=\frac{\dot{\phi}^{2}-2V(\phi)}{\dot{\phi}^{2}+2V(\phi)}\in \left[-1;1\right]\,.
\end{align}
In the spatially flat model, one may obtain the following equations of motion:
\begin{align}
    H^{2} & =\frac{\kappa^{2}}{3}\left[\frac{1}{2}\dot{\phi}^{2}+V(\phi)+\rho_{M}\right]\,, \\ \label{I-Friedmann}
    \dot{H} & =-\frac{\kappa^{2}}{2}\left[\dot{\phi}^{2}+\rho_{M}\left(1+\omega_{M}\right)\right]\,, \\ \label{II-Friedmann}
    \ddot{\phi} & = -3H \dot{\phi}-V'(\phi)\,.
\end{align}

During radiation/matter domination epoch the total matter energy density $\rho_{M}$ dominates over the quintessence one:
\begin{equation}
    \rho_{M}\gg \rho_{\phi}\,,
\end{equation}
so that, $\rho_{\phi}$ tracks the $\rho_{M}$, and therefore, the DE density emerges at late times (as appears to be the case in reality).
\subsection{Tracking behavior}
Due to obvious reasons, the tracking behavior of the SF (quintessence) depends on the specific form of the potential, namely, if one considers a steep $V(\phi)$, then:
\begin{equation}
    \frac{1}{2}\dot{\phi}^{2}\gg V(\phi) \quad\implies\quad \omega_{\phi}\simeq 1\,,
\end{equation}
and this yields a \textbf{stiff matter/kination} behavior of the quintessence energy density:
\begin{equation}
    \rho_{\phi}\propto a^{-6}\,,
\end{equation}
that \textit{decreases much faster than the considered background fluid energy density}.

In order to obtain the late-time cosmic acceleration, we must assume the following conditions:
\begin{equation}
    \omega_{\phi}<-\frac{1}{3} \quad\implies\quad \frac{1}{2}\dot{\phi}^{2}<V(\phi)\,.
\end{equation}
This means that the self-interaction potential, $V(\phi)$, must be shallow enough for the slow-roll conditions to be satisfied (see \eqref{slow-roll-parameters}). It is also useful to introduce the notion of a new parameter which relates the second derivative of the SF and the Hubble friction term:
\begin{equation}
    \zeta\equiv\frac{\ddot{\phi}}{3H\dot{\phi}}\,,
\end{equation}
therefore, one may write:
\begin{equation}
    3H\dot{\phi}\left(1+\zeta\right)=-V'(\phi) \quad\implies\quad \dot{\phi}=-\frac{V'(\phi)}{3H\left(1+\zeta\right)}\,.
\end{equation}
Another useful identity that can be obtained using \eqref{rho-p-EoS} is of the following form:
\begin{equation}
    1+\omega_{\phi}=\frac{\rho_{\phi}+p_{\phi}}{\rho_{\phi}}=\frac{\dot{\phi}^{2}}{\rho_{\phi}}\,,
\end{equation}
and implies a compact form for the quintessence EoS parameter:
\begin{equation}\label{EoS-q}
    \omega_{\phi}=-1+\frac{V'(\phi)^{2}}{9H^{2}\left(1+\zeta\right)^{2}\rho_{\phi}}\,.
\end{equation}
Moreover, during the DE-dominated slow-roll phase, the EoS parameter \eqref{EoS-q} takes a much more compact form:
\begin{equation}
    \begin{dcases}
    \left|\zeta\right|\ll 1 \\
    \rho_{\phi}\simeq V(\phi) \\
    H^{2}\simeq\frac{\kappa^{2}}{3}V(\phi)
\end{dcases}
\quad\implies\quad
\boxed{\omega_{\phi}\simeq -1+\frac{M_{\mathrm{Pl}}^{2}}{3}\left(\frac{V'(\phi)}{V(\phi)}\right)^{2}\equiv -1+\frac{2}{3}\epsilon_{V}}\,.
\end{equation}

\subsection{Types of quintessence potentials}
The literature identifies two main types of SF potential in quintessence models \cite{Caldwell:2005tm}:
\begin{examplebox}[Types of quintessence potentials]
    \begin{enumerate}[label=(\Roman*)]
        \item \textit{\textbf{'Freezing' models}} - in the past, the SF was rolling along $V(\phi)$, but this movement gradually slowed down after entering the phase of late-time cosmic accelerated expansion:
        \begin{enumerate}[label=(\alph*)]
            \item \textbf{Fermion condensate model as a dynamical SUSY-breaking mechanism} \cite{Binetruy:1998rz} (there is no minimum of the SF potential, so that the SF rolls down the $V(\phi)$ toward infinity \cite{Ratra:1987rm,Zlatev:1998tr}):
            \begin{equation}\label{inverse-power-law-potential}
                V(\phi)=M^{4+n}\phi^{-n} \quad\land\quad n>0\,;
            \end{equation}
            \item \textbf{SUGRA models} \cite{Brax:1999gp} (there is a minimum at which the SF could be trapped, namely, $\omega_{\phi}=-1$):
            \begin{equation}
                V(\phi)=M^{4+n}\phi^{-n}\,\exp{\left[\alpha\left(\frac{\phi}{m_{\mathrm{Pl}}}\right)^{2}\right]} \quad\land\quad \alpha>0\,,
            \end{equation}
            where:
            \begin{equation}
                \phi_{\mathrm{min}}=m_{\mathrm{Pl}}\sqrt{\frac{n}{2\alpha}} \quad\land\quad m_{\phi}^{2}\equiv V''\left(\phi_{\mathrm{min}}\right)=\frac{4\alpha}{m_{\mathrm{Pl}}^{2}}M^{4+n}\left(\frac{2\alpha}{n\,m_{\mathrm{Pl}}}\right)^{n/2}e^{n/2}\,;
            \end{equation}
        \end{enumerate}
        \item \textit{\textbf{'Thawing' models}} - until recently, the SF was frozen by Hubble friction, but it has started to evolve as the Hubble parameter has dropped below the SF mass $m_{\phi}$:
        \begin{enumerate}[label=(\alph*)]
            \item \textbf{Chaotic inflation-like models} ($n=\{2,4\}$ and $V_{0}=0$ \cite{Linde:1983gd}) but with different mass scale $M$:
            \begin{equation}
                V(\phi)=V_{0}+M^{4-n}\phi^{n} \quad\land\quad n>0\,.
            \end{equation}
            Originally, the model with $n=1$ was designed to replace the cosmological constant with a slowly varying scalar field \cite{Linde:1990ta} (whose potential could take negative values \cite{Kallosh:2003bq}).
            \item \textbf{Pseudo-Nambu-Goldstone boson (pNGB) model} \cite{Frieman:1995pm} (the SF is almost frozen at the maximum of its potential when $m_{\phi}<H$, but it begins to roll down at the present value of $m_{\phi}\simeq H_{0}$):
            \begin{equation}
                V(\phi)=M^{4}\cos^{2}{\left(\frac{\phi}{f}\right)}\,.
            \end{equation}
        \end{enumerate}
    \end{enumerate}
\end{examplebox}
\begin{examplebox}[Deviation from $\omega_{\phi}\simeq -1$ for the explicit SF potentials]
    \begin{enumerate}[label=(\Roman*)]
    \item \textit{\textbf{'Freezing' models}}:
        \begin{enumerate}[label=(\alph*)]
            \item \textbf{Inverse power-law} (\textit{Ratra-Peebles}):
            \begin{equation}
                \frac{V'(\phi)}{V(\phi)}=-\frac{n}{\phi} \quad\implies\quad \epsilon_{V}=\frac{1}{2\kappa^{2}}\left(\frac{n}{\phi}\right)^{2}
            \end{equation}
            yields:
            \begin{equation}
                \boxed{\omega_{\phi}\simeq -1+\frac{n^{2}}{3\kappa^{2}\phi^{2}}=-1+\frac{n^{2}}{3}\left(\frac{M_{\mathrm{Pl}}}{\phi}\right)^{2}}
            \end{equation}
            $\implies$ As the SF rolls to larger values the slope decreases and the EoS parameter $\omega_{\phi}$ freezes toward the value of -1;
            \item \textbf{SUGRA-corrected inverse power-law} (\textit{Brax-Martin}):
            \begin{equation}
                \frac{V'(\phi)}{V(\phi)}=-\frac{n}{\phi}+2\alpha\frac{\phi}{M_{\mathrm{Pl}}^{2}} \quad\implies\quad \epsilon_{V}=\frac{1}{2\kappa^{2}}\left(-\frac{n}{\phi}+2\alpha\frac{\phi}{M_{\mathrm{Pl}}^{2}}\right)^{2}
            \end{equation}
            yields:
            \begin{equation}
                \boxed{\omega_{\phi}\simeq -1+\frac{M_{\mathrm{Pl}}^{2}}{3}\left(-\frac{n}{\phi}+2\alpha\frac{\phi}{M_{\mathrm{Pl}}^{2}}\right)^{2}}\,.
            \end{equation}
            \begin{itemize}
                \item $V(\phi)$ has a minimum at:
                \begin{equation}
                    \phi_{\mathrm{min}}^{2}=\frac{n}{2\alpha}M_{\mathrm{Pl}}^{2}\,,
                \end{equation}
                \item Expanding around the minimum:
                \begin{equation}
                    \phi=\phi_{\mathrm{min}}+\delta\phi
                \end{equation}
                one can find that:
                \begin{equation}
                    \frac{V'(\phi)}{V(\phi)}\simeq\frac{V''(\phi)}{V(\phi)}\Bigg|_{\phi=\phi_{\mathrm{min}}}\delta\phi=\frac{4\alpha}{M_{\mathrm{Pl}}^{2}}\delta\phi\,,
                \end{equation}
                therefore:
                \begin{equation}
                    \omega_{\phi}\simeq -1+\frac{16\alpha^{2}}{3M_{\mathrm{Pl}}^{2}}\left(\delta\phi\right)^{2}
                \end{equation}
            \end{itemize}
            $\implies$ Exponential factor pushes $\omega_{\phi}$ closer to -1 than the pure inverse power-law;
        \end{enumerate}
        \item \textit{\textbf{'Thawing' models}}:
        \begin{enumerate}[label=(\alph*)]
            \item \textbf{Polynomial}:
            \begin{equation}
                \frac{V'(\phi)}{V(\phi)}=\frac{n\,M^{4-n}\phi^{n-1}}{V_{0}+M^{4-n}\phi^{n}} \quad\implies\quad \epsilon_{V}=\frac{1}{2\kappa^{2}}\left(\frac{n\,M^{4-n}\phi^{n-1}}{V_{0}+M^{4-n}\phi^{n}}\right)^{2}
            \end{equation}
            yields:
            \begin{equation}
                \boxed{\omega_{\phi}\simeq -1+\frac{M_{\mathrm{Pl}}^{2}}{3}\frac{n^{2}M^{2(4-n)}\phi^{2(n-1)}}{\left(V_{0}+M^{4-n}\phi^{n}\right)^{2}}}\,.
            \end{equation}
            Near the frozen regime:
                \begin{equation}
                    \left|\phi\right|\ll \left(\frac{V_{0}}{M^{4-n}}\right)^{1/n}\,,
                \end{equation}
            so that, for:
            \begin{itemize}
                \item $n=1$:
                \begin{equation}
                    \boxed{\omega_{\phi}\simeq -1+\frac{M_{\mathrm{Pl}}^{2}}{3}\frac{M^{6}}{V_{0}^{2}}=\mathrm{const}}\,,
                \end{equation}
                \item $n=2$:
                \begin{equation}
                    \boxed{\omega_{\phi}\simeq -1+\frac{4}{3}\frac{M_{\mathrm{Pl}}^{2}\,M^{4}}{V_{0}^{2}}\phi^{2}}
                \end{equation}
            \end{itemize}
            $\implies$ Standard thawing behavior;
            \item \textbf{pNGB}:
            \begin{equation}
                \frac{V'(\phi)}{V(\phi)}=-\frac{2}{f}\tan{\left(\frac{\phi}{f}\right)} \quad\implies\quad \epsilon_{V}=\frac{2}{\kappa^{2}f^{2}}\tan^{2}{\left(\frac{\phi}{f}\right)}
            \end{equation}
            yields:
            \begin{equation}
                \boxed{\omega_{\phi}\simeq -1+\frac{4}{3\kappa^{2}f^{2}}\tan^{2}{\left(\frac{\phi}{f}\right)}=-1+\frac{4}{3}\left(\frac{M_{\mathrm{Pl}}}{f}\right)^{2}\tan^{2}{\left(\frac{\phi}{f}\right)}}
            \end{equation}
            $\implies$ Standard thawing evolution.
        \end{enumerate}
    \end{enumerate}
\end{examplebox}

\subsection{Dynamical systems approach}
The mathematical formalism associated with \textbf{\textit{dynamical systems}}\footnote{See \cite{Wainwright1997,Coley2003,Bahamonde:2017ize} for more details on the application of dynamical systems in cosmology.} \cite{Perko2001,Wiggins2003} is a powerful research tool for determining the evolution of the physical system under consideration (in our case, a cosmological model) \textit{without explicitly analyzing solutions for specific initial conditions}. It allows us to obtain qualitative statements for all possible solutions.
\begin{examplebox}[Dynamical system - formal definition]
    \textit{\textbf{Dynamical system}} is any abstract system consisting of:
            \begin{itemize}
                \item A space (\textit{state space} or \textit{phase space});
                \item A mathematical rule describing the evolution of points in that space.
            \end{itemize}
    \textit{\textbf{\underline{Remark:}}}\\
    In many cases, it is challenging to identify a state space that has a \textit{clear physical interpretation} and allows for accurate mathematical analysis.
\end{examplebox}
\begin{examplebox}[Dynamical system - mathematical definition]
    The dynamical system is defined by the following vector differential equation \cite{Perko2001,Wiggins2003}:
    \begin{equation}
        \boxed{\dot{\vec{x}}=\vec{f}\left(\vec{x}\right)\quad\land\quad \vec{f}:X\to X}\,,
    \end{equation}
    where:
    \begin{equation}
        \vec{x}=\left(x_{1},x_{2},\ldots,x_{n}\right)
    \end{equation}
    is the element of the state space $X\subseteq\mathbb{R}^{n}$ and:
    \begin{equation}
        \dot{(\ )}\equiv\frac{d}{dt}\quad\land\quad t\in\mathbb{R}\,,
    \end{equation}
    and $\vec{f}$ is a \textit{\textbf{vector field}} on $\mathbb{R}^{n}$ such that:
    \begin{equation}\label{fx}
    \vec{f}\left(\vec{x}\right)=\Bigl(f_{1}\left(\vec{x}\right),f_{2}\left(\vec{x}\right),\ldots,f_{n}\left(\vec{x}\right)\Bigr)\,.
    \end{equation}
    $\implies$ \underline{The system of $n$ equations which describe the dynamical behavior of the $n$ variables.}\\
    \textit{\textbf{\underline{Remark:}}}\\
    \textit{In general, $t$ variable does not have to be related to the \textbf{physical time}!}
\end{examplebox}
This dissertation will not focus strictly on the stability analysis of individual cosmological models, as this has already been carried out in one of the authors' previous studies \cite{Postolak:2025qmv}. Details of the stability analysis and the methods employed can be found in \cite{Perko2001,Wiggins2003,Wainwright1997,Coley2003,Bahamonde:2017ize} and references therein. For our analysis, we will rely on the values of the dynamical variables and cosmological parameters that determine the behavior of each model.

First of all, we need to introduce the notion of the so-called \textit{\textbf{expansion normalized (EN) variables}} \cite{Wainwright1997}:
\begin{examplebox}[EN variables]
    \begin{equation}\label{EN-variables}
        x\equiv\frac{\kappa\,\dot{\phi}}{\sqrt{6}H} \quad\land\quad y\equiv\frac{\kappa\,\sqrt{V}}{\sqrt{3}H}
    \end{equation}
\end{examplebox}
together with the cosmological density parameters:
\begin{examplebox}[Cosmological density parameters]
    \begin{equation}\label{cosmological-density-parameters}
        \Omega_{M}\equiv\frac{\kappa^{2}\rho_{M}}{3H^{2}}=1-x^{2}-y^{2} \quad\land\quad \Omega_{\phi}\equiv\frac{\kappa^{2}\rho_{\phi}}{3H^{2}}=x^{2}+y^{2}
    \end{equation}
\end{examplebox}
with the constraint condition (for a spatially flat FLRW Universe):
\begin{equation}
    \Omega_{M}+\Omega_{\phi}=1\,.
\end{equation}
From the Friedmann equations \eqref{I-Friedmann} and \eqref{II-Friedmann} one may obtain:
\begin{equation}
    \frac{\dot{H}}{H^{2}}=-3x^{2}+\frac{3}{2}\left(1+\omega_{M}\right)\left(x^{2}+y^{2}-1\right)\,.
\end{equation}
Moreover, the effective EoS parameter becomes:
\begin{equation}
    \omega_{\mathrm{eff}}\equiv\frac{p_{\mathrm{tot}}}{\rho_{\mathrm{tot}}}=\omega_{M}\Omega_{M}+\omega_{\phi}\Omega_{\phi}=\omega_{M}+x^{2}\left(1-\omega_{M}\right)-y^{2}\left(1+\omega_{M}\right)
\end{equation}
and the EoS for the DE (quintessence) is of the form:
\begin{equation}
    \omega_{\phi}\equiv\frac{p_{\phi}}{\rho_{\phi}}=\frac{x^{2}-y^{2}}{x^{2}+y^{2}}\,.
\end{equation}
Introducing the notion of number of $e$-folds (our 'time' variable in the dynamical system):
\begin{equation}
    N\equiv\ln{a} \quad\implies\quad \frac{d}{dN}=\frac{1}{H}\frac{d}{dt}
\end{equation}
we get the dynamical equation for the $x$ variable (kinetic energy of the SF):
\begin{equation}
    \frac{dx}{dN}=\frac{\kappa}{\sqrt{6}}\left(\frac{\ddot{\phi}}{H^{2}}-\frac{\dot{\phi}}{H}\frac{\dot{H}}{H^{2}}\right)\,,
\end{equation}
and for $y$ variable (potential energy of the SF):
\begin{equation}
    \frac{dy}{dN}=\frac{\kappa}{\sqrt{3}}\left(\frac{\dot{\phi}\,V'(\phi)}{2\sqrt{V(\phi)}H^{2}}-\frac{\sqrt{V(\phi)}\dot{H}}{H^{3}}\right)\,.
\end{equation}
Note that the dynamical system formed by the EN variables \eqref{EN-variables} and the cosmological density parameters \eqref{cosmological-density-parameters} \textbf{\textit{will not be closed}} unless an additional variable describing the slope of the scalar field potential is introduced, along with a new parameter:
\begin{examplebox}[Closure of the autonomous dynamical system]
    \begin{itemize}
        \item New variable:
        \begin{equation}\label{lambda-phi-variable}
            \lambda_{\phi}\equiv -\frac{V'(\phi)}{\kappa\,V(\phi)}
        \end{equation}
        satisfies the following dynamical equation:
        \begin{equation}
            \frac{d\lambda_{\phi}}{dN}=-\sqrt{6}\,\lambda_{\phi}^{2}\,x\,\bigl(\Gamma_{\phi}-1\bigr)\,,
        \end{equation}
        where:
        \begin{equation}
            \Gamma_{\phi}\equiv\frac{V''(\phi)\,V(\phi)}{V'(\phi)^{2}}=\Gamma_{\phi}\left(\lambda_{\phi}\right)
        \end{equation}
        is an additional parameter;
        \item If the function:
        \begin{equation}
            \lambda_{\phi}=\lambda_{\phi}(\phi)
        \end{equation}
        is invertible, i.e. one can obtain
        \begin{equation}
            \phi=\phi\left(\lambda_{\phi}\right)\,,
        \end{equation}
        so that, we can write $\Gamma_{\phi}=\Gamma_{\phi}\left(\lambda_{\phi}\right)$:
        \begin{equation}
            \Gamma_{\phi}=\Gamma_{\phi}\bigl[\phi\left(\lambda_{\phi}\right)\bigr]\,.
        \end{equation}
        $\implies$ \underline{\textit{\textbf{Closed autonomous dynamical system.}}}
    \end{itemize}
\end{examplebox}
By using the variable $\lambda_{\phi}$ and the parameter $\Gamma_{\phi}$, we can derive the final form of the equations that describe the cosmological dynamical system under consideration:
\begin{examplebox}[Final form of the cosmological dynamical system]
    \begin{equation}\label{final-DS}
    \begin{dcases}
        \frac{dx}{dN}= -3x-\frac{3}{2}x\Bigl[x^{2}\left(\omega_{M}-1\right)+\left(y^{2}-1\right)\left(\omega_{M}+1\right)\Bigr]+\sqrt{\frac{3}{2}}\,\lambda_{\phi}\,y^{2} \\
        \frac{dy}{dN}= -\frac{1}{2}y\left[\sqrt{6}\,\lambda_{\phi}\,x+3x^{2}\left(\omega_{M}-1\right)+3\left(y^{2}-1\right)\left(\omega_{M}+1\right)\right] \\
        \frac{d\lambda_{\phi}}{dN}= -\sqrt{6}\,\lambda_{\phi}^{2}\,x\,\bigl(\Gamma_{\phi}-1\bigr) \,.
    \end{dcases}
\end{equation}
\end{examplebox}
\begin{examplebox}[Specific case - constant $\lambda_{\phi}$]
    In the case of a constant value of the $\lambda_{\phi}$ variable:
    \begin{equation}
        \lambda_{\phi}=\lambda\equiv\mathrm{const} \quad\implies\quad \kappa\,\lambda=-\frac{V'(\phi)}{V(\phi)}
    \end{equation}
    one obtains an \textbf{exponential form of the SF potential}:
    \begin{equation}
        V(\phi)=V_{0}\,e^{-\kappa\lambda\phi}\,.
    \end{equation}
\end{examplebox}
\begin{examplebox}[Remark]
\begin{itemize}
    \item The SF potentials characterized by:
    \begin{equation}
        \lambda_{\phi}\neq\mathrm{const}
    \end{equation}
    can be classified as a \textbf{"freezing" models with no minimum of the SF potential};
    \item The fixed points derived under the assumption of:
    \begin{equation}
        \lambda_{\phi}=\mathrm{const}
    \end{equation}
    may be interpreted as \textbf{'instantaneous' fixed points evolving over time}, as long as the \textit{timescale of variation of} $\lambda_{\phi}$ \textit{is much smaller than the Hubble time}, $H^{-1}$ \cite{delaMacorra:1999ff,Ng:2001hs}.
\end{itemize}
\end{examplebox}

\subsection{Scaling and tracking solutions}
In order to relate the dynamical system variables with the observational parameters, we introduce the notion of a new variable (the ratio of KE and PE of the SF/quintessence) \cite{Steinhardt:1999nw}:
\begin{equation}\label{X-definition}
    X\equiv\frac{1+\omega_{\phi}}{1-\omega_{\phi}}=\frac{\dot{\phi}^{2}}{2V(\phi)}\,.
\end{equation}
Differentiating $X$ with respect to the number of $e$-folds $N$:
\begin{equation}
    \frac{dX}{dN}=\frac{1}{6}\frac{d\ln{X}}{dN}+1=-\frac{V'(\phi)}{V(\phi)}\frac{\dot{\phi}^{2}+2V(\phi)}{6H\dot{\phi}}
\end{equation}
and using the fact that \cite{Amendola_Tsujikawa_2010q}:
\begin{equation}
    \sqrt{\frac{\left(1+\omega_{\phi}\right)}{\Omega_{\phi}}}=2\sqrt{3}\,H\frac{\left|\dot{\phi}\right|}{\left(\dot{\phi}^{2}+2V(\phi)\right)}
\end{equation}
gives the following identity:
\begin{equation}\label{-lambda-phi}
    \frac{V'(\phi)}{\kappa\,V(\phi)}\equiv -\lambda_{\phi}=\pm \sqrt{\frac{3\left(\omega_{\phi}+1\right)}{\Omega_{\phi}}}\left(\frac{1}{6}\frac{d\ln{X}}{dN}+1\right)\,.
\end{equation}
Differentiating \eqref{-lambda-phi} with respect to the SF produces:
\begin{equation}\label{Gamma-phi-minus-one}
    \Gamma_{\phi}-1=\frac{V(\phi)}{V'(\phi)}\frac{H}{\dot{\phi}}\left[\frac{1}{2}\frac{\omega'_{\phi}}{\left(1+\omega_{\phi}\right)}-\frac{1}{2}\frac{\Omega'_{\phi}}{\Omega_{\phi}}+\frac{Y''}{6+Y'}\right]\,,
\end{equation}
where:
\begin{equation}
    \omega'_{\phi}\equiv\frac{d\omega_{\phi}}{dN} \quad\land\quad \Omega'_{\phi}\equiv\frac{d\Omega_{\phi}}{dN} \quad\land\quad Y'\equiv\frac{d\ln{X}}{dN}\,.
\end{equation}
Using the following identities \cite{Amendola_Tsujikawa_2010q}:
\begin{equation}
    \frac{\omega'_{\phi}}{\left(1+\omega_{\phi}\right)}=\frac{Y}{\left(X+1\right)} \quad\land\quad \frac{\Omega'_{\phi}}{\Omega_{\phi}}=3\left(1-\Omega_{\phi}\right)\frac{\bigl[\omega_{M}\left(X+1\right)-X+1\bigr]}{\left(X+1\right)}
\end{equation}
in the equation \eqref{Gamma-phi-minus-one} yields the final form of the relation for the $\Gamma_{\phi}$ parameter in terms of observational quantities:
\begin{equation}
    \boxed{\Gamma_{\phi}=1+3\frac{\left(\omega_{M}-\omega_{\phi}\right)\left(1-\Omega_{\phi}\right)}{\left(6+Y'\right)\left(1+\omega_{\phi}\right)}-\frac{Y'}{\left(1+X\right)\left(6+Y'\right)\left(1+\omega_{\phi}\right)}-2\frac{Y''}{\left(6+Y'\right)^{2}\left(1+\omega_{\phi}\right)}}\,.
\end{equation}
During radiation domination (RD) and matter domination (MD) epochs the quintessence density parameter is negligible:
\begin{equation}
    \Omega_{\phi}\ll 1\,.
\end{equation}
If we further assume that the parameter $\Gamma_{\phi}$ \textit{varies slowly over time}, then the EoS parameter for quintessence approaches a nearly constant value:
\begin{equation}
    \omega_{\phi}\simeq\mathrm{const}
\end{equation}
due to the fact that:
\begin{equation}
    Y'\to 0 \quad\land\quad Y''\to 0\,.
\end{equation}
The explicit formula for $\omega_{\phi}$ becomes the following:
\begin{equation}\label{EoS-quintessence}
    \boxed{\omega_{\phi}=-\frac{2\left(1-\Gamma_{\phi}\right)+\omega_{M}\left(1-\Omega_{\phi}\right)}{1+\Omega_{\phi}-2\Gamma_{\phi}}\simeq\frac{2\left(1-\Gamma_{\phi}\right)+\omega_{M}}{2\Gamma_{\phi}-1}=\frac{\omega_{M}-2\left(\Gamma_{\phi}-1\right)}{1+2\left(\Gamma_{\phi}-1\right)}\simeq\mathrm{const}}\,.
\end{equation}
\begin{examplebox}[Scaling solution]
\begin{itemize}
    \item For the exponential self-interaction potentials for the quintessence one obtains:
    \begin{equation}
        \Gamma_{\phi}=1\,,
    \end{equation}
    so that, the EoS parameter \eqref{EoS-quintessence} takes the same value as for the dominant/matter fluid:
    \begin{equation}
        \boxed{\omega_{\phi}=\omega_{M}}
    \end{equation}
    $\equiv$ \textbf{Scaling solution} \cite{Copeland:1997et,Ferreira:1997hj,Copeland:2006wr};
    \item Physical implications of the scaling solutions:
    \begin{itemize}
        \item Reduction of sensitivity to initial conditions;
        \item Energy density of quintessence naturally becomes subdominant over long periods of cosmic evolution;
        \item Help to address aspects of the coincidence problem;
        \item They emerge as attractors in a variety of quintessence and scalar-tensor models.
    \end{itemize}
\end{itemize}
\end{examplebox}
\begin{examplebox}[Tracking solution]
\begin{itemize}
    \item For the other types of the SF/quintessence potentials we are dealing with the following values of the $\Gamma_{\phi}$ parameter:
    \begin{equation}
        \Gamma_{\phi}>1\,,
    \end{equation}
    so that:
    \begin{equation}
        \boxed{\omega_{\phi}<\omega_{M}}
    \end{equation}
    $\implies$ The quintessence energy density evolves slowly compared to the background (matter) energy density, provided that the value of the parameter $\Gamma_{\phi}$ remains nearly constant \cite{Steinhardt:1999nw}:
    \begin{equation}
        \boxed{\left|\frac{d\left(\Gamma_{\phi}-1\right)}{dN}\right|\ll\left|\left(\Gamma_{\phi}-1\right)\right|}
    \end{equation}
    $\equiv$ \textbf{Tracking solution} \cite{Steinhardt:1999nw,Zlatev:1998tr,Copeland:2006wr};
    \item Tracking solutions address the following physical aspects:
    \begin{itemize}
        \item Sensitivity to initial conditions;
        \item Fine-tuning problems;
        \item Coincidence problem.
    \end{itemize}
\end{itemize}
\end{examplebox}
\begin{examplebox}[Example of tracking solution - Inverse power-law SF potential]
    \begin{itemize}
        \item The inverse power-law potential of the form \eqref{inverse-power-law-potential} yields:
        \begin{equation}
            \Gamma_{\phi}=\frac{n+1}{n}=1+n^{-1}>1\,;
        \end{equation}
        \item The late-time accelerated expansion is possible if:
        \begin{equation}
            \lambda_{\phi}^{2}<2 \quad\implies\quad \lambda_{\phi}\in\left(-\sqrt{2};\sqrt{2}\right) \quad\implies\quad \phi>\frac{n}{4\sqrt{\pi}}m_{\mathrm{Pl}}
        \end{equation}
        $\implies$ \textbf{The condition is independent of the mass scale} $\mathbf{M}$;
        \item For:
        \begin{equation}
            n\sim\mathcal{O}(1) \quad\implies\quad \phi\sim m_{\mathrm{Pl}}\,;
        \end{equation}
        \item The present value of the quintessence PE is:
        \begin{equation}
            V\left(\phi_{0}\right)\simeq H_{0}^{2}\,m_{\mathrm{Pl}}^{2} \quad\land\quad \phi_{0}\simeq m_{\mathrm{Pl}},
        \end{equation}
        so that, the mass scale $M$ can be estimated as:
        \begin{equation}
            M\simeq\left(\frac{H_{0}}{m_{\mathrm{Pl}}}\right)^{2/(4+n)}m_{\mathrm{Pl}}\simeq 10^{-(46-19n)/(4+n)}\,\mathrm{GeV}\,,
        \end{equation}
        where:
        \begin{equation}
            H_{0}\simeq 10^{-42}\,\mathrm{GeV}\,;
        \end{equation}
        \item The explicit values of $M$ for specific values of $n$:
        \begin{equation}
            M\simeq
            \begin{dcases}
                3.98\times 10^{-6}\,\mathrm{GeV}\simeq 4\,\mathrm{keV} \qquad (n=1) \\
                4.64\times 10^{-2}\,\mathrm{GeV}\simeq 46\,\mathrm{MeV} \qquad (n=2) \\
                3.73\times 10^{1}\,\mathrm{GeV}\simeq 37\,\mathrm{GeV} \qquad (n=3) \\
                5.62\times 10^{3}\,\mathrm{GeV}\simeq 5.6\,\mathrm{TeV} \qquad (n=4) \\
            \end{dcases}
        \end{equation}
    \end{itemize}
    $\implies$ $\boldsymbol{n\gtrsim3}$ \textbf{leads to scales close to the electroweak or beyond the Standard Model scales.}
\end{examplebox}

\subsection{Tracker solutions}
\begin{examplebox}[Tracker solutions]
\begin{itemize}
    \item The \textit{\textbf{tracker solutions}} are \textbf{special trajectories in the phase space that 'attract' other trajectories};
    \item They are characterized by:
    \begin{equation}
        \omega_{\phi}\simeq\mathrm{const} \quad\land\quad \Omega_{\phi}\simeq\mathrm{const}\,;
    \end{equation}
    \item A broad spectrum of initial conditions ultimately align to a shared, universal evolutionary trajectory.
\end{itemize}
\end{examplebox}
Let us consider once again the inverse power-law (Ratra-Peebles) SF potential \eqref{inverse-power-law-potential} which yields:
\begin{equation}\label{observation1}
    \begin{dcases}
        V'(\phi)<0 \\
        \dot{\phi}>0
    \end{dcases}
    \quad\land\quad \phi>0\,.
\end{equation}
In order to make our cosmological model more physical, let us consider a cosmological model with radiation (relativistic matter), non-relativistic matter and the SF (quintessence). In such a case the total energy density and pressure of cosmological fluid becomes:
\begin{equation}
    \rho_{M}\equiv\rho_{r}+\rho_{m} \quad\land\quad p_{M}\equiv p_{r}+p_{m}=\underbrace{\omega_{r}}_{=\frac{1}{3}}\rho_{r}+\underbrace{\omega_{m}}_{=0}\rho_{m}=\frac{1}{3}\rho_{r}\,.
\end{equation}
Now, we must introduce a new dynamical variable responsible for the evolution of the radiation in our dynamical system:
\begin{equation}
    z\equiv\frac{\kappa\sqrt{\rho_{r}}}{\sqrt{3}H}\,,
\end{equation}
therefore, the density parameters take the following forms:
\begin{equation}
    \Omega_{\phi}=x^{2}+y^{2} \quad\land\quad \Omega_{r}=z^{2} \quad\land\quad \Omega_{m}=1-x^{2}-y^{2}-z^{2}
\end{equation}
and the effective EoS parameter becomes:
\begin{equation}
    \omega_{\mathrm{eff}}\equiv\omega_{r}\Omega_{r}+\omega_{m}\Omega_{m}+\omega_{\phi}\Omega_{\phi}=\frac{1}{3}z^{2}+x^{2}-y^{2}\,.
\end{equation}
The resulting dynamical system is now $4D$:
\begin{equation}
    \begin{dcases}
        \frac{dx}{dN}= \frac{1}{2}\left[3x^{3}+x\left(z^{2}-3y^{2}-3\right)+\sqrt{6}\,\lambda_{\phi}\,y^{2}\right] \\
        \frac{dy}{dN}= \frac{1}{2}y\left[3\left(x^{2}-y^{2}+1\right)+z^{2}-\sqrt{6}\,\lambda_{\phi}\,x\right] \\
        \frac{dz}{dN}= \frac{1}{2}z\left[3\left(x^{2}-y^{2}\right)+z^{2}-1\right] \\
        \frac{d\lambda_{\phi}}{dN}= -\frac{\sqrt{6}}{n}x\,\lambda_{\phi}^{2}\,.
    \end{dcases}
\end{equation}
By using the relation \eqref{observation1} one can deduce that:
\begin{equation}
    \lambda_{\phi}>0 \quad\land\quad \frac{d\lambda_{\phi}}{dN}<0\,,
\end{equation}
which means that the variable $\lambda_{\phi}$ decreases with the time parameter ($N$).

Moreover, the use of the relationship for the quintessence EoS parameter \eqref{EoS-quintessence} \textit{in the tracking regime} produces the general relation of the form:
\begin{equation}
    \omega_{\phi}\simeq\frac{n\,\omega_{M}-2}{n+2}
\end{equation}
that takes the following explicit forms in the specific cosmological epochs:
\begin{enumerate}[label=(\alph*)]
    \item RD:
    \begin{equation}
        \omega_{M}=\omega_{r}=\frac{1}{3} \quad\implies\quad \boxed{\omega_{\phi}\simeq \frac{n-6}{3\left(n+2\right)}}\,;
    \end{equation}
    \item MD:
    \begin{equation}
        \omega_{M}=\omega_{m}=0 \quad\implies\quad \boxed{\omega_{\phi}\simeq -\frac{2}{n+2}}\,.
    \end{equation}
\end{enumerate}
Furthermore, from \eqref{-lambda-phi} we can identify that \cite{Amendola_Tsujikawa_2010q}:
\begin{equation}
    \frac{1}{6}\frac{d\ln{X}}{dN}=\Delta(t)-1 \quad\land\quad \Delta(t)\equiv\lambda_{\phi}\sqrt{\frac{\Omega_{\phi}}{3\left(1+\omega_{\phi}\right)}}
\end{equation}
and from the definition of $X$ variable \eqref{X-definition} the above relation can be rewritten as follows \cite{Amendola_Tsujikawa_2010q}:
\begin{equation}
    \frac{1}{6}\frac{d\ln{X}}{dN}=\frac{1}{3\left(1-\omega_{\phi}^{2}\right)}\omega'_{\phi}\,.
\end{equation}
We know that for tracker solutions:
\begin{equation}
    \omega_{\phi}\simeq\mathrm{const}\,,
\end{equation}
so that:
\begin{equation}
    \Delta(t)\simeq 1\,.
\end{equation}
Finally, we determine the final form of the quintessence density parameter for the tracker solution:
\begin{examplebox}[Quintessence density parameter for the tracker solution]
    \begin{equation}
        \boxed{\Omega_{\phi}\simeq\frac{3\left(1+\omega_{\phi}\right)}{\lambda_{\phi}^{2}}}\,.
    \end{equation}
\end{examplebox}
\begin{examplebox}[BBN bound on quintessence]
    The BBN bound on the tracking/scaling SF (quintessence) is \cite{Bean:2001wt}:
    \begin{equation}
        \boxed{\Omega_{\phi}^{(\mathrm{BBN})}<0.045}\,.
    \end{equation}
\end{examplebox}

\section{Reconstructing quintessence from observational data}
\begin{examplebox}[Differences between the CC and quintessence]
    From an observational perspective, the following two aspects that distinguish the cosmological constant from quintessence should be noted:
    \begin{itemize}
        \item Unlike the cosmological constant, the \textit{\textbf{equation of state for quintessence may vary during the evolution of the Universe}};
        \item \textit{\textbf{Quintessence may fluctuate and cluster}} if the SF has non-zero mass.
    \end{itemize}
\end{examplebox}
For the aforementioned reasons, it is possible to determine the nature of dark energy, establishing whether it remains constant over time (i.e. the cosmological constant) or evolves over time (e.g. quintessence).

\subsection{Statefinder diagnosis}
\begin{figure}[htbp]
    \centering
    \includegraphics[width=1\linewidth]{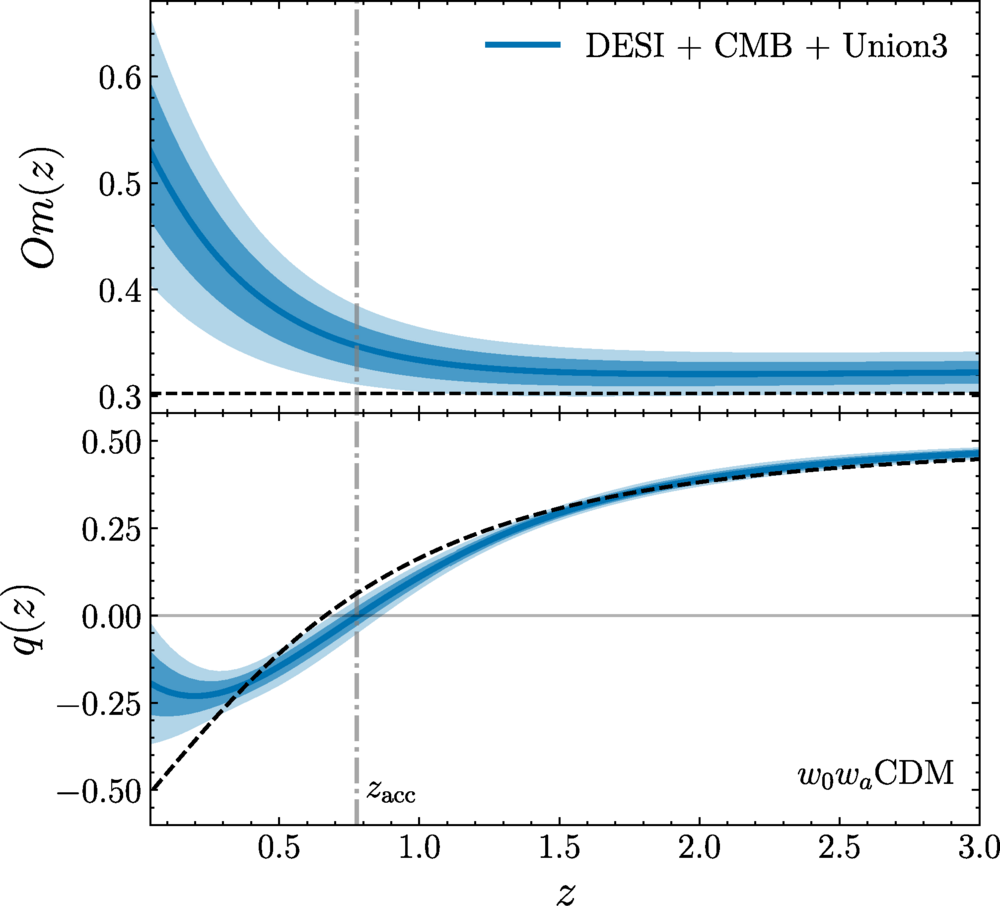}
    \caption{Evolution of the deceleration parameter (diagram at the bottom) obtained during the DESI mission (DESI DR2) \cite{DESI:2025fii}.}
    \label{fig:q-diagram-DESI}
\end{figure}

From the observational data, we know that the deceleration parameter defined via:
\begin{equation}
    q\equiv\frac{\ddot{a}}{a H^{2}}=\frac{d\ln{H}}{d\ln{(1+z)}}-1
\end{equation}
has changed the sign from positive to negative at the redshift/scale factor of the value \cite{DESI:2025fii} (see Fig.~\ref{fig:q-diagram-DESI}):
\begin{equation}
    z_{\mathrm{acc}}\simeq 0.8 \quad\iff\quad a_{\mathrm{acc}}\simeq 0.56\,.
\end{equation}
It has been found that the parameter $q$ can be used to construct new parameters that distinguish the LCDM (CC) model from other dark energy models. The parameters are known as the \textit{\textbf{'Statefinder' parameters}} \cite{Sahni:2002fz}:
\begin{examplebox}['Statefinder' parameters]
    \begin{equation}\label{statefinder-parameters}
        r\equiv\frac{\dddot{a}}{a H^{3}} \quad\land\quad s\equiv\frac{r-1}{3\left(q-\frac{1}{2}\right)}\,.
    \end{equation}
    \textit{\textbf{\underline{Remark:}}}\\
    Notably, since they are constructed purely from geometrical quantities, they can also be applied to MG models.
\end{examplebox}
In order to verify this criterion, let us consider a \textbf{cosmological model incorporating non-relativistic matter and dark energy}. In such a case, the Friedmann equations take the form:
\begin{align}\label{Friedmann-DE}
    H^{2} & = \frac{\kappa^{2}}{3}\bigl(\rho_{m}+\rho_{DE}\bigr)\,, \\
    \dot{H} & = -\frac{\kappa^{2}}{2}\Bigl[\rho_{m}+\rho_{DE}\bigl(1+\omega_{DE}\bigr)\Bigr]\,,\label{II-Friedmann-DE}
\end{align}
and the continuity equations are:
\begin{align}\label{continuity-m}
    \dot{\rho}_{m} & =-3H\rho_{m}\,, \\
    \dot{\rho}_{DE} & =-3H\rho_{DE}\bigl(1+\omega_{DE}\bigr)\,.\label{continuity-DE}
\end{align}
Using \eqref{statefinder-parameters} together with \eqref{Friedmann-DE} produces:
\begin{equation}
    q=\frac{1}{2}+\frac{3}{2}\omega_{DE}\,\Omega_{DE} \quad\land\quad r=1+3\frac{\dot{H}}{H}+\frac{\ddot{H}}{H^{3}}\,,
\end{equation}
where:
\begin{equation}
    \Omega_{DE}\equiv\frac{\kappa^{2}\rho_{DE}}{3H^{2}}
\end{equation}
is the density parameter for the DE fluid. In the next step, differentiating \eqref{II-Friedmann-DE} with the use of \eqref{continuity-m} and \eqref{continuity-DE} gives:
\begin{equation}\label{ddot-H-statefinder}
    \begin{aligned}
        \ddot{H} & = -\frac{\kappa^{2}}{2}\Bigl[\dot{\rho}_{m}+\rho_{DE}\,\dot{\omega}_{DE}+\dot{\rho}_{DE}\bigl(1+\omega_{DE}\bigr)\Bigr] \\
        & = -\frac{\kappa^{2}}{2}\Bigl[-3H\rho_{m}+\rho_{DE}\,\dot{\omega}_{DE}-3H\bigl(1+\omega_{DE}\bigr)^{2}\rho_{DE}\Bigr] \\
        & = \frac{\kappa^{2}}{2}\Bigl[3H\rho_{m}+3H\bigl(1+\omega_{DE}\bigr)^{2}\Bigr]-\frac{\kappa^{2}}{2}\rho_{DE}\,\dot{\omega}_{DE}\,.
    \end{aligned}
\end{equation}
Dividing the last expression in \eqref{ddot-H-statefinder} by $H^{3}$ and expressing everything in terms of the density parameters implies:
\begin{equation}
    \frac{\ddot{H}}{H^{3}}=\frac{9}{2}\Bigl[\Omega_{m}+\Omega_{DE}\bigl(1+\omega_{DE}\bigr)^{2}\Bigr]-\frac{3}{2}\Omega_{DE}\frac{\dot{\omega}_{DE}}{H}\,.
\end{equation}
Furthermore, using the fact that:
\begin{equation}
    3\frac{\dot{H}}{H^{2}}=-\frac{9}{2}\bigl[1+\omega_{DE}\,\Omega_{DE}\bigr]
\end{equation}
gives the relation for the parameter $r$:
\begin{equation}
    r=1+\frac{9}{2}\Omega_{DE}\left[\omega_{DE}\bigl(1+\omega_{DE}\bigr)-\frac{1}{3}\frac{\dot{\omega}_{DE}}{H}\right]\,,
\end{equation}
so that the statefinder parameters take the more compact form:
\begin{equation}
    r=1+\frac{9}{2}\omega_{DE}\,\Omega_{DE}\,s \quad\land\quad s=1+\omega_{DE}-\frac{1}{3}\frac{\dot{\omega}_{DE}}{H}\,.
\end{equation}
Equivalently, by using the barotropic EoS for the DE fluid/component:
\begin{equation}
    p_{DE}=\omega_{DE}\,\rho_{DE}\,,
\end{equation}
one can write:
\begin{equation}
    \dot{p}_{DE}=\dot{\omega}_{DE}\,\rho_{DE}+\omega_{DE}\,\dot{\rho}_{DE}=\rho_{DE}\Bigl[\dot{\omega}_{DE}-3H\omega_{DE}\bigl(1+\omega_{DE}\bigr)\Bigr]\,,
\end{equation}
and, in addition, the following ratio:
\begin{equation}\label{ratio-p-rho-DE}
    \frac{\dot{p}_{DE}}{\dot{\rho}_{DE}}=\omega_{DE}-\frac{\dot{\omega}_{DE}}{3H\bigl(1+\omega_{DE}\bigr)}\,.
\end{equation}
Multiplying the ratio \eqref{ratio-p-rho-DE} by $(\omega_{DE}+1)/\omega_{DE}$ yields the final formula for the parameter $s$:
\begin{equation}
    \boxed{s=\left(\frac{1+\omega_{DE}}{\omega_{DE}}\right)\left(\frac{\dot{p}_{DE}}{\dot{\rho}_{DE}}\right)=\left(\frac{1}{\omega_{DE}}+1\right)\left(\frac{\dot{p}_{DE}}{\dot{\rho}_{DE}}\right)}\,.
\end{equation}
\begin{examplebox}[$\Lambda$CDM model]
    In the $\Lambda$CDM model we are dealing with the cosmological constant characterized by:
    \begin{equation}
        \omega_{DE}=\omega_{\Lambda}\equiv -1=\mathrm{const}\,,
    \end{equation}
    and therefore, we can identify this model with a \textbf{\textit{specific point}} in the $(r,s)$ phase space:
    \begin{equation}
        \boxed{\left(r,s\right)=(1,0)}\,.
    \end{equation}
\end{examplebox}
\begin{examplebox}[Remark]
    \textbf{For the general DE models the trajectories evolve in the} $\boldsymbol{\left(r,s\right)}$ \textbf{plane} so \textbf{one could distinguish them from the corcondance LCDM (CC) model}.
\end{examplebox}

\subsubsection{Quintessence with constant EoS}
\begin{figure}[htbp]
    \centering
    \includegraphics[width=1\linewidth]{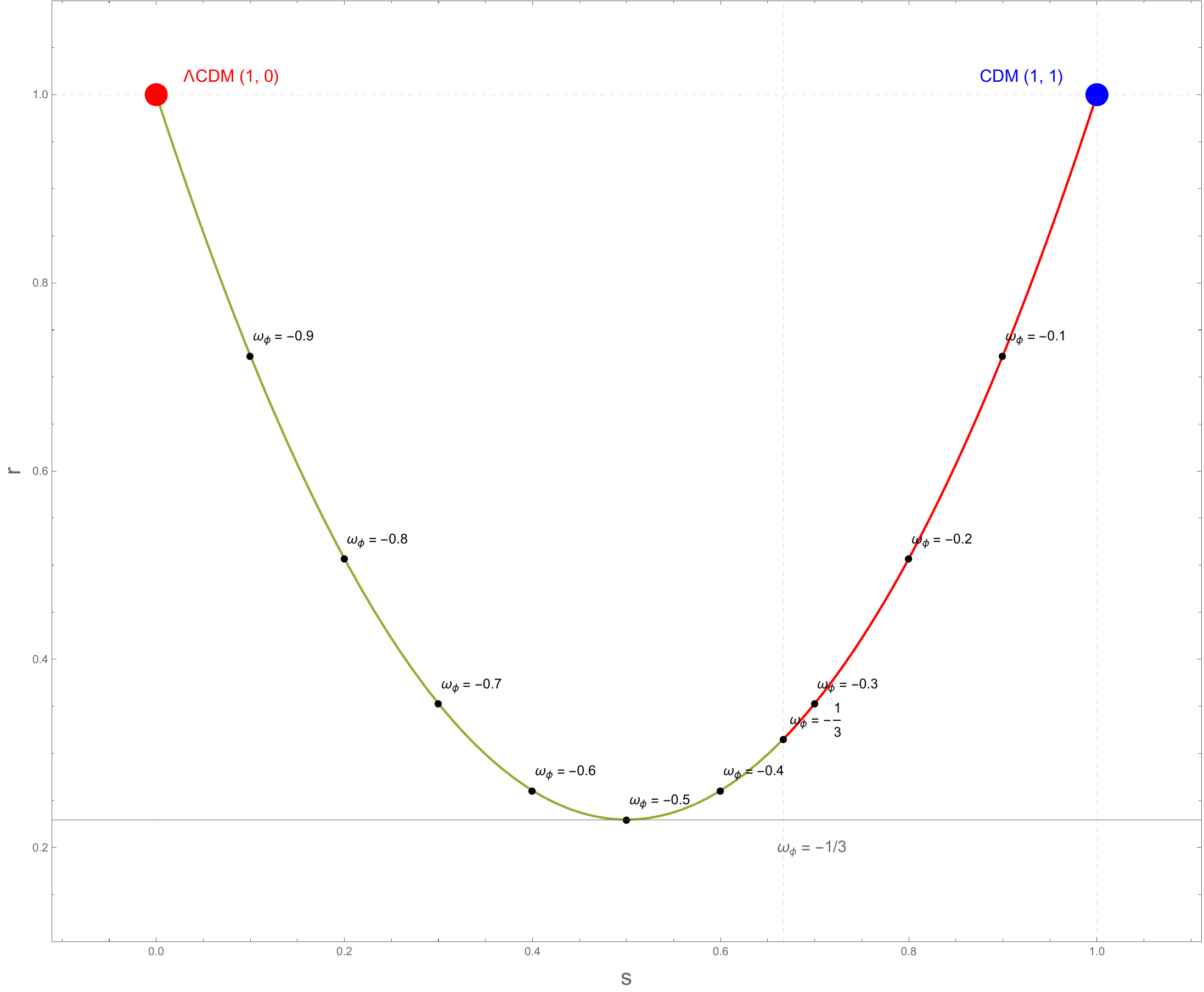}
    \caption{The statefinder phase space for a present value of the DE density parameter with the assumption of different constant values of the quintessence EoS. The limiting cases are: LCDM model $(1,0)$ and CDM $(1,1)$. The green segment of the curve represents the accelerated expansion of the Universe.}
    \label{fig:Statefinder-constant-EoS}
\end{figure}

The simplest form of the quintessence equation of state is:
\begin{equation}
    \omega_{\phi}=\mathrm{const}\in\left(-1;0\right) \quad\implies\quad \dot{\omega}_{\phi}=0\,,
\end{equation}
which yields the following forms of the statefinder parameters:
\begin{align}
    r & = 1+\frac{9}{2}\omega_{\phi}\,\Omega_{\phi}\,s=1+\frac{9}{2}\omega_{\phi}\,\Omega_{\phi}\left(1+\omega_{\phi}\right)\,, \\
    s & = 1+\omega_{\phi}=\mathrm{const}\,.\label{vertical-line}
\end{align}
Therefore, as the parameter $\Omega_{\phi}$ ranges from 0 to 1, the parameter $r$ decreases from 1 to:
\begin{equation}
    r\to 1+\frac{9}{2}\omega_{\phi}\left(1+\omega_{\phi}\right)\,,
\end{equation}
so that one obtains a vertical line at \eqref{vertical-line}. Moreover, in the limiting case of:
\begin{equation}
    \begin{dcases}
        \Omega_{\phi}\to 1 \\
        \omega_{\phi}=s-1
    \end{dcases}
    \quad\implies\quad
    r=1+\frac{9}{2}s\left(s-1\right)\,,
\end{equation}
we obtain the final relationship in the statefinder plane/phase space that is of the form:
\begin{examplebox}[Statefinder parameters for quintessence with constant EoS]
    \begin{equation}
        \left(r,s\right)=\left(1+\frac{9}{2}s\left(s-1\right),1+\omega_{\phi}\right)\,.
    \end{equation}
\end{examplebox}
Fig.~\ref{fig:Statefinder-constant-EoS} illustrates the statefinder parameters space for a case characterized by the current value of the dark energy density parameter, i.e. \cite{Planck:2018vyg}:
\begin{equation}
    \Omega_{\phi}=0.6847\,.
\end{equation}

\subsubsection{Reconstruction of the quintessence potential with a constant EoS}
\begin{figure}[htbp]
    \centering
    \includegraphics[width=1\linewidth]{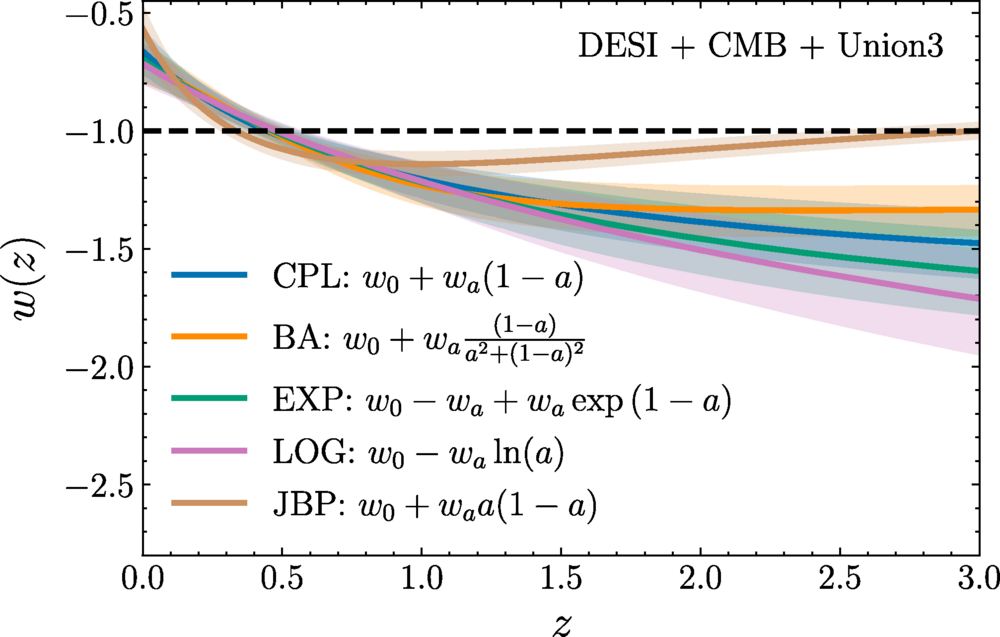}
    \caption{Evolution of the DE EoS for different alternative parametrizations \cite{DESI:2025fii}.}
    \label{fig:Alternative-EoS-DESI}
\end{figure}

The next step in our analysis is to attempt to \textit{\textbf{reconstruct the quintessence potential}}. The starting point is the continuity equation for dark energy, namely the SF/quintessence:
\begin{equation}
    \dot{\rho}_{DE}+3H\rho_{DE}\bigl(1+\omega_{DE}\bigr)=0\,.
\end{equation}
Dividing by $\rho_{DE}$ and separating the variables gives:
\begin{equation}\label{dln-rho-DE-dt}
    \frac{d\ln{\rho_{DE}}}{dt}=-3H\bigl(1+\omega_{DE}\bigr)\,.
\end{equation}
By using the definition of the Hubble parameter\footnote{Also known as the \textbf{Hubble-Lema\^{i}tre parameter} \cite{Lemaitre:1927zz,Hubble:1929ig,Lemaitre:1931zza,Lemaitre:1931zz,Lemaitre:1933gd,Nussbaumer:2011ew}. See \cite{Lemaitre1984,Kragh01102007,Holder:2012Lemaitre,Kragh:2018jrz} for more details on the influence of Georges Lema\^{i}tre's contributions to modern cosmology.}\footnote{An attempt to determine its current value is one of the so-called \textit{\textbf{cosmological tensions}} \cite{Perivolaropoulos:2021jda,Abdalla:2022yfr,CosmoVerseNetwork:2025alb}, known as the \textit{\textbf{Hubble tension}} \cite{Verde:2019ivm,Riess:2019qba,DiValentino:2021izs,Schoneberg:2021qvd,DiValentino:2024yew}.}:
\begin{equation}
    H\equiv\frac{\dot{a}}{a} \quad\implies\quad dt=\frac{d\ln{a}}{H}\,
\end{equation}
one can rewrite \eqref{dln-rho-DE-dt} as:
\begin{equation}
    \frac{d\ln{\rho_{DE}}}{d\ln{a}}=-3\bigl[1+\omega_{DE}(a)\bigr]\,.
\end{equation}
Integrating this relation from today, namely, with the current initial conditions:
\begin{equation}
    a_{0}=1 \quad\land\quad \rho_{DE}\left(a_{0}\right)=\rho_{DE}^{(0)}
\end{equation}
implies:
\begin{equation}\label{rho-DE-a}
    \boxed{\rho_{DE}(a)=\rho_{DE}^{(0)}\exp{\left[-3\int_{1}^{a}\frac{1+\omega_{DE}\left(a'\right)}{a'}da'\right]}=\rho_{DE}^{(0)}\exp{\left[-3\int_{0}^{N}\bigl[1+\omega_{DE}\left(N'\right)\bigr]dN'\right]}}\,.
\end{equation}
In terms of redshift:
\begin{equation}
    1+z\equiv\frac{a_{0}}{a} \quad\land\quad a_{0}=1
\end{equation}
one may write:
\begin{equation}
    d\ln{\left(1+z\right)}=-d\ln{a}=-dN \quad\land\quad dt=-\frac{dz}{H\left(1+z\right)}\,,
\end{equation}
which produces:
\begin{equation}\label{dln-rho-dz}
    \frac{d\ln{\rho_{DE}}}{dz}=3\frac{\left[1+\omega_{DE}(z)\right]}{\left(1+z\right)}\,.
\end{equation}
Integrating \eqref{dln-rho-dz} from $z=0$ to some value of $z$ yields:
\begin{equation}\label{rho-DE-z}
    \boxed{\rho_{DE}(z)=\rho_{DE}^{(0)}\exp{\left[\int_{0}^{z}\frac{3\bigl[1+\omega_{DE}\left(z'\right)\bigr]}{1+z'}dz'\right]}}\,.
\end{equation}
\begin{examplebox}[Parameterizations of the EoS for DE]
    \begin{enumerate}[label=(\alph*)]
        \item \textit{\textbf{CPL (Chevallier-Polarski-Linder)}} \cite{Chevallier:2000qy,Linder:2002et} - \textit{can, but does not have to, be interpreted} as a Taylor expansion (up to the 1st order) of an unknown function \cite{LinderStambul2025}, $\omega_{DE}(a)$, around $a_{0}=1$:
        \begin{equation}
            \boxed{\omega_{DE}^{(\mathrm{CPL})}(a)=\omega_{0}+\omega_{a}\left(1-a\right) \quad\iff\quad \omega_{DE}^{(\mathrm{CPL})}(z)=\omega_{0}+\omega_{a}\frac{z}{\left(1+z\right)}}\,,
        \end{equation}
        with:
        \begin{equation}
            \omega_{0}\equiv\omega_{DE}(1)=\mathrm{const} \quad\land\quad \omega_{a}\equiv -\frac{d\omega_{DE}(a)}{da}\Biggl|_{a=a_{0}}=\mathrm{const}\,,
        \end{equation}
        by \eqref{rho-DE-a} and \eqref{rho-DE-z} implies:
        \begin{align}
            \rho_{DE}^{(\mathrm{CPL})}(a) & = \rho_{DE}^{(0)}\,a^{-3\left(1+\omega_{0}+\omega_{a}\right)}\,\exp{\bigl[3\omega_{a}a\bigr]} \\
            \rho_{DE}^{(\mathrm{CPL})}(z) & = \rho_{DE}^{(0)}\,\left(1+z\right)^{3\left(1+\omega_{0}+\omega_{a}\right)}\exp{\left[\frac{3\omega_{a}}{1+z}\right]}\,;
        \end{align}
        \item \textit{\textbf{BA (Barboza-Alcaniz)}} \cite{Barboza:2008rh,Efstathiou:1999tm}:
        \begin{equation}
            \boxed{\omega_{DE}^{(\mathrm{BA})}(a)=\omega_{0}+\omega_{a}\frac{\left(1-a\right)}{\left[a^{2}+\left(1-a\right)^{2}\right]} \quad\iff\quad \omega_{DE}^{(\mathrm{BA})}(z)=\omega_{0}+\omega_{a}\frac{z\left(1+z\right)}{\left(1+z^{2}\right)}}\,,
        \end{equation}
        yields:
        \begin{align}
            \rho_{DE}^{(\mathrm{BA})}(a) & = \rho_{DE}^{(0)}\,a^{-3\left(1+\omega_{0}+\omega_{a}\right)}\bigl[1+2a\left(a-1\right)\bigr]^{\frac{3}{2}\omega_{a}}  \,, \\
            \rho_{DE}^{(\mathrm{BA})}(z) & = \rho_{DE}^{(0)}\,\left(1+z\right)^{3\left(1+\omega_{0}\right)}\left(1+z^{2}\right)^{\frac{3}{2}\omega_{a}}  \,;
        \end{align}
        \item \textbf{\textit{EXP (exponential)}} \cite{Dimakis:2016mip,Pan:2019brc}:
        \begin{equation}
            \boxed{\omega_{DE}^{(\mathrm{EXP})}(a)=\left(\omega_{0}-\omega_{a}\right)+\omega_{a}\exp{\left[1-a\right]} \iff \omega_{DE}^{(\mathrm{EXP})}(z)=\omega_{0}+\omega_{a}\left(\exp{\left[\frac{z}{1+z}\right]}-1\right)}\,,
        \end{equation}
        produces:
        \begin{align}
            \rho_{DE}^{(\mathrm{EXP})}(a) & = \rho_{DE}^{(0)}\,a^{-3\left(1+\omega_{0}-\omega_{a}\right)}\,\exp{\bigl[-3e\,\omega_{a}\,\Ei{\left(-a\right)}\bigr]}  \,, \\
            \rho_{DE}^{(\mathrm{EXP})}(z) & = \rho_{DE}^{(0)}\,\left(1+z\right)^{3\left(1+\omega_{0}-\omega_{a}\right)}\,\exp{\left[-3e\,\omega_{a}\,\Ei{\left(-\frac{1}{1+z}\right)}\right]}  \,,
        \end{align}
        where:
        \begin{equation}
            \Ei(x)\equiv -\int_{-x}^{\infty}\frac{e^{-t}}{t}dt=\int_{-\infty}^{x}\frac{e^{t}}{t}dt \quad\land\quad x\neq 0\in\mathbb{R}
        \end{equation}
        is the \textit{exponential integral function} \cite{Temme:2010ExpLogSinCosIntegrals};
        \item \textit{\textbf{LOG (logarithmic)}} \cite{Efstathiou:1999tm}:
        \begin{equation}
            \boxed{\omega_{DE}^{(\mathrm{LOG})}(a)=\omega_{0}-\omega_{a}\ln{a} \quad\iff\quad \omega_{DE}^{(\mathrm{LOG})}(z)=\omega_{0}-\omega_{a}\ln{\left[\frac{1}{1+z}\right]}}\,,
        \end{equation}
        generates:
        \begin{align}
            \rho_{DE}^{(\mathrm{LOG})}(a) & = \rho_{DE}^{(0)}\,\exp{\left[\frac{3}{2}\frac{\left(1+\omega_{0}-\omega_{a}\ln{a}\right)^{2}}{\omega_{a}}\right]}  \,, \\
            \rho_{DE}^{(\mathrm{LOG})}(z) & = \rho_{DE}^{(0)}\,\exp{\left[\frac{3}{2}\frac{\Bigl(1+\omega_{0}-\omega_{a}\ln{\left[\frac{1}{1+z}\right]}\Bigr)}{\omega_{a}}\right]}  \,;
        \end{align}
        \item \textit{\textbf{JBP (Jassal-Bagla-Padmanabhan)}} \cite{Jassal:2005qc}:
        \begin{equation}
            \boxed{\omega_{DE}^{(\mathrm{JBP})}(a)=\omega_{0}+\omega_{a}a\left(1-a\right) \quad\iff\quad \omega_{DE}^{(\mathrm{JBP})}(z)=\omega_{0}+\omega_{a}\frac{z}{\left(1+z\right)^{2}}}\,,
        \end{equation}
        leads to:
        \begin{align}
            \rho_{DE}^{(\mathrm{JBP})}(a) & = \rho_{DE}^{(0)}\,a^{-3\left(1+\omega_{0}\right)}\,\exp{\left[\frac{3}{2}\omega_{a}a\left(a-2\right)\right]}  \,, \\
            \rho_{DE}^{(\mathrm{JBP})}(z) & = \rho_{DE}^{(0)}\,\left(1+z\right)^{3\left(1+\omega_{0}\right)}\,\exp{\left[-\frac{3}{2}\omega_{a}\frac{\left(1+2z\right)}{\left(1+z\right)^{2}}\right]}  \,.
        \end{align}
    \end{enumerate}
    Fig.~\ref{fig:Alternative-EoS-DESI} presents the evolution curves obtained from the aforementioned parameterizations of the DE EoS as part of the DESI mission.\footnote{See \cite{Liu:2025myr,Carloni:2025dqt,Li:2026hwq} for other possible approaches.}
\end{examplebox}

\begin{figure}[htbp]
    \centering
    \includegraphics[width=1\linewidth]{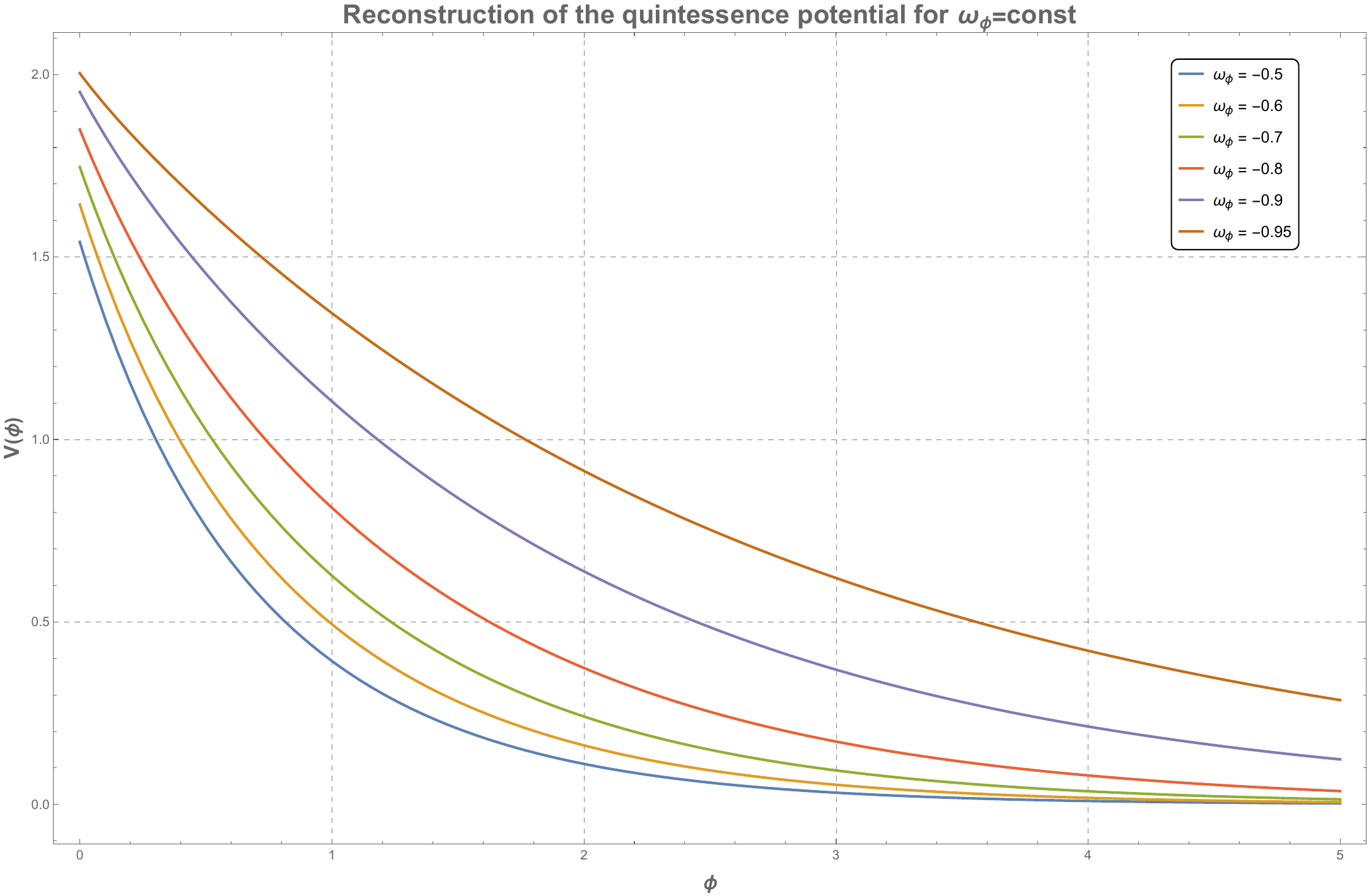}
    \caption{Reconstructed quintessence SF potential \eqref{reconstructed-quintessence-potential} for a different constant values of $\omega_{\phi}$. It was assumed that: $\kappa^{2}=H_{0}^{2}=1$, $\phi_{0}=0$ and $\Omega_{M}^{(0)}=0.315$.}
    \label{fig:reconstruction-quintessence}
\end{figure}

Using \eqref{rho-DE-z} one may write that:
\begin{equation}\label{E-squared-z}
    E^{2}(z)\equiv\frac{H^{2}(z)}{H_{0}^{2}}=\Omega_{M}^{(0)}\left(1+z\right)^{3}+\underbrace{\left(1-\Omega_{M}^{(0)}\right)}_{\Omega_{DE}^{(0)}}\left(1+z\right)^{3\left(1+\omega_{\phi}\right)}\,.
\end{equation}
From the Einstein field equations \eqref{I-Friedmann} and \eqref{II-Friedmann} one could obtain the following:
\begin{equation}\label{dot-phi-squared-V}
    \dot{\phi}^{2}=-\frac{2}{\kappa^{2}}\dot{H}-\rho_{M} \quad\land\quad \kappa^{2}\,V(\phi)=3H^{2}+\dot{H}-\frac{\kappa^{2}}{2}\rho_{M}\,.
\end{equation}
The change of variable from the the cosmological time $t$ to the redshift $z$ yields:
\begin{equation}
    \frac{dz}{dt}=-\left(1+z\right)H \quad\land\quad \frac{dt}{dz}=-\frac{1}{\left(1+z\right)H}
\end{equation}
for any function $X(z)\equiv X$ gives:
\begin{equation}
    \frac{dX}{dt}=\frac{dX}{dz}\frac{dz}{dt}=-\left(1+z\right)H\frac{dX}{dz}\,.
\end{equation}
Applying this relationship into $H^{2}$ produces:
\begin{equation}\label{dH-squared}
    \frac{dH^{2}}{dt}=2H\dot{H}=-\left(1+z\right)H\frac{dH^{2}}{dz} \quad\implies\quad \dot{H}=-\frac{1}{2}\left(1+z\right)\frac{dH^{2}}{dz}\,,
\end{equation}
and for $\dot{\phi}$:
\begin{equation}\label{dot-phi-z}
    \dot{\phi}=-\left(1+z\right)H\frac{d\dot{\phi}}{dz} \quad\implies\quad \dot{\phi}^{2}=\left(1+z\right)^{2}H^{2}\left(\frac{d\phi}{dz}\right)^{2}\,.
\end{equation}
Substituting \eqref{dH-squared} into $\dot{\phi}^{2}$ from \eqref{dot-phi-squared-V} implies:
\begin{equation}
    \dot{\phi}^{2}=\frac{1+z}{\kappa^{2}}\frac{dH^{2}}{dz}-\rho_{M}\,.
\end{equation}
Moreover, using \eqref{dot-phi-z} and dividing by $\left(1+z\right)^{2}H^{2}$ gives:
\begin{equation}
    \left(\frac{d\phi}{dz}\right)^{2}=\frac{1}{\kappa^{2}\left(1+z\right)H^{2}}\frac{dH^{2}}{dz}-\frac{\rho_{M}}{\left(1+z\right)^{2}H^{2}}\,.
\end{equation}
Furthermore, one can rewrite the last term as:
\begin{equation}
    \frac{\rho_{M}}{\left(1+z\right)^{2}H^{2}}=3\frac{H_{0}^{2}}{\kappa^{2}}\Omega_{M}^{(0)}\frac{\left(1+z\right)}{H^{2}}=3\frac{\Omega_{M}^{(0)}}{\kappa^{2}}\frac{\left(1+z\right)}{E^{2}(z)}\,.
\end{equation}
and the first term as:
\begin{equation}
    \frac{1}{H^{2}}\frac{dH^{2}}{dz}=\frac{d\ln{H^{2}}}{dz}=2\frac{d\ln{E}}{dz}\,,
\end{equation}
therefore:
\begin{equation}\label{dphi-dz-squared}
    \frac{\kappa^{2}}{2}\left(\frac{d\phi}{dz}\right)^{2}=\frac{1}{\left(1+z\right)}\frac{d\ln{E}}{dz}-\frac{3}{2}\Omega_{M}^{(0)}\frac{\left(1+z\right)}{E^{2}(z)}\,.
\end{equation}
Next, using the 2nd relation from \eqref{dot-phi-squared-V} with \eqref{dH-squared} generates new relationship for the SF potential:
\begin{equation}
    \kappa^{2}\,V(\phi)=3H^{2}-\frac{\left(1+z\right)}{2}\frac{dH^{2}}{dz}-\frac{\kappa^{2}}{2}\rho_{M}\,
\end{equation}
and dividing this formula by $3H_{0}^{2}$ and rewriting in terms of $E(z)$ yields:
\begin{align}
    \frac{\kappa^{2}\,V(\phi)}{3H_{0}^{2}} & = \underbrace{\frac{H^{2}}{H_{0}^{2}}}_{E^{2}(z)}-\frac{\left(1+z\right)}{6}\frac{d}{dz}\underbrace{\left(\frac{H^{2}}{H_{0}^{2}}\right)}_{E^{2}(z)}-\frac{1}{2}\Omega_{M}^{(0)}\left(1+z\right)^{3} \\
    & = E^{2}(z)-\frac{\left(1+z\right)}{6}\frac{d E^{2}(z)}{dz}-\frac{1}{2}\Omega_{M}^{(0)}\left(1+z\right)^{3}\,.
\end{align}
Inserting \eqref{E-squared-z} into \eqref{dphi-dz-squared} and simplifying gives:
\begin{equation}\label{dphi-dz-squared-final}
    \frac{\kappa^{2}}{2}\left(\frac{d\phi}{dz}\right)^{2}=\frac{3\left(1+\omega_{\phi}\right)}{2}\frac{\left(1-\Omega_{M}^{(0)}\right)\left(1+z\right)^{-2}}{\left(1-\Omega_{M}^{(0)}\right)+\Omega_{M}^{(0)}\left(1+z\right)^{-3\omega_{\phi}}}\,.
\end{equation}
Integrating \eqref{dphi-dz-squared-final} with the following substitution \cite{Amendola_Tsujikawa_2010q}:
\begin{equation}
    u\equiv\sqrt{\frac{\Omega_{M}^{(0)}}{1-\Omega_{M}^{(0)}}}\left(1+z\right)^{-\frac{3}{2}\omega_{\phi}}
\end{equation}
leads to a relationship between the SF and the redshift, expressed as follows:
\begin{equation}
    \phi(z)-\phi_{0}=\sgn{\left(\omega_{\phi}\right)}\sqrt{\frac{4\left(1+\omega_{\phi}\right)}{3\kappa^{2}}}\frac{1}{\left|\omega_{\phi}\right|}\ln{\left(\frac{\sqrt{1+u^{2}}+1}{u}\right)} \quad\land\quad \phi_{0}=\mathrm{const}\,,
\end{equation}
which could be rewritten as:
\begin{equation}
    u=\sinh^{-1}{\left[\left|\omega_{\phi}\right|\sqrt{\frac{3\kappa^{2}}{4\left(1+\omega_{\phi}\right)}}\left(\phi-\phi_{0}\right)\right]}\,.
\end{equation}
Since:
\begin{equation}
    u^{2}\propto\left(1+z\right)^{-3\omega_{\phi}}\,,
\end{equation}
substituting back into $V(z)$ and eliminating $z$ yields the \textit{\textbf{final form of the reconstructed SF/quintessence self-interaction potential}} \cite{Amendola_Tsujikawa_2010q}:
\begin{examplebox}[Final form of the reconstructed quintessence potential for a constant EoS parameter]
    \begin{equation}\label{reconstructed-quintessence-potential}
        \boxed{V(\phi)=\frac{3H_{0}^{2}\left(1-\omega_{\phi}\right)}{2\kappa^{2}}\frac{\left(1-\Omega_{M}^{(0)}\right)^{1/\left|\omega_{\phi}\right|}}{\left(\Omega_{M}^{(0)}\right)^{\beta}}\sinh^{-2\beta}{\left[\left|\omega_{\phi}\right|\sqrt{\frac{3\kappa^{2}}{4\left(1+\omega_{\phi}\right)}}\left(\phi-\phi_{0}+\phi_{1}\right)\right]}}\,,
    \end{equation}
    where:
    \begin{equation}
        \beta\equiv\frac{\left(1+\omega_{\phi}\right)}{\left|\omega_{\phi}\right|} \quad\land\quad \phi_{1}\equiv\sqrt{\frac{4\left(1+\omega_{\phi}\right)}{3\kappa^{2}}}\frac{1}{\left|\omega_{\phi}\right|}\ln{\left(\frac{1+\sqrt{1-\Omega_{M}^{(0)}}}{\sqrt{\Omega_{M}^{(0)}}}\right)}\,.
    \end{equation}
\end{examplebox}
Fig.~\ref{fig:reconstruction-quintessence} illustrates behavior of the reconstructed quintessence potential for a different values of $\omega_{\phi}$.

\subsubsection{Freezing vs. thawing models in the $(r,s)$ plane}
Now let us consider the general behavior of evolution in the statefinder phase space for the two types of quintessence model: freezing and thawing.

To illustrate freezing-type models, we will use a model with an inverse power-law SF potential:
\begin{equation}
    V(\phi)=M^{4+n}\phi^{-n}\,.
\end{equation}
In such a case, one obtains the following matter tracking:
\begin{equation}
    \omega_{\phi}\simeq -\frac{2}{\left(2+n\right)}\,,
\end{equation}
so that initially:
\begin{equation}
    \Omega_{\phi}\ll 1
\end{equation}
implies the following formulae for the statefinder parameters:
\begin{equation}
    r\simeq 1 \quad\land\quad s=1+\omega_{\phi}=\frac{n}{2+n}>0\,.
\end{equation}
In the late-time epoch, the quintessence density parameter approaches:
\begin{equation}
    \Omega_{\phi}\to 1\,,
\end{equation}
and follows the curve of the form:
\begin{equation}
    r=1+\frac{9}{2}s\left(s-1\right)
\end{equation}
towards:
\begin{equation}
    \left(r,s\right)=\left(1,0\right)\,.
\end{equation}
Furthermore, from the fact that:
\begin{equation}
    s>0 \quad\land\quad \omega_{\phi}<0
\end{equation}
one may obtain:
\begin{equation}\label{statefinder-relation-1}
    s=1+\omega_{\phi}-\frac{\dot{\omega}_{\phi}}{3\omega_{\phi}H}>0 \quad\implies\quad \frac{\dot{\omega}_{\phi}}{H}>3\omega_{\phi}\left(1+\omega_{\phi}\right)\,.
\end{equation}

On the other hand, the thawing models (e.g. pNGB) start their evolution near the point:
\begin{equation}
    \left(r,s\right)\simeq\left(1,0\right)
\end{equation}
with:
\begin{equation}
    \omega_{\phi}\in \left(-1;0\right) \quad\land\quad \dot{\omega}_{\phi}>0
\end{equation}
that implies:
\begin{equation}
    r<1 \quad\land\quad s>0\,.
\end{equation}
Consequently, the \textbf{thawing trajectories in the statefinder phase space evolve in the opposite direction with respect to the freezing models}.

\begin{figure}[htbp]
    \centering
    \includegraphics[width=1\linewidth]{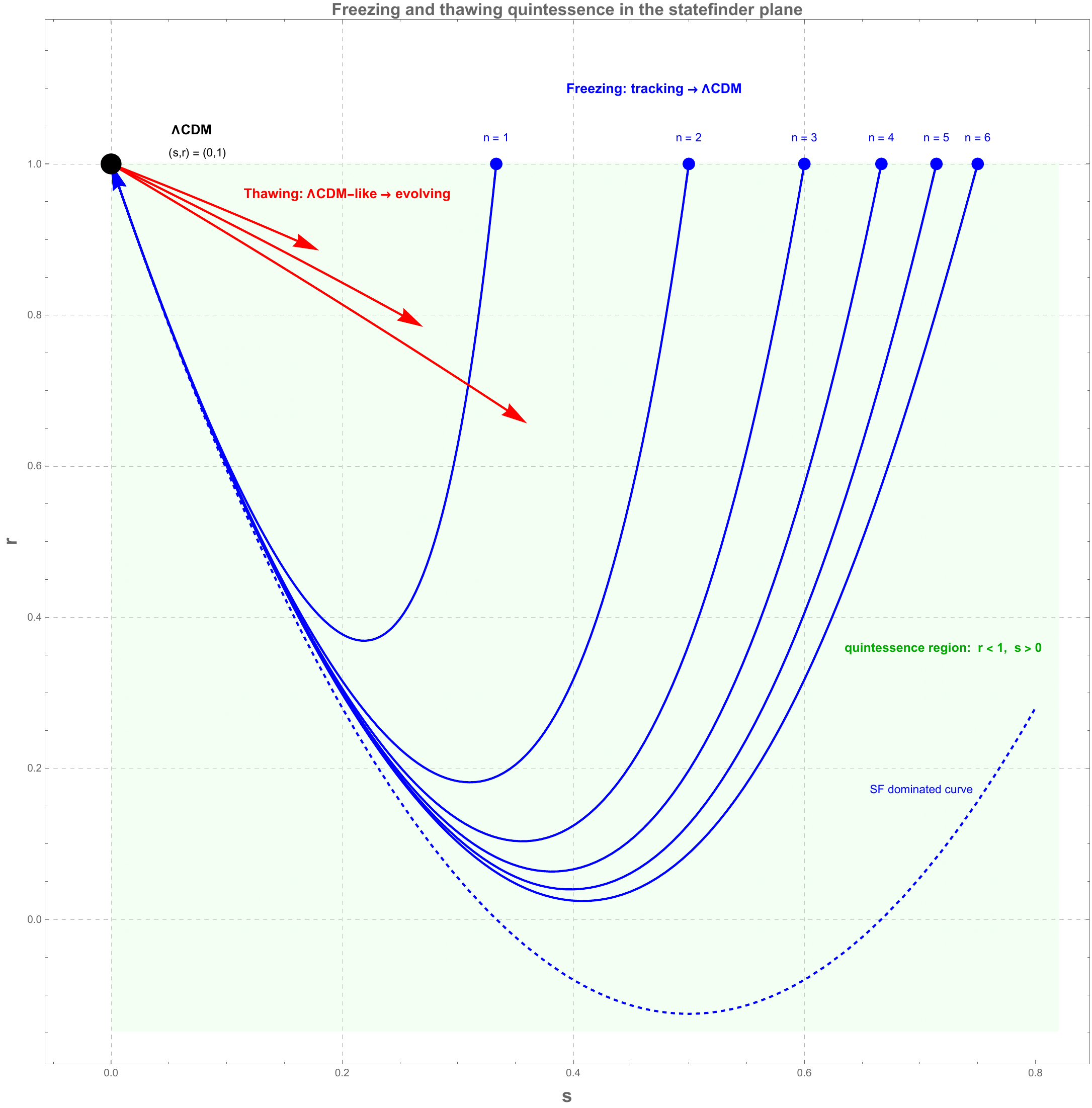}
    \caption{Thawing and freezing quintessence in the statefinder phase space. The diagram illustrates the main conclusion drawn from the analysis of the models: thawing models begin at the LCDM regime and evolve from there, whereas freezing models evolve into the LCDM region.}
    \label{fig:Freezing-thawing-statefinder}
\end{figure}

However, it is important to note that the allowed parameter space for both quintessence classes is characterized by a condition of the form:
\begin{examplebox}[Allowed parameter space for thawing and freezing quintessence models]
    \begin{equation}
        \boxed{r<1 \quad\land\quad s>0}\,.
    \end{equation}
\end{examplebox}
Fig.~\ref{fig:Freezing-thawing-statefinder} illustrates the schematic behavior of the thawing and freezing quintessence models in the statefinder plane.

\subsection{$(\omega_{\phi},\omega'_{\phi})$ plane}
For the purposes of our analysis, let us introduce a quantity representing the kinetic energy of the SF (quintessence):
\begin{equation}
    K\equiv\frac{1}{2}\dot{\phi}^{2}
\end{equation}
which implies:
\begin{equation}
    \rho_{\phi}=K+V(\phi) \quad\land\quad p_{\phi}=K-V(\phi)\,,
\end{equation}
so that:
\begin{equation}
    \omega_{\phi}=\frac{K-V(\phi)}{K+V(\phi)}\,.
\end{equation}
\subsubsection{Exact evolution equation}
Differentiation of the formula:
\begin{equation}
    1+\omega_{\phi}=\frac{\rho_{\phi}+p_{\phi}}{\rho_{\phi}}=\frac{\dot{\phi}^{2}}{\rho_{\phi}}=\frac{2K}{\rho_{\phi}}
\end{equation}
with:
\begin{equation}
    K'\equiv\frac{dK}{dN}=\frac{\dot{K}}{H}=\frac{\dot{\phi}\ddot{\phi}}{H}=2K\xi \quad\land\quad \xi\equiv\frac{\ddot{\phi}}{H\dot{\phi}}
\end{equation}
and:
\begin{equation}
    \rho'_{\phi}\equiv\frac{d\rho_{\phi}}{dN}=\frac{\dot{\rho}_{\phi}}{H}=-3\left(\rho_{\phi}+p_{\phi}\right)=-6K=-3\left(1+\omega_{\phi}\right)\rho_{\phi}
\end{equation}
gives:
\begin{align}
    \left(1+\omega_{\phi}\right)'\equiv\frac{d\left(1+\omega_{\phi}\right)}{dN} & = 2\left(\frac{K'}{\rho_{\phi}}-\frac{K\rho'_{\phi}}{\rho_{\phi}^{2}}\right)=2\left[\frac{2K\xi}{\rho_{\phi}}+\frac{K}{\rho_{\phi}}3\left(1+\omega_{\phi}\right)\right] \\
    & = \left(1+\omega_{\phi}\right)\Bigl[2\xi+3\left(1+\omega_{\phi}\right)\Bigr]\,.
\end{align}
Since:
\begin{equation}
    \left(1+\omega_{\phi}\right)'=\omega'_{\phi}\,,
\end{equation}
therefore, the \textit{\textbf{exact evolution equation for the quintessence EoS}} becomes \cite{Linder:2006sv}:
\begin{examplebox}[Exact evolution equation for the quintessence EoS]
    \begin{equation}\label{exact-EoS-evolution-quintessence}
        \boxed{\omega'_{\phi}=\left(1+\omega_{\phi}\right)\Bigl[2\xi+3\left(1+\omega_{\phi}\right)\Bigr]}\,.
    \end{equation}
    $\implies$ \textbf{The sign of} $\boldsymbol{\xi}$ \textbf{parameter (acceleration vs. deceleration of the SF) controls the sign of} $\boldsymbol{\omega'_{\phi}}$.
\end{examplebox}
The \textbf{boundary between acceleration and deceleration of quintessence} is equivalent to the following case:
\begin{equation}
    \left(\ddot{\phi}=0 \iff \xi=0\right) \quad\implies\quad \boxed{\omega'_{\phi}=3\left(1+\omega_{\phi}\right)^{2}}\,.
\end{equation}
Equivalently, one could rewrite \eqref{exact-EoS-evolution-quintessence} by eliminating the $\xi$ parameter with the use of $\lambda_{\phi}$ and $\Omega_{\phi}$ which produce:
\begin{equation}
    \boxed{\omega'_{\phi}=-3\left(1-\omega_{\phi}^{2}\right)+\lambda_{\phi}\left(1-\omega_{\phi}\right)\sqrt{3\left(1+\omega_{\phi}\right)\,\Omega_{\phi}}}\,.
\end{equation}

\subsubsection{Thawing models}
In the case of thawing models, the \textbf{SF is initially frozen by Hubble friction during MD epoch}:
\begin{equation}
    \omega'_{\phi}>0 \quad\land\quad \omega_{\phi}\simeq -1 \quad\land\quad \left|\xi\right|\ll 1\,,
\end{equation}
therefore:
\begin{equation}
    1+\omega_{\phi}\ll 1 \quad\implies\quad 3\left(1+\omega_{\phi}\right)^{2}\ll 2\xi\left(1+\omega_{\phi}\right)\,,
\end{equation}
which implies the following formula:
\begin{equation}
    \omega'_{\phi}\simeq 2\xi\left(1+\omega_{\phi}\right)\,.
\end{equation}
The physical bound (upper limit) on how fast $\dot{\phi}$ could grow goes from the MD epoch, namely:
\begin{equation}
    \left|\ddot{\phi}\right|\lesssim\frac{\dot{\phi}}{t} \quad\land\quad H(t)\simeq\frac{2}{3t}\,,
\end{equation}
so that the constraint on the $\xi$ parameter becomes:
\begin{equation}
    \xi\lesssim\frac{3}{2}\,.
\end{equation}
Nevertheless, for the pNGB model \cite{Caldwell:2005tm}, the authors obtain the lower bound of the form:
\begin{equation}
    \xi\gtrsim\frac{1}{2}
\end{equation}
once the SF starts to move significantly. Hence, the \textbf{thawing models can be described by the region}:
\begin{examplebox}[Thawing quintessence EoS phase-space region]
    \begin{equation}
        \boxed{1+\omega_{\phi}\lesssim\omega'_{\phi}\lesssim 3\left(1+\omega_{\phi}\right)}\,.
    \end{equation}
\end{examplebox}
\subsubsection{Freezing models}
\begin{figure}[htbp]
    \centering
    \includegraphics[width=1\linewidth]{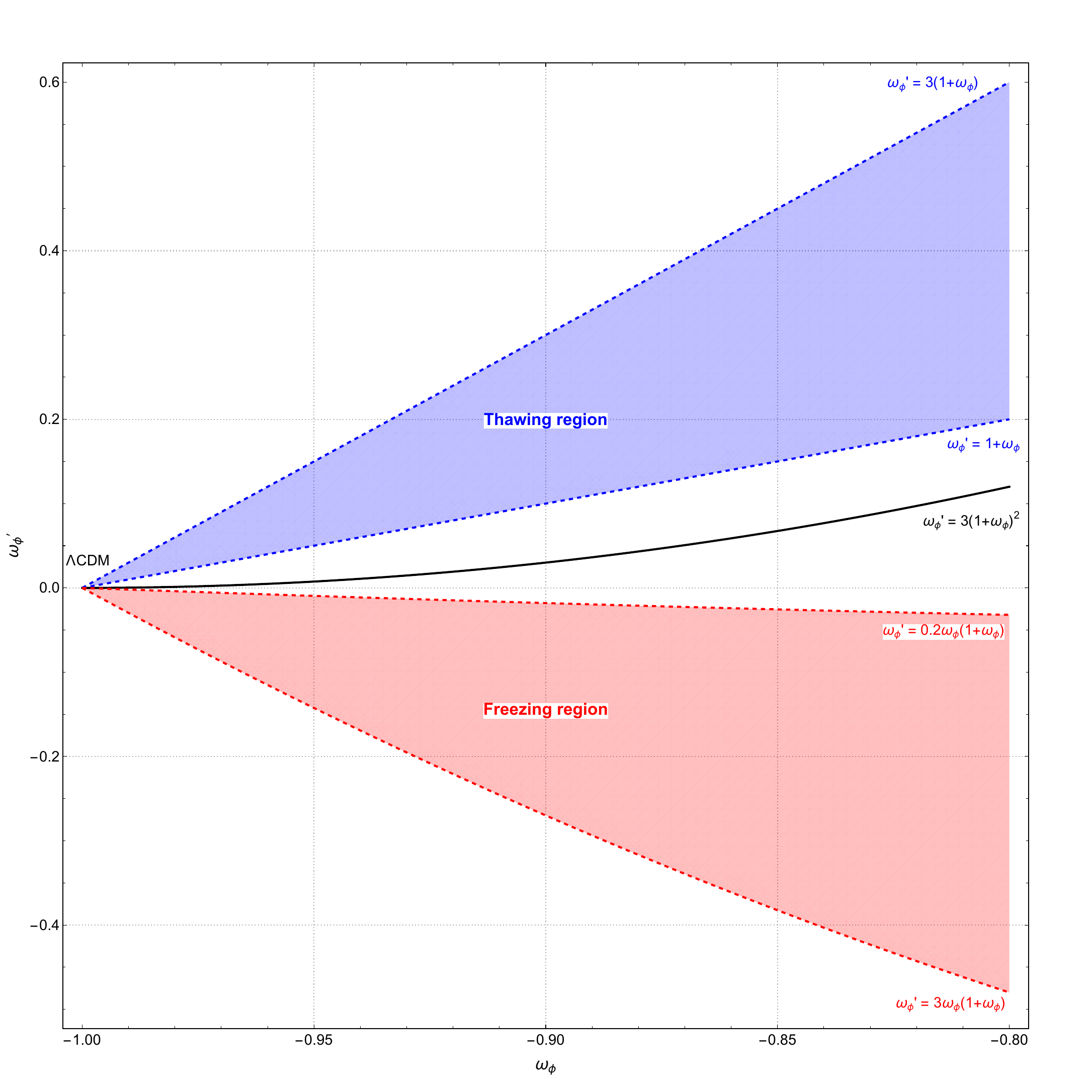}
    \caption{The allowed EoS phase-space for thawing and freezing quintessence models. The black curve shows the border between the acceleration and deceleration of the SF.}
    \label{fig:thawing-freezing-EoS-region}
\end{figure}

The freezing models of the quintessence decelerate:
\begin{equation}
    \omega'_{\phi}<0 \quad\land\quad \ddot{\phi}<0 \implies \xi<0
\end{equation}
as the SF potential shallows, therefore $\omega_{\phi}$ decreases towards $-1$ value. The lower bound comes from a statefinder relation \eqref{statefinder-relation-1} and gives:
\begin{equation}
    \omega'_{\phi}>3\omega_{\phi}\left(1+\omega_{\phi}\right)\,.
\end{equation}
On the other hand, the upper bound can be determined by the fact that the freezing fields cannot change too rapidly once DE is appreciable ($z\lesssim1$) \cite{Caldwell:2005tm}:
\begin{equation}
    \omega'_{\phi}\lesssim 0.2\,\omega_{\phi}\left(1+\omega_{\phi}\right)\,.
\end{equation}
Therefore, the acceptable region in the EoS phase space for freezing quintessence models takes the following form:
\begin{examplebox}[Freezing quintessence EoS phase-space region]
    \begin{equation}
        \boxed{3\omega_{\phi}\left(1+\omega_{\phi}\right)<\omega'_{\phi}\lesssim 0.2\,\omega_{\phi}\left(1+\omega_{\phi}\right)}\,.
    \end{equation}
\end{examplebox}
Fig.~\ref{fig:thawing-freezing-EoS-region} illustrates the allowed phase space that describes the evolution of the quintessence EoS for thawing and freezing models.

\subsubsection{Zones of avoidance}
\begin{figure}[htbp]
    \centering
    \includegraphics[width=1\linewidth]{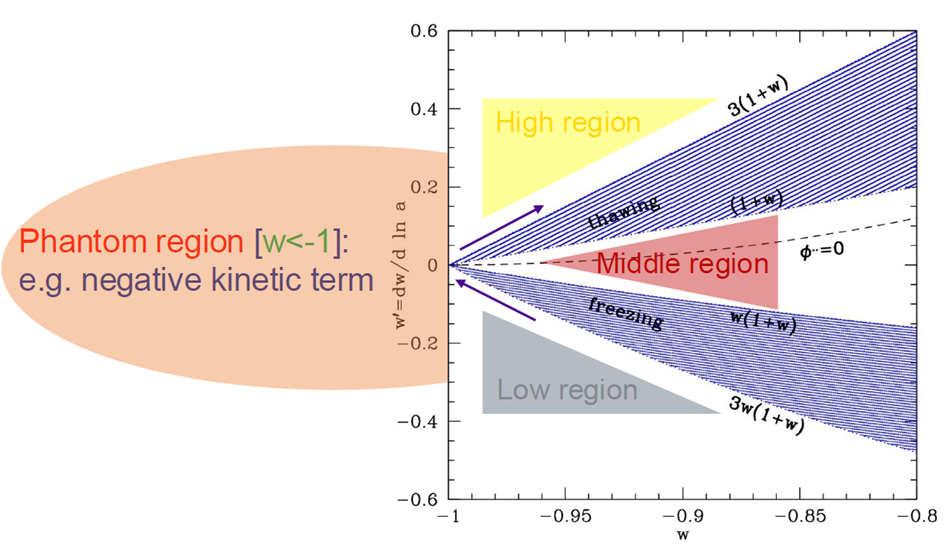}
    \caption{Zones of avoidance for the quintessence models \cite{Linder:2024rdj,LinderStambul2025}.}
    \label{fig:zones-of-avoidance}
\end{figure}

In addition to the above limitations arising from the design of specific classes of quintessence models, there are others originating from the fundamental laws of physics - \textit{\textbf{zones of avoidance}}\footnote{As stated in \cite{Linder:2024rdj}, zones of Avoidance are, in fact, a test of the physics of the SF (i.e., checking the Klein-Gordon equation itself).} (see Fig.~\ref{fig:zones-of-avoidance}) \cite{Linder:2024rdj,LinderStambul2025}:
\begin{examplebox}[Zones of avoidance for quintessence scenarios]
    \begin{enumerate}[label=(\alph*)]
        \item \textbf{High region} - violation of the early-times RD epoch and MD era:
        \begin{equation}
            \boxed{\omega'_{\phi}>3\,\omega_{\phi}\left(1+\omega_{\phi}\right)}\,;
        \end{equation}
        \item \textbf{Middle region} - \textit{fine tuning} (balance of the Hubble friction and the slope of the SF potential at the present):
        \begin{equation}
            \ddot{\phi}\simeq 0 \quad\implies\quad \boxed{1+\omega_{\phi}<\omega'_{\phi}<\omega_{\phi}\left(1+\omega_{\phi}\right)}\,;
        \end{equation}
        \item \textbf{Low region} - non-canonical evolution of the SF, namely, rolling upslope (e.g. k-essence):
        \begin{equation}
            \boxed{\omega'_{\phi}<3\,\omega_{\phi}\left(1+\omega_{\phi}\right)}\,;
        \end{equation}
        \item \textbf{Phantom region} (e.g. negative kinetic term of the SF in the action):
        \begin{equation}
            \boxed{\omega_{\phi}<-1}\,.
        \end{equation}
    \end{enumerate}
    \underline{\textit{\textbf{Conclusions:}}}
     \begin{enumerate}[label=(\roman*)]
         \item Current data suggests that \textbf{DE (quintessence) may indeed behave as if it lies in the high region zone of avoidance};
         \item The \textbf{mirage DE model best fits the current data}. It has several unusual properties, including \textit{phantom crossing} and a \textit{non-de Sitter future state}. Moreover, it \textit{exhibits rapid evolution, which violates the idea of a long period of evolution during MD epoch};
         \item The data prefers the \textbf{phantom crossing being a real, not merely apparent, characteristic by} $\boldsymbol{\gtrsim 3\sigma}$.
     \end{enumerate}
\end{examplebox}

\subsection{\texorpdfstring{$Om(z)$}{Om(z)} diagnosis}

\begin{examplebox}[$Om(z)$ diagnosis]
    \begin{itemize}
        \item The $Om(z)$ diagnosis \cite{Sahni:2008xx,Zunckel:2008ti,Shafieloo:2009hi}:
    \begin{equation}
        Om(z)\equiv\frac{h^{2}(z)-1}{\left(1+z\right)^{3}-1}\,,
    \end{equation}
    where:
    \begin{equation}
        h(z)\equiv\frac{H(z)}{H_{0}}
    \end{equation}
    is a \textbf{null test of the spatially flat LCDM model} because for DE in the form of CC:
    \begin{equation}
        Om(z)=\frac{\Omega_{M}^{(0)}\left[\left(1+z\right)^{3}-1\right]}{\left(1+z\right)^{3}-1}=\Omega_{M}^{(0)}=\mathrm{const} \quad\land\quad \Omega_{M}^{(0)}=\Omega_{DM}^{(0)}+\Omega_{BM}^{(0)}\,;
    \end{equation}
    \item If:
    \begin{equation}
        Om(z)\neq\mathrm{const}\,,
    \end{equation}
    then this indicates a \textbf{deviation from} $\boldsymbol{\Lambda}$\textbf{CDM model} (see the diagram at the top of Fig.~\ref{fig:q-diagram-DESI} for the results from DESI DR2 \cite{DESI:2025fii});
    \item For DE with $\omega_{DE}=\mathrm{const}$:
    \begin{enumerate}[label=(\alph*)]
        \item \textbf{Quintessence-like behavior} corresponds to:
        \begin{equation}
            \omega_{DE}>-1 \quad\implies\quad \frac{d\,Om(z)}{dz}<0\,,
        \end{equation}
        \item \textbf{Phantom-like behavior} coincides with:
        \begin{equation}
            \omega_{DE}<-1 \quad\implies\quad \frac{d\,Om(z)}{dz}>0\,.
        \end{equation}
    \end{enumerate}
    \end{itemize}
    \underline{\textit{\textbf{Conclusion:}}}\\
    The $Om(z)$ diagnosis can \textbf{distinguish quintessence-like and phantom-like expansion histories without directly reconstructing the DE EoS}, $\omega_{DE}(z)$.
\end{examplebox}
The quantities $q(z)$ and $Om(z)$ are \textit{\textbf{sensitive indicators of new physics}} because they only respond to the 'shape' of the normalized expansion history, $h(z)$. They are \textit{not affected by degeneracies between DE and matter (BM and DM) densities at the background level} \cite{Wasserman:2002gb,Kunz:2007rk,Shafieloo:2011zv}.

\subsection{DESI DR2}
\label{sec:DESI-DR2-ch4}
\begin{figure}[htbp]
    \centering
    \includegraphics[width=1\linewidth]{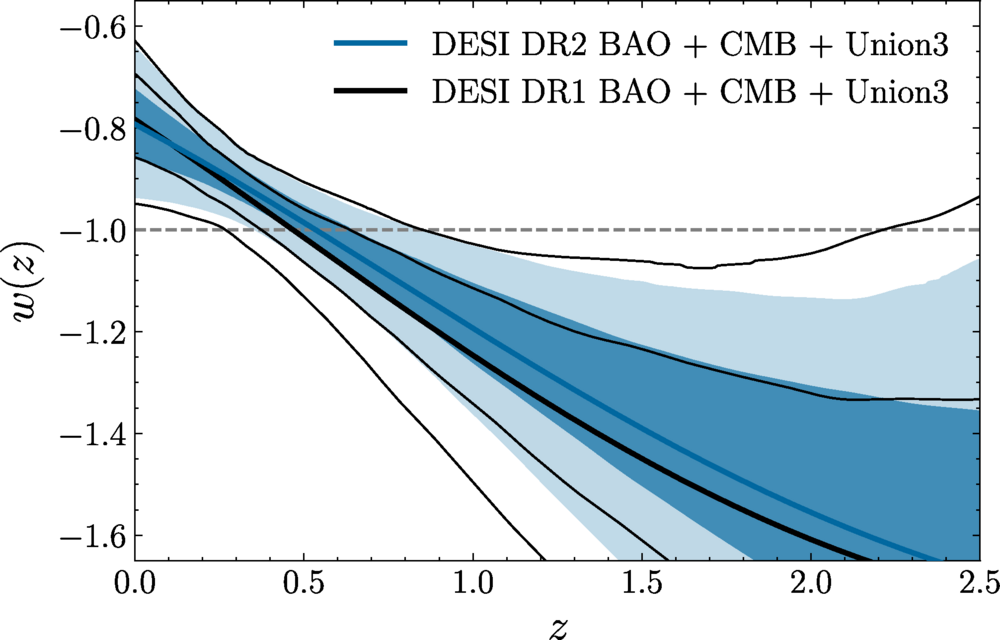}
    \caption{Comparison of the DE EoS evolution obtained via Gaussian process (GP) reconstruction in the case of DESI DR1 and DESI DR2 \cite{DESI:2025fii}.}
    \label{fig:DESI-DR2-DESI-DR1-comparison}
\end{figure}

\begin{figure}[htbp]
    \centering
    \includegraphics[width=1\linewidth]{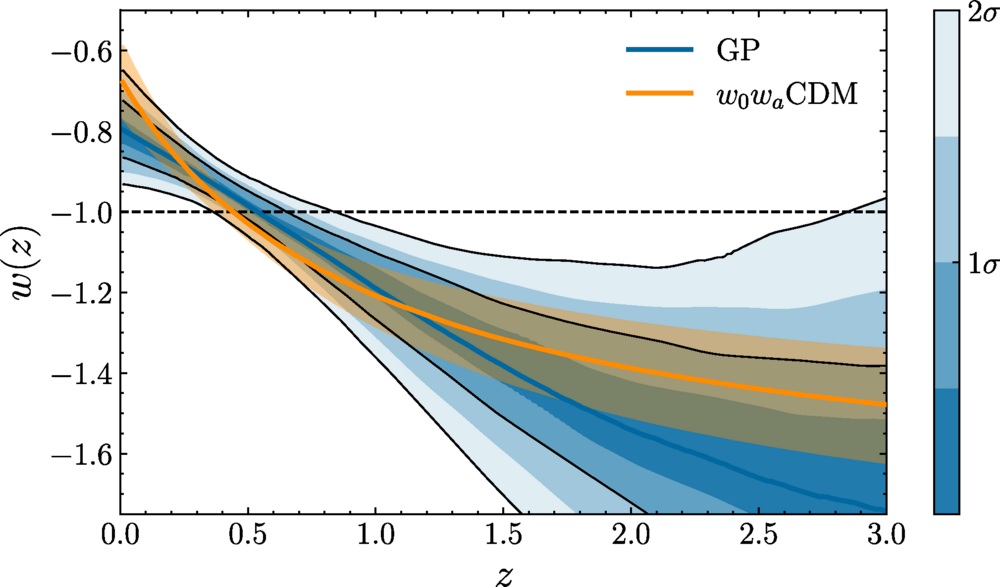}
    \caption{Comparison of GP reconstruction of the DE EoS with the $\omega_{0}\omega_{a}$ parametrization \cite{DESI:2025fii}.}
    \label{fig:DESI-DR2-omega0-omegaa}
\end{figure}

The \textit{\textbf{Dark Energy Spectroscopic Instrument (DESI)}}\footnote{\href{https://www.desi.lbl.gov/}{https://www.desi.lbl.gov/.}} mission may help determine the true nature of dark energy by providing clues as to whether it is a cosmological constant or a quantity that evolves over time along with the Universe \cite{Linder:2024rdj,Rezaei:2025vhb,Capozziello:2025qmh,Chaudhary:2025uzr}. This fundamental question can possibly be answered by examining a parameter of the EoS for dark energy. Reconstructing its evolution (or constancy, in the case of the cosmological constant) across different cosmological epochs can yield significant insights (see Fig.~\ref{fig:DESI-DR2-DESI-DR1-comparison} for a comparison between the results from DESI DR1 and DESI DR2, and Fig.~\ref{fig:DESI-DR2-omega0-omegaa} for a non-parametrized EoS for DE vs. $\omega_{0}\omega_{a}$ parametrization).


\subsubsection{$\omega_{0}\omega_{a}$ results}
\begin{figure}[htbp]
    \centering
    \includegraphics[width=1\linewidth]{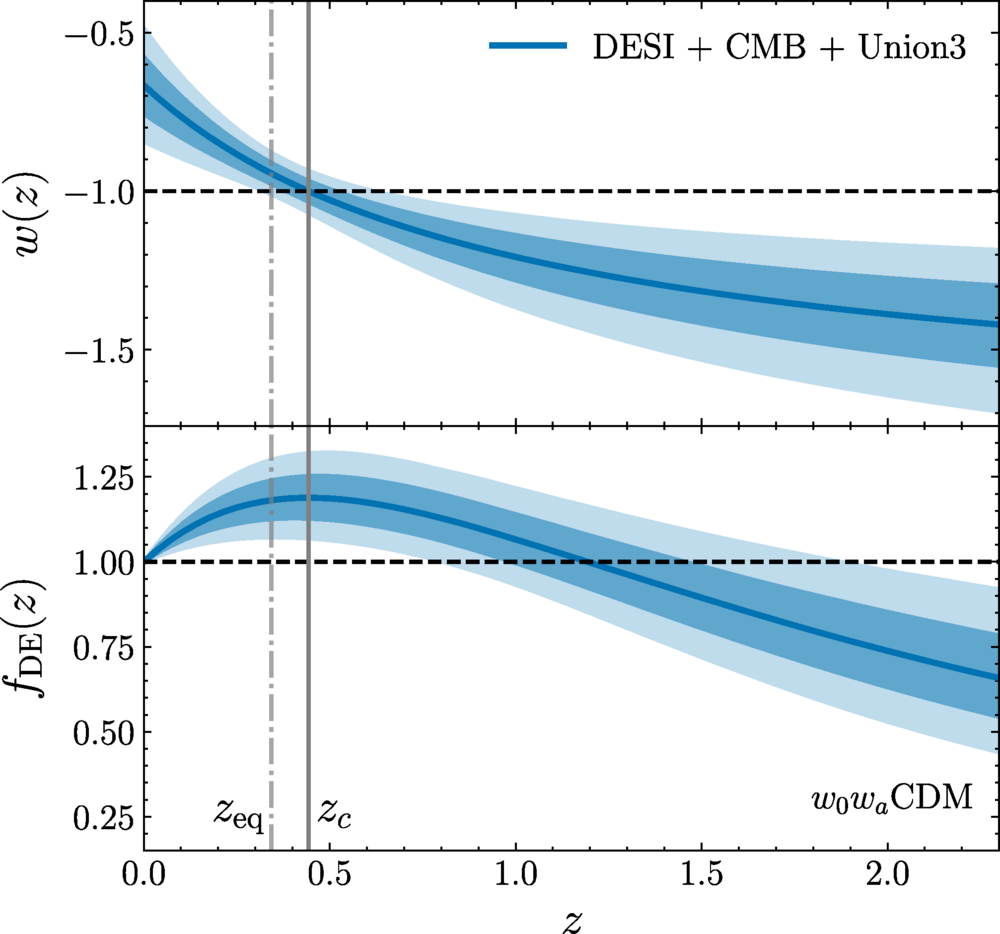}
    \caption{The DE EoS parameter and normalized DE density $f_{DE}(z)\equiv\rho_{DE}(z)/\rho_{DE}^{(0)}$ evolution for the $\omega_{0}\omega_{a}$ parametrization \cite{DESI:2025fii}.}
    \label{fig:DE-EoS-DE-normalized-density}
\end{figure}

\begin{examplebox}[$\omega_{0}\omega_{a}$ results]
    \begin{itemize}
        \item The combined data (DESI+CMB+Union3, DESI DR1 BAO, DESI DR2 BAO) favor the region of the parameters characterized by \cite{DESI:2025fii}:
        \begin{equation}
            \omega_{0}>-1 \quad\land\quad \omega_{a}<0\,,
        \end{equation}
        that is, \textbf{\textit{away from the CC}}. This means that the EoS of DE was phantom-like:
        \begin{equation}
            \omega_{DE}(z)<-1
        \end{equation}
        in the distant past and has evolved to (\textit{crossed the PDL}):
        \begin{equation}
            \omega_{DE}(z)>-1
        \end{equation}
        at the present (see Fig.~\ref{fig:DE-EoS-DE-normalized-density});
        \item From a physical point of view, a phantom EoS indicates for an increase of the energy density with the expansion of the Universe:
        \begin{equation}
            \frac{d\rho_{DE}}{da}>0\,,
        \end{equation}
        before reaching its maximum at \cite{DESI:2025fii}:
        \begin{equation}
            z_{c}\simeq 0.45 \quad\iff\quad a_{c}\simeq 0.69\,;
        \end{equation}
        \item This is a \textit{'naive' suggestion} that points to \textit{\textbf{'phantom crossing' at high redshift}} \cite{Caldwell:1999ew}, which would in turn imply a \textbf{violation of the NEC} \cite{Visser:1999de,Cattoen:2006yh,Curiel2017,MartinMoruno2017,Kontou:2020bta,Hawking_Ellis_2023ECs,Ford:1994bj,Tipler:1978zz,Cattoen:2005dx,Visser:1997qk,Visser:1997tq,Carroll_2019chapter}:
        \begin{equation}
            \rho+p\geq 0\,;
        \end{equation}
        \item If confirmed, the \textbf{phantom crossing would have profound implications not only for cosmology, but also for fundamental physics} \cite{Curiel2017,Kontou:2020bta}. This would also suggest that the dark sector of the Universe is \textit{far more complex than we previously thought}.
    \end{itemize}
    \underline{\textbf{\textit{Remark:}}}\\
    \textbf{The} $\boldsymbol{\omega_{0}\omega_{a}}$ \textbf{parametrization is an effective description that may not accurately approximate the true behavior of the DE EoS,} $\boldsymbol{\omega_{DE}(z)}$.
\end{examplebox}

\begin{figure}[htbp]
    \centering
    \includegraphics[width=1\linewidth]{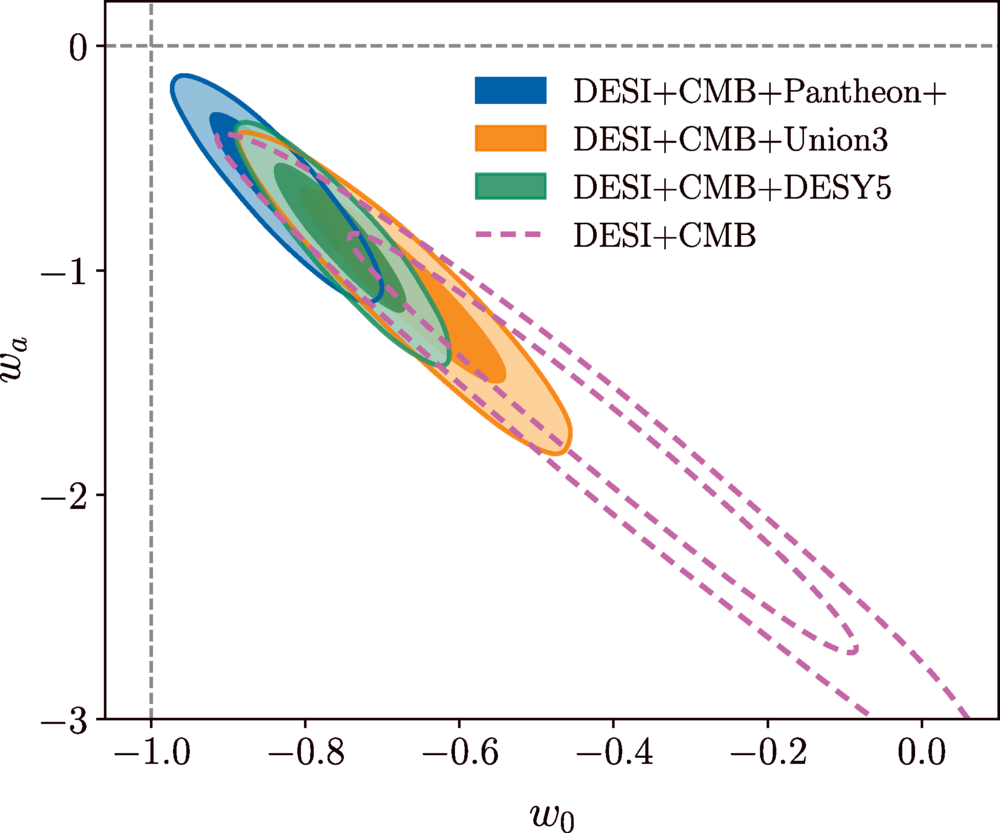}
    \caption{The posterior distributions of $\omega_{0}$ and $\omega_{a}$ parameters for the $\omega_{0}\omega_{a}$CDM model \cite{DESI:2025zgx}.}
    \label{fig:CPL-distribution}
\end{figure}

\begin{examplebox}[DESI DR2 and the preference for the $\omega_{0}\omega_{a}$CDM model over the $\Lambda$CDM model]
    The \textbf{DESI DR2 BAO data} \cite{DESI:2025zgx,DESI:2025fii} has led to an \textit{\textbf{increase in the evidence for a departure from LCDM in the form of evolving dark energy}} (see Fig.~\ref{fig:CPL-distribution}):
    \begin{itemize}
        \item \textbf{\textit{DESI + CMB + Pantheon}}$\boldsymbol{+}$\\
        ($\omega_{0}\omega_{a}$CDM model over $\Lambda$CDM at the $\boldsymbol{2.8\sigma}$ level):
        \begin{equation}
            \omega_{0}=-0.838 \pm 0.055\quad\land\quad \omega_{a}=-0.62_{-0.19}^{+0.22}\,;
        \end{equation}
        \item \textbf{\textit{DESI + CMB + Union3}}\\
        ($\omega_{0}\omega_{a}$CDM model over $\Lambda$CDM at the $\boldsymbol{3.8\sigma}$ level):
        \begin{equation}
            \omega_{0}=-0.667 \pm 0.088\quad\land\quad \omega_{a}=-1.09_{-0.27}^{+0.31}\,;
        \end{equation}
        \item \textbf{\textit{DESI + CMB + DESY5}}\\
        ($\omega_{0}\omega_{a}$CDM model over $\Lambda$CDM at the $\boldsymbol{4.2\sigma}$ level):
        \begin{equation}
            \omega_{0}=-0.752 \pm 0.057\quad\land\quad \omega_{a}=-0.86_{-0.20}^{+0.23}\,.
        \end{equation}
        \end{itemize}
\end{examplebox}

\subsubsection{Interactions between DE and DM as an alternative to a phantom crossing}

\begin{examplebox}[Interactions between DE and DM]
    \begin{itemize}
        \item  The interacting DE (IDE) models \cite{Linder:2025zxb,LinderStambul2025,Caldera-Cabral:2008yyo,vanderWesthuizen:2025vcb,vanderWesthuizen:2025mnw,vanderWesthuizen:2025rip,Pan:2025qwy,Guedezounme:2025wav,Li:2026xaz,Amendola:1999er} are characterized by the following modified conservation law (in this case energy flows from DE to DM sector):
        \begin{equation}
            \nabla_{\nu}T^{\nu}{}_{\mu\,(DE)}=-\nabla_{\nu}T^{\nu}{}_{\mu\,(DM)}=Q_{\mu}\equiv -Q\,u_{\mu}
        \end{equation}
        that implies:
        \begin{align}
            \dot{\rho}_{DE}+3H\rho_{DE}\left(1+\omega_{DE}\right) & =-Q \,, \\
            \dot{\rho}_{DM}+3H\rho_{DM} & = +Q\,,
        \end{align}
        so that:
        \begin{align}
            \omega_{DE}^{\mathrm{eff}} & = \omega_{DE}+\frac{Q}{3H\rho_{DE}} \,, \\
            \omega_{DM}^{\mathrm{eff}} & = -\frac{Q}{3H\rho_{DM}}\,,
        \end{align}
        where $Q$ denotes the energy transfer rate;
        \item One can explicitly show impact of the interactions on the deceleration parameter \cite{Linder:2025zxb}:
        \begin{align}
            q\equiv -1+\frac{\dot{H}}{H^{2}} & = \frac{1}{2}\sum_{i}\left(1+3\omega_{i}\right)\,\Omega_{i}(z) \\
            & = \frac{1}{2}\Bigl[1+3\Bigl(\omega_{DM}^{\mathrm{eff}}\,\Omega_{DM}(z)+\omega_{DE}^{\mathrm{eff}}\,\Omega_{DE}(z)\Bigr)\Bigr] \\
            & = \frac{1}{2}\left[1-\frac{Q}{H}\frac{\kappa^{2}}{3H^{2}}+3\omega_{DE}\,\Omega_{DE}+\frac{Q}{H}\frac{\kappa^{2}}{3H^{2}}\right] \\
            & = \frac{1}{2}\Bigl[1+3\omega_{DE}\,\Omega_{DE}(z)\Bigr]\,.
        \end{align}
        $\implies$ \underline{\textbf{Two interesting implications}:}
        \begin{enumerate}[label=(\roman*)]
            \item The \textbf{\textit{interaction term}} $\boldsymbol{Q}$ \textbf{\textit{does not appear explicitly}} - \textbf{it is hidden in the evolution of the effective DE density};
            \item If we consider the acceleration at a given cosmological epoch, the data can be used to determine the values of $\Omega_{DE}^{(0)}$ and $q^{(0)}$. This allows us to determine the bare quantity before interaction (rather than the 'measured' $\omega_{DE}^{\mathrm{eff}(0)}$):
            \begin{equation}
                \omega_{0}\equiv\omega_{DE}^{(0)}\,.
            \end{equation}
        \end{enumerate}
    \end{itemize}
\end{examplebox}

\begin{figure}[htbp]
    \centering
    \includegraphics[width=1\linewidth]{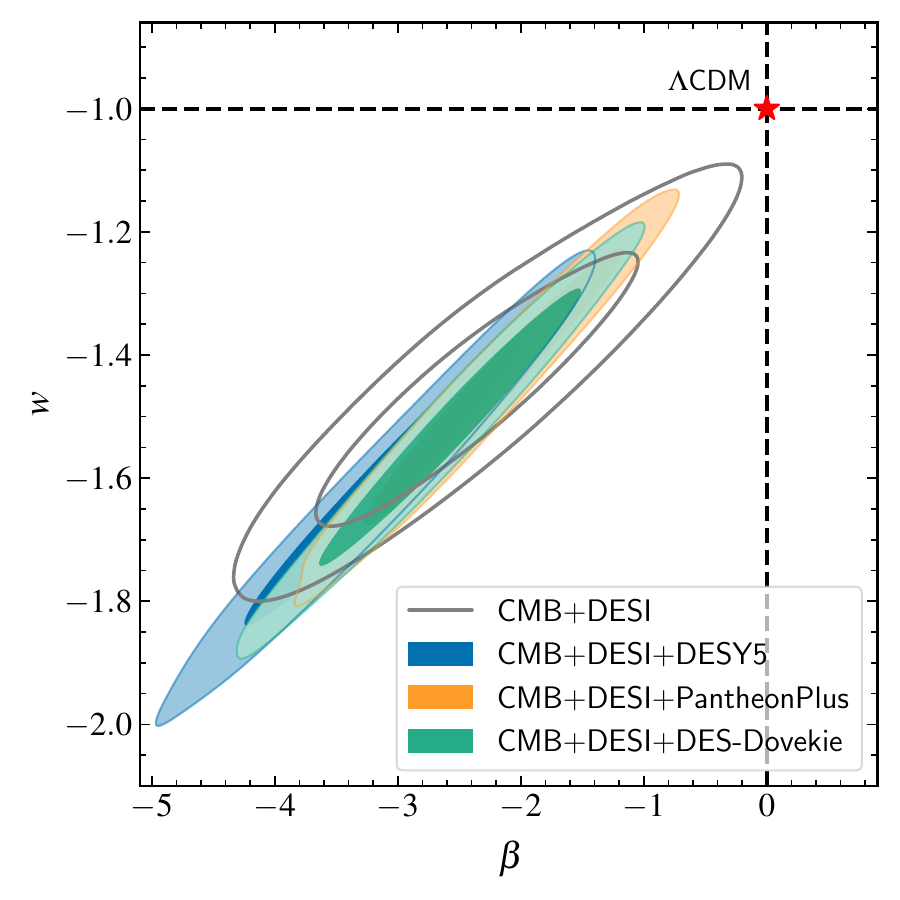}
    \caption{Constraints on the free parameters used in the IDE model with \eqref{IDE-Q} \cite{Li:2026xaz}.}
    \label{fig:IDE-constraints}
\end{figure}

\begin{examplebox}[Example of IDE model]
    \begin{itemize}
        \item For the energy transfer rate form favored by observational data \cite{Li:2024qso,Li:2025muv}:
        \begin{equation}\label{IDE-Q}
            Q=\beta H_{0}\,\rho_{DE}\propto\rho_{DE}
        \end{equation}
        and a constant EoS parameter $\omega$ for the DE fluid in \cite{Li:2026xaz} the following results were obtained for different datasets (see Fig.~\ref{fig:IDE-constraints} for an illustrated constraints on the free parameters):
        \begin{enumerate}[label=(\alph*)]
            \item \textbf{CMB+DESI DR2}:
            \begin{equation}
                \omega=-1.45^{+0.13}_{-0.15} \quad\land\quad \beta=-2.37^{+0.82}_{-0.93} \,;
            \end{equation}
            \item \textbf{CMB+DESI DR2+Pantheon}$\boldsymbol{+}$:
            \begin{equation}
                \omega=-1.46^{+0.16}_{-0.13} \quad\land\quad \beta=-2.25^{+0.73}_{-0.63} \,;
            \end{equation}
            \item \textbf{CMB+DESI DR2+DESY5}:
            \begin{equation}
                \omega=-1.60\pm 0.15 \quad\land\quad \beta=-3.08\pm 0.70 \,;
            \end{equation}
            \item \textbf{CMB+DESI DR2+DES-Dovekie}:
            \begin{equation}
                \omega=-1.52^{+0.16}_{-0.13} \quad\land\quad \beta=-2.59^{+0.76}_{-0.62} \,;
            \end{equation}
        \end{enumerate}
        \item The authors of \cite{Li:2026xaz} obtained a \textbf{strong preference for a non-vanishing coupling}, reaching the $5\sigma$ level, and  their data \textbf{favor a phantom DE component accompanied by energy transfer to the DM sector}.
    \end{itemize}
    \underline{\textbf{\textit{Conclusion:}}}\\
    \textbf{Physically distinct interacting realizations of the dark sector can produce similar late-time expansion histories.}
\end{examplebox}

\subsubsection{Quintessence}
The quintessence scenario \cite{Caldwell:2005tm,Linder:2006sv,Wetterich:1987fm,Ratra:1987rm,Cahn:2008gk} is one of the types of DE considered in the data analysis \cite{DESI:2025zgx,DESI:2025fii}.

A general analysis of the BAO data regarding the behavior of the DE EoS indicates that \textit{\textbf{'freezing' models}} characterized by:
\begin{equation}
    \omega_{a}>0
\end{equation}
\textit{\textbf{"are not favored by observations"}} \cite{DESI:2025fii}, therefore, the main findings of the analyses concern the \textbf{'thawing' quintessence} model.

\subsubsection{'Thawing' quintessence}
\begin{figure}[htbp]
    \centering
    \includegraphics[width=1\linewidth]{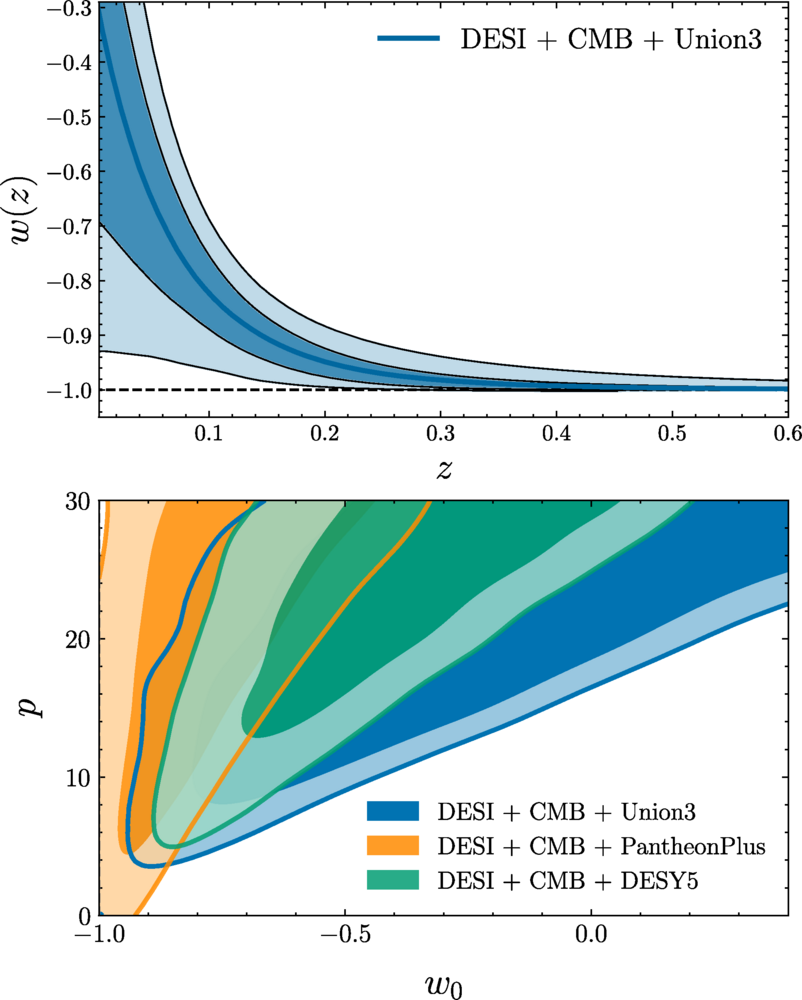}
    \caption{Constraints on the algebraic thawing functional form \eqref{algebraic-formula} and evolution of the non-phantom DE EoS obtained from DESI DR2 analysis \cite{DESI:2025fii}.}
    \label{fig:Constraints-algebraic}
\end{figure}

\begin{examplebox}['Thawing' quintessence and DESI DR2]
    \begin{itemize}
        \item The phase-space dynamics of the 'thawing' quintessence scenarios \cite{Caldwell:2005tm,Ferreira:1997hj,Scherrer:2007pu} such as aforementioned \textbf{pNGB models} \cite{Frieman:1995pm}:
        \begin{equation}
            V(\phi)=M^{4}\left[1+\cos{\left(\frac{\phi}{f}\right)}\right]\,,
        \end{equation}
        or simple\textbf{ monomial SF potentials} very well known from a \textbf{high-energy physics} phenomenology \cite{Copeland:2006wr}:
        \begin{equation}
            V(\phi)\propto m^{2}\phi^{2} \quad\land\quad V(\phi)\propto\lambda\,\phi^{4}
        \end{equation}
        may be well approximated by the explicit EoS parametrization of the $\omega_{0}\omega_{a}$ form \cite{dePutter:2008wt}:
        \begin{equation}
            \omega_{a}\simeq -1.58 \left(1+\omega_{0}\right)\,,
        \end{equation}
        and allows the EoS for DE to cross the PDL\footnote{Phantom divide line.} ($\omega_{DE}=-1$) which is \textbf{impossible from the physical point of view in the case of quintessence models} \cite{Caldwell:1999ew,Cline:2003gs,Vikman:2004dc}\footnote{For more information on phantom/ghost SF cosmological models, see \cite{Carroll:2003st,Caldwell:2003vq,Dabrowski:2003jm,Sami:2003xv,Nojiri:2005sx,Nojiri:2005sr,Sbisa:2014pzo,Ludwick:2017tox}.};
        \item The thawing dynamics guaranteeing that $\omega_{DE}>-1$ at all times can be expressed by the following algebraic formula \cite{Linder:2007wa,Crittenden:2007yy,Linder:2015zxa}:
        \begin{equation}\label{algebraic-formula}
            1+\omega_{DE}(a)=\left(1+\omega_{0}\right)\,a^{p}\left(\frac{1+b}{1+b\,a^{-3}}\right)^{1-p/3} \quad\land\quad b=0.5\,, 
        \end{equation}
        where $\omega_{0}$ and $p$ are \textit{free parameters}. Fig.~\ref{fig:Constraints-algebraic} shows the constraints obtained for the space of free parameters appearing in the formula \eqref{algebraic-formula}.
    \end{itemize}
\end{examplebox}
\subsubsection{Conclusions}
\begin{examplebox}[Main findings from DESI DR2]
    \begin{itemize}
        \item Statistical analyses associated with the DESI DR2 \textbf{suggest a departure from the} $\boldsymbol{\Lambda}$\textbf{CDM model at the} $\boldsymbol{2.8-4.2\sigma}$ \textbf{level} which means that the \textit{\textbf{LCDM model is challenged}};
        \item Taking the best fit at face value seem to indicate the so-called \textit{\textbf{mirage DE}} scenario \cite{Linder:2007ka,Mattsson:2009mxq} that consists of \cite{Linder:2024rdj,LinderStambul2025}:
        \begin{enumerate}[label=(\alph*)]
            \item \textbf{Phantom behavior}:
            \begin{equation}
                \omega_{DE}<-1
            \end{equation}
            at $z\gtrsim 1$;
            \item \textbf{'Superevolving' faster than the Hubble friction from MD epoch would allow};
            \item \textbf{Crossing of the PDL}:
            \begin{equation}
                \omega_{DE}=-1\,;
            \end{equation}
            \item  \textbf{Recent evolving toward}:
            \begin{equation}
                \omega_{DE}>-1\,,
            \end{equation}
            i.e. \textbf{away from the CC}.
        \end{enumerate}
        $\implies$ \textit{\textbf{We understand the physics behind each of the properties, but they generally do not all go together!}}
    \end{itemize}
    $\implies$ \underline{\textbf{\textit{New directions(?):}}}
    \begin{enumerate}[label=(\alph*)]
        \item \textbf{Non-canonical kinetic term of the SF} \cite{Chiba:1999ka,Armendariz-Picon:2000nqq,Kobayashi:2011nu,Arkani-Hamed:2003juy};
        \item \textbf{Multi-field scenarios} \cite{Feng:2004ad,Guo:2004fq,Hu:2004kh,Cai:2009zp,Cai:2025mas};
        \item \textbf{Non-standard vacuum, e.g. negative SF potentials, phase transitions} etc. \cite{Kallosh:2003bq,Perivolaropoulos:2004yr,Lykkas:2015kls,Felder:2002jk,Bassett:2002qu,Mortonson:2009qq,Parker:1999td,Parker:2003as,DiValentino:2017zyq}.
    \end{enumerate}
\end{examplebox}

\section{Summary}
\begin{examplebox}[Main conclusions]
    \begin{enumerate}[label=(\alph*)]
        \item Quintessence is the simplest dynamical DE model based on a minimally coupled canonical SF;
        \item Canonical quintessence satisfies:
        \begin{equation}
            \rho_{\phi}+p_{\phi}=\dot{\phi}^{2}\geq 0\,,
        \end{equation}
        and therefore cannot cross the phantom divide line:
        \begin{equation}
            \omega=-1\,;
        \end{equation}
        \item Scaling and tracking solutions provide mechanisms that reduce the dependence of the late-time SF evolution on initial conditions;
        \item Statefinder diagnostics, the $(\omega_{\phi},\omega'_{\phi})$ plane and the $Om(z)$ diagnostic provide complementary ways of distinguishing $\Lambda$CDM, thawing quintessence, freezing quintessence and phantom-like expansion histories;
        \item DESI DR2 suggests that a time-dependent effective DE EoS may provide a better phenomenological description of the late-time expansion history than a strict cosmological constant;
        \item The DESI DR2 preference for phantom-crossing-like behavior cannot be straightforwardly interpreted as minimal canonical quintessence. It points instead toward a broader theoretical landscape, including non-canonical fields, interacting dark sectors, multi-field models and modified gravity.
    \end{enumerate}
\end{examplebox}
The overall conclusion is that quintessence remains a fundamental reference model for dynamical dark energy. Even if the simplest canonical realization is too restrictive to explain all possible observational indications, especially an effective phantom crossing, it provides the baseline against which more general theories should be compared. For this reason, quintessence is not only a model of late-time acceleration, but also a diagnostic framework for organizing deviations from $\Lambda$CDM and for identifying which theoretical extensions of the dark-energy sector are physically required.
\begin{savequote}
"The world, an entity out of everything, was created by neither gods nor men, but was, is and will be eternally living fire, regularly becoming ignited and regularly becoming extinguished."
\qauthor{\textbf{Heraclitus}}
"[\ldots] The prospect that inflation did not occur deserves serious consideration. If we step back, there seem to be two logical possibilities. Either the universe had a beginning, which we commonly dub the “big bang,” or there was no beginning and what has been called the big bang was actually a “big bounce,” a transition from some preceding cosmological phase to the present expanding phase."
\qauthor{\textbf{Anna (Ijjas) Rosenzweig, Paul J. Steinhardt, Abraham Loeb} \cite{Ijjas2017SciAm}}
\end{savequote}

\chapter{Ekpyrotic and Cyclic Universe}
\label{Sec:ekpyrotic}
\ifpdf
    \graphicspath{{Chapter6/Figs/Raster/}{Chapter6/Figs/PDF/}{Chapter6/Figs/}}
\else
    \graphicspath{{Chapter6/Figs/Vector/}{Chapter6/Figs/}}
\fi

\section{Introduction}
We understand the Universe's evolution pretty well from the BBN until the present. We also know that the \textbf{early Universe must have been in a very special state} characterized by \textit{homogeneity}, \textit{spatial flatness} and \textit{tiny curvature perturbations}. Cosmic inflation \cite{Starobinsky:1980te,Guth:1980zm,Linde1982,Linde:1983psb,Linde:1983gd,Goldwirth:1991rj,Liddle2000,Brandenberger2000,Burgess:2002ub,Gordon:2000hv,Bartolo:2004if,Bassett:2005xm,Mukhanov_2005a,Mukhanov_2005b,Lemoine2007,Chen:2006nt,Baumann:2009ds,Baumann:2009ni,Chen:2010xka,Guth:2013sya,delCampo2014chapter,Baumann:2014nda,Baumann:2014cja,Senatore:2016aui,Guzzetti:2016mkm,Maggiore-inflation-PGWs,Guth1984SciAm,Martin:2013tda,Nojiri:2017ncd,Planck:2018jri,Planck:2019kim,Chowdhury:2019otk,Baumann_2022inflation,Ketov2024chapter} offers a possible answer for such a very early-times state by the mechanism of rapid exponential expansion. Nevertheless, there are significant issues and concerns regarding this formalism \cite{Penrose:1988mg,Borde:1993xh,Borde:1996pt,Earman1999,Brandenberger2000,Martin:2000xs,Borde:2001nh,Gibbons:2006pa,Steinhardt2011,Brandenberger:2011eq,Brandenberger:2012aj,Ijjas:2013vea,Ijjas:2014nta,Ijjas:2015hcc,Ijjas2017,DiTucci:2019xcr,Martin2020,Dawid:2023mmv,Garfinkle:2023vzf,Postolak:2024xtm,Martin:2024qnn,Ferreira:2025lrd}. Moreover, based on observations, it can be concluded that, regardless of its true nature, dark energy \cite{Peebles:2002gy,Linder:2005ne,Copeland:2006wr,Frieman:2008sn,dePutter:2008wt,Amendola_Tsujikawa_2010book,Linder:2024rdj,DESI:2025wyn,Rezaei:2025vhb,Capozziello:2025qmh,DESI:2025fii,LinderStambul2025,Linder:2025zxb} currently has a dominant influence on the Universe.

\begin{figure}[htbp]
    \centering
    \includegraphics[width=0.55\linewidth]{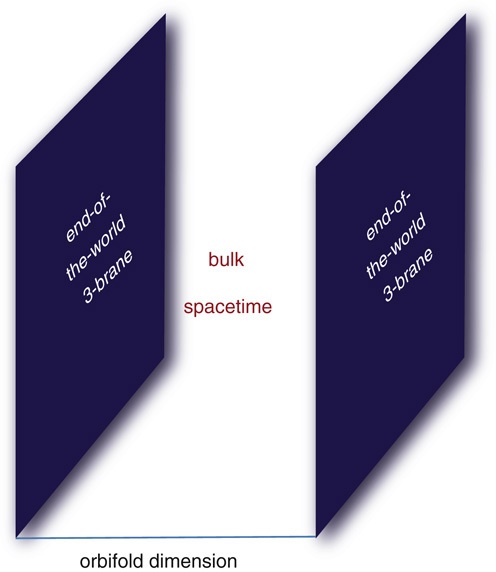}
    \caption{The braneworld picture of our Universe - two end-of-the-world 3-branes and the orbifold dimension embedded in $5D$ bulk spacetime \cite{Lehners:2008vx}.}
    \label{fig:braneworld-picture}
\end{figure}

The \textbf{\textit{ekpyrotic}}\footnote{The name comes from the Ancient Greek word '\textit{ekpyrosis}' (conflagration) \cite{Lapidge1978StoicCosmology,Salles_2025book,SteinhardtTurok2007EndlessUniverse}.} and \textbf{\textit{cyclic (ekpyrotic) cosmology}} \cite{SteinhardtTurok2007EndlessUniverse,Khoury:2001wf,Steinhardt:2002ih,Steinhardt:2001st,Khoury:2001bz,Khoury:2001zk,Khoury:2003rt,Khoury:2003vb,Boyle:2003km,Erickson:2003zm,Khoury:2004xi,Steinhardt:2004gk,Lehners:2007ac,Garfinkle:2008ei,Ijjas:2018qbo,Nastase2019ekpyrotic,Nastase2019cyclic,Ijjas:2014fja,Levy:2015awa,Ijjas:2016tpn,Ijjas:2016vtq,Ijjas:2019pyf,Cook:2020oaj,Ijjas:2020dws,Ijjas:2021zwv,Ijjas:2021gkf,Kist:2022mew,Lehners:2008vx,Ijjas:2015hcc}  belong to the class of \textit{bouncing cosmological models} \cite{Tolman:1931fei,Tolman1934book,Barrow1995,Dabrowski:1995ae,Clifton:2007tn,Battefeld:2014uga,Brandenberger:2016vhg,Novello:2008ra,Nojiri:2017ncd} and draw the main inspiration from the \textit{braneworld model} of the cosmos (\textit{brane cosmology}) \cite{Horava:1996ma,Horava:1995qa,Langlois:2002bb,Papantonopoulos2002,Brax:2003fv,Maartens2005chapter,Maartens:2010ar}. In such a formalism spacetime is effectively $5D$ with one dimension being a line segment. There are two $(3+1)D$ boundary branes at the endpoints of the line segment (orbifold) - Fig.~\ref{fig:braneworld-picture}.
All matter and forces, except gravity, are localized on branes. However, gravity can propagate throughout the entire spacetime. From this perspective, our Universe is one of the boundary branes that can only interact with one another through gravitational force (as long as the branes are far apart). The key assumption is that an attractive force acts on the branes, causing them to move closer together during the ekpyrotic phase. This, in turn, flattens the branes. Ultimately, the branes collide and then bounce off each other. In this picture, the collision could be viewed as a \textbf{Big Bang}. Furthermore, because the collision is \textit{inelastic}, \textit{matter and radiation are produced on the branes}. Because of quantum fluctuations, the branes are slightly rippled and do not collide at the same time everywhere, generating temperature fluctuations. Furthermore, the distance between the boundary branes stabilizes shortly after the collision, and the attractive interaction begins to take effect again. After a significant amount of time, which will be estimated in the next section of this chapter, another collision occurs, and the model becomes cyclic.

\section{Ekpyrotic phase}
The ekpyrotic phase demonstrates how a contracting phase could resolve known cosmological issues \cite{Khoury:2001wf,Erickson:2006wc}. In order to demonstrate this, let us consider the general form of the FLRW metric:
\begin{equation}\label{FLRW-with-k}
    ds^{2}=-dt^{2}+a^{2}(t)\left[\frac{dr^{2}}{1-k\,r^{2}}+r^{2}\,d\Omega_{2}^{2}\right]\,,
\end{equation}
where:
\begin{equation}
    d\Omega_{2}^{2}\equiv d\theta^{2}+\sin^{2}{\left(\theta\right)}\,d\varphi^{2}
\end{equation}
is the line element on the unit two-sphere. We will also assume that the Universe is filled with a \textbf{perfect fluid}:
\begin{equation}
    T_{\mu\nu}=\left(\rho+p\right)\,u_{\mu}u_{\nu}+p\,g^{\mu\nu}\,,
\end{equation}
where the 4-velocities are given by:
\begin{equation}
    u_{\mu}=\left(-1,0,0,0\right) \quad\land\quad u^{\mu}=\left(1,0,0,0\right) \,,
\end{equation}
that is described by an energy-momentum tensor of the following form:
\begin{equation}\label{perfect-fluid-tensor}
    T_{\nu}{}^{\mu}=\diag\bigl[-\rho,p,p,p\bigr]\,.
\end{equation}
Using \eqref{FLRW-with-k} and \eqref{perfect-fluid-tensor} one may derive the Friedmann equations:
\begin{align}
    \left(\frac{\dot{a}}{a}\right)^{2}\equiv H^{2} & = \frac{\kappa^{2}}{3}\rho-\frac{k}{a^{2}} \,,\label{I-FE-k} \\
     \frac{\ddot{a}}{a}\equiv\dot{H}+H^{2} & = -\frac{\kappa^{2}}{6}\left(\rho+3p\right)\,, \label{II-FE-k}
\end{align}
and also the Raychaudhuri equation:
\begin{equation}
    \dot{H}=-\frac{\kappa^{2}}{2}\left(\rho+p\right)+\frac{k}{a^{2}}\,.
\end{equation}
The Bianchi identity:
\begin{equation}
    \nabla^{\nu}G_{\mu\nu}=0
\end{equation}
leads to the covariant conservation law for the stress-energy tensor:
\begin{equation}
    \nabla^{\nu}T_{\mu\nu}=0\,,
\end{equation}
so that one gets the continuity equation in cosmology:
\begin{equation}
    \dot{\rho}+3H\left(\rho+p\right)=0\,.
\end{equation}
Next, by assuming the constant barotropic EoS for the cosmological fluid:
\begin{equation}
    \omega\equiv\frac{p}{\rho}=\mathrm{const}
\end{equation}
one obtains the standard form of the energy density in terms of the scale factor:
\begin{equation}\label{rho-a}
    \rho(a)\propto a^{-3\left(1+\omega\right)}\,.
\end{equation}
Now we may introduce the notion of the cosmological density parameter:
\begin{equation}
    \Omega(t)\equiv\frac{\rho(t)}{\rho_{\mathrm{crit}}(t)} \quad\land\quad \rho_{\mathrm{crit}}\equiv 3H^{2}\,.
\end{equation}
After combining it with the 1st Friedmann equation \eqref{I-FE-k}, we find that it satisfies the following relation:
\begin{equation}\label{minus-Omega-k}
    \Omega-1=\frac{k}{\left(a H\right)^{2}}\equiv -\Omega_{k}\,.
\end{equation}
At the present epoch \cite{Planck:2018vyg}:
\begin{equation}
    \left|\Omega-1\right|_{0}\equiv\left|\Omega-1\right|\Big|_{t=t_{0}}\equiv\left|\Omega_{k}\right|\lesssim 10^{-3}
\end{equation}
and the extrapolation back in time for the Planck time yields the following ratio:
\begin{equation}
    \frac{\left|\Omega-1\right|_{Pl}}{\left|\Omega-1\right|_{0}}=\frac{\left(a H\right)^{2}_{0}}{\left(a H\right)^{2}_{Pl}}\,.
\end{equation}
For the RD epoch:
\begin{equation}
    \omega_{r}=\frac{1}{3} \quad\implies\quad
    \begin{dcases}
        a(t)\propto t^{1/2} \\
        \rho(a)\propto a^{-4}\propto t^{-2}
    \end{dcases}
\end{equation}
one gets:
\begin{equation}\label{Omega-minus-one-Planck}
    \left|\Omega-1\right|_{Pl}\lesssim\frac{t_{Pl}}{t_{0}}\times 10^{-3}\simeq\frac{5.39\times 10^{-44}\,\mathrm{s}}{4.35\times 10^{17}\,\mathrm{s}}\times 10^{-3}\sim 10^{-64}\,.
\end{equation}
\begin{examplebox}[Flatness problem]
    \textbf{The Universe must have been extremely flat in its early stages!}
\end{examplebox}
\subsection{The most general form of the Friedmann equations and Raychaudhuri equation}

Now, let us consider the \textbf{most general case of the FLRW-like Universe} by taking into account \textit{spatial curvature}, \textit{non-relativistic matter/cosmological dust (BM and DM)}, \textit{radiation (relativistic matter)}, \textit{homogeneous shear anisotropy} and the \textit{canonical SF}:
\begin{equation}\label{general-action-ekpyrotic}
    S=\int d^{4}x\,\sqrt{-g}\left[\frac{1}{2\kappa^{2}}R-\frac{1}{2}g^{\mu\nu}\partial_{\mu}\phi\,\partial_{\nu}\phi-V(\phi)\right]+S_{M}\bigl[g_{\mu\nu},\chi\bigr]
\end{equation}
Using \eqref{general-action-ekpyrotic} one can \textbf{generalize the Friedmann equations} \eqref{I-FE-k}-\eqref{II-FE-k} to obtain:
\begin{examplebox}[Generalized Friedmann equations]
    \begin{align}
        \left(\frac{\dot{a}}{a}\right)^{2}\equiv H^{2} & = \frac{\kappa^{2}}{3}\left[\rho_{r}^{(0)}a^{-4}+\rho_{m}^{(0)}a^{-3}+\rho_{\phi}^{(0)}a^{-3\left(1+\omega_{\phi}\right)}\right]+\frac{1}{3}\sigma_{(0)}^{2}a^{-6}-k\,a^{-2} \,,\label{I-FE-k-generalized} \\
        \frac{\ddot{a}}{a}\equiv\dot{H}+H^{2} & = -\frac{\kappa^{2}}{6}\left[2\rho_{r}^{(0)}a^{-4}+\rho_{m}^{(0)}a^{-3}+\left(1+3\omega_{\phi}\right)\rho_{\phi}^{(0)}a^{-3\left(1+\omega_{\phi}\right)}\right]-\frac{2}{3}\sigma_{(0)}^{2}a^{-6} \,, \label{II-FE-k-generalized}
    \end{align}
\end{examplebox}
where:
\begin{equation}\label{EoS-parameters-ekp}
    \omega_{r}=\frac{1}{3} \quad\land\quad \omega_{m}=0 \quad\land\quad \omega_{\phi}\equiv\frac{p_{\phi}}{\rho_{\phi}}=\frac{\frac{1}{2}\dot{\phi}^{2}-V(\phi)}{\frac{1}{2}\dot{\phi}^{2}+V(\phi)}\Bigg|_{\frac{1}{2}\dot{\phi}^{2}\ll V(\phi)}\simeq -1\,,
\end{equation}
and $\sigma$ denotes the energy density of shear anisotropy. Combining \eqref{I-FE-k-generalized} with \eqref{II-FE-k-generalized} yields the \textbf{generalized Raychaudhuri equation}:
\begin{examplebox}[Generalized Raychaudhuri equation]
    \begin{equation}
        \dot{H}=-\frac{\kappa^{2}}{2}\left[\frac{4}{3}\rho_{r}^{(0)}a^{-4}+\rho_{m}^{(0)}a^{-3}+\left(1+\omega_{\phi}\right)\rho_{\phi}^{(0)}a^{-3\left(1+\omega_{\phi}\right)}\right]-\sigma_{(0)}^{2}a^{-6}+k\,a^{-2}\,.
    \end{equation}
\end{examplebox}

\subsection{Negative SF potential}
In the framework of STTs the ekpyrotic model is equipped with \textit{\textbf{very steep negative exponential SF self-interaction potential}} \cite{Khoury:2001wf,Steinhardt:2001st}:
\begin{examplebox}[Ekpyrotic SF potential]
    \begin{equation}\label{ekpyrotic-SF-potential}
        \boxed{V(\phi)=-V_{0}\,e^{-c \phi}}\,.
    \end{equation}
\end{examplebox}
The \textbf{justification for the negative potential}\footnote{For further details on negative SF potentials in cosmology, see \cite{Felder:2002jk,Heard:2002dr}.} \textbf{of the SF} is the following:
\begin{enumerate}[label=(\alph*)]
    \item Achieving a \textit{\textbf{turnaround}} (\textit{non-singular transition from expansion to contraction}) implies the condition:
    \begin{equation}
        H=0\,,
    \end{equation}
    so that the negative $V(\phi)$ may cancel the positive kinetic energy of matter content of the Universe;
    \item In order to \textit{\textbf{dominate over the anisotropy/shear}}:
    \begin{equation}
        \rho_{\sigma}\propto\sigma^{2}\,a^{-6}
    \end{equation}
    from \eqref{EoS-parameters-ekp} one can observe that the requirement of:
    \begin{equation}
        \omega>1
    \end{equation}
    should be satisfied.
\end{enumerate}

\subsection{Multi-field model}
In the case of a spatially flat FLRW Universe with multiple SFs, the Einstein-SF equations take the following form:
\begin{align}
    H^{2} & = \frac{\kappa^{2}}{3}\left[\frac{1}{2}\sum_{i}\dot{\phi}_{i}^{2}+\sum_{i}V_{i}\left(\phi_{i}\right)\right]\,, \label{I-FE-multifield} \\
    \dot{H}+H^{2} & = \frac{\kappa^{2}}{3}\left[\sum_{i}V_{i}\left(\phi_{i}\right)-\sum_{i}\dot{\phi}_{i}^{2}\right]\,, \label{II-FE-multifield} \\
    \ddot{\phi}_{i} & = -3H\dot{\phi}_{i}-V'_{i}\left(\phi_{i}\right)\,, \label{KG-multifield}
\end{align}
where:
\begin{equation}
    V'_{i}\left(\phi_{i}\right)\equiv\frac{d V_{i}\left(\phi_{i}\right)}{d \phi_{i}}\,,
\end{equation}
and the Raychaudhuri equation becomes:
\begin{equation}\label{Raychaudhuri-multifield}
    \dot{H}=-\frac{\kappa^{2}}{2}\sum_{i}\dot{\phi}_{i}^{2}\,.
\end{equation}
Given that all of the SFs have negative exponential self-interaction potentials of the form:
\begin{equation}
    V_{i}\left(\phi_{i}\right)=-V_{i}\,e^{-c_{i}\phi_{i}} \quad\land\quad c_{i}\gg 1\,,
\end{equation}
then, the Einstein-scalar EoM satisfy the scaling solution of the form\footnote{We assume that the ekpyrotic phase (slow contraction) occurs for negative values of cosmological time ($t<0$) and that the bounce takes place at $t=0$.} \cite{Heard:2002dr,Guo:2003eu,Koyama:2007mg,Koyama:2007ag,Lehners:2008vx}:
\begin{examplebox}[Ekpyrotic multi-field scaling solution]
    \begin{equation}\label{scaling-ekpyrotic}
        \boxed{a(t)=(-t)^{p} \quad\land\quad \phi_{i}=\frac{2}{c_{i}}\ln{\left(-\sqrt{\frac{c_{i}^{2}\,V_{i}}{2}}t\right)} \quad\land\quad p\equiv\sum_{i}\frac{2}{c_{i}^{2}}}\,.
    \end{equation}
\end{examplebox}
Consequently, we obtain a \textit{\textbf{very slowly contracting Universe with a constant EoS}}:
\begin{examplebox}[Ekpyrotic EoS]
    \begin{equation}\label{ekpyrotic-EoS}
        \boxed{\omega=\frac{2}{3p}-1\gg 1}\,.
    \end{equation}
    $\implies$ The additional term with $\omega\gg 1$ comes to dominate the cosmological evolution over the curvature and shear/anisotropy contributions:
    \begin{equation}
        \Omega_{k}\equiv -\frac{k}{a^{2}H^{2}}\propto\left(a H\right)^{-2} \quad\land\quad \Omega_{\sigma}\equiv\frac{\sigma^{2}}{3a^{6}H^{2}}\propto a^{-6}H^{-2}\,.
    \end{equation}
    $\implies$ If we neglect quantum effects, the \textit{\textbf{Universe becomes exponentially flat and isotropic}} as it approaches the \textit{\textbf{Big Crunch}} \cite{Tolman1934book,Dicus1983,Davies1997LastThreeMinutes,Cattoen:2005dx}.
\end{examplebox}

\subsection{Ekpyrosis as a solution for the flatness problem}
There are \textbf{two main conditions} that must be met to \textbf{resolve the flatness problem}:
\begin{enumerate}[label=(\alph*)]
    \item \textbf{Ekpyrotic EoS}:
    \begin{equation}
        \omega\gg 1
    \end{equation}
    requires a \textbf{sufficiently steep negative SF potential}. In a \textit{single-field scenario} \eqref{ekpyrotic-EoS} and \eqref{scaling-ekpyrotic} imply the \textbf{condition on the free parameter} of the form:
    \begin{equation}
        \omega=\frac{2}{3p}-1=\frac{c^{2}}{3}-1\gg 1 \quad\implies\quad \boxed{c \gg\sqrt{6}}\,;
    \end{equation}
    \item Scaling solution \eqref{scaling-ekpyrotic}, which during ekpyrosis satisfies:
    \begin{equation}
        a(t)\simeq\mathrm{const} \quad\land\quad H(t)=\frac{p}{t}\propto t^{-1}\,,
    \end{equation}
    and together with \eqref{minus-Omega-k} and \eqref{Omega-minus-one-Planck} yields that the \textbf{comoving Hubble scale} $\boldsymbol{\left|aH\right|}$ ($H<0$ during ekpyrotic phase) \textbf{must grow by at least 60} $\boldsymbol{e}$\textbf{-folds in magnitude}:
    \begin{equation}\label{60-efolds}
        \boxed{\ln{\left[\frac{\left|a H\right|_{\mathrm{end}}}{\left|a H\right|_{\mathrm{beg}}}\right]}\gtrsim 60 \quad\implies\quad \left|t_{\mathrm{beg}}\right|\gtrsim e^{60}\left|t_{\mathrm{end}}\right|}\,.
    \end{equation}
\end{enumerate}
Moreover, as we will show in Sec.~\ref{Sec:perturbations} the following condition is also necessary in order to explain the \textit{observed amplitude of cosmological perturbations}:
\begin{equation}
    t_{\mathrm{end}}\simeq -10^{3}\MPl^{-1}\,,
\end{equation}
so that the \textit{\textbf{minimal duration of the ekpyrotic phase}} is given by \cite{Lehners:2008vx}:
\begin{examplebox}[Minimum duration of the ekpyrotic phase]
    \begin{equation}
        \boxed{\left|t_{\mathrm{beg}}\right|\gtrsim 10^{30}\MPl^{-1}\simeq 10^{-13}\,\mathrm{s}}\,.
    \end{equation}
\end{examplebox}
\subsubsection{Cyclic scenario}
From the \textit{cyclic model perspective}, the SF potential $V(\phi)$ should \textit{interpolate between the GUT and DE scale}, therefore from the scaling solution \eqref{scaling-ekpyrotic} we get:
\begin{equation}
    V\bigl[\phi(t)\bigr]=-\frac{2}{c^{2}\,t^{2}}\propto -t^{-2} \quad\land\quad \boxed{\left|t_{\mathrm{beg}}\right|=\sqrt{\frac{\left|V_{\mathrm{end}}\right|}{\left|V_{\mathrm{beg}}\right|}}\left|t_{\mathrm{end}}\right|\simeq\sqrt{10^{112}}\,10^{3}\MPl^{-1}\simeq 10^{16}\,\mathrm{s}\simeq 10^{8}\,\mathrm{yr}} \,.
\end{equation}
In such a case, the \textbf{ekpyrosis lasts for hundreds of millions of years}.

\subsubsection{Higher-dimensional perspective}

From a \textit{higher-dimensional perspective}, the \textit{\textbf{radion}} \cite{Binetruy:2001tc,Goldberger:1999uk,Goldberger:1999un,Csaki:1999mp,
Brax:2003fv,Brax:2004xh,Maartens:2010ar}, which \textit{determines the distance between the branes}, is one of the scalar fields. Moreover, as previously mentioned, $V(\phi)$ manifests an attractive force between the end-of-the-world branes \cite{Khoury:2001wf,Khoury:2001bz,Steinhardt:2001st,
Steinhardt:2004gk,Lehners:2008vx}. The crucial point is that the \textit{\textbf{brane scale factors do not diverge at the collision}} (see Sec.~\ref{Sec:Crunch}) and due to the \textit{inelastic nature of the collision}, \textit{\textbf{matter is produced at a finite temperature}} \cite{Steinhardt:2001st,Steinhardt:2004gk,Turok:2004gb,Takamizu:2004rq,
Lehners:2008vx}. It is also worth noting that \textbf{no topological defects will form if the temperature is below the GUT scale}. Therefore, the \textit{\textbf{slow contraction phase (ekpyrosis) seems to solve all basic cosmological issues}}.

\section{Towards the Big Crunch}\label{Sec:Crunch}

In this section, we will demonstrate that the phase of \textbf{ekpyrotic contraction} ($\omega>1$) \textbf{prevents BKL chaotic mixmaster scenario} \cite{Belinsky:1970ew} and \textbf{produces a nearly scale-invariant spectrum of scalar perturbations}. These perturbations then serve as seeds for the large scale structure (LSS) of the Universe.

\subsection{Avoiding chaos}
In the case of the contraction phase, chaos is the problem we are faced with. It was shown in \cite{Belinsky:1970ew} that if all matter components have an EoS satisfying:
\begin{equation}
    \omega<1\,,
\end{equation}
then, the contracting Universe becomes unstable when subjected to small perturbations. The \textbf{metric becomes highly anisotropic}, taking on a \textit{\textbf{Kasner form}} \cite{Kasner:1921zz}. That is, \textbf{all spatial dimensions except one shrink away}. However, the \textbf{metric repeatedly jumps from one Kasner form to another}. These \textit{BKL oscillations} are referred to as \textbf{\textit{chaotic mixmaster behavior}} \cite{Misner:1969hg,MisnerThorneWheeler1973MixmasterUniverse}.
\begin{examplebox}[Kasner model and mixmaster Universe]
    \begin{itemize}
        \item The \textit{\textbf{Kasner model}} for an anisotropic Universe \cite{Kasner:1921zz} is the \textbf{Bianchi type I}\footnote{Further details on the Bianchi classification can be found in \cite{Ellis:1968vb,Jantzen:1979Bianchi,Jantzen:2001me,Belinski_Henneaux_2017Bianchi,Plebanski_Krasinski_2024Banchi,Nastase_2025Bianchi}.} cosmological model:
        \begin{equation}
            ds^{2}=-dt^{2}+e^{2\alpha(t)}\left(e^{2\beta(t)}\right)_{ij}\,dx^{i}\,dx^{j}
        \end{equation}
        $\equiv$ The \textit{simplest anisotropic model with spatially flat $t=\mathrm{const}$ hypersurfaces with different expansion rates (scale factors) in three orthogonal directions}:
        \begin{equation}\label{Kasner-metric}
            \begin{aligned}
                ds^{2} & =-dt^{2}+t^{2p_{1}}\,dx^{2}+t^{2p_{2}}\,dy^{2}+t^{2p_{3}}\,dz^{2} \\
                & = -dt^{2}+a_{1}^{2}(t)\,dx^{2}+a_{2}^{2}(t)\,dy^{2}+a_{3}^{2}(t)\,dz^{2}\,,
            \end{aligned}
        \end{equation}
        where:
        \begin{equation}\label{Kasner-condition-1}
            p_{1}+p_{2}+p_{3}=p_{1}^{2}+p_{2}^{2}+p_{3}^{2}=1\,,
        \end{equation}
        and one of the $p_{i}$'s must be non-positive, for instance:
        \begin{equation}\label{Kasner-condition-2}
            -\frac{1}{3}\leq p_{1} \leq 0\,.
        \end{equation}
        In fact, one could observe that such a kind of model describes an \textit{expanding Universe}, since the volume element increases linearly:
        \begin{equation}
            \sqrt{-g}=\sqrt{g^{(3)}}=t\,;
        \end{equation}
        \item In the case of \textit{\textbf{generalized Kasner model}} \cite{MisnerThorneWheeler1973MixmasterUniverse} we allow for a \textit{more general time-dependence} while preserve some of the simplicity of the condition \eqref{Kasner-condition-1} on $p_{i}$'s. In fact, the exponents satisfy the relation, e.g.:
        \begin{equation}
            p_{2}\equiv\frac{d\ln{g_{22}}}{d\ln{g}}\,,
        \end{equation}
        so that the $3\times 3$ spatial metric can be parametrized as \cite{MisnerThorneWheeler1973MixmasterUniverse}:
        \begin{equation}
            g_{ij}=e^{2\alpha}\left(e^{2\beta}\right)_{ij} \quad\iff\quad \left(\ln{g}\right)_{ij}=2\alpha\,\delta_{ij}+2\beta_{ij}\,,
        \end{equation}
        where $\beta_{ij}$ is a traceless $3\times 3$ symmetric matrix, and the exponential is a matrix power series:
        \begin{equation}
            \det{\left(e^{2\beta}\right)}=1 \quad\land\quad \sqrt{g}=e^{3\alpha}\,.
        \end{equation}
        Moreover, using \eqref{Kasner-metric} and \eqref{Kasner-condition-1} one can define:
        \begin{equation}
            p_{ij}\equiv\frac{d\left(\ln{g}\right)_{ij}}{d\ln{\left[\det{\left(g\right)}\right]}}=\frac{1}{3}\left(\delta_{ij}+\frac{d\beta_{ij}}{d\alpha}\right)
        \end{equation}
        and the 1st Kasner condition becomes:
        \begin{equation}\label{generalized-Kasner-condition-1}
            1=\sum_{i}p_{i}\equiv\tr{\left(p_{ij}\right)}=1+\frac{1}{3}\tr{\left(\frac{d\beta}{d\alpha}\right)}\,.
        \end{equation}
        The relation \eqref{generalized-Kasner-condition-1} is an identity due to the fact that:
        \begin{equation}
            \tr{\left(\beta_{ij}\right)}=0\,.
        \end{equation}
        The 2nd Kasner condition in \eqref{Kasner-condition-1} takes the general form of:
        \begin{equation}\label{Kasner-condition-generalized-3}
            \tr{\left(p^{2}\right)}=1 \quad\implies\quad \left(\frac{d\beta_{ij}}{d\alpha}\right)^{2}=6\,.
        \end{equation}
        For the diagonal $\beta_{ij}$ we have only two independent components (since there is only one condition on the three dagonal components) that may be parametrized as \cite{MisnerThorneWheeler1973MixmasterUniverse}:
        \begin{equation}
            \beta_{11}=\beta_{+}+\sqrt{3}\,\beta_{-} \quad\land\quad \beta_{22}=\beta_{+}-\sqrt{3}\,\beta_{-} \quad\land\quad \beta_{33}=-2\beta_{+}
        \end{equation}
        and imply another form of the generalized 2nd Kasner condition \eqref{Kasner-condition-generalized-3}:
        \begin{equation}\label{Kasner-condition-generalized-4}
            \left(\frac{d\beta_{+}}{d\alpha}\right)^{2}+\left(\frac{d\beta_{-}}{d\alpha}\right)^{2}=1\,,
        \end{equation}
        where:
        \begin{equation}
            \frac{d\beta_{+}}{d\alpha}=\frac{1}{2}\bigl(1-3p_{3}\bigr) \quad\land\quad \frac{d\beta_{-}}{d\alpha}=\frac{1}{2}\sqrt{3}\bigl(p_{1}-p_{2}\bigr)\,.
        \end{equation}
        \item By introducing the \textbf{spatial curvature} (left-invariant one-forms on $SU(2)$):
        \begin{equation}
            \begin{dcases}
                \sigma^{1}=\cos{\psi}\,d\theta+\sin{\psi}\,\sin{\theta}\,d\phi \\
                \sigma^{2}=\sin{\psi}\,d\theta-\cos{\psi}\,\sin{\theta}\,d\phi \\
                \sigma^{3}=d\psi+\cos{\theta}\,d\phi
            \end{dcases}
        \end{equation}
        we obtain the model of \textit{\textbf{mixmaster Universe}} \cite{Misner:1969hg,MisnerThorneWheeler1973MixmasterUniverse} which is the \textbf{Bianchi type IX} model $\equiv$ \textit{Closed, homogeneous, anisotropic model with spatial topology related to} $S^{3}\simeq SU(2)$:
        \begin{equation}
            \boxed{ds^{2}=-N^{2}(t)dt^{2}+e^{2\alpha(t)}\left(e^{2\beta(t)}\right)_{ij}\,\sigma^{i}\sigma^{j}}\,,
        \end{equation}
        where:
        \begin{equation}
            \beta_{ij}=\diag{\left(\beta_{+}+\sqrt{3}\,\beta_{-},\beta_{+}-\sqrt{3}\,\beta_{-},-2\beta_{+}\right)} \quad\land\quad \sum_{k=1}^{3}\beta_{k}(t)=0\,,
        \end{equation}
        and:
        \begin{equation}
            \begin{dcases}
                a_{1}(t)=e^{\alpha+\beta_{+}+\sqrt{3}\,\beta_{-}} \\
                a_{2}(t)=e^{\alpha+\beta_{+}-\sqrt{3}\,\beta_{-}} \\
                a_{3}(t)=e^{\alpha-2\beta_{+}} 
            \end{dcases}
            \quad\land\quad
            \begin{dcases}
                \beta_{+}=\frac{\beta_{1}+\beta_{2}}{2}=-\frac{\beta_{3}}{2} \\
                \beta_{-}=\frac{\beta_{1}-\beta_{2}}{2\sqrt{3}}\,.
            \end{dcases}
        \end{equation}
        \textbf{\textit{Physical interpretation}}:
        \begin{enumerate}[label=(\alph*)]
            \item The term $\boldsymbol{e^{\alpha}}$ measures the \textbf{average scale factor};
            \item The terms $\boldsymbol{\beta_{+}}$ and $\boldsymbol{\beta_{-}}$ describe \textbf{anisotropic distortions}.
        \end{enumerate}
    \end{itemize}
\end{examplebox}
Now, we want to understand how the ekpyrotic phase might prevent chaos. This requires us to consider cosmological perturbations and, in particular, to examine the evolution of anisotropies. In the \textit{\textbf{synchronous gauge}}, the metric becomes of the form \cite{Kodama:1984ziu,Mukhanov:1990me,Malik:2008im}:
\begin{equation}
    ds^{2}=-dt^{2}+g_{ij}\left(x^{\mu}\right)\,dx^{i}\,dx^{j}\,.
\end{equation}
We also know that, during the contraction phase, spatial gradients quickly become irrelevant when compared to time gradients \cite{Belinsky:1970ew}, thus, one can consider the simplified form of the line element (of the Kasner form):
\begin{equation}\label{Kasner-ekpyrotic-metric}
    ds^{2}=-dt^{2}+a^{2}(t)\sum_{i}e^{2\beta_{i}(t)} \quad\land\quad \sum_{i}\beta_{i}(t)=0\,.
\end{equation}
This means that the dynamics is \textit{\textbf{ultralocal}} (\textit{near the crunch/singularity, the time evolution of each spatial point becomes effectively independent of neighboring points}) and can be described by the following EoM:
\begin{equation}
    \begin{dcases}
        3H^{2}=\frac{1}{2}\sum_{i}\dot{\beta}_{i}^{2}(t)+\ldots \\
        \ddot{\beta}_{i}(t)+3H\dot{\beta}_{i}(t)=0
    \end{dcases}
\end{equation}
for which the solution is a \textit{growing mode} satisfying the relation:
\begin{equation}
    \dot{\beta}_{i}(t)\propto a^{-3}(t) \quad\implies\quad 3H^{2}=\sigma_{(0)}^{2}\,a^{-6}+\ldots\,.
\end{equation}
Therefore, as stated previously, the anisotropies scale as a stiff matter fluid ($\omega=1$) and would quickly dominate the evolution of the Universe if not for the presence of the ekpyrotic SF, for which the energy density grows faster, namely:
\begin{equation}
    \rho_{\phi}\propto a^{-2/p}\,,
\end{equation}
so that:
\begin{equation}\label{ekpyrotic-growing-mode}
    \beta_{i}(t)\propto \left(-t\right)^{1-3p}\,.
\end{equation}
An important conclusion follows from equation \eqref{ekpyrotic-growing-mode}: the exponents in the metric \eqref{Kasner-ekpyrotic-metric} approach constant values in the ekpyrotic phase if:
\begin{equation}
    p<\frac{1}{3}\,.
\end{equation}
Therefore, there exists a \textit{\textbf{cosmic no-hair theorem for ekpyrosis}} \cite{Erickson:2003zm}:
\begin{examplebox}[No-hair theorem for ekpyrosis]
    An anisotropic and inhomogeneous contracting Universe converges to a homogeneous, flat, and isotropic Universe if it contains an energy density with the EoS parameter that satisfies the condition of:
    \begin{equation}
        \boxed{\omega>1}\,.
    \end{equation}
\end{examplebox}
The 1st Friedmann equation \eqref{I-FE-k-generalized} and the ratio \eqref{60-efolds} imply that during the slow contraction phase:
\begin{equation}
    \frac{\left(a^{6}H^{2}\right)_{\mathrm{beg}}}{\left(a^{6}H^{2}\right)_{\mathrm{end}}}\geq e^{120}\,.
\end{equation}
This means that the \textbf{relative importance of anisotropies drops rapidly} and \textit{\textbf{at the end of the ekpyrosis, the size of the pre-ekpyrotic anisotropies and spatial curvatures has decreased exponentially!}}

\subsubsection{Kinetic phase}
The ekpyrotic SF potential \eqref{ekpyrotic-SF-potential} turns off and becomes irrelevant near the Big Crunch, thus, the Universe enters a KE-dominated phase with the EoS \cite{Khoury:2003rt,Steinhardt:2004gk,Lehners:2008vx}:
\begin{equation}
    \omega=1\,.
\end{equation}
In such a case, the field equations take the form:
\begin{equation}
    3H^{2}=\frac{1}{2}\dot{\phi}^{2}=-\dot{H} \quad\land\quad \ddot{\phi}+3H\dot{\phi}=0
\end{equation}
and lead to the simple relation between a scale factor and the SF:
\begin{equation}
    a(\phi)\propto\exp{\left(\frac{\phi}{\sqrt{6}}\right)}
\end{equation}
with the explicit form of the solution:
\begin{equation}
    a(t)=a_{0}(-t)^{1/3} \quad\land\quad \phi(t)=\phi_{0}+\sqrt{\frac{2}{3}}\ln{\left(-t\right)}\,.
\end{equation}
As stated previously, we will show in Sec.~\ref{Sec:perturbations} that the ekpyrotic phase lasts until:
\begin{equation}
    t\sim 10^{3}\,t_{Pl}
\end{equation}
before the Big Crunch. Moreover, the \textbf{kinetic phase becomes relevant until the QG regime is reached} \cite{Khoury:2003rt,Steinhardt:2004gk,Lehners:2008vx}:
\begin{equation}
    t\sim -\MPl^{-1}
\end{equation}
and continues in the first epoch after the Big Bang. Furthermore, one can show that during kinetic phase:
\begin{equation}
    \dot{\phi}^{2}\propto a^{-6}\,,
\end{equation}
which means that the \textbf{SF energy density scales exactly with the anisotropies}. This suggests that the \textbf{relative importance of anisotropy remains constant during the KE-dominated epoch}. In addition, it was shown in \cite{Erickson:2003zm} that also the curvature terms are not dangerous during kinetic domination. Nevertheless, in the presence of $p$-forms (higher-rank antisymmetric gauge fields) \textbf{dangerous modes could grow} (as a power of cosmic time and not longer than about 7 $e$-folds) and they depend on their coupling to the SF \cite{Erickson:2003zm}. An important example is the \textbf{heterotic M-theory} \cite{Lukas:1998tt} in which critical coupling \cite{Erickson:2003zm} allows for a chaotic evolution towards the Big Crunch when there is sufficient energy density in the $p$-form modes. The conclusion is the following:
\begin{examplebox}[Conclusion]
    The \textit{\textbf{ekpyrotic phase could prevent the chaotic evolution into the Big Crunch}} because the \textit{energy density of potentially harmful modes is diluted so much during ekpyrosis that they do not have enough time to dominate before the Crunch occurs} \cite{Erickson:2003zm}. Additionally, there is a \textit{suppression by the topology of the internal manifold} \cite{Wesley:2005bd}. Thus, \textit{\textbf{chaos is prevented by delaying it until we enter the QG regime, where the field equations of GR break down}}.
\end{examplebox}

\subsection{Milne Universe}
\begin{figure}[htbp]
    \centering
    \includegraphics[width=1\linewidth]{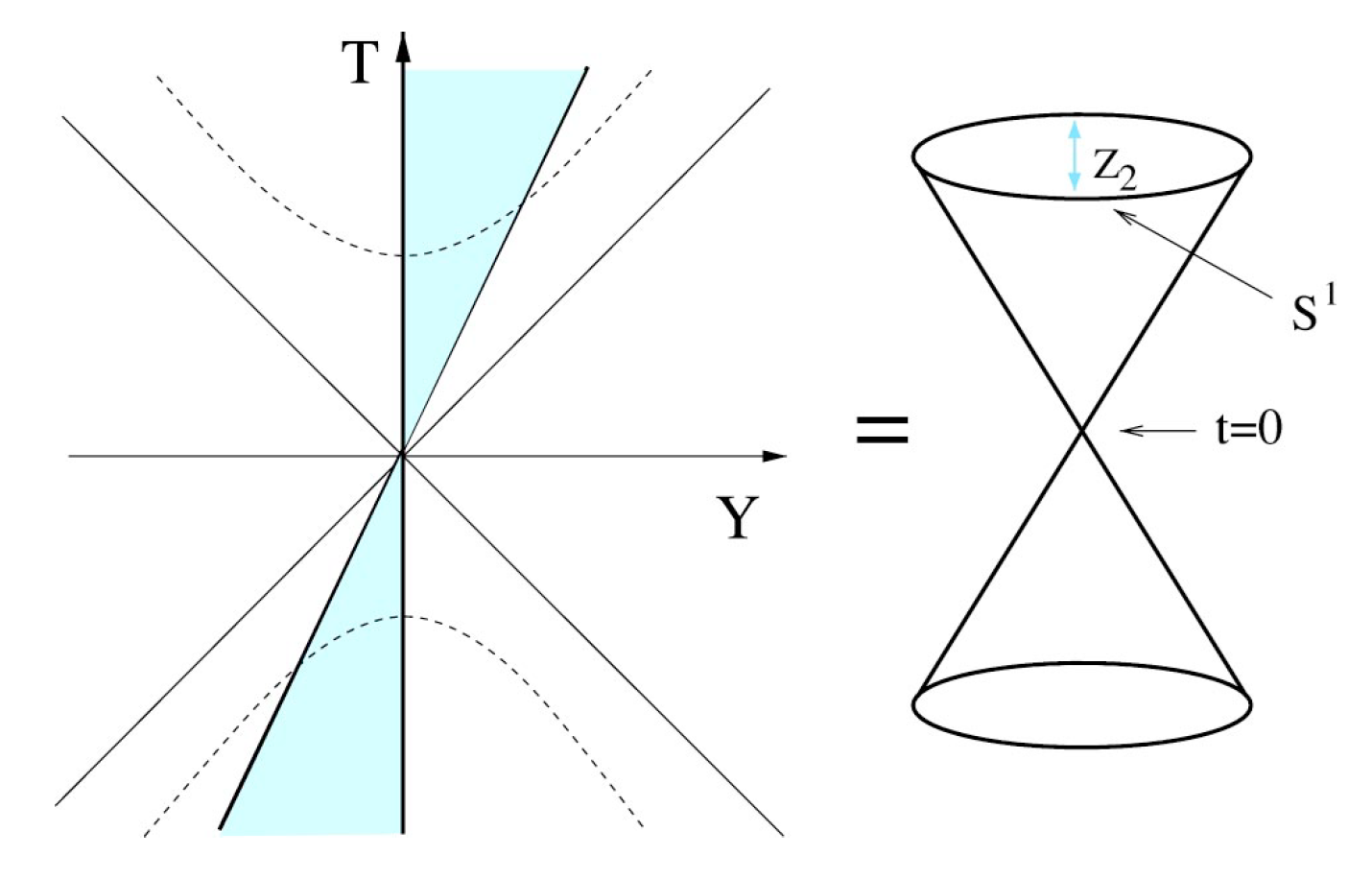}
    \caption{Embedding of the collision of two branes in Minkowski spacetime \cite{Tolley:2003nx}. In their notation $u\to T$, $v\to Y$, $T\to t$, and dashed lines denote $t=\mathrm{const}$ Lorentz-invariant coordinate.}
    \label{fig:brane-collision-Minkowski}
\end{figure}

In the $4D$ formalism, approaching the Big Crunch (KE-domination) indicates a \textit{\textbf{singularity}}:
\begin{examplebox}[Big Crunch singularity]
    The \textit{\textbf{Big Crunch singularity}}\footnote{More information on \textit{spacetime singularities} can be found in \cite{Penrose:1964wq,Hawking:1970zqf,Tipler:1978zz,Hawking1996nature,Joshi2014chapter,Berger2014chapter,Hawking_Ellis_2023Sing1,Hawking_Ellis_2023Sing2,Hawking_Ellis_2023Sing3,Penrose:1980ge,deHaro:2023lbq,Rendall2005}.} is a \textbf{future spacelike cosmological singularity reached after a phase of contraction}. In the FLRW formalism it corresponds to \cite{Cattoen:2005dx}:
    \begin{equation}
        a(t)\to 0 \quad\land\quad \left|H(t)\right|\to \infty
        \quad\land\quad \rho(t)\to \infty \quad\land\quad
        t\to t_{\mathrm{crunch}}^{-}\,.
    \end{equation}
\end{examplebox}
However, near $t=0$, the higher-dimensional ($5D$) formalism must be taken into account to fully describe cosmological evolution \cite{Khoury:2001wf,Lehners:2008vx}. Specifically, we must consider \textit{\textbf{gravity in}} $\boldsymbol{5D}$ \textit{\textbf{spacetime}}, where \textit{one spatial dimension is a line segment} (\textbf{Ho\v{r}ava-Witten theory} \cite{Horava:1995qa,Horava:1996ma}). The line segment can be interpreted as a circle that has undergone a $\mathbb{Z}_{2}$ reflection symmetry along an axis. In such a case, the metric takes the following form \cite{Khoury:2001bz}:
\begin{equation}\label{Horava-Witten-metric}
    ds_{5}^{2}=\exp{\left(-\sqrt{\frac{2}{3}}\phi\right)}\,ds_{4}^{2}+\exp{\left(2\sqrt{\frac{2}{3}}\phi\right)}\,dy^{2}\,,
\end{equation}
where:
\begin{equation}\label{orbifold-coordinate}
    y\in \left(-y_{0},y_{0}\right)
\end{equation}
is the \textbf{orbifold coordinate} and $\phi$ denotes the \textbf{radion field} (\textit{parametrizes the size of the orbifold}). The SF-dependent prefactor in front of $ds_{4}^{2}$ \textit{preserves the canonical SF presence} after dimensional reduction. Therefore, one can apply the scaling solution \eqref{scaling-ekpyrotic} inside the $5D$ metric \eqref{Horava-Witten-metric} and obtain:
\begin{equation}
    ds_{5}^{2}=(-t)^{-2/3}\left[-dt^{2}+(-t)^{2/3}\,dx_{3}^{2}\right]+(-t)^{4/3}\,dy^{2}\,.
\end{equation}
Introducing a new time coordinate \cite{Lehners:2008vx}:
\begin{equation}
    T=(-t)^{2/3}
\end{equation}
yields into the metric of \textit{\textbf{'compactified Milne mod}} $\boldsymbol{\mathbb{Z}_{2}}$' $\boldsymbol{\times\,\mathbb{R}_{3}}$ \textit{\textbf{Universe}} \cite{Tolley:2003nx}\footnote{The Milne model corresponds to the empty open FLRW Universe ($k=-1$, $\rho=0$, $\Lambda=0$, $a(t)\propto t$). It is equivalent to a patch of Minkowski spacetime written in hyperbolic coordinates \cite{Milne1935,Mukhanov_2005Milne}.}:
\begin{examplebox}[Compactified Milne mod $\mathbb{Z}_{2}$ $\times\,\mathbb{R}_{3}$ metric]
    \begin{equation}
        ds_{5}^{2}=-dT^{2}+T^{2}\,dy^{2}+dx_{3}^{2}\,.
    \end{equation}
    $\equiv$ \textbf{Two orbifold planes in the form of} $\boldsymbol{(3+1)D}$ \textbf{boundaries of the line segment that approach each other, collide, and move away from each other.}
\end{examplebox}
Notably, another change of coordinates:
\begin{equation}
    u=T\cosh{\left(y\right)} \quad\land\quad v=T\sinh{\left(y\right)}\,,
\end{equation}
resulted in \textbf{Minkowski space becoming an embedded spacetime}:
\begin{equation}
    ds_{5}^{2}=-du^{2}+dv^{2}+dx_{3}^{2}\,.
\end{equation}
This means that \textit{\textbf{spacetime is flat except at the moment of collision, providing a solution to any gravitational theory without CC, and higher derivative corrections become small for a Milne-like spacetime.}}
\begin{examplebox}[Conclusion]
    \textbf{The simplest model for the collision of branes is the compactified Milne space.}
\end{examplebox}
Fig.~\ref{fig:brane-collision-Minkowski} demonstrates that, locally, the collision of two branes can be embedded in Minkowski spacetime. It requires a two-step procedure \cite{Tolley:2003nx}:
\begin{enumerate}[label=(\roman*)]
    \item Compactification of the $y$ coordinate by identifying:
    \begin{equation}
        y\to y+2y_{0}
    \end{equation}
    in order to produce the double-conical spacetime (on the right);
    \item Orbifolding of the circular sections of the cones by the $\mathbb{Z}_{2}$ symmetry:
    \begin{equation}
        y\to 2y_{0}-y\,,
    \end{equation}
    so that the two fixed points of the $\mathbb{Z}_{2}$ symmetry become two tensionless branes (on the left) that move at a relative velocity of:
    \begin{equation}
        v_{\mathrm{rel}}=\tanh{\left(y_{0}\right)}
    \end{equation}
    which collide and pass through one another (at the moment of collision).
\end{enumerate}
\begin{examplebox}[Scale factor behavior during the Big Crunch/collision]
    \begin{itemize}
        \item The Big Crunch singularity, as seen from the $5D$ perspective, is the \textit{\textbf{temporary contraction of the orbifold dimension}} (\textit{with other dimensions being finite});
        \item Using \eqref{Horava-Witten-metric} with \eqref{scaling-ekpyrotic} gives the value of the $4D$ scale factor at the collision:
        \begin{equation}
            a\to 0\,;
        \end{equation}
        \item However, from the metric \eqref{Horava-Witten-metric} one can deduce that the brane scale factors are actually given by:
        \begin{equation}
        \begin{aligned}
            a_{5} & =a\,\exp{\left(-\frac{1}{2}\sqrt{\frac{2}{3}}\phi\right)}=a\,\exp{\left(-\frac{\phi}{\sqrt{6}}\right)}=a_{0}\left(-t\right)^{1/3}\,\exp{\left[-\left(\frac{\phi_{0}+\sqrt{\frac{2}{3}}\ln{(-t)}}{\sqrt{6}}\right)\right]} \\
            & = a_{0}\exp{\left(-\frac{\phi_{0}}{\sqrt{6}}\right)}\left(-t\right)^{1/3}\left(-t\right)^{-1/3}=a_{0}\exp{\left(-\frac{\phi_{0}}{\sqrt{6}}\right)}=\mathrm{const}\,.
        \end{aligned}
        \end{equation}
    \end{itemize}
    \underline{\textit{\textbf{Conclusion:}}}\\
    \textbf{The matter density and temperature on the branes both remain finite at the collision!}
\end{examplebox}
The above considerations reveal the general idea behind the ekpyrotic model:
\begin{examplebox}[General idea of the ekpyrotic model]
    The \textbf{ekpyrotic phase produces an exponentially flat and isotropic Universe} (in $4D$ spacetime). Meanwhile, the \textbf{kinetic phase and the transition from the Big Crunch to the Big Bang could be described by the compactified Milne Universe} (in $5D$ spacetime)\footnote{More details concerning the Milne spacetime in the vicinity of branes collision can be found in \cite{Turok:2004gb,Niz:2006ef}.}.
\end{examplebox}
\begin{examplebox}[Remark]
    Since one would need \textit{\textbf{QG to fully describe the transition}}, we could only \textit{\textbf{effectively describe the spacetime up to the Planck time before and after the collision}}.
\end{examplebox}
\subsubsection{Matching conditions}
At this point in our analysis, we are confronted with the issue of how the Big Crunch can be matched to the Big Bang:
\begin{examplebox}['Matching conditions']
    In the literature, there are two (different) main approaches to the \textit{\textbf{'matching conditions'}} (\textit{between the Big Crunch and the Big Bang}) \cite{Tolley:2003nx,Brandenberger:2001bs,Hwang:2001zt,Lyth:2001pf,Martin:2001ue,Durrer:2002jn,Tolley:2002cv,Battefeld:2004mn,Creminelli:2004jg,Martin:2004pm}:
    \begin{enumerate}[label=(\alph*)]
        \item \textbf{\textit{The spacetime metric could remain unchanged after a collision}} \cite{Creminelli:2004jg} - based on the following justification:
        \begin{itemize}
            \item Due to ekpyrosis, the Universe is isotropic (up to exponentially small terms);
            \item Quantum effects create small anisotropies in the metric (see Sec.~\ref{Sec:perturbations} for more details);
            \item In the synchronous gauge:
            \begin{equation}
                g_{0\mu}=\eta_{0\mu}\,,
            \end{equation}
            so that the metric becomes of the form \cite{Lehners:2008vx}:
            \begin{equation}
                ds^{2}=-dt^{2}+a^{2}(t)\,e^{2\xi\left(x^{\mu}\right)}\,e^{2h_{ij}\left(x^{\mu}\right)}\,dx^{i}\,dx^{j}\,,
            \end{equation}
            where $\xi\left(x^{\mu}\right)$ denotes \textit{scalar curvature perturbations} and $h_{ij}\left(x^{\mu}\right)$ indicates \textit{tensor perturbations};
        \end{itemize}
        $\implies$ Locally, the \textit{small long-wavelength perturbations} could be \textbf{gauged away}. This could be done by \textit{local rescaling of the spatial coordinates} (locally, these fluctuations are integration constants);\\
        $\implies$ The \textbf{Universe has the same history at every point in space}, \textit{up to exponentially small terms};\\
        $\implies$ The \textbf{metric would reemerge from the transition with the same perturbations} (\textit{the same applies not only to linear perturbations, but also to the full nonlinear metric});
        \item \textbf{\textit{Dynamics of the transition is important}} \cite{Tolley:2002cv,Tolley:2003nx} - based on the following reasoning:
        \begin{itemize}
            \item By considering free fields in a compactified Milne spacetime, we can analytically continue the evolution of the fields around the singularity (matching perturbations on surfaces of constant energy density \cite{Khoury:2001zk,Turok:2004yx});
        \end{itemize}
        $\implies$ \textbf{Mixing of different perturbations modes};\\
        $\implies$ The \textbf{history of the Universe relies significantly on exponentially small perturbation terms};
        \begin{itemize}
            \item Such an approach \textbf{breaks down if one takes interactions into account};
            \item \textit{Mixing of modes may occur when the 4D effective formalism is no longer applicable} (near the collision) \cite{McFadden:2005mq};
        \end{itemize}
    \end{enumerate}
\end{examplebox}
Regarding matching conditions, we will use the first approach for the purposes of this study.

\subsection{New ekpyrotic scenarios}
\begin{figure}[htbp]
    \centering
    \includegraphics[width=1\linewidth]{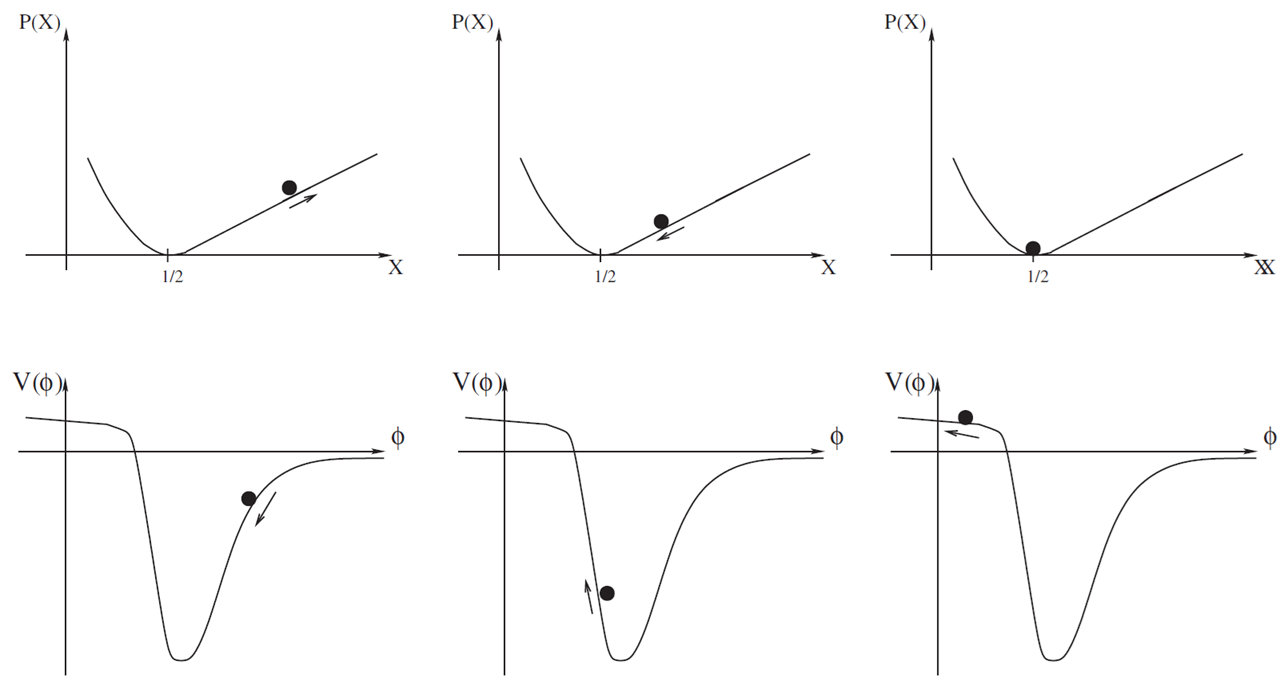}
    \caption{Dynamical evolution of the SF in the new ekpyrotic cosmology \cite{Buchbinder:2007ad}.}
    \label{fig:dynamical-evolution-new-ekpyrotic}
\end{figure}

\begin{figure}[htbp]
    \centering
    \includegraphics[width=1\linewidth]{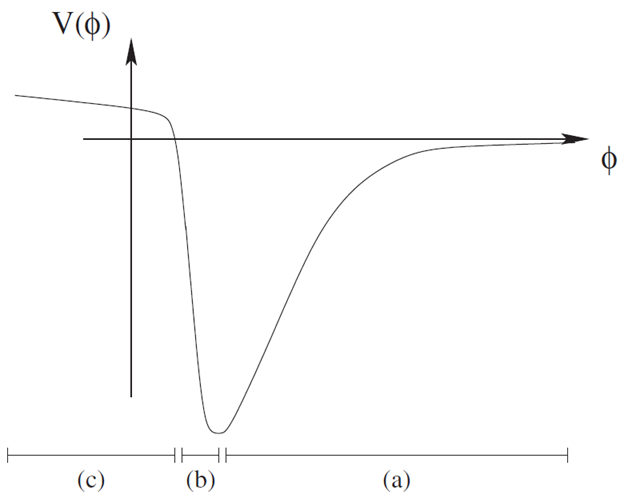}
    \caption{Generic shape of the ekpyrotic SF potential \cite{Buchbinder:2007ad}.}
    \label{fig:generic-ekpyrotic-potential}
\end{figure}

In the \textit{\textbf{new ekpyrotic models}} \cite{Buchbinder:2007ad,Buchbinder:2007tw,Creminelli:2007aq}, the contracting and expanding stages are connected by a \textit{\textbf{smooth, non-singular bounce}}. This bounce can be fully described by a $4D$ EFT, in which the \textit{\textbf{energy does not reach the Planck scale}}, and \textit{\textbf{cosmological perturbations can evolve unambiguously through the bounce}}. Nevertheless, a \textbf{smooth cosmological bounce would require the violation of one or more of the assumptions of the Penrose-Hawking singularity theorems}:
\begin{examplebox}[Penrose–Hawking singularity theorems]
    \begin{itemize}
        \item The classical Hawking-Penrose singularity theorems \cite{Penrose:1964wq,Hawking:1970zqf,Hawking1966a,Hawking1966b,Hawking1967a}\footnote{For a review, see also \cite{Senovilla:2014gza,Landsman:2022hrn}.} demonstrate that \textit{\textbf{spacetime singularities are not just a result of exact FLRW symmetry, rather, they are a general consequence of classical GR under certain geometric and physical conditions}};
        \item The main assumptions can be summarized as follows:
        \begin{enumerate}[label=(\roman*)]
            \item \textbf{Einstein gravity}:
            The spacetime $(\mathcal{M},g_{\mu\nu})$ is described by the classical Einstein field equations:
            \begin{equation}
                G_{\mu\nu}=\kappa^{2}\,T_{\mu\nu}\,;
            \end{equation}
            \item \textbf{Causal regularity}:
            The spacetime satisfies an appropriate causality condition, such as the absence of closed timelike curves or global hyperbolicity. Alternatively, it satisfies a related global causal assumption;
            \item \textbf{Energy/convergence condition}:
            Due to the Einstein equations, the matter satisfies an energy condition, which implies the focusing of geodesic congruences. For timelike geodesics, this condition is expressed as a timelike convergence:
            \begin{equation}
                R_{\mu\nu}\,u^{\mu}\,u^{\nu}\geq 0\,,
            \end{equation}
            for every timelike vector $u^{\mu}$. Moreover, for null geodesics one can use the null convergence condition of the form:
            \begin{equation}
                R_{\mu\nu}k^{\mu}k^{\nu}\geq 0\,,
            \end{equation}
            for every null vector $k^{\mu}$. In the case of Einstein gravity these are related to the SEC:
            \begin{equation}
                \rho+p \geq 0 \quad\land\quad \rho+3p \geq 0
            \end{equation}
            and NEC:
            \begin{equation}
                \rho+p \geq 0\,,
            \end{equation}
            respectively;
            \item \textbf{Generic curvature condition}:
            The spacetime has nonzero tidal curvature along the relevant causal geodesics. Therefore, the focusing predicted by the Raychaudhuri equation cannot be avoided through exact degeneracy;
            \item \textbf{Initial/trapping condition}:
            The convergence of geodesics is forced by a certain condition. In the case of gravitational collapse, this condition is usually the existence of a closed trapped surface. For cosmological cases, this condition can be replaced by a global expansion or contraction condition on a spacelike hypersurface;
        \end{enumerate}
        \item Based on these assumptions, the \textbf{Raychaudhuri equation indicates that initially converging timelike or null geodesic congruences will converge in a finite amount of proper time or affine parameter}. Therefore, \textit{\textbf{spacetime cannot be extended indefinitely along all causal geodesics}};
        \item Thus, in the \textit{precise geometric sense}, a \textit{\textbf{spacetime singularity indicates geodesic incompleteness, not necessarily the divergence of a specific curvature scalar}}. In cosmology, theorems imply that a \textbf{classical expanding Universe that satisfies the above assumptions is typically incomplete in the past}, whereas a \textbf{classical contracting Universe that satisfies the corresponding time-reversed assumptions is typically incomplete in the future, leading to a Big Crunch-type singularity};
    \end{itemize}
    \underline{\textit{\textbf{Conclusion:}}}\\
    For a \textit{\textbf{nonsingular bounce or regular transition through a Crunch to occur, at least one of the theorems' assumptions must fail}}. This can happen through a \textbf{violation of the relevant energy condition}, \textbf{QG effects}, \textbf{modified gravity}, \textbf{extra-dimensional dynamics}, or a \textbf{change in the global causal structure}.
\end{examplebox}
Additionally, the Hubble parameter must satisfy the following condition for the non-singular bounce (transition from contraction to expansion - minimum of the scale factor):
\begin{examplebox}[Condition for a (non-singular) cosmological bounce]
    \begin{equation}\label{bounce-condition-H}
        \boxed{a(t_{B})=a_{B}>0 \quad\land\quad H(t_{B})=0 \quad\land\quad \dot{H}(t_{B})>0}\,.
    \end{equation}
\end{examplebox}
The new ekpyrotic scenarios employ a mechanism known as a \textit{\textbf{ghost condensate}} \cite{Arkani-Hamed:2003pdi}\footnote{For critical comments on this formalism, see \cite{Kallosh:2007ad}.}:
\begin{examplebox}[Ghost condensate]
    \begin{itemize}
        \item The \textbf{effective Lagrangian for a ghost condensate} model is of the form \cite{Arkani-Hamed:2003pdi}:
        \begin{equation}\label{ghost-condensate}
            \boxed{\mathcal{L}=\sqrt{-g}\,M^{4} P(X) \quad\land\quad X\equiv -\frac{1}{2m^{4}}\left(\partial\phi\right)^{2}=-\frac{1}{2m^{4}}g^{\mu\nu}\partial_{\mu}\phi\,\partial_{\nu}\phi}\,,
        \end{equation}
        where $P(X)$ is a function to be specified as well as $m$ and $M$. The later are two mass scales that must be specified by the fundamental microscopic theory. As we will see, they must also satisfy certain consistency conditions;
        \item This type of theory allows for a constant shift symmetry \cite{Arkani-Hamed:2003pdi}:
        \begin{equation}
            \phi\to \phi+\mathrm{const}\,.
        \end{equation}
    \end{itemize}
\end{examplebox}
to \textbf{violate the NEC}:
\begin{examplebox}[Violation of the NEC]
    \begin{equation}
        \boxed{\left(\rho+p<0 \iff \omega\equiv\frac{p}{\rho}<-1\right) \quad\implies\quad \dot{H}=-\frac{\kappa^{2}}{2}\left(\rho+p\right)=-\frac{\kappa^{2}}{2}\rho\left(1+\omega\right)>0}\,.
    \end{equation}
\end{examplebox}
Note, however, that it is not yet clear what kind of theories may realize such a formalism \cite{Adams:2006sv,Kallosh:2007ad}. Taking into account a homogeneous SF:
\begin{equation}
    \phi=\phi(t) \quad\implies\quad \left(\partial\phi\right)^{2}=-\dot{\phi}^{2}
\end{equation}
one obtains that:
\begin{equation}
    X=\frac{\dot{\phi}^{2}}{2m^{4}}\,.
\end{equation}
Varying the action:
\begin{equation}
    S=\int d^{4}x\,\sqrt{-g}\,M^{4} P(X)
\end{equation}
with respect to the SF:
\begin{equation}
    \delta P(X)=P'(X)\,\delta X \quad\land\quad \delta X=-\frac{1}{m^{4}}g^{\mu\nu}\partial_{\mu}\phi\,\partial_{\nu}\delta\phi=-\frac{1}{m^{4}}\partial^{\mu}\phi\,\partial_{\mu}\delta\phi
\end{equation}
yields:
\begin{equation}
    \delta S=-\frac{M^{4}}{m^{4}}\int d^{4}x\,\sqrt{-g}\,P'(X)\,\partial^{\mu}\phi\,\partial_{\mu}\delta\phi=\frac{M^{4}}{m^{4}}\int d^{4}x\,\sqrt{-g}\,\nabla_{\mu}\bigl[P'(X)\,\partial^{\mu}\phi\bigr]\delta\phi\,.
\end{equation}
Therefore, the \textbf{EoM for a ghost condensate} takes the form of:
\begin{examplebox}[EoM for a ghost condensate]
    \begin{equation}\label{ghost-condensate-EoM}
        \nabla_{\mu}\bigl[P'(X)\,\partial^{\mu}\phi\bigr]=0 \quad\implies\quad \boxed{\frac{d}{dt}\Bigl[a^{3}P'(X)\,\dot{\phi}\Bigr]=0}\,.
    \end{equation}
\end{examplebox}
If one supposes that the function $P(X)$ has an extremum at:
\begin{equation}
    X=X_{0} \quad\land\quad P'\left(X_{0}\right)=0\,,
\end{equation}
the EoM \eqref{ghost-condensate-EoM} is automatically solved:
\begin{equation}
    a^{3}P'(X)\,\dot{\phi}=0\,.
\end{equation}
It is convenient to use the value of:
\begin{equation}
    X_{0}=\frac{1}{2}
\end{equation}
in order to cancel out the factors of $1/2$:
\begin{equation}
    X=\frac{\dot{\phi}^{2}}{2m^{4}}=\frac{1}{2} \quad\implies\quad \dot{\phi}^{2}=m^{4}\,,
\end{equation}
so that:
\begin{equation}\label{dot-phi-ghost}
    \dot{\phi}=\pm m^{2}\,.
\end{equation}
Choosing the negative branch of \eqref{dot-phi-ghost} gives:
\begin{equation}
    \dot{\phi}=-m^{2} \quad\implies\quad \phi(t)=-m^{2}\,t+\phi_{0}\,.
\end{equation}
Moreover, because of the shift symmetry one may set:
\begin{equation}
    \phi_{0}=0\,,
\end{equation}
therefore, the time-dependent expectation value\footnote{The background SF takes the form that spontaneously selects a preferred time direction while preserving the shift symmetry up to time translations.} (EV) takes the form:
\begin{equation}
    \langle\phi\rangle=\phi(t)=-m^{2}\,t\,.
\end{equation}
The \textbf{energy-momentum tensor for a ghost condensate} \cite{Lehners:2008vx}:
\begin{equation}
    T_{\mu\nu}=M^{4} P(X)\,g_{\mu\nu}+\left(\frac{M}{m}\right)^{4}P'(X)\,\partial_{\mu}\phi\,\partial_{\nu}\phi\,,
\end{equation}
where:
\begin{equation}\label{ghost-rho-p}
    \rho=M^{4}\bigl[2P'(X)\,X-P(X)\bigr] \quad\land\quad p=M^{4}P(X)\,,
\end{equation}
indicates that \textit{ghost condensate itself does not yet violate the NEC}, \textbf{at the extremum \textit{it has the same EoS as CC}}:
\begin{equation}
    \omega\equiv\frac{p}{\rho}=\frac{M^{4}P(X)}{-M^{4}P(X)}=-1\,.
\end{equation}
\begin{examplebox}[Remark]
    The \textit{\textbf{NEC might be violated by fluctuations around the extremum}} \cite{Lehners:2008vx,Arkani-Hamed:2003pdi,Buchbinder:2007ad}:
    \begin{equation}
        \phi=-m^{2}\,t+\pi\left(t,\vec{x}\right) \quad\land\quad \mathcal{L}_{\mathrm{eff}}\propto 2X_{0}\,P''\left(X_{0}\right)\dot{\pi}^{2}\,.
    \end{equation}
    $\implies$ \textbf{Correct sign of KE for the extremum being a minimum (no ghosts around the ghost condensate).}
\end{examplebox}
Expanding the energy density, $\rho$, in \eqref{ghost-rho-p} up to the 1st order in fluctuations gives:
\begin{equation}
    \rho\simeq -K M^{4}\frac{\dot{\pi}}{m^{2}} \quad\land\quad K\equiv P''\left(X_{0}\right)>0\,.
\end{equation}
Such fluctuations could arise due to the presence of the SF potential:
\begin{equation}
    \rho\simeq -K \left(\frac{M}{m}\right)^{2}\dot{\pi}+V(\phi) \quad\land\quad p\simeq -V(\phi)\,.
\end{equation}
Now, it is crucial to observe that the Friedmann equation:
\begin{equation}
    \dot{H}=-\frac{\kappa^{2}}{2}\bigl(\rho+p\bigr)
\end{equation}
immediately implies:
\begin{equation}\label{dotH-dotpi}
    \boxed{\dot{H}\simeq\frac{\kappa^{2}}{2}K\frac{M^{4}}{m^{2}}\dot{\pi}} \quad\implies\quad \pi(t)\simeq \frac{2}{\kappa^{2}}\frac{m^{2}}{K M^{4}}H(t)+\pi_{0}\,.
\end{equation}
\begin{examplebox}[Conclusion]
    \textit{\textbf{There is a possibility of the NEC violation}} for:
    \begin{equation}
        \boxed{\dot{\pi}>0 \quad\implies\quad \dot{H}>0}\,.
    \end{equation}
\end{examplebox}
However, a \textit{\textbf{bounce is still difficult to achieve}} because the modes that violate the NEC \textit{scale with a positive power of the scale factor} ($\omega<-1$ in \eqref{rho-a}). \textbf{They tend to become subdominant in a contracting phase and dominant in an expanding one} (\textit{\textbf{this is exactly what we do not want here!}}). The problem can be solved by assuming that \textit{\textbf{the same SF is responsible for both ekpyrosis and the cosmological bounce}}. During the slow contraction phase, the function $P(X)$ should be \textit{linear}, and near the minimum (\textit{\textbf{ghost phase}}) it should be \textit{quadratic} \cite{Buchbinder:2007ad}:
\begin{equation}
    P(X)=\frac{1}{2}K\bigl(X-X_{0}\bigr)^{2}\,.
\end{equation}
Furthermore, the potential $V(\phi)$ needs to be \textit{positive} at the time of the bounce to push the EoS parameter to values below $-1$ that violate the NEC \cite{Buchbinder:2007ad} (Fig.~\ref{fig:dynamical-evolution-new-ekpyrotic}). Following the cosmological bounce, the \textbf{NEC-violating fluid decays into matter and radiation, thus reheating the universe} \cite{Buchbinder:2007ad}.
Based on the above information, the \textit{\textbf{general behavior of the SF within the ekpyrotic potential can be summarized as follows}}:
\begin{examplebox}[Generic ekpyrotic SF potential]
    The \textit{\textbf{generic ekpyrotic SF potential}}:
    \begin{equation}
        V(\phi)=-V_{0}\,e^{-c\phi} \quad\land\quad c\gg 1
    \end{equation}
    consists of \textbf{three main regions} (Fig.~\ref{fig:generic-ekpyrotic-potential}) \cite{Buchbinder:2007ad}:
    \begin{enumerate}[label=(\alph*)]
        \item Region (a) - \textbf{\textit{steep and negative region}}:
        \begin{itemize}
            \item The \textbf{field rolls down the potential} and \textbf{admits the scaling solution} (attractor) \eqref{scaling-ekpyrotic};
            \item As the modes exit the horizon, there is a \textbf{generation of large-scale density fluctuations};
            \item As we will show in Sec.~\ref{Sec:perturbations}, one may obtain a \textit{\textbf{nearly scale-invariant spectrum}} if the \textit{\textbf{'fast-roll' conditions}} are satisfied \cite{Khoury:2003rt}:
            \begin{equation}\label{fast-roll-conditions}
                \boxed{\epsilon\ll 1 \quad\land\quad \left|\eta\right|\ll 1}\,,
            \end{equation}
            where:
            \begin{equation}\label{fast-roll-parameters}
                \boxed{\epsilon\equiv \MPl^{-2}\left(\frac{V(\phi)}{V'(\phi)}\right)^{2} \quad\land\quad \eta\equiv 1-\frac{V''(\phi)\,V(\phi)}{V'(\phi)^{2}}}\,;
            \end{equation}
            \item Stage of the \textit{\textbf{very slow contraction}};
        \end{itemize}
        \item Region (b) - \textit{\textbf{minimum of the SF potential}}:
        \begin{itemize}
            \item Required to \textbf{avoid being left with a large negative vacuum energy};
        \end{itemize}
        \item Region (c) - \textit{\textbf{flat and positive region}} (\textit{in the case of ghost condensation mechanism for the bounce}):
        \begin{equation}
            V(\phi)\simeq M^{4}\left(1-\beta\frac{M^{2}}{m^{2}}\phi\right) \quad\land\quad \beta>0\,;
        \end{equation}
        \begin{itemize}
            \item \textbf{Reversion from contraction to expansion (\textit{non-singular bounce})}.
        \end{itemize}
    \end{enumerate}
\end{examplebox}

\subsubsection{Instabilities of the models}
The expansion of the ghost condensate action \eqref{ghost-condensate} around the NEC-violating background gives the following formula for the SF:
\begin{equation}
    \phi=-m^{2}\,t+\pi_{0}(t)+\pi\left(t,\vec{x}\right)\,.
\end{equation}
In a consequence, one may obtain the EoM \cite{Buchbinder:2007ad}:
\begin{equation}
    \ddot{\pi}+3H\dot{\pi}=-\frac{V'(\phi)}{K}\left(\frac{m}{M}\right)\,.
\end{equation}
Moreover, from equation \eqref{dotH-dotpi} one could observe that:
\begin{equation}
    \dot{H}\sim\dot{\pi}\,,
\end{equation}
so that the effective Lagrangian becomes \cite{Creminelli:2006xe}:
\begin{equation}\label{effective-Lagrangian-gradient-ghost}
    \mathcal{L}_{\pi}\propto\frac{1}{2}\dot{\pi}^{2}+\frac{1}{K M^{4}}\dot{H}\left(\vec{\nabla}\pi\right)^{2}-\frac{1}{2}\frac{m^{4}}{K M^{4} \left(M'\right)^{2}}\left(\vec{\nabla}^{2}\pi\right)^{2}\,,
\end{equation}
where the quantity:
\begin{equation}
    M'\equiv\frac{M^{2}}{\bar{M}}
\end{equation}
denotes an \textit{effective mass scale associated with the higher-derivative stabilizing operator} $(\vec{\nabla}^{2}\pi)^2$ and is introduced in order to connect the analysis with the convention of \cite{Creminelli:2006xe}.

The form of \eqref{effective-Lagrangian-gradient-ghost} immediately implies that for:
\begin{equation}
    \dot{H}\neq 0
\end{equation}
we must deal with a \textbf{generation of a non-zero gradient term for the fluctuations} (\textit{with a wrong sign on the NEC-violating background}). Using the \textit{explicit forms of the operators} in the \textit{\textbf{Fourier space}}:
\begin{equation}
    \nabla^{2}\to -k^{2} \quad\land\quad \nabla^{4}\to k^{4}\,,
\end{equation}
one obtains the dispersion relation:
\begin{examplebox}[Ghost condensate (new ekpyrotic) dispersion relation]
    \begin{equation}\label{dispersion-ghost}
        \boxed{\omega^{2}=-2\frac{1}{K M^{4}}\dot{H}\,k^{2}+\frac{m^{4}}{K M^{4} \left(M'\right)^{2}}k^{4}}
    \end{equation}
\end{examplebox}
that has a \textit{\textbf{gradient instability at long wavelengths}} (1st term on the RHS of \eqref{dispersion-ghost} is negative for $\dot{H}>0$). Moreover, one may observe also that the 2nd term on the RHS of \eqref{dispersion-ghost} \textbf{stabilizes the UV sector}.

Rewriting the dispersion relation \eqref{dispersion-ghost} as:
\begin{equation}
    \omega^{2}(k)=-2A\,k^{2}+B\,k^{4}\,,
\end{equation}
where:
\begin{equation}
    A\equiv\frac{\dot{H}}{K M^{4}} \quad\land\quad B\equiv\frac{m^{4}}{K M^{4} \left(M'\right)^{2}}
\end{equation}
one can find the most negative value:
\begin{equation}
    \frac{\partial\omega^{2}}{\partial\left(k^{2}\right)}=-2A+2B\,k^{2}=0 \quad\implies\quad k_{*}^{2}=\frac{A}{B}=\dot{H}\frac{\left(M'\right)^{2}}{m^{4}}
\end{equation}
which corresponds to:
\begin{equation}
    \omega_{\mathrm{grad}}^{2}=\left|\omega^{2}\left(k_{*}\right)\right|=\frac{A^{2}}{B}=\left(\dot{H}\frac{M'}{\sqrt{K} M^{2} m^{2}}\right)^{2}\,.
\end{equation}
Such \textbf{instabilities are absent} if \textbf{their rate is less than the Hubble rate}:
\begin{equation}
    \left|\omega_{\mathrm{grad}}\right|\lesssim\left|H\right|\,,
\end{equation}
which yields the bound of the following form:
\begin{equation}
    \frac{\dot{H}}{H}\lesssim\frac{\sqrt{K} M^{2} m^{2}}{M'}\,.
\end{equation}

The second type of instability could arise when one considers mixing of fluctuations with gravity - \textit{\textbf{Jeans instability}} (\textit{long-wavelength instability of the ghost condensate fluctuation}, $\pi$)\footnote{Sufficiently long wavelength modes grow instead of oscillating.}. In such a case, one obtains the following growth rate \cite{Arkani-Hamed:2003pdi,Arkani-Hamed:2005teg}:
\begin{equation}
    \omega_{J}\sim\frac{\sqrt{K}M^{2} m^{2}}{M'}\,,
\end{equation}
that is harmless under the \textit{sub-Hubble condition} of:
\begin{equation}
    \omega_{J}\lesssim H\,.
\end{equation}
Thus, considering those two types of possible instabilities yields the double inequality condition of the following form:
\begin{examplebox}[Condition for avoiding (gradient and Jeans) instabilities in the new ekpyrotic models]
    \begin{equation}\label{instabilities-condition}
        \boxed{\frac{\dot{H}}{H}\lesssim\frac{\sqrt{K}M^{2} m^{2}}{M'}\lesssim H}\,.
    \end{equation}
    \underline{\textit{\textbf{Remark:}}}\\
    The condition \eqref{instabilities-condition} becomes a parametric window only if:
    \begin{equation}\label{instabilities-condition-1}
        \dot{H}\ll H^{2}\,.
    \end{equation}
    In other case, \textit{at least one instability grows faster than the Hubble friction}.
\end{examplebox}

\subsubsection{Constraint on a successful bounce}
Approaching the cosmological (non-singular) bounce implies:
\begin{equation}
    \left|H\right|\to 0\,,
\end{equation}
so that the condition \eqref{instabilities-condition-1} is not satisfied. The inastabilities are of the Jeans type, and their effects could be alleviated if the entire period of the NEC violation spans approximately one $e$-fold (the bounce phase is rather quick) \cite{Buchbinder:2007ad}:
\begin{equation}
    H\Delta t\lesssim 1\,.
\end{equation}
Furthermore, the value of $V(\phi)$ at the \textit{end of ekpyrosis} must be appropriate in order for the ekpyrotic phase to \textit{\textbf{be long enough}} \cite{Buchbinder:2007ad}:
\begin{equation}
    \boxed{m^{4}\,e^{2\mathcal{N}_{\mathrm{ekp}}}\ll\left|V_{\mathrm{end}}\right|\ll \frac{1}{2}c^{2}M^{4}K}\,,
\end{equation}
where:
\begin{equation}
    \boxed{\mathcal{N}_{\mathrm{ekp}}\equiv\frac{1}{2}\ln{\left|\frac{V_{\mathrm{end}}}{V_{\mathrm{beg}}}\right|}}
\end{equation}
is the \textbf{number of} $\boldsymbol{e}$\textbf{-folds of the ekpyrotic phase}. In fact, it means that the \textbf{scale} $\boldsymbol{M}$ \textbf{must be exponentially larger than} $\boldsymbol{m}$.

\subsubsection{Conclusions}
The question remains: \textit{\textbf{How could this large hierarchy in EFT arise from a more fundamental theory?}} However, that is not the focus of our discussion here. Rather, the \textit{\textbf{new ekpyrotic models aim to provide proof-of-concept that a cosmological history could be described entirely within the framework of a}} $\boldsymbol{4D}$ \textit{\textbf{EFT}}. This would allow us to follow the evolution of all fields, especially cosmological perturbations, without ambiguity through the bounce.

\section{Cosmological perturbations}\label{Sec:perturbations}
The \textit{\textbf{ekpyrotic phase addresses standard cosmological puzzles and provides a new mechanism for generating cosmological perturbations}} \cite{Khoury:2001wf,Khoury:2001zk,Gratton:2003pe,Boyle:2004gv,Brandenberger2004perturbations,Novello:2008ra,Lehners:2008vx,Battefeld:2014uga,Brandenberger:2016vhg}. Although both scalar and tensor perturbations are formed, their properties turn out to be very different, as we will explain it in this section.

\subsection{Scalar perturbations}
\begin{figure}[htbp]
    \centering
    \includegraphics[width=1\linewidth]{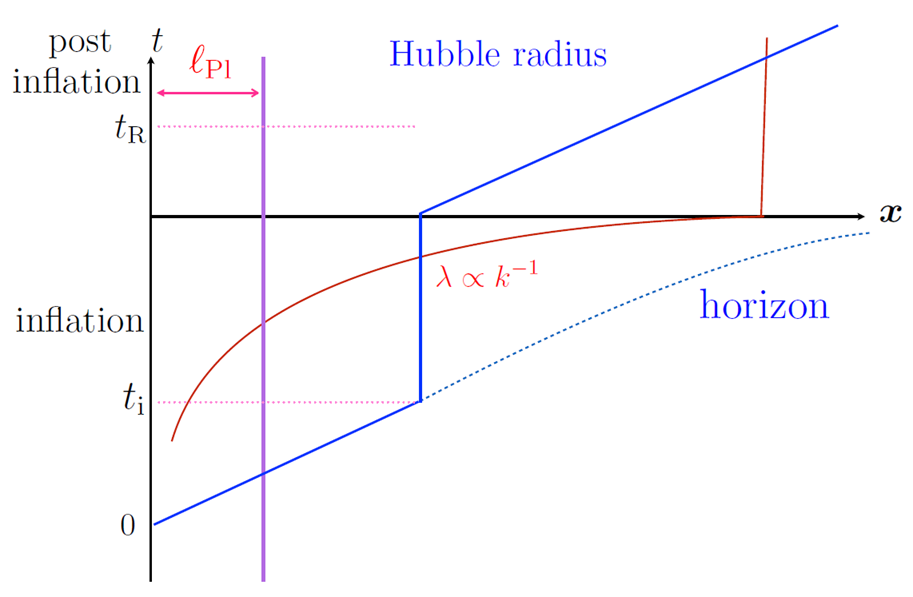}
    \caption{Space-time sketch of inflationary cosmology with emphasis on the trans-Planckian problem \cite{Brandenberger:2016vhg}.}
    \label{fig:Inflation-sketch}
\end{figure}

\begin{figure}[htbp]
    \centering
    \includegraphics[width=1\linewidth]{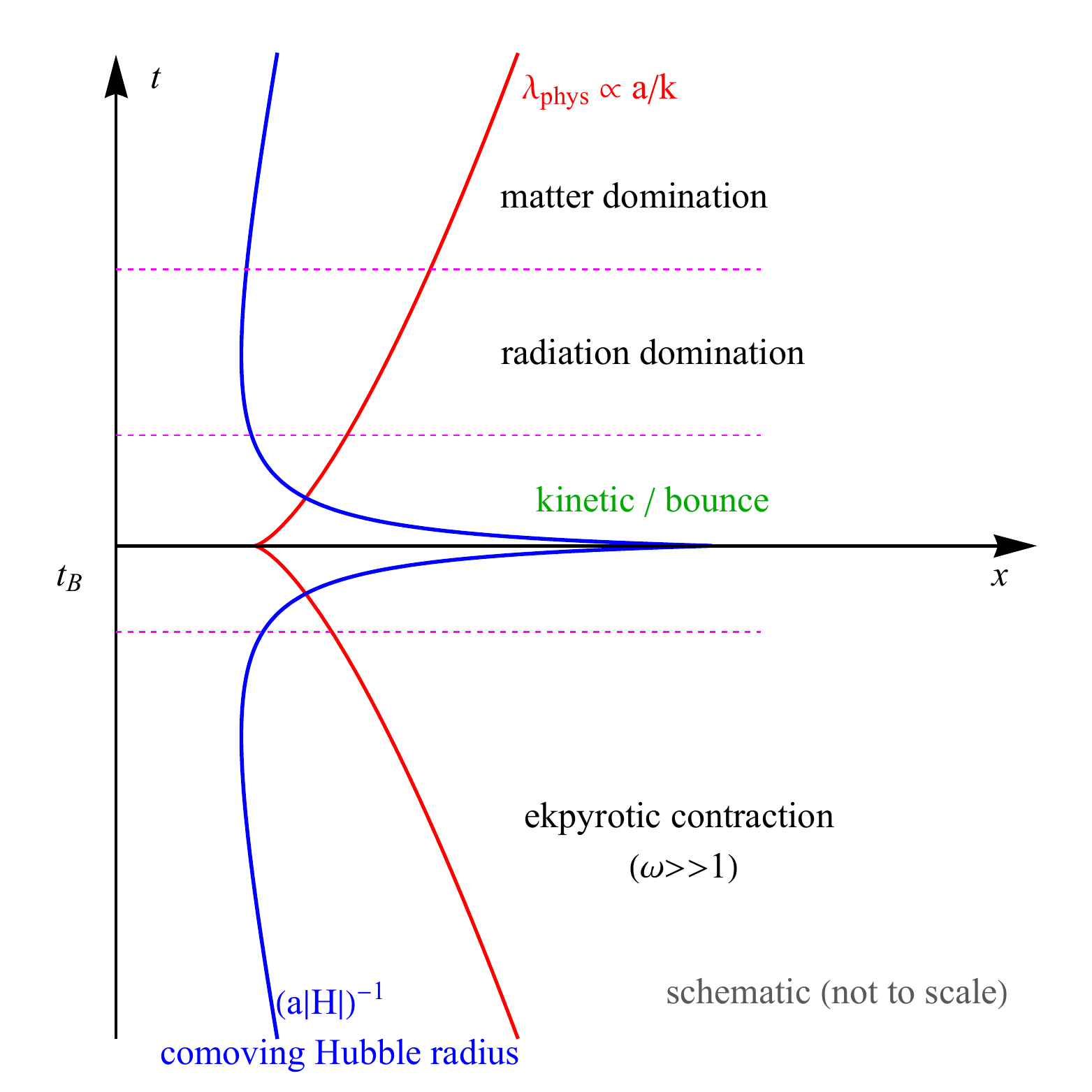}
    \caption{Space-time sketch of ekpyrotic cosmology.}
    \label{fig:Ekpyrotic-sketch}
\end{figure}

The angular anisotropy power spectrum of the CMB encodes primordial scalar perturbations whose wavelengths exited the effective horizon in the early Universe and later re-entered the Hubble radius, leaving imprints on the photon-baryon plasma and on the surface of last scattering. This suggests a period of cosmological evolution during which the \textit{\textbf{modes were initially in causal contact}}. There are two possible mechanisms for achieving this:
\begin{enumerate}[label=(\roman*)]
    \item \textbf{Cosmological inflation} - the \textit{modes are stretched beyond the approximately constant horizon} (Fig.~\ref{fig:Inflation-sketch});
    \item Ekpyrosis - the \textit{horizon shrinks rapidly due to increasing ultralocality as we approach the Big Crunch, while the fluctuation modes remain roughly constant} (Fig.~\ref{fig:Ekpyrotic-sketch}).
\end{enumerate}


\subsubsection{Single SF model}
The simple heuristic argument to obtain a \textit{\textbf{scale-invariant spectrum of perturbations}} comes from considering a \textit{single SF with a steep negative potential} which corresponds to a \textit{very slow contracting Universe in the ekpyrotic phase}. In such a case, we can initially ignore gravity \cite{Lehners:2007ac}. Therefore, we can consider a SF in Minkowski spacetime:
\begin{equation}\label{Minkowski-action}
    S=\int d^{4}x\left[-\frac{1}{2}\left(\partial\phi\right)^{2}+V_{0}\,e^{-c\phi}\right] \quad\land\quad \left(\partial\phi\right)^{2}\equiv\eta^{\mu\nu}\partial_{\mu}\phi\,\partial^{\nu}\phi=\partial_{\mu}\phi\,\partial^{\mu}\phi\,.
\end{equation}
It can be seen that the PE is unbounded from below, and that the SF diverges to ${}^{-}\infty$ in finite time.
\begin{examplebox}[Remark]
    The $V_{0}$ parameter \textit{\textbf{is not physical}} because it is possible to \textbf{shift the SF and alter its value arbitrarily}.
\end{examplebox}
In the ekpyrotic and cyclic models the SF potential is expected to run up towards zero at large negative values of the field (heterotic M-theory limit - vanishing string coupling \cite{Steinhardt:2001st}). The \textit{\textbf{quantum fluctuations of the background SF acquire a scale-invariant spectrum}} (\textit{to leading order in} $\hslash$) as it rolls down towards ${}^{-}\infty$.

\begin{figure}[htbp]
    \centering
    \includegraphics[width=1\linewidth]{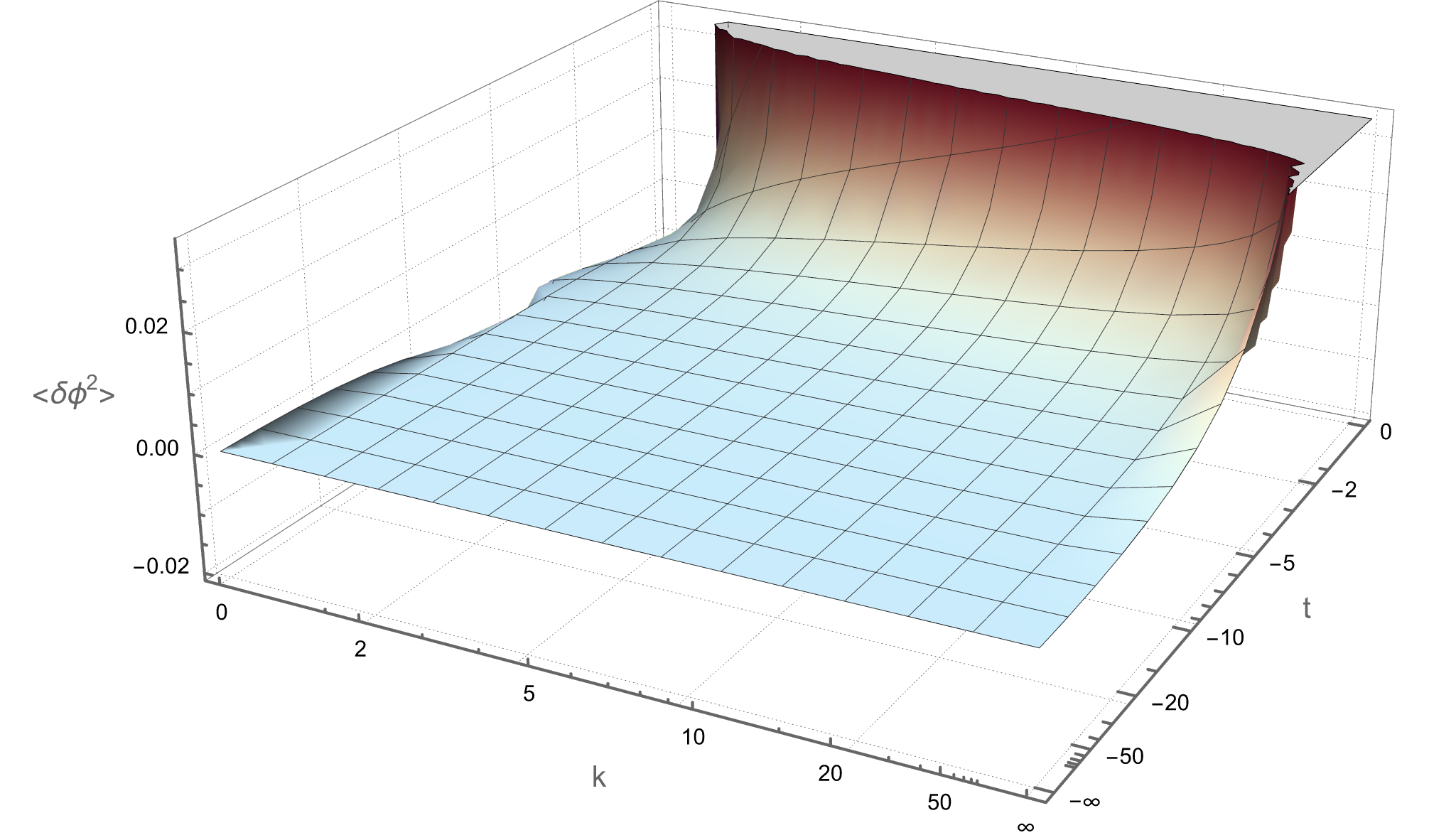}
    \caption{Variance of the quantum fluctuation.}
    \label{fig:quantum-variance}
\end{figure}

The following aspects are responsible for this:
\begin{enumerate}[label=(\roman*)]
    \item The action \eqref{Minkowski-action} \textit{\textbf{is scale-variant in the classical sense}} - shifting the field:
    \begin{equation}
        \phi\to \phi+\psi
    \end{equation}
    and rescaling coordinates by:
    \begin{equation}
        x^{\mu}\to x^{\mu}\,e^{\frac{1}{2}c\psi}
    \end{equation}
    rescales the action \eqref{Minkowski-action} by a factor $e^{c\psi}$, and is therefore a symmetry of the CFT;
    \item Rescaling the SF:
    \begin{equation}
        \phi\to\frac{\phi}{c}
    \end{equation}
    and redefining the $V_{0}$ constant, the parameter $c$ may be taken out to the front of the action and absorbed into Dirac constant:
    \begin{equation}
        \hslash\to\frac{\hslash}{c^{2}}
    \end{equation}
    in the formula $i\frac{S}{\hslash}$ governing the quantum theory;
    \item The SF $\phi$ has dimension of mass in $4D$ framework.
\end{enumerate}
The spatially homogeneous background solution represents a \textit{'zero energy density'} in the SF sector. In cyclic models, the 'zero energy' condition provides a \textit{\textbf{reasonable initial state for perturbations}}, as the phase of perturbation generation is followed by a phase of very low energy density and an extended stage of accelerated expansion \cite{Lehners:2007ac}. The classical EoM are time-translation invariant, so that the allowed perturbation is of the form of spatially homogeneous time-delay \cite{Lehners:2007ac,Lehners:2008vx}:
\begin{equation}
    \phi=\frac{2}{c}\ln{\bigl[-A\left(t+\delta t\right)\bigr]}
\end{equation}
and the quantum fluctuations scale as:
\begin{equation}
    \delta\phi=\frac{2}{c}\ln{\left(1+\frac{\delta t}{t}\right)}\simeq\frac{2}{c}\frac{\delta t}{t} \propto t^{-1}\,.
\end{equation}
For long wavelengths (modes effectively frozen by causality):
\begin{equation}\label{frozen-in}
    \left|k\,t\right|\ll 1\,,
\end{equation}
one could expect that the perturbations follow the same behavior. It means that the quantum variance becomes of the form:
\begin{equation}\label{quantum-variance-Minkowski}
    \left\langle\delta\phi^{2}\right\rangle\propto\hslash\,t^{-2}
\end{equation}
and the restoration of constants:
\begin{equation}
    \phi\to c\,\phi \quad\land\quad \hslash\to c^{2}\,\hslash
\end{equation}
left the result unchanged. Moreover, $\delta\phi$ has the dimension of $t^{-1}$ in $4D$ spacetime, which means that the constant of proportionality in quantum variance \eqref{quantum-variance-Minkowski} is dimensionless. Specifically, the perturbations $\delta\phi$ have a \textit{\textbf{scale-invariant spectrum of spatial fluctuations}}. We will now demonstrate this through explicit mathematical calculations:
\begin{examplebox}[Derivation of the scale-invariant spectrum of spatial fluctuations]
    \begin{itemize}
        \item First of all, let us expand the SF as:
        \begin{equation}
            \phi=\phi_{B}(t)+\delta\phi\left(t,\vec{x}\right)\,,
        \end{equation}
        which produces the following linear EoM:
        \begin{equation}
            \ddot{\delta\phi}+V''(\phi)\,\delta\phi=\vec{\nabla}^{2}\delta\phi\,;
        \end{equation}
        \item The zero energy condition for the classical background gives:
        \begin{equation}
            V''(\phi)=c^{2}\,V(\phi)=-\frac{1}{2}c^{2}\,\dot{\phi}_{B}^{2}(t)=-\frac{2}{t^{2}}\,;
        \end{equation}
        \item The next step is the Fourier decomposition:
        \begin{equation}
            \delta\phi\left(t,\vec{x}\right)=\sum_{\vec{k}}\left[a_{\vec{k}}\,\chi_{k}(t)\,e^{i\vec{k}\cdot\vec{x}}+a_{\vec{k}}^{\dagger}\,\chi_{k}^{*}(t)\,e^{-i\vec{k}\cdot\vec{x}}\right]\,,
        \end{equation}
        with the mode functions obeying the following EoM:
        \begin{equation}
            \ddot{\delta{\chi}}_{k}=\frac{2\chi_{k}}{t^{2}}-k^{2}\,\chi_{k}\,;
        \end{equation}
        \item Moreover, the \textit{incoming Minkowski vacuum state} is the \textbf{Bunch-Davies vacuum} \cite{Bunch:1978yq,Birrell_Davies_1982}:
        \begin{equation}\label{Bunch-Davies-Minkowski}
            \chi_{k}(t)=\frac{1}{\sqrt{2k}}e^{-ikt}\left(1-\frac{i}{k\,t}\right)\,,
        \end{equation}
        which for large $|k\,t|$ tends to the Minkowski positive frequency mode and for $|k\,t|\to 0$ enters the growing time-delay solution:
        \begin{equation}
            \chi_{k}\propto t^{-1}\,;
        \end{equation}
        \item The variance of the quantum fluctuation becomes:
        \begin{equation}\label{quantum-variance-chi-squared}
            \left\langle\delta\phi^{2}\right\rangle=\hslash\int\frac{d^{3}k}{\left(2\pi\right)^{3}}\left|\chi_{k}\right|^{2}\,,
        \end{equation}
        where the use of \eqref{Bunch-Davies-Minkowski} yields:
        \begin{equation}\label{chi-squared-Minkowski}
            \left|\chi_{k}\right|^{2}=\frac{1}{2k}\left(1+\frac{1}{k^{2}\,t^{2}}\right)\,;
        \end{equation}
        \item The 1st term in \eqref{chi-squared-Minkowski} is the usual \textit{\textbf{UV divergence}} that could be subtract by keeping only the \textit{physical (long-wavelength)} growth:
        \begin{equation}\label{chi-squared-Minkowski-long-wavelength}
            \left|\chi_{k}\right|^{2}\to\frac{1}{2}\frac{1}{k^{3}\,t^{2}}\,;
        \end{equation}
        \item Inserting \eqref{chi-squared-Minkowski-long-wavelength} into the formula for the quantum variance \eqref{quantum-variance-chi-squared} and integrating:
        \begin{equation}
            d^{3}k=4\pi\,k^{2}dk
        \end{equation}
        produces:
        \begin{equation}\label{quantum-variance-final-Minkowski}
            \boxed{\left\langle\delta\phi^{2}\right\rangle=\hslash\int\frac{4\pi\,k^{2}dk}{\left(2\pi\right)^{3}}\frac{1}{2k^{3}\,t^{2}}=\hslash\int\frac{k^{2}dk}{4\pi^{2}}\frac{1}{k^{3}\,t^{2}}=\frac{\hslash}{4\pi^{2}}\frac{\ln{k}}{t^{2}}}\,.
        \end{equation}
    \end{itemize}
    $\implies$ \textit{\textbf{Scale-invariance of the spectrum of growing SF perturbations (each logarithmic $k$-interval contributes equally)}} - Fig.~\ref{fig:quantum-variance}.
    \begin{itemize}
        \item From a physical point of view: the integral \eqref{quantum-variance-final-Minkowski} is over the modes that have '\textit{frozen in}' \eqref{frozen-in} and follow the time-delay solution.
    \end{itemize}
\end{examplebox}
\begin{examplebox}[Remark]
    We know that \textit{\textbf{observational data from the CMB spectrum is strictly related to the curvature perturbation of spacetime geometry}}. Therefore, even though its effects seem to be small corrections \cite{Creminelli:2004jg}, \textit{\textbf{we must take into account the influence of gravity}}.
\end{examplebox}
Let us consider an \textbf{\textit{ekpyrotic SF that is minimally coupled to gravity (GR)}}:
\begin{equation}\label{MC-SF-GR}
    \mathcal{L}=\sqrt{-g}\left[\frac{1}{2\kappa^{2}}R-\frac{1}{2}g^{\mu\nu}\partial_{\mu}\phi\,\partial_{\nu}\phi+V_{0}\,e^{-c\phi}\right]
\end{equation}
characterized by a \textit{\textbf{'fast-roll' parameter}}:
\begin{equation}
    \epsilon\equiv\frac{3}{2}\left(1+\omega\right)\gg 1\,.
\end{equation}
One can also observe that the scaling solution \eqref{scaling-ekpyrotic} implies that:
\begin{equation}
    \epsilon=\frac{1}{p}\,.
\end{equation}
From now on, we will perform the analysis using the notion of conformal time $\tau$:
\begin{equation}
    dt=a\,d\tau \quad\land\quad \frac{dX}{d\tau}\equiv X'\,.
\end{equation}
In terms of conformal time, the scaling solution takes the form:
\begin{examplebox}[Ekpyrotic scaling solution in terms of conformal time]
    \begin{equation}
        \boxed{a(\tau)=(-\tau)^{p/(1-p)} \quad\land\quad \phi_{i}(\tau)=\frac{2}{c_{i}\left(1-p\right)}\ln{\left(-\tau\right)} \quad\land\quad p\equiv\sum_{i}\frac{2}{c_{i}^{2}}}\,.
    \end{equation}
\end{examplebox}
The \textit{\textbf{single-field scalar perturbation theory for ekpyrotic models}} can be described by the following metric\footnote{We will use the notation from \cite{Boyle:2004gv}.} \cite{Khoury:2001wf,Khoury:2001zk,Boyle:2004gv}:
\begin{equation}\label{perturbations-metric}
    \boxed{\frac{ds^{2}}{a^{2}}=-\bigl(1+2A\,Y\bigr)d\tau^{2}-2B\,Y_{i}\,d\tau\,dx^{i}+\Bigl[\bigl(1+2H_{L}\,Y\bigr)\delta_{ij}+2H_{T}\,Y_{ij}\Bigr]dx^{i}\,dx^{j}}
\end{equation}
and the perturbed SF:
\begin{equation}
    \phi=\phi_{0}(\tau)+\delta\phi(\tau)\,Y\,,
\end{equation}
where:
\begin{equation}
    Y=Y\left(\vec{x}\right) \quad\land\quad Y_{i}=Y_{i}\left(\vec{x}\right) \quad\land\quad Y_{ij}=Y_{ij}\left(\vec{x}\right)
\end{equation}
are scalar harmonics, and:
\begin{equation}
    A=A(\tau) \quad\land\quad B=B(\tau) \quad\land\quad H_{L}=H_{L}(\tau) \quad\land\quad H_{T}=H_{T}(\tau)
\end{equation}
are \textit{\textbf{time-dependent scalar metric perturbation amplitudes} for a given Fourier/scalar harmonic mode} $Y\left(\vec{x}\right)$ \cite{Kodama:1984ziu,Mukhanov:1990me,Malik:2008im,Bardeen:1980kt,Ma:1995ey}:
\begin{itemize}
    \item $A$ - \textit{\textbf{lapse perturbation}} (\textit{scalar perturbation of the time-time metric component} $g_{00}$);
    \item $B$ - \textit{\textbf{scalar shift perturbation}} (\textit{scalar part of the mixed time-space metric component} $g_{0i}$);
    \item $H_{L}$ - \textit{\textbf{trace/isotropic scalar perturbation of the spatial metric}} \textit{(perturbs the local spatial scale factor and is related to the scalar curvature perturbation});
    \item $H_{T}$ - \textit{\textbf{traceless scalar perturbation of the spatial metric}} (\textit{represents the scalar shear part of the spatial metric perturbation}).
\end{itemize}
\begin{examplebox}[Scalar harmonics]
    The scalar harmonics are defined by the following relations:
    \begin{equation}
        \nabla^{2}Y=-k^{2}\,Y \quad\land\quad Y_{i}=-\frac{1}{k}\partial_{i}Y \quad\land\quad Y_{ij}=\frac{1}{k^{2}}\partial_{i}\partial_{j}Y+\frac{1}{3}\delta_{ij}Y\,,
    \end{equation}
    \begin{equation}
        \delta^{ij}Y_{ij}=0\,.
    \end{equation}
\end{examplebox}
The perturbed Einstein field equations take the form of:
\begin{equation}
    \delta G_{\nu}^{\mu}=\delta T_{\nu}^{\mu}
\end{equation}
provides a \textit{connection between the metric and matter perturbations}. In the considered case (scalar perturbations in a spatially flat FLRW Universe with a MC SF), \textbf{scalar perturbations can be fully described by a single gauge-invariant variable} (however, the \textit{\textbf{choice of gauge is not unique}}).
\begin{examplebox}[Newtonian gauge]
    \begin{itemize}
        \item The \textit{\textbf{Newtonian gauge}} is \textit{convenient for considering the Newtonian (gauge-invariant) potential}, $\Phi$, and is defined by the following set of the scalar metric perturbation amplitudes:
    \begin{equation}
        \boxed{B=H_{T}=0 \quad\land\quad A=\Phi=-H_{L}}\,,
    \end{equation}
    so that the line element \eqref{perturbations-metric} becomes:
    \begin{equation}
        \boxed{ds^{2}=a^{2}(\tau)\Bigl[-\bigl(1+2\Phi\bigr)d\tau^{2}+\bigl(1-2\Phi\bigr)d\vec{x}^{2}\Bigr]}\,;
    \end{equation}
    \item Background energy density and pressure of the SF are given by:
    \begin{equation}
        \rho_{\phi}\equiv\frac{\phi_{0}'^{2}}{2a^{2}}+V\left(\phi_{0}\right) \quad\land\quad p_{\phi}\equiv\frac{\phi_{0}'^{2}}{2a^{2}}-V\left(\phi_{0}\right)\,;
    \end{equation}
    \item Linear matter perturbations:
    \begin{subequations}
        \begin{align}
            \delta\rho_{\phi} & =\frac{\phi'_{0}}{a^{2}}\delta\phi'-\frac{\phi_{0}'^{2}}{a^{2}}\Phi+V_{,\phi}\,\delta\phi \,,\label{delta-rho-phi-Newtonian} \\
            \delta p_{\phi} & =\frac{\phi'_{0}}{a^{2}}\delta\phi'-\frac{\phi_{0}'^{2}}{a^{2}}\Phi-V_{,\phi}\,\delta\phi  \,,\label{delta-p-phi-Newtonian}
        \end{align}
    \end{subequations}
    and momentum density:
    \begin{equation}
        \delta T^{0}{}_{i}=-\bigl(\rho_{\phi}+p_{\phi}\bigr)\partial_{i}v \quad\land\quad \bigl(\rho_{\phi}+p_{\phi}\bigr)v=\frac{\phi'_{0}}{a^{2}}\delta\phi \,;
    \end{equation}
    \item $0i$ field equation (momentum constraint):
    \begin{equation}\label{0i-Newtonian}
        \Phi'+\mathcal{H}\Phi=\frac{1}{2}\phi'_{0}\,\delta\phi \quad\land\quad \mathcal{H}\equiv\frac{a'}{a}\,;
    \end{equation}
    \item $00$ and the trace of $ij$ field equations:
    \begin{subequations}
        \begin{align}
            -3\mathcal{H}\bigl(\Phi'+\mathcal{H}\Phi\bigr)+k^{2}\Phi & = \frac{1}{2}a^{2}\delta\rho_{\phi}\,,\label{00-Newtonian} \\
            \Phi''+3\mathcal{H}\Phi'+\bigl(2\mathcal{H}'+\mathcal{H}^{2}\bigr)\Phi & = \frac{1}{2}a^{2}\delta p_{\phi}\,;
            \label{trace-ij-Newtonian}
        \end{align}
    \end{subequations}
    \item The background equations:
    \begin{subequations}
        \begin{align}
            \mathcal{H}^{2} & = \frac{1}{3}a^{2}\rho_{\phi}\,,\label{IFE-Newtonian} \\
            \mathcal{H}' & = \mathcal{H}^{2}-\frac{1}{2}a^{2}\bigl(\rho_{\phi}+p_{\phi}\bigr)\,,\label{IIFE-Newtonian} \\
            \phi''_{0} & = -2\mathcal{H}\phi'_{0}-a^{2}\,V_{,\phi}\,;
            \label{KG-Newtonian}
        \end{align}
    \end{subequations}
    \item Taking the relation for $\delta\phi$ from \eqref{0i-Newtonian}, plugging into \eqref{delta-rho-phi-Newtonian} and \eqref{delta-p-phi-Newtonian}, and subtracting the pressure and density combinations appearing in \eqref{00-Newtonian} and \eqref{trace-ij-Newtonian} eliminate $\delta\phi'$. After some rather simple algebra gives the \textit{\textbf{final form of the EoM}}:
    \begin{equation}\label{Newtonian-gauge-EoM}
        \boxed{\Phi''+2\left(\mathcal{H}-\frac{\phi''_{0}}{\phi'_{0}}\right)\Phi'+\left(k^{2}+2\mathcal{H}'-2\mathcal{H}\frac{\phi''_{0}}{\phi'_{0}}\right)\Phi=0}\,,
    \end{equation}
    where:
    \begin{equation}
        k\equiv\left|\vec{k}\right|
    \end{equation}
    is the magnitude of the comoving Fourier 3-vector.
    \end{itemize}
\end{examplebox}
\begin{examplebox}[Comoving gauge]
    \begin{itemize}
        \item The \textit{\textbf{comoving gauge}} is \textit{convenient for considering the (gauge-invariant) curvature perturbation (on spatial hypersurfaces)}, $\xi$, and is defined by the following set of the scalar metric perturbation amplitudes:
        \begin{equation}
            \boxed{H_{T}=\delta T_{i}^{0}=0 \implies\delta\phi=0 \quad\land\quad \xi=H_{L}}\,,
        \end{equation}
        therefore, the perturbed metric in \textit{\textbf{ADM}}\footnote{Arnowitt-Deser-Misner.} \textit{\textbf{variables}} becomes of the form \cite{Arnowitt:1959ah,Arnowitt:1960,Arnowitt:1962hi,Nastase_2025ADM,Yi_2026ADM}:
        \begin{equation}
            \boxed{ds^{2}=-N^{2}d\tau^{2}+h_{ij}\bigl(dx^{i}+N^{i}d\tau\bigr)\bigl(dx^{j}+N^{j}d\tau\bigr)}\,,
        \end{equation}
        where:
        \begin{equation}
            h_{ij}=a^{2}(\tau)\,e^{2\xi}\,\delta_{ij}\,,
        \end{equation}
        with lapse:
        \begin{equation}
            N=a(1+\alpha)
        \end{equation}
        and shift:
        \begin{equation}
            N_{i}=a^{2}\partial_{i}B \,;
        \end{equation}
        \item Linearized \eqref{MC-SF-GR} up to 1st order in $\alpha$ and $B$ together with the linearized momentum constraint fixes the following:
        \begin{equation}\label{alpha-B-comoving}
            \alpha=\frac{\xi'}{\mathcal{H}} \quad\land\quad \partial^{2}B=-\frac{\partial^{2}\xi}{\mathcal{H}}+\epsilon\,\xi'\,,
        \end{equation}
        where:
        \begin{equation}\label{epsilon-comoving}
            \epsilon\equiv\frac{\phi_{0}'^{2}}{2\mathcal{H}^{2}}
        \end{equation}
        \item Substituting $\alpha$ and $B$ from \eqref{alpha-B-comoving} and expanding to 2nd order gives:
        \begin{equation}\label{2nd-order-action-comoving}
            S^{(2)}=\frac{1}{2}\int d\tau\,d^{3}x\,z^{2}\Bigl[\xi'^{2}-\bigl(\nabla\xi\bigr)^{2}\Bigr]\,,
        \end{equation}
        where:
        \begin{equation}
            z\equiv\frac{a\,\phi'_{0}}{\mathcal{H}}=\frac{a^{2}\,\phi'_{0}}{a'}\,;
        \end{equation}
        \item Varying the action \eqref{2nd-order-action-comoving} with respect to $\xi$ yields the following forms of the EoM:
        \begin{subequations}
            \begin{align}
                \frac{d}{d\tau}\bigl(z^{2}\,\xi'\bigr)-z^{2}\,\nabla^{2}\xi & =0 \,,\\
                \xi''+2\frac{z'}{z}\xi'+k^{2}\xi & = 0\,.\label{Comoving-gauge-EoM}
            \end{align}
        \end{subequations}
    \end{itemize}
\end{examplebox}
Using the constraints, one could relate $\Phi$ and $\xi$:
\begin{equation}
    \xi=\Phi+\frac{1}{\epsilon}\left(\frac{\Phi'}{\mathcal{H}}+\Phi\right) \quad\land\quad \Phi=-\epsilon\frac{\mathcal{H}}{k^{2}}\xi'\,.
\end{equation}
Furthermore, it is convenient to use \textbf{new variables}\footnote{They have the same $k$-dependence (the same spectral properties) as $\Phi$ and $\xi$.} of the form \cite{Mukhanov:1985rz}:
\begin{equation}\label{Mukhanov-variables}
    u\equiv\frac{a}{\phi'_{0}}\Phi \quad\land\quad v\equiv z\,\xi\,.
\end{equation}
Using the definitions from \eqref{Mukhanov-variables} in \eqref{Newtonian-gauge-EoM} and \eqref{Comoving-gauge-EoM} produces the following EoM:
\begin{align}
    u''+\left[k^{2}-\frac{\left(z^{-1}\right)''}{z^{-1}}\right]u & =0\,,\label{u-EoM-Mukhanov} \\
    v''+\left(k^{2}-\frac{z''}{z}\right)v & =0\,.\label{v-EoM-Mukhanov}
\end{align}
Moreover, differentiating the definitions in \eqref{Mukhanov-variables} gives the explicit relations between $u$ and $v$ of the form:
\begin{align}
    k\,v & = 2k\left(u'+\frac{z'}{z}u\right)\,, \\
    -k\,u & = \frac{1}{2k}\left(v'+\frac{\left(z^{-1}\right)'}{z^{-1}}v\right)\,.
\end{align}
For a constant EoS parameter (scaling background):
\begin{equation}\label{scaling-conformal-time}
    a(\tau)\propto(-\tau)^{q} \quad\land\quad \mathcal{H}=\frac{q}{\tau}\,,
\end{equation}
where:
\begin{equation}
    q=\frac{1}{\epsilon-1}\,.
\end{equation}
From \eqref{epsilon-comoving} one obtains:
\begin{equation}
    \phi'_{0}=\pm\sqrt{2\epsilon}\mathcal{H}\,,
\end{equation}
and therefore:
\begin{equation}
    z=\frac{a\,\phi'_{0}}{\mathcal{H}}=\pm\sqrt{2\epsilon}\,a\propto a\,.
\end{equation}
Using the scaling solution \eqref{scaling-conformal-time} this generates the following relations:
\begin{align}
    \frac{z''}{z} & =\frac{a''}{a}=\frac{q\left(q-1\right)}{\tau^{2}}=\frac{2-\epsilon}{\left(\epsilon-1\right)^{2}\,\tau^{2}}\,, \\
    \frac{\left(z^{-1}\right)''}{z^{-1}} & = \frac{q\left(q+1\right)}{\tau^{2}}=\frac{\epsilon}{\left(\epsilon-1\right)^{2}\,\tau^{2}}\,.
\end{align}
\begin{examplebox}[Conclusion]
    At the early Universe (for a large $|\tau|$) the $k^{2}$ term dominates over the $(z^{-1})''/z^{-1}$ and $z''/z$ terms in the brackets of \eqref{u-EoM-Mukhanov} and \eqref{v-EoM-Mukhanov}.\\
    $\implies$ \textit{\textbf{The relevant modes remain well within the horizon, satisfying asymptotically oscillatory solutions.}}
\end{examplebox}
Conversely, on scales smaller than the horizon, the curvature of spacetime is negligible, and the boundary conditions corresponding to the adiabatic Minkowski-like vacuum selected in the asymptotic subhorizon regime (\textit{\textbf{Bunch-Davies vacuum}}) can be applied \cite{Bunch:1978yq}:
\begin{examplebox}[Bunch-Davies initial conditions]
    \begin{equation}\label{Bunch-Davies-ics}
        \boxed{u\to\frac{i}{\left(2k\right)^{3/2}}e^{-ik\tau} \quad\land\quad v\to\frac{1}{\sqrt{2k}}e^{-ik\tau} \quad\land\quad \left|k\,\tau\right|\to\infty}\,.
    \end{equation}
\end{examplebox}
In fact, the EoM for $u$ and $v$ variables now become of the \textit{\textbf{Bessel equations}} form:
\begin{align}
    u''+\left[k^{2}-\frac{\epsilon}{\left(\epsilon-1\right)^{2}\,\tau^{2}}\right]u & = 0\,, \\
    v''+\left[k^{2}-\frac{2-\epsilon}{\left(\epsilon-1\right)^{2}\,\tau^{2}}\right]v & = 0\,,
\end{align}
where the exact solutions are of the forms \cite{King_Billingham_Otto_2003Bessel,Prosperetti_2011Bessel,Whittaker_Watson_2021Bessel}:
\begin{examplebox}[Bessel solutions]
    \begin{align}
        u(x) & = \sqrt{x}\left[A_{1}\,H_{\alpha}^{(1)}(x)+A_{2}\,H_{\alpha}^{(2)}(x)\right]\,, \\
        v(x) & = \sqrt{x}\left[B_{1}\,H_{\beta}^{(1)}(x)+B_{2}\,H_{\beta}^{(2)}(x)\right]\,,
    \end{align}
    where:
    \begin{equation}
        x\equiv\left|k\,\tau\right|\,,
    \end{equation}
    $A_{1,2}$, $B_{1,2}$ are constants, and $H_{\alpha,\beta}^{(1,2)}$ denotes the Hankel functions with the following orders:
    \begin{equation}
        \boxed{\alpha\equiv\sqrt{\frac{\left(z^{-1}\right)''}{z^{-1}}\tau^{2}+\frac{1}{4}}=\frac{1}{2}\left|\frac{\epsilon+1}{\epsilon-1}\right| \quad\land\quad \beta\equiv\sqrt{\frac{z''}{z}\tau^{2}+\frac{1}{4}}=\frac{1}{2}\left|\frac{\epsilon-3}{\epsilon-1}\right|}\,.
    \end{equation}
\end{examplebox}
In the \textbf{distant past}:
\begin{equation}
    x\to\infty\,,
\end{equation}
and the Hankel functions \textit{asymptotically} behave as \cite{King_Billingham_Otto_2003Bessel,Prosperetti_2011Bessel,Whittaker_Watson_2021Bessel}:
\begin{equation}
    H_{s}^{(1,2)}(x)\xrightarrow[x\to\infty]{}\sqrt{\frac{2}{\pi\,x}}\exp{\left[\pm i\left(x-s\frac{\pi}{2}-\frac{\pi}{4}\right)\right]}\,,
\end{equation}
and in such a case the boundary conditions determine the solutions: $\tau<0$ during the ekpyrotic contraction, so that:
\begin{equation}
    x=-k\,\tau\,,
\end{equation}
and using the Bunch-Davies initial conditions \eqref{Bunch-Davies-ics} it yields:
\begin{equation}
    u(\tau)\xrightarrow[\tau\to{}^{-}\infty]{}\frac{i}{\left(2k\right)^{3/2}}e^{-ik\tau}=\frac{i}{\left(2k\right)^{3/2}}e^{ix} \quad\land\quad v(\tau)\xrightarrow[\tau\to{}^{-}\infty]{}\frac{1}{\sqrt{2k}}e^{-ik\tau}=\frac{1}{\sqrt{2k}}e^{ix}\,.
\end{equation}
Thus, only the $H_{s}^{(1)}$ branch is allowed, which means that:
\begin{equation}
    A_{2}=B_{2}=0\,,
\end{equation}
and the solutions become of the form:
\begin{equation}
    \boxed{u(x)=\frac{1}{2k}\mathcal{P}_{1}\sqrt{\frac{\pi\,x}{4k}}H_{\alpha}^{(1)}(x) \quad\land\quad v(x)=\mathcal{P}_{2}\sqrt{\frac{\pi\,x}{4k}}H_{\beta}^{(1)}(x)}\,,
\end{equation}
where:
\begin{equation}
    \mathcal{P}_{1}\equiv\exp{\left[\frac{1}{4}i\bigl(2\alpha+3\bigr)\pi\right]} \quad\land\quad \mathcal{P}_{2}\equiv\exp{\left[\frac{1}{4}i\bigl(2\beta+1\bigr)\pi\right]}
\end{equation}
are \textit{phase factors}.

\begin{figure}[htbp]
    \centering
    \includegraphics[width=1\linewidth]{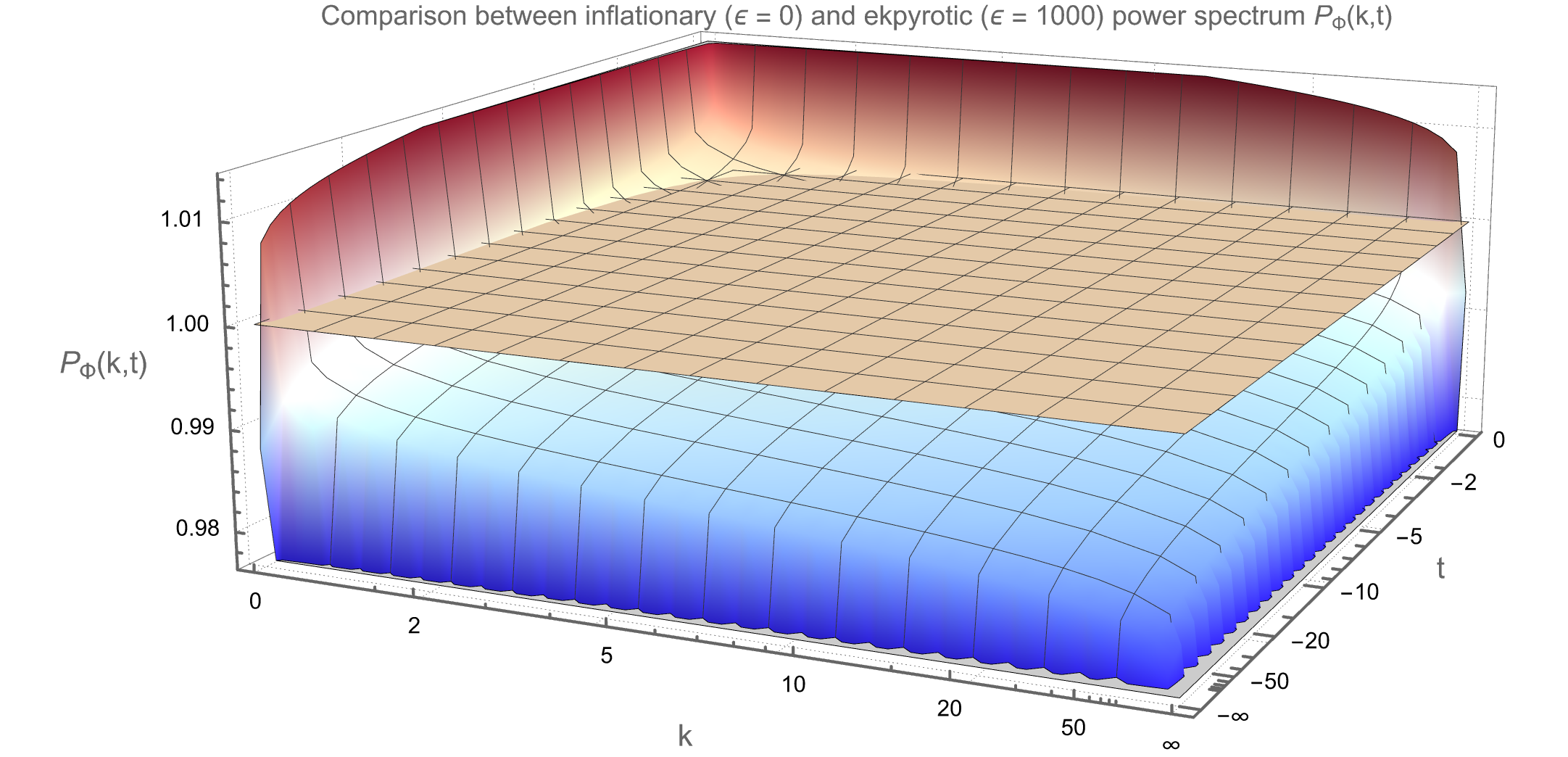}
    \caption{Comparison between inflationary and ekpyrotic power spectrum for Newtonian potential.}
    \label{fig:Newtonian-spectrum-comparison}
\end{figure}

\begin{figure}[htbp]
    \centering
    \includegraphics[width=1\linewidth]{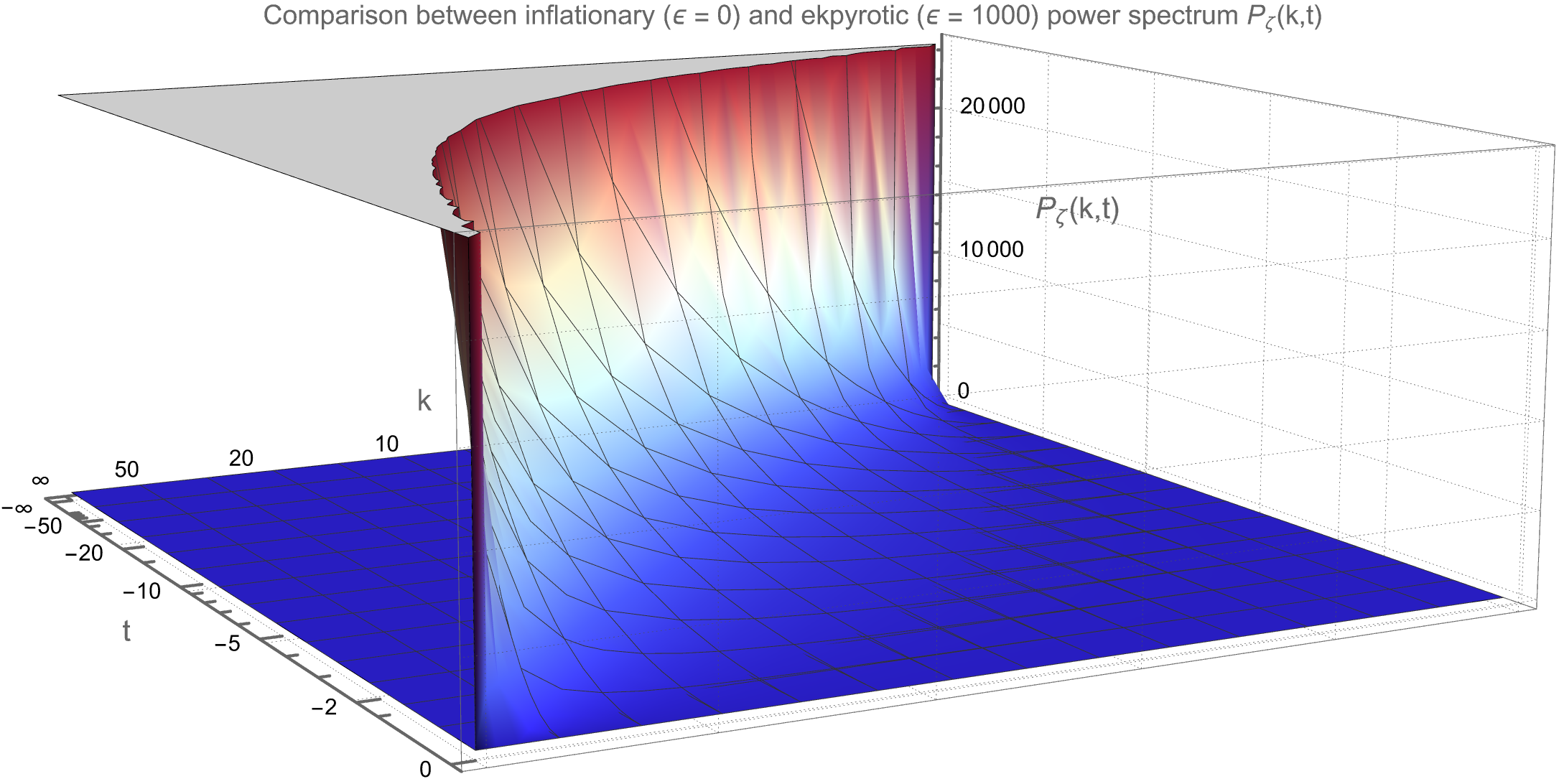}
    \caption{Comparison between inflationary and ekpyrotic power spectrum for curvature perturbation.}
    \label{fig:curvature-spectrum-comparison}
\end{figure}

The power spectra of $\Phi$ and $\xi$ can be determined by considering the \textit{\textbf{super-Hubble (late time) limit}}, in which comoving scales are outside the Hubble radius, namely \cite{King_Billingham_Otto_2003Bessel,Prosperetti_2011Bessel,Whittaker_Watson_2021Bessel}:
\begin{equation}\label{Bessel-late-time}
    H_{s}^{(1)}(x)\xrightarrow[x\to 0]{}-\frac{i}{\pi}\Gamma(s)\left(\frac{x}{2}\right)^{-s} \quad\land\quad s>0\,,
\end{equation}
where $\Gamma(s)$ denotes the \textit{Euler gamma function}. The use of \eqref{Bessel-late-time} implies:
\begin{equation}
    u(x)\propto x^{1/2-\alpha} \quad\land\quad v(x)\propto x^{1/2-\beta}\,.
\end{equation}
Moreover, by taking into account the fact that:
\begin{equation}
    \Phi=\frac{\phi'_{0}}{a}u \quad\land\quad \xi=\frac{v}{z} \quad\land\quad z\propto a\,,
\end{equation}
one can write the \textit{\textbf{power spectra for}} $\boldsymbol{\Phi}$ and $\boldsymbol{\xi}$ explicitly:
\begin{examplebox}[Power spectra]
    \begin{enumerate}[label=(\alph*)]
        \item The \textbf{power spectrum for the Newtonian potential} takes the form of:
        \begin{equation}\label{Newtonian-power-spectrum}
            \boxed{P_{\Phi}(k)\equiv\frac{k^{3}}{2\pi^{2}}\frac{\phi_{0}'^{2}}{a^{2}}\left|u\right|^{2}\propto x^{1-2\alpha}}\,;
        \end{equation}
        \item The \textbf{power spectrum for the curvature perturbation} becomes:
        \begin{equation}\label{curvature-power-spectrum}
            \boxed{P_{\xi}(k)\equiv\frac{k^{3}}{2\pi^{2}}\frac{\left|v\right|^{2}}{z^{2}}\propto x^{3-2\beta}}\,.
        \end{equation}
    \end{enumerate}
\end{examplebox}
In addition, the \textit{\textbf{spectral indices}} take the form:
\begin{examplebox}[Spectral indices]
    \begin{enumerate}[label=(\alph*)]
        \item The \textbf{spectral index for} $\boldsymbol{P_{\Phi}(k)}$:
        \begin{equation}
            \boxed{n_{\Phi}-1\equiv\frac{d\ln{P_{\Phi}(k)}}{d\ln{k}}\Bigg|_{\tau}=1-2\alpha=1-\left|\frac{\epsilon+1}{\epsilon-1}\right|}\,,
        \end{equation}
        is \textit{\textbf{invariant under the transformation}}:
        \begin{equation}
            \epsilon\to\frac{1}{\epsilon}
        \end{equation}
        $\implies$ Manifestation (at the linear level) of \textbf{duality between cosmological inflation ($\epsilon\to 0$) and ekpyrosis ($\epsilon\to\infty$)} \cite{Boyle:2004gv};\\
        $\implies$ \textit{\textbf{In both limits the spectrum is nearly scale-invariant}} (see Fig.~\ref{fig:Newtonian-spectrum-comparison});
        \item The \textbf{spectral index for} $\boldsymbol{P_{\xi}(k)}$:
        \begin{equation}
            \boxed{n_{\xi}-1\equiv\frac{d\ln{P_{\xi}(k)}}{d\ln{k}}\Bigg|_{\tau}=3-2\beta=3-\left|\frac{\epsilon-3}{\epsilon-1}\right|}
        \end{equation}
        in the ekpyrotic limit becomes:
        \begin{equation}
            \begin{dcases}
                \epsilon\gg 1 \\
                \beta\to\frac{1}{2}
            \end{dcases}
            \quad\implies\quad
            \boxed{n_{\xi}-1\to 2}\,.
        \end{equation}
        $\implies$ \textit{\textbf{A very blue spectrum (much more power on smaller scales) that is in disagreement with observations!}} \cite{Planck:2018vyg,Planck:2018jri,Planck:2019kim} (see Fig.~\ref{fig:curvature-spectrum-comparison})
    \end{enumerate}
\end{examplebox}
The results obtained for the \textbf{case of a constant EoS can be generalized to the case of a slowly varying EoS} as follows \cite{Gratton:2003pe}:
\begin{examplebox}[Slowly varying EoS case]
    \begin{itemize}
        \item Let us define the \textbf{dimensionless steepness of the SF potential}:
        \begin{equation}
            c(\phi)\equiv -\frac{V_{,\phi}}{V}\,,
        \end{equation}
        and the \textbf{ekpyrotic fast-roll parameters}:
        \begin{equation}
            \bar{\epsilon}\equiv\left(\frac{V}{V_{,\phi}}\right)^{2}=\frac{1}{c^{2}} \quad\land\quad \bar{\eta}\equiv 1-\frac{V\,V_{,\phi\phi}}{V_{,\phi}^{2}}\,;
        \end{equation}
        \item Using once again the background EoS variable:
        \begin{equation}
            \epsilon\equiv\frac{\phi'^{2}}{2\mathcal{H^{2}}}=\frac{3}{2}\bigl(1+\omega\bigr)
        \end{equation}
        in the case of ekpyrotic scaling solution with a (locally) exponential SF potential one finds:
        \begin{equation}
            \epsilon=\frac{1}{2}c^{2} \quad\implies\quad \frac{1}{\epsilon}=\frac{2}{c^{2}}=2\bar{\epsilon}\,;
        \end{equation}
        \item The EoM now becomes of the form:
        \begin{equation}
            \boxed{u''+\left[k^{2}-\frac{\left(z^{-1}\right)''}{z^{-1}}\right]u=0 \quad\land\quad z\equiv\frac{a\,\phi'_{0}}{\mathcal{H}}=\pm a\sqrt{2\epsilon}=a\,c}\,,
        \end{equation}
        therefore:
        \begin{equation}
            \frac{1}{z}=\frac{1}{a\,c}\,;
        \end{equation}
        \item For a slowly varying backgrounds, the 'mass term' can be approximated by:
        \begin{equation}
            \frac{\left(z^{-1}\right)''}{z^{-1}}=\frac{\nu_{u}^{2}-\frac{1}{4}}{\tau^{2}}\,,
        \end{equation}
        so that:
        \begin{equation}
            u(x)=\sqrt{x}\,H_{\nu_{u}}^{(1)}(x) \quad\land\quad x\equiv k\left|\tau\right|\,;
        \end{equation}
        \item From \eqref{Newtonian-power-spectrum} we can deduce that:
        \begin{equation}
            \boxed{P_{\Phi}(k)\propto x^{1-2\nu_{u}} \quad\land\quad n_{\Phi}-1=1-2\nu_{u}}\,;
        \end{equation}
        \item Now, we must explicitly compute the 'mass term'. To do this, let us introduce the notion of a new variable:
        \begin{equation}
            y\equiv\frac{1}{z}\,,
        \end{equation}
        in order to expand $y''/y$ in small departures from the scaling solution:
        \begin{align}
            \frac{y'}{y} & = -\left(\frac{a'}{a}\right)-\left(\frac{c'}{c}\right)=-\mathcal{H}-\frac{c'}{c} \,, \\
            \frac{y''}{y} & = \left(\frac{y'}{y}\right)'+\left(\frac{y'}{y}\right)^{2} \,;
        \end{align}
        \item Together with the background identity:
        \begin{equation}
            \mathcal{H}'=\mathcal{H}^{2}\bigl(1-\epsilon\bigr)
        \end{equation}
        one gets:
        \begin{equation}
            \frac{y''}{y}=\mathcal{H}^{2}\epsilon-\left(\frac{c'}{c}\right)'+2\mathcal{H}\left(\frac{c'}{c}\right)+\left(\frac{c'}{c}\right)^{2}\,;
        \end{equation}
        \item Changing the derivatives:
        \begin{equation}
            \frac{d}{d\tau}=\mathcal{H}\frac{d}{dN} \quad\land\quad N\equiv\ln{a}
        \end{equation}
        gives:
        \begin{equation}
            \frac{c'}{c}=\mathcal{H}\frac{d\ln{c}}{dN}\equiv\mathcal{H}\Delta \quad\land\quad \left(\frac{c'}{c}\right)'=\mathcal{H}'\Delta+\mathcal{H}^{2}\frac{d\Delta}{dN}\,,
        \end{equation}
        where:
        \begin{equation}
            \Delta\equiv\frac{d\ln{c}}{dN}\,;
        \end{equation}
        \item The \textit{\textbf{slow variation}} means that:
        \begin{equation}
            \Delta=\mathcal{O}\bigl(\bar{\eta}\,\epsilon\bigr) \quad\land\quad \frac{d\Delta}{dN}=\mathcal{O}\bigl(\bar{\eta}^{2}\,\epsilon^{2}\bigr)\,;
        \end{equation}
        \item Up to a 1st order (dropping $\frac{d\Delta}{dN}$ and $(c'/c)^{2}$):
        \begin{equation}
            \frac{y''}{y}\simeq\mathcal{H}^{2}\Bigl[\epsilon-\bigl(1-\epsilon\bigr)\Delta+2\Delta\Bigr]=\mathcal{H}^{2}\Bigl[\epsilon+\bigl(1+\epsilon\bigr)\Delta\Bigr]\,;
        \end{equation}
        \item In order to express $\Delta$ in terms of $\bar{\eta}$, we note that:
        \begin{equation}
            \bar{\eta}\equiv 1-\frac{V\,V_{,\phi\phi}}{V_{,\phi}^{2}}=\frac{c_{,\phi}}{c^{2}}\,,
        \end{equation}
        which implies:
        \begin{equation}
            \Delta\equiv\frac{d\ln{c}}{dN}=\frac{c_{,\phi}}{c}\frac{d\phi}{dN}=\frac{c_{,\phi}}{c}\frac{\phi'}{\mathcal{H}}=\frac{c_{,\phi}}{c}\sqrt{2\epsilon}=c\,\bar{\eta}\sqrt{2\epsilon}=2\epsilon\,\bar{\eta}\,;
        \end{equation}
        \item Putting altogether gives:
        \begin{equation}
            \frac{\left(z^{-1}\right)''}{z^{-1}}=\frac{y''}{y}\simeq\mathcal{H}^{2}\Bigl[\epsilon+2\bigl(1+\epsilon\bigr)\epsilon\,\bar{\eta}\Bigr]\,;
        \end{equation}
        \item In the case of a quasi-power-law contraction:
        \begin{equation}
            \mathcal{H}^{2}\simeq\frac{1}{\bigl(\epsilon-1\bigr)^{2}\,\tau^{2}}
        \end{equation}
        we get:
        \begin{equation}
            \frac{\left(z^{-1}\right)''}{z^{-1}}\simeq\frac{1}{\bigl(\epsilon-1\bigr)^{2}\,\tau^{2}}\Bigl[\epsilon+2\epsilon\bigl(1+\epsilon\bigr)\bar{\eta}\Bigr]\,;
        \end{equation}
        \item For a large $\epsilon$ and small $\bar{\eta}$:
        \begin{equation}
            \frac{\left(z^{-1}\right)''}{z^{-1}}=\frac{\nu_{u}^{2}-\frac{1}{4}}{\tau^{2}}\sim\frac{1}{\tau^{2}}\frac{1+2\epsilon\,\bar{\eta}}{\epsilon}\,;
        \end{equation}
        \item Expansion of $\nu_{u}$ around $1/2$ yields:
        \begin{equation}
            \nu_{u}-\frac{1}{2}\simeq\frac{1}{\epsilon}+2\bar{\eta}=2\bigl(\bar{\epsilon}+\bar{\eta}\bigr)\,;
        \end{equation}
        \item Finally, the spectral index takes the form:
        \begin{equation}
            \boxed{n_{\Phi}-1=1-2\nu_{u}\simeq -4\bigl(\bar{\epsilon}+\bar{\eta}\bigr)}\,,
        \end{equation}
        and in the \textbf{\textit{special case of an exact exponential SF potential}}:
        \begin{equation}
            c=\mathrm{const} \quad\implies\quad \bar{\eta}=0
        \end{equation}
        reduces into:
        \begin{equation}
            \boxed{n_{\Phi}-1\simeq -4\bar{\epsilon}}\,;
        \end{equation}
        \item In fact, one can show that for a large $\epsilon$:
        \begin{equation}
            \boxed{1-\frac{\epsilon+1}{\epsilon-1}\simeq -\frac{2}{\epsilon}=-4\bar{\epsilon}}\,.
        \end{equation}
    \end{itemize}
\end{examplebox}
Our analysis of the generation of scalar perturbations in the single-field ekpyrotic model led us to the following conclusions:
\begin{examplebox}[Conclusions for a single-field model]
    \begin{enumerate}[label=(\roman*)]
        \item A \textit{\textbf{single-field ekpyrotic model cannot produce a scale-invariant spectrum of cosmological perturbations}} \textit{unless Newtonian potential and curvature perturbations mix at the bounce}. In this case, the \textit{scale-invariant component of} $\Phi$ \textit{could dominate the blue intrinsic spectrum of} $\xi$ on large scales;
        \item The \textit{\textbf{mixing process depends on the specifics of the bounce dynamics and is a model-specific prediction}}. The examples include \cite{Lehners:2008vx}:
        \begin{itemize}
            \item \textbf{Higher-dimensional effects} \cite{Tolley:2003nx,McFadden:2005mq};
            \item \textit{\textbf{Breakdown of}} $\boldsymbol{4D}$ \textit{\textbf{EFT formalism}} $\implies$ \textit{\textbf{New degrees of freedom become relevant near the brane collision epoch}} $\implies$ \textit{\textbf{A more complete EFT that takes into account more fields}}.
        \end{itemize}
    \end{enumerate}
\end{examplebox}

\subsubsection{Two fields model}
\begin{figure}[htbp]
    \centering
    \includegraphics[width=1\linewidth]{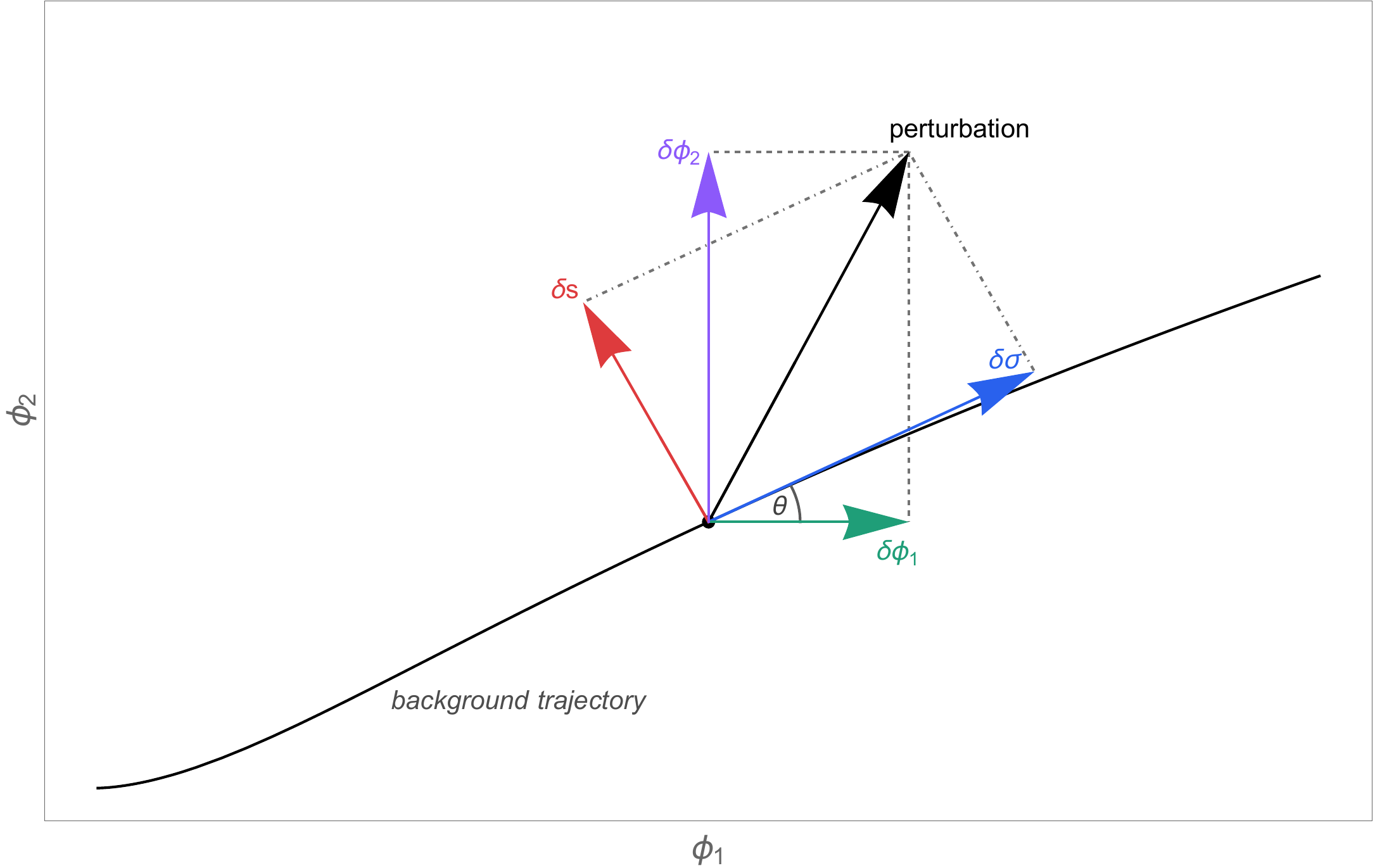}
    \caption{Decomposition of an arbitrary perturbation into an adiabatic ($\delta\sigma$) and entropy ($\delta s$) component.}
    \label{fig:decomposition-adiabatic-entropy}
\end{figure}

As we emphasized in the previous section of this study, the natural extension of the single-field EFT formalism is the \textit{\textbf{multi-field scenario}} \cite{Koyama:2007mg}. From a phenomenological point of view, this extension appears justified by the higher-dimensional formalism, which posits the existence of at least \textbf{two additional degrees of freedom in the form of scalar fields} \cite{Lehners:2008vx,Becker:2006dvp,Khoury:2001wf,Brax:2003fv,Maartens:2010ar,Lukas:1998tt,Garriga:2001ar,Ovrut:2002hi,Kobayashi:2002pw,Grana:2005jc,Lehners:2006ir,Cicoli:2023opf}:
\begin{enumerate}[label=(\alph*)]
    \item \textit{\textbf{Radion} field} (it specifies the \textit{separation between the end-of-the-world branes});
    \item \textit{\textbf{Volume modulus} field} (of the internal manifold).
\end{enumerate}
The presence of two SFs allows one to \textbf{decompose the perturbations
into an adiabatic mode, parallel to the background trajectory in field space, and an entropy, or isocurvature, mode, orthogonal to it} (Fig.~\ref{fig:decomposition-adiabatic-entropy}). In multi-field ekpyrotic/collapsing backgrounds, the entropy perturbation can contain a growing mode during the contracting phase and may subsequently source the curvature perturbation through a conversion mechanism \cite{Notari:2002yc,Koyama:2007if,Lehners:2008vx}.
The curvature perturbation can be sourced by entropy perturbations, which, if they acquire a nearly scale-invariant spectrum, can generate nearly scale-free curvature perturbations just before the bounce. These then transform into growing-mode perturbations during the subsequent expansion phase.

Our scenario involves an attractive force between the boundary branes and may be described by the following potential for the two scalar fields in $4D$ EFT (\textbf{two SFs MC to gravity/GR}) \cite{Lehners:2008vx,Lehners:2007ac,Koyama:2007mg,Koyama:2007if,Lehners:2010fy}:
\begin{examplebox}[Two-field ekpyrotic SF potential]
    \begin{equation}\label{ekpyrotic-multifield-potential}
        \boxed{V\left(\phi_{1},\phi_{2}\right)=-V_{1}\exp{\left[-\int c_{1}\left(\phi_{1}\right)d\phi_{1}\right]}-V_{2}\exp{\left[-\int c_{2}\left(\phi_{2}\right)d\phi_{2}\right]} \quad\land\quad V_{1},V_{2}>0}\,.
    \end{equation}
\end{examplebox}
If we further assume that the functions $c_{1}\left(\phi_{1}\right)$ and $c_{2}\left(\phi_{2}\right)$ vary slowly, then the SF potentials in \eqref{ekpyrotic-multifield-potential} are locally exponential. Furthermore, for scaling solutions, the two SFs diverge simultaneously into:
\begin{equation}
    \phi_{i}\to{}^{-}\infty\,.
\end{equation}
To address this issue, we introduce a new variable (\textit{\textbf{path length along the background SF trajectory}}), $\sigma$, defined by the relation \cite{Gordon:2000hv}:
\begin{equation}
    \dot{\sigma}\equiv\sqrt{\dot{\phi}_{1}^{2}+\dot{\phi}_{2}^{2}} \,,
\end{equation}
where $\dot{\sigma}$ denotes the adiabatic speed. Moreover, one can also introduce the angle of the background trajectory given by (see Fig.~\ref{fig:decomposition-adiabatic-entropy} again):
\begin{equation}\label{sin-cos-theta}
    \sin{\theta}\equiv\frac{\dot{\phi}_{2}}{\dot{\sigma}} \quad\land\quad \cos{\theta}\equiv\frac{\dot{\phi}_{1}}{\dot{\sigma}}\,.
\end{equation}
In such a case, the \textbf{fast-roll background parameter} becomes:
\begin{equation}\label{epsilon-dot-sigma}
    \epsilon\equiv\frac{3}{2}\bigl(1+\omega\bigr)=\frac{\dot{\sigma}^{2}}{2H^{2}}\,,
\end{equation}
and is valid for any spatially flat FLRW background dominadet by (MC) canonical SFs.

During the generation of entropic perturbations, we assume that the \textbf{background trajectory in SF space is straight}:
\begin{equation}\label{theta-zero}
    \dot{\theta}=0\,.
\end{equation}
Equivalently, the \textbf{adiabatic and entropy directions are fixed during this stage}. Hence, the \textbf{entropy perturbation is generated during the ekpyrotic contracting phase}, while its \textbf{conversion into the curvature perturbation is postponed until a later bending of the trajectory} \cite{Lehners:2008vx,Gordon:2000hv,Lehners:2007ac,Koyama:2007mg,Notari:2002yc}, for which:
\begin{equation}
    \dot{\theta}\neq 0\,.
\end{equation}
It is convenient to parametrize a straight line in the field space using the following relation:
\begin{equation}
    \dot{\phi}_{2}\equiv\gamma\dot{\phi}_{1} \quad\land\quad \gamma\equiv\tan{\theta}=\mathrm{const}\,.
\end{equation}
Using the general identity:
\begin{equation}
    \dot{\theta}=\frac{\dot{\phi}_{2}\,V_{,\phi_{1}}-\dot{\phi}_{1}\,V_{,\phi_{2}}}{\dot{\phi}_{1}^{2}+\dot{\phi}_{2}^{2}}
\end{equation}
for a straight line \eqref{theta-zero} requires:
\begin{equation}
    \frac{\dot{\phi}_{2}}{\dot{\phi}_{1}}=\frac{V_{,\phi_{2}}}{V_{,\phi_{1}}}\,,
\end{equation}
and for separable negative exponential potentials \eqref{ekpyrotic-multifield-potential}:
\begin{equation}
    V_{,\phi_{i}}=c_{i}\left(\phi_{i}\right)\,V_{i}\exp{\left[-\int c_{i}\left(\phi_{i}\right)d\phi_{i}\right]}
\end{equation}
yields:
\begin{equation}
    \gamma\equiv\frac{\dot{\phi}_{2}}{\dot{\phi}_{1}}=\frac{V_{,\phi_{2}}}{V_{,\phi_{1}}}=\frac{c_{2}\left(\phi_{2}\right)\,V_{2}\exp{\left[-\int c_{2}\left(\phi_{2}\right)d\phi_{2}\right]}}{c_{1}\left(\phi_{1}\right)\,V_{1}\exp{\left[-\int c_{1}\left(\phi_{1}\right)d\phi_{1}\right]}}\,.
\end{equation}
The scaling solution requires the following ratio to be a constant:
\begin{equation}\label{R-ratio}
    R\equiv\frac{V_{2}\exp{\left[-\int c_{2}\left(\phi_{2}\right)d\phi_{2}\right]}}{V_{1}\exp{\left[-\int c_{1}\left(\phi_{1}\right)d\phi_{1}\right]}}=\mathrm{const}\,.
\end{equation}
Taking a time derivative gives:
\begin{equation}
    \frac{\dot{R}}{R}=-c_{2}\,\dot{\phi}_{2}+c_{1}\,\dot{\phi}_{1}=\bigl(c_{1}-\gamma\,c_{2}\bigr)\dot{\phi}_{1}\,,
\end{equation}
and yields:
\begin{equation}\label{c1-gamma-c2}
    c_{1}=\gamma\,c_{2}\,.
\end{equation}
In such a case, the SF potential \eqref{ekpyrotic-multifield-potential} could be written as:
\begin{equation}\label{straight-line-SF-potential}
    \boxed{V\left(\phi_{1},\phi_{2}\right)=\tilde{V}\left(\phi_{1}\right)+\gamma^{2}\,\tilde{V}\left(\frac{\phi_{2}}{\gamma}\right)}
\end{equation}
for some function $\tilde{V}$.

\begin{figure}[htbp]
    \centering
    \includegraphics[width=1\linewidth]{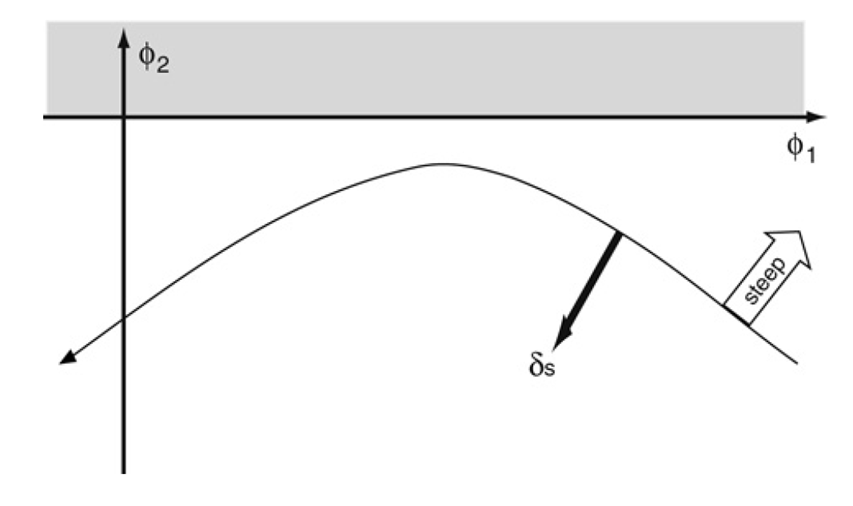}
    \caption{Background trajectory following a ridge in the SF potential during the ekpyrotic phase \cite{Lehners:2008vx}.}
    \label{fig:ridge-SF-potential}
\end{figure}

Along the straight trajectory \eqref{theta-zero}:
\begin{equation}
    \sin{\theta}=\frac{\gamma}{\sqrt{1+\gamma^{2}}} \quad\land\quad \cos{\theta}=\frac{1}{\sqrt{1+\gamma^{2}}}\,,
\end{equation}
and defining the slope along the adiabatic direction:
\begin{equation}
    V_{,\sigma}\equiv V_{,\phi_{1}}\cos{\theta}+V_{,\phi_{2}}\sin{\theta}
\end{equation}
together with \eqref{c1-gamma-c2} lead into the simple relation for the ratio \eqref{R-ratio}:
\begin{equation}
    R=\gamma^{2}\,.
\end{equation}
This yields into the following form of the effective exponent:
\begin{equation}
    c_{\mathrm{eff}}\equiv -\frac{V_{,\sigma}}{V}=\frac{c_{2}\,\gamma}{\sqrt{1+\gamma^{2}}}=\frac{c_{1}\,c_{2}}{\sqrt{c_{1}^{2}+c_{2}^{2}}}\,,
\end{equation}
so that the \textit{\textbf{effective adiabatic SF potential}} \textit{locally} becomes of the form:
\begin{equation}
    \boxed{V(\sigma)=-V_{0}\,e^{-c_{\mathrm{eff}}\sigma}}\,.
\end{equation}
For an exponential with (constant) $c_{\mathrm{eff}}$, the exact ekpyrotic scaling solution obeys:
\begin{equation}
    \epsilon=\frac{1}{2}c_{\mathrm{eff}}^{2}\,,
\end{equation}
and by using \eqref{c1-gamma-c2} gives:
\begin{equation}
    \epsilon_{\mathrm{ekp}}=\frac{1}{2}c_{\mathrm{eff}}^{2}=\frac{1}{2}\frac{c_{1}^{2}}{\left(1+\gamma^{2}\right)}=\frac{1}{2}\frac{\left|\gamma\,c_{1}\,c_{2}\right|}{\left(1+\gamma^{2}\right)}\,,
\end{equation}
where the last equality comes from the fact that:
\begin{equation}
    c_{1}^{2}=\gamma^{2}\,c_{2}^{2}=\left|\gamma\,c_{1}\,c_{2}\right|\,.
\end{equation}
\begin{examplebox}[Conclusion]
    \begin{itemize}
        \item An \textbf{ekpyrotic phase is obtained if the factor of} $\boldsymbol{c_{1}\,c_{2}}$ \textbf{is large}, namely:
        \begin{equation}
            c_{\mathrm{eff}}^{2}>6 \quad\implies\quad \boxed{\frac{c_{1}^{2}\,c_{2}^{2}}{c_{1}^{2}+c_{2}^{2}}>6}\,;
        \end{equation}
        \item For the symmetric case:
        \begin{equation}
            c_{1}=c_{2}=c
        \end{equation}
        one has:
        \begin{equation}
            c_{\mathrm{eff}}^{2}=\frac{1}{2}c^{2} \quad\implies\quad \boxed{c^{2}>12 \quad\implies\quad c>2\sqrt{3}}\,.
        \end{equation}
    \end{itemize}
\end{examplebox}
\begin{examplebox}[The value of $\gamma$ parameter]
    \begin{itemize}
        \item In phenomenological two-field ekpyrotic models, one usually assumes:
        \begin{equation}
            \gamma\sim\mathcal{O}(1)\,;
        \end{equation}
        \item \textit{\textbf{Heterotic M-theory}} (colliding branes solution with empty branes) - during ekpyrotic phase \cite{Lehners:2006ir}:
        \begin{equation}
            \gamma= -\frac{1}{\sqrt{3}}\,;
        \end{equation}
        \item The specific value of the parameter $\gamma$ \textbf{\textit{does not qualitatively affect the results}} \cite{Lehners:2008vx,Lehners:2008my}.
    \end{itemize}
\end{examplebox}
In the case of the contracting Universe, a growing mode is represented by the entropy perturbation $\delta s$ (relative fluctuation in the two SFs), which at linear order takes the form \cite{Gordon:2000hv}:
\begin{equation}\label{entropy-perturbation-linear}
    \boxed{\delta s\equiv\frac{\dot{\phi}_{1}\delta\phi_{2}-\dot{\phi}_{2}\delta\phi_{1}}{\dot{\sigma}}}\,.
\end{equation}
It represents the \textit{\textbf{gauge-invariant perturbation that is perpendicular to the background SF trajectory}} (Fig.~\ref{fig:decomposition-adiabatic-entropy}). The EoM can be derived as follows \cite{Gordon:2000hv}:
\begin{examplebox}[Derivation of the (linear order) EoM for the entropy perturbation]
    \begin{itemize}
        \item Let us introduce the notion of \textbf{unit vector tangent to the background trajectory}:
        \begin{equation}
            e_{\sigma}^{i}=\left(\cos{\theta},\sin{\theta}\right)\,,
        \end{equation}
        and the \textbf{unit vector orthogonal to the trajectory}:
        \begin{equation}
            e_{s}^{i}=\left(-\sin{\theta},\cos{\theta}\right)\,;
        \end{equation}
        \item The adiabatic and entropy perturbations can be obtained by \textit{projecting the field perturbations along these two directions}:
        \begin{subequations}
            \begin{align}
                \delta\sigma & = e_{\sigma}^{i}\,\delta\phi_{i}= \cos{\theta}\,\delta\phi_{1}+\sin{\theta}\,\delta\phi_{2} \,,\label{delta-sigma} \\
                \delta s & = e_{s}^{i}\,\delta\phi_{i}= -\sin{\theta}\,\delta\phi_{1}+\cos{\theta}\,\delta\phi_{2} \,;\label{delta-s}
            \end{align}
        \end{subequations}
        \item Using \eqref{sin-cos-theta} gives the relation for the entropy perturbation \eqref{entropy-perturbation-linear}:
        \begin{equation}
            \delta s=-\frac{\dot{\phi}_{2}}{\dot{\sigma}}\delta\phi_{1}+\frac{\dot{\phi}_{1}}{\dot{\sigma}}=\frac{\dot{\phi}_{1}\,\delta\phi_{2}-\dot{\phi}_{2}\,\delta\phi_{1}}{\dot{\sigma}}\,;
        \end{equation}
        \item This combination is in fact gauge-invariant (at linear oder) because under a small time shift:
        \begin{equation}
            \delta\phi_{i}\to\delta\phi_{i}-\dot{\phi}_{i}\,\delta t\,,
        \end{equation}
        the extra terms cancel in the numerator:
        \begin{equation}
            \dot{\phi}_{1}\left(-\dot{\phi}_{2}\,\delta t\right)-\dot{\phi}_{2}\left(-\dot{\phi}_{1}\,\delta t\right)=0\,;
        \end{equation}
        \item The background EoM are of the form:
        \begin{equation}\label{background-EoM-SF-i}
            \ddot{\phi}_{i}+3H\dot{\phi}_{i}+V_{,\phi_{i}}=0\,;
        \end{equation}
        \item Projecting \eqref{background-EoM-SF-i} along the adiabatic direction produces:
        \begin{equation}
            \ddot{\sigma}+3H\dot{\sigma}+V_{,\sigma}=0 \quad\land\quad V_{,\sigma}\equiv e_{\sigma}^{i}V_{,\phi_{i}}\,;
        \end{equation}
        \item On the other hand, projecting orthogonally yields the \textit{\textbf{bending equation}}:
        \begin{equation}\label{turning-rate}
            \dot{\theta}=-\frac{V_{,s}}{\dot{\sigma}} \quad\land\quad V_{,s}\equiv e_{s}^{i}V_{,\phi_{i}}
        \end{equation}
        $\implies$ \textbf{The} $\boldsymbol{\dot{\theta}}$ \textbf{measures the bending of the background trajectory in the field space};
        \item Linear perturbation EoM in the Newtonian gauge are of the form:
        \begin{equation}\label{linear-EoM-Newtonian-perturbation}
            \ddot{\delta\phi}_{i}+3H\dot{\delta\phi}_{i}+\frac{k^{2}}{a^{2}}\delta\phi_{i}+V_{,\phi_{i}\phi_{j}}\delta\phi_{j}=4\dot{\phi}_{i}\dot{\Phi}-2V_{,\phi_{i}}\Phi\,;
        \end{equation}
        \item In order to obtain the entropy equation, we must project \eqref{linear-EoM-Newtonian-perturbation} along $e_{s}^{i}$. \textbf{During this projection the basis vectors themselves are time-dependent}:
        \begin{equation}
            \dot{e}_{\sigma}^{i}=\dot{\theta}\,e_{s}^{i} \quad\land\quad \dot{e}_{s}^{i}=-\dot{\theta}\,e_{\sigma}^{i}
        \end{equation}
        $\implies$ Additional terms proportional to $\dot{\theta}$ and $\dot{\theta}^{2}$;
        \item The use of background EoM and the constraint (1st Friedmann equation) eliminate the adiabatic term:
        \begin{equation}\label{entropy-EoM-perturbation-general}
            \boxed{\ddot{\delta s}+3H\dot{\delta s}+\left(\frac{k^{2}}{a^{2}}+V_{ss}+3\dot{\theta}^{2}\right)\delta s=4k^{2}\frac{\dot{\theta}}{a^{2}\,\dot{\sigma}}\Phi}\,,
        \end{equation}
        where:
        \begin{equation}
            V_{ss}\equiv e_{s}^{i}\,e_{s}^{j}\,V_{,\phi_{i}\phi_{j}}=\dot{\sigma}^{2}\left(\dot{\phi}_{1}^{2}\,V_{,\phi_{1}\phi_{2}}-2\dot{\phi}_{1}\dot{\phi}_{2}V_{,\phi_{1}\phi_{2}}+\dot{\phi}_{2}^{2}\,V_{,\phi_{1}\phi_{2}}\right)\,;
        \end{equation}
        \item For the straight trajectory phase:
        \begin{equation}
            \dot{\theta}=0 \quad\implies\quad \boxed{\ddot{\delta s}+3H\dot{\delta s}+\left(\frac{k^{2}}{a^{2}}+V_{ss}\right)\delta s=0}
        \end{equation}
        $\implies$ \textbf{\textit{Relevant during the generation of entropy perturbations in the ekpyrotic contracting phase}} ($\delta s$ \textit{evolves independently of} $\Phi$);
        \item \textit{\textbf{A later bending of the trajectory is needed to convert the entropy perturbation into the curvature one.}}
    \end{itemize}
\end{examplebox}
On \textbf{super-Hubble scales}:
\begin{equation}
    \frac{k}{a\left|H\right|}\ll 1\,,
\end{equation}
and \eqref{entropy-EoM-perturbation-general} takes the simple form of:
\begin{equation}\label{super-Hubble-EoM-entropy}
    \ddot{\delta s}+3H\dot{\delta s}+V_{ss}\delta s=0\,.
\end{equation}
In terms of conformal time, $\tau$, and the canonical variable:
\begin{equation}
    \delta S\equiv a\,\delta s\,,
\end{equation}
the EoM \eqref{super-Hubble-EoM-entropy} becomes:
\begin{equation}
    \boxed{\delta S''+\left(-\frac{a''}{a}+a^{2}\,V_{ss}\right)\delta S=0}\,,
\end{equation}
so that the \textit{\textbf{'effective potential'}} governing $\delta S$ is:
\begin{examplebox}['Effective potential' for the entropy perturbation]
    \begin{equation}
        \boxed{\mathcal{U}(\tau)\equiv\frac{a''}{a}-a^{2}\,V_{ss}}\,,
    \end{equation}
\end{examplebox}
and yields the EoM:
\begin{examplebox}[The EoM for entropy perturbation]
    \begin{equation}\label{deltaS-EoM}
        \boxed{\delta S''+\left(k^{2}-\mathcal{U}\right)\delta S=0}\,.
    \end{equation}
\end{examplebox}
\begin{examplebox}[Specific form of entropy mass in the straight line approximation]
\begin{itemize}
    \item The general form of the 2nd derivative becomes:
    \begin{equation}\label{Vss}
        V_{ss}=\sin^{2}{\left(\theta\right)}\,V_{,\phi_{1}\phi_{1}}-2\sin{\left(\theta\right)}\cos{\left(\theta\right)}\,V_{,\phi_{1}\phi_{2}}+\cos^{2}{\left(\theta\right)}\,V_{,\phi_{2}\phi_{2}}\,;
    \end{equation}
    \item In the case of a straight scaling trajectory, the specific form of the SF potential \eqref{straight-line-SF-potential} yields:
    \begin{equation}
        V_{,\phi_{1}\phi_{2}}=0 \quad\land\quad V_{,\phi_{1}\phi_{1}}=\tilde{V}''\left(\phi_{1}\right)\,;
    \end{equation}
    \item Moreover:
    \begin{equation}
        V_{,\phi_{2}\phi_{2}}=\frac{d^{2}}{d\phi_{2}^{2}}\left[\gamma^{2}\,\tilde{V}\left(\frac{\phi_{2}}{\gamma}\right)\right]=\tilde{V}''\left(\frac{\phi_{2}}{\gamma}\right)\,;
    \end{equation}
    \item Along the straight scaling trajectory \eqref{c1-gamma-c2} implies relation between the SFs of the form:
    \begin{equation}
        \phi_{2}=\gamma\,\phi_{1} \quad\implies\quad \phi_{1}=\frac{\phi_{2}}{\gamma}\,,
    \end{equation}
    so that:
    \begin{equation}
        V_{,\phi_{2}\phi_{2}}=V_{,\phi_{1}\phi_{1}}\,;
    \end{equation}
    \item In a consequence, \eqref{Vss} can be written as follows:
    \begin{equation}
        \boxed{V_{ss}=V_{,\phi_{1}\phi_{1}}\Bigl[\underbrace{\sin^{2}{\left(\theta\right)}+\cos^{2}{\left(\theta\right)}}_{1}\Bigr]=V_{,\phi_{1}\phi_{1}}}\,.
    \end{equation}
\end{itemize}
\end{examplebox}
For the entropy mode, the canonically normalized Fourier mode obeys the \textit{\textbf{Mukhanov-Sasaki equation}}:
\begin{equation}\label{Mukhanov-Sasaki-EoM}
    y''+\bigl[k^{2}-\mathcal{U}(\tau)\bigr]y=0
\end{equation}
with:
\begin{equation}
    y=\delta S
\end{equation}
for entropy field (two-field ekpyrosis) or:
\begin{equation}
    y=v
\end{equation}
for curvature, and:
\begin{equation}
    \mathcal{U}=\frac{z''}{z}
\end{equation}
in the single-field analogue. If the effective potential $\mathcal{U}(\tau)$ scales exactly like $\tau^{-2}$:
\begin{equation}
    \mathcal{U}(\tau)=\frac{\nu^{2}-\frac{1}{4}}{\tau^{2}}\,,
\end{equation}
then the EoM \eqref{Mukhanov-Sasaki-EoM} (and also \eqref{deltaS-EoM}) is form-invariant under the dilation:
\begin{equation}
    \left(\tau,\vec{x}\right)\to\left(\lambda\tau,\lambda\vec{x}\right)
\end{equation}
provided that:
\begin{equation}
    k\to\frac{k}{\lambda}\,.
\end{equation}
In such a case, the EoM depends only on the dimensionless combination:
\begin{equation}
    x\equiv k\left|\tau\right|\,,
\end{equation}
and turns the mode equation into a Bessel equation with Hankel solutions:
\begin{equation}
    y(\tau)\propto \sqrt{-\tau}\,H_{\nu}^{(1)}\left(x\right)\,,
\end{equation}
so that on super-Hubble scales, one obtains the universal power-law behavior:
\begin{equation}
    \left|y\right|\propto k^{-\nu} \quad\implies\quad \boxed{P(k)\propto k^{3-2\nu}}\,.
\end{equation}
Thus, the\textbf{\textit{ spectral tilt is constant}}:
\begin{equation}
    \boxed{n-1=3-2\nu}\,.
\end{equation}
\begin{examplebox}[Conclusion]
    \begin{itemize}
        \item The scale-invariance corresponds to:
        \begin{equation}\label{scale-invariance-2-fields-Ueff}
            \boxed{\nu=\frac{3}{2} \quad\implies\quad \mathcal{U}(\tau)\,\tau^{2}=2}
        \end{equation}
        $\implies$ If $\mathcal{U}(\tau)$ \textit{\textbf{were not proportional to}}:
        \begin{equation}
            \mathcal{U}(\tau)\propto\tau^{-2}\,,
        \end{equation}
        \textit{\textbf{the EoM would introduce an extra time scale!}}\\
        $\implies$ The solution \textit{\textbf{would no longer be a pure function of}} $\boldsymbol{x}$, the \textit{\textbf{super-Hubble amplitude would pick up additional}} $\boldsymbol{\tau}$\textit{\textbf{-dependence}}, and the spectrum \textit{\textbf{would not be a clean power-law (or would have strong running/features).}}
    \end{itemize}
\end{examplebox}
Our task now is to check whether the effective potential, $\mathcal{U}(\tau)$, satisfies the condition for a scale-invariant spectrum \eqref{scale-invariance-2-fields-Ueff}. We will make the necessary calculations following the details from \cite{Lehners:2007ac}:
\begin{examplebox}[Detailed calculations for the spectrum of entropy perturbation]
    \begin{itemize}
        \item Starting from general relations:
        \begin{equation}
            \mathcal{H}\equiv\frac{a'}{a}=a H \quad\land\quad a'=a\mathcal{H}=a^{2}H\,,
        \end{equation}
        and taking 1st derivative yields:
        \begin{equation}
            \mathcal{H}'=\left(a H\right)'=a'H+a H'=a^{2}H^{2}+a^{2}\dot{H}=a^{2}\left(H^{2}+\dot{H}\right)\,,
        \end{equation}
        so that:
        \begin{equation}
            \frac{a''}{a}=\mathcal{H}'+\mathcal{H}^{2}=a^{2}\left(H^{2}+\dot{H}\right)+a^{2}H^{2}=a^{2}\left(2H^{2}+\dot{H}\right)\,;
        \end{equation}
        \item Using the fact that:
        \begin{equation}
            \epsilon\equiv -\frac{\dot{H}}{H^{2}} \quad\implies\quad \dot{H}=-\epsilon H^{2}
        \end{equation}
        gives:
        \begin{equation}
            \frac{a''}{a}=2a^{2}H^{2}\left(1-\frac{\epsilon}{2}\right)\,;
        \end{equation}
        \item Now, considering the Klein-Gordon (KG) equation:
        \begin{equation}
            \ddot{\phi}_{1}+3H\dot{\phi}_{1}+V_{,\phi_{1}}=0
        \end{equation}
        produces the formula for the derivative of the SF potential:
        \begin{equation}\label{V-phi1-derivative}
            V_{,\phi_{1}}=-\left(\ddot{\phi}_{1}+3H\dot{\phi}_{1}\right)\,;
        \end{equation}
        \item Differentiating \eqref{V-phi1-derivative} with respect to cosmological time:
        \begin{equation}
            \frac{d}{dt}\bigl(V_{,\phi_{1}}\bigr)=V_{,\phi_{1}\phi_{1}}\dot{\phi}_{1}=-\left(\dddot{\phi}_{1}+3\dot{H}\dot{\phi}_{1}+3H\ddot{\phi}_{1}\right)\,,
        \end{equation}
        and dividing by $\dot{\phi}_{1}$ (with the straight trajectory condition $\dot{\theta}$=0) yields:
        \begin{equation}\label{V-phi1-phi1-derivative}
            V_{,\phi_{1}\phi_{1}}=-\frac{\dddot{\sigma}}{\dot{\sigma}}-3\dot{H}-3H\frac{\ddot{\sigma}}{\dot{\sigma}}\,,
        \end{equation}
        where we used the relations:
        \begin{equation}
            \dot{\phi}_{1}=\dot{\sigma}\,\cos{\theta} \quad\land\quad \ddot{\phi}_{1}=\ddot{\sigma}\,\cos{\theta} \quad\land\quad \dddot{\phi}_{1}=\dddot{\sigma}\,\cos{\theta}\,;
        \end{equation}
        \item In order to express the adiabatic derivatives by $\epsilon$, we use \eqref{epsilon-dot-sigma} in the form of:
        \begin{equation}
            \dot{\sigma}^{2}=2\epsilon H^{2}\,,
        \end{equation}
        and rewrite the last segment from the RHS of \eqref{V-phi1-phi1-derivative}:
        \begin{equation}\label{ddot-sigma-dot-sigma}
            \frac{\ddot{\sigma}}{\dot{\sigma}}=\frac{1}{2}\frac{\dot{\epsilon}}{\epsilon}+\frac{\dot{H}}{H}=\frac{1}{2}H\frac{\epsilon_{,N}}{\epsilon}-\epsilon H \quad\land\quad \epsilon_{,N}\equiv\frac{d\epsilon}{dN}\,;
        \end{equation}
        \item Now, we can differentiate \eqref{ddot-sigma-dot-sigma} and add the square in order to obtain the 1st part on the RHS of \eqref{V-phi1-phi1-derivative}:
        \begin{equation}
            \frac{\dddot{\sigma}}{\dot{\sigma}}=\frac{d}{dt}\left(\frac{\ddot{\sigma}}{\dot{\sigma}}\right)+\left(\frac{\ddot{\sigma}}{\dot{\sigma}}\right)^{2}\,;
        \end{equation}
        \item Keeping the terms through $\mathcal{O}\left(\epsilon^{2}\right)$ and $\mathcal{O}\left(\epsilon_{,N}\right)$, and neglecting $\epsilon_{,NN}$ with $\epsilon_{,N}^{2}$ (slow variation), one finds that:
        \begin{equation}
            \frac{\dddot{\sigma}}{\dot{\sigma}}\simeq H^{2}\left(2\epsilon^{2}-\frac{5}{2}\epsilon_{,N}\right) \quad\land\quad \frac{\ddot{\sigma}}{\dot{\sigma}}\simeq H\left(-\epsilon+\frac{1}{2}\frac{\epsilon_{,N}}{\epsilon}\right)\,;
        \end{equation}
        \item Inserting this into \eqref{V-phi1-phi1-derivative} produces:
        \begin{equation}
            V_{,\phi_{1}\phi_{1}}\simeq -H^{2}\left(2\epsilon^{2}-\frac{5}{2}\epsilon_{,N}\right)+3\epsilon H^{2}-3H^{2}\left(-\epsilon+\frac{1}{2}\frac{\epsilon_{,N}}{\epsilon}\right)\,;
        \end{equation}
        \item Combining the last two pieces gives:
        \begin{equation}\label{3-epsilon-H}
            3\epsilon H^{2}-3H^{2}\left(-\epsilon+\frac{1}{2}\frac{\epsilon_{,N}}{\epsilon}\right)=6\epsilon H^{2}-\frac{3}{2}H^{2}\frac{\epsilon_{,N}}{\epsilon}\,;
        \end{equation}
        \item The last term in \eqref{3-epsilon-H} is \textit{\textbf{suppressed by}} $\boldsymbol{\epsilon^{-1}}$ \textit{\textbf{and belongs to the subleading}} $\boldsymbol{\mathcal{O}\left(\epsilon^{0}\right)}$\,;
        \item Thus, the enhanced terms are:
        \begin{equation}
            \boxed{a^{2}\,V_{,\phi_{1}\phi_{1}}=-a^{2}H^{2}\left(2\epsilon^{2}-6\epsilon-\frac{5}{2}\epsilon_{,N}\right)+\mathcal{O}\left(\epsilon^{0}\right)}\,;
        \end{equation}
        \item We can also observe that:
        \begin{equation}
            \mathcal{H}'=\mathcal{H}^{2}\bigl(1-\epsilon\bigr)\,,
        \end{equation}
        which yields the relation:
        \begin{equation}
            \mathcal{H}^{-1}=\int_{0}^{\tau}\bigl(\epsilon-1\bigr)d\tau'\,;
        \end{equation}
        \item Using the identity:
        \begin{equation}
            1=\frac{d(\tau)}{d\tau}
        \end{equation}
        and integrating by parts gives the relation:
        \begin{equation}
            \int_{0}^{\tau}\bigl(\epsilon-1\bigr)d\tau'=\Bigl[\bigl(\epsilon-1\bigr)\tau'\Bigr]\Bigg|_{0}^{\tau}-\int_{0}^{\tau}\epsilon'\tau'd\tau'=\epsilon\,\tau\left(1-\frac{1}{\epsilon}\right)-\int_{0}^{\tau}\epsilon'\tau'd\tau'\,;
        \end{equation}
        \item Writing the last integral in a form convenient for another integration by parts yields:
        \begin{equation}
            \bigl(\epsilon\,\tau\bigr)^{-1}\int_{0}^{\tau}\epsilon'\tau'd\tau'=\frac{\epsilon'}{\epsilon}\tau-\bigl(\epsilon\,\tau\bigr)^{-1}\int_{0}^{\tau}\frac{d}{d\tau}\bigl(\epsilon'\tau'\bigr)\tau'd\tau'\,;
        \end{equation}
        \item Now, we can use the fact that:
        \begin{equation}
            \epsilon'=\mathcal{H}\,\epsilon_{,N}\,,
        \end{equation}
        and to leading order:
        \begin{equation}
            \mathcal{H}\tau=\epsilon^{-1}
        \end{equation}
        in the scaling solution;
        \item The 2nd terms becomes:
        \begin{equation}
            -\bigl(\epsilon\,\tau\bigr)^{-1}\int_{0}^{\tau}\frac{d}{d\tau}\bigl(\epsilon'\tau'\bigr)\tau'd\tau'=-\bigl(\epsilon\,\tau\bigr)^{-1}\int_{0}^{\tau}\frac{1}{\epsilon}{\biggl(\frac{\epsilon_{,N}}{\epsilon}\biggr)}_{,N}d\tau'=\mathcal{O}\left(\epsilon^{-2}\right)
        \end{equation}
        $\implies$ \textit{\textbf{Negligible at our order}};
        \item The remaining part is:
        \begin{equation}
            -\bigl(\epsilon\,\tau\bigr)^{-1}\int_{0}^{\tau}\epsilon'\tau'd\tau'\simeq-\frac{\epsilon'}{\epsilon}\tau=-\mathcal{H}\tau\frac{\epsilon_{,N}}{\epsilon}\simeq-\frac{\epsilon_{,N}}{\epsilon^{2}}\,;
        \end{equation}
        \item Taking everything altogether gives:
        \begin{equation}\label{conformal-H-minus1-epsilon}
            \mathcal{H}^{-1}\simeq\epsilon\,\tau\left(1-\frac{1}{\epsilon}-\frac{\epsilon_{,N}}{\epsilon^{2}}\right)\,;
        \end{equation}
        \item Equivalently, inverting \eqref{conformal-H-minus1-epsilon}:
        \begin{subequations}
        \begin{align}
            \mathcal{H}\tau & = \frac{1}{\epsilon}\left(1+\frac{1}{\epsilon}+\frac{\epsilon_{,N}}{\epsilon^{2}}\right)+\mathcal{O}\left(\epsilon^{-3}\right) \,,\label{mathcalH-tau} \\
            \left(\mathcal{H}\tau\right)^{2} & = \frac{1}{\epsilon^{2}}\left(1+\frac{2}{\epsilon}+2\frac{\epsilon_{,N}}{\epsilon^{2}}\right)\,;\label{mathcalH-tau-squared}
        \end{align}
        \end{subequations}
        \item Combining everything yields:
        \begin{equation}\label{entropy-mass-final}
        \begin{aligned}
            \tau^{2}\left(\frac{a''}{a}-a^{2}\,V_{,\phi_{1}\phi_{1}}\right) & =\tau^{2}\mathcal{H}^{2}\left[\bigl(2-\epsilon\bigr)+\left(2\epsilon^{2}-6\epsilon-\frac{5}{2}\epsilon_{,N}\right)\right] \\
            & =\tau^{2}\mathcal{H^{2}}\left(2\epsilon^{2}-7\epsilon+2-\frac{5}{2}\epsilon_{,N}\right)\,;
        \end{aligned}
        \end{equation}
        \item Now, we must use \eqref{mathcalH-tau-squared} in the last expression from \eqref{entropy-mass-final}. The dominant piece comes from:
        \begin{equation}
            2\epsilon^{2}\bigl(\mathcal{H}\tau\bigr)^{2}\simeq 2\epsilon^{2}\cdot \epsilon^{-2}=2\,,
        \end{equation}
        and the next to the leading segments are the following:
        \begin{subequations}
            \begin{align}
                2\epsilon^{2}\bigl(\mathcal{H}\tau\bigr)^{2} & \simeq 2\left(\frac{2}{\epsilon}+2\frac{\epsilon_{,N}}{\epsilon^{2}}\right) \,, \\
                -7\epsilon\bigl(\mathcal{H}\tau\bigr)^{2} & \simeq -\frac{7}{\epsilon}\,, \\
                -\frac{5}{2}\epsilon_{,N}\bigl(\mathcal{H}\tau\bigr)^{2} & \simeq -\frac{5}{2}\frac{\epsilon_{,N}}{\epsilon^{2}}\,;
            \end{align}
        \end{subequations}
        \item Keeping the controlled terms, the \textit{\textbf{final result}} is:
        \begin{equation}
            \boxed{\tau^{2}\left(\frac{a''}{a}-a^{2}\,V_{,\phi_{1}\phi_{1}}\right)\simeq 2\left(1-\frac{3}{2\epsilon}+\frac{3}{4}\frac{\epsilon_{,N}}{\epsilon^{2}}\right)}\,.
        \end{equation}
    \end{itemize}
    \underline{\textit{\textbf{Conclusion:}}}\\
    \textbf{The leading terms are the sought} $\boldsymbol{\simeq 2}$ \textbf{with the first two controlled corrections.}
\end{examplebox}
The mode equation can be written as:
\begin{equation}
    \delta S''+\left[k^{2}-\frac{\nu^{2}-\frac{1}{4}}{\tau^{2}}\right]\delta S=0\,,
\end{equation}
where:
\begin{equation}
    \boxed{\nu^{2}-\frac{1}{4}=2\left(1-\frac{3}{2\epsilon}+\frac{3}{4}\frac{\epsilon_{,N}}{\epsilon^{2}}\right)}\,.
\end{equation}
The expansion of $\nu$ at large $\epsilon$ produces:
\begin{equation}
    \nu=\frac{3}{2}-\frac{1}{\epsilon}+\frac{1}{2}\frac{\epsilon_{,N}}{\epsilon^{2}}+\ldots\,.
\end{equation}
\textit{On super-Hubble scales}, the \textit{\textbf{power spectrum for entropic perturbations}} is of the form:
\begin{examplebox}[Power spectrum for entropic perturbations]
    \begin{equation}
        \boxed{P_{\delta s}(k)\propto k^{3-2\nu}}\,,
    \end{equation}
    so that the spectral index becomes:
    \begin{equation}\label{entropic-spectral-index}
        \boxed{n_{\delta s}-1\equiv\frac{d\ln{P_{\delta s}}}{d\ln{k}}=3-2\nu=\underbrace{\frac{2}{\epsilon}}_{\text{gravitational}}-\underbrace{\frac{\epsilon_{,N}}{\epsilon^{2}}}_{\text{non-gravitational}}}\,.
    \end{equation}
    \underline{\textit{\textbf{Conclusion:}}}\\
    \begin{itemize}
        \item The \textbf{1st term} is often interpreted as the \textit{\textbf{gravitational contribution to the tilt}}: it is already \textbf{present for an exact exponential ekpyrotic potential}:
        \begin{equation}
            \epsilon_{,N}=0\,,
        \end{equation}
        and \textbf{originates from the finite gravitational background evolution} encoded in $a''/a$ and $\mathcal{H}\tau$. It gives a \textit{\textbf{small blue tilt}};
        \item The \textbf{2nd term} is associated with the \textbf{slow variation of the ekpyrotic steepness parameter}, or equivalently with \textbf{deviations from an exact exponential potential}. It is therefore a \textit{\textbf{potential shape}}, or \textit{\textbf{non-gravitational, contribution}}. For:
        \begin{equation}
            \epsilon_{,N}>0\,,
        \end{equation}
        it gives a \textit{\textbf{red contribution and can compensate the gravitational blue tilt}}.
    \end{itemize}
    $\implies$ \underline{\textit{\textbf{Nearly scale-invariant spectrum of entropy perturbations in a contracting Universe!}}}
\end{examplebox}
\begin{examplebox}[Remark]
The \textbf{growth of entropy perturbations} in the two-field ekpyrotic mechanism is closely related to the transverse instability of the background trajectory in field space \cite{Lehners:2007ac,Tolley:2007nq}. In the tachyonic entropic mechanism, the background trajectory lies along a ridge of the SF potential, while the entropy direction corresponds to an unstable transverse direction \cite{Koyama:2007mg}. Equivalently, during the straight trajectory phase, the \textit{\textbf{entropy perturbation is governed by an effective negative mass squared}}:
\begin{equation}
    V_{ss}<0\,,
\end{equation}
which allows $\delta s$ to grow. To obtain a \textit{\textbf{sufficiently long ekpyrotic phase}}, the trajectory must therefore remain sufficiently close to the crest of the ridge for long enough before the subsequent conversion of entropy perturbations into curvature perturbations (Fig.~\ref{fig:ridge-SF-potential}).
\end{examplebox}
In the case of the two-field model, it is also worth noting that the following \textit{\textbf{theoretical and observational requirements}} apply \cite{Lehners:2008vx,Buchbinder:2007ad,Boyle:2003km,Lehners:2007ac,Levy:2015awa,Koyama:2007mg,Buchbinder:2007tw,Koyama:2007if,Lehners:2010fy,Boyle:2005se,Lehners:2008qe,Boyle:2007zx}:
\begin{examplebox}[Constraints on the two-field ekpyrosis]
    \begin{enumerate}[label=(\alph*)]
        \item Amplitude normalization of scalar curvature perturbations;
        \item Spectral tilt vs. ekpyrotic fast-roll history;
        \item Non-Gaussianity;
        \item Reheating temperature window and GWs BBN bound;
        \item Timing of RD vs. crossing the potential well;
        \item Duration of ekpyrosis;
        \item Instability ridge and initial conditions.
    \end{enumerate}
\end{examplebox}
Using the turning rate/bending equation \eqref{turning-rate} of the background trajectory and the comoving curvature perturbation:
\begin{equation}
    \mathcal{R}=\frac{H}{\dot{\sigma}}Q_{\sigma}\,,
\end{equation}
where $Q_{\sigma}$ is the gauge-invariant adiabatic field perturbation, one
finds that \textit{\textbf{on large scales the curvature perturbation is sourced by the entropy perturbation}} \cite{Lehners:2008vx,Gordon:2000hv,Lehners:2007ac,Koyama:2007mg}:
\begin{equation}\label{comoving-curvature-evolution}
    \dot{\mathcal{R}}=2\frac{H}{\dot{\sigma}}\dot{\theta}\,\delta s\,.
\end{equation}
In the spatially flat gauge:
\begin{equation}
    Q_{\sigma}=\delta\sigma \quad\implies\quad \xi\equiv\mathcal{R}=\frac{H}{\dot{\sigma}}\delta\sigma\,.
\end{equation}
Integrating the large scale evolution \eqref{comoving-curvature-evolution} over the short conversion interval $\Delta t$ with:
\begin{equation}
    \dot{\theta}\neq 0
\end{equation}
yields the relation for \textbf{\textit{linear curvature perturbation}}:
\begin{equation}\label{conversion-integral}
    \boxed{\xi_{L}\left(t_{f}\right)-\xi_{L}\left(t_{i}\right)=2\int_{t_{i}}^{t_{f}}\frac{H}{\dot{\sigma}}\dot{\theta}\,\delta s^{(1)}dt} \quad\land\quad H<0\,.
\end{equation}
Thus, the \textbf{conversion is only efficient while the trajectory bends}. Furthermore, \textit{\textbf{curvature can be generated on super-Hubble scales by entropy perturbations}}, whereas the \textit{\textbf{converse is not true in the long-wavelength limit}}.

\textbf{During the conversion}, the \textit{\textbf{(instantaneous) fast-roll parameter}} is of the form:
\begin{equation}
    \epsilon_{c}\equiv\frac{\dot{\sigma}^{2}}{2H^{2}}\,,
\end{equation}
so that:
\begin{equation}
    \frac{2H}{\dot{\sigma}}=\frac{2H}{\sqrt{2\epsilon_{c}}\left|H\right|}=\sqrt{\frac{2}{\epsilon_{c}}}\frac{H}{\left|H\right|}=
    \begin{dcases}
        +\sqrt{\frac{2}{\epsilon_{c}}},\quad H>0\quad\text{(expansion)}\,, \\
        -\sqrt{\frac{2}{\epsilon_{c}}},\quad H<0\quad\text{(contraction)}\,.
    \end{dcases}
\end{equation}
In the contracting phase, the conversion integral \eqref{conversion-integral} becomes:
\begin{equation}\label{xi-L-ti-tf}
    \xi_{L}\left(t_{f}\right)-\xi_{L}\left(t_{i}\right)=-\int_{t_{i}}^{t_{f}}\sqrt{\frac{2}{\epsilon_{c}}}\dot{\theta}\,\delta s^{(1)}dt\,,
\end{equation}
therefore:
\begin{equation}
    \boxed{\left|\xi_{L}\right|\propto\frac{1}{\sqrt{\epsilon_{c}}}}\,.
\end{equation}
Thus, the \textit{\textbf{EoS during the conversion process affects the efficiency of entropy-to-curvature conversion}}: \textit{larger} $\epsilon_{c}$ \textit{suppresses} the generated curvature perturbation, while \textit{smaller} $\epsilon_c$ \textit{enhances} it. This dependence also influences the magnitude of the 2nd order correction.

\begin{figure}[htbp]
    \centering
    \includegraphics[width=1\linewidth]{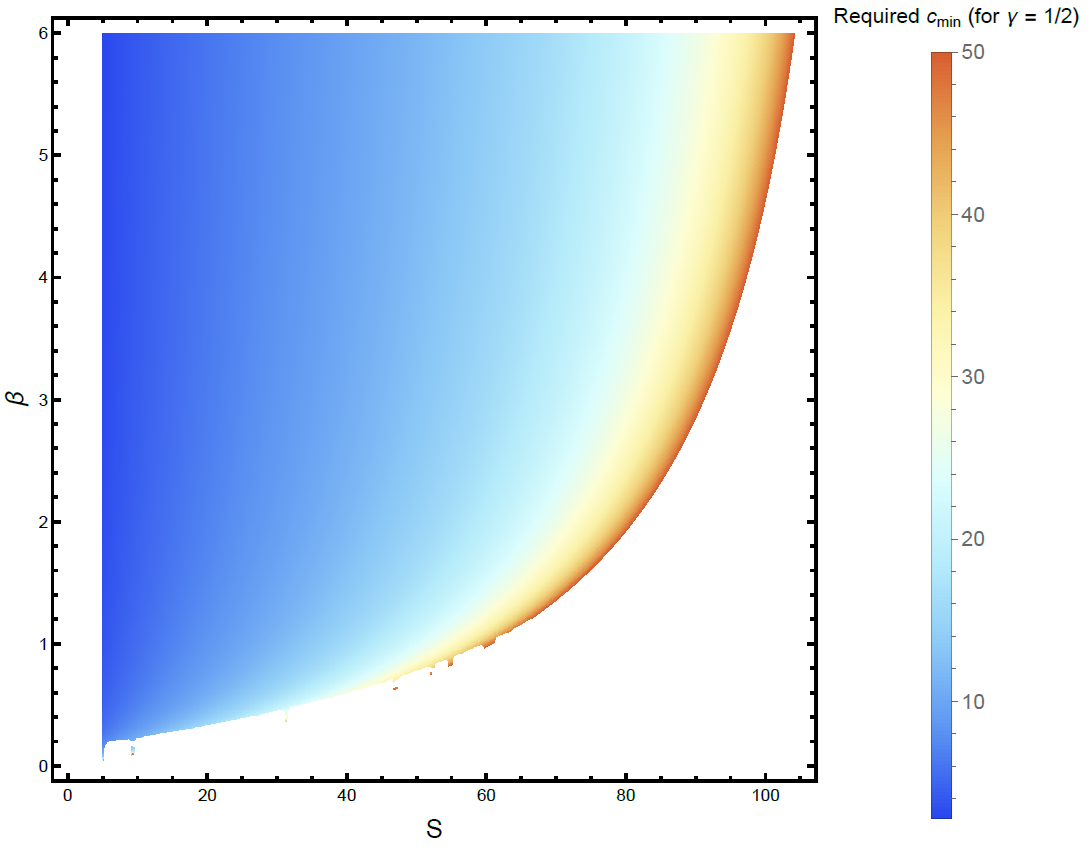}
    \caption{Parameter space $\left(S,\beta\right)$ consistent with Planck 2018 data for a fixed value of $\gamma=1/2$ (required $c_{\mathrm{min}}$).}
    \label{fig:parameter-space-S-beta}
\end{figure}

Additionally, the \textbf{background trajectory could bend due to conversion}:
\begin{examplebox}[Types of the conversion mechanism]
    \begin{enumerate}[label=(\alph*)]
        \item After the ekpyrotic epoch during kination \cite{Lehners:2007ac};
        \item During ekpyrosis \cite{Koyama:2007ag,Koyama:2007mg,Koyama:2007if} (\textit{\textbf{ruled out by observations}} \cite{Planck:2018jri,Planck:2019kim});
        \item During the ghost condensate transition \cite{Buchbinder:2007ad,Creminelli:2007aq};
        \item After the Big Bang by modulated preheating \cite{Battefeld:2007st}.
    \end{enumerate}
\end{examplebox}
\begin{examplebox}[Conversion during kination]
    \begin{itemize}
        \item In the original ekpyrotic and cyclic scenarios, the epoch of SF potential domination ends at:
        \begin{equation}
            t=t_{\mathrm{end}}<0
        \end{equation}
        before the possible cosmological bounce (Big Crunch to Big Bang transition) at:
        \begin{equation}
            t_{B}=0\,;
        \end{equation}
        \item After this comes an era of KE domination and a change in the EoS behavior:
        \begin{equation}
            \omega\gg 1\to\omega=1 \quad\iff\quad \epsilon_{\mathrm{ekp}}\gg 1\to\epsilon=3\,;
        \end{equation}
        \item This phase is when the conversion from entropy to curvature perturbations takes place \cite{Lehners:2007ac};
        \item The conversion arises naturally in the heterotic M-theory embedding of the cyclic scenario due to a negative-tension brane bouncing off a spacetime singularity \cite{Lehners:2007nb}\\
        $\implies$ A \textit{\textbf{bend of the field space trajectory}} in the $4D$ EFT before the collision with the positive-tension brane;
        \item $4D$ EFT $\implies$ There are two SF moduli fields ($\phi_{1}$ and $\phi_{2}$), that exist on the half-plane given by:
        \begin{equation}
            \phi_{1}\in\left({}^{-}\infty,{}^{+}\infty\right) \quad\land\quad \phi_{2}\in\left({}^{-}\infty,0\right)\,,
        \end{equation}
        with the boundary at:
        \begin{equation}\label{moduli-boundary}
            \phi_{2}=0\,;
        \end{equation}
        \item The cosmological solution reaches the moduli boundary \eqref{moduli-boundary} and is reflected off of it at time:
        \begin{equation}
            t=t_{\mathrm{ref}}\,;
        \end{equation}
        \item In terms of the general framework, the \textit{\textbf{SF trajectory smoothly bends during kinetic domination}}\\
        $\implies$ $N$ SFs with a generic \textit{\textbf{K{\"a}hler metric}}\footnote{For a complex SF or moduli space, the kinetic sector is naturally described by a K{\"a}hler metric, which locally can be written in terms of a K{\"a}hler potential $K$ as $G_{i\bar{j}}=\partial_i\partial_{\bar{j}}K$. This structure is standard in complex/K{\"a}hler geometry and appears naturally in supersymmetric and string compactification effective theories \cite{Huybrechts2005Kahler,Moroianu_2007Kahler,Voisin_2002Kahler}.} $g_{ij}(\phi)$ on the SF space \cite{Lehners:2007ac}:
        \begin{equation}\label{dot-xi-Kahler}
            \boxed{\dot{\xi}=\frac{H}{\dot{H}}g_{ij}(\phi)\frac{D^{2}\phi^{i}}{Dt^{2}}s^{j}}\,,
        \end{equation}
        where the $N-1$ entropy perturbations expressed as:
        \begin{equation}\label{s-i}
            s^{i}\equiv\delta\phi^{i}-\dot{\phi}^{i}\frac{g_{jk}(\phi)\dot{\phi}^{j}\delta\phi^{k}}{g_{lm}(\phi)\dot{\phi}^{l}\dot{\phi}^{m}}
        \end{equation}
        are the $\delta\phi^{i}$ components orthogonal to the background trajectory, and $D^{2}/Dt^{2}$ denotes the geodesic operator on the SF space;
        \item In our case, the SF space is flat, therefore:
        \begin{equation}
            g_{ij}=\delta_{ij}\,,
        \end{equation}
        so that:
        \begin{equation}
            \frac{D}{Dt}\to\frac{d}{dt}\,;
        \end{equation}
        \item In the case of two-field model, \eqref{s-i} becomes:
        \begin{equation}
            s^{1}=-\frac{\dot{\phi}_{2}\,\delta s}{\dot{\sigma}} \quad\land\quad s^{2}=\frac{\dot{\phi}_{1}\,\delta s}{\dot{\sigma}}\,;
        \end{equation}
        \item Assuming that the SF trajectory reflects off the boundary at \eqref{moduli-boundary}, the SF trajectory takes the simple form:
        \begin{equation}
            \dot{\phi}_{2}=
            \begin{dcases}
                -\tilde{\gamma}\dot{\phi}_{1}\,,\quad t<t_{\mathrm{ref}}\,,\\
                +\tilde{\gamma}\dot{\phi}_{1}\,,\quad t>t_{\mathrm{ref}}\,,
            \end{dcases}
        \end{equation}
        where:
        \begin{equation}
            \dot{\phi}_{1}\simeq\mathrm{const}<0
        \end{equation}
        near the bounce;
        \item The cosmological bounce yields into the \textit{Dirac delta function} on the RHS of \eqref{dot-xi-Kahler}:
        \begin{equation}
            \frac{D^{2}\phi_{2}}{Dt^{2}}=2\delta_{D}\left(t-t_{\mathrm{ref}}\right)\dot{\phi}_{2}\,t_{\mathrm{ref}}^{+}=2\tilde{\gamma}\,\dot{\phi}_{1}\delta_{D}\left(t-t_{\mathrm{ref}}\right)\,;
        \end{equation}
        \item From the higher-dimensional perspective, the time $t_{\mathrm{ref}}$ corresponds to the time of the bounce at the negative-tension brane;
        \item In fact, the relation \eqref{dot-xi-Kahler} means that there are scale-invariant entropy perturbations by the time of $t_{\mathrm{ref}}$\\
        $\implies$ \textit{\textbf{Instantaneous conversion into the same (scale-invariant) long-wavelength spectrum of curvature perturbations}};
        \item The integrated turn angle is given by:
        \begin{equation}\label{Delta-theta}
            \Delta\theta=\int\dot{\theta}\,dt=\frac{\dot{\phi}_{1}}{\dot{\sigma}^{2}}\Delta\dot{\phi}_{2}=\frac{\dot{\phi}_{1}}{\dot{\sigma}^{2}}\left(2\tilde{\gamma}\,\dot{\phi}_{1}\right)=\frac{2\tilde{\gamma}}{1+\tilde{\gamma}^{2}}\,;
        \end{equation}
        \item Plugging \eqref{Delta-theta} into \eqref{xi-L-ti-tf} and treating $\delta s$ as:
        \begin{equation}\label{delta-s-constant}
            \delta s=\mathrm{const}
        \end{equation}
        during an idealized instantaneous turn gives the following conversion rule:
        \begin{equation}
            \boxed{\xi_{L}=\sqrt{\frac{2}{\epsilon_{c}}}\Delta\theta\,\delta s^{(1)}\left(t_{\mathrm{ref}}\right)=\sqrt{\frac{2}{\epsilon_{c}}}\frac{2\tilde{\gamma}}{\left(1+\tilde{\gamma}^{2}\right)}\delta s^{(1)}\left(t_{\mathrm{ref}}\right)}\,,
        \end{equation}
        which is maximized at:
        \begin{equation}
            \tilde{\gamma}=1\,,
        \end{equation}
        with the maximum of $1$ - \textit{\textbf{the most efficient}} $\boldsymbol{2D}$ \textit{\textbf{bend for the conversion}};
        \item The background evolution during the conversion is given by:
        \begin{equation}
            a(t)\propto (-t)^{1/3} \quad\land\quad H(t)=\frac{1}{3t} \quad\land\quad \epsilon_{c}=3\,;
        \end{equation}
        \item With the assumption of straight trajectory and negligible SF potential, the evolution of the entropy mode between $t_{\mathrm{end}}$ and $t_{\mathrm{ref}}$ (large scale entropy) can be described by the simple formula:
        \begin{equation}
            \ddot{\delta s}+\frac{1}{t}\dot{\delta s}=0
        \end{equation}
        with the solution:
        \begin{equation}
            \delta s(t)=A+B\ln{(-t)}\,;
        \end{equation}
        \item Matching to the growing mode:
        \begin{equation}
            \delta s(t)\propto t^{-1}
        \end{equation}
        in the ekpyrotic phase at:
        \begin{equation}
            t=t_{\mathrm{end}}
        \end{equation}
        yields the following relation at the reflection time:
        \begin{equation}
            \delta s\left(t_{\mathrm{ref}}\right)=\delta s\left(t_{\mathrm{end}}\right)\left[1+\ln{\left(\frac{\left|t_{\mathrm{end}}\right|}{\left|t_{\mathrm{ref}}\right|}\right)}\right]\,;
        \end{equation}
        \item Moreover, the ekpyrotic phase ends at the time given by:
        \begin{equation}
            \left|V_{\mathrm{end}}\right|\simeq\frac{2}{c^{2}\,t_{\mathrm{end}}^{2}} \quad\implies\quad \boxed{t_{\mathrm{end}}^{2}\simeq\frac{2}{c^{2}\left|V_{\mathrm{end}}\right|}}\,;
        \end{equation}
        \item Quantizing $\delta S$ in the ekpyrotic vacuum gives (at the leading order):
        \begin{equation}
            \boxed{\Delta^{2}_{\delta S}(k)\Big|_{t=t_{\mathrm{end}}}=\frac{\hslash}{8\pi^{2}}\frac{c_{1}^{2}\left|V_{\mathrm{end}}\right|}{\MPl^{2}}}\,;
        \end{equation}
        \item Combining with the previous calculations gives:
        \begin{equation}
            \xi_{L}=\sqrt{\frac{2}{\epsilon_{c}}}\frac{2\tilde{\gamma}}{\left(1+\tilde{\gamma}^{2}\right)}\delta s\left(t_{\mathrm{ref}}\right)=\sqrt{\frac{2}{\epsilon_{c}}}\frac{2\tilde{\gamma}}{\left(1+\tilde{\gamma}^{2}\right)}\delta s\left(t_{\mathrm{end}}\right)\left[1+\ln{\left(\frac{\left|t_{\mathrm{end}}\right|}{\left|t_{\mathrm{ref}}\right|}\right)}\right]\,,
        \end{equation}
        therefore:
        \begin{equation}\label{Delta-xi-squared-KE}
            \boxed{\Delta_{\xi}^{2}(k)=\frac{2}{\epsilon_{c}}\frac{4\tilde{\gamma}^{2}}{\left(1+\tilde{\gamma}^{2}\right)^{2}}\Delta_{\delta s}^{2}(k)\Bigg|_{t=t_{\mathrm{end}}}\cdot\left[1+\ln{\left(\frac{\left|t_{\mathrm{end}}\right|}{\left|t_{\mathrm{ref}}\right|}\right)}\right]^{2}}\,;
        \end{equation}
        \item Thus, the variance is of the form:
        \begin{equation}\label{variance-xi-KE}
            \left\langle\xi^{2}\right\rangle=\int\frac{dk}{k}\Delta_{\xi}^{2}(k)\,;
        \end{equation}
        \item The formulae \eqref{Delta-xi-squared-KE} and \eqref{variance-xi-KE} imply that the \textbf{result depends only logarithmically on the time} $\boldsymbol{t_{\mathrm{ref}}}$, and the \textbf{main dependence is on the minimum value of the effective SF potential and} $\boldsymbol{c_{1}}$ \textbf{parameter};
        \item Cosmological observations from the CMB \cite{Planck:2018vyg} (at the $68\%$ CL):
        \begin{equation}\label{Planck-ns-As}
            \boxed{n_{s}=0.9649\pm 0.0042 \quad\land\quad A_{s}\equiv\Delta_{\xi}^{2}(k)\Big|_{k_{*}=0.05\,\mathrm{Mpc}^{-1}}\simeq 2.1\times 10^{-9}}
        \end{equation}
        requires:
        \begin{equation}
            \boxed{c_{1}\sqrt{\left|V_{\mathrm{end}}\right|}\simeq 10^{-3}\MPl}
        \end{equation}
        $\simeq$ \textit{\textbf{GUT scale}} \\
        $\implies$ \textbf{The ekpyrotic phase ends at a time of}:
        \begin{equation}
            \boxed{t\simeq 10^{3}t_{Pl}}
        \end{equation}
        before the Big Crunch.
    \end{itemize}
\end{examplebox}
Since we have established that \textit{\textbf{curvature perturbations with the correct amplitude can arise from a multiple-field ekpyrotic phase}}, it is important to discuss the \textbf{spectral index of these perturbations}, as it is a \textit{\textbf{measurable quantity}}.
\begin{examplebox}[Predictions for the spectral index]
    \begin{enumerate}[label=(\alph*)]
        \item \textbf{Model-independent estimating procedure} \cite{Khoury:2003vb}:
        \begin{itemize}
            \item Rewriting the spectral index \eqref{entropic-spectral-index} in terms of number of $e$-folds before the end of ekpyrosis, $\mathcal{N}$:
            \begin{equation}\label{d-mathcal-N}
                d\mathcal{N}=\bigl(\epsilon-1\bigr)dN\simeq\epsilon\,dN \quad\land\quad \epsilon\gg 1
            \end{equation}
            yields the following relation:
            \begin{equation}
                \frac{\epsilon_{,N}}{\epsilon^{2}}\simeq\frac{1}{\epsilon}\frac{d\epsilon}{d\mathcal{N}}=\frac{d\ln{\epsilon}}{d\mathcal{N}}\,,
            \end{equation}
            so that the \textit{\textbf{'model-independent estimator'}} takes the form:
            \begin{equation}
                \boxed{n_{s}-1=\frac{2}{\epsilon}-\frac{d\ln{\epsilon}}{d\mathcal{N}}}\,;
            \end{equation}
        \end{itemize}
        \item \textbf{Power-law ekpyrotic EoS}:
        \begin{itemize}
            \item Imposing:
            \begin{equation}
                \boxed{\epsilon\left(\mathcal{N}\right)\simeq\mathcal{N}^{\alpha} \quad\land\quad \alpha>0}
            \end{equation}
            gives:
            \begin{equation}
                \frac{2}{\epsilon}=\frac{2}{\mathcal{N}^{\alpha}} \quad\land\quad \frac{d\ln{\epsilon}}{d\mathcal{N}}=\frac{\alpha}{\mathcal{N}}\,;
            \end{equation}
            \item This produces the spectral index of the form:
            \begin{equation}
                \boxed{n_{s}-1\simeq\frac{2}{\mathcal{N}^{\alpha}}-\frac{\alpha}{\mathcal{N}}}\,;
            \end{equation}
            \item The sign of the tilt is sensitive to the parameter $\alpha$;
            \item For:
            \begin{equation}
                \mathcal{N}\sim 60\,,
            \end{equation}
            one can obtain:
            \begin{equation}
                \boxed{n_{s}\simeq
                \begin{dcases}
                    1.01667\,,\quad \alpha=1\quad\text{(slightly blue)}\,,\\
                    0.96722\,,\quad \alpha=2\quad\text{(moderately red)}\,;
                \end{dcases}}
            \end{equation}
            \item The limiting value separating the blue spectrum from the red one comes from the condition:
            \begin{equation}
                \frac{\alpha}{\mathcal{N}}=2\mathcal{N}^{-\alpha} \quad\implies\quad \boxed{\alpha=\frac{W\bigl[2\mathcal{N}\ln{\left(\mathcal{N}\right)}\bigr]}{\ln{\left(\mathcal{N}\right)}}}\,,
            \end{equation}
            where $W[x]$ denotes the \textit{Lambert} $W$ \textit{function} (\textit{product logarithm});
            \item For:
            \begin{equation}
                \mathcal{N}\sim 60
            \end{equation}
            we get:
            \begin{equation}
                \alpha\simeq 1.13777\,;
            \end{equation}
            \item The band for the entropic mechanism becomes:
            \begin{equation}
                \boxed{0.97\lesssim n_{s}\lesssim 1.02}\,;
            \end{equation}
            \item In the case of the Newtonian potential:
            \begin{equation}
                n_{\Phi}-1=-\frac{2}{\epsilon}-\frac{\epsilon_{,N}}{\epsilon^{2}} \quad\implies\quad n_{\Phi}-1=-\frac{2}{\epsilon}-\frac{d\ln{\epsilon}}{d\mathcal{N}}\,,
            \end{equation}
            therefore:
            \begin{equation}
                n_{\Phi}-1\simeq -\left(\frac{2}{\mathcal{N}^{\alpha}}+\frac{\alpha}{\mathcal{N}}\right)\,;
            \end{equation}
            \item For $\mathcal{N}\sim 60$ one obtains that the \textit{\textbf{red spectrum}}:
            \begin{equation}
                \boxed{n_{\Phi}\simeq
                \begin{dcases}
                    0.95\,,\quad \alpha=1\,,\\
                    0.96611\,,\quad \alpha=2\,,
                \end{dcases}}
            \end{equation}
            that has a \textit{maximum} corresponding to:
            \begin{equation}
                n_{\Phi}\simeq 0.970703 \quad\land\quad \alpha\simeq 1.51358\,,
            \end{equation}
            with the bound of:
            \begin{equation}
                \boxed{0.95\lesssim n_{\Phi}\lesssim 0.97}
            \end{equation}
            $\implies$ \textbf{\textit{In agreement with the observational constraints}} \cite{Planck:2018jri};
            \item At the time of \textbf{exit from the horizon}:
            \begin{equation}
                k=a\left|H\right|=\left|\mathcal{H}\right|\,,
            \end{equation}
            therefore:
            \begin{equation}
                \frac{d}{d\ln{k}}=-\frac{d}{d\mathcal{N}}
            \end{equation}
            $\implies$ \textbf{Running of the spectral tilt} has the form:
            \begin{equation}
                \boxed{\frac{d n_{s}}{d\ln{k}}=-\frac{d n_{s}}{d\mathcal{N}}=\frac{2\alpha}{\mathcal{N}^{\alpha+1}}-\frac{\alpha}{\mathcal{N}^{2}}}\,;
            \end{equation}
            \item For $\mathcal{N}\sim 60$:
            \begin{equation}\label{running-power-law}
                \boxed{\frac{d n_{s}}{d\ln{k}}\simeq
                \begin{dcases}
                    2.78\times 10^{-4}\,,\quad \alpha=1\,,\\
                    -5.37\times 10^{-4}\,,\quad \alpha=2\,;
                \end{dcases}}
            \end{equation}
            \item Observational data from the Planck mission (CMB constraints) \cite{Planck:2018jri}:
            \begin{equation}\label{Planck-running}
                \boxed{\frac{d n_{s}}{d\ln{k}}=-4.5\times 10^{-3}\pm 6.7\times 10^{-3}}
            \end{equation}
            are \textit{\textbf{in agreement with the obtained result}} \eqref{running-power-law};
            \item \textbf{Running of the running of the spectral index} takes the form:
            \begin{equation}
                \frac{d^{2}n_{s}}{d\left(\ln{k}\right)^{2}}=\frac{2\alpha\left(\alpha+1\right)}{\mathcal{N}^{\alpha+2}}-\frac{2\alpha}{\mathcal{N}^{3}}\,;
            \end{equation}
            \item For $\mathcal{N}\sim 60$:
            \begin{equation}\label{running-running-power-law}
                \boxed{\frac{d^{2}n_{s}}{d\left(\ln{k}\right)^{2}}\simeq
                \begin{dcases}
                    9.26\times 10^{-6}\,,\quad \alpha=1\,,\\
                    -1.76\times 10^{-5}\,,\quad \alpha=2\,;
                \end{dcases}}
            \end{equation}
            \item The Planck 2018 observational bound \cite{Planck:2018jri}:
            \begin{equation}\label{Planck-running-running}
                \boxed{\frac{d^{2}n_{s}}{d\left(\ln{k}\right)^{2}}=9\times 10^{-3}\pm 1.2\times 10^{-2}}
            \end{equation}
            is \textit{\textbf{in agreement with}} \eqref{running-running-power-law};
            \item For the \textbf{Newtonian potential}, the \textit{\textbf{running}} and \textit{\textbf{running of the running}} become:
            \begin{equation}
                \boxed{\frac{d n_{\Phi}}{d\ln{k}}=-\left(\frac{2\alpha}{\mathcal{N}^{\alpha+1}}+\frac{\alpha}{\mathcal{N}^{2}}\right) \quad\land\quad \frac{d^{2}n_{\Phi}}{d\left(\ln{k}\right)^{2}}=-\left(\frac{2\alpha\left(\alpha+1\right)}{\mathcal{N}^{\alpha+2}}+\frac{2\alpha}{\mathcal{N}^{3}}\right)}\,;
            \end{equation}
            \item For $\mathcal{N}\sim 60$:
            \begin{equation}
                \boxed{\frac{d n_{\Phi}}{d\ln{k}}\simeq
                \begin{dcases}
                    -8.33\times 10^{-4}\,,\quad \alpha=1\,,\\
                    -5.74\times 10^{-4}\,,\quad \alpha=2\,,
                \end{dcases}}
            \end{equation}
            and:
            \begin{equation}
                \boxed{\frac{d^{2}n_{\Phi}}{d\left(\ln{k}\right)^{2}}\simeq
                \begin{dcases}
                    -2.78\times 10^{-5}\,,\quad \alpha=1\,,\\
                    -1.94\times 10^{-5}\,,\quad \alpha=2\,,
                \end{dcases}}
            \end{equation}
            which is \textit{\textbf{in agreement with the observational data}} (\eqref{Planck-running} and \eqref{Planck-running-running});
        \end{itemize}
        \item \textbf{Ekpyrotic SF potential}:
        \begin{itemize}
            \item The two ekpyrotic SFs satisfy the relation:
            \begin{equation}
                \dot{\phi}_{2}=\gamma\,\dot{\phi}_{1}
            \end{equation}
            together with the steep SF potential of the form:
            \begin{equation}
                V\left(\phi_{1}\right)=-V_{0}\,\exp{\left[-\int c\left(\phi_{1}\right)d\phi_{1}\right]}\,;
            \end{equation}
            \item The adiabatic field along the trajectory becomes:
            \begin{equation}
                \dot{\sigma}=\sqrt{\dot{\phi}_{1}^{2}+\dot{\phi}_{2}^{2}}=\left|\dot{\phi}_{1}\right|\sqrt{1+\gamma^{2}} \quad\land\quad e_{\sigma}^{i}=\frac{\dot{\phi}_{i}}{\dot{\sigma}}=\frac{\left(1,\gamma\right)}{\sqrt{1+\gamma^{2}}}\,;
            \end{equation}
            \item The directional derivative of the SF potential along the path is given by:
            \begin{equation}
                V_{,\sigma}=e_{\sigma}^{i}\,V_{,\phi_{i}}\simeq e_{\sigma}^{1}\,V_{,\phi_{1}}=\frac{1}{\sqrt{1+\gamma^{2}}}V_{,\phi_{1}}\,,
            \end{equation}
            and therefore, restoring the $\MPl$ (knowing that $c\left(\phi_{1}\right)$ has a dimension of $M^{-1}$):
            \begin{equation}
                -\frac{V_{,\sigma}}{V}\equiv C_{\sigma}=\frac{c\left(\phi_{1}\right)}{\sqrt{1+\gamma^{2}}}
            \end{equation}
            $\implies$ The \textbf{\textit{effective steepness along the trajectory}} has the following form:
            \begin{equation}
                \boxed{c_{\sigma}\equiv \MPl\,C_{\sigma}=\MPl\frac{c\left(\phi_{1}\right)}{\sqrt{1+\gamma^{2}}}}\,;
            \end{equation}
            \item For a locally exponential ekpyrotic SF potential, the ekpyrotic scaling solution gives:
            \begin{equation}\label{epsilon-c-sigma}
                \boxed{\epsilon=\frac{1}{2}C_{\sigma}^{2}\,\MPl^{2}=\frac{c^{2}\left(\phi_{1}\right)}{2\left(1+\gamma^{2}\right)}\MPl^{2}}
            \end{equation}
            $\equiv$ \textit{\textbf{Two-field generalization of the single-field result}};
            \item The \textbf{gravitational contribution is positive (blue) and explicitly suppressed by a factor of} $\boldsymbol{\MPl^{-2}}$:
            \begin{equation}
                \boxed{\frac{2}{\epsilon}=\frac{2}{\frac{c^{2}\left(\phi_{1}\right)\MPl^{2}}{2\left(1+\gamma^{2}\right)}}=4\frac{\left(1+\gamma^{2}\right)}{c^{2}\left(\phi_{1}\right)\MPl^{2}}}\,;
            \end{equation}
            \item For the non-gravitational contribution:
            \begin{equation}
                \gamma=\mathrm{const}\,,
            \end{equation}
            so that the SF-dependence of $\epsilon$ is entirely carried by the function $c\left(\phi_{1}\right)$:
            \begin{equation}
                \ln{\epsilon}=\ln{\left[\frac{c^{2}\left(\phi_{1}\right)\MPl^{2}}{2\left(1+\gamma^{2}\right)}\right]}=2\ln{\bigl[c\left(\phi_{1}\right)\bigr]}+\mathrm{const} \quad\land\quad \frac{d\ln{\epsilon}}{d\phi_{1}}=2\frac{c_{,\phi_{1}}}{c}\,;
            \end{equation}
            \item Converting to derivative with respect to $\mathcal{N}$:
            \begin{equation}
                \frac{d\phi_{1}}{d\mathcal{N}}=\frac{\dot{\phi}_{1}}{H}=\frac{\dot{\sigma}}{H}\frac{1}{\sqrt{1+\gamma^{2}}}=\frac{\sqrt{2\epsilon}H}{H}\frac{1}{\sqrt{1+\gamma^{2}}}=\sqrt{\frac{2\epsilon}{1+\gamma^{2}}}\,,
            \end{equation}
            and using \eqref{d-mathcal-N} yields:
            \begin{equation}
                \frac{d\phi_{1}}{d\mathcal{N}}=\frac{1}{\epsilon}\frac{d\phi_{1}}{dN}=\frac{1}{\epsilon}\sqrt{\frac{2\epsilon}{1+\gamma^{2}}}=\sqrt{\frac{2}{\epsilon\left(1+\gamma^{2}\right)}}\,;
            \end{equation}
            \item Eliminating $\epsilon$ by \eqref{epsilon-c-sigma} gives:
            \begin{equation}
                \frac{d\phi_{1}}{d\mathcal{N}}=\frac{2}{c\left(\phi_{1}\right)}\,,
            \end{equation}
            so that:
            \begin{equation}
                \frac{d\ln{\epsilon}}{d\mathcal{N}}=\frac{d\ln{\epsilon}}{d\phi_{1}}\frac{d\phi_{1}}{d\mathcal{N}}=2\left(\frac{c_{,\phi_{1}}}{c}\right)\left(\frac{2}{c}\right)=4\frac{c_{,\phi_{1}}}{c^{2}}\,;
            \end{equation}
            \item Collecting the two pieces together imply:
            \begin{equation}
                \boxed{n_{s}-1=4\left[\frac{\left(1+\gamma^{2}\right)}{c^{2}\MPl^{2}}-\frac{c_{,\phi_{1}}}{c^{2}}\right]}\,;
            \end{equation}
            \item In the case of a pure exponential SF potential:
            \begin{equation}
                c_{,\phi_{1}}=0 \quad\implies\quad \boxed{n_{s}-1=4\frac{\left(1+\gamma^{2}\right)}{c^{2}\MPl^{2}}>0}
            \end{equation}
            $\implies$ \textbf{\textit{Slightly blue spectrum}};
            \item For instance, if:
            \begin{equation}
                c=\frac{20}{\MPl} \quad\land\quad \gamma=\frac{1}{2}
            \end{equation}
            the spectral tilt becomes:
            \begin{equation}
                n_{s}-1\simeq 0.0125\,;
            \end{equation}
            \item In the case of a realistic cyclic scenario, the SF potential must flatten in order to end the ekpyrotic phase:
            \begin{equation}
                c_{,\phi_{1}}>0
            \end{equation}
            $\implies$ \textbf{For} $\boldsymbol{c\gg 1}$, the \textbf{non-gravitational term is often dominant and drives the tilt red}.
        \end{itemize}
    \end{enumerate}
\end{examplebox}
\begin{examplebox}[Power-law $c\left(\phi_{1}\right)$ dependence]
    \begin{itemize}
        \item For a power-law form of the $c(\phi)$ function:
        \begin{equation}
            c(\phi)=A\,\phi^{\beta} \quad\land\quad \beta\neq -1\,,
        \end{equation}
        the \textit{\textbf{'integrated steepness'}} takes the form:
        \begin{equation}
            S=\int c(\phi)\,d\phi\,;
        \end{equation}
        \item Therefore:
        \begin{equation}
            S=\frac{A}{\beta+1}\phi^{\beta+1} \quad\implies\quad A=\frac{\left(\beta+1\right)}{\phi^{\beta+1}}S\,;
        \end{equation}
        \item The derivatives become:
        \begin{equation}
            c_{,\phi}=\beta\,A\,\phi^{\beta-1} \quad\land\quad \frac{c_{,\phi}}{c^{2}}=\frac{\beta}{\left(\beta+1\right)S}\,;
        \end{equation}
        \item If:
        \begin{equation}
            c\gg 1\,,
        \end{equation}
        then the gravitational term is small, and the dominant contribution is:
        \begin{equation}\label{ns-beta}
            n_{s}-1\simeq -\frac{4}{S}\frac{\beta}{1+\beta}\,;
        \end{equation}
        \item For:
        \begin{equation}
            S\sim 100\,,
        \end{equation}
        the spectral tilt \eqref{ns-beta} gives:
        \begin{equation}
            n_{s}-1\simeq -0.04\frac{\beta}{\left(1+\beta\right)}
        \end{equation}
        $\implies$ $n_{s}$ \textbf{agrees with the data} (\textit{slightly red}):
        \begin{equation}
            0.96\lesssim n_{s}\lesssim 1
        \end{equation}
        for the range of:
        \begin{equation}
            0<\beta<{}^{+}\infty\,.
        \end{equation}
    \end{itemize}
\end{examplebox}
\begin{examplebox}[General analysis for the $c\left(\phi_{1}\right)$ function]
    \begin{itemize}
        \item Let us introduce the notion of a \textit{\textbf{gravitational part variable}}:
        \begin{equation}
            B\equiv 4\frac{\left(1+\gamma^{2}\right)}{c^{2}\MPl^{2}}\geq 0\,,
        \end{equation}
        and set the numerical parameter:
        \begin{equation}
            y\equiv\frac{\beta}{1+\beta}\in\left(0,1\right)
        \end{equation}
        that produce the new form of the spectral index:
        \begin{equation}
            \boxed{n_{s}=1+B-\frac{4}{S}y}\,;
        \end{equation}
        \item For a fixed $B$ value, the minimum tilt (maximally red) at a given $S$ is described by:
        \begin{equation}
            n_{s}^{\mathrm{min}}(S)=1+B-\frac{4}{S}\,,
        \end{equation}
        and acquired as:
        \begin{equation}
            \beta\to{}^{+}\infty \quad\implies\quad y\to 1\,;
        \end{equation}
        \item On the other hand, the maximum tilt is:
        \begin{equation}
            n_{s}^{\mathrm{max}}=1+B
        \end{equation}
        for:
        \begin{equation}
            \beta\to 0 \quad\implies\quad y\to 0\,;
        \end{equation}
        \item Therefore, the observational bound on $n_{s}$ \eqref{Planck-ns-As} gives:
        \begin{equation}
            \boxed{n_{s}^{\mathrm{low}}=0.9607 \quad\land\quad n_{s}^{\mathrm{high}}=0.9691}\,;
        \end{equation}
        \item Analytic allowed region in the $\left(S,\beta\right)$ parameter space becomes:
        \begin{equation}
            \boxed{n_{s}^{\mathrm{low}}\leq 1+B-\frac{4}{S}y \leq n_{s}^{\mathrm{high}}}\,;
        \end{equation}
    \end{itemize}
    $\implies$
    \begin{enumerate}[label=(\alph*)]
        \item \textbf{Existence} (\textit{upper bound on} $S$) - \textbf{the reddest point}:
        \begin{equation}
            \boxed{S\leq S_{\mathrm{max}}(B)\equiv\frac{4}{1+B-n_{s}^{\mathrm{high}}}}\,;
        \end{equation}
        \begin{enumerate}[label=(\roman*)]
            \item $B\to 0$:
            \begin{equation}
                S_{\mathrm{max}}(0)\simeq 129.4\,;
            \end{equation}
            \item $c=20/\MPl$ and $\gamma=1/2$:
            \begin{equation}
                B=0.0125 \quad\implies\quad S_{\mathrm{max}}\simeq 92.2\,;
            \end{equation}
            \item $c=50/\MPl$ and $\gamma=1/2$:
            \begin{equation}
                B=0.002 \quad\implies\quad S_{\mathrm{max}}\simeq 121.6
            \end{equation}
        \end{enumerate}
        $\implies$ The \textbf{enormous integrated steepnesses} (e.g. $S\sim 10^{3}$) \textbf{are ruled out} in this minimal approach (they are \textit{too blue even at} $y=1$);
        \item Allowed range of $\beta$ at a fixed $S$ and $B$:
        \begin{itemize}
            \item Let us introduce:
            \begin{equation}
                y_{\mathrm{min}}=\frac{S}{4}\Bigl(1+B-n_{s}^{\mathrm{high}}\Bigr) \quad\land\quad y_{\mathrm{max}}=\frac{S}{4}\Bigl(1+B-n_{s}^{\mathrm{low}}\Bigr)\,;
            \end{equation}
            \item Then:
            \begin{equation}
                \beta_{\mathrm{min}}\left(S,B\right)=\frac{y_{\mathrm{min}}}{1-y_{\mathrm{min}}} \quad\land\quad \beta_{\mathrm{max}}\left(S,B\right)=\frac{y_{\mathrm{max}}}{1-y_{\mathrm{max}}}\,,
            \end{equation}
            with the assumption that if $y_{\mathrm{min}}\leq 0$ one takes $\beta_{\mathrm{min}}=0$ and if $y_{\mathrm{max}}\geq 1$ the interval extends to $\beta\to{}^{+}\infty$;
            \item The examples (with $B=0$, for simplicity):
            \begin{enumerate}[label=(\roman*)]
                \item $S=60$:
                \begin{equation}
                    y\in\left[0.4635,0.5895\right] \quad\implies\quad \beta\in\left[0.864,1.436\right]\,;
                \end{equation}
                \item $S=90$:
                \begin{equation}
                    y\in\left[0.6953,0.8843\right] \quad\implies\quad \beta\in\left[2.282,7.639\right]\,;
                \end{equation}
                \item $S\to S_{\infty}(0)=101.8$:
                \begin{equation}
                    y\to 1 \quad\implies\quad \beta_{\mathrm{max}}\to{}^{+}\infty\,,
                \end{equation}
                where:
                \begin{equation}
                    S_{\infty}(B)\equiv\frac{4}{1+B-n_s^{\mathrm{low}}}\,.
                \end{equation}
            \end{enumerate}
        \end{itemize}
    \end{enumerate}
    \underline{\textit{\textbf{Conclusion:}}}\\
    A larger $S$, which \textit{suppresses the red contribution} $-4y/S$, \textbf{must be compensated} by a larger $\beta$, i.e. by a \textit{faster local variation of the steepness function}, \textbf{in order to reproduce the observed slightly red tilt} (Fig.~\ref{fig:parameter-space-S-beta}).
\end{examplebox}

\subsection{Non-Gaussianity}
In the ekpyrotic models, the SF potential is very steep, and therefore, the self-interactions of the SFs can be important \cite{Koyama:2007ag,Lehners:2007ac,Lehners:2010fy,Lehners:2008my}. In particular, in the entropic mechanism, the same steepness that is needed to generate a nearly scale-invariant entropy spectrum also enhances non-linear interactions in the entropy direction. In a consequence, \textit{\textbf{many canonical two-field ekpyrotic models predict a sizeable local-type non-Gaussianity}} \cite{Buchbinder:2007ad,Lehners:2007ac,Koyama:2007ag,Lehners:2008my, Lehners:2010fy}. This feature \textbf{distinguishes them from minimal single-field slow-roll inflation, where the non-Gaussian signal is slow-roll suppressed} \cite{Bartolo:2004if,Maldacena:2002vr,Acquaviva:2002ud}, although more general inflationary scenarios can also generate observable non-Gaussianity \cite{Bartolo:2004if,Chen:2010xka}.

\textit{\textbf{Non-Gaussian observables are therefore a particularly important discriminator between the early Universe scenarios}}. Some simple ekpyrotic conversion mechanisms, especially \textbf{conversion during the ekpyrotic phase, are strongly constrained or excluded by current bounds on primordial non-Gaussianity} \cite{Planck:2019kim,Koyama:2007ag}, whereas other realizations can give values close to the sensitivity of present or future cosmological surveys \cite{Meerburg:2019qqi,Lehners:2010fy,Alvarez:2014vva}. Thus, the \textit{\textbf{non-Gaussian predictions provide a crucial test of the viability of ekpyrotic and cyclic models}} and offer a direct probe of the non-linear physics operating in the very early Universe.

\subsubsection{Perturbations in the two-field model}
The general form of the EoM for the entropy perturbation (up to a 2nd order) on large scales is given by \cite{Langlois:2006vv}\footnote{On large scales, $\delta\varepsilon_{\mathrm{com}}^{(1)}$ is gradient-suppressed. At the 2nd order, $\delta\varepsilon_{\mathrm{com}}^{(2)}$ can also be neglected
when the non-local momentum contribution is negligible. However, in a contracting background this should be regarded as an approximation rather than an automatic decay argument.}:
\begin{equation}\label{EoM-delta-s-general}
    \begin{aligned}
        & \ddot{\delta s}+3H\dot{\delta s}+\left(V_{ss}+3\dot{\theta}^{2}\right)\delta s+\frac{\dot{\theta}}{\dot{\sigma}}\left(\dot{\delta s}^{(1)}\right)^{2}+\frac{2}{\dot{\sigma}}\left(\ddot{\theta}+\dot{\theta}\frac{V_{\sigma}}{\dot{\sigma}}-\frac{3}{2}H\dot{\theta}\right)\dot{\delta s}^{(1)}\delta s^{(1)} \\
        & + \left(\frac{1}{2}V_{sss}-5\frac{\dot{\theta}}{\dot{\sigma}}V_{ss}-9\frac{\dot{\theta}^{3}}{\dot{\sigma}}\right)\left(\delta s^{(1)}\right)^{2}+2\frac{\dot{\theta}}{\dot{\sigma}}\delta\varepsilon_{\mathrm{com}}^{(2)}=0\,,
    \end{aligned}
\end{equation}
where:
\begin{equation}
    \delta\varepsilon_{\mathrm{com}}^{(1)}=\delta\rho^{(1)}-\frac{\dot{\rho}}{\dot{\sigma}}\delta\sigma^{(1)}
\end{equation}
and
\begin{equation}
    \delta\varepsilon_{\mathrm{com}}^{(2)}=\delta\rho^{(2)}-\frac{\dot{\rho}}{\dot{\sigma}}\delta\sigma^{(2)}-\frac{\delta\sigma}{\dot{\sigma}}\left[\dot{\delta\varepsilon}_{\mathrm{com}}^{(1)}+\frac{1}{2}\dot{\left(\frac{\dot{\rho}}{\dot{\sigma}}\right)}\delta\sigma+\frac{\dot{\rho}}{\dot{\sigma}}\dot{\theta}\delta s\right]
\end{equation}
denote the comoving energy density perturbations.

During the generation of entropy perturbations, the background trajectory is
assumed to be straight:
\begin{equation}
    \dot{\theta}=0\,,
\end{equation}
so that entropy perturbations are not yet converted into curvature perturbations. Expanding the entropy potential around the background trajectory up to a 2nd order in $ \delta s$, on large cosmological scales, one obtains \cite{Lehners:2010fy}:
\begin{equation}\label{delta-s-EoM-2nd-order}
    \ddot{\delta s}+3H\dot{\delta s}+V_{ss}\delta s+\frac{1}{2}V_{sss}\left(\delta s^{(1)}\right)^{2}=0\,.
\end{equation}
Here $V_{ss}$ and $V_{sss}$ denote derivatives of the SF potential in the entropy direction, evaluated on the background trajectory. The quadratic
term proportional to $V_{sss}$ is responsible for the \textit{\textbf{intrinsic non-linear evolution of the entropy perturbation and contributes to the local-type non-Gaussianity generated in the entropic ekpyrotic mechanism}} \cite{Buchbinder:2007ad,Lehners:2007ac,Gordon:2000hv,Koyama:2007ag,Lehners:2010fy,Lehners:2008my}.

In the case of constant exponents in the SF potential:
\begin{equation}
    c_{i}=\mathrm{const}\,,
\end{equation}
one gets that:
\begin{equation}
    V_{ss}=-\frac{2}{t^{2}} \quad\land\quad \frac{1}{2}V_{sss}=\frac{\left(1-\gamma^{2}\right)}{\gamma\sqrt{\left(1+\gamma^{2}\right)}\,t^{2}}\sqrt{\left|\gamma\,c_{1}\,c_{2}\right|}\,.
\end{equation}
The expansion of the entropy perturbation (up to a 2nd order) becomes \cite{Koyama:2007if,Lehners:2007wc}:
\begin{equation}\label{delta-s-expansion}
    \delta s(t)=\delta s^{(1)}(t)+\delta s^{(2)}(t)=\delta s_{\mathrm{end}}\frac{t_{\mathrm{end}}}{t}+\tilde{c}\left(\delta s_{\mathrm{end}}\frac{t_{\mathrm{end}}}{t}\right)^{2}\,,
\end{equation}
with:
\begin{equation}\label{tilde-c}
    \tilde{c}\equiv\frac{\left(\gamma^{2}-1\right)}{4\gamma\sqrt{\left(1+\gamma^{2}\right)}}\sqrt{\left|\gamma\,c_{1}\,c_{2}\right|}=\frac{\left(\gamma^{2}-1\right)}{4\gamma}\sqrt{2\epsilon_{\mathrm{ekp}}}\,.
\end{equation}
\begin{examplebox}[Conclusions]
\begin{itemize}
    \item The \textit{\textbf{intrinsic non-Gaussianity associated with ekpyrotic models}} (2nd term in \eqref{delta-s-expansion}) is of $\mathcal{O}\left(\sqrt{\epsilon_{\mathrm{ekp}}}\right)$ and \textit{\textbf{arises from the steepness of the SF potentials and the SFs' self-interactions}};
    \item The situation is \textbf{completely different when it comes to inflation, where the intrinsic non-Gaussianity is very small because the SF potential is flat}:
    \begin{equation}
        \epsilon_{\mathrm{infl}}\ll1\,.
    \end{equation}
\end{itemize}
\end{examplebox}
Nevertheless, rather than measuring the non-Gaussianity present in the entropy perturbation itself, we measure the non-Gaussianity imprinted on the curvature perturbation. Therefore, it is important to understand how strongly this intrinsic non-Gaussianity is transferred to the curvature perturbation. Therefore, let us consider the evolution of the curvature perturbation at the long-wavelength limit \cite{Langlois:2006vv}:
\begin{equation}\label{non-linear-xi}
    \boxed{\dot{\xi}=2\frac{H}{\dot{\sigma}}\dot{\theta}\,\delta s-\frac{H}{\dot{\sigma}^{2}}\left(V_{ss}+4\dot{\theta}^{2}\right)\left(\delta s^{(1)}\right)^{2}}\,.
\end{equation}
Here, the 1st term describes the linear source of curvature perturbations by entropy perturbations whenever the background trajectory bends ($\dot{\theta}\neq 0$), while the 2nd term describes the non-linear transfer of entropy perturbations into the curvature perturbation.

\begin{figure}[htbp]
            \centering
            \includegraphics[width=1\linewidth]{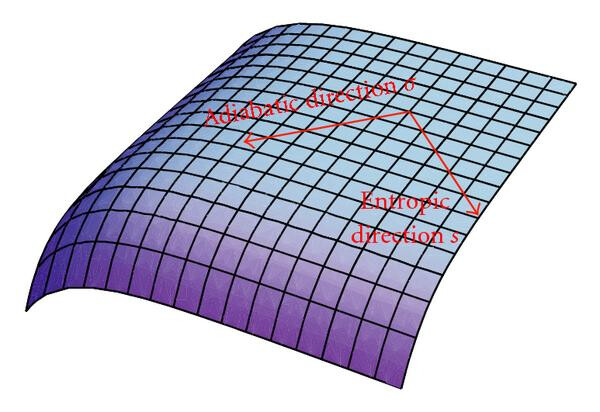}
            \caption{The two-field ekpyrotic potential combines an ekpyrotic/'steep' ($\sigma$) direction with a transverse tachyonic/'shallow' ($s$) direction \cite{Lehners:2010fy}.}
            \label{fig:potential-steep-shallow}
\end{figure}

\begin{figure}[htbp]
            \centering
            \includegraphics[width=1\linewidth]{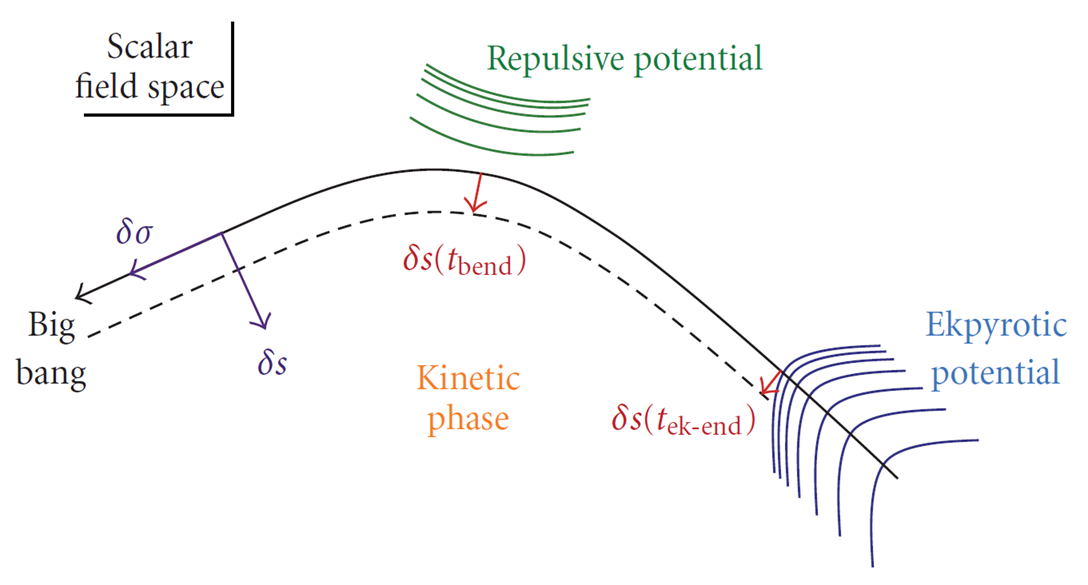}
            \caption{Schematic field-space picture of the entropy-to-curvature conversion in the two-field ekpyrotic scenario. After the ekpyrotic phase, the background SF trajectory enters a kinetic phase and is bent by a repulsive potential. This bending converts entropy perturbations into adiabatic perturbations \cite{Lehners:2010fy}.}
            \label{fig:bending-repulsive-potential}
\end{figure}

\begin{figure}[htbp]
    \centering
    \includegraphics[width=1\linewidth]{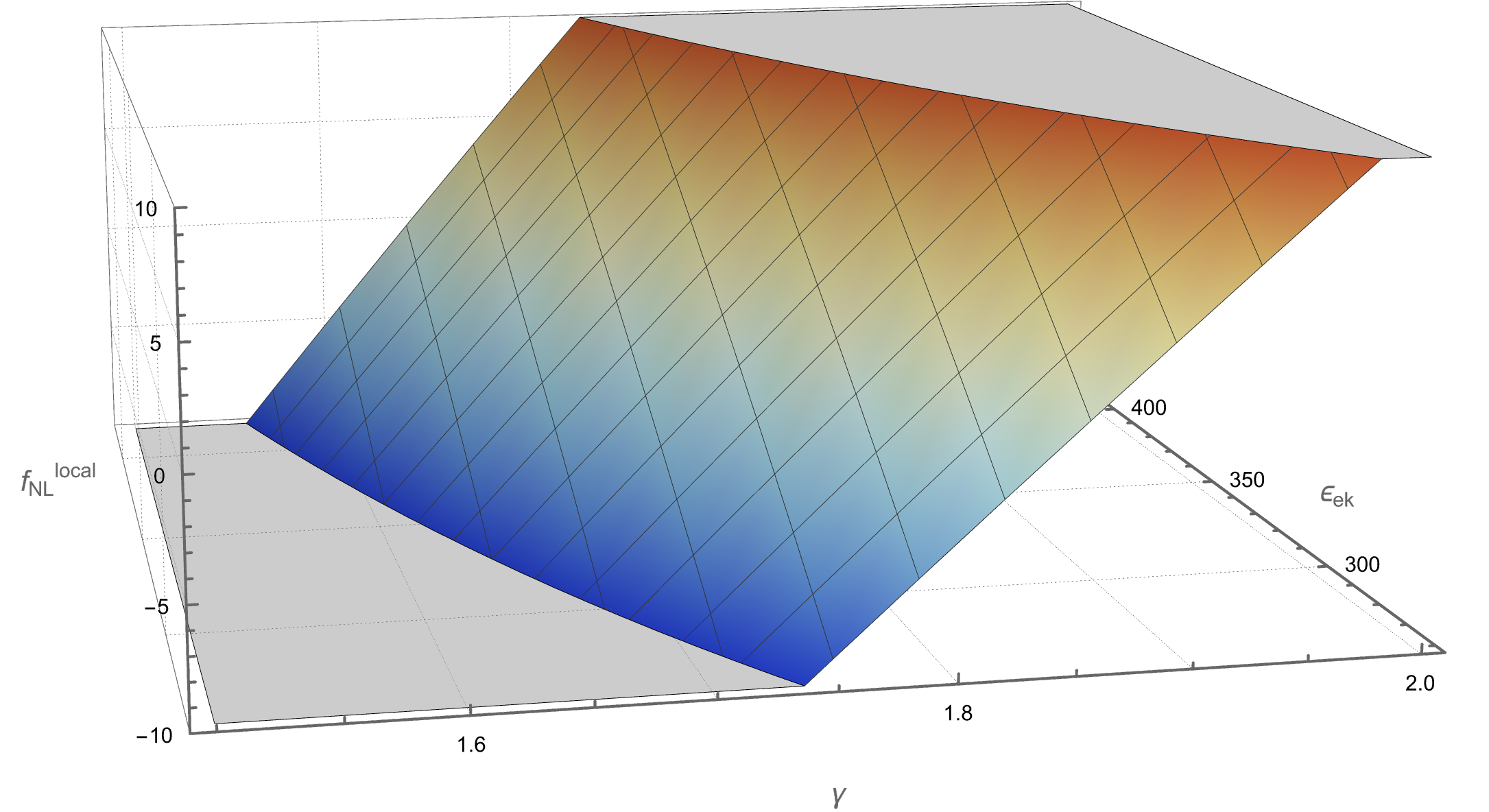}
    \caption{The allowed $\left(\gamma,\epsilon_{\mathrm{ekp}}\right)$ parameter space (for $\gamma>0$) based on the assumption of $\left|f_{NL}^{\mathrm{local}}\right|\lesssim 10$.}
    \label{fig:fNL-positive-gamma}
\end{figure}

\begin{figure}[htbp]
    \centering
    \includegraphics[width=1\linewidth]{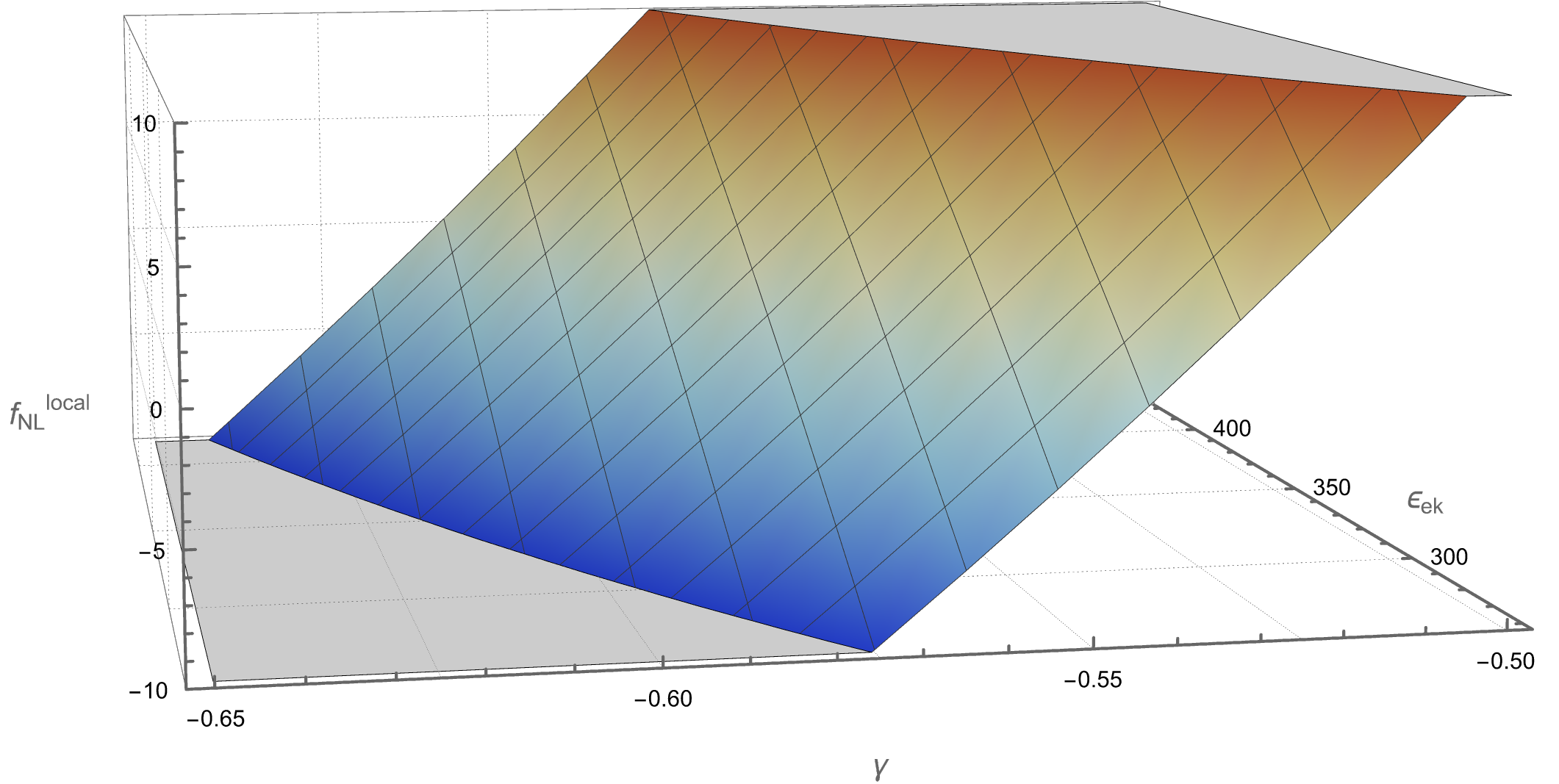}
    \caption{The allowed $\left(\gamma,\epsilon_{\mathrm{ekp}}\right)$ parameter space (for $\gamma<0$) based on the assumption of $\left|f_{NL}^{\mathrm{local}}\right|\lesssim 10$.}
    \label{fig:fNL-negative-gamma}
\end{figure}

In such a case, the local wavelength-independent non-Gaussian contribution to $\xi$ could be described in terms of linear Gaussian curvature perturbation $\xi_{L}$ as follows \cite{Komatsu:2001rj}:
\begin{equation}
    \xi=\xi_{L}+\frac{3}{5}f_{NL}\,\xi_{L}^{2}\,.
\end{equation}
The two-field formalism described by \eqref{delta-s-EoM-2nd-order} yields a useful decomposition of the local non-Gaussianity generated during the entropy-to-curvature conversion \cite{Lehners:2010fy,Lehners:2008my,Lehners:2007wc}:
\begin{align}
    f_{NL}^{\mathrm{intrinsic}} & = \frac{5}{3\xi_{L}^{2}}\int 2\frac{H}{\dot{\sigma}}\dot{\theta}\,\delta s^{(2)}dt \,,\label{fNL-intrinsic}\\
    f_{NL}^{\mathrm{reflection}} & = -\frac{5}{3\xi_{L}^{2}}\int\frac{H}{\dot{\sigma}^{2}}\left(V_{ss}+4\dot{\theta}^{2}\right)\left(\delta s^{(1)}\right)^{2}dt \,,\label{fNL-reflection}\\
    f_{NL}^{\mathrm{integrated}} & = \frac{5}{6\xi_{L}^{2}}\left[\delta s^{(1)}\left(t_{\mathrm{end}}\right)\right]^{2} \,,\label{fNL-integrated}
\end{align}
with the \textbf{total non-Gaussianity parameter}:
\begin{equation}
    \boxed{f_{NL}=f_{NL}^{\mathrm{intrinsic}}+f_{NL}^{\mathrm{reflection}}+f_{NL}^{\mathrm{integrated}}}\,.
\end{equation}
\begin{examplebox}[Physical interpretation of specific contributions to non-Gaussianity]
    \begin{enumerate}[label=(\alph*)]
        \item $f_{NL}^{\mathrm{intrinsic}}$ contribution - results from the \textit{\textbf{direct conversion of the intrinsic non-linearity associated with the entropy perturbation into a corresponding non-linearity in the curvature perturbation}};
        \item $f_{NL}^{\mathrm{reflection}}$ contribution - comes from the \textit{\textbf{non-linear source term}} in \eqref{non-linear-xi} and is \textbf{generated during the conversion};
        \item $f_{NL}^{\mathrm{integrated}}$ contribution - arises from the \textit{\textbf{2nd order sourcing of the curvature perturbation by the entropy perturbation}} in \eqref{non-linear-xi} and is \textbf{produced during the ekpyrotic phase before the conversion}.
    \end{enumerate}
\end{examplebox}
Combining \eqref{xi-L-ti-tf}, \eqref{delta-s-expansion}, \eqref{tilde-c} and \eqref{fNL-intrinsic} implies the \textit{\textbf{estimation for the intrinsic part of non-Gaussianity}}:
\begin{examplebox}[Estimation for the intrinsic part of non-Gaussianity]
    \begin{equation}
        \boxed{f_{NL}^{\mathrm{intrinsic}}\simeq\pm\frac{5}{3\xi_{L}^{2}}\int\sqrt{\frac{2}{\epsilon_{c}}}\,\dot{\theta}\,\tilde{c}\,\left(\delta s^{(1)}\right)^{2}dt\sim\frac{\sqrt{\epsilon_{c}\,\epsilon_{\mathrm{ekp}}}}{\Delta\theta}\Bigg|_{\Delta\theta\sim\mathcal{O}(1)}\sim\sqrt{\epsilon_{c}\,\epsilon_{\mathrm{ekp}}}}
    \end{equation}
    $\equiv$ \textbf{The geometric mean of the} $\boldsymbol{\epsilon}$ \textbf{parameters during the generation and conversion processes}.
    \begin{itemize}
        \item \underline{\textbf{Provides a basic estimate for} $\boldsymbol{\left|f_{NL}\right|}$};
        \item Explains why the \textit{\textbf{non-Gaussianity in ekpyrotic (cyclic) models exceeds that in simple inflationary models by more than an order of magnitude}}, and \textit{\textbf{how the EoS during the conversion could significantly impact the predictions}}.
    \end{itemize}
\end{examplebox}
In order to improve upon this qualitative estimate and make it more precise, we need to consider the details of the conversion mechanism. As results independent of a specific model implementation can only be obtained when a conversion occurs while the KE of the SF is dominant, our discussion will focus exclusively on this scenario.
\begin{examplebox}[Conversion during the kination]
    \begin{itemize}
        \item During the kinetic-domination stage:
        \begin{equation}
            \epsilon_{c}=3 \quad\implies\quad \boxed{f_{NL}\sim\sqrt{\epsilon_{\mathrm{ekp}}}\sim\mathcal{O}\left(c_{1}\right)}\,;
        \end{equation}
        \item If the exponents in the potential for ekpyrotic SFs are not equal:
        \begin{equation}
            c_{1}\neq c_{2}\,,
        \end{equation}
        then, during the ekpyrotic phase, the potential decreases more rapidly on one side of the background SF trajectory ('ridge') than on the other\\
        $\implies$ \textit{\textbf{'Steep' and 'shallow' directions}} (Fig.~\ref{fig:potential-steep-shallow});
        \item The bend may be understood as a situation in which one of the SFs is reflected, while the other is not significantly altered;
        \item Without loss of generality, one can select trajectories that bend towards the 'shallow' direction (e.g. cyclic models based on heterotic M-theory with $\gamma=-1/\sqrt{3}$) \cite{Lehners:2008vx} (Fig.~\ref{fig:ridge-SF-potential}):
        \begin{equation}
            \gamma<0 \quad\land\quad \dot{\theta}>0\,,
        \end{equation}
        so that:
        \begin{equation}
            -1<\gamma<0 \quad\land\quad \tilde{c}>0\,;
        \end{equation}
        \item Analogously, choosing:
        \begin{equation}
            \gamma<-1 \quad\land\quad \tilde{c}<0
        \end{equation}
        corresponds to bending towards the 'steep' direction;
        \item As shown in \cite{Lehners:2008my}, the assumption that only the field $\phi_{2}$ reflects \textbf{does not change the result};
        \item In the context of heterotic M-theory, the reflection takes place because the field $\phi_{2}$ approaches the boundary of moduli space \eqref{moduli-boundary} and experiences a bounce \cite{Lehners:2007nb};
        \item The reflection is likely due to a \textit{\textbf{'reflection/repulsive potential'}}, $V^{R}\left(\phi_{2}\right)$, which is a \textit{\textbf{separate contribution that is not associated with the exponential SFs potentials}} \cite{Lehners:2008vx} (Fig.~\ref{fig:bending-repulsive-potential});
        \item For a pure kination before the conversion of the perturbations, the background SF trajectory is also a straight line during the kinetic phase before the bend, therefore, the EoM becomes:
        \begin{equation}
            \ddot{\delta s}+\frac{1}{t}\dot{\delta s}=0\,;
        \end{equation}
        \item Matching with the ekpyrotic solution \eqref{delta-s-expansion} at $t=t_{\mathrm{end}}$ yields:
        \begin{equation}
            \boxed{\delta s(t)=\delta s_{\mathrm{end}}\left[1+\ln{\left(\frac{\left|t_{\mathrm{end}}\right|}{\left|t\right|}\right)}\right]+\tilde{c}\,\delta s_{\mathrm{end}}^{2}\left[1+2\ln{\left(\frac{\left|t_{\mathrm{end}}\right|}{\left|t\right|}\right)}\right]}\,;
        \end{equation}
        \item Moreover, one can also deduce that during kination:
        \begin{equation}
            t\frac{\dot{\delta s}}{\delta s}\simeq -\frac{1}{1+\ln{\left(\frac{\left|t_{\mathrm{end}}\right|}{\left|t\right|}\right)}}\,,
        \end{equation}
        and during the ekpyrosis:
        \begin{equation}
            t\,\dot{\delta s}\sim\delta s\,;
        \end{equation}
        \item During the conversion, the KE of the SF is dominant, but $V^{R}\left(\phi_{2}\right)$ also significantly influences the evolution of entropy perturbations;
        \item For a \textit{\textbf{'gradual bending' approximation}} \cite{Buchbinder:2007at}:
        \begin{equation}
            \dot{\theta}=\mathrm{const}\neq 0 \quad\land\quad t\in\left[t_{\mathrm{ref}},t_{\mathrm{ref}}+\Delta t\right]\,,
        \end{equation}
        with the assumption:
        \begin{equation}
            \Delta\theta\sim\mathcal{O}(1)\,\mathrm{radian}\,,
        \end{equation}
        yields the \textbf{total bending angle} of the form:
        \begin{equation}
            \left|\dot{\theta}\right|\simeq\frac{1}{\left|t_{\mathrm{ref}}\right|}\,;
        \end{equation}
        \item In this case, we can relate the 2nd derivative of the repulsive potential with respect to $s$ to the rate of change of the angle $\dot{\theta}$:
        \begin{equation}\label{V-R-ss}
            V^{R}_{ss}=-2\dot{\theta}^{2}+\frac{\dot{\theta}}{\gamma\,t}\,;
        \end{equation}
        \item Furthermore, in the context of 'gradual' conversion:
        \begin{equation}
            \Delta t\sim \left|t_{\mathrm{ref}}\right|\,,
        \end{equation}
        so that one could ignore higher-derivatives of the bending angle;
        \item Furthermore, \textbf{sharp transitions} result in values of the parameter $f_{NL}$ that are \textbf{too high from an observational perspective} \cite{Lehners:2007wc};
        \item At the linear order, the EoM \eqref{EoM-delta-s-general} simplifies into:
        \begin{equation}
            \ddot{\delta s}^{(1)}+3H\dot{\delta s}^{(1)}+\left(\dot{\theta}^{2}+\frac{\dot{\theta}}{\gamma\,t}\right)\delta s^{(1)}=0\,;
        \end{equation}
        \item Using the relation for a 1st order \eqref{delta-s-constant} one could set:
        \begin{equation}
            \dot{\delta s}^{(1)}=0\,,
        \end{equation}
        therefore, one can neglect the (Hubble) damping piece in the EoM:
        \begin{equation}\label{delta-s-1-without-damping}
            \ddot{\delta s}^{(1)}+\left(\dot{\theta}^{2}+\frac{\dot{\theta}}{\gamma\,t}\right)\delta s^{(1)}=0\,;
        \end{equation}
        \item Introducing a time-dependent coefficient:
        \begin{equation}
            Q(t)\equiv\dot{\theta}^{2}+\frac{\dot{\theta}}{\gamma\,t}\,,
        \end{equation}
        allows for use of the 'frozen-coefficient' approximation;
        \item Evaluation of $Q(t)$ at the midpoint:
        \begin{equation}
            t_{*}\equiv t_{\mathrm{ref}}+\frac{1}{2}\Delta t
        \end{equation}
        gives:
        \begin{equation}
            Q(t)\simeq Q\left(t_{*}\right)\equiv\omega^{2}=\dot{\theta}^{2}+\frac{\dot{\theta}}{\gamma\,t_{*}}\,,
        \end{equation}
        so that:
        \begin{equation}
            \omega\equiv\dot{\theta}\sqrt{1+\frac{1}{\dot{\theta}\,\gamma\,t_{*}}}\,;
        \end{equation}
        \item For gradual-like reflections with:
        \begin{equation}
            \left|\dot{\theta}\,\gamma\,t_{*}\right|\sim\mathcal{O}(1)
        \end{equation}
        this yields:
        \begin{equation}
            \omega\simeq\mathcal{O}(1)\dot{\theta}\,;
        \end{equation}
        \item Thus, the EoM \eqref{delta-s-1-without-damping} becomes of a form for an \textbf{oscillator with a constant frequency}:
        \begin{equation}
            \ddot{\delta s}^{(1)}+\omega^{2}\delta s^{(1)}\simeq 0 \,,
        \end{equation}
        with an explicit solution:
        \begin{equation}\label{delta-s-1-solution}
            \boxed{\delta s^{(1)}(t)=\delta s^{(1)}\left(t_{\mathrm{ref}}\right)\cos{\Bigl[\omega\bigl(t-t_{\mathrm{ref}}\bigr)\Bigr]}}
        \end{equation}
        $\implies$ \textit{\textbf{Rather than undergoing continuous logarithmic growth, the entropic perturbation can suppress the amplitude during the conversion}};
        \item It is important to note that the \textit{\textbf{conversion from entropy to curvature perturbations is not as effective as one might naively expect}}:
        \begin{equation}
            \xi_{L}=-\sqrt{\frac{2}{3}}\dot{\theta}\int\delta s^{(1)}dt=-\sqrt{\frac{2}{3}}\frac{\dot{\theta}}{\omega}\delta s^{(1)}\left(t_{\mathrm{ref}}\right)\sin{\left(\omega\Delta t\right)}\,,
        \end{equation}
        where:
        \begin{equation}
            \omega\Delta t\simeq 3 \quad\implies\quad \sin{\left(\omega\Delta t\right)}\simeq\frac{1}{3}
        \end{equation}
        in the case of subdominant reflections;
        \item The \textbf{2nd order entropy perturbation} satisfies the EoM of the form:
        \begin{equation}
            \ddot{\delta s}^{(2)}+\omega^{2}\left[\delta s^{(2)}-\frac{\dot{\theta}}{\dot{\sigma}}\left(\delta s^{(1)}\right)^{2}\right]=0
        \end{equation}
        $\implies$ \textit{\textbf{The 2nd order entropy perturbation is sourced by the square of the linear entropy perturbation during the bending phase}};
        \item Using \eqref{delta-s-1-solution} with:
        \begin{equation}
            \alpha\equiv\frac{\dot{\theta}}{\dot{\sigma}}=\mathrm{const}\simeq\mathcal{O}(1)
        \end{equation}
        at the beginning of the reflection with the initial condition:
        \begin{equation}
            \dot{\delta s}^{(2)}\left(t_{\mathrm{ref}}\right)\simeq 0
        \end{equation}
        gives the following EoM:
        \begin{equation}
            \boxed{\ddot{\delta s}^{(2)}+\omega^{2}\delta s^{(2)}=\omega^{2}\alpha\left(\delta s^{(1)}\right)^{2}}\,,
        \end{equation}
        with the solution of the form:
        \begin{equation}
            \boxed{\begin{aligned}
                \delta s^{(2)}(t)= &\,\delta s^{(2)}\left(t_{\mathrm{ref}}\right)\cos{\Bigl[\omega\bigl(t-t_{\mathrm{ref}}\bigr)\Bigr]} \\
                & +\alpha\left(\delta s^{(1)}\left(t_{\mathrm{ref}}\right)\right)^{2}\Biggl\{\frac{1}{2}-\frac{1}{3}\cos{\Bigl[\omega\bigl(t-t_{\mathrm{ref}}\bigr)\Bigr]}-\frac{1}{6}\cos{\Bigl[2\omega\bigl(t-t_{\mathrm{ref}}\bigr)\Bigr]}\Biggr\}
            \end{aligned}}\,;
        \end{equation}
        \item For large values of $\epsilon_{\mathrm{ekp}}$ ($|\tilde{c}|$) the initial intrinsic contribution:
        \begin{equation}
            \delta s^{(2)}\left(t_{\mathrm{ref}}\right)\sim\tilde{c}\left[\delta s^{(1)}\left(t_{\mathrm{ref}}\right)\right]^2
        \end{equation}
        dominates over the additional source term generated during the reflection. In a consequence, the 2nd order entropy perturbation approximately follows the same oscillatory behavior as the 1st order perturbation during the bending phase;
        \item Nevertheless, \textbf{small values of} $\boldsymbol{|\tilde{c}|}$ \textbf{generate significant corrections};
        \item In order to evaluate all of the contributions for $f_{NL}$ we must also compute the following integral:
        \begin{equation}
        \begin{aligned}
            \left[\delta s^{(1)}\left(t_{\mathrm{ref}}\right)\right]^{-2}\int\delta s^{(2)}dt= &\,\tilde{c}\frac{1+2\ln{\left(\frac{\left|t_{\mathrm{end}}\right|}{\left|t_{\mathrm{ref}}\right|}\right)}}{\left[1+\ln{\left(\frac{\left|t_{\mathrm{end}}\right|}{\left|t_{\mathrm{ref}}\right|}\right)}\right]^{2}}\frac{\sin{\left(\omega\Delta t\right)}}{\omega} \\
            & +\alpha\left[\frac{1}{2}\Delta t-\underbrace{\left(\frac{\sin{\left(\omega\Delta t\right)}}{3\omega}+\frac{\sin{\left(2\omega\Delta t\right)}}{12\omega}\right)}_{\to 0}\right]\,;
        \end{aligned}
        \end{equation}
        \item Now, the \textit{\textbf{intrinsic part of non-Gaussianity}} becomes:
        \begin{equation}
            \boxed{f_{NL}^{\mathrm{intrinsic}}\simeq A_{\mathrm{conv}}\sqrt{\epsilon_{\mathrm{ekp}}}+B_{\mathrm{conv}}}\,,
        \end{equation}
        with:
        \begin{equation}
            X\equiv\omega\Delta t \quad\land\quad L\equiv\ln{\left(\frac{\left|t_{\mathrm{end}}\right|}{\left|t_{\mathrm{ref}}\right|}\right)}\,,
        \end{equation}
        where:
        \begin{equation}
            \boxed{A_{\mathrm{conv}}\equiv -\frac{5}{4\sqrt{3}}\frac{\left(\gamma^{2}-1\right)}{\gamma}\frac{\omega}{\dot{\theta}}\frac{1}{\sin{X}}\frac{\left(1+2L\right)}{\left(1+L\right)^{2}}}\,,
        \end{equation}
        and:
        \begin{equation}
            \boxed{B_{\mathrm{conv}}\equiv -\frac{5}{2}\sqrt{\frac{2}{3}}\alpha\frac{\omega}{\dot{\theta}}\frac{\frac{1}{2}X-\frac{1}{3}\sin{X}-\frac{1}{12}\sin{\left(2X\right)}}{\sin^{2}{X}}}\,;
        \end{equation}
        $\implies$ \textit{\textbf{The crucial contribution is proportional to}} $\boldsymbol{\sqrt{\epsilon_{\mathrm{ekp}}}}$. For a \textit{\textbf{generic asymmetric two-field ekpyrotic SF potential, this contribution can therefore exceed the non-Gaussianity expected in minimal single-field slow-roll inflation by more than an order of magnitude!}}
        \item The $A_{\mathrm{conv}}$ contribution comes entirely from the part of the integral proportional to $\tilde{c}$. Since: \begin{equation}
            \tilde{c}=\frac{\gamma^2-1}{4\gamma}\sqrt{2\epsilon_{\mathrm{ekp}}}\,,
        \end{equation}
        this contribution vanishes for a symmetric transverse potential, corresponding to:
        \begin{equation}
            \gamma^2=1 \quad\implies\quad \tilde{c}=0\,.
        \end{equation}
        In this case there is no intrinsic 2nd order entropy perturbation generated by the asymmetry of the entropy direction at the beginning of the conversion;
        \item The $B_{\mathrm{conv}}$ contribution arises from the part of the 2nd order entropy perturbation generated during the conversion itself, i.e. from the non-linear sourcing by the square of the linear entropy mode. It is independent of the ekpyrotic steepness parameter $\epsilon_{\mathrm{ekp}}$, but it depends on the details of the conversion process through: $\alpha$,$\omega/\dot{\theta}$, and $X$;
        \item Therefore:
        \begin{equation}
            \boxed{f_{NL}^{\mathrm{intrinsic}}\sim\mathcal{O}\bigl(\sqrt{\epsilon_{\mathrm{ekp}}}\bigr)+\mathcal{O}(1)}\,;
        \end{equation}
        \item The \textbf{reflection part of non-Gaussianity} can be derived explicitly from \eqref{fNL-reflection} and \eqref{V-R-ss} with the use of:
        \begin{equation}
            \gamma\,t_{\mathrm{ref}}\,\dot{\theta}\simeq 1 \quad\land\quad \int t\sin^{2}{\Bigl[\omega\bigl(t-t_{\mathrm{ref}}\bigr)\Bigr]}dt\simeq \frac{1}{2}\Delta t\left(t_{\mathrm{ref}}+\frac{1}{2}\Delta t\right)\,,
        \end{equation}
        so that \cite{Lehners:2008vx}:
        \begin{equation}
            \boxed{f_{NL}^{\mathrm{reflection}}\simeq\frac{5}{6\xi_{L}^{2}}\int\left(2\dot{\theta}^{2}t+\frac{\dot{\theta}}{\gamma}\right)\left(\delta s^{(1)}\right)^{2}dt\simeq -\frac{15\omega^{2}}{8\dot{\theta}^{2}\sin^{2}{\left(\omega\Delta t\right)}}\frac{\left|t_{\mathrm{ref}}+\frac{1}{2}\Delta t\right|}{\Delta t}<0}
        \end{equation}
        $\implies$ The $f_{NL}^{\mathrm{reflection}}$ \textbf{\textit{is independent of}} $\epsilon_{\mathrm{ekp}}$ and \textit{\textbf{depends only on the kinematics of the bend!}}
        \item The integrated piece becomes:
        \begin{equation}
            \boxed{f_{NL}^{\mathrm{integrated}}=\frac{5}{6\xi_{L}^{2}}\left[\delta s^{(1)}\left(t_{\mathrm{end}}\right)\right]^{2}\simeq\frac{5\omega^{2}}{4\dot{\theta}^{2}\left[1+\ln{\left(\frac{\left|t_{\mathrm{end}}\right|}{\left|t_{\mathrm{ref}}\right|}\right)}\right]^{2}\sin^{2}{\left(\omega\Delta t\right)}}}
        \end{equation}
        $\implies$ The $f_{NL}^{\mathrm{integrated}}$ \textbf{\textit{is also independent of}} $\epsilon_{\mathrm{ekp}}$ and \textit{\textbf{also depends only on the kinematics of the bend!}}
    \end{itemize}
\end{examplebox}
\begin{examplebox}[Estimation for the total non-Gaussianity in the ekpyrotic/cyclic models]
    The assumption of:
    \begin{equation}
        \ln{\left(\frac{\left|t_{\mathrm{end}}\right|}{\left|t_{\mathrm{ref}}\right|}\right)}\sim\mathcal{O}(1) \quad\land\quad t_{*}\sim\mathcal{O}(1)
    \end{equation}
    yield \cite{Lehners:2008vx}:
    \begin{equation}
        \begin{dcases}
            f_{NL}^{\mathrm{intrinsic}}\simeq 12\tilde{c}+50 \\
            f_{NL}^{\mathrm{reflection}}\simeq -150 \\
            f_{NL}^{\mathrm{integrated}}\simeq 15
        \end{dcases}
        \quad\implies\quad
        \boxed{f_{NL}\simeq 12\tilde{c}-85\simeq 4\frac{\left(\gamma^{2}-1\right)}{\gamma}\sqrt{\epsilon_{\mathrm{ekp}}}-85\sim\mathcal{O}(10)}
    \end{equation}
\end{examplebox}
\begin{examplebox}[Observational constraints on non-Gaussianity ($68\%\,CL$)]
    \begin{enumerate}[label=(\alph*)]
        \item \textbf{CMB (Planck 2018)} \cite{Planck:2019kim}:
        \begin{equation}
            \boxed{f_{NL}^{\mathrm{local}}=-0.9\pm 5.1}\,;
        \end{equation}
        \item \textbf{PR4 re-analysis} \cite{Jung:2025nss}:
        \begin{equation}
            \boxed{f_{NL}^{\mathrm{local}}=-0.1\pm 5.0}\,;
        \end{equation}
        \item \textbf{LSS (DESI DR1)} \cite{Chaussidon:2024qni}:
        \begin{equation}
            \boxed{f_{NL}^{\mathrm{local}}=-3.6_{-9.1}^{+9.0}}\,;
        \end{equation}
        \item \textbf{DESI LRG and Planck lensing} \cite{Bermejo-Climent:2024bcb}:
        \begin{equation}
            \boxed{f_{NL}^{\mathrm{local}}=39_{-38}^{+40}}\,;
        \end{equation}
    \end{enumerate}
    \underline{\textit{\textbf{Conclusion:}}}\\
    The results of all current probes are consistent with a \textit{\textbf{nearly Gaussian primordial distribution}}, with an effective bound for $f_{NL}^{\mathrm{local}}$ very close to zero:
    \begin{equation}
        \boxed{\left|f_{NL}^{\mathrm{local}}\right|\lesssim 10}\,.
    \end{equation}
\end{examplebox}
\begin{examplebox}[Allowed parameter space]
    The allowed parameter space require a combination of \cite{Lehners:2008vx,Lehners:2010fy}:
    \begin{enumerate}[label=(\roman*)]
        \item \textbf{Smooth conversion} in order to \textit{reflection and integrated contributions become under control};
        \item Range of $\tilde{c}\left(\gamma,\epsilon_{\mathrm{ekp}}\right)$ where the large $\sqrt{\epsilon_{\mathrm{ekp}}}$ term \textbf{does not overshoot the observational bound} (Fig.~\ref{fig:fNL-positive-gamma} and Fig.~\ref{fig:fNL-negative-gamma}).
    \end{enumerate}
\end{examplebox}

\subsection{Tensor perturbations}
The ekpyrotic phase involves the generation of GWs, but with a \textit{\textbf{spectrum that differs radically from the prediction of inflation}} \cite{Guzzetti:2016mkm,Caprini:2018mtu,Kamionkowski:2015yta}. Therefore, \textit{\textbf{GWs offer an additional method of distinguishing between models of the early Universe}}. However, it is currently difficult to predict \textit{if our measurement technology will be sensitive enough to detect the GW background in the future}.

This section will provide an in-depth analysis of the key studies on the generation of ekpyrotic GWs \cite{Khoury:2001wf,Boyle:2003km,Boyle:2004gv}.

In the transverse-traceless (TT) gauge, the tensor perturbations are characterized by the following metric \cite{Kodama:1984ziu,Mukhanov:1990me,Malik:2008im}:
\begin{equation}
    \frac{ds^{2}}{a^{2}}=-d\tau^{2}+\left[\delta_{ij}+2h_{T}\left(\tau,k\right)Y_{ij}^{(2)}\left(\vec{x}\right)\right]dx^{i}\,dx^{j}\,,
\end{equation}
where $Y_{ij}^{(2)}\left(\vec{x}\right)$ denotes tensor harmonics on $\mathbb{R}^{3}$ satisfying:
\begin{equation}
    \nabla^{2}Y_{ij}^{(2)}=-k^{2}Y_{ij}^{(2)}\,,
\end{equation}
and being transverse and traceless:
\begin{equation}
    \partial_{i}Y_{ij}^{(2)}=\delta^{ij}Y_{ij}^{(2)}=0\,.
\end{equation}
The quantity $h_{T}(\tau,k)$ denotes the mode amplitude of one of the two tensor polarizations. The TT description is convenient because tensor perturbations are gauge-invariant at first order: scalar and vector gauge transformations do not generate a transverse-traceless tensor component \cite{Bardeen:1980kt,Kodama:1984ziu,Mukhanov:1990me,Malik:2008im}.

\begin{figure}[htbp]
    \centering
    \includegraphics[width=0.9\linewidth]{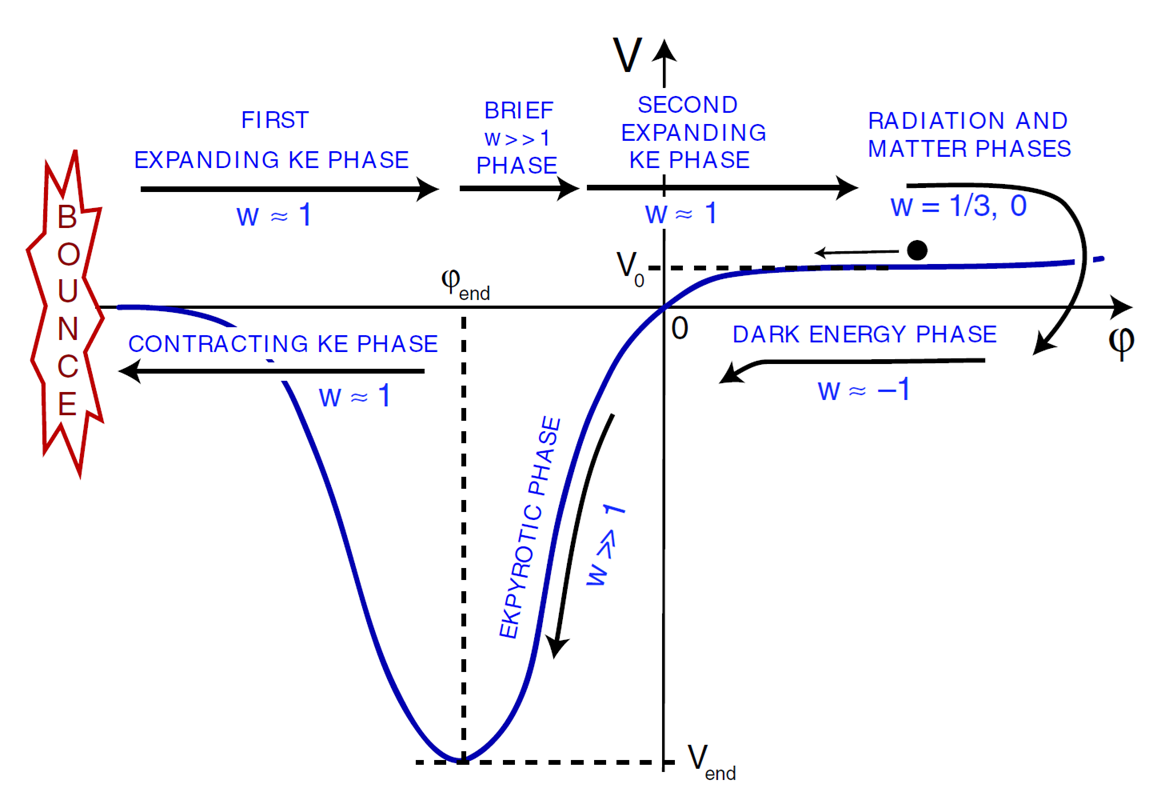}
    \caption{SF potential for the cyclic ekpyrotic Universe model \cite{Erickson:2006wc}.}
    \label{fig:cyclic-SF-potential}
\end{figure}

\begin{figure}[htbp]
    \centering
    \includegraphics[width=0.7\linewidth]{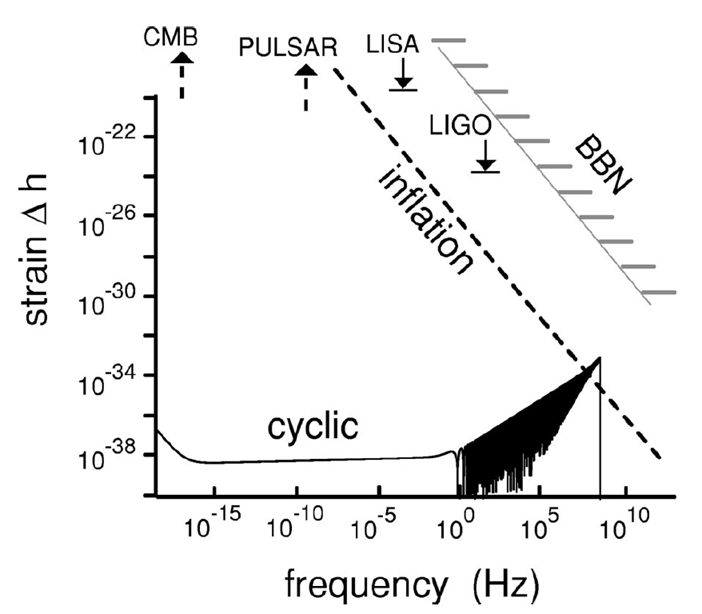}
    \caption{Present day dimensionless strain $\Delta h\left(k,\tau_{0}\right)$ for $T_{r}=10^{7}$ GeV and $V_{end}^{1/4}=10^{14}$ GeV \cite{Boyle:2003km}.}
    \label{fig:present-day-strain-cyclic-model}
\end{figure}

The expansion in Fourier modes and polarizations gives:
\begin{equation}
    h_{ij}\left(\tau,\vec{x}\right)=\sum_{\lambda=+,\times}\int\frac{d^{3}k}{\left(2\pi\right)^{3/2}}e^{i\vec{k}\cdot\vec{x}}h_{\lambda}(\tau,k)\,e_{ij}^{(\lambda)}\left(\hat{k}\right)\,,
\end{equation}
with a single tensor harmonic mode:
\begin{equation}
    h_{ij}=2h_{T}\,Y_{ij}^{(2)}\,,
\end{equation}
where the polarization tensors $e_{ij}^{(\lambda)}$ satisfy the following relations:
\begin{equation}
    k^{i}e_{ij}^{(\lambda)}=\delta^{ij}e_{ij}^{(\lambda)}=0 \quad\land\quad e_{ij}^{(\lambda)}\,e_{\left(\lambda'\right)}^{ij}=\delta_{\lambda\lambda'}\,.
\end{equation}
Starting from the Einstein-Hilbert action:
\begin{equation}
    S_{EH}=\frac{1}{2}\MPl^{2}\int d^{4}x\,\sqrt{-g}\,R
\end{equation}
and expanding up to a 2nd order in $h_{ij}$ gives:
\begin{equation}
    g_{00}=-a^{2}(\tau) \quad\land\quad g_{ij}=a^{2}(\tau)\bigl[\delta_{ij}+h_{ij}\bigr]\,.
\end{equation}
Imposing the TT conditions, the quadratic action for tensor perturbations becomes \cite{Mukhanov:1990me,Maggiore:2007ulw,Maggiore:2018sht}:
\begin{equation}
    S_{T}^{(2)}=\frac{1}{8}\MPl^{2}\int d\tau\,d^{3}x\,a^{2}(\tau)\left[\left(h'_{ij}\right)^{2}-\left(\partial_{l}h_{ij}\right)^{2}\right]\,.
\end{equation}
The use of the mode expansion and orthogonality of the tensor harmonics yields (for each polarization):
\begin{equation}
    S_{T}^{(2)}=\frac{1}{8}\MPl^{2}\sum_{\lambda}\int d\tau\,d^{3}k\,a^{2}(\tau)\left[\left|h'_{\lambda}(\tau,k)\right|^{2}-k^{2}\left|h_{\lambda}(\tau,k)\right|^{2}\right]\,,
\end{equation}
so that the EoM takes the form:
\begin{equation}
    h''_{\lambda}+2\mathcal{H}h'_{\lambda}+k^{2}h_{\lambda}=0\,,
\end{equation}
and therefore:
\begin{equation}\label{hT-EoM-1}
    h''_{T}+2\mathcal{H}h'_{T}+k^{2}h_{T}=0\,.
\end{equation}
Now, we can introduce a new rescaled variable of the form:
\begin{equation}
    f_{T}(\tau,k)\equiv a(\tau)\,h_{T}(\tau,k)\,.
\end{equation}
The explicit forms of the derivatives become:
\begin{equation}\label{fT-derivatives}
    f'_{T}=a'\,h_{T}+a\,h'_{T} \quad\land\quad f''_{T}=a''\,h_{T}+2a'\,h'_{T}+a\,h''_{T}\,.
\end{equation}
Moreover, using \eqref{hT-EoM-1} and the 2nd relation from \eqref{fT-derivatives} implies:
\begin{equation}
    f''_{T}=a''\,h_{T}+2a'\,h'_{T}+a\left(-2\mathcal{H}h'_{T}-k^{2}h_{T}\right)=a''\,h_{T}+2a'\,h'_{T}-2a'\,h'_{T}-a\,k^{2}h_{T}= a''\,h_{T}-a\,k^{2}h_{T}\,.
\end{equation}
Thus, the EoM becomes of the well-known form:
\begin{equation}
    \boxed{f''_{T}+\left(k^{2}-\frac{a''}{a}\right)f_{T}=0}\,.
\end{equation}
Imposing the Bunch-Davies initial conditions yields analogy with the scalar perturbations, therefore:
\begin{equation}
    \boxed{f_{T}(\tau,k)=\mathcal{P}_{2}\sqrt{\frac{\pi\,x}{4k}}H_{\beta}^{(1)}(x)} \quad\land\quad x\equiv\left|k\tau\right|\,,
\end{equation}
where:
\begin{equation}
    \beta^{2}=\frac{\left(\epsilon-3\right)^{2}}{4\left(\epsilon-1\right)^{2}} \quad\implies\quad \boxed{\beta=\frac{1}{2}\frac{\left|\epsilon-3\right|}{\left|\epsilon-1\right|}}
\end{equation}
and:
\begin{equation}
    \mathcal{P}_{2}=\exp{\left[i\left(\beta\frac{\pi}{2}+\frac{\pi}{4}\right)\right]}\,.
\end{equation}
In such a case, the \textbf{dimensionless tensor power spectrum} of $f_{T}$ is defined as:
\begin{equation}
    \mathcal{P}_{T}(k)\propto k^{3}\left|h_{T}\right|^{2}=k^{3}\left|\frac{f_{T}}{a}\right|^{2}\,.
\end{equation}
The \textbf{tensor spectral index} (in the \textit{super-horizon regime}) becomes of the form:
\begin{equation}
    n_{T}\equiv\frac{d\ln{\left(k^{3}\left|h_{T}\right|^{2}\right)}}{d\ln{k}}=3-2\beta\,,
\end{equation}
therefore:
\begin{equation}
    \boxed{n_{T}=3-\left|\frac{\epsilon-3}{\epsilon-1}\right|}\,.
\end{equation}
In the \textbf{ekpyrotic scenario} under consideration:
\begin{examplebox}[Ekpyrotic tensor spectral index]
    \begin{equation}
        \boxed{\epsilon\gg 1 \quad\implies\quad n_{T}=3-\left(1-\frac{2}{\epsilon-1}\right)=2+\frac{2}{\epsilon-1}\simeq 2}
    \end{equation}
    $\implies$ \textit{\textbf{Tensor spectrum is very blue!}}
\end{examplebox}
\begin{examplebox}[GW background in the cyclic ekpyrotic scenario]
    The cyclic ekpyrotic scenario corresponds to the following form of the SF potential (Fig.~\ref{fig:cyclic-SF-potential}):
    \begin{equation}
        \boxed{V(\phi)=V_{0}\left(1-e^{-c\phi}\right)F(\phi) \quad\lor\quad V(\phi)=V_{0}\left(e^{b\phi}-e^{-c\phi}\right)F(\phi)}\,,
    \end{equation}
    where:
    \begin{equation}
        b\ll 1 \quad\land\quad c\gg 1
    \end{equation}
    and:
    \begin{equation}
        F(\phi)\xrightarrow[\phi>\phi_{\mathrm{end}}]{}1 \quad\land\quad F(\phi)\xrightarrow[\phi<\phi_{\mathrm{end}}]{}0\,.
    \end{equation}
    In such a type of models, the \textit{\textbf{present day GW spectrum can be divided into three regimes}} (Fig.~\ref{fig:present-day-strain-cyclic-model}) \cite{Boyle:2003km}:
    \begin{enumerate}[label=(\roman*)]
        \item \textbf{Low frequency (LF) regime} - \textit{long-wavelength modes that re-enter the horizon after M-R equality}:
        \begin{equation}
            k<k_{\mathrm{eq}} \quad\implies\quad \boxed{\Delta h\left(k,\tau_{0}\right)\simeq\frac{\sqrt{\Gamma}\,k_{0}^{2}}{\pi\,\MPl H_{r}^{\alpha}}k^{-1+\alpha}}\,;
        \end{equation}
        \item \textbf{Medium frequency (MF) regime} - \textit{modes which re-enter the horizon between M-R equality and the onset of RD}:
        \begin{equation}
            k_{eq}<k<k_{r} \quad\implies\quad \boxed{\Delta h\left(k,\tau_{0}\right)\simeq\frac{\sqrt{\Gamma}\,k_{0}^{2}}{\pi\,\MPl H_{r}^{\alpha}}\left(\frac{k^{\alpha}}{k_{eq}}\right)}\,;
        \end{equation}
        \item \textbf{High frequency (HF) regime} - modes which exit the horizon during the ekpyrotic phase and re-enter during the (expanding) kination phase:
        \begin{equation}
            k_{r}<k<k_{end} \quad\implies\quad \boxed{\Delta h\left(k,\tau_{0}\right)\simeq\left(\frac{\sqrt{2}}{\pi}\right)^{3/2}\frac{\left(\Gamma H_{r}\right)^{1/2-\alpha}k_{0}^{2}}{\MPl\,k_{eq}\,k_{r}}\left|\cos{\left(k\,\tau_{r}-\frac{\pi}{4}\right)}\right|k^{1/2+\alpha}}\,,
        \end{equation}
    \end{enumerate}
    where:
    \begin{equation}
        k_{0}\equiv a_{0}H_{0} \quad\land\quad k_{eq}\equiv a_{eq}H_{eq}  \quad\land\quad k_{r}\equiv a_{r}H_{r} \quad\land\quad k_{end}\equiv a_{end}\left|H_{end}\right|\,,
    \end{equation}
    and:
    \begin{equation}
        \alpha\equiv\frac{2}{c^{2}-2}\ll 1 \quad\land\quad \Gamma\equiv\left|\frac{\tau_{r}}{\tau_{end}}\right|\,.
    \end{equation}
\end{examplebox}

\subsubsection{Physical interpretation and comparison with inflation}
For a mode with comoving wavenumber $k$, the dimensionless tensor power spectrum could be written as:
\begin{equation}
    \mathcal{P}_{T}(k)\sim\left(\frac{H_{*}}{\MPl}\right)^{2}\left(\frac{k}{k_{*}}\right)^{n_{T}}\,,
\end{equation}
where $H_{*}$ denotes the characteristic Hubble parameter at the time when the mode of scale $k_{*}$ is generated (e.g. at the end of ekpyrosis or when the mode exits the horizon during inflation):
\begin{enumerate}[label=(\alph*)]
    \item Cosmological inflation\footnote{$r$ denotes the \textit{\textbf{tensor-to-scalar ratio}}: $r\equiv\frac{A_{T}}{A_{s}}$.}:
    \begin{equation}
        \boxed{n_{T}\simeq -2\epsilon_{\mathrm{infl}}\simeq -\frac{r}{8} \quad\implies\quad \mathcal{P}_{T}(k)\simeq\left(\frac{H_{*}}{\MPl}\right)^{2}}
    \end{equation}
    $\implies$ \textit{\textbf{Nearly scale-invariant spectrum}};\\
    A mode of the wavelength comparable to the present Hubble radius ($H_{0}^{-1}$) has amplitude of the order:
    \begin{equation}
        h\left(k_{0}\right)\sim\frac{H_{*}}{\MPl}\,;
    \end{equation}
    \item Ekpyrosis:
    \begin{equation}
        \boxed{n_{T}\simeq 2 \quad\implies\quad \mathcal{P}_{T}(k)\propto k^{2}}\,;
    \end{equation}
    Normalizing at some high frequency scale $k_{B}$ associated with the end of the ekpyrotic phase, with the amplitude:
    \begin{equation}
        \mathcal{P}_{T}\left(k_{B}\right)\sim\left(\frac{H_{B}}{\MPl}\right)^{2}
    \end{equation}
    implies:
    \begin{equation}
        \mathcal{P}_{T}\left(k\right)=\mathcal{P}_{T}\left(k_{B}\right)\left(\frac{k}{k_{B}}\right)^{2}\,;
    \end{equation}
    The present Hubble scale ($\sim 60\,e$-folds in comoving scale):
    \begin{equation}
        k_{0}\sim e^{-60}k_{B}
    \end{equation}
    yields:
    \begin{equation}
        \boxed{\frac{\mathcal{P}_{T}\left(k_{0}\right)}{\mathcal{P}_{T}\left(k_{B}\right)}\sim\left(\frac{k_{0}}{k_{B}}\right)^{2}\sim e^{-120}}
    \end{equation}
    $\implies$ The root mean square amplitude on the present horizon scales is suppressed by \cite{Khoury:2001wf}:
    \begin{equation}
        \boxed{h\left(k_{0}\right)\sim\frac{H_{B}}{\MPl}e^{-60}}
    \end{equation}
    $\implies$ \textit{\textbf{In the simplest ekpyrotic models, the primordial tensor signal is exponentially suppressed and effectively unobservable!}}
\end{enumerate}

\subsubsection{Higher-dimensional perspective}
From a higher-dimensional perspective, tensor perturbations correspond to fluctuations of the higher-dimensional metric. In the $4D$ effective description these appear as PGWs, while in the higher-dimensional picture they are associated with bulk gravitational degrees of freedom, including the massless graviton zero mode and, in general, massive Kaluza-Klein excitations \cite{Maartens:2010ar,Lehners:2008vx}:
\begin{examplebox}[PGWs from the ekpyrotic model as the bulk modes]
    \begin{itemize}
        \item \textbf{Gravitons can propagate throughout the full higher-dimensional spacetime} because they are \textit{\textbf{excitations of the bulk metric itself}}. \textbf{They are not confined to the visible brane} \cite{Lehners:2008vx,Maartens:2010ar,Khoury:2001wf};
        \item \textit{Following the ekpyrotic phase, the visible brane enters a kination phase}, and according to the complete higher-dimensional model, the \textit{\textbf{geometry near the brane collision is a compactified Milne model}} (i.e. locally Minkowski space) \cite{Lehners:2008vx,Tolley:2003nx,Khoury:2001bz,Turok:2004gb,Lehners:2006pu}.
    \end{itemize}
    $\implies$ \underline{\textit{\textbf{Important consequences:}}}
    \begin{enumerate}[label=(\alph*)]
        \item \textit{\textbf{No amplification in the Minkowski-like regime}}:\\
        The wavelengths of modes shorter than the characteristic scale of the compactified Milne region (usually the GUT scale) are such that they do not exit during the $4D$ ekpyrotic phase. \textbf{In the Minkowski-like regime, their EoM is effectively that of free waves in a flat space}; that is to say, \textbf{they are not amplified by the background and remain in their vacuum state} \cite{Lehners:2008vx,Tolley:2003nx,Khoury:2001bz,Turok:2004gb,Lehners:2006pu};
        \item \textit{\textbf{High-frequency cut-off}}:\\
        Modes that leave the horizon during the ekpyrotic contracting phase (when 4D GR is a good approximation) are subject to the mode amplification described by the 4D effective equation for $f_T$. This results in a \textbf{maximum comoving wave number}, $k_{\mathrm{max}}$, (or \textbf{maximum frequency}), \textbf{beyond which the primordial tensor spectrum is cut off}. Therefore, modes with $k>k_{\mathrm{max}}$, where:
        \begin{equation}
            k_{\mathrm{max}}\sim k_{\mathrm{end}}\equiv a_{\mathrm{end}}\left|H_{\mathrm{end}}\right|
        \end{equation}
        \textit{never undergo ekpyrotic amplification} \cite{Boyle:2003km}. Therefore, the \textit{\textbf{high-frequency modes just below this cut-off carry most of the GW energy density because of their very blue spectrum}}. These modes are most tightly constrained by BBN \cite{Caprini:2018mtu,Boyle:2007zx,Maggiore:1999vm}.
    \end{enumerate}
\end{examplebox}

\subsubsection{Observational implications}
The primordial tensor spectrum (PGWs) in ekpyrotic models is extremely blue and suppressed on large scales, which implies the following:
\begin{examplebox}[Observational implications for ekpyrotic PGWs]
    \begin{itemize}
        \item The direct GW background from the ekpyrotic phase is negligible on CMB and LSS scales;
        \item On these scales, the dominant tensor contribution is expected to come from 2nd order (induced) GWs, which are sourced by 1st order scalar perturbations \cite{Baumann:2007zm}.
    \end{itemize}
    $\implies$ \underline{\textit{\textbf{Notable consequences:}}}
    \begin{enumerate}[label=(\alph*)]
        \item This is in \textbf{direct contrast to the simple slow-roll inflation model, whose robust prediction is a nearly scale-invariant primordial tensor spectrum, with the amplitude being directly related to the inflationary energy scale by the so called \textit{Lyth bound}} \cite{Lyth:1996im}:
        \begin{equation}
            \boxed{\frac{H}{\MPl}=\pi\,\sqrt{A_{s}}\sqrt{\frac{r}{2}}\simeq\sqrt{\frac{r}{0.01}}\times 10^{-5}}\,,
        \end{equation}
        so that:
        \begin{equation}
            \boxed{E_{\mathrm{infl}}\equiv\left(3H^{2}\MPl^{2}\right)^{1/4}=5\times 10^{-3}\left(\frac{r}{0.01}\right)^{1/4}\MPl \quad\land\quad r\propto E_{\mathrm{infl}}^{4}}\,.
        \end{equation}
        In this sense, a measurement of $r$ would directly determine the energy scale of inflation. The associated Lyth bound implies that an observable tensor-to-scalar ratio typically requires a super-Planckian field-space excursion in canonical single-field slow-roll models \cite{Lyth:1996im,Baumann:2009ds}\footnote{A related field-space distance budget for bouncing and cyclic cosmologies was developed in \cite{Postolak:2026okk}.};
        \item The observation of \textit{\textbf{primordial B-mode polarization with a nearly scale-invariant spectrum would strongly support the theory of inflation}} \cite{Seljak:1996gy,Kamionkowski:1996ks,Kamionkowski:2015yta} and \textit{\textbf{strongly disfavor the simplest four-dimensional ekpyrotic/cyclic models.}};
        \item On the other hand, \textit{\textbf{if PGWs stay undetected at very low values of the tensor-to-scalar ratio, it is fully compatible with ekpyrotic models}}. However, it \textbf{does not uniquely select the ekpyrotic framework over the inflationary one} \cite{Baumann:2009ds,Planck:2018jri,Lyth:1996im}.
    \end{enumerate}
\end{examplebox}
\begin{examplebox}[Integral constraint and $\Delta N_{\mathrm{eff}}$]
    \begin{itemize}
        \item The \textbf{GW contribution to the radiation density} can be expressed as a \textit{\textbf{change in the effective number of relativistic degrees of freedom beyond photons}}, as given by the following formula:
        \begin{equation}
            \rho_{r}=\rho_{\gamma}\left[1+\frac{7}{8}\left(\frac{4}{11}\right)^{4/3}N_{\mathrm{eff}}\right]\,,
        \end{equation}
        where \cite{Planck:2018vyg}:
        \begin{equation}
            N_{\mathrm{eff}}=3.046+\Delta N_{\mathrm{eff}}\,,
        \end{equation}
        and $\rho_{\gamma}$ denotes the photon energy density;
        \item Therefore, the \textbf{\textit{GW contribution}} may be written as \cite{Maggiore:1999vm,Maggiore:2007ulw,Caprini:2018mtu}:
        \begin{equation}
            \boxed{\Delta N_{\mathrm{eff}}^{(\mathrm{GW})}=\frac{8}{7}\left(\frac{11}{4}\right)^{4/3}\frac{1}{\Omega_{\gamma}^{(0)}}\int d\ln{(f)}\,\Omega_{\mathrm{GW}}^{(0)}(f)}\,;
        \end{equation}
        \item Using the observational bounds from CMB and BBN \cite{Planck:2018vyg,Planck:2018jri}:
        \begin{equation}
            \Omega_{\gamma}^{(0)}h^{2}\simeq 2.47\times 10^{-5} \quad\land\quad \Delta N_{\mathrm{eff}}\lesssim\mathcal{O}(0.3)
        \end{equation}
        yields the integral constraint of the form:
        \begin{equation}
            \boxed{\int d\ln{(f)}\,\Omega_{\mathrm{GW}}^{(0)}(f)\,h^{2}\lesssim\mathcal{O}\left(10^{-6}\right)}\,;
        \end{equation}
        \item For a blue power-law spectrum with an UV cut-off, $f_{\mathrm{max}}$:
        \begin{equation}
            \Omega_{\mathrm{GW}}^{(0)}(f)=\Omega_{\mathrm{max}}\left(\frac{f}{f_{\mathrm{max}}}\right)^{n_{T}} \quad\land\quad f_{\mathrm{BBN}}<f\leq f_{\mathrm{max}}\,,
        \end{equation}
        one finds for $f_{\mathrm{max}}\gg f_{\mathrm{BBN}}$ and $n_{T}>0$ that:
        \begin{equation}
            \int_{f_{\mathrm{BBN}}}^{f_{\mathrm{max}}}d\ln{(f)}\,\Omega_{\mathrm{GW}}^{(0)}(f)=\frac{\Omega_{\mathrm{max}}}{n_{T}}\left[1-\left(\frac{f_{\mathrm{BBN}}}{f_{\mathrm{max}}}\right)^{n_{T}}\right]\simeq\frac{\Omega_{\mathrm{max}}}{n_{T}}\,;
        \end{equation}
        \item The ekpyrotic scenario with $n_{T}\simeq 2$ implies the BBN bound of the form:
        \begin{equation}
            \boxed{\Omega_{\mathrm{max}}\,h^{2}\lesssim n_{T}\times 10^{-6}\sim 2\times 10^{-6}}
        \end{equation}
        $\implies$ \textit{\textbf{Explanation for why the high-frequency modes just below the cut-off dominate the constraint.}}
    \end{itemize}
\end{examplebox}
\begin{examplebox}[CMB tensor-to-scalar ratio]
    \begin{itemize}
        \item The tensor-to-scalar ratio:
        \begin{equation}
            r\left(k_{*}\right)\equiv\frac{\mathcal{P}_{T}\left(k_{*}\right)}{\mathcal{P}_{\xi}\left(k_{*}\right)}
        \end{equation}
        current observational constraint is \cite{Tristram:2020wbi,Campeti:2022vom}:
        \begin{equation}
            \boxed{r_{0.05}<0.036}\,;
        \end{equation}
        \item Therefore, for ekpyrosis, if the CMB pivot mode is $N$ $e$-folds larger in wavelength than the brane-collision scale, means that:
        \begin{equation}
            \frac{k_{*}}{k_{B}}=e^{-N}\,,
        \end{equation}
        so that:
        \begin{equation}
            \mathcal{P}_{T}\left(k_{*}\right)=A_{T}\left(\frac{k_{*}}{k_{B}}\right)^{n_{T}}=A_{T}\,e^{-n_{T}N}\,;
        \end{equation}
        \item Taking the values of:
        \begin{equation}
            n_{T}\simeq 2 \quad\land\quad N\simeq 60
        \end{equation}
        yields:
        \begin{equation}
            e^{-n_{T}N}=e^{-120}\sim 10^{-53}\,,
        \end{equation}
        thus, even for a high ekpyrotic scale, such as:
        \begin{equation}
            A_{T}\sim 10^{-10}\,,
        \end{equation}
        this produces:
        \begin{equation}
            \boxed{\mathcal{P}_{T}\left(k_{*}\right)\sim 10^{-62} \quad\land\quad r\left(k_{*}\right)\sim 10^{-53}}
        \end{equation}
        $\implies$ \textit{\textbf{Completely negligible compared to current and upcoming CMB sensitivities!}}
    \end{itemize}
    \underline{\textit{\textbf{Conclusion:}}}\\
    \textbf{Primordial gravitational waves are a crucial observational discriminator between inflationary and standard ekpyrotic/cyclic scenarios.} \textbf{A detection of a nearly scale-invariant primordial tensor spectrum would strongly support an inflationary origin and would strongly disfavor the simplest four-dimensional ekpyrotic/cyclic models}, whereas a \textbf{continued non-detection of PGWs remains compatible with ekpyrosis but does not uniquely select it over low-scale inflation}.
\end{examplebox}

\section{Cyclic ekpyrotic cosmology}
The minimal version of the cyclic (ekpyrotic) cosmological model can be described using GR with minimally coupled SF:
\begin{equation}
    S=\int d^{4}x\,\sqrt{-g}\left[\frac{1}{2}R-\frac{1}{2}g^{\mu\nu}\partial_{\mu}\phi\,\partial_{\nu}\phi-V(\phi)\right]\,,
\end{equation}
where the cyclic ekpyrotic SF potential contains a regions of: a shallow positive plateau, a steep negative region, and a region in which the SF potential is effectively cut-off. Therefore, it may take one of the following forms \cite{Steinhardt:2001st,Lehners:2008vx}:
\begin{equation}
    V(\phi)=V_{0}\left(1-e^{-c\phi}\right)F(\phi) \quad\lor\quad V(\phi)=V_{0}\left(e^{b\phi}-e^{-c\phi}\right)F(\phi)\,,
\end{equation}
with:
\begin{equation}
    V_{0}>0 \quad\land\quad b\ll 1 \quad\land\quad c\gg 1\,.
\end{equation}
Moreover, the function describing the cut-off must satisfy the following conditions (as mentioned previously):
\begin{itemize}
    \item $F(\phi)\to 1$ on the plateau and the ekpyrotic slope;
    \item $F(\phi)\to 0$ deep in the well of the SF potential.
\end{itemize}

\subsection{Structure of one cycle}
\begin{figure}[htbp]
    \centering
    \includegraphics[width=1\linewidth]{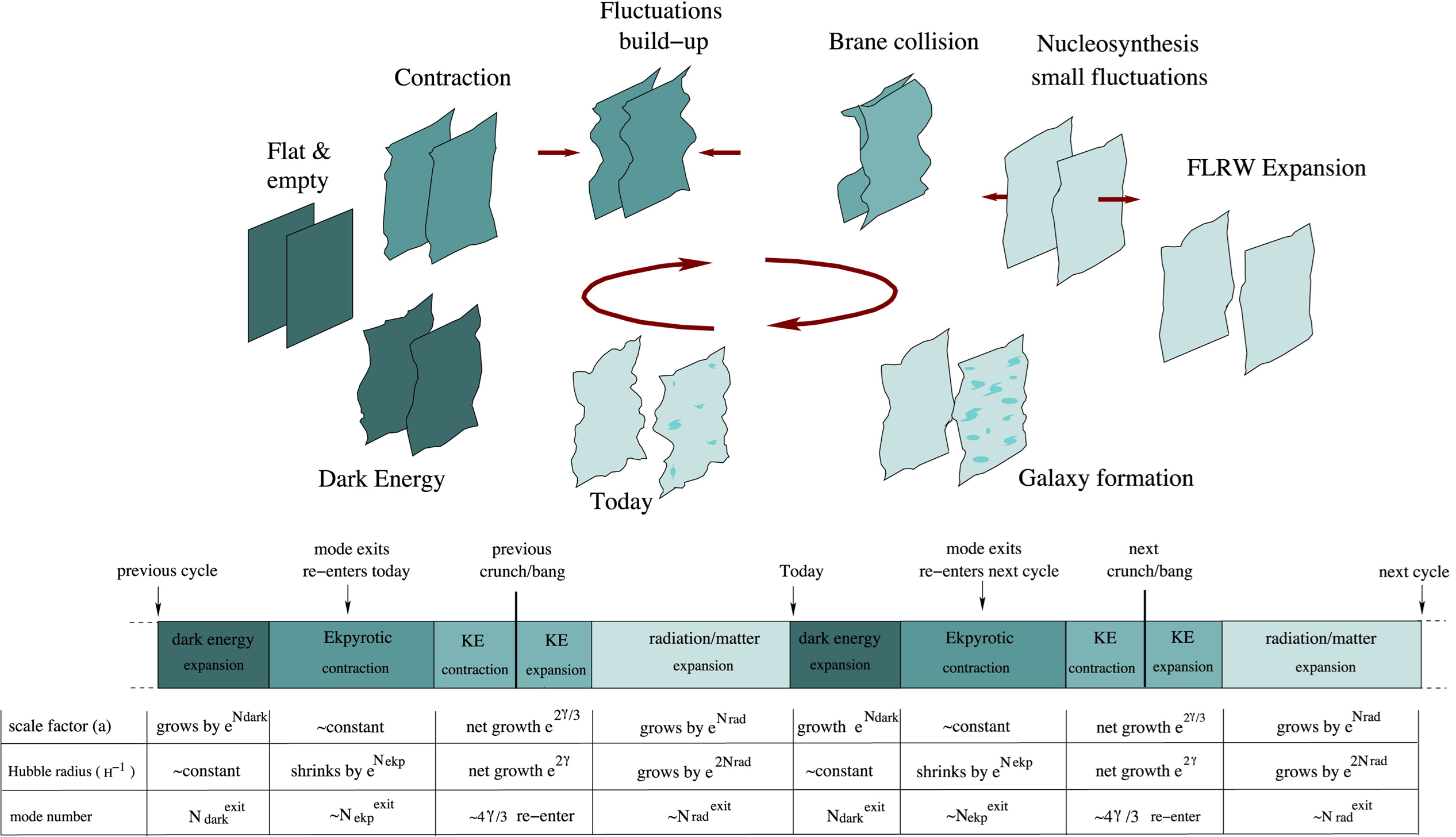}
    \caption{Schematic of the cyclic Universe \cite{Battefeld:2014uga}.}
    \label{fig:schematic-cyclic-Universe}
\end{figure}

A single cycle can be divided into the following cosmological phases (see Fig.~\ref{fig:schematic-cyclic-Universe}) \cite{Steinhardt:2001st,Ijjas:2019pyf,Lehners:2008vx}:
\begin{examplebox}[Structure of the cosmological cycle]
    \begin{enumerate}[label=(\roman*)]
        \item \textbf{DE/quintessence phase} - the SF slowly evolves on the positive plateau:
        \begin{equation}
            V(\phi)\simeq V_{0}>0 \quad\land\quad \omega_{\phi}\simeq -1\,,
        \end{equation}
        and the scale factor grows;
        \item \textbf{Ekpyrotic slow contraction} - the field rolls down a steep negative potential:
        \begin{equation}
            V(\phi)\simeq -V_{0}\,e^{-c\phi}\,,
        \end{equation}
        and generates an ultra-stiff EoS:
        \begin{equation}
            \omega\gg 1\,;
        \end{equation}
        \item \textbf{KE-dominated contraction} - the SF potential is negligible:
        \begin{equation}
            V(\phi)\simeq 0 \quad\land\quad \rho_{\phi}\simeq\frac{1}{2}\dot{\phi}^{2}\,,
        \end{equation}
        so that the KE of the field becomes a dominant contribution with the EoS:
        \begin{equation}
            \omega_{\phi}\simeq 1\,;
        \end{equation}
        \item \textbf{Cosmological bounce/collision of the branes} - the Universe passes from contraction to expansion. In the original higher-dimensional cyclic picture this transition is interpreted as a collision of boundary branes, whereas in modern effective descriptions it is often replaced by a non-singular bounce mechanism. In both cases, radiation and matter are produced at the transition \cite{Khoury:2001bz,Turok:2004gb,Lehners:2006pu,Lehners:2008vx,Ijjas:2019pyf};
        \item \textbf{KE-dominated expansion, RD, and MD} - immediately after the bounce/collision the scalar KE may still dominate for a short period, after which RD and MD reproduce the usual hot Big Bang cosmological evolution;
        \item \textbf{Return to the plateau of the SF potential} - after the radiation and matter epochs, the SF evolves back toward the positive plateau of the potential. The Universe then enters another \textbf{DE/quintessence phase}, which prepares the onset of the next ekpyrotic contraction.
    \end{enumerate}
    \underline{\textit{\textbf{Remark:}}}\\
    The \textbf{original cyclic model employs a singular brane collision}. However, \textbf{modern effective cyclic scenarios typically replace this with a non-singular bounce} \cite{Khoury:2001bz,Turok:2004gb,Ijjas:2016vtq,Nojiri:2017ncd}.
\end{examplebox}

\subsection{Growth of a scale factor and Hubble parameter}
It is now useful to define the following parameter related to the magnitude of the SF potential at the beginning and end of ekpyrosis, namely:
\begin{equation}
    V_{\mathrm{beg}}\equiv\left|V\left(t_{\mathrm{beg}}\right)\right| \quad\land\quad V_{\mathrm{end}}\equiv\left|V\left(t_{\mathrm{end}}\right)\right| \quad\land\quad V_{\mathrm{beg}}<V_{\mathrm{end}}\,.
\end{equation}
During an ekpyrotic contraction the SF potential becomes of the form:
\begin{equation}\label{ekpyrotic-scaling-potential}
    V(\phi)\simeq -V_{0}\,e^{-c\phi}\,,
\end{equation}
and one may use the scaling solution given by:
\begin{equation}
    a(t)=a_{0}(-t)^{p} \quad\land\quad \phi(t)=q\,\ln{(-t)}+\phi_{0}\,,
\end{equation}
where:
\begin{equation}\label{p-c-relation}
    t<0 \quad\land\quad p=\frac{2}{c^{2}}>0\,,
\end{equation}
and the EoS and fast-roll parameter are defined as:
\begin{equation}
   \omega_{\mathrm{ekp}}=\frac{c^{2}}{3}-1\gg 1 \quad\land\quad \epsilon\equiv -\frac{\dot{H}}{H^{2}}=\frac{3}{2}\bigl(1+\omega\bigr) \quad\land\quad \epsilon_{\mathrm{ekp}}=\frac{1}{p}=\frac{c^{2}}{2}\gg 1\,. 
\end{equation}
This implies:
\begin{equation}\label{scaling-H}
    H(t)=\frac{p}{t} \quad\land\quad \dot{H}(t)=-\frac{p}{t^{2}} \quad\land\quad \dot{\phi}(t)=\frac{q}{t} \quad\land\quad \ddot{\phi}(t)=-\frac{q}{t^{2}}\,,
\end{equation}
therefore:
\begin{equation}
    V(\phi)=-\tilde{V}(-t)^{-cq} \quad\land\quad \tilde{V}\equiv V_{0}\,e^{-c\phi_{0}}>0\,.
\end{equation}
Furthermore, the 1st derivative of the potential \eqref{ekpyrotic-scaling-potential} with respect to $\phi$ is:
\begin{equation}
    V_{,\phi}=-c\,V(\phi)=c\,\tilde{V}(-t)^{-cq}\,.
\end{equation}
The use of the Klein-Gordon equation:
\begin{equation}
    \ddot{\phi}+3H\dot{\phi}+V_{,\phi}=0
\end{equation}
gives the following relation:
\begin{equation}
    \frac{q\left(3p-1\right)}{t^{2}}+c\,\tilde{V}(-t)^{-cq}=0\,,
\end{equation}
so that the time-dependence of the ekpyrotic SF potential becomes:
\begin{equation}
    c\,q=2 \quad\implies\quad V(\phi)=-\tilde{V}(-t)^{-2} \quad\implies\quad \left|V\right|\propto t^{-2}\,.
\end{equation}
In such a case, the ratio of $V_{\mathrm{end}}$ and $V_{\mathrm{beg}}$ takes the form:
\begin{equation}\label{V-ratio}
    \frac{V_{\mathrm{end}}}{V_{\mathrm{beg}}}=\left(\frac{\left|t_{\mathrm{beg}}\right|}{\left|t_{\mathrm{end}}\right|}\right)^{2}\,.
\end{equation}
One may also immediately find (using \eqref{scaling-H}) the relation between the ratio \eqref{V-ratio} and the ratio of the Hubble parameters at the end and beginning of the ekpyrotic phase:
\begin{equation}
    \frac{\left|H_{\mathrm{end}}\right|}{\left|H_{\mathrm{beg}}\right|}=\frac{\left|t_{\mathrm{beg}}\right|}{\left|t_{\mathrm{end}}\right|}=\left(\frac{V_{\mathrm{end}}}{V_{\mathrm{beg}}}\right)^{1/2}\,.
\end{equation}
The number of ekpyrotic $e$-folds can now be defined as:
\begin{equation}
    \boxed{N_{\mathrm{ekp}}\equiv\ln{\left(\frac{\left|H_{\mathrm{end}}\right|}{\left|H_{\mathrm{beg}}\right|}\right)}=\frac{1}{2}\ln{\left(\frac{V_{\mathrm{end}}}{V_{\mathrm{beg}}}\right)}}\,,
\end{equation}
and the scale factor changes by a value of:
\begin{equation}
    \boxed{\frac{a_{\mathrm{end}}}{a_{\mathrm{beg}}}=\left(\frac{\left|t_{\mathrm{end}}\right|}{\left|t_{\mathrm{beg}}\right|}\right)^{p}=\left(\frac{V_{\mathrm{end}}}{V_{\mathrm{beg}}}\right)^{-p/2}}\,,
\end{equation}
or, equivalently:
\begin{equation}\label{Delta-ln-a-ekp}
    \boxed{\Delta\ln{a_{\mathrm{ekp}}}=-p\,N_{\mathrm{ekp}}}\,.
\end{equation}
We can observe that for $p\ll 1$, such a contraction is small even if $N_{\mathrm{ekp}}\gg 1$ \cite{Lehners:2008vx}.

\subsection{Energy and bounce conditions}
\subsubsection{Classical energy conditions}
For a perfect fluid, described by the following energy-momentum tensor:
\begin{equation}
    T_{\mu\nu}=\bigl(\rho+p\bigr)u_{\mu}u_{\nu}+p\,g_{\mu\nu}\,,
\end{equation}
the \textit{\textbf{pointwise (classical) energy conditions}} become of the form \cite{Curiel2017,Kontou:2020bta,Hawking_Ellis_2023ECs}:
\begin{examplebox}[Classical energy conditions]
    \begin{enumerate}[label=(\alph*)]
        \item \textbf{Null energy condition (NEC)} - \textit{energy density measured along any null direction is non-negative}:
        \begin{equation}
            \boxed{\rho+p\geq 0}\,;
        \end{equation}
        \item \textbf{Weak energy condition (WEC)} - \textit{physical observer measures a non-negative local energy density}:
        \begin{equation}
            \boxed{\rho\geq 0 \quad\land\quad \rho+p\geq 0}\,;
        \end{equation}
        \item \textbf{Strong energy condition (SEC)} - attractive gravity and geodesic focusing in the Raychaudhuri equation (\textit{violated for an accelerated expansion of the Universe}):
        \begin{equation}
            \boxed{\rho+3p\geq 0 \quad\land\quad \rho+p\geq 0}\,;
        \end{equation}
        \item \textbf{Dominant energy condition (DEC)} - \textit{energy flux measured by any physical observer is non-spacelike and future-directed}:
        \begin{equation}
            \boxed{\rho\geq 0 \quad\land\quad \left|p\right|\leq\rho}\,.
        \end{equation}
    \end{enumerate}
\end{examplebox}
For a canonical SF considered in the standard ekpyrotic scenario:
\begin{equation}
    \rho_{\phi}+p_{\phi}=\dot{\phi}^{2}>0\,,
\end{equation}
so that the NEC is satisfied. Also the WEC and SEC are satisfied during the ekpyrotic phase:
\begin{equation}
    \rho_{\phi}+3p_{\phi}=\bigl(1+3\omega_{\phi}\bigr)\rho_{\phi}>0\,.
\end{equation}
Nevertheless, the \textbf{DEC is violated} due to the fact that:
\begin{equation}
    \boxed{\omega_{\phi}>1 \quad\implies\quad \left|p_{\phi}\right|=\omega_{\phi}\rho_{\phi}>\rho_{\phi}}\,.
\end{equation}

\subsubsection{Bounce conditions}
As we stated before, the non-singular bounce requires:
\begin{examplebox}[Non-singular bounce condition]
    \begin{equation}\label{bounce-cyclic-condition}
        \boxed{a\left(t_{B}\right)=a_{B}>0 \quad\land\quad H\left(t_{B}\right)=0 \quad\land\quad \dot{H}\left(t_{B}\right)>0}\,.
    \end{equation}
\end{examplebox}
The use of 1st Friedmann equation (with spatial curvature):
\begin{equation}
    H^{2}=\frac{\rho}{3}-\frac{k}{a^{2}}
\end{equation}
for a 2nd condition in \eqref{bounce-cyclic-condition} implies that:
\begin{equation}
    \rho_{B}=3\frac{k}{a_{B}^{2}}\,.
\end{equation}
On the other hand, the 2nd Friedmann equation:
\begin{equation}
    \dot{H}=-\frac{1}{2}\bigl(\rho+p\bigr)+\frac{k}{a^{2}}
\end{equation}
for a 3rd condition in \eqref{bounce-cyclic-condition} yields:
\begin{equation}
    \dot{H}_{B}=-\frac{1}{2}\bigl(\rho_{B}+p_{B}\bigr)+\frac{k}{a_{B}^{2}}\,,
\end{equation}
so that:
\begin{equation}
    \rho_{B}+p_{B}<2\frac{k}{a_{B}^{2}}\,.
\end{equation}
This results in \textit{\textbf{three cases, depending on the value of spatial curvature}}:
\begin{examplebox}[Bounce conditions for different spatial curvatures]
    \begin{enumerate}[label=(\alph*)]
    \item \textbf{Spatially flat Universe} ($k=0$):
    \begin{equation}
        \boxed{\rho_{B}=0 \quad\land\quad \rho_{B}+p_{B}<0}
    \end{equation}
    requires \textit{\textbf{violation of the NEC}} - \textit{\textbf{main obstacle in a flat ekpyrotic Universe}};
    \item \textbf{Closed Universe} ($k=1$):
    \begin{equation}
        \rho_{B}=\frac{3}{a_{B}^{2}}>0 \quad\land\quad \rho_{B}+p_{B}<\frac{2}{a_{B}^{2}}=\frac{2}{3}\rho_{B}
    \end{equation}
    in order to satisfy the NEC requires:
    \begin{equation}
        \boxed{-\rho_{B}\leq p_{B}<-\frac{\rho_{B}}{3}}\,,
    \end{equation}
    but violates the SEC:
    \begin{equation}
        \boxed{p_{B}<-\frac{\rho_{B}}{3} \quad\implies\quad \rho_{B}+3p_{B}<0}\,.
    \end{equation}
    Thus, in the closed FLRW universe, the \textbf{NEC violation could be avoided when obtaining the cosmological non-singular bounce};
    \item \textbf{Open Universe} ($k=-1$):
    \begin{equation}
        \rho_{B}=-\frac{3}{a_{B}^{2}}<0\,,
    \end{equation}
    so that the \textbf{non-singular bounce is impossible for a positive total energy density content}.
\end{enumerate}
\underline{\textit{\textbf{Conclusion:}}}\\
\textbf{In the case of a standard spatially flat ekpyrotic formalism, one must take into account either a singular brane collision, a controlled NEC-violating phase, or a MG framework.}
\end{examplebox}

\subsection{Geodesic incompleteness and singularity resolution}
Ekpyrotic contraction is a smoothing mechanism, but in the ($4D$) \textit{\textbf{Einstein frame effective description it is not, by itself, a non-singular or geodesically complete cosmology}}\footnote{For related discussions of past completeness in inflationary spacetimes, see \cite{Borde:2001nh,Lesnefsky:2022fen}.} \cite{Kinney:2021imp,Lesnefsky:2022fen,Pavlovic:2023mke}. \textbf{Geodesic completeness means that every maximal causal geodesic, timelike or null, has an unbounded parameter domain}: proper time for timelike geodesics and affine parameter for null geodesics \cite{Hawking_Ellis_2023Sing1,Hawking_Ellis_2023Sing3,Clarke_1994book}. Conversely, the existence of even one inextendible causal geodesic with finite proper time or finite affine parameter is sufficient to establish geodesic incompleteness. This distinction is important: ekpyrosis can dynamically suppress spatial curvature, shear anisotropies and inhomogeneities relative to the ekpyrotic energy density, while the $4D$ spacetime may still approach a Big Crunch-type boundary unless it is supplemented by a well-defined non-singular bounce or by an embedding in a geodesically complete completion.

The 4-velocity of a comoving observer is given by:
\begin{equation}
    u^{\mu}=\delta_{0}^{\mu}\,,
\end{equation}
and its proper time is simply the cosmological time:
\begin{equation}
    d\tau_{\mathrm{prop}}=dt\,.
\end{equation}
Thus, the \textbf{proper time remaining before the Big Crunch} at $t=0$ \textbf{is a finite quantity}:
\begin{equation}\label{tau-prop-infty}
    \boxed{\Delta\tau_{\mathrm{prop}}=\int_{t_{i}}^{0}dt=\left|t_{i}\right|<\infty}\,.
\end{equation}
Therefore, the \textbf{comoving time-like geodesics are incomplete in the} $\boldsymbol{4D}$ \textbf{effective geometry (the same applies to radial null geodesics)}. In fact, using the conformal time:
\begin{equation}
    dt=a\,d\tau\,,
\end{equation}
one can show that their affine parameter satisfies:
\begin{equation}
    d\lambda\propto a^{2}d\tau=a\,dt\,,
\end{equation}
and near the bounce/Crunch (for $t\to{0}^{-}$):
\begin{equation}\label{Delta-lambda}
    \boxed{\Delta\lambda\propto\int_{t_{i}}^{0}a(t)\,dt\propto\int_{t_{i}}^{0}\left|t\right|^{p}dt=\frac{\left|t_{i}\right|^{p+1}}{p+1}<\infty}\,.
\end{equation}
This indicates for a \textbf{null geodesic incompleteness}. The relations \eqref{tau-prop-infty} and \eqref{Delta-lambda} constitute the \textit{time-reversed analogue of the usual FLRW singularity behavior}.

Moreover, the \textbf{incompleteness is associated with the divergence of curvature invariants}. For the general FLRW metric, the Ricci scalar takes the following form:
\begin{equation}
    R=6\left(2H^{2}+\dot{H}+\frac{k}{a^{2}}\right)\,.
\end{equation}
Using the ekpyrotic scaling solution \eqref{scaling-H} yields:
\begin{equation}
    2H^{2}+\dot{H}=\frac{p\left(2p-1\right)}{t^{2}} \quad\land\quad \frac{k}{a^{2}}\propto\left|t\right|^{-2p}\,,
\end{equation}
so that the SF contribution proportional to $t^{-2}$ diverges more rapidly than the curvature contribution. Thus:
\begin{equation}
    R\simeq 6\frac{p\left(2p-1\right)}{t^{2}} \quad\land\quad t\to 0^{-}\,.
\end{equation}
In similar manner, the \textit{Kretschmann scalar} may be given by:
\begin{equation}
    K\equiv R_{\mu\nu\rho\sigma}R^{\mu\nu\rho\sigma}=12\left[\left(H^{2}+\dot{H}\right)^{2}+\left(H^{2}+\frac{k}{a^{2}}\right)^{2}\right]\,,
\end{equation}
so that, in the ekpyrotic scaling regime it becomes of the form:
\begin{equation}
    K\propto t^{-4} \quad\land\quad t\to 0^{-}\,.
\end{equation}
The above means that the $4D$ \textbf{ekpyrotic solution approaches a genuine curvature singularity}. Equivalently:
\begin{equation}
    3\left(H^{2}+\frac{k}{a^{2}}\right)=\rho\,,
\end{equation}
the total energy density diverges as:
\begin{equation}
    \rho\propto 3\frac{p^{2}}{t^{2}}\,.
\end{equation}
This demonstrates that the \textit{\textbf{ekpyrotic smoothing phase and the singularity-resolution mechanism are distinct concepts}}. The ultra-stiff phase suppresses anisotropy and spatial curvature, however, it does not convert the contracting branch into an expanding branch in the context of standard $4D$ GR. \textbf{A complete cyclic model must therefore specify how the Universe is extended through the Crunch/bounce surface}. In the original ekpyrotic and cyclic brane scenarios, the $4D$ Crunch is interpreted as a collision of boundary branes or orbifold planes. Near the collision, the higher-dimensional geometry is often approximated by compactified Milne spacetime, which is locally flat but still requires a prescription for evolving fields and perturbations through the orbifold collision. Hence, the \textit{\textbf{compactified Milne picture may soften the interpretation of the singularity in the higher-dimensional geometry, but it does not by itself provide a universally accepted}} $\boldsymbol{4D}$ \textit{\textbf{geodesically complete solution}}.

\textit{\textbf{In non-singular effective cyclic models, the singular brane collision is replaced by a smooth bounce}} \cite{Lehners:2008vx,Buchbinder:2007ad,Koehn:2015vvy}. In spatially flat GR, this requires a transient NEC-violating phase \cite{Novello:2008ra,Koehn:2015vvy}. In contrast, MG and LQG  models can change the effective Friedmann equation in such a way that a bounce occurs without violating the matter sector NEC \cite{Battefeld:2014uga,Ijjas:2016vtq,Nojiri:2017ncd,Novello:2008ra,Singh:2006im,Ashtekar:2006wn,Ashtekar:2011ni}. In order to extend the solutions of the field equations via a cosmological bounce, it must also be ensured that the curvature invariants are finite quantities. It should also be noted that cosmological perturbations must be under control, with no ghost or gradient instabilities \cite{Battefeld:2014uga,Ijjas:2016vtq}.

The issue is also connected with \textbf{past eternity}. Even if the local background undergoes repeated ekpyrotic cycles, \textit{\textbf{global geodesic completeness does not occur automatically}}. Entropy production, net volume growth, the validity range of the $4D$ effective theory, and the physics of the bounce must all be controlled in each cycle. For this reason, \textbf{\textit{modern cyclic models are best viewed as providing a local smoothing and recycling mechanism, and the question of how to achieve global, geodesically complete realization remains an open problem dependent on the model}} \cite{Lehners:2008vx,Ijjas:2019pyf,Ijjas:2021zwv}.

\subsection{Kinetic phase, reheating and closure of the cycle}
\subsubsection{Kinetic phase}
Once the field has exited the steep negative part of the SF potential, the potential becomes negligible, and the KE of the SF dominates. During this stage, the KE dominates over spatial curvature:
\begin{equation}
    \rho_{\phi}\gg 3\frac{\left|k\right|}{a^{2}} \quad\implies\quad \Bigl(\omega=1 \quad\land\quad \rho_{\phi}\propto a^{-6}\Bigr)\,,
\end{equation}
and implies the following solution:
\begin{equation}
    a(t)\propto\left|t\right|^{1/3} \quad\land\quad H(t)=\frac{1}{3t} \quad\land\quad \phi(t)=\phi_{*}\pm\sqrt{\frac{2}{3}}\ln{\left|t\right|}\,.
\end{equation}
The spatial curvature remains subdominant in the contracting kination phase if it was already suppressed by ekpyrosis. During the expanding kinetic stage, the relative curvature fraction can increase in relation to the scalar kinetic energy due to the fact that:
\begin{equation}\label{rho-phi-rho-k}
    \rho_{\phi}\propto a^{-6} \quad\land\quad \left|\rho_{k}\right|\propto a^{-2}\,.
\end{equation}
Nevertheless, evolution is not spoiled in viable cyclic solutions because the curvature contribution has already been strongly suppressed by the preceding ekpyrotic phase \cite{Lehners:2008vx,Battefeld:2014uga}.

Now, let us denote the time at which the kination epoch has started by $t_{\mathrm{end}}$, which is characterized by the following relation:
\begin{equation}
    \rho_{\phi}\left(t_{\mathrm{end}}\right)\simeq V_{\mathrm{end}}\,,
\end{equation}
and the time, labeled by $t_r$, at which the radiation contribution equals the SF KE after reheating of the Universe:
\begin{equation}
    \rho_{r}\left(t_{r}\right)\simeq\rho_{\phi}\left(t_{r}\right)\simeq T_{r}^{4}\,,
\end{equation}
where $T_{r}$ denotes radiation temperature scale. Using the behavior of the SF energy density given by \eqref{rho-phi-rho-k}, one can obtain the following ratio of the scale factors:
\begin{equation}
    \boxed{\frac{a_{r}}{a_{\mathrm{end}}}=\left(\frac{V_{\mathrm{end}}}{T_{r}^{4}}\right)^{1/6}=\left(\frac{V_{\mathrm{end}}^{1/4}}{T_{r}}\right)^{2/3} \quad\land\quad E_{\mathrm{end}}\equiv V_{\mathrm{end}}^{1/4}}\,.
\end{equation}
Moreover, by defining the following parameter - \textit{dimensionless measure of the duration of the kination epoch}:
\begin{equation}
    \gamma_{\mathrm{KE}}\equiv\ln{\left(\frac{V_{\mathrm{end}}^{1/4}}{T_{r}}\right)}
\end{equation}
provides the relationship of the form:
\begin{equation}\label{Delta-ln-a-KE}
    \boxed{\Delta\ln{a_{\mathrm{KE}}}=\ln{\left(\frac{a_{r}}{a_{\mathrm{end}}}\right)}=\frac{2}{3}\gamma_{\mathrm{KE}}}\,.
\end{equation}
Moreover, using the fact that:
\begin{equation}
    H^{2}\propto\rho\,,
\end{equation}
gives:
\begin{equation}\label{Hr-Hend}
    \boxed{\frac{\left|H_{r}\right|}{\left|H_{\mathrm{end}}\right|}=\left(\frac{T_{r}^{4}}{V_{\mathrm{end}}}\right)^{1/2}=e^{-2\gamma_{\mathrm{KE}}}}\,.
\end{equation}

\subsubsection{RD and MD}
After reheating, the temperature of the Universe approximately scales as:
\begin{equation}
    T\propto a^{-1}\,,
\end{equation}
so that the logarithmic growth of the scale factor from RD to the present epoch becomes:
\begin{equation}
    N_{\mathrm{rad}}\equiv\ln{\left(\frac{T_{r}}{T_{0}}\right)}\,,
\end{equation}
and therefore:
\begin{equation}\label{ln-H0-Hr}
    \boxed{\ln{\left(\frac{\left|H_{0}\right|}{\left|H_{r}\right|}\right)}\simeq -2N_{\mathrm{rad}}}\,.
\end{equation}
During the pure RD:
\begin{equation}
    H\propto T^{2}\,,
\end{equation}
the formula \eqref{ln-H0-Hr} is exact. For a \textit{realistic model}, both \textit{MD and late DE domination introduce corrections of} $\mathcal{O}(1)$ \textit{to the logarithmic estimate}. These corrections do not change the leading result for the cycle-counting arguments \cite{Lehners:2008vx,Steinhardt:2001st}.

\subsubsection{Net expansion per cycle}
\begin{figure}[htbp]
    \centering
    \includegraphics[width=0.9\linewidth]{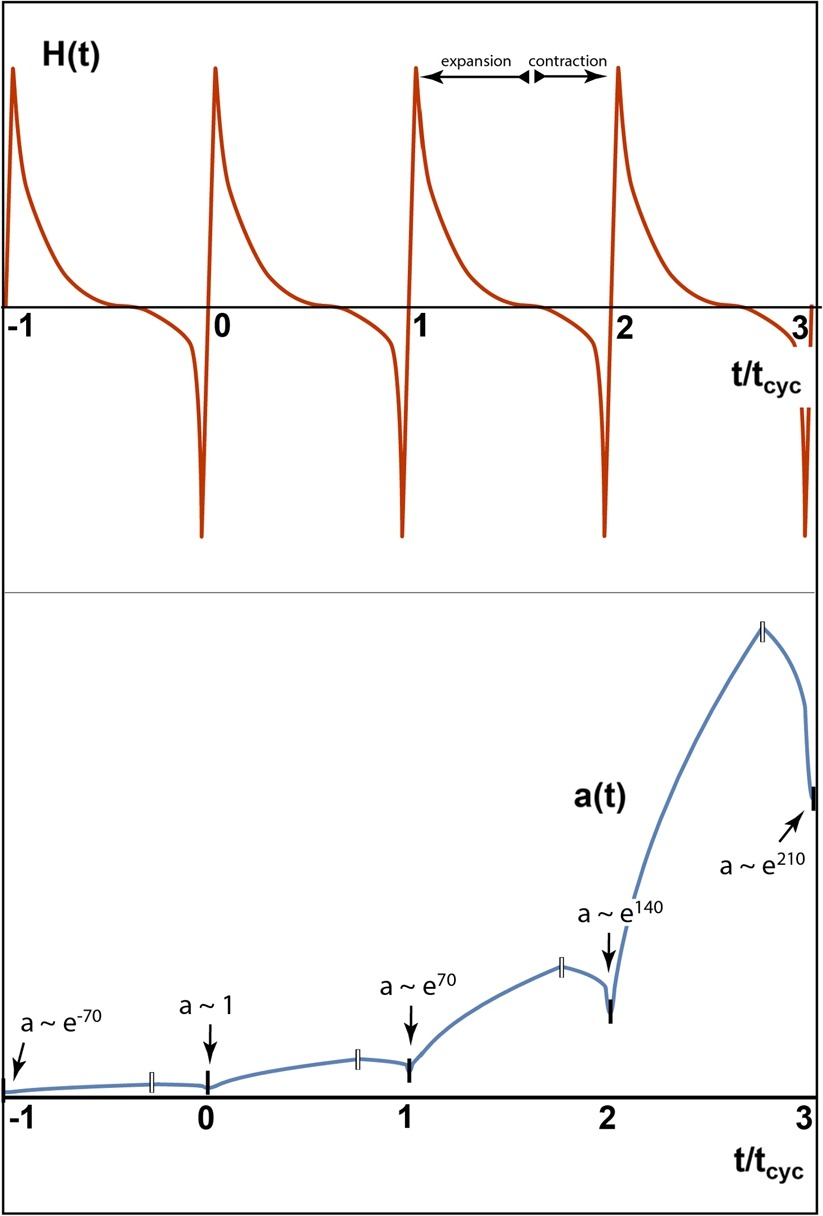}
    \caption{Evolution of the Hubble parameter (with an amplitude of $10^{10}$ GeV) and the scale factor in the cyclic (ekpyrotic) model of the Universe \cite{Ijjas:2019pyf}.}
    \label{fig:cyclic-model-a-H-evolution}
\end{figure}

Considering the main contributions from the cosmological epochs in the cyclic model produces the following \textit{\textbf{net change in the scale factor over one cycle}}:
\begin{equation}
    \Delta\ln{a_{\mathrm{cycle}}}\equiv N_{\mathrm{DE}}+\Delta\ln{a_{\mathrm{ekp}}}+\Delta\ln{a_{\mathrm{KE}}}+N_{\mathrm{rad}}\,,
\end{equation}
and with the use of \eqref{Delta-ln-a-ekp} and \eqref{Delta-ln-a-KE} gives:
\begin{equation}
    \Delta\ln{a_{\mathrm{cycle}}}=N_{\mathrm{DE}}-\underbrace{p N_{\mathrm{ekp}}}_{\to 0\,(p\ll 1)}+\frac{2}{3}\gamma_{\mathrm{KE}}+N_{\mathrm{rad}}
\end{equation}
which produces the final form given by the relation:
\begin{equation}
    \boxed{\Delta\ln{a_{\mathrm{cycle}}}\simeq N_{\mathrm{DE}}+\frac{2}{3}\gamma_{\mathrm{KE}}+N_{\mathrm{rad}}}\,.
\end{equation}
\begin{examplebox}[Conclusion]
    \textbf{Cyclic ekpyrotic cosmology is not exactly periodic in physical volume because the scale factor increases significantly from one cycle to the next} (Fig.~\ref{fig:cyclic-model-a-H-evolution}) \cite{Steinhardt:2001st,Lehners:2008vx,Ijjas:2019pyf,Ijjas:2021zwv}.
\end{examplebox}

\subsubsection{Closure condition for the Hubble parameter}
Tracking $|H|$ across the ekpyrotic, kinetic, and radiation phases provides an approximate cyclic closure condition for the Hubble scale. In the case of ekpyrotic phase, one may write that:
\begin{equation}
    \left|H_{\mathrm{end}}\right|=\left|H_{\mathrm{beg}}\right|e^{N_{\mathrm{ekp}}}\,,
\end{equation}
so that, together with \eqref{Hr-Hend} and \eqref{ln-H0-Hr} we obtain the relation:
\begin{equation}
    \boxed{\ln{\left(\frac{\left|H_{0}\right|}{\left|H_{\mathrm{beg}}\right|}\right)}\simeq N_{\mathrm{ekp}}-2\bigl(\gamma_{\mathrm{KE}}+N_{\mathrm{rad}}\bigr)}\,.
\end{equation}
Assuming that:
\begin{equation}
    \left|H_{0}\right|\simeq\left|H_{\mathrm{beg}}\right|\,,
\end{equation}
implies the \textit{\textbf{final form of the closure condition}} \cite{Steinhardt:2001st,Lehners:2008vx}:
\begin{examplebox}[Closure condition for the Hubble parameter]
    \begin{equation}
        \boxed{N_{\mathrm{ekp}}\simeq 2\bigl(\gamma_{\mathrm{KE}}+N_{\mathrm{rad}}\bigr)}\,.
    \end{equation}
\end{examplebox}
\begin{examplebox}[Numerical example]
    For the following values of the parameters/quantities (\textit{GUT-like high-energy scale}, \textit{high reheating/radiation temperature}, \textit{ultra-slow ekpyrotic contraction parameter}, \textit{present CMB temperature scale}) \cite{Steinhardt:2001st,Ijjas:2019pyf}:
    \begin{equation}
        E_{\mathrm{end}}\equiv V_{\mathrm{end}}^{1/4}=10^{15}\,\mathrm{GeV} \quad\land\quad T_{r}=10^{12}\,\mathrm{GeV} \quad\land\quad p=10^{-3} \quad\land\quad T_{0}\simeq 10^{-4}\,\mathrm{eV}
    \end{equation}
    one obtains that:
    \begin{equation}
        \gamma_{\mathrm{KE}}=\ln{\left(\frac{10^{15}}{10^{12}}\right)}=\ln{\left(10^{3}\right)}\simeq 6.9 \quad\land\quad N_{\mathrm{rad}}=\ln{\left(\frac{10^{12}\,\mathrm{GeV}}{10^{-4}\,\mathrm{eV}}\right)}=\ln{\left(10^{25}\right)}\simeq 57.6\,,
    \end{equation}
    so that:
    \begin{equation}
        N_{\mathrm{ekp}}\simeq 2\bigl(6.9+57.6\bigr)=129\,.
    \end{equation}
    Moreover, by taking into account:
    \begin{equation}
        N_{\mathrm{DE}}=0\,,
    \end{equation}
    yields:
    \begin{equation}
        \boxed{\Delta\ln{a_{\mathrm{cycle}}}\simeq\frac{2}{3}\times 6.9+57.6\simeq 62.2}
    \end{equation}
    $\implies$ \textbf{\textit{The physical volume increases significantly from cycle to cycle}} \cite{Steinhardt:2001st,Lehners:2008vx}.
\end{examplebox}

\subsection{Radiation must not become dominant too early}
The cyclic model predicts that radiation will not dominate until the SF has climbed out of the negative SF potential well and reaches the positive plateau. Let us denote by $t_\mathrm{cross}$ the time at which the SF crosses from the negative region to the plateau scale $V_0$, and $t_r$ the time at which RD begins. The requirement is:
\begin{equation}\label{RD-too-early-condition}
    t_{r}>t_\mathrm{cross}\,.
\end{equation}
During the kinetic phase:
\begin{equation}
    \phi(t)=\phi_{\mathrm{end}}+s\sqrt{\frac{2}{3}}\ln{\left(\frac{\left|t\right|}{\left|t_{\mathrm{end}}\right|}\right)} \quad\land\quad s=\pm 1\,.
\end{equation}
If the SF potential is approximately exponential as the field climbs out of the well, then:
\begin{equation}
    \left|V(t)\right|\propto\left(\frac{\left|t\right|}{\left|t_{\mathrm{end}}\right|}\right)^{-sc\sqrt{2/3}}\,,
\end{equation}
and choosing the sign corresponding to a climb from $V_{\mathrm{end}}$ to $V_{0}$ gives:
\begin{equation}
    \frac{V_{\mathrm{end}}}{V_{0}}=\left(\frac{\left|t_{\mathrm{cross}}\right|}{\left|t_{\mathrm{end}}\right|}\right)^{c\sqrt{2/3}}\,.
\end{equation}
Therefore:
\begin{equation}
    \left|t_{\mathrm{cross}}\right|=\left|t_{\mathrm{end}}\right|\left(\frac{V_{\mathrm{end}}}{V_{0}}\right)^{\sqrt{3/2}/c}\,,
\end{equation}
and using the relationship between $p$ and $c$ given by \eqref{p-c-relation} implies the following equality:
\begin{equation}
    \frac{1}{c}\sqrt{\frac{3}{2}}=\frac{1}{2}\sqrt{3p} \quad\implies\quad c=\sqrt{\frac{2}{p}}\,.
\end{equation}
Furthermore, one can observe that:
\begin{equation}
    \left|t_{\mathrm{end}}\right|\sim V_{\mathrm{end}}^{-1/2} \quad\land\quad t_{r}\sim H_{r}^{-1}\sim T_{r}^{-2}\,,
\end{equation}
so that, the condition \eqref{RD-too-early-condition} becomes:
\begin{equation}\label{RD-too-early-condition-1}
    \boxed{T_{r}\lesssim V_{\mathrm{end}}^{1/4}\left(\frac{V_{0}}{V_{\mathrm{end}}}\right)^{\sqrt{3p/16}}=E_{\mathrm{end}}\left(\frac{V_{0}}{V_{\mathrm{end}}}\right)^{\sqrt{3p/16}}}\,.
\end{equation}
In the case of physically realistic values:
\begin{equation}
    V_{0}\ll V_{\mathrm{end}} \quad\land\quad p\ll 1\,,
\end{equation}
so that the exponent is small and the \textbf{upper bound} \eqref{RD-too-early-condition-1} \textbf{is a relatively weak condition} \cite{Lehners:2008vx}.

\subsection{Semi-classical consistency of the cosmological bounce}

The discussion above shows that a \textit{\textbf{non-singular cyclic completion requires a mechanism beyond the classical}} ($4D$) \textit{\textbf{ekpyrotic scaling solution}}. However, \textbf{\textit{specifying such a mechanism is not sufficient by itself}}: one must also \textbf{verify that the corresponding semi-classical or MG description is used within its domain of validity}. This point is particularly important because the bounce occurs close to the regime in which the curvature, energy density, and time derivatives of the background can become large.

A useful way to organize the possible corrections is through an effective gravitational action of the following schematic form:
\begin{equation}
    S_{\mathrm{eff}}=\int d^{4}x\,\sqrt{-g}\left[\frac{\MPl^{2}}{2}R+\alpha_{1}R^{2}+\alpha_{2}R_{\mu\nu}R^{\mu\nu}+\alpha_{3}R_{\mu\nu\rho\sigma}R^{\mu\nu\rho\sigma}+\ldots\right]+S_{\phi}+S_{m}\,.
\end{equation}
The higher-curvature terms should be regarded as an expansion in powers of a cutoff scale, $\Lambda$. At the level of dimensional estimates one expects that:
\begin{equation}
    \alpha_{i}\sim\frac{\MPl^{2}}{\Lambda^{2}}\,,
\end{equation}
so that the effective expansion is under perturbative control only if:
\begin{equation}
    \frac{\left|R\right|}{\Lambda^{2}}\ll 1 \quad\land\quad \frac{\left|R_{\mu\nu}R^{\mu\nu}\right|}{\Lambda^{4}}\ll 1 \quad\land\quad \frac{\left|R_{\mu\nu\rho\sigma}R^{\mu\nu\rho\sigma}\right|}{\Lambda^{4}}\ll 1\,.
\end{equation}
In a consequence, if the bounce takes place at curvatures comparable to the cut-off, the effective higher-derivative theory cannot be treated as a controlled low-energy expansion. In that case, the bounce must be interpreted as a genuinely UV-sensitive phenomenon, rather than as a robust prediction of the low-energy effective theory \cite{Donoghue:1994dn,Burgess_2020,Nojiri:2017ncd,Shankaranarayanan:2022wbx}.

In a semi-classical treatment, the background equations may be written as:
\begin{equation}
    G_{\mu\nu}+\Delta G_{\mu\nu}^{\mathrm{HD}}=\frac{1}{\MPl^{2}}\left(T_{\mu\nu}^{\mathrm{cl}}+\left\langle T_{\mu\nu}\right\rangle_{\mathrm{q}}\right)\,,
\end{equation}
where $\Delta G_{\mu\nu}^{\mathrm{HD}}$ denotes higher-derivative gravitational corrections, $T_{\mu\nu}^{\mathrm{cl}}$ is the classical matter contribution, and $\left\langle T_{\mu\nu}\right\rangle_{\mathrm{q}}$ is the renormalized stress-energy tensor of quantum fields on the time-dependent background \cite{Birrell_Davies_1982,Hu:2008rga,Parker_Toms_2009book}. A self-consistent semi-classical bounce requires that the quantum backreaction does not become uncontrollably large:
\begin{equation}
    \left|\left\langle T_{\mu\nu}\right\rangle_{\mathrm{q}}\right|\lesssim\left|T_{\mu\nu}^{\mathrm{cl}}\right|+\MPl^{2}\left|\Delta G_{\mu\nu}^{\mathrm{HD}}\right|\,.
\end{equation}
If this condition fails, then the background solution cannot be trusted without solving the full backreaction problem.

Another important diagnostic is \textit{\textbf{adiabaticity}}. In the vicinity of a smooth bounce, the time dependence of the scale factor and the effective masses of quantum fields can result in gravitational particle production. This is a standard effect in QFT in curved spacetime. When the background undergoes a non-adiabatic evolution, the notion of positive frequency modes before and after the transition differs. The corresponding vacua are then related by a Bogoliubov transformation \cite{Birrell_Davies_1982,Parker_Toms_2009book}.

For a canonically normalized field mode $v_{k}$, one may write the mode equation in the form:
\begin{equation}
    v''_{k}+\omega_{k}^{2}(\tau)v_{k}=0\,,
\end{equation}
where a prime denotes differentiation with respect to conformal time. The leading WKB/adiabaticity condition is:
\begin{equation}
    \left|\frac{\omega'_{k}}{\omega_{k}^{2}}\right|\ll 1\,.
\end{equation}
More generally, adiabatic evolution requires a hierarchy of conditions involving higher derivatives of $\omega_k$, such as \cite{Birrell_Davies_1982,Parker_Toms_2009book}:
\begin{equation}
    \left|\frac{\omega''_{k}}{\omega_{k}^{3}}\right|\ll 1
    \quad\land\quad
    \left|\frac{\omega_{k}'^{\,2}}{\omega_{k}^{4}}\right|\ll 1\,.
\end{equation}
If these conditions are violated, positive and negative frequency modes mix, and particles are produced. The late-time mode functions can be written as:
\begin{equation}
    v_{k}^{\mathrm{out}}=\alpha_{k}v_{k}^{\mathrm{in}}+\beta_{k}v_{k}^{\mathrm{in}\,*}\,,
\end{equation}
where $\alpha_{k}$ and $\beta_{k}$ are the \textit{Bogoliubov coefficients} and:
\begin{equation}
    n_{k}=|\beta_{k}|^{2}
\end{equation}
is the occupation number of produced particles \cite{Birrell_Davies_1982,Parker_Toms_2009book}.

For a field whose dynamics is described in terms of a comoving frequency $\omega_{k}(\tau)$, the corresponding energy density can be estimated as:
\begin{equation}
    \rho_{\mathrm{prod}}\sim\frac{1}{a^{4}}\int\frac{d^{3}k}{\left(2\pi\right)^{3}}\,\omega_{k}\,\left|\beta_{k}\right|^{2}\,.
\end{equation}
The factor of $a^{-4}$ reflects the conversion from comoving momentum and comoving frequency to physical energy density.

Therefore, a controlled bounce requires:
\begin{equation}
    \rho_{\mathrm{prod}}\ll\rho_{\mathrm{bg}}\,,
\end{equation}
where $\rho_{\mathrm{bg}}$ denotes the energy scale associated with the background sector responsible for the bounce, rather than necessarily $3M_{\mathrm{Pl}}^{2}H^{2}$ at the exact bouncing point. If this condition is not satisfied, the produced particles backreact on the geometry and the assumed background solution is no longer self-consistent. This issue is part of the broader consistency problem of non-singular bouncing cosmologies \cite{Novello:2008ra,Battefeld:2014uga,Brandenberger:2016vhg,Koehn:2015vvy}.

The only exception is the case in which particle production is part of the intended reheating mechanism. Then the produced energy density must not be neglected, but included self-consistently in the post-bounce evolution and in the matching to the hot expanding phase \cite{Albrecht:1982mp,Allahverdi:2010xz,Amin:2014eta,Turok:2004gb,Takamizu:2004rq}.
```

Another issue concerns \textit{\textbf{matching}}. In a genuinely non-singular bounce, no separate matching prescription should be necessary: the \textit{\textbf{background and perturbation variables should evolve continuously through the bounce according to the EoM}}. This is conceptually different from a singular brane collision, where one must prescribe matching rules for the fields and perturbations at the collision surface. Therefore, a successful non-singular effective model should determine all physically relevant quantities without introducing an additional singular transition rule. \textit{\textbf{This requirement is especially important for the curvature perturbation, because its evolution determines whether the pre-bounce smoothing and perturbation generation mechanisms survive into the expanding branch}} \cite{Battefeld:2014uga,Koehn:2015vvy,Ijjas:2016vtq}.

Therefore, the semi-classical or MG sector must satisfy several consistency requirements in addition to the existence of a regular background solution:
\begin{examplebox}[Consistency requirements]
    \begin{enumerate}[label=(\roman*)]
        \item The curvature expansion must remain below the cutoff scale of the effective theory;
        \item The quantum stress-energy backreaction must remain subdominant or be included self-consistently;
        \item Particle production must not destabilize the background unless it is part of the intended reheating mechanism;
        \item Perturbations must pass through the bounce without requiring an \textit{ad hoc singular matching rule};
        \item The post-bounce state must connect smoothly to a hot expanding FLRW phase.
\end{enumerate}
\end{examplebox}
Thus, \textit{\textbf{the role of the semi-classical or MG completion is not only to replace the singularity by a cosmological (non-singular) bounce but also to make the bounce predictive}}. \textbf{Without these additional consistency requirements, the bounce may remove the background singularity formally while leaving the physical evolution uncontrolled.}

\subsection{Entropy and local cyclicity}
The fact that:
\begin{equation}
    \Delta\ln a_{\mathrm{cycle}}>0
\end{equation}
has an \textbf{important thermodynamic interpretation}. It means that the \textbf{cyclic scenario is not globally periodic in the sense of Tolman's oscillatory Universe}. \textit{In Tolman's picture, entropy production from one cycle to the next implies that later cycles become larger and longer, whereas earlier cycles become smaller and shorter}. Therefore, an \textit{\textbf{exactly periodic sequence of global cycles is incompatible with the second law of thermodynamics}} \cite{Tolman:1931fei,Tolman1934book}.

Modern ekpyrotic/cyclic cosmology avoids this conclusion by changing what is meant by a cycle. The \textit{\textbf{scale factor is not required to return to its previous value}}. Instead, the \textit{\textbf{local thermodynamic and geometrical conditions may approximately repeat, while the global volume increases}}. One may express this as:
\begin{equation}
    H(t+T)\simeq H(t) \quad\land\quad \rho(t+T)\simeq\rho(t) \quad\land\quad T_{\mathrm{temp}}(t+T)\simeq T_{\mathrm{temp}}(t)\,,
\end{equation}
but:
\begin{equation}
    a(t+T)\simeq e^{\Delta N_{\mathrm{cycle}}}\,a(t) \quad\land\quad \Delta N_{\mathrm{cycle}}>0\,.
\end{equation}
Thus, the \textbf{model is cyclic in local physical conditions, not in the global scale factor} \cite{Steinhardt:2001st,Lehners:2008vx,Ijjas:2019pyf,Ijjas:2021zyf}.

The entropy accounting can be stated in a simple quantitative form. For a relativistic plasma:
\begin{equation}
    \rho_{r}=\frac{\pi^{2}}{30}g_{*}T^{4} \quad\land\quad s=\frac{2\pi^{2}}{45}g_{*s}T^{3}\,,
\end{equation}
where $s$ is the entropy density. Therefore, the comoving entropy is of the form:
\begin{equation}
    S_{\mathrm{com}}=s\,a^{3}\,.
\end{equation}
In the case of an adiabatic phase:
\begin{equation}
g_{*s}T^{3}a^{3}=\mathrm{const}\,,
\end{equation}
and, if $g_{*s}$ is approximately constant, then:
\begin{equation}
    T\propto a^{-1}\,.
\end{equation}
However, irreversible processes during a cycle can increase the total entropy. Let us assume that the total entropy in a large physical region grows by a factor of:
\begin{equation}
\frac{S_{n+1}}{S_{n}}=e^{\Delta S_{N}}\,,
\end{equation}
while the physical volume grows by:
\begin{equation}
    \frac{V_{n+1}}{V_{n}}=\left(\frac{a_{n+1}}{a_n}\right)^{3}=e^{3\Delta N_{\mathrm{cycle}}}\,.
\end{equation}
Thus, the entropy density changes as:
\begin{equation}
    \frac{s_{n+1}}{s_n}=\frac{\frac{S_{n+1}}{V_{n+1}}}{\frac{S_n}{V_n}}=\exp{\bigl(\Delta S_N-3\Delta N_{\mathrm{cycle}}\bigr)}\,.
\end{equation}
Therefore, the total entropy may increase:
\begin{equation}
S_{n+1}>S_n\,,
\end{equation}
while the entropy density remains bounded or approximately cyclic provided that:
\begin{equation}
    \Delta S_N \lesssim 3\Delta N_{\mathrm{cycle}}\,.
\end{equation}
\textit{\textbf{This is the thermodynamic meaning of local cyclicity: although entropy is not destroyed, it becomes diluted as physical volume grows.}}

This interpretation is especially important for black holes (BHs) due to the fact that they carry a large entropy given by the \textit{Bekenstein–Hawking formula} \cite{Bekenstein:1973ur,Hawking:1975vcx}:
\begin{equation}
    S_{\mathrm{BH}}=\frac{A}{4G}=\frac{M^{2}}{2\MPl^{2}}\,,
\end{equation}
for a Schwarzschild BH characterized by its mass, $M$. In a globally periodic model, such entropy would accumulate and obstruct the return to an identical state. In a local cyclic model, high-entropy objects can be carried outside the patch that participates in the next ekpyrotic phase. The second law is not violated, because the entropy is not erased globally; rather, the next cycle begins in a smooth local region whose entropy density has been diluted by expansion \cite{Ijjas:2021zyf}.

The \textbf{ekpyrotic phase complements this thermodynamic dilution by suppressing gravitational disorder}. Matter entropy is not the only relevant quantity in cosmology: anisotropy, inhomogeneity, Weyl curvature and BHs also contribute to the gravitational arrow of time. The \textbf{ultra-stiff ekpyrotic phase dynamically selects smooth contracting patches and prevents anisotropies from dominating the approach to the bounce}. Hence, the \textit{\textbf{entropy problem and the smoothing problem are related, but not identical: volume growth dilutes entropy density, whereas ekpyrotic contraction suppresses the geometrical sources of gravitational disorder}}.

It is useful to \textit{\textbf{distinguish this local entropy resolution from the stronger question of global past completeness}}. A \textbf{model can dilute entropy density from cycle to cycle and still fail to be geodesically complete in the past}. Conversely, \textbf{attempts to construct geodesically complete cyclic cosmologies must address entropy production separately}. \textit{\textbf{This is why the entropy problem, the geodesic-completeness problem and the bounce stability problem should be treated as related but logically distinct consistency conditions}} \cite{Kinney:2021imp,Pavlovic:2023mke}.

\subsection{Advantages and open issues}
The main theoretical advantages of cyclic (ekpyrotic) models are the following:
\begin{examplebox}[Advantages of cyclic ekpyrotic models]
    \begin{enumerate}[label=(\alph*)]
        \item \textbf{Suppression of anisotropy and curvature} - the ultra-stiff EoS $\omega\gg 1$ ensures that $\rho_{\mathrm{ekp}}$ grows faster than shear, ordinary matter an the spatial curvature during the contraction phase \cite{Erickson:2003zm,Lehners:2008vx,Cook:2020oaj};
        \item \textbf{Causal smoothing without accelerated expansion} - the comoving Hubble radius decreases during the ekpyrotic contraction, which allows large-scale modes to exit the horizon despite the fact that the Universe is contracting \cite{Khoury:2001wf,Lehners:2008vx};
        \item \textbf{Generation of perturbations in multi-field models} - The entropy perturbations perpendicular to the background SF trajectory may possess an almost scale-invariant spectrum and can subsequently transform into curvature perturbations \cite{Lehners:2008my,Koyama:2007mg,Lehners:2008vx};
        \item Local entropy reset - the net growth in physical volume per cycle can dilute the entropy density in the next observable patch, avoiding locally the naive Tolman problem \cite{Steinhardt:2001st,Lehners:2008vx,Ijjas:2021zwv}.
    \end{enumerate}
\end{examplebox}
On the other hand, the main open issues are:
\begin{examplebox}[Open issues]
    \begin{enumerate}[label=(\alph*)]
        \item \textbf{Cosmological bounce and violation of NEC} - the non-singular bounce in spatially flat ($4D$) FLRW Universe requires NEC violation. Nevertheless, stable and UV-complete NEC-violating phases remain difficult to construct \cite{Battefeld:2014uga,Ijjas:2016vtq,Kontou:2020bta};
        \item \textbf{Singularity resolution} - the original framework of the collision of branes is singular in the $4D$ ST EFT and needs a higher-dimensional resolution (which is \textit{highly speculative}) \cite{Khoury:2001bz,Turok:2004gb,Lehners:2006pu};
        \item \textbf{Anisotropy through the bounce} - ekpyrosis suppresses anisotropy before the bounce/crunch, but they can re-grow during NEC-violating or MG bounce phases \cite{Battefeld:2014uga};
        \item \textbf{Matching of perturbations} - matching the perturbations through the possible bounce/crunch is a subtle issue, and is a model-dependent aspect of the considered formalism \cite{Tolley:2003nx,Lehners:2008vx,Battefeld:2014uga};
        \item \textbf{Non-Gaussianity} - multi-field ekpyrotic models may generate significant local non-Gaussianity due to the steep (negative) SF potential and non-linear entropy interactions \cite{Lehners:2008my,Lehners:2008vx};
        \item \textbf{Global non-periodicity} - the entropy and volume increase from one cycle to the another, so that, the Universe is not globally periodic (even if local patches undergo repeated cycles) \cite{Steinhardt:2001st,Tolman1934book,Ijjas:2021zwv}.
    \end{enumerate}
\end{examplebox}

\section{Conclusions}
The ekpyrotic contraction is a powerful smoothing mechanism. It suppresses anisotropy, ordinary matter and spatial curvature dynamically while satisfying the NEC, WEC and SEC. However, since the EoS parameter $\omega_{\mathrm{ekp}}$ is much greater than one, the DEC is violated. Furthermore, the same canonical ekpyrotic phase cannot produce a non-singular bounce in a spatially flat $4D$ STT. A complete cyclic model requires the ekpyrotic mechanism to be supplemented by a higher-dimensional collision of end-of-the-world branes, a carefully controlled NEC-violating EFT, a closed Universe bounce, or a modification of GR.

The $e$-fold approach demonstrates how a cycle can be internally consistent. Namely, ekpyrosis generates significant growth in the number of $e$-folds in $|H|$, while the scale factor changes only slightly. The kination and RD/MD epochs result in significant net growth of the scale factor.

Moreover, the entropy considerations do not forbid such a local cyclicity. The Tolman's argument excludes exactly periodic entropy-producing Universe, but cyclic ekpyrotic cosmology is not exactly periodic in a global sense. Instead, the total entropy and physical volume increase, while the entropy density in the relevant local patch can be reset from one cycle to the another. Construction of a fully stable, geodesically and UV-complete cyclic ekpyrotic model remains an open problem at the intersection of GR, EFT, higher-dimensional gravity and quantum cosmology.
\begin{savequote}
"My life amounts to no more than one drop in a limitless ocean. Yet what is any ocean, but a multitude of drops?"
\qauthor{\textbf{David Mitchell}, \textit{Cloud Atlas}}
"Maybe in the excitement of the development of a cosmology that could be tested and even established, we have inadvertently put aside something interesting to be dealt with later, and then forgotten it. Maybe as the cosmological tests are tightened, they will at last reveal the failure of the extrapolation of standard physics."
\qauthor{\textbf{Philip James Edwin Peebles} \cite{Peebles:2022bya}}
\end{savequote}

\chapter{Conclusions and Perspectives}
\label{Sec:conclusions}
\ifpdf
    \graphicspath{{Chapter7/Figs/Raster/}{Chapter7/Figs/PDF/}{Chapter7/Figs/}}
\else
    \graphicspath{{Chapter7/Figs/Vector/}{Chapter7/Figs/}}
\fi

\label{ch:conclusions-perspectives}

\section{Conclusions}

The purpose of this dissertation was to analyze scalar field descriptions of the dark sector of the Universe, with emphasis on scalar-tensor cosmology, effective field theory, dynamical dark energy, and ekpyrotic/cyclic alternatives to inflation. The central message is that scalar fields offer a unifying mathematical framework for addressing various seemingly distinct problems in cosmology: the origin of primordial perturbations, the nature of dark energy, possible interactions in the dark sector, modifications of gravity, and non-singular or cyclic scenarios for the early Universe.

The standard $\Lambda$CDM model remains the reference framework for observational cosmology. It is simple, predictive and remarkably successful. However, it is also phenomenological: it contains a cold dark matter component and a cosmological constant whose microscopic origins remain unknown. The first lesson of this dissertation is therefore that one should distinguish between the empirical success of $\Lambda$CDM and the physical interpretation of its dark components. This point was also the motivation of \cite{Borowiec:2023kmq}, where the separation between baryonic and dark matter within the $\Lambda$CDM formalism was examined. At the level of the homogeneous expansion, baryons and cold dark matter scale in the same way, whereas their physical distinction appears through interactions, perturbations, clustering, and non-gravitational properties.

The second lesson is that scalar-tensor theories provide a minimal and theoretically well-motivated way of going beyond GR. In such theories, gravity is mediated by the metric tensor and by at least one scalar degree of freedom. The scalar may be interpreted as a varying gravitational coupling, a remnant of extra dimensions, a dilaton-like field, a DE component, or an effective degree of freedom generated by a more fundamental theory \cite{Brans:1961sx,Dicke:1961gz,Faraoni2004chapter,Fujii_Maeda_2003,Clifton:2011jh}. The formalism developed in Chapter \ref{Sec:STTs} shows that scalar-tensor gravity is not a single model, but a class of theories characterized by coupling functions, scalar potentials and matter-coupling prescriptions.

A key conceptual point is the distinction between the Jordan and Einstein conformal frames. At the mathematical level, the two frames may be related by a regular conformal transformation and a scalar field redefinition. At the physical level, the interpretation depends on how matter couplings, units and observables are treated. The pragmatic conclusion adopted in this thesis is that classical cosmological observables should be expressed in a frame-consistent or frame-invariant way, while off-shell quantum quantities and renormalization-group flows may retain frame dependence \cite{Faraoni:2006fx,Catena:2006bd,Chiba:2013mha,Domenech:2016yxd}. This viewpoint is particularly important when comparing scalar-tensor models with observations.

The third lesson concerns inflation and its ultraviolet sensitivity. Inflation is a powerful mechanism for explaining the large-scale smoothness, flatness and primordial perturbations of the observable Universe \cite{Starobinsky:1980te,Guth:1980zm,Linde1982,Baumann:2009ds,Baumann_2022inflation}. Nevertheless, the EFT analysis of Chapter \ref{Sec:inflation} shows that the success of inflation at the level of observations does not by itself guarantee UV completeness. In the EFT language, unknown high-energy physics generates higher-dimensional operators suppressed by a cutoff scale. For example, Planck-suppressed corrections can generate shifts of the slow-roll parameter of the schematic form:
\begin{equation}
    \Delta\eta_{V}\sim c\left(\frac{M_{\mathrm{Pl}}}{\Lambda}\right)^{2}\,,
\end{equation}
which is dangerous when $\Lambda\lesssim M_{\mathrm{Pl}}$ and $c\sim\mathcal{O}(1)$. This is the essence of the eta problem. Therefore, a complete inflationary model requires not only agreement with CMB data, but also a mechanism protecting the inflaton potential against radiative corrections and Planck-suppressed operators \cite{Cheung:2007st,Weinberg:2008hq,Donoghue:1994dn,Burgess_2020,Kallosh:1995hi,Harlow:2018tng}.

The fourth lesson concerns dark energy. Quintessence is the simplest dynamical alternative to the cosmological constant. For a canonical minimally coupled scalar field:
\begin{equation}
    \rho_{\phi}=\frac{1}{2}\dot{\phi}^{2}+V(\phi) \quad\land\quad p_{\phi}=\frac{1}{2}\dot{\phi}^{2}-V(\phi)\,,
\end{equation}
and therefore:
\begin{equation}
    \rho_{\phi}+p_{\phi}=\dot{\phi}^{2}\geq 0\,.
\end{equation}
Consequently, canonical quintessence cannot cross the phantom divide line $\omega=-1$. This makes quintessence both attractive and restrictive: it is the minimal dynamical DE model, but it cannot realize all possible effective expansion histories. The DESI DR2 results motivate a careful treatment of time-dependent dark energy, because combined analyses indicate that the late-time expansion history may prefer an effective evolving equation of state over a strict cosmological constant \cite{DESI:2025zgx,DESI:2025fii}. However, if the reconstructed effective equation of state crosses the phantom divide, then minimal canonical quintessence is not sufficient. One must consider non-canonical models, interacting dark sectors, multi-field scenarios, or modified gravity.

The fifth lesson concerns ekpyrotic and cyclic cosmology. Ekpyrotic contraction provides a powerful smoothing mechanism because an ultra-stiff scalar field with:
\begin{equation}
    \omega_{\mathrm{ekp}}\gg 1
\end{equation}
dynamically suppresses anisotropy, spatial curvature and ordinary matter during contraction \cite{Khoury:2001wf,Lehners:2008vx,Battefeld:2014uga}. This makes ekpyrosis an important alternative to inflationary smoothing. However, the same canonical ekpyrotic phase cannot by itself generate a non-singular bounce in a spatially flat four-dimensional GR or scalar-tensor formalism. A complete cyclic model requires an additional mechanism: a higher-dimensional brane collision, a controlled NEC-violating EFT, a (spatially) closed Universe cosmological bounce, or a modification of gravity.

The preprint \cite{Postolak:2026okk} developed a phase-resolved approach to field-space distance bounds in ekpyrotic, bouncing and cyclic cosmologies. The important point is that the field-space distance should not be treated as a single undifferentiated number. Instead, different phases of a cosmological cycle contribute differently. In the ekpyrotic smoothing phase, one finds approximately that:
\begin{equation}
    d_{\mathrm{ek}}\simeq\frac{\sqrt{2\epsilon_{\mathrm{ek}}}}{\epsilon_{\mathrm{ek}}-1}N_{\mathrm{sm}}\,,
\end{equation}
and in the ultra-stiff limit:
\begin{equation}
    d_{\mathrm{ek}}\simeq\sqrt{\frac{2}{\epsilon_{\mathrm{ek}}}}N_{\mathrm{sm}}\,.
\end{equation}
Thus, a large amount of smoothing can remain compatible with a sub-Planckian field-space distance if $\epsilon_{\mathrm{ek}}$ is sufficiently large. This illustrates how ekpyrotic cosmology can be constrained using not only perturbations and background dynamics, but also EFT and field-space consistency conditions.

\begin{examplebox}[Main conclusions of the dissertation]
    \begin{enumerate}[label=(\alph*)]
        \item The dark sector should be treated as a physical problem, not merely as a set of phenomenological density parameters;
        \item Scalar fields provide a unifying language for inflation, quintessence, scalar field dark matter, scalar-tensor gravity and ekpyrotic/cyclic cosmology;
        \item Scalar-tensor theories are among the most economical extensions of GR and provide a natural framework for testing whether the dark sector is modified matter, modified gravity, or an interaction between both descriptions;
        \item The Jordan and Einstein frames are mathematically related when the conformal transformation is regular, but their physical interpretation requires a consistent treatment of matter couplings, units and observables;
        \item Inflation is observationally successful, but EFT analysis shows that it is UV sensitive. Slow roll requires protection against radiative corrections and Planck-suppressed operators;
        \item Canonical quintessence is the minimal dynamical DE model, but it cannot cross the phantom divide. DESI DR2 therefore motivates broader scalar field and modified gravity interpretations;
        \item Ekpyrotic contraction provides a robust smoothing mechanism, but a complete cyclic scenario still requires a controlled bounce, a consistent treatment of perturbations, and a solution to geodesic-completeness and entropy issues;
        \item The author’s articles and preprints \cite{Borowiec:2023kmq,Postolak:2024xtm,Postolak:2025qmv,Postolak:2026okk,CosmoVerseNetwork:2025alb} form a coherent research line: from the interpretation of the dark sector, through scalar-tensor dynamics, to early Universe alternatives and cosmological tensions.
    \end{enumerate}
\end{examplebox}

\section{Perspectives}
The results of this dissertation suggest several directions for future work.

\subsection{Scalar-tensor dark sector phenomenology}

The first direction is the further development of non-minimally coupled scalar field dark sector models. The case study \cite{Postolak:2025qmv} shows that Einstein-frame scalar-tensor cosmology can be analyzed using dynamical systems methods. The next step is to connect this phase-space analysis more directly with data. This requires implementing the model at the level of background evolution, linear perturbations and parameter inference.

A realistic observational analysis should include:
\begin{enumerate}[label=(\alph*)]
    \item CMB constraints from Planck and future CMB experiments;
    \item BAO constraints from DESI and future surveys;
    \item SNe Ia data;
    \item Weak lensing and large-scale-structure constraints;
    \item Local bounds on fifth forces and PPN parameters;
    \item Gravitational wave constraints on the propagation of tensor modes.
\end{enumerate}
Such a program would make it possible to determine whether a non-minimally coupled scalar field can simultaneously satisfy local gravity constraints and produce observable cosmological effects.

\subsection{Dark sector separation and cosmological tensions}

The second direction is connected with the interpretation of cosmological tensions. The CosmoVerse White Paper \cite{CosmoVerseNetwork:2025alb} emphasizes that tensions may arise from observational systematics, astrophysical modeling, or new fundamental physics. From the viewpoint of this dissertation, the most important theoretical question is whether the dark sector should be split into separately conserved components or treated as an interacting effective sector.

The work \cite{Borowiec:2023kmq} motivates this question already within $\Lambda$CDM, while DESI DR2 motivates it from the DE side. Interacting dark energy, decaying dark matter, scalar-mediated fifth forces and scalar-tensor modifications can produce similar background expansion histories. Therefore, future work should focus on observables that break this degeneracy. The simultaneous use of background, perturbative and gravitational wave observables is necessary to distinguish modified matter from modified gravity.

\subsection{UV control of inflation}

The third direction concerns inflation. The EFT analysis of Chapter \ref{Sec:inflation} shows that inflationary model building is not only a question of fitting the scalar spectral index $n_{s}$ and the tensor-to-scalar ratio $r$. It is also a question of radiative stability, symmetry protection and UV completion. Future work should therefore focus on identifying which inflationary operators are technically natural and which require fine tuning.

Particularly important problems include:
\begin{enumerate}[label=(\alph*)]
    \item The relation between approximate shift symmetries and quantum gravity constraints;
    \item The role of additional light or heavy fields during inflation;
    \item The observational signatures of non-standard sound speed and non-Gaussianity;
    \item The consistency of reheating with the EFT assumptions used during inflation;
    \item The connection between the trans-Planckian problem and the cutoff of the inflationary EFT.
\end{enumerate}
In this sense, \textit{\textbf{inflation remains a highly successful phenomenological framework, but not a completed fundamental theory}}.

\subsection{Ekpyrotic, bouncing and cyclic cosmology}

The fourth direction is the development of controlled ekpyrotic and cyclic models. The analysis of Chapter \ref{Sec:ekpyrotic} and the preprint \cite{Postolak:2026okk} show that ekpyrosis can provide efficient smoothing while keeping the field-space distance under control. However, the most difficult part of the scenario is not the ekpyrotic contraction itself, but the transition through the bounce and the global completion of the cyclic history.

Future work should address:
\begin{enumerate}[label=(\alph*)]
    \item The construction of stable NEC-violating or modified gravity bounces;
    \item The matching of scalar and tensor perturbations through the bounce;
    \item The control of anisotropy and gradient instabilities near the bounce;
    \item The relation between field-space distance bounds and swampland-inspired constraints;
    \item The entropy budget and the distinction between local cyclicity and global periodicity;
    \item The geodesic completeness of cyclic cosmological spacetimes.
\end{enumerate}
A fully satisfactory cyclic model must solve all these problems simultaneously. It is not enough to remove the background singularity formally; the bounce must also be stable, predictive and compatible with perturbation theory.

\subsection{Observational prospects}

The next decade will provide increasingly precise tests of the dark sector. DESI, Euclid, Rubin, Roman, next-generation CMB experiments, gravitational wave standard sirens and 21 cm cosmology will improve constraints on the background expansion, growth of structure and gravitational dynamics. These data will test whether the late-time Universe is consistent with a cosmological constant or whether a dynamical DE component is required.

From the perspective of this dissertation, the most important future tests are not only measurements of a single parameter such as $\omega_{0}$. Rather, the decisive tests will involve consistency relations between different sectors. A scalar-tensor theory that mimics $\Lambda$CDM at the background level may still produce deviations in perturbations or in the relation between matter and metric potentials. Conversely, an apparent dynamical DE signal may be absorbed into a more general modified gravity or interacting dark sector description. This is why the future of dark sector cosmology lies in combined, multi-probe consistency tests.

\section{Final remarks}

The main conclusion of this dissertation is that scalar fields are not merely convenient mathematical tools. They are a natural bridge between cosmology, gravity, quantum field theory and high-energy physics. They appear in inflation, quintessence, scalar-tensor gravity, string-inspired theories and ekpyrotic cosmology. At the same time, they expose the deepest theoretical difficulties of modern cosmology: naturalness, UV sensitivity, the meaning of the dark sector, the interpretation of cosmological tensions, the role of conformal frames and the possibility of resolving the initial singularity.

The thesis therefore supports a cautious but constructive viewpoint. The $\Lambda$CDM model should remain the benchmark. Inflation should remain the leading early-Universe framework. Quintessence should remain the minimal reference model for dynamical dark energy. Ekpyrotic and cyclic models should remain important alternatives that clarify what inflation does and does not explain. However, none of these frameworks should be treated as final. Each of them is best understood as an effective description whose domain of validity must be tested.

In this sense, the dark sector is not only a missing piece of the cosmic inventory. It is a window into the limits of our present theories of matter, gravity, and the early Universe. The scalar field approach developed in this dissertation provides one possible route toward that window: mathematically controlled, phenomenologically testable and open to future revision.


\begin{spacing}{0.9}


\bibliographystyle{elsarticle-num}
\cleardoublepage
\bibliography{References/references} 



\end{spacing}





\printthesisindex 

\end{document}